\documentclass[11pt,a4paper]{article}
\usepackage[T1]{fontenc}
\usepackage{lmodern}
\usepackage[margin=24mm,bottom=26mm,top=24mm]{geometry}
\usepackage{graphicx}
\usepackage{microtype}
\usepackage{float}
\usepackage{xcolor}
\usepackage{booktabs}
\usepackage{array}
\usepackage{longtable}
\usepackage{fancyhdr}
\usepackage{titlesec}
\usepackage[font=small,labelfont=bf,skip=6pt]{caption}
\usepackage[hidelinks,bookmarks=true,bookmarksnumbered=true]{hyperref}
\hypersetup{
    colorlinks=true, linkcolor=[rgb]{0.16,0.47,0.84},
    urlcolor=[rgb]{0.16,0.47,0.84},
  pdftitle={The state of nuclear data for fusion activation simulations},
  pdfsubject={FNS decay heat benchmark, four evaluated nuclear data libraries}}
\titleformat{\section}{\LARGE\bfseries}{}{0pt}{}
\titlespacing*{\section}{0pt}{0pt}{12pt}
\titleformat{\subsection}{\Large\bfseries}{}{0pt}{}
\titleformat{\subsubsection}{\normalsize\bfseries}{}{0pt}{}
\titlespacing*{\subsubsection}{0pt}{10pt}{2pt}
\begin{document}
\begin{titlepage}\vspace*{4cm}
{\Huge\bfseries The state of nuclear data\\[6pt]for fusion activation simulations\par}
\vspace{14mm}{\large TENDL-2025, ENDF/B-VIII.1, JEFF-4.0, JENDL-5\par}
\vspace{4mm}{\large against the IAEA CoNDERC FNS decay-heat benchmark\par}
\vspace{14mm}{\large J. Shimwell\par}
\vfill\end{titlepage}
\tableofcontents\clearpage
\section{Summary}
This report compares decay-heat calculations made with four nuclear data libraries, TENDL-2025, ENDF/B-VIII.1, JEFF-4.0 and JENDL-5, against the JAEA FNS decay-heat benchmark. The main comparison uses the 2000 campaign, in which 73 foils were irradiated for five minutes in a mostly 14 MeV neutron field and their decay heat was measured for about an hour. The 1996 seven-hour campaign, with 29 foils and cooling times of up to 400 days, is compared separately and is not included in the rankings.

Over the 69 elemental foils solved by all four libraries, the median deviation from the measurement (the mean absolute C/E deviation of each foil) is JENDL-5 17 percent, TENDL-2025 19 percent, JEFF-4.0 22 percent and ENDF/B-VIII.1 44 percent. No library is best everywhere: JENDL-5 on 24, JEFF-4.0 on 18, TENDL-2025 on 17 and ENDF/B-VIII.1 on 10 of the foils. For each element the report breaks the heat down by product and lists the nuclear data on which the libraries differ, and the final sections propose measurements, including enriched-isotope irradiations, that would resolve the largest differences.

\section{The benchmark}
The \href{https://nds.iaea.org/conderc/fusion/}{JAEA Fusion Neutronics Source (FNS) decay-heat experiments} are distributed by the IAEA through \href{https://nds.iaea.org/conderc/fusion/}{CoNDERC}. Samples were irradiated with D-T neutrons, moved to a calorimeter, and their decay heat was measured as they cooled. The benchmark covers 73 foils, 132 irradiations and four irradiation positions, each with a measured neutron spectrum.

The calculations use the published spectrum for each position directly, so no neutron transport is needed. Cross sections are collapsed with the spectrum to one-group reaction rates, and the Bateman equations are solved over the irradiation and cooling history. The main comparison uses the 2000 five-minute campaign because it is the only one that includes all 73 foils, all irradiated at position 3 and measured at the same cooling times.

The 1996 seven-hour campaign irradiated 29 foils at position 7 and followed some of them for up to 400 days, so it is more sensitive to long-lived products. Its results are shown separately and are not used in the periodic tables, win counts or distributions. The campaigns differ in spectrum, irradiation history and measurement, so they are not pooled. Iron shows why: with the same libraries, its median calculated heat is 6 to 9 percent above the 2000 five-minute measurements and 5 to 8 percent below the 1996 five-minute measurements.

\begin{figure}[H]
\centering
\includegraphics[width=\textwidth,height=0.82\textheight,keepaspectratio]{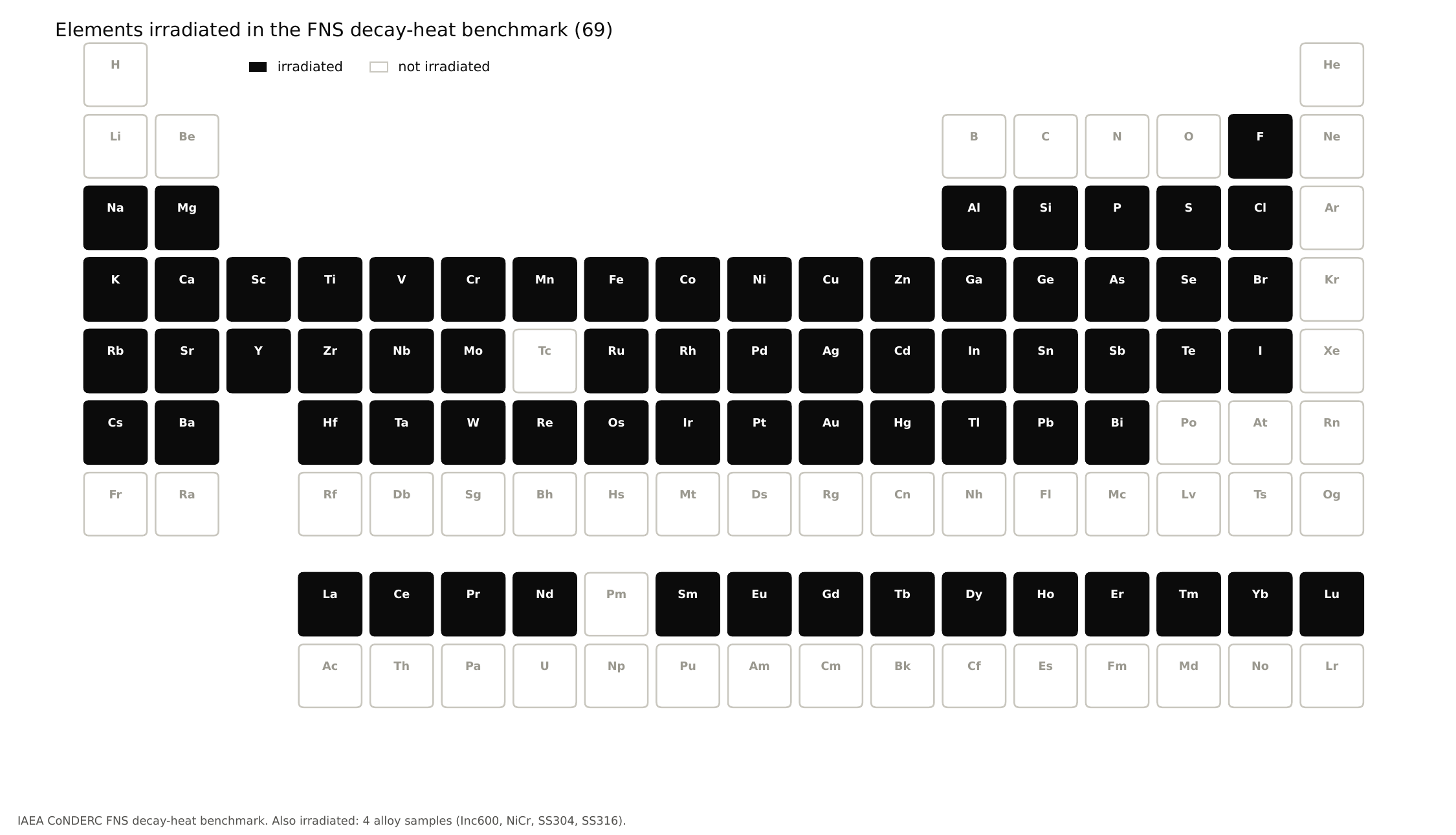}
\caption{Elements irradiated in the benchmark. Four alloy samples were also irradiated.}
\end{figure}
\clearpage

\section{Library configurations}
Each calculation takes its neutron cross sections, reaction channels and isomeric branching from one library, and uses that library's decay data and fission yields where it has them. TENDL-2025 has neither, so the TENDL calculations use UKDD-2020 decay data and UKFY-4.2 fission yields, the UKAEA libraries FISPACT-II pairs with TENDL. These are published as ENDF-6 files in the FISPACT-II nuclear data repository and were converted for this work. A difference between two curves can therefore come from the neutron data, the decay data or both. The table in the next section lists the source of every input.

All data are used as published. Values that look wrong are left in place, because adjusting them to match the benchmark would hide the effect being tested.

\section{Previous validation}
The \href{https://www-nds.iaea.org/media/docs/fusion/UKAEA-CCFE-RE-20-004.pdf}{2020 UKAEA validation} analysed this benchmark with FISPACT-II and TENDL-2019, ENDF/B-VIII.0, JEFF-3.3, EAF2010 and, where it had coverage, IRDFF-II. All of those libraries have since been superseded.

This report repeats the comparison with TENDL-2025, ENDF/B-VIII.1, JEFF-4.0 and JENDL-5, and goes further for each element. It breaks the heat down by product, compares the inputs behind each product, separates reaction rates from isomeric branching, and plots the energy-dependent cross sections or branching ratios where the libraries differ. Covariance results are given with the fraction of production they cover. This narrows a disagreement down to particular evaluated data, but it cannot say which library is right. That needs differential measurements, such as those in EXFOR, or direct measurements of isomeric ratios.

\section{Interpreting C/E values}
A C/E value applies to the spectrum, irradiation time and cooling times of its campaign. Agreement in a field dominated by 14 MeV neutrons says little about spectra in which low-energy reactions dominate. The median experimental uncertainty is about 7 percent and larger for some foils, so differences smaller than this are not significant. Calculation uncertainties matter too. When all four libraries miss the measurement by a similar amount, the measurement may be at fault, but so may data or methods that the libraries share.

\clearpage
\section{Nuclear data used}
The table gives the source of each input for the 408 calculations, covering 73 five-minute and 29 seven-hour foils including alloys. Each column is one calculation configuration. Cross sections, reaction channels and isomeric branching come from the named library; the decay data and fission-yield sources are read from each result's recorded metadata, and where results record different sources all of them are listed. Failed calculations, if any, are listed in their own section.
\begingroup\small\setlength{\tabcolsep}{2.5pt}
\renewcommand{\arraystretch}{1.3}
\begin{tabular}{@{}p{0.27\linewidth}*{4}{p{0.17\linewidth}}@{}}\toprule
Input & TENDL-2025 & ENDF/B-VIII.1 & JEFF-4.0 & JENDL-5 \\ \midrule
Neutron cross sections & TENDL-2025 & ENDF/B-VIII.1 & JEFF-4.0 & JENDL-5 \\
Neutron reaction channels & TENDL-2025 & ENDF/B-VIII.1 & JEFF-4.0 & JENDL-5 \\
Isomeric branching & TENDL-2025 & ENDF/B-VIII.1 & JEFF-4.0 & JENDL-5 \\
Decay half-lives and modes & UKDD-2020$^{*}$ & ENDF/B-VIII.1 & JEFF-4.0 & JENDL-5 \\
Decay energies and photon emission & UKDD-2020$^{*}$ & ENDF/B-VIII.1 & JEFF-4.0 & JENDL-5 \\
Fission yields & UKFY-4.2$^{*}$ & ENDF/B-VIII.1 & JEFF-4.0 & JENDL-5 \\
Cross-section covariance & TENDL-2025 & ENDF/B-VIII.1 & JEFF-4.0 & JENDL-5 \\
\bottomrule\end{tabular}\endgroup\par
An asterisk marks an input taken from a different library. TENDL-2025 has no decay data or fission yields, so it uses UKDD-2020 decay data and UKFY-4.2 fission yields from the FISPACT-II nuclear data repository, converted locally. The transmutation chain combines the listed reaction channels, decay modes and yields. A listed covariance source does not mean that every nuclide or channel has covariance data. The other uncertainties come with the decay, branching or yield data they belong to.
Every calculation uses a measured FNS neutron spectrum: position 3 for the five-minute campaign and position 7 for the seven-hour campaign. No photonuclear data are used. The energy of decay gammas is included in the decay heat.
\subsection*{Data releases}
{\small\begin{longtable}{lll}\toprule
Data library & Release & SHA-256 prefix \\ \midrule
\endfirsthead\toprule Data library & Release & SHA-256 prefix \\ \midrule\endhead
ENDF/B-VIII.1 & 2026-11-13 & \texttt{e04637b8bc40} \\
JEFF-4.0 & 2026-11-13 & \texttt{0226481b9d95} \\
JENDL-5 & 2026-11-13 & \texttt{6e47e06d399b} \\
TENDL-2025 & 2026-11-14 & \texttt{66deceec1b2d} \\
UKDD-2020 & converted from decay2020.tar.bz2 & \texttt{f2ff9501f6d2} \\
UKDD-2020 & processed locally & \texttt{50f4881de3ee} \\
UKFY-4.2 & converted from UKFY42data.tar.bz2 & \texttt{e7a2505c32ef} \\
UKFY-4.2 & processed locally & \texttt{25f0c6193f02} \\
\bottomrule\end{longtable}}
The dates are those of the processed data releases, not of the evaluations. For a published release the hash is that of its manifest, which identifies the exact files used. For a library converted locally it is that of the archive downloaded from the FISPACT-II repository. The full hashes are stored with each result.
\clearpage
\section{Analysis method}
For each element the calculated heat is compared with the measurement and then broken down by activation product. For the products that carry the heat, the reaction rates, isomeric branching, half-lives and decay energies are compared across the libraries. Where a reaction rate or branching ratio differs, its energy dependence is plotted to show where the difference lies.

A product missing from one library's heat breakdown may point to a missing reaction channel, but it can also come from a different state assignment or a reporting threshold. Likewise, agreement between the libraries does not rule out an error they share, so a deviation common to all of them calls for a check of both the shared data and the experiment.

An input that differs between libraries is a possible explanation, not a proven cause. The product-removal test subtracts a product's heat from the total without re-solving the inventory, so it shows how large the contribution is, not what corrected data would give.

\begin{figure}[H]
\centering
\includegraphics[width=\textwidth,height=0.82\textheight,keepaspectratio]{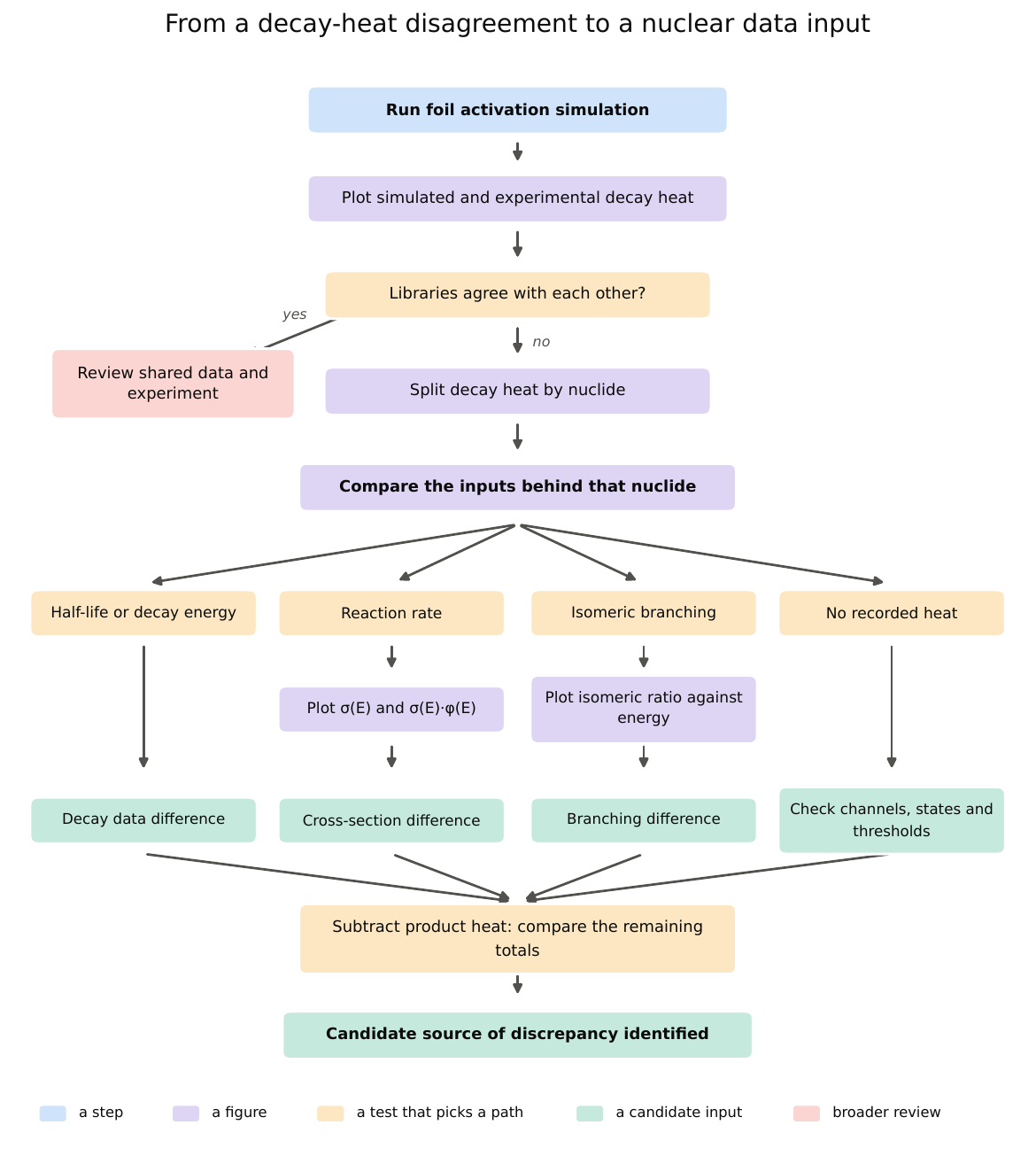}
\caption{The steps from a decay-heat disagreement to a single nuclear data input.}
\end{figure}
\clearpage

\section{Neutron spectrum}
\begin{figure}[H]
\centering
\includegraphics[width=\textwidth,height=0.82\textheight,keepaspectratio]{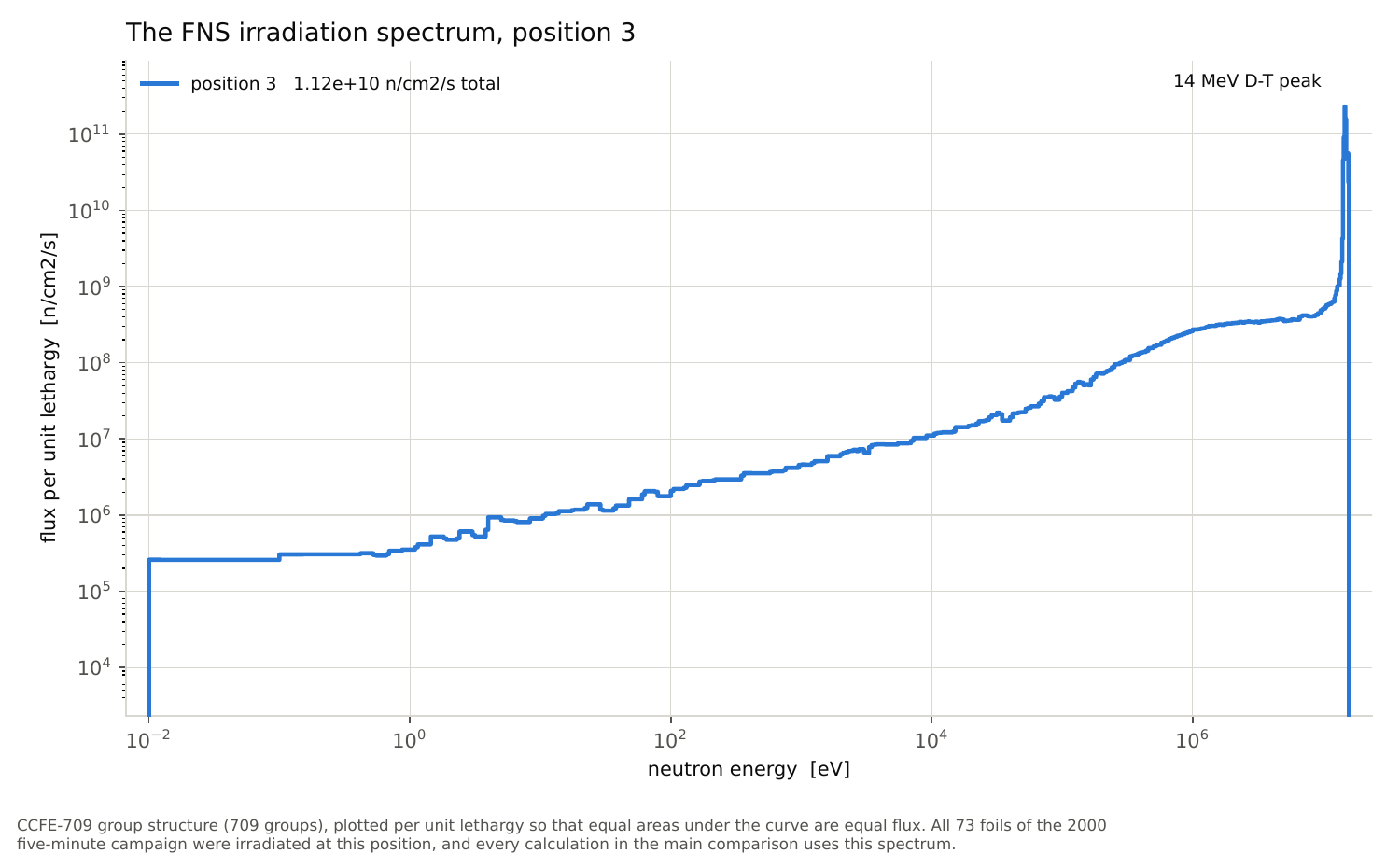}
\caption{The measured neutron spectrum at irradiation position 3, used for every calculation in the main comparison.}
\end{figure}
\clearpage

\section{Results by element}
Elements are ordered alphabetically. Each has three figures: the calculated decay heat against the measurement, with C/E below; the share of the heat from each activation product during cooling; and the inputs for those products, each relative to its median across the libraries. Cross-section and branching plots are added where the libraries differ. A short note then lists the differences and suggests measurements. Where a long-lived product matters and the foil was in the 1996 seven-hour campaign, that comparison is also shown, but it is not included in the rankings.

\clearpage
\subsection{Ag: Silver}\label{foil:Ag}
\begin{figure}[H]
\centering
\includegraphics[width=\textwidth,height=0.82\textheight,keepaspectratio]{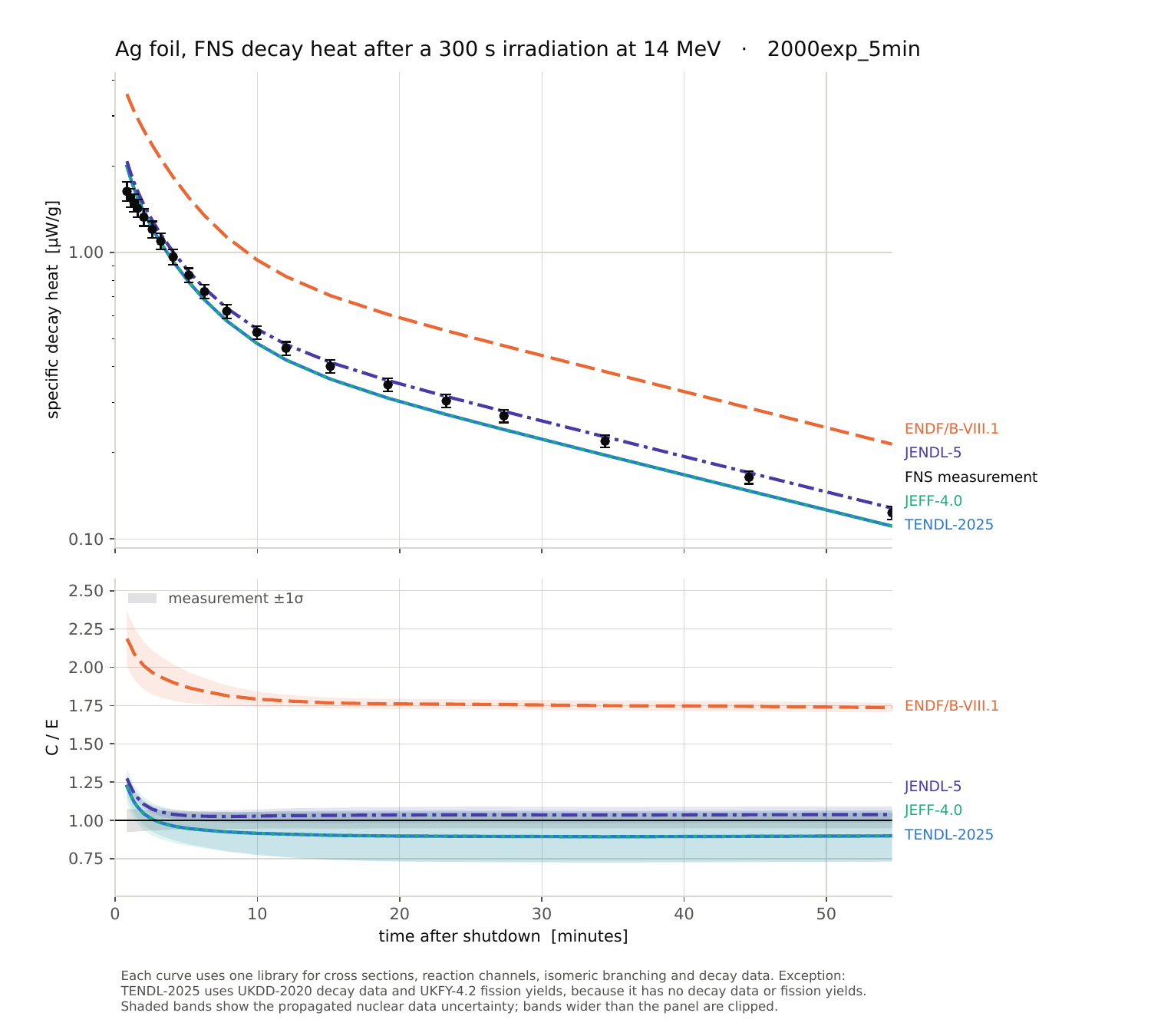}
\caption{Ag: decay heat against the measurement, with C/E beneath. Mean $|C/E-1|$ by library: JENDL-5 7\%, JEFF-4.0 9\%, TENDL-2025 9\%, ENDF/B-VIII.1 88\%.}
\end{figure}
\clearpage

\begin{figure}[H]
\centering
\includegraphics[width=\textwidth,height=0.82\textheight,keepaspectratio]{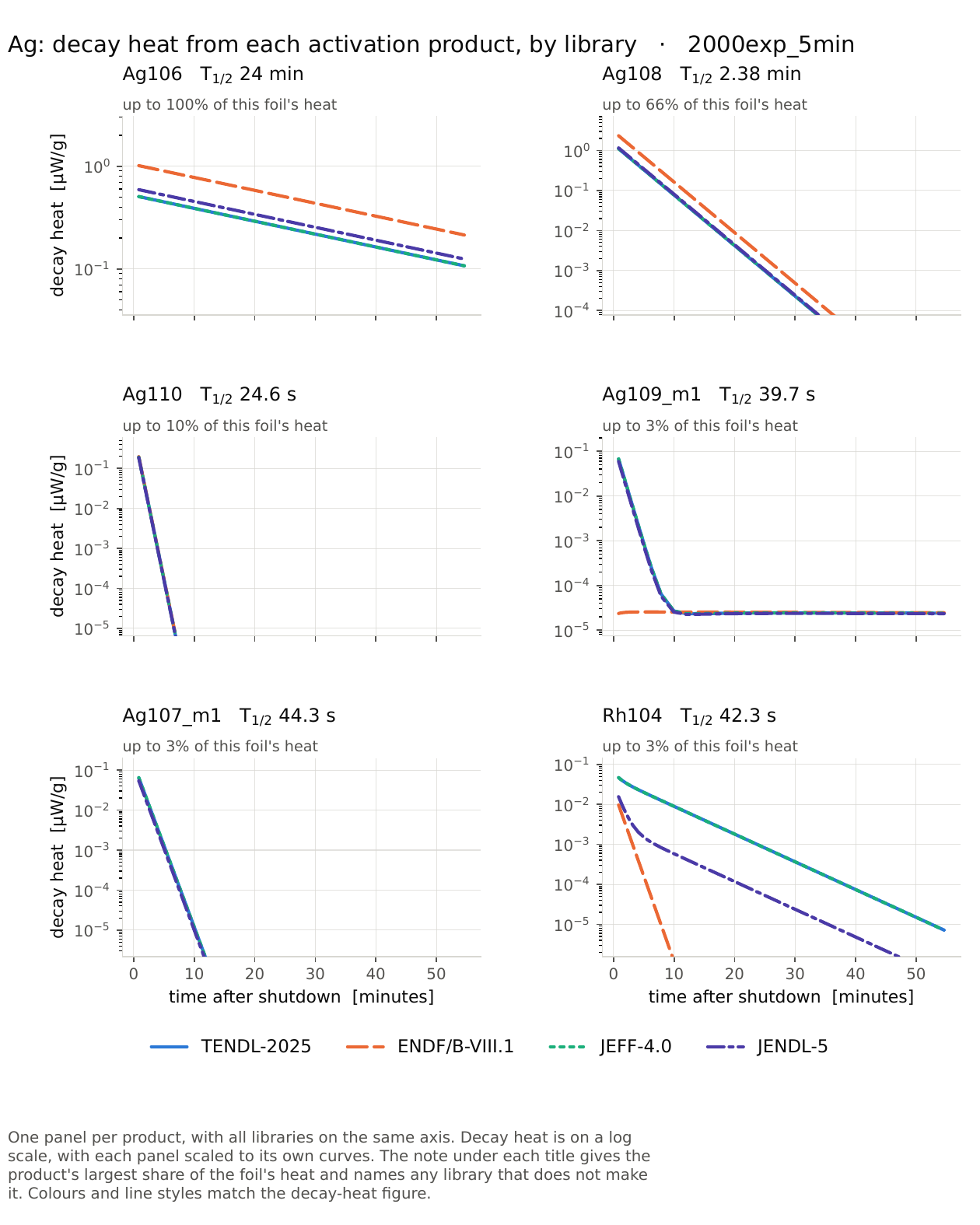}
\caption{Ag: share of the decay heat from each activation product during cooling.}
\end{figure}
\clearpage

\begin{figure}[H]
\centering
\includegraphics[width=\textwidth,height=0.82\textheight,keepaspectratio]{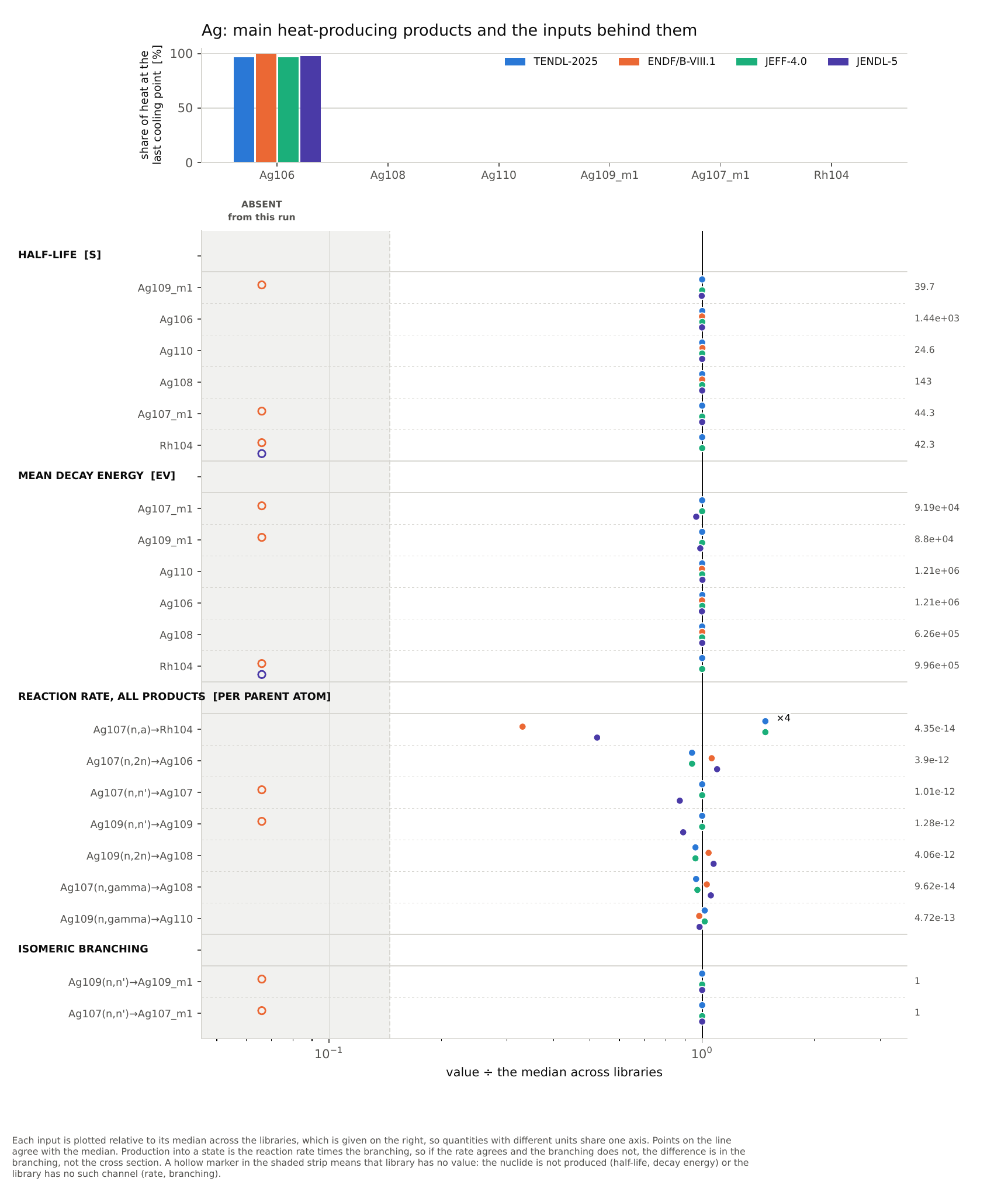}
\caption{Ag: inputs for the main heat-producing products, each relative to its median across the available libraries.}
\end{figure}
\clearpage

\begin{figure}[H]
\centering
\includegraphics[width=\textwidth,height=0.82\textheight,keepaspectratio]{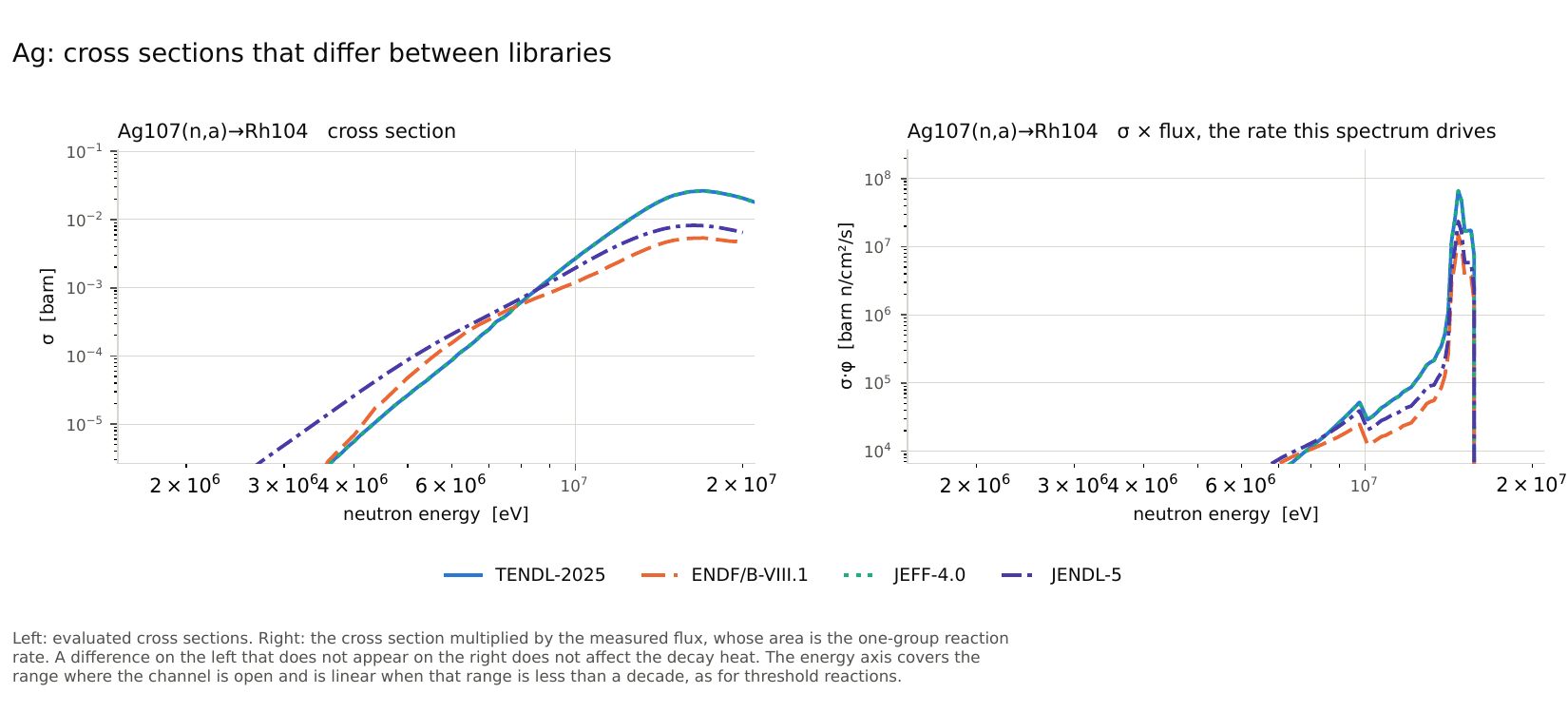}
\caption{Ag: cross sections of the channels whose spectrum-averaged rates differ, with the neutron spectrum and their product.}
\end{figure}
\clearpage

ENDF/B-VIII.1 stands out at 88 percent from the measurement, against 9 percent for the next furthest library and 7 percent for JENDL-5. The median measurement uncertainty is 6 percent.

\subsubsection*{Differences in the nuclear data}
\begin{itemize}\setlength{\itemsep}{6pt}
\item ENDF/B-VIII.1 gives the total reaction rate of Ag107(n,a)→Rh104 as 1.435e-14, against 6.421e-14 for the median of the others, a factor of 4.\\[2pt]{\itshape This is the total rate of the channel over all its products, so the difference is in the cross section, not in the split between states.}
\end{itemize}
\subsubsection*{Suggested measurements}
\begin{itemize}\setlength{\itemsep}{4pt}
\item Natural Ag has 2 isotopes, so the measured heat combines several parents. Enriched single-isotope targets would distinguish their contributions. The candidate parent is Ag107. Enrichment alone does not isolate a single reaction channel.
\item This foil makes Ag106\textbackslash{}\_m1 (half-life 8.46 days), which gives up to 2 percent of the calculated heat. A five-minute irradiation and 55 minute measurement only see the start of this contribution. There is no seven-hour FNS measurement of this foil. A longer irradiation followed for days to months would be needed to test this product's production and decay.
\item The reaction rates are averaged over the measured spectrum. Activation measurements at several well-defined neutron energies, including below 14 MeV, would show which energy range the differences come from.
\end{itemize}
\clearpage
\subsection{Al: Aluminum}\label{foil:Al}
\begin{figure}[H]
\centering
\includegraphics[width=\textwidth,height=0.82\textheight,keepaspectratio]{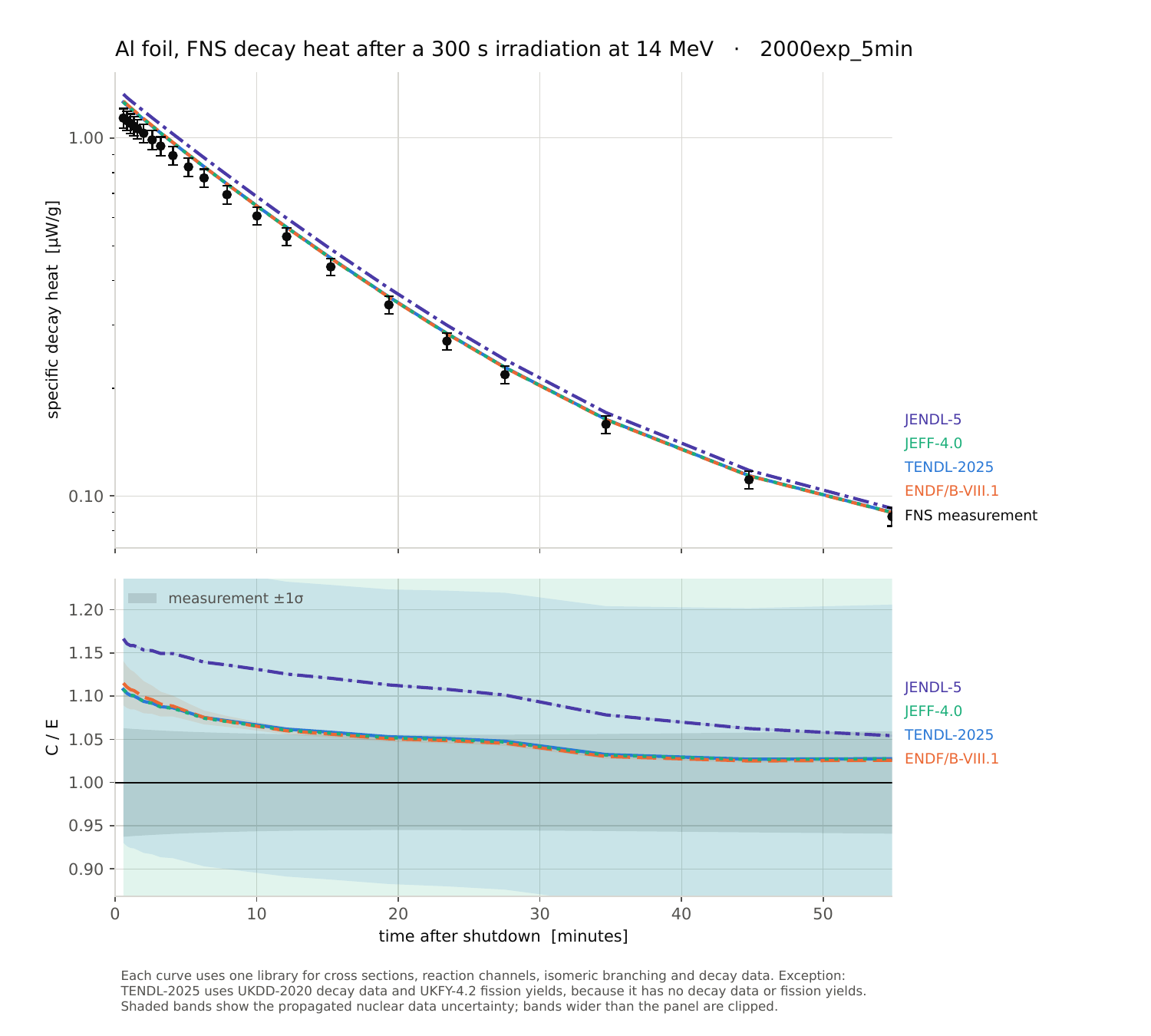}
\caption{Al: decay heat against the measurement, with C/E beneath. Mean $|C/E-1|$ by library: JEFF-4.0 7\%, TENDL-2025 7\%, ENDF/B-VIII.1 7\%, JENDL-5 13\%.}
\end{figure}
\clearpage

\begin{figure}[H]
\centering
\includegraphics[width=\textwidth,height=0.82\textheight,keepaspectratio]{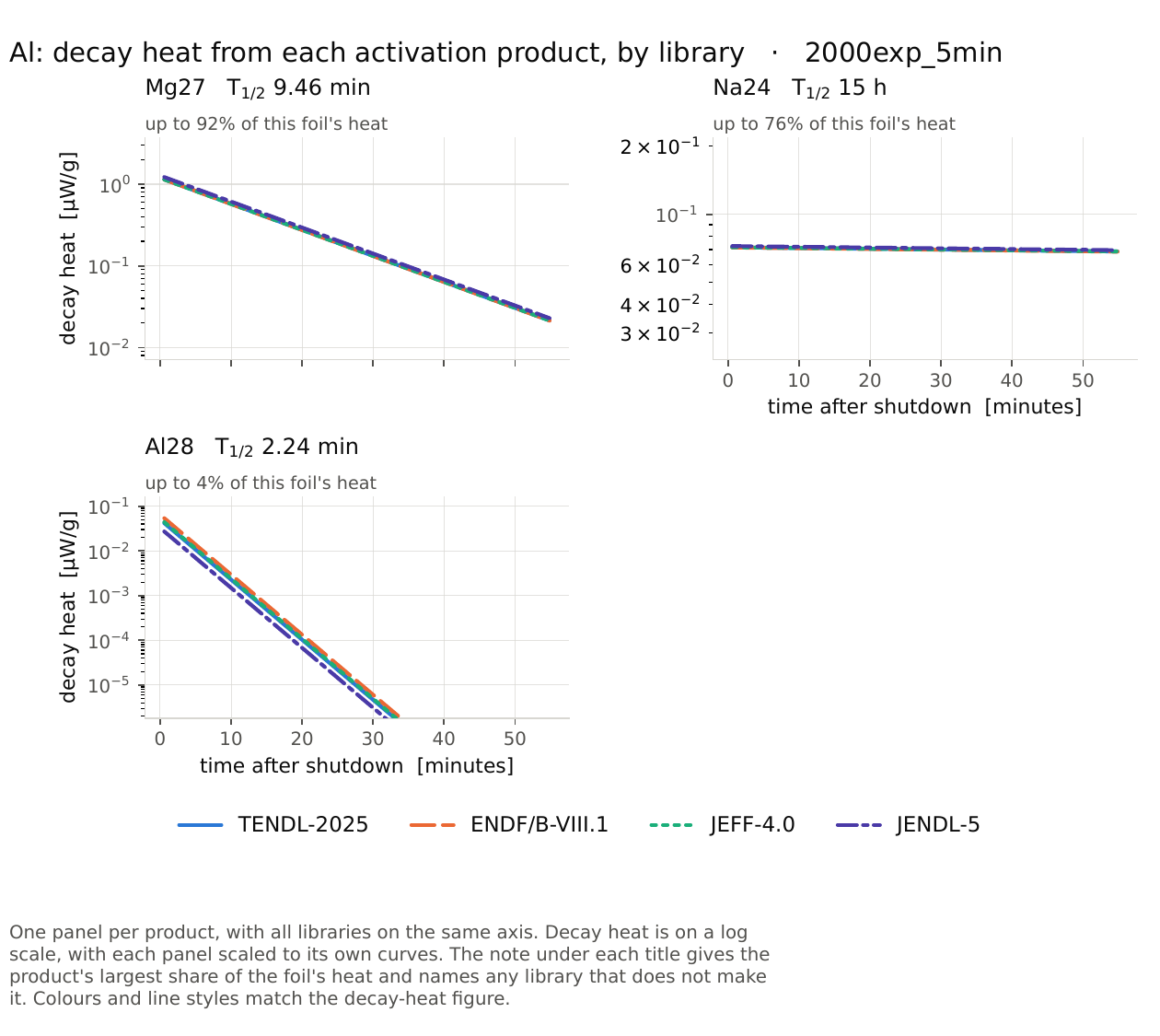}
\caption{Al: share of the decay heat from each activation product during cooling.}
\end{figure}
\clearpage

\begin{figure}[H]
\centering
\includegraphics[width=\textwidth,height=0.82\textheight,keepaspectratio]{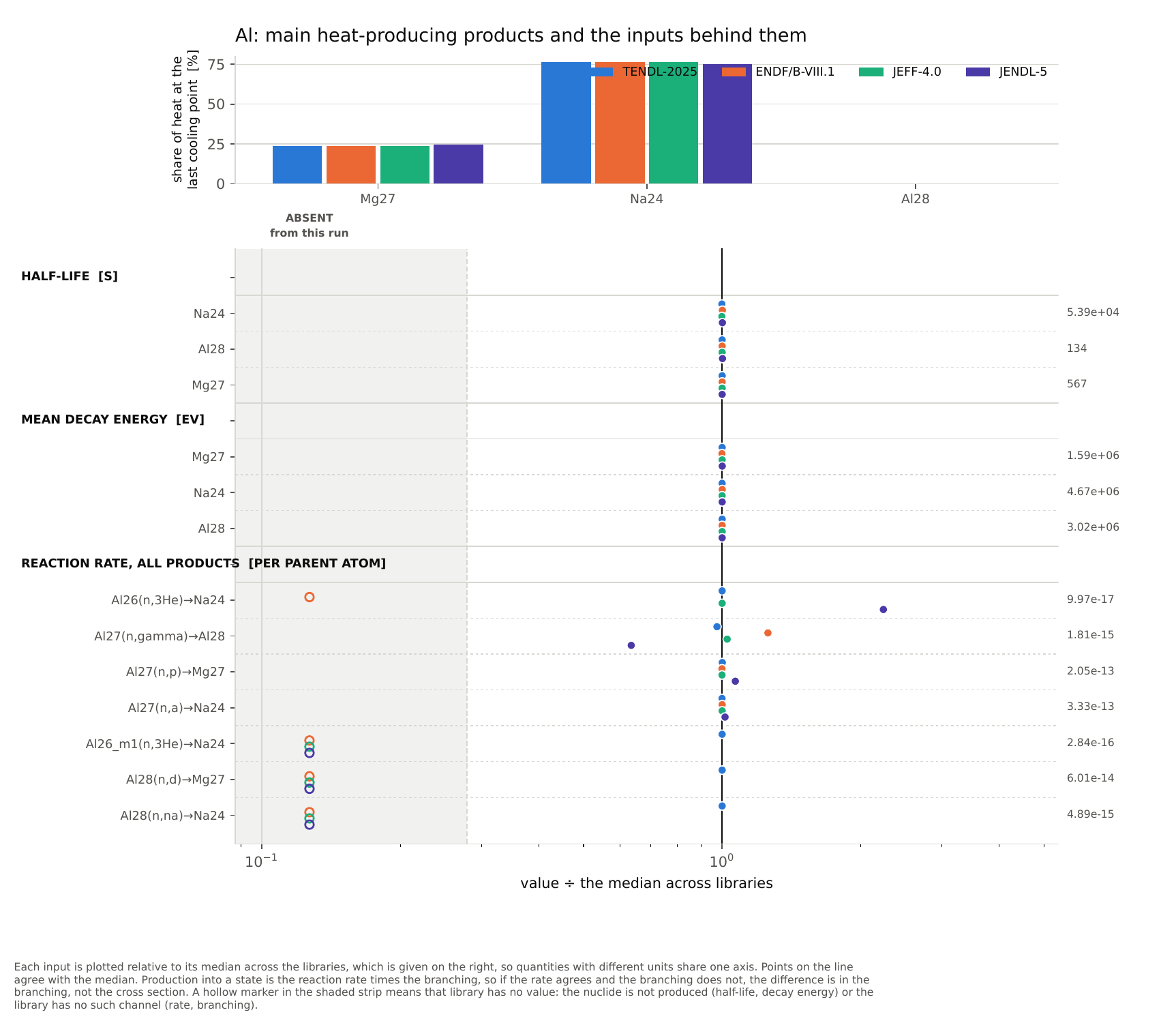}
\caption{Al: inputs for the main heat-producing products, each relative to its median across the available libraries.}
\end{figure}
\clearpage

\begin{figure}[H]
\centering
\includegraphics[width=\textwidth,height=0.82\textheight,keepaspectratio]{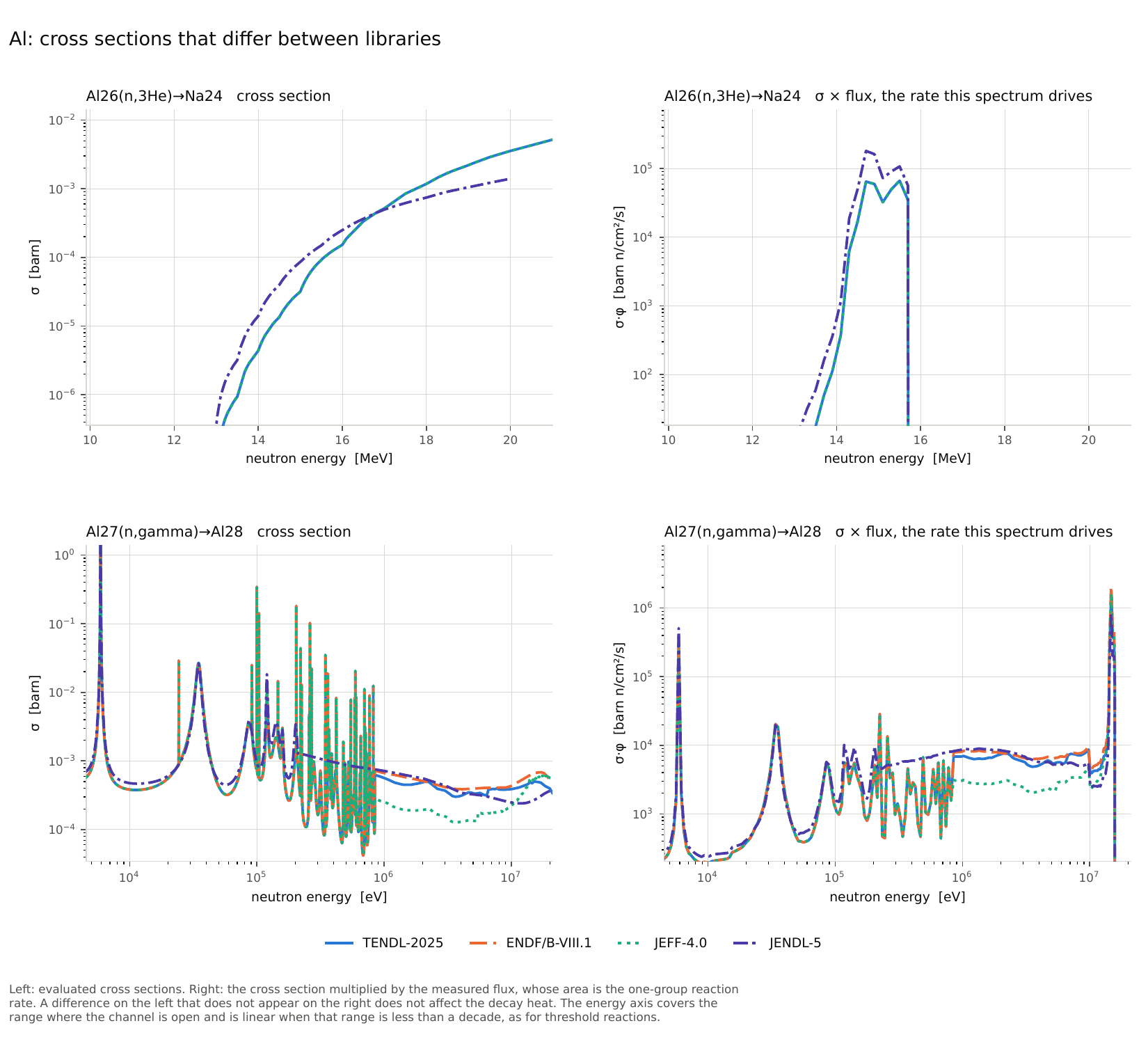}
\caption{Al: cross sections of the channels whose spectrum-averaged rates differ, with the neutron spectrum and their product.}
\end{figure}
\clearpage

Mean absolute C/E deviations range from 7 to 13 percent. The spread of 6 percentage points is smaller than the median measurement uncertainty of 6 percent, so the measurement cannot separate the libraries.

\subsubsection*{Differences in the nuclear data}
\begin{itemize}\setlength{\itemsep}{6pt}
\item JENDL-5 gives the total reaction rate of Al26(n,3He)→Na24 as 2.233e-16, against 9.968e-17 for the median of the others, a factor of 2.\\[2pt]{\itshape This is the total rate of the channel over all its products, so the difference is in the cross section, not in the split between states.}
\end{itemize}
\subsubsection*{Suggested measurements}
\begin{itemize}\setlength{\itemsep}{4pt}
\item Al has a single stable isotope, so the foil is already a single-isotope target and enrichment would not help.
\item The reaction rates are averaged over the measured spectrum. Activation measurements at several well-defined neutron energies, including below 14 MeV, would show which energy range the differences come from.
\end{itemize}
\clearpage
\subsection{As: Arsenic}\label{foil:As}
\begin{figure}[H]
\centering
\includegraphics[width=\textwidth,height=0.82\textheight,keepaspectratio]{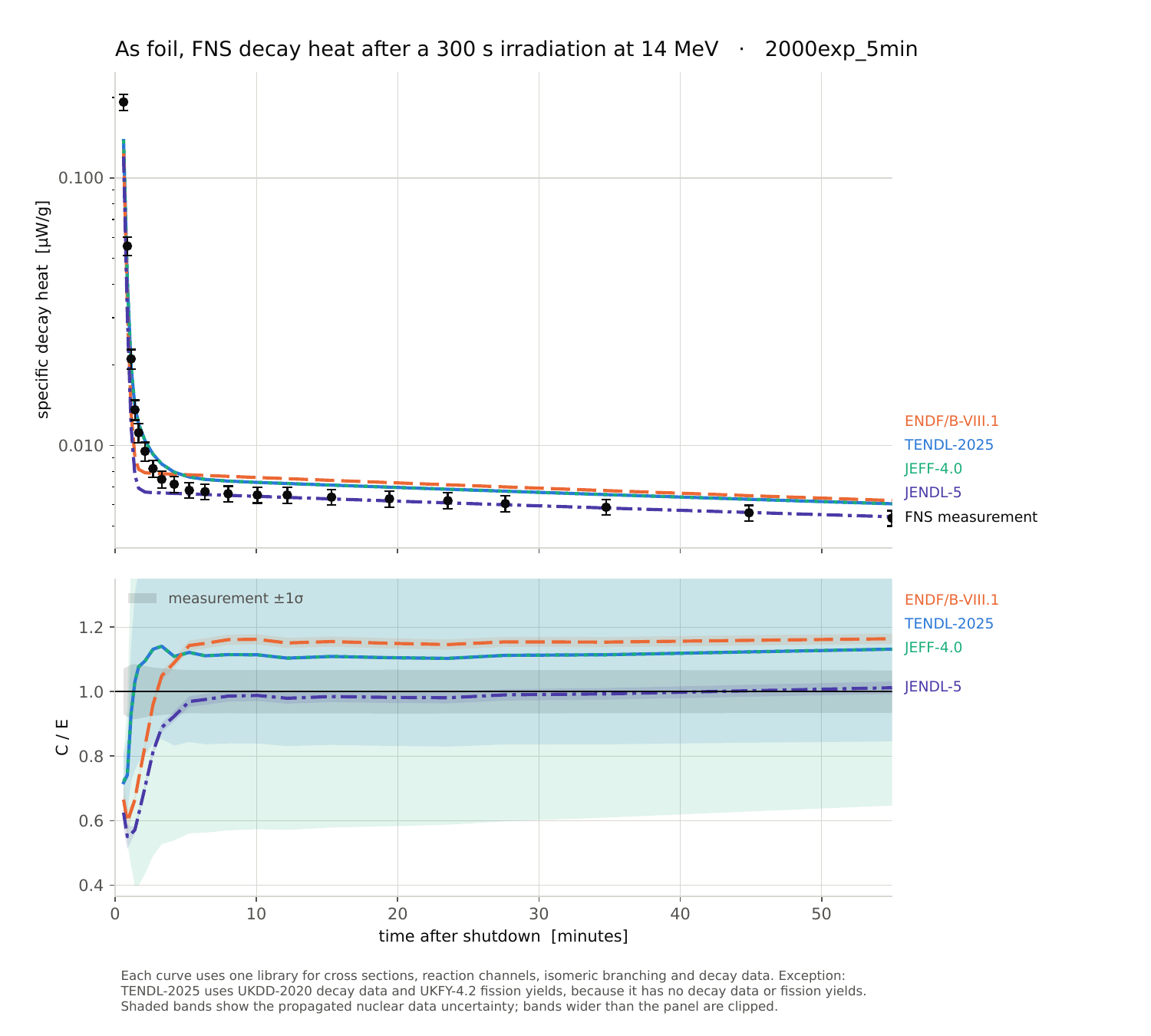}
\caption{As: decay heat against the measurement, with C/E beneath. Mean $|C/E-1|$ by library: JEFF-4.0 12\%, TENDL-2025 12\%, JENDL-5 14\%, ENDF/B-VIII.1 19\%.}
\end{figure}
\clearpage

\begin{figure}[H]
\centering
\includegraphics[width=\textwidth,height=0.82\textheight,keepaspectratio]{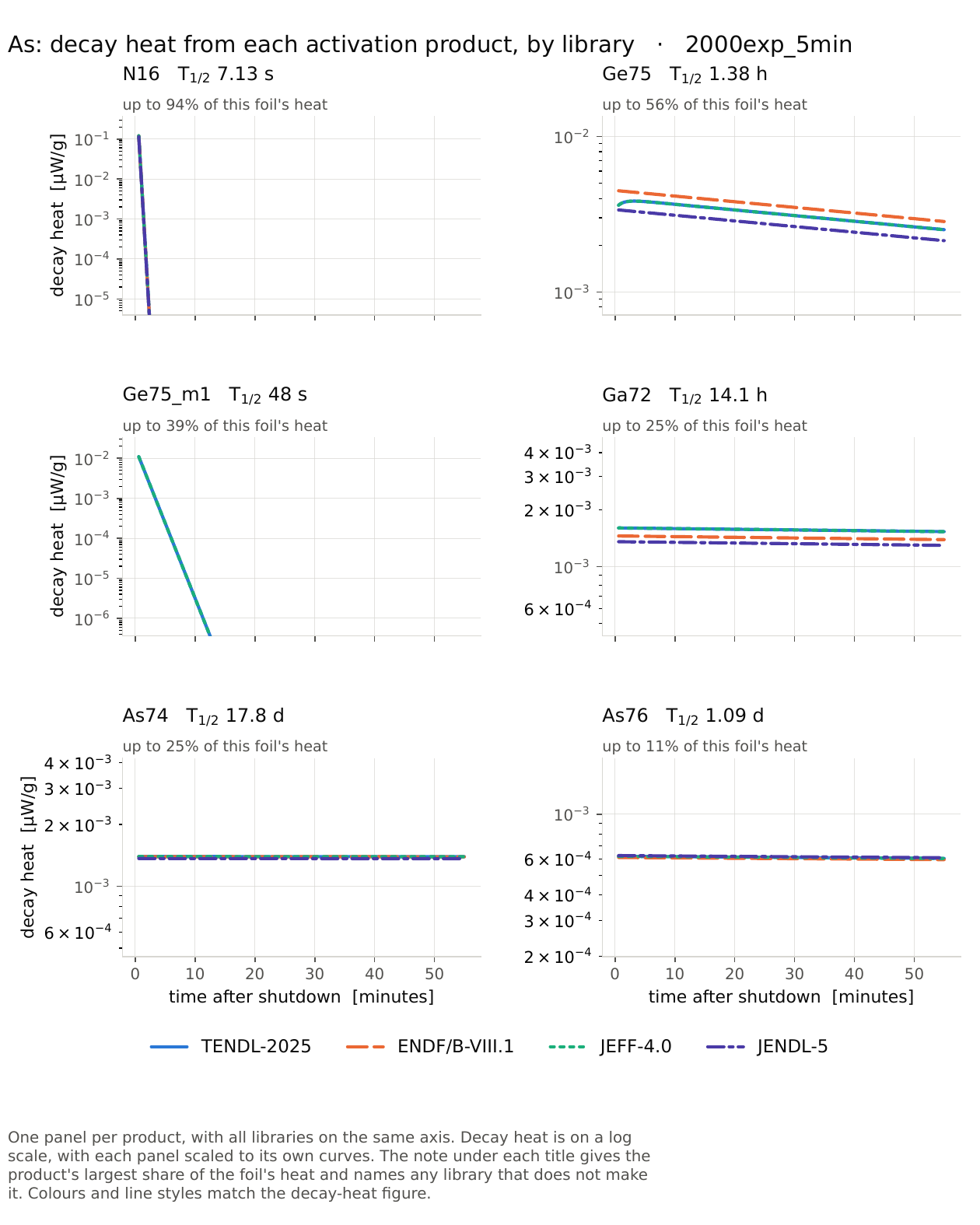}
\caption{As: share of the decay heat from each activation product during cooling.}
\end{figure}
\clearpage

\begin{figure}[H]
\centering
\includegraphics[width=\textwidth,height=0.82\textheight,keepaspectratio]{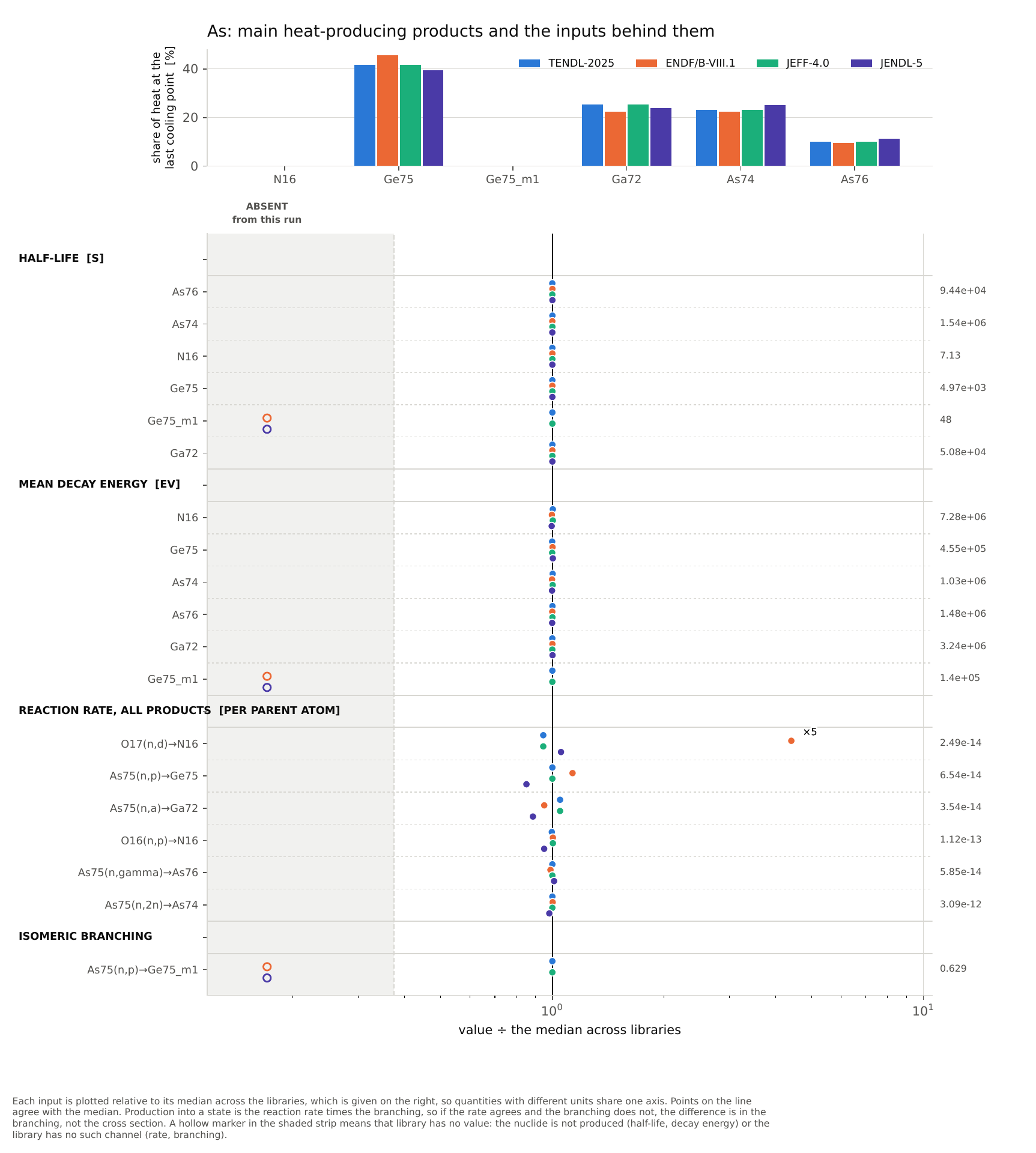}
\caption{As: inputs for the main heat-producing products, each relative to its median across the available libraries.}
\end{figure}
\clearpage

\begin{figure}[H]
\centering
\includegraphics[width=\textwidth,height=0.82\textheight,keepaspectratio]{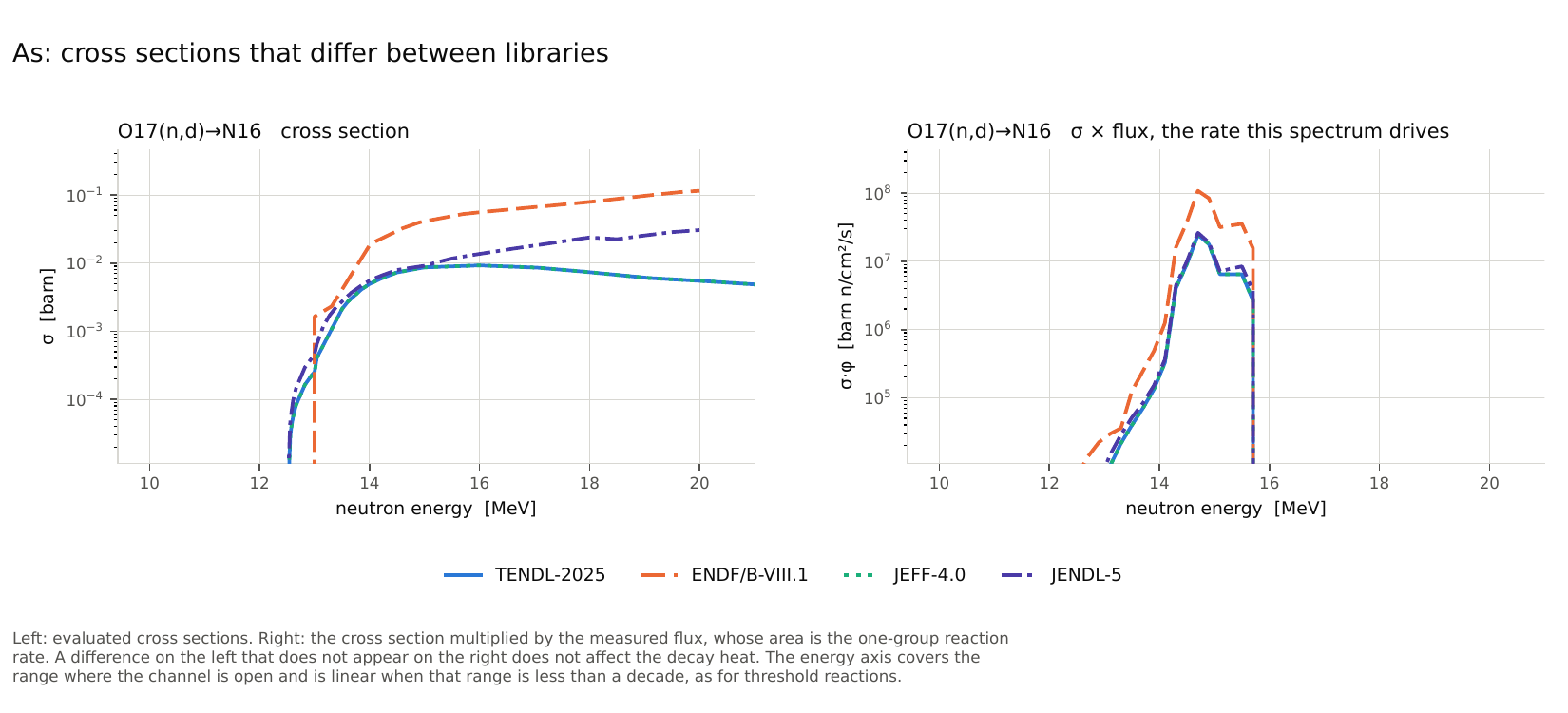}
\caption{As: cross sections of the channels whose spectrum-averaged rates differ, with the neutron spectrum and their product.}
\end{figure}
\clearpage

Mean absolute C/E deviations range from 12 to 19 percent. The spread of 6 percentage points is smaller than the median measurement uncertainty of 7 percent, so the measurement cannot separate the libraries.

\subsubsection*{Differences in the nuclear data}
\begin{itemize}\setlength{\itemsep}{6pt}
\item ENDF/B-VIII.1 gives the total reaction rate of O17(n,d)→N16 as 1.096e-13, against 2.351e-14 for the median of the others, a factor of 5.\\[2pt]{\itshape This is the total rate of the channel over all its products, so the difference is in the cross section, not in the split between states.}
\end{itemize}
\subsubsection*{Suggested measurements}
\begin{itemize}\setlength{\itemsep}{4pt}
\item As has a single stable isotope, so the foil is already a single-isotope target and enrichment would not help.
\item The sample is a compound that also contains O. Their activation adds to the measured heat; a metal or purer sample would remove it.
\item This foil makes As74 (half-life 17.8 days), which gives up to 25 percent of the calculated heat. A five-minute irradiation and 55 minute measurement only see the start of this contribution. There is no seven-hour FNS measurement of this foil. A longer irradiation followed for days to months would be needed to test this product's production and decay.
\item The reaction rates are averaged over the measured spectrum. Activation measurements at several well-defined neutron energies, including below 14 MeV, would show which energy range the differences come from.
\end{itemize}
\clearpage
\subsection{Au: Gold}\label{foil:Au}
\begin{figure}[H]
\centering
\includegraphics[width=\textwidth,height=0.82\textheight,keepaspectratio]{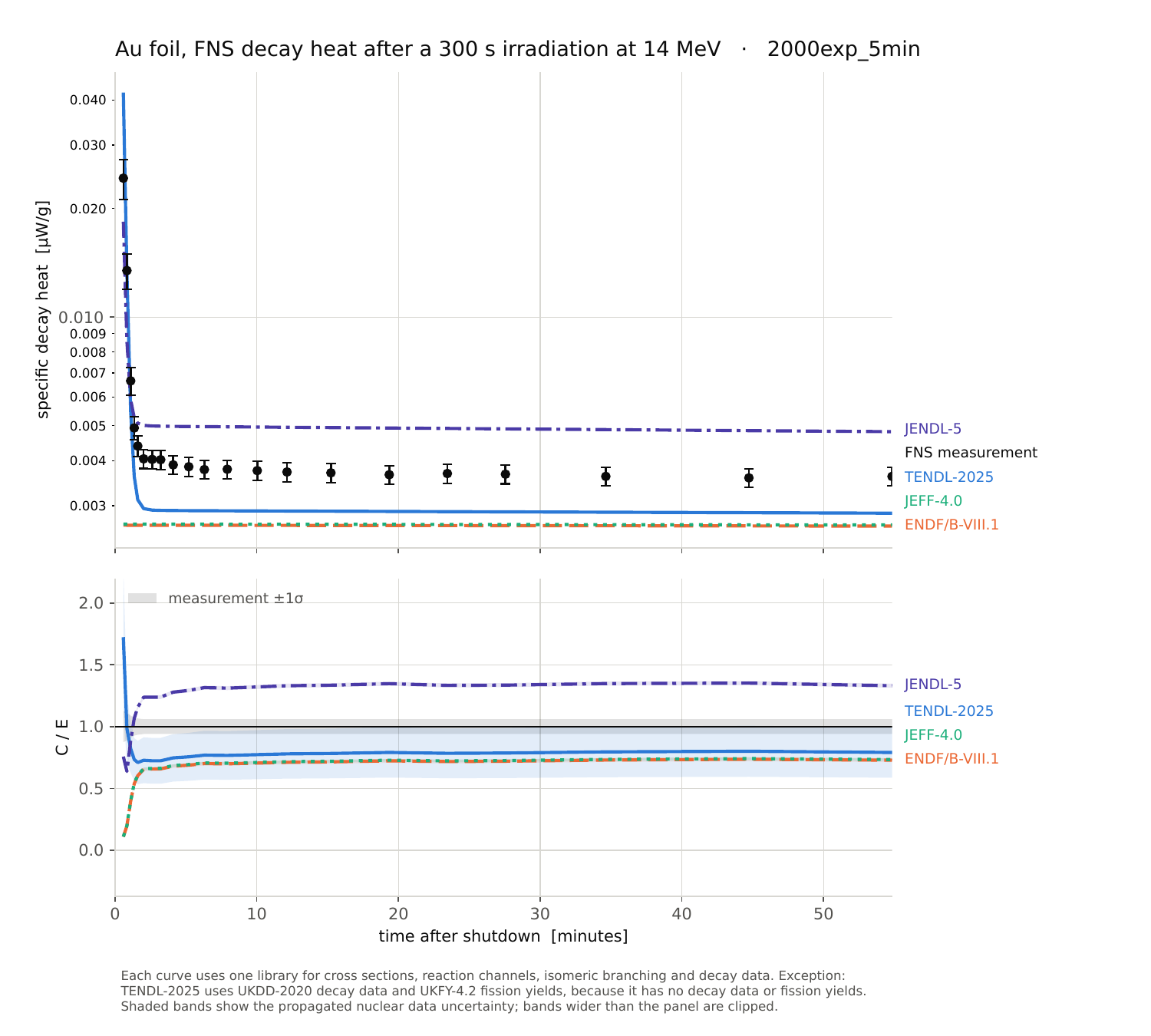}
\caption{Au: decay heat against the measurement, with C/E beneath. Mean $|C/E-1|$ by library: TENDL-2025 25\%, JENDL-5 28\%, JEFF-4.0 37\%, ENDF/B-VIII.1 38\%.}
\end{figure}
\clearpage

\begin{figure}[H]
\centering
\includegraphics[width=\textwidth,height=0.82\textheight,keepaspectratio]{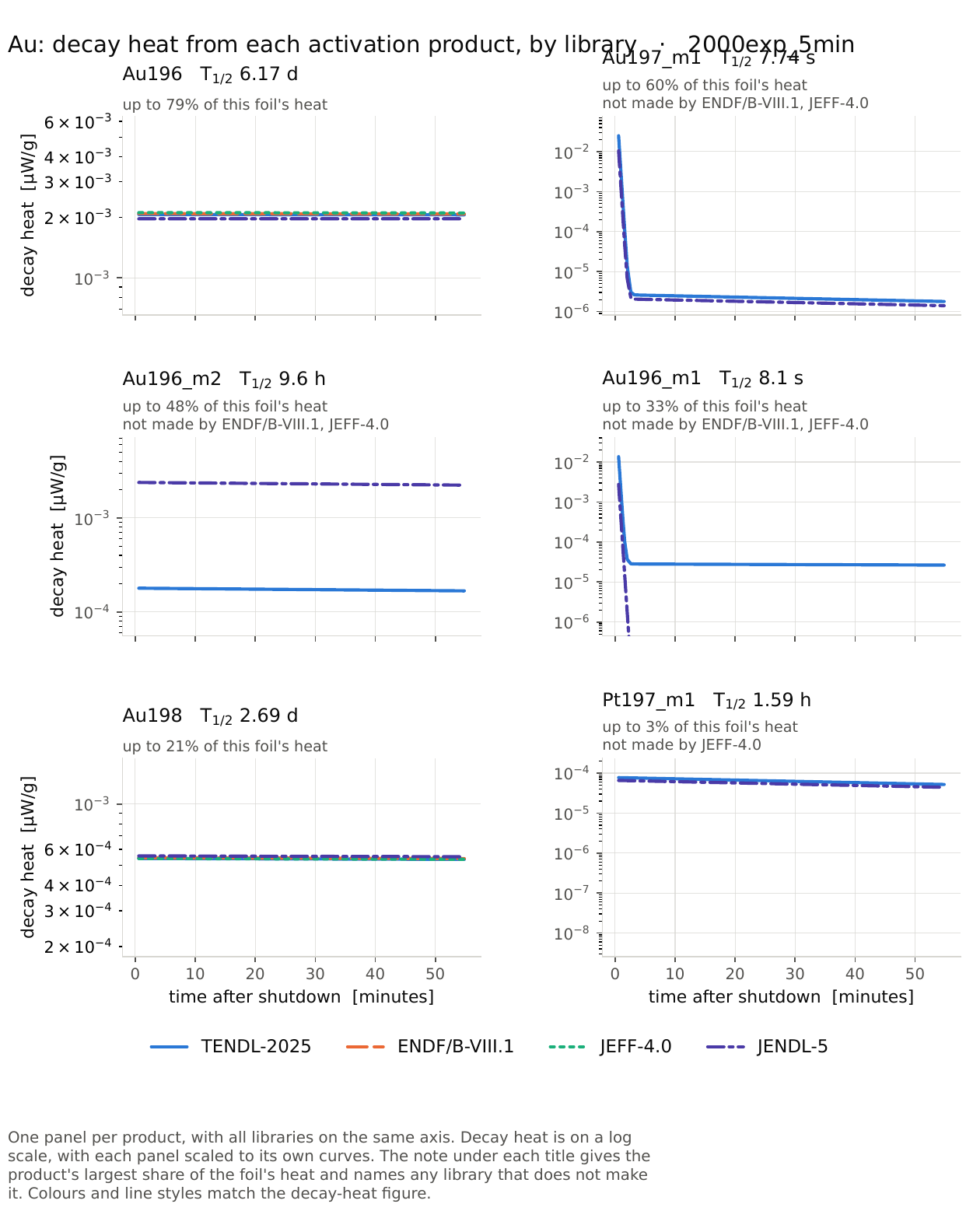}
\caption{Au: share of the decay heat from each activation product during cooling.}
\end{figure}
\clearpage

\begin{figure}[H]
\centering
\includegraphics[width=\textwidth,height=0.82\textheight,keepaspectratio]{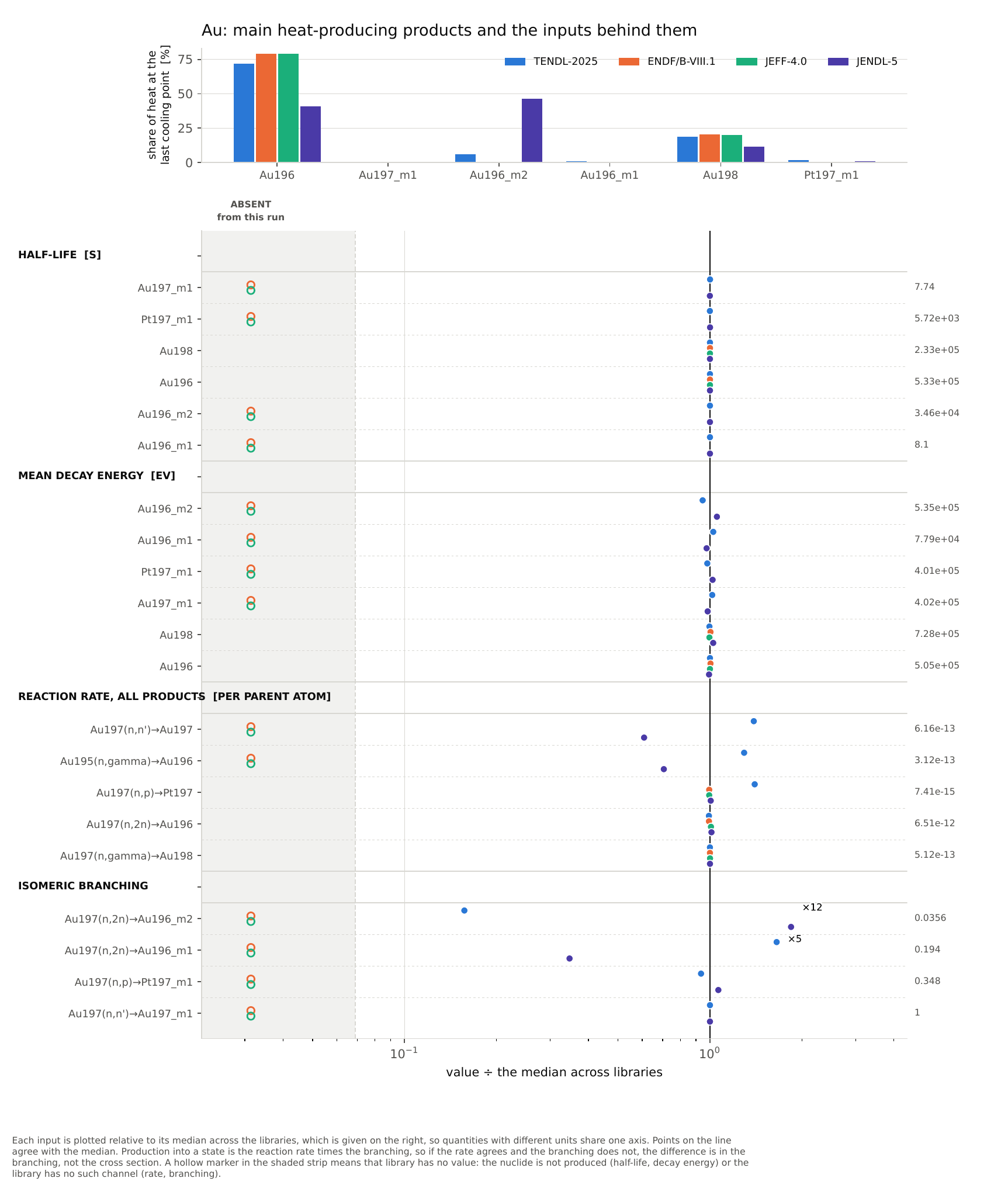}
\caption{Au: inputs for the main heat-producing products, each relative to its median across the available libraries.}
\end{figure}
\clearpage

\begin{figure}[H]
\centering
\includegraphics[width=\textwidth,height=0.82\textheight,keepaspectratio]{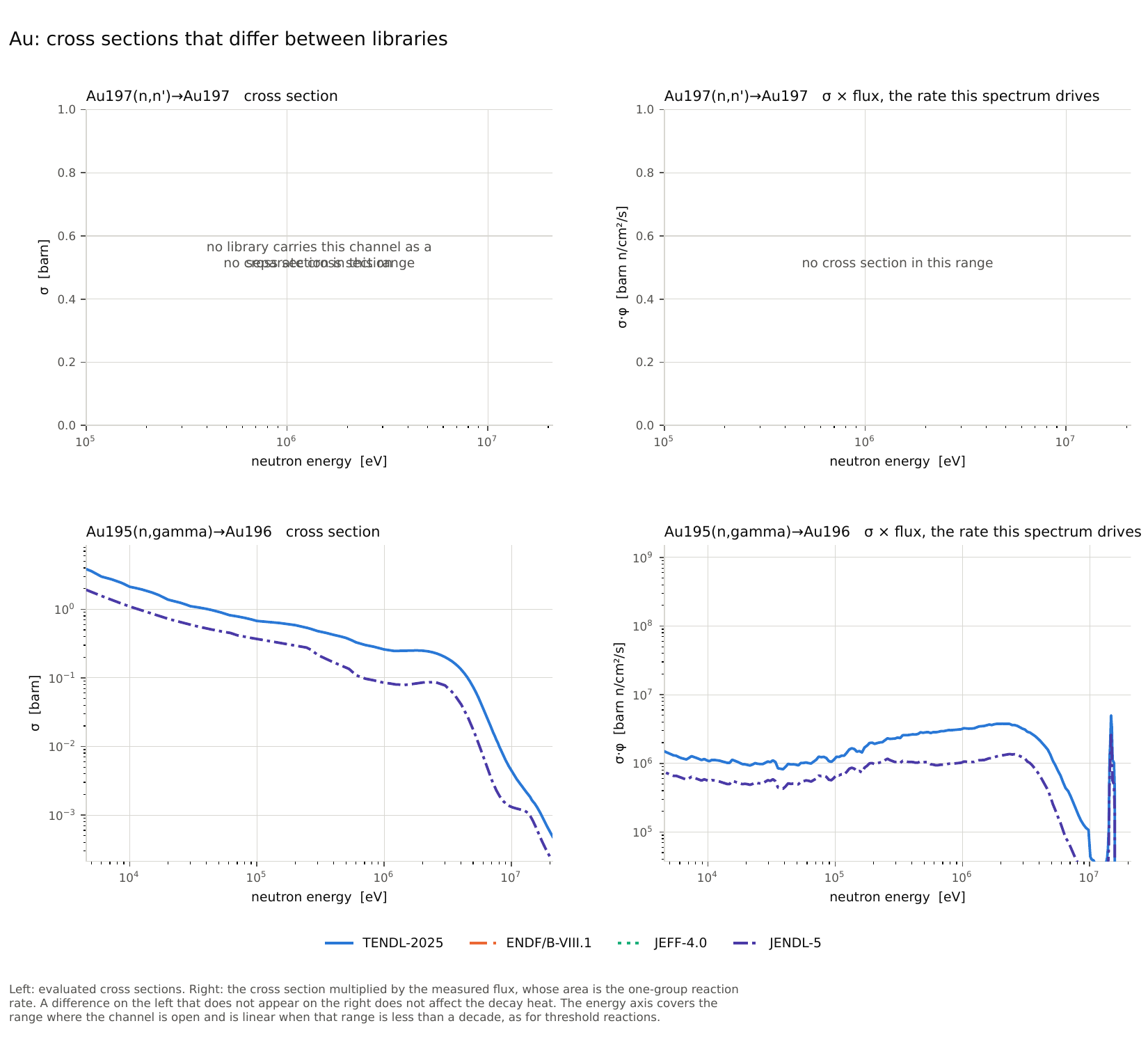}
\caption{Au: cross sections of the channels whose spectrum-averaged rates differ, with the neutron spectrum and their product.}
\end{figure}
\clearpage

\begin{figure}[H]
\centering
\includegraphics[width=\textwidth,height=0.82\textheight,keepaspectratio]{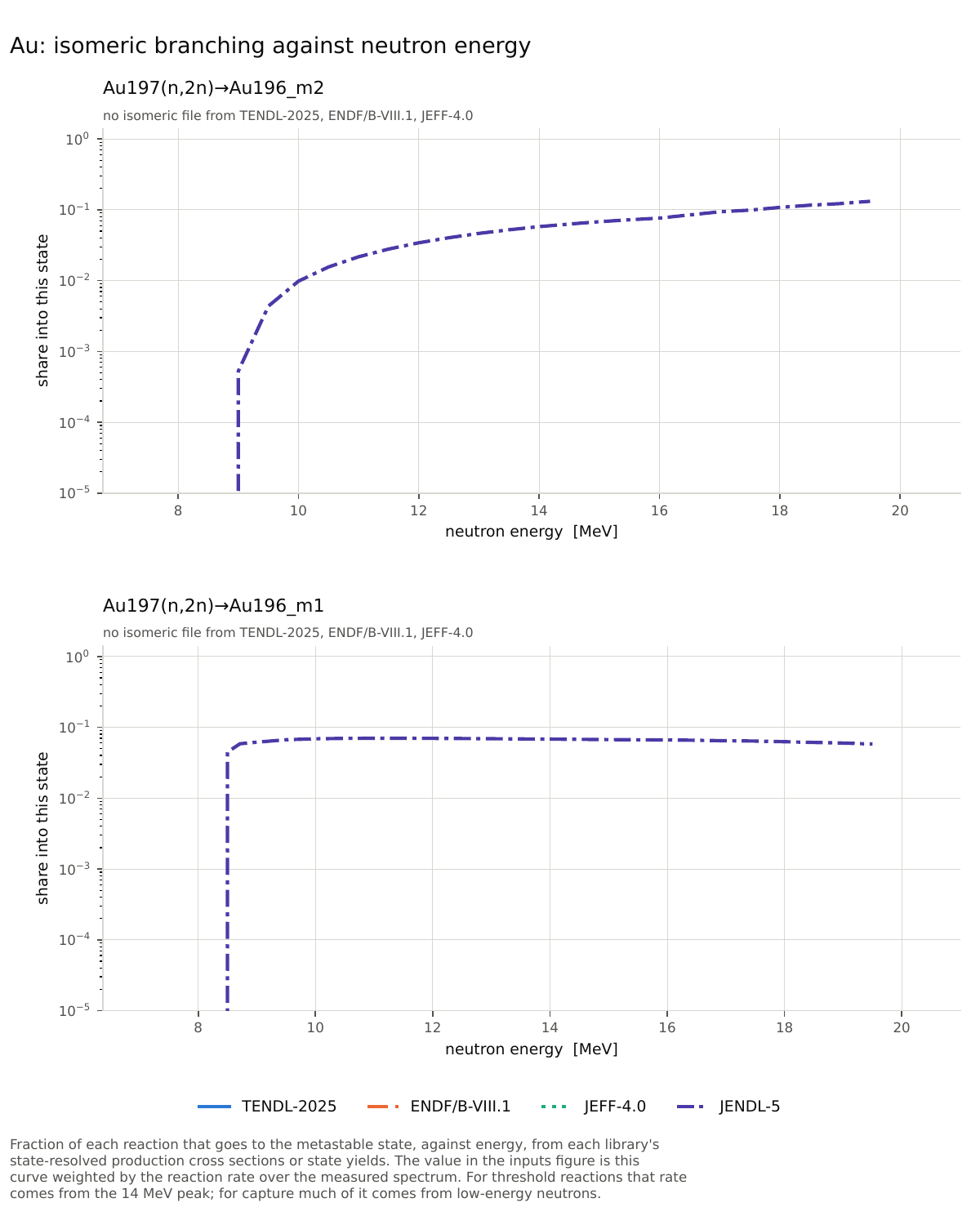}
\caption{Au: isomeric branching ratio against neutron energy for the channels whose spectrum-averaged ratio differs between libraries.}
\end{figure}
\clearpage

The libraries are between 25 and 38 percent from the measurement, which has a median uncertainty of 6 percent. TENDL-2025 is closest and ENDF/B-VIII.1 furthest.

\subsubsection*{Differences in the nuclear data}
\begin{itemize}\setlength{\itemsep}{6pt}
\item Only JENDL-5 and TENDL-2025 have the isomeric branching of Au197(n,2n)→Au196\_m2, and they differ by a factor of 12: 0.06557 against 0.005586.\\[2pt]{\itshape Isomeric branching sets how a channel's production is split between the ground and metastable states. It should be checked together with the total rate of the channel.}
\item Only TENDL-2025 and JENDL-5 have the isomeric branching of Au197(n,2n)→Au196\_m1, and they differ by a factor of 5: 0.3209 against 0.06735.
\item Only TENDL-2025 and JENDL-5 have the total reaction rate of Au197(n,n')→Au197, and they differ by a factor of 2: 8.564e-13 against 3.746e-13.\\[2pt]{\itshape This is the total rate of the channel over all its products, so the difference is in the cross section, not in the split between states.}
\item ENDF/B-VIII.1, JEFF-4.0 give no heat from Au197\_m1; the other libraries do.\\[2pt]{\itshape The heat breakdown alone does not show why. Possible causes are a missing reaction channel, a different state assignment, different decay data or a reporting threshold.}
\item ENDF/B-VIII.1, JEFF-4.0 give no heat from Au196\_m2; the other libraries do.
\item ENDF/B-VIII.1, JEFF-4.0 give no heat from Au196\_m1; the other libraries do.
\end{itemize}
\subsubsection*{Suggested measurements}
\begin{itemize}\setlength{\itemsep}{4pt}
\item Au has a single stable isotope, so the foil is already a single-isotope target and enrichment would not help.
\item This foil makes Au196 (half-life 6.17 days), which gives up to 79 percent of the calculated heat. A five-minute irradiation and 55 minute measurement only see the start of this contribution. There is no seven-hour FNS measurement of this foil. A longer irradiation followed for days to months would be needed to test this product's production and decay.
\item The reaction rates are averaged over the measured spectrum. Activation measurements at several well-defined neutron energies, including below 14 MeV, would show which energy range the differences come from.
\item Measuring the activity of each state separately, for example by gamma spectrometry where the states have distinct lines, would constrain the isomeric ratio. This needs corrections for feeding between states, detector efficiency and irradiation history.
\end{itemize}
\clearpage
\subsection{Ba: Barium}\label{foil:Ba}
\begin{figure}[H]
\centering
\includegraphics[width=\textwidth,height=0.82\textheight,keepaspectratio]{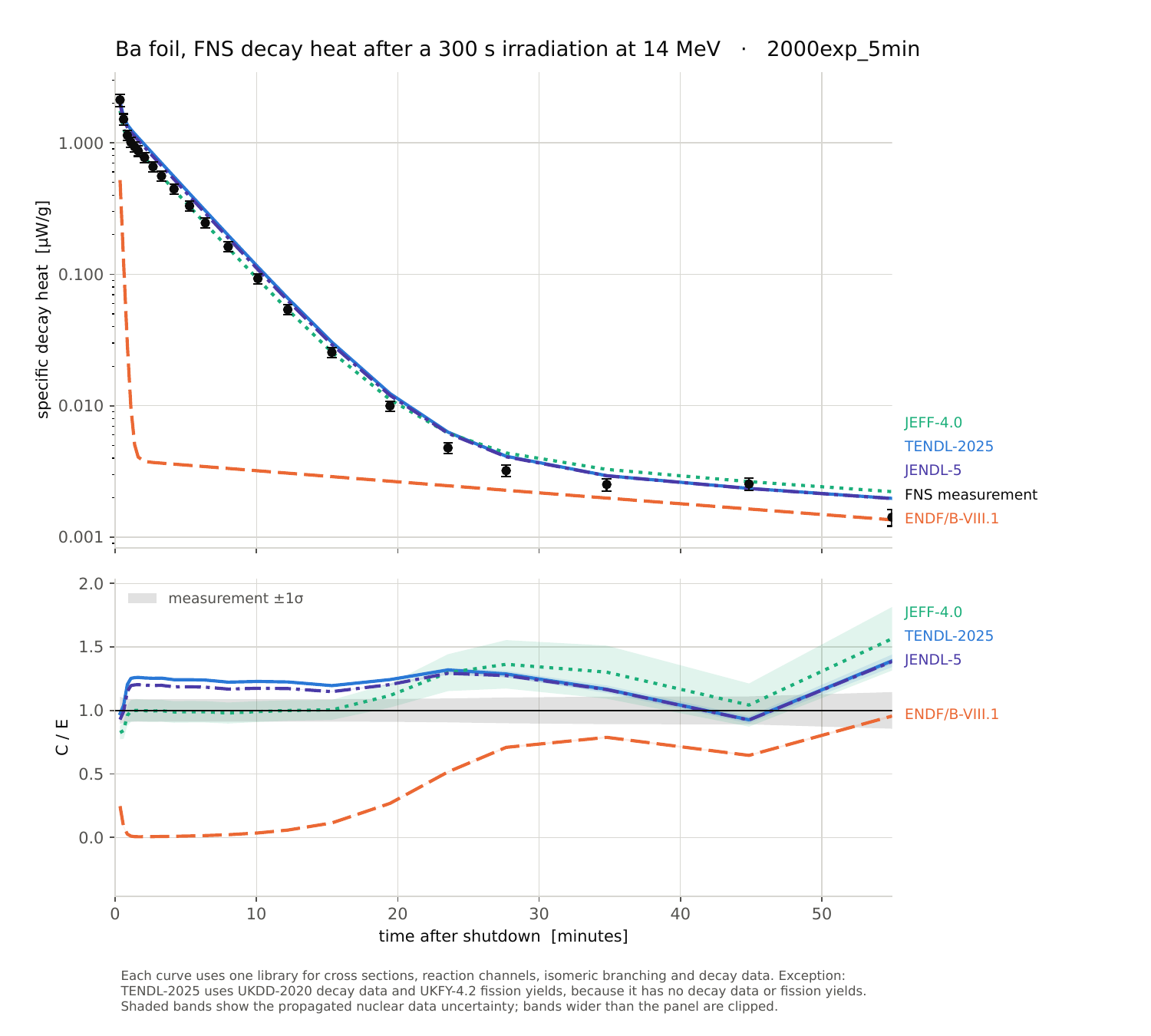}
\caption{Ba: decay heat against the measurement, with C/E beneath. Mean $|C/E-1|$ by library: JEFF-4.0 10\%, JENDL-5 18\%, TENDL-2025 22\%, ENDF/B-VIII.1 79\%.}
\end{figure}
\clearpage

\begin{figure}[H]
\centering
\includegraphics[width=\textwidth,height=0.82\textheight,keepaspectratio]{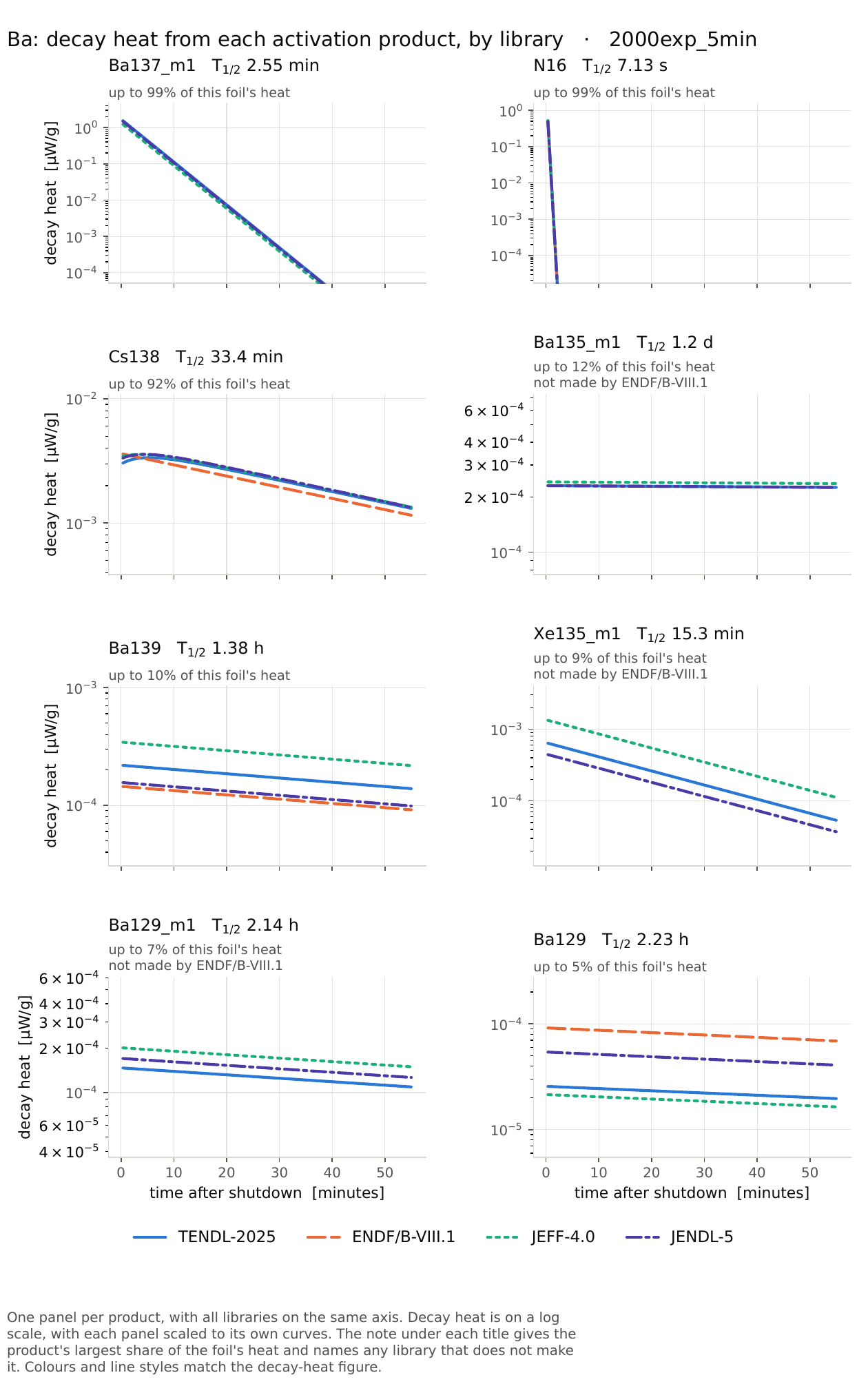}
\caption{Ba: share of the decay heat from each activation product during cooling.}
\end{figure}
\clearpage

\begin{figure}[H]
\centering
\includegraphics[width=\textwidth,height=0.82\textheight,keepaspectratio]{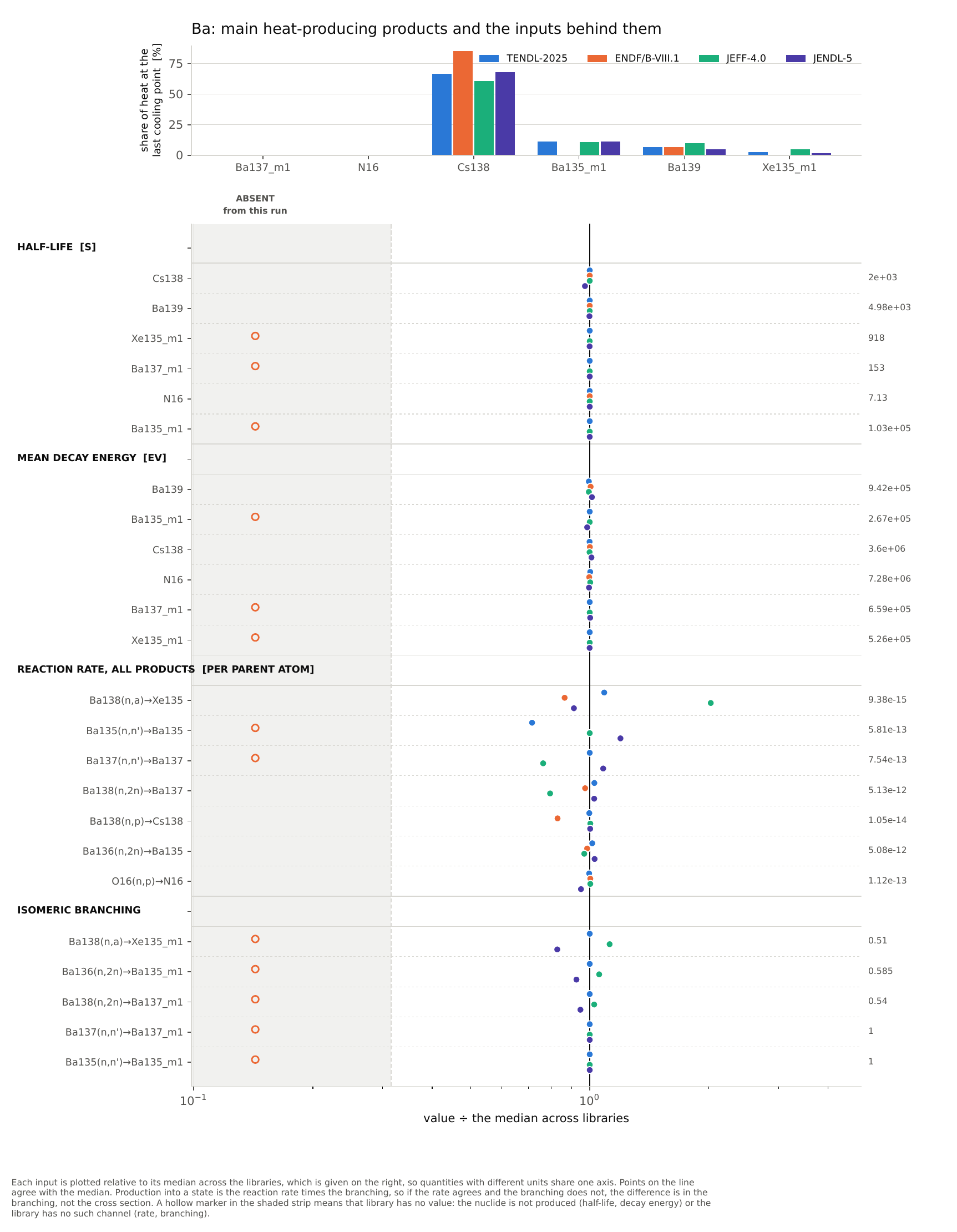}
\caption{Ba: inputs for the main heat-producing products, each relative to its median across the available libraries.}
\end{figure}
\clearpage

\begin{figure}[H]
\centering
\includegraphics[width=\textwidth,height=0.82\textheight,keepaspectratio]{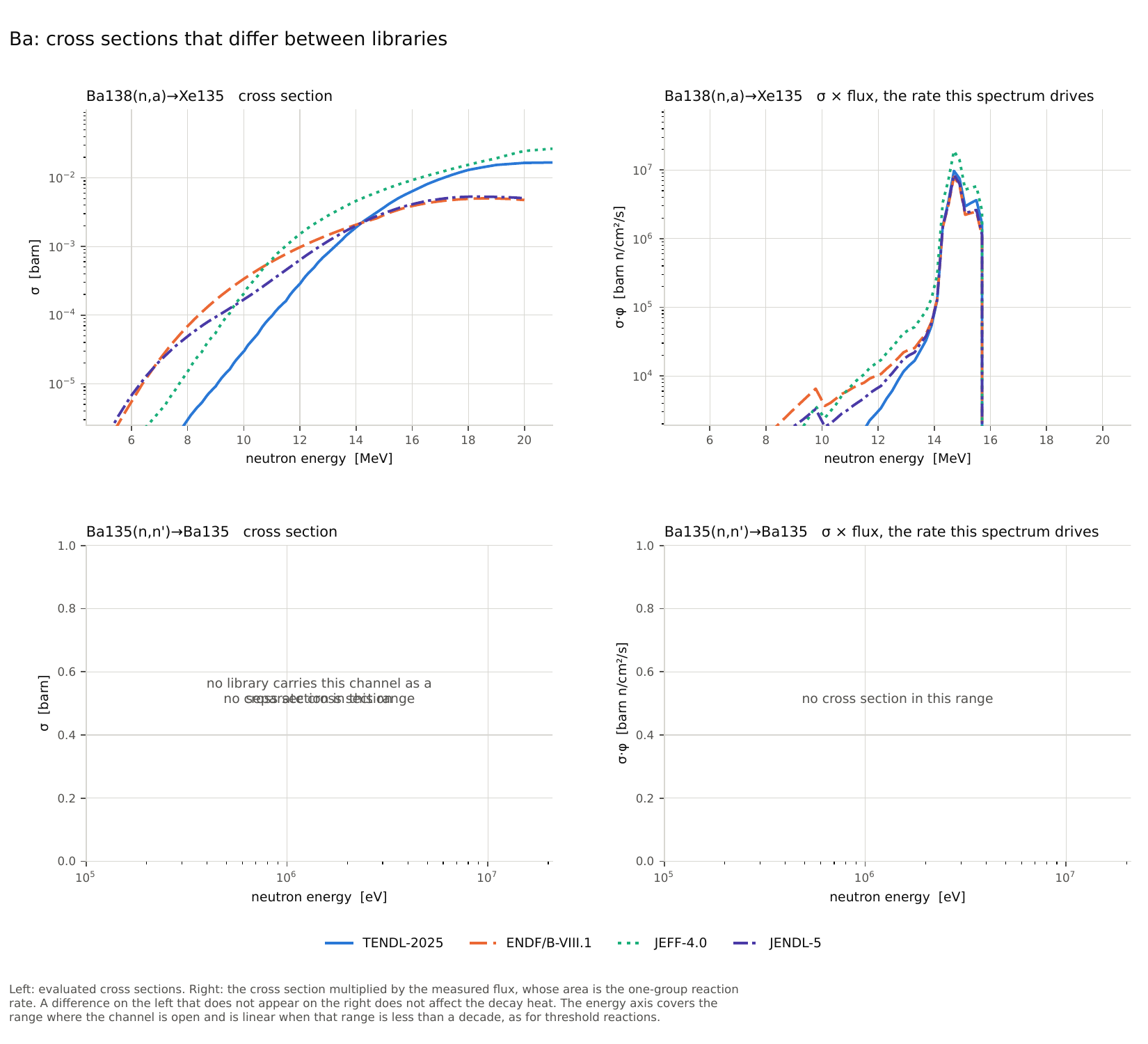}
\caption{Ba: cross sections of the channels whose spectrum-averaged rates differ, with the neutron spectrum and their product.}
\end{figure}
\clearpage

\subsubsection*{1996 seven-hour campaign}
The 1996 campaign irradiated this foil for 7.17 hours at position 7 and measured it out to 24.18 days. In the five-minute calculation Ba135\_m1 (half-life 1.2 days) gives up to 12 percent of the heat, and the longer irradiation and cooling test its production and decay data. TENDL-2025 is closest. This campaign uses a different spectrum, so it is not included in the main rankings.

\begin{figure}[H]
\centering
\includegraphics[width=\textwidth,height=0.82\textheight,keepaspectratio]{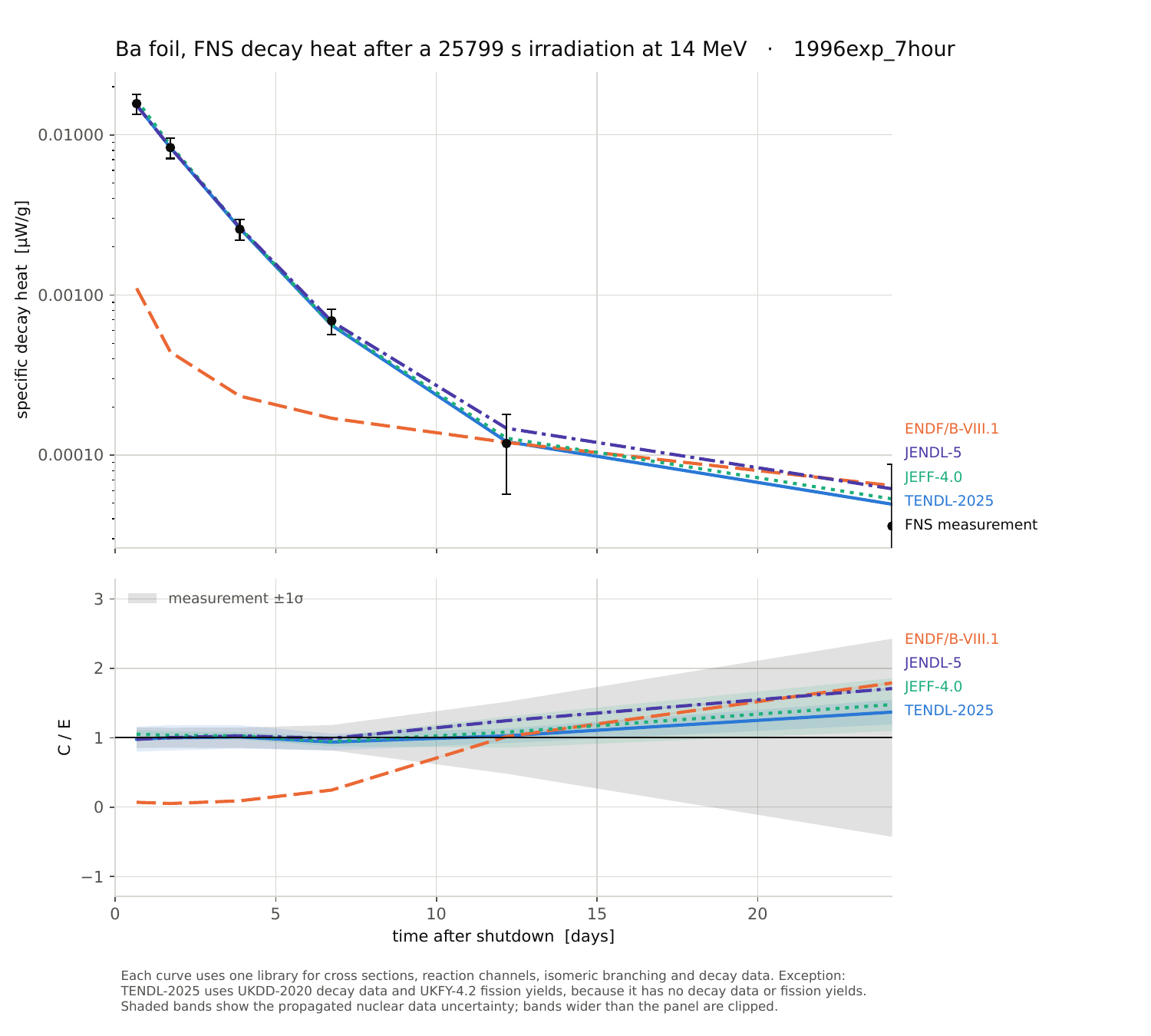}
\caption{Ba: 1996 seven-hour campaign, irradiation position 7. Mean $|C/E-1|$ by library: TENDL-2025 8\%, JEFF-4.0 12\%, JENDL-5 17\%, ENDF/B-VIII.1 72\%. }
\end{figure}
\clearpage

ENDF/B-VIII.1 stands out at 79 percent from the measurement, against 22 percent for the next furthest library and 10 percent for JEFF-4.0. The median measurement uncertainty is 9 percent.

\subsubsection*{Differences in the nuclear data}
\begin{itemize}\setlength{\itemsep}{6pt}
\item JEFF-4.0 gives the total reaction rate of Ba138(n,a)→Xe135 as 1.896e-14, against 8.554e-15 for the median of the others, a factor of 2.\\[2pt]{\itshape This is the total rate of the channel over all its products, so the difference is in the cross section, not in the split between states.}
\item ENDF/B-VIII.1 gives no heat from Ba135\_m1; the other libraries do.\\[2pt]{\itshape The heat breakdown alone does not show why. Possible causes are a missing reaction channel, a different state assignment, different decay data or a reporting threshold.}
\item ENDF/B-VIII.1 gives no heat from Xe135\_m1; the other libraries do.
\end{itemize}
\subsubsection*{Suggested measurements}
\begin{itemize}\setlength{\itemsep}{4pt}
\item Natural Ba has 7 isotopes, so the measured heat combines several parents. Enriched single-isotope targets would distinguish their contributions. The candidate parent is Ba138. Enrichment alone does not isolate a single reaction channel.
\item The sample is a compound that also contains C, O. Their activation adds to the measured heat; a metal or purer sample would remove it.
\item The reaction rates are averaged over the measured spectrum. Activation measurements at several well-defined neutron energies, including below 14 MeV, would show which energy range the differences come from.
\end{itemize}
\clearpage
\subsection{Bi: Bismuth}\label{foil:Bi}
\begin{figure}[H]
\centering
\includegraphics[width=\textwidth,height=0.82\textheight,keepaspectratio]{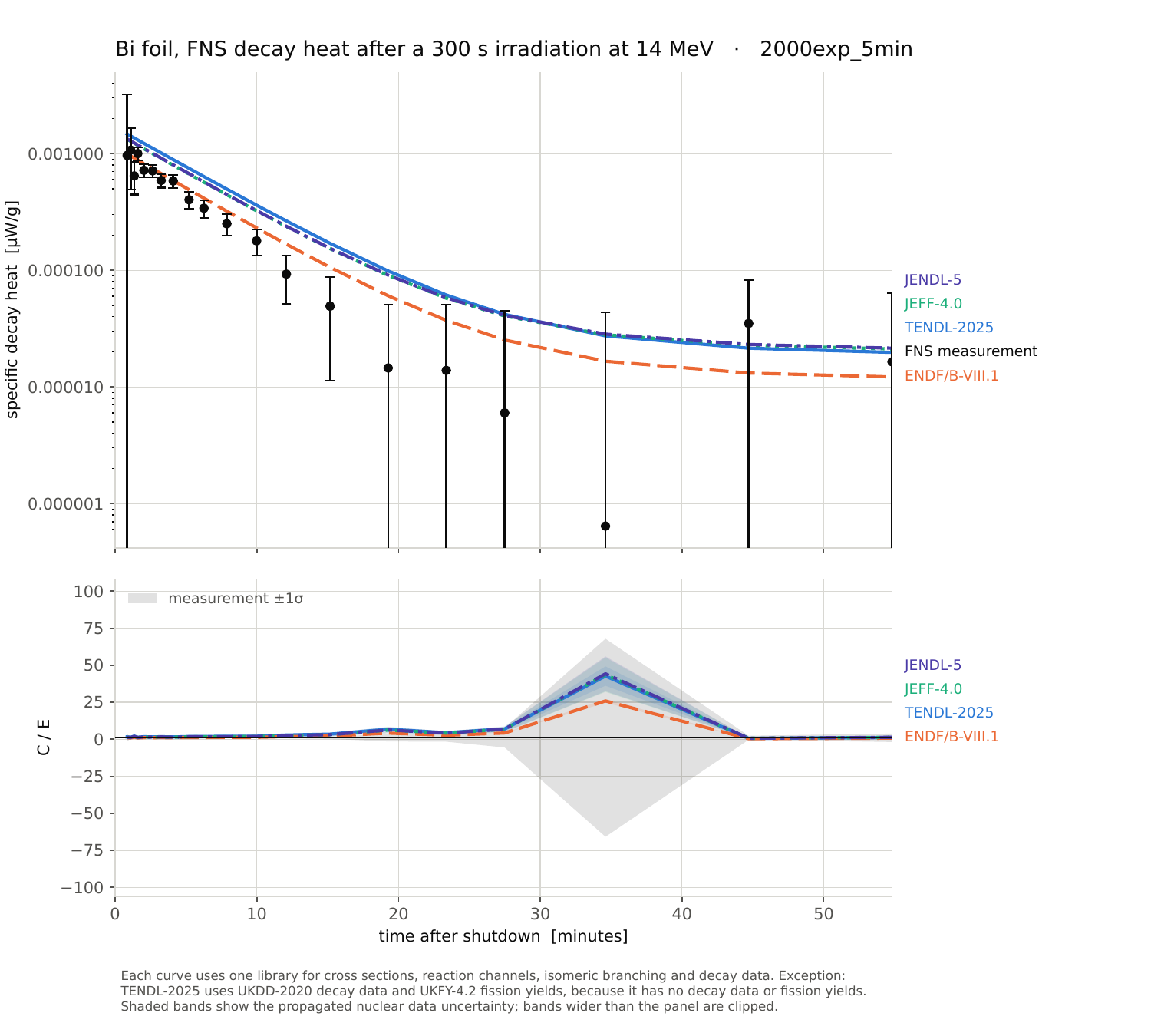}
\caption{Bi: decay heat against the measurement, with C/E beneath. Mean $|C/E-1|$ by library: ENDF/B-VIII.1 189\%, JEFF-4.0 337\%, JENDL-5 342\%, TENDL-2025 351\%.}
\end{figure}
\clearpage

\begin{figure}[H]
\centering
\includegraphics[width=\textwidth,height=0.82\textheight,keepaspectratio]{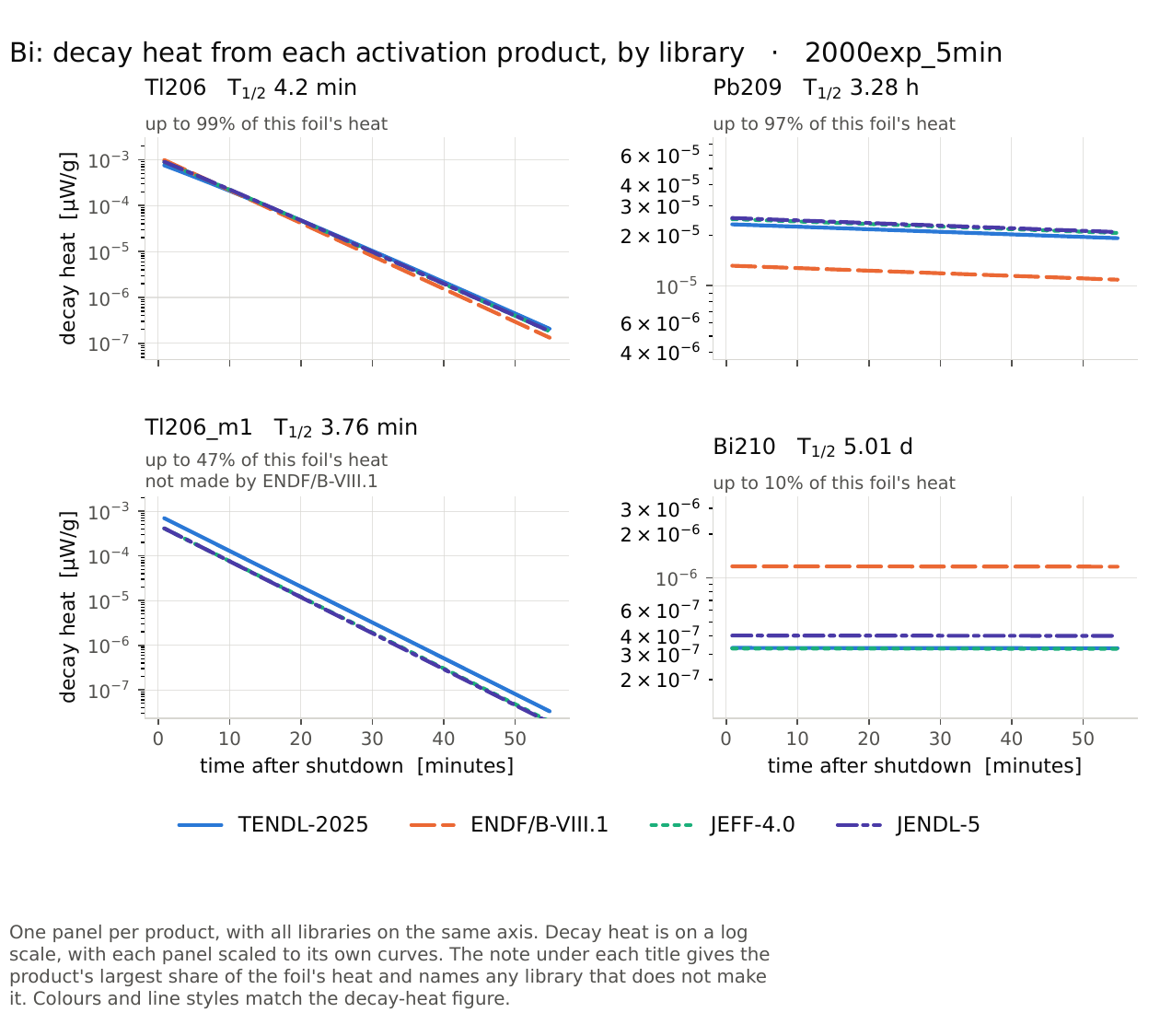}
\caption{Bi: share of the decay heat from each activation product during cooling.}
\end{figure}
\clearpage

\begin{figure}[H]
\centering
\includegraphics[width=\textwidth,height=0.82\textheight,keepaspectratio]{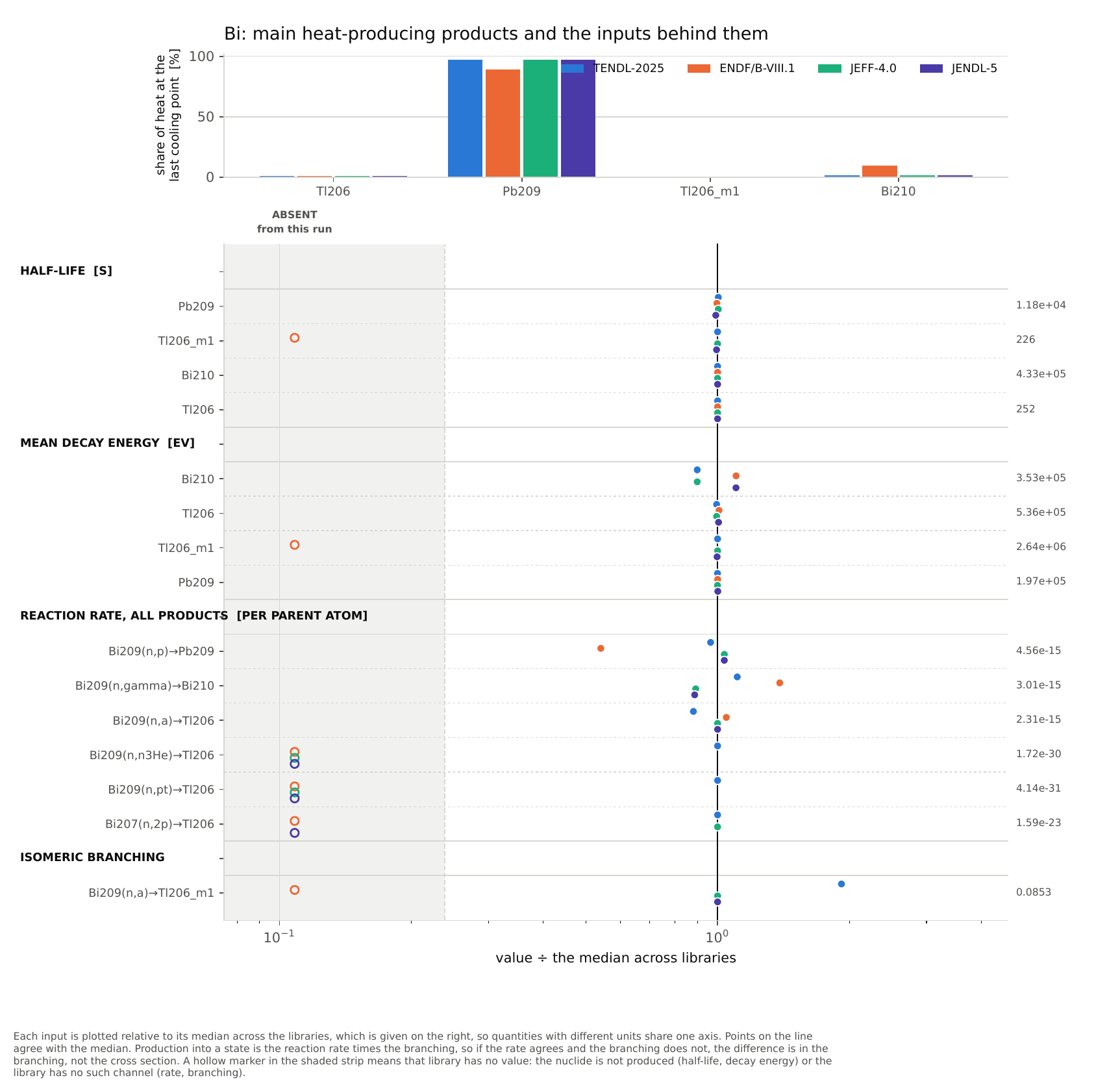}
\caption{Bi: inputs for the main heat-producing products, each relative to its median across the available libraries.}
\end{figure}
\clearpage

\begin{figure}[H]
\centering
\includegraphics[width=\textwidth,height=0.82\textheight,keepaspectratio]{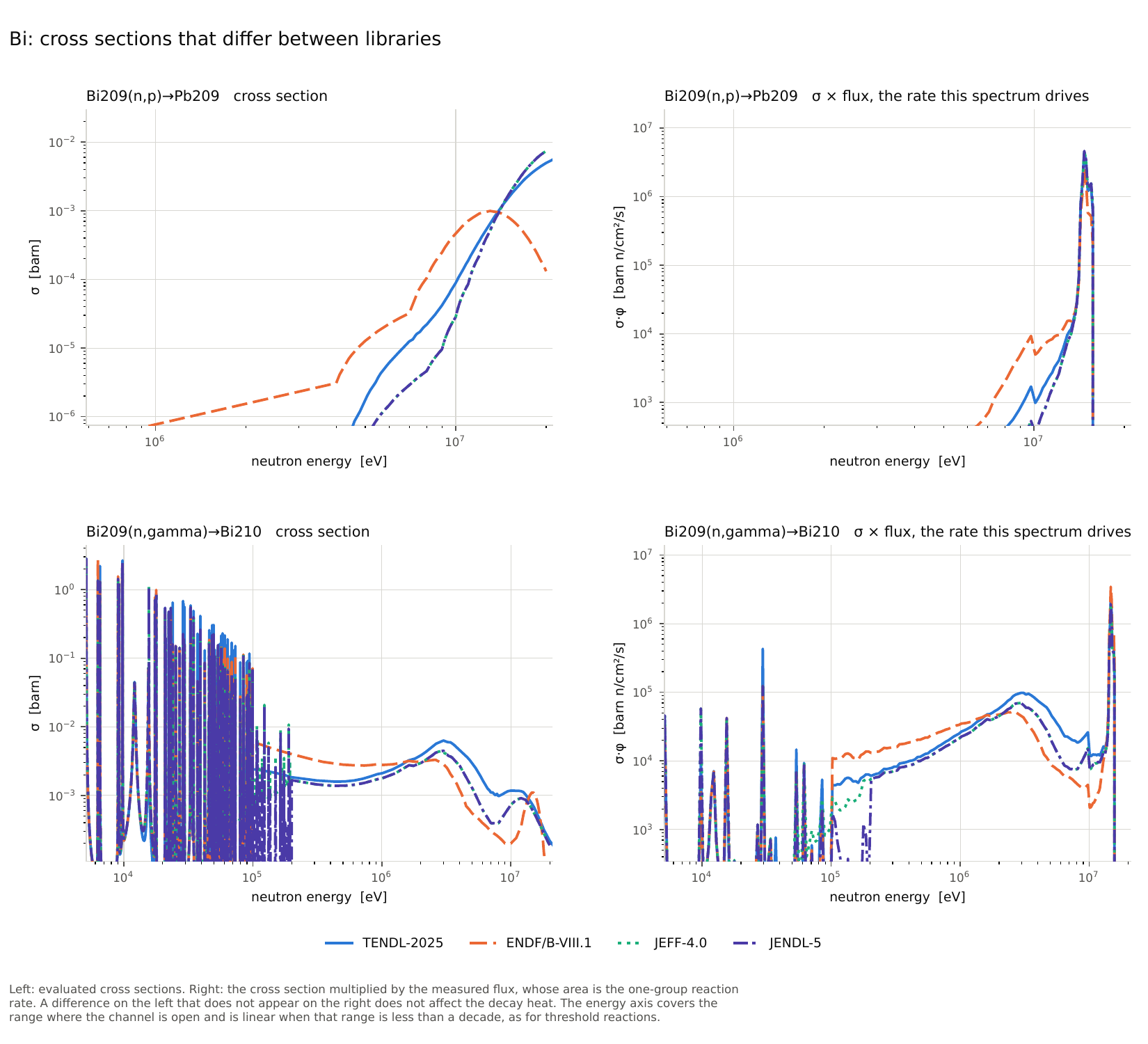}
\caption{Bi: cross sections of the channels whose spectrum-averaged rates differ, with the neutron spectrum and their product.}
\end{figure}
\clearpage

\begin{figure}[H]
\centering
\includegraphics[width=\textwidth,height=0.82\textheight,keepaspectratio]{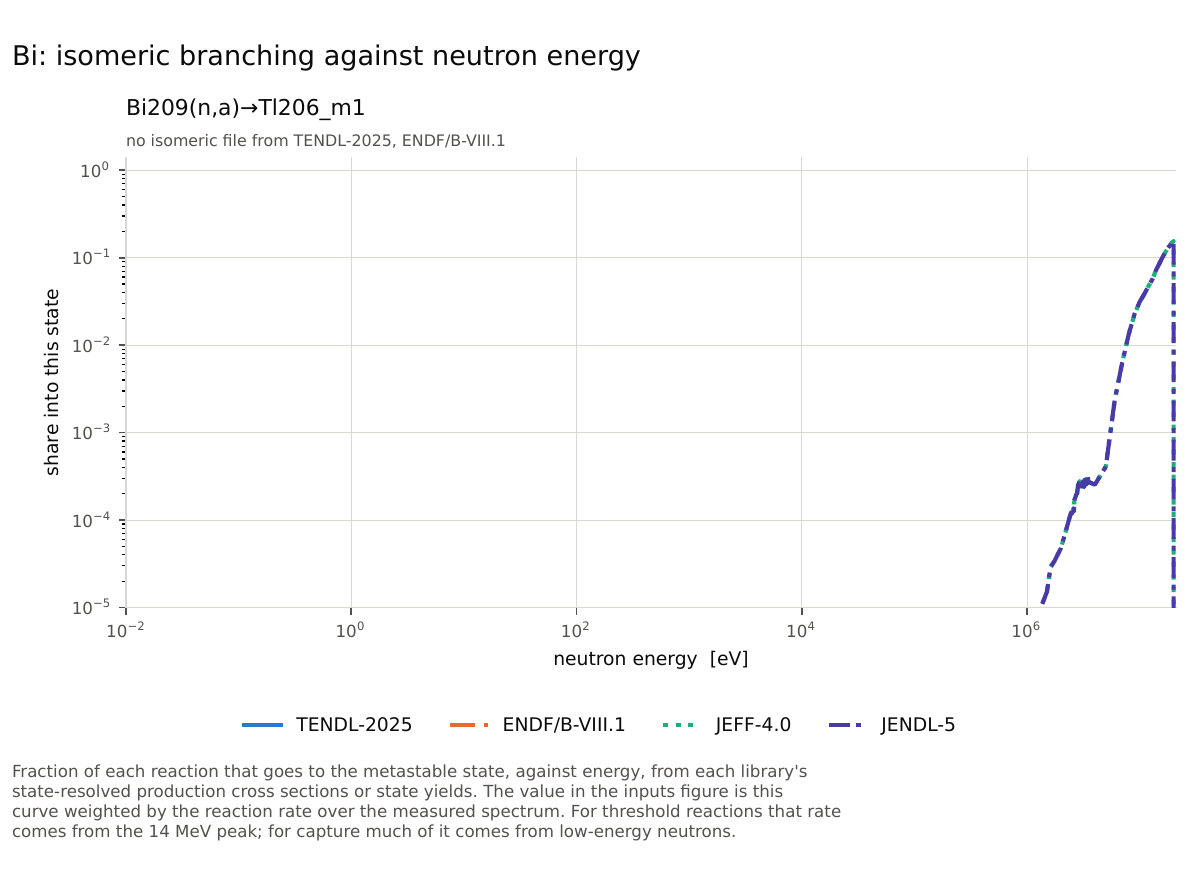}
\caption{Bi: isomeric branching ratio against neutron energy for the channels whose spectrum-averaged ratio differs between libraries.}
\end{figure}
\clearpage

\subsubsection*{1996 seven-hour campaign}
The 1996 campaign irradiated this foil for 7.17 hours at position 7 and measured it out to 12.17 days. In the five-minute calculation Bi210 (half-life 5.01 days) gives up to 10 percent of the heat, and the longer irradiation and cooling test its production and decay data. TENDL-2025 is closest. This campaign uses a different spectrum, so it is not included in the main rankings.

\begin{figure}[H]
\centering
\includegraphics[width=\textwidth,height=0.82\textheight,keepaspectratio]{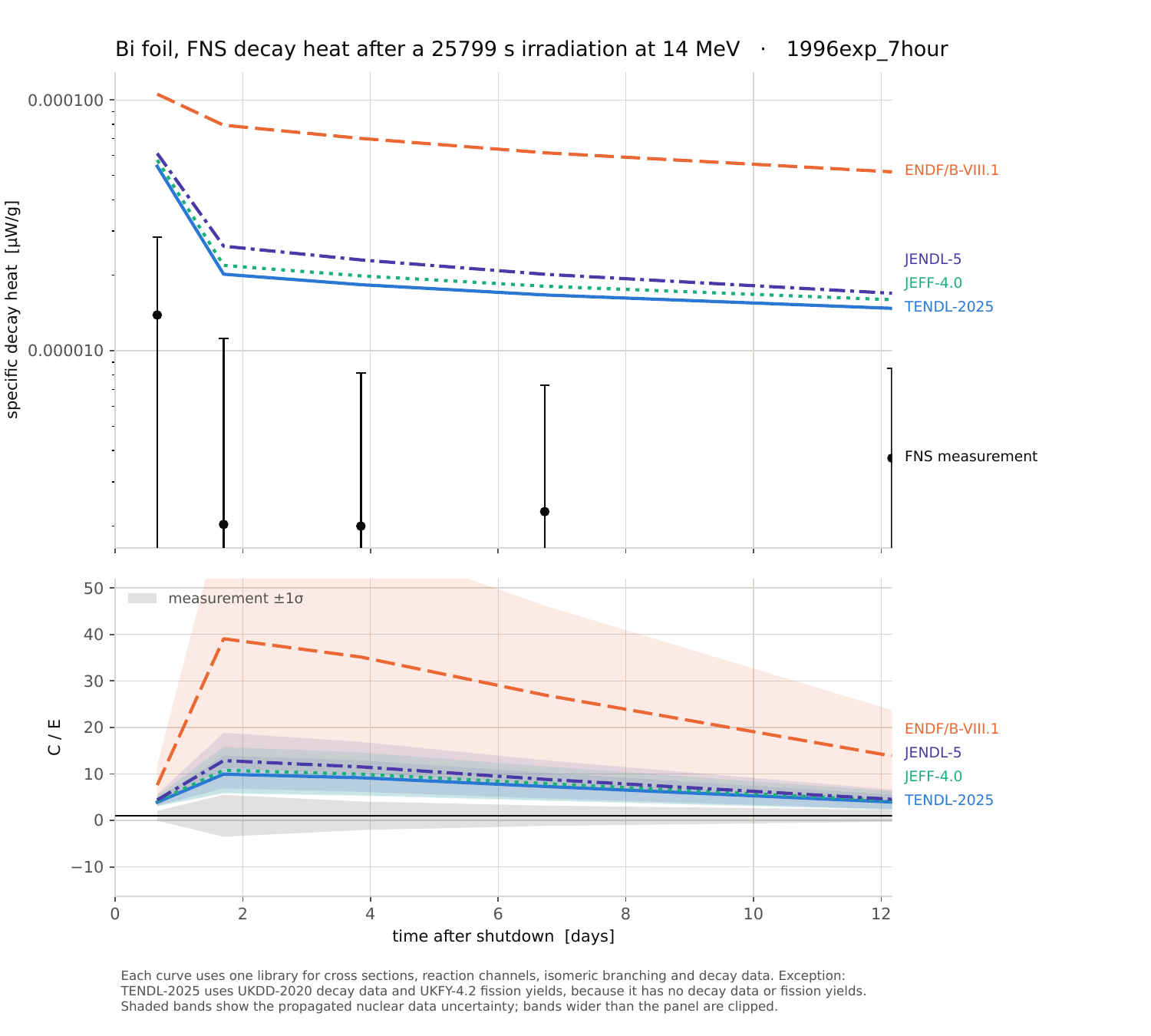}
\caption{Bi: 1996 seven-hour campaign, irradiation position 7. Mean $|C/E-1|$ by library: TENDL-2025 586\%, JEFF-4.0 642\%, JENDL-5 744\%, ENDF/B-VIII.1 2352\%. }
\end{figure}
\clearpage

The libraries are between 189 and 351 percent from the measurement, which has a median uncertainty of 38 percent. ENDF/B-VIII.1 is closest and TENDL-2025 furthest.

\subsubsection*{Differences in the nuclear data}
\begin{itemize}\setlength{\itemsep}{6pt}
\item ENDF/B-VIII.1 gives no heat from Tl206\_m1; the other libraries do.\\[2pt]{\itshape The heat breakdown alone does not show why. Possible causes are a missing reaction channel, a different state assignment, different decay data or a reporting threshold.}
\end{itemize}
\subsubsection*{Suggested measurements}
\begin{itemize}\setlength{\itemsep}{4pt}
\item Bi has a single stable isotope, so the foil is already a single-isotope target and enrichment would not help.
\end{itemize}
\clearpage
\subsection{Br: Bromine}\label{foil:Br}
\begin{figure}[H]
\centering
\includegraphics[width=\textwidth,height=0.82\textheight,keepaspectratio]{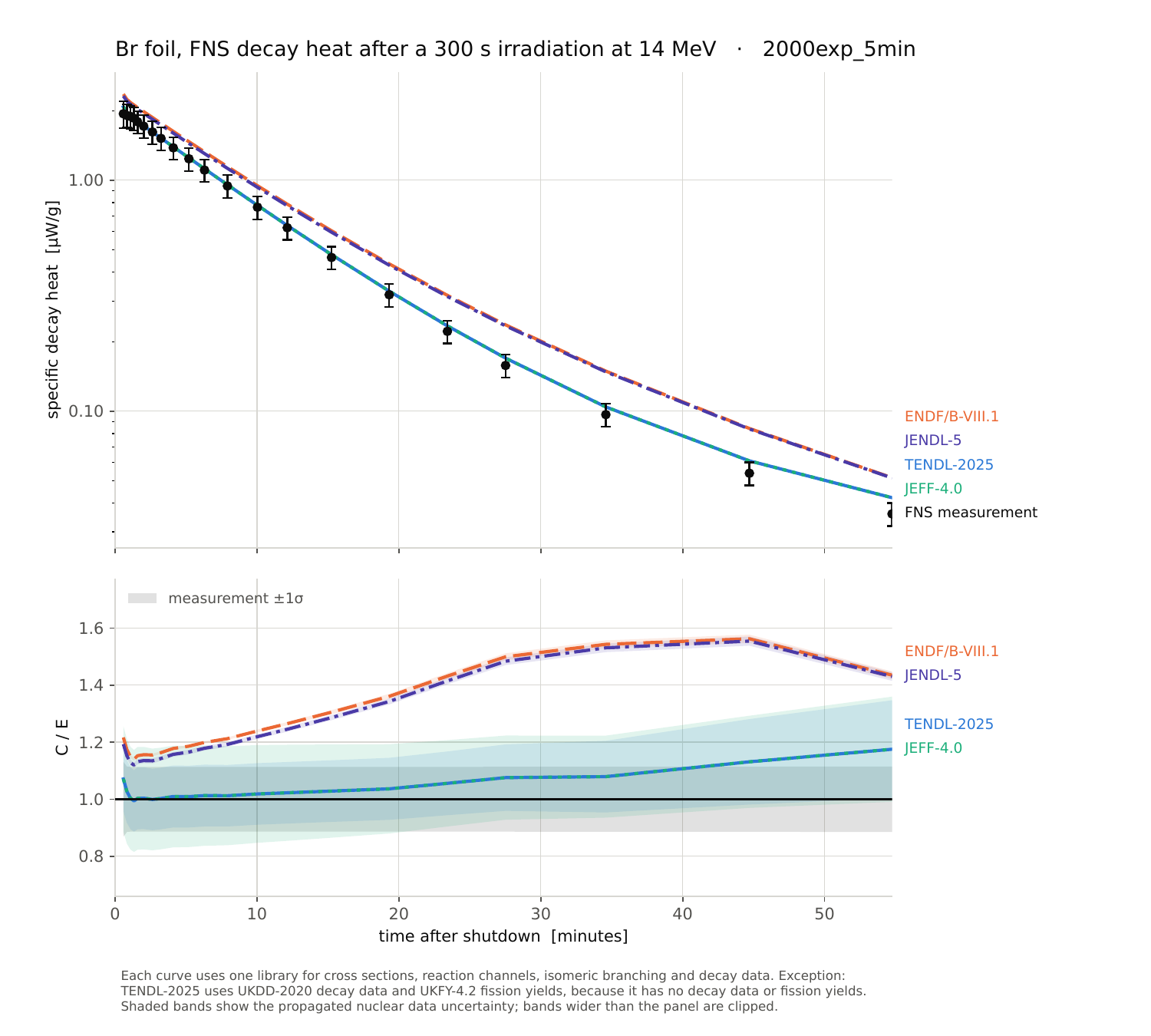}
\caption{Br: decay heat against the measurement, with C/E beneath. Mean $|C/E-1|$ by library: JEFF-4.0 4\%, TENDL-2025 4\%, JENDL-5 25\%, ENDF/B-VIII.1 27\%.}
\end{figure}
\clearpage

\begin{figure}[H]
\centering
\includegraphics[width=\textwidth,height=0.82\textheight,keepaspectratio]{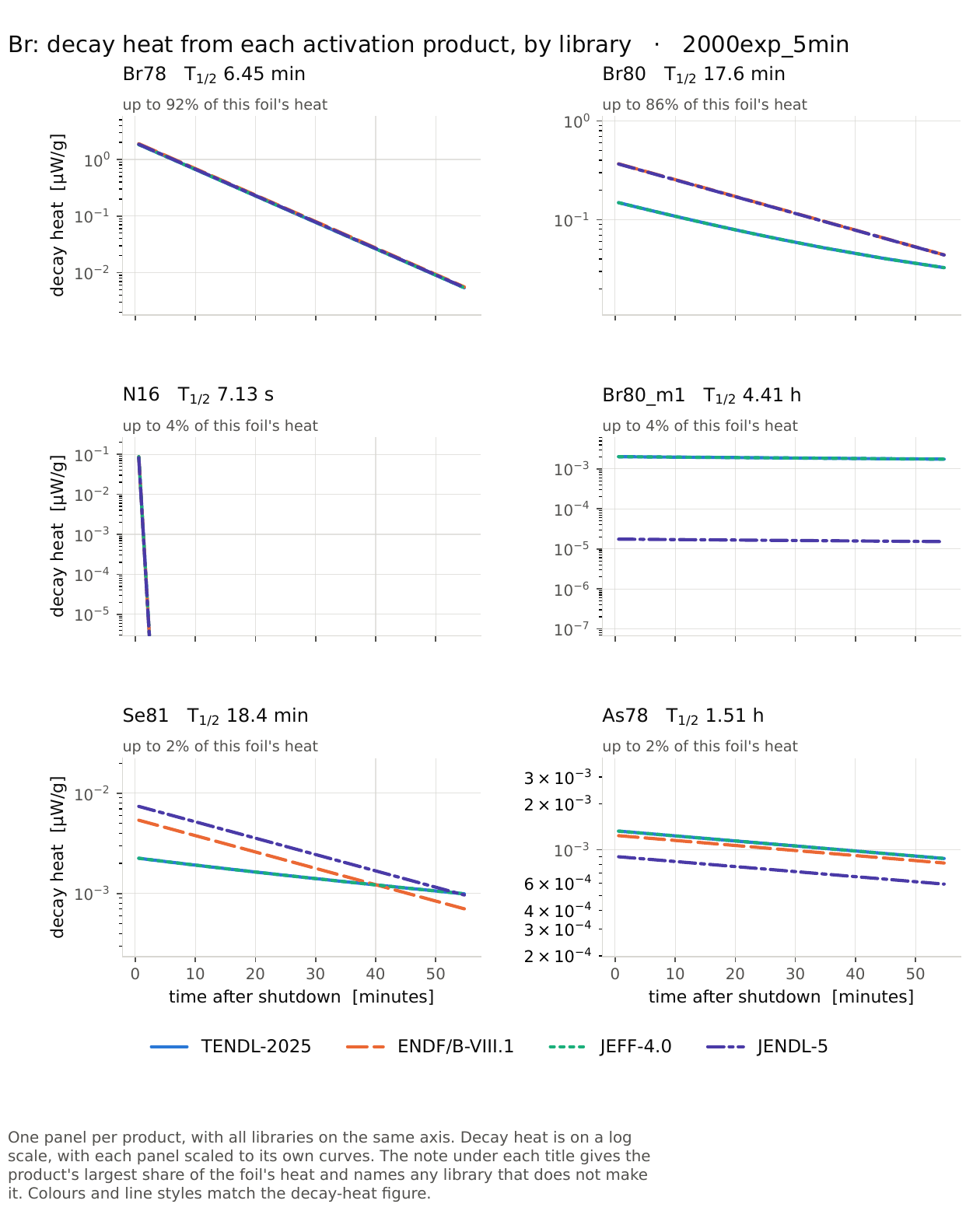}
\caption{Br: share of the decay heat from each activation product during cooling.}
\end{figure}
\clearpage

\begin{figure}[H]
\centering
\includegraphics[width=\textwidth,height=0.82\textheight,keepaspectratio]{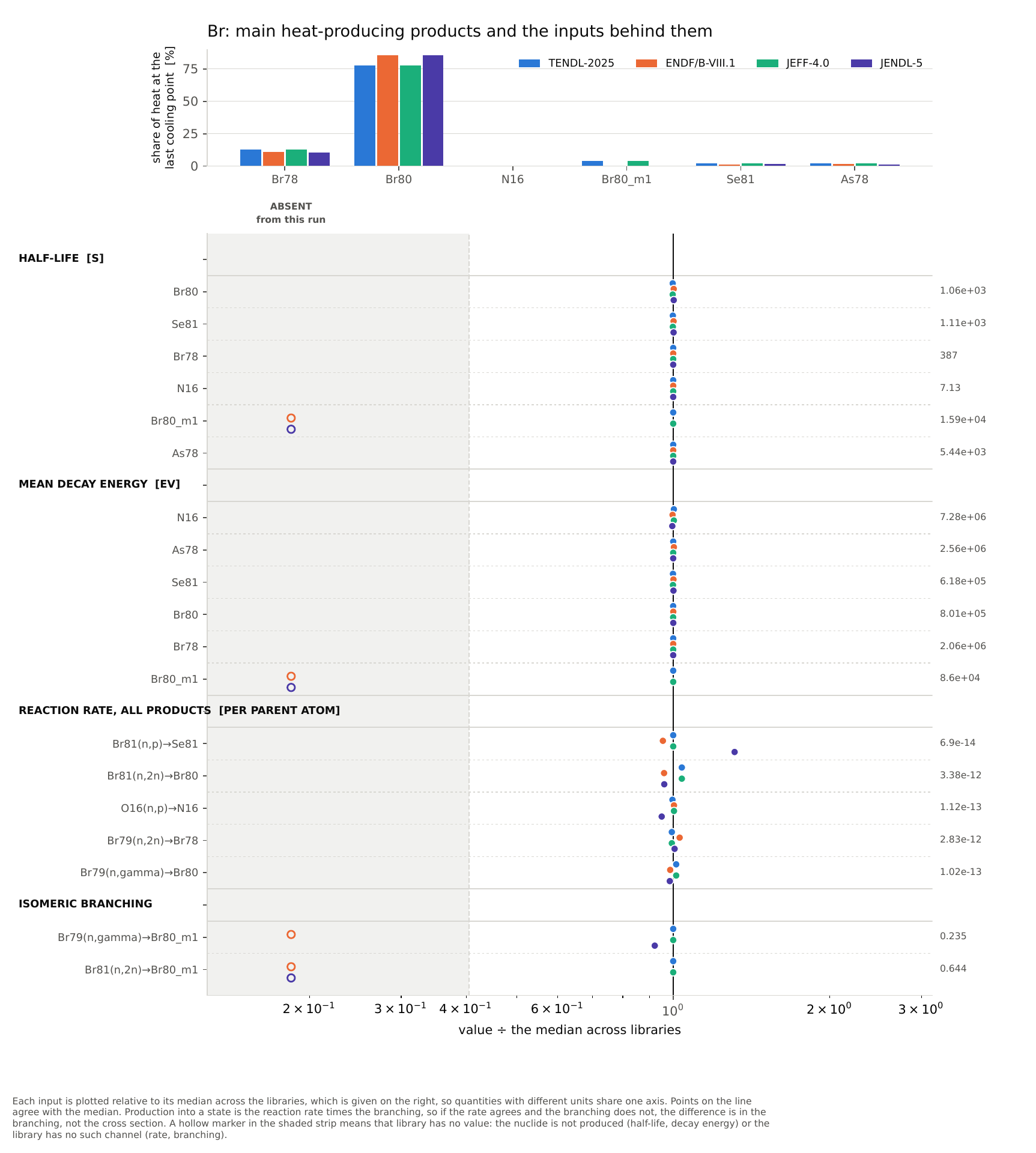}
\caption{Br: inputs for the main heat-producing products, each relative to its median across the available libraries.}
\end{figure}
\clearpage

The libraries are between 4 and 27 percent from the measurement, which has a median uncertainty of 11 percent. TENDL-2025 is closest and ENDF/B-VIII.1 furthest.

\subsubsection*{Suggested measurements}
\begin{itemize}\setlength{\itemsep}{4pt}
\item No input differs enough between the libraries to be listed. An error shared by all of them, or a smaller difference, is still possible.
\end{itemize}
\clearpage
\subsection{Ca: Calcium}\label{foil:Ca}
\begin{figure}[H]
\centering
\includegraphics[width=\textwidth,height=0.82\textheight,keepaspectratio]{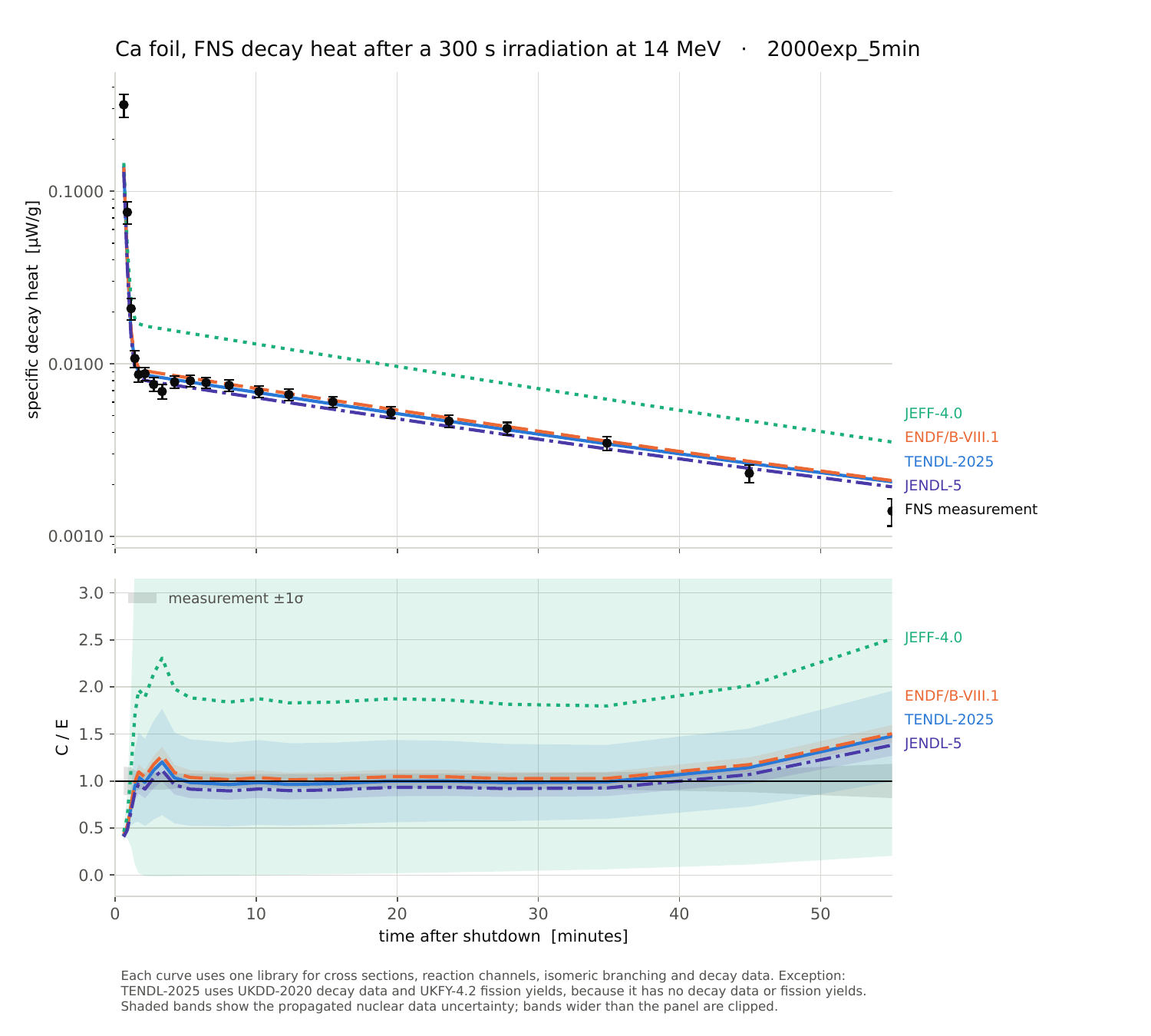}
\caption{Ca: decay heat against the measurement, with C/E beneath. Mean $|C/E-1|$ by library: TENDL-2025 12\%, ENDF/B-VIII.1 14\%, JENDL-5 15\%, JEFF-4.0 86\%.}
\end{figure}
\clearpage

\begin{figure}[H]
\centering
\includegraphics[width=\textwidth,height=0.82\textheight,keepaspectratio]{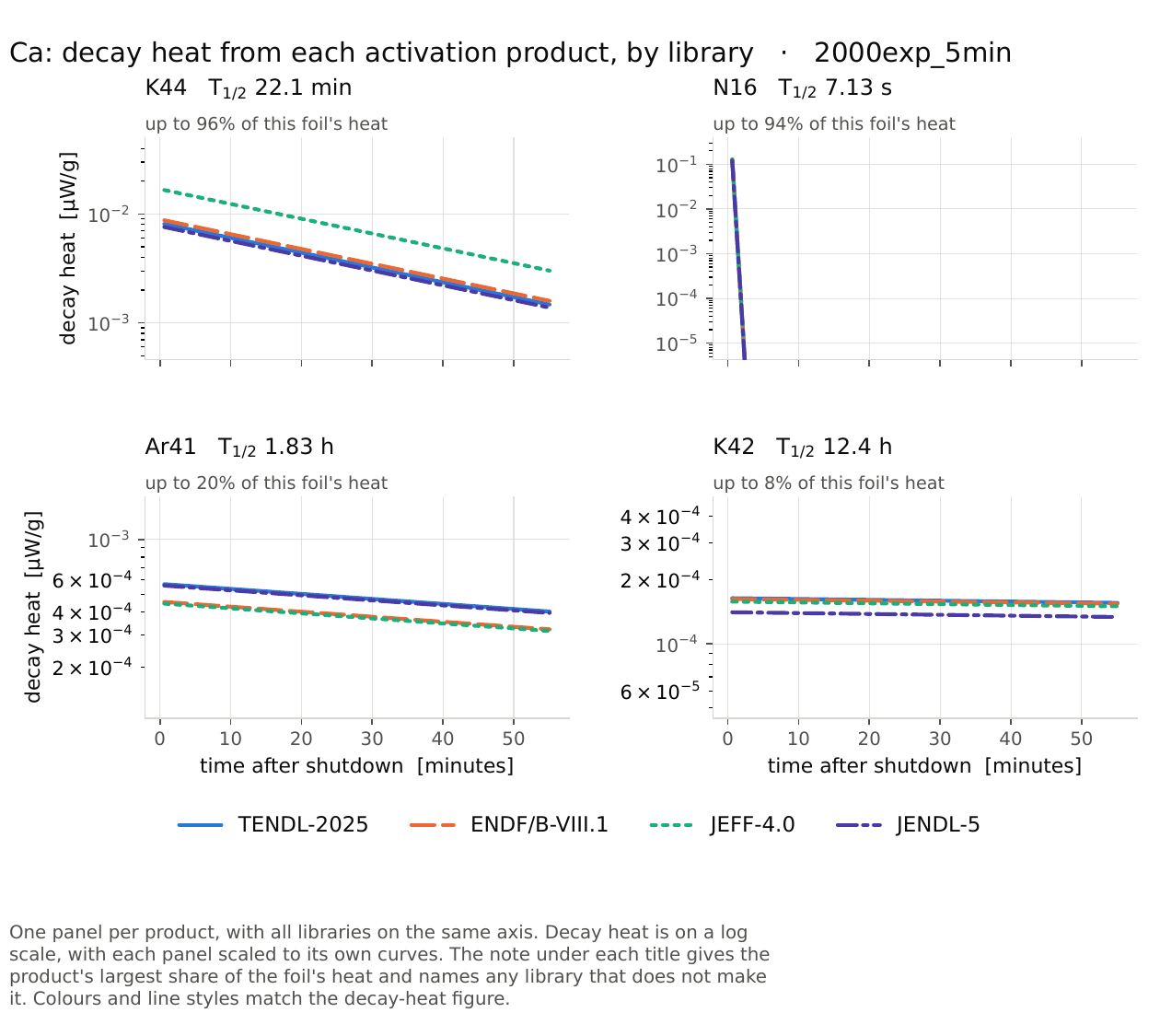}
\caption{Ca: share of the decay heat from each activation product during cooling.}
\end{figure}
\clearpage

\begin{figure}[H]
\centering
\includegraphics[width=\textwidth,height=0.82\textheight,keepaspectratio]{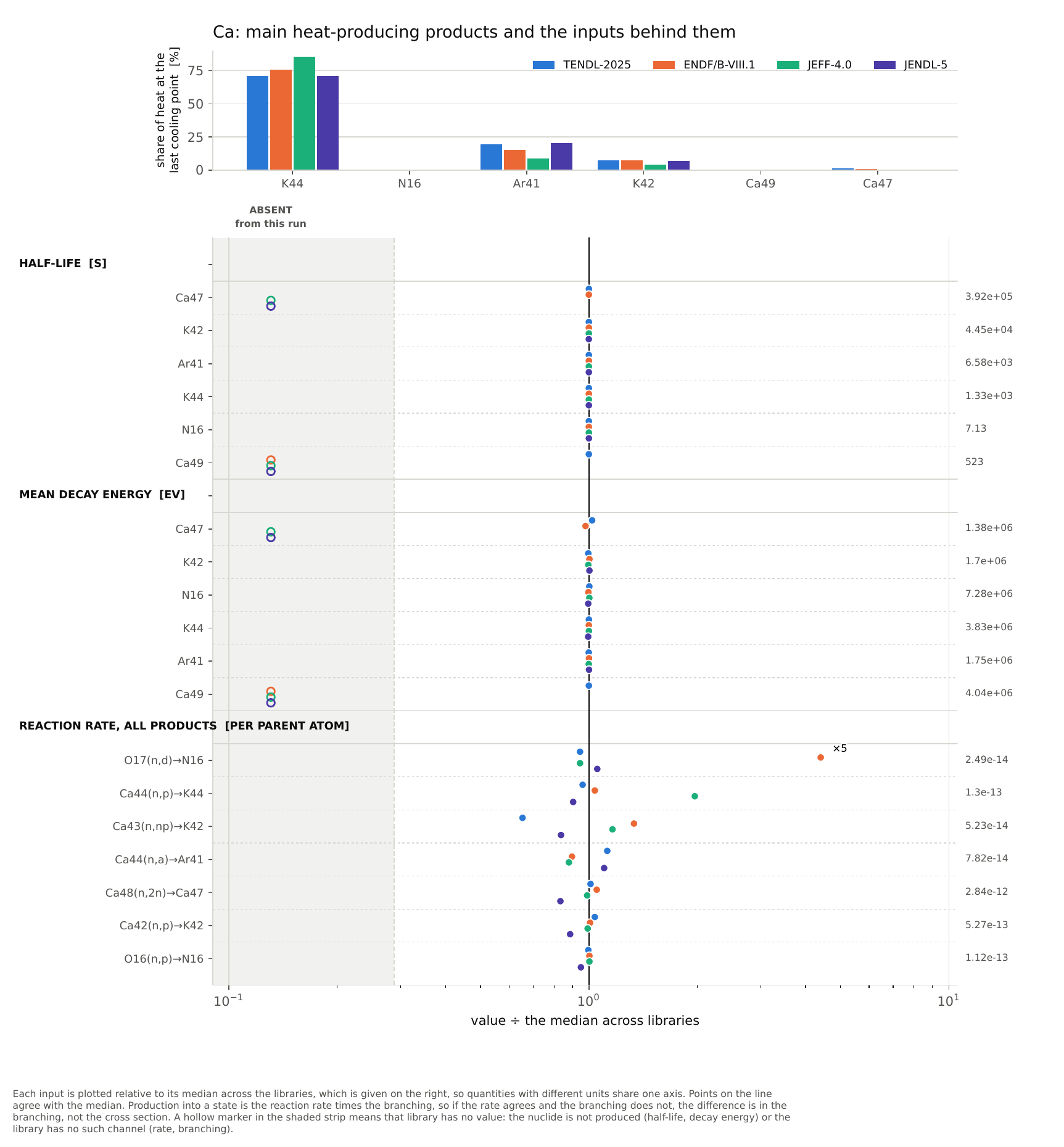}
\caption{Ca: inputs for the main heat-producing products, each relative to its median across the available libraries.}
\end{figure}
\clearpage

\begin{figure}[H]
\centering
\includegraphics[width=\textwidth,height=0.82\textheight,keepaspectratio]{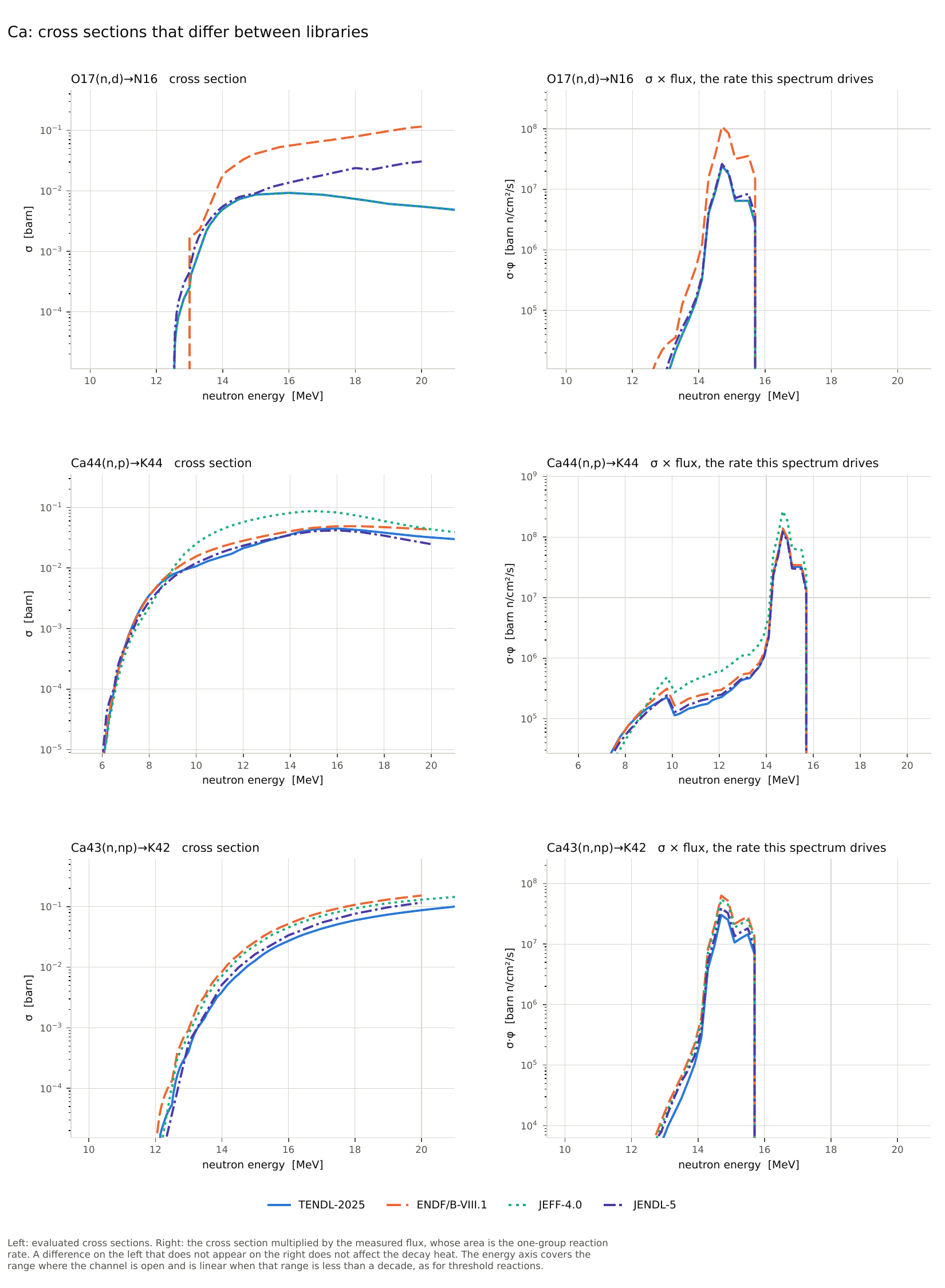}
\caption{Ca: cross sections of the channels whose spectrum-averaged rates differ, with the neutron spectrum and their product.}
\end{figure}
\clearpage

\subsubsection*{1996 seven-hour campaign}
The 1996 campaign irradiated this foil for 7.17 hours at position 7 and measured it out to 24.19 days. In the five-minute calculation Ca47 (half-life 4.54 days) gives up to 1 percent of the heat, and the longer irradiation and cooling test its production and decay data. ENDF/B-VIII.1 is closest. This campaign uses a different spectrum, so it is not included in the main rankings.

\begin{figure}[H]
\centering
\includegraphics[width=\textwidth,height=0.82\textheight,keepaspectratio]{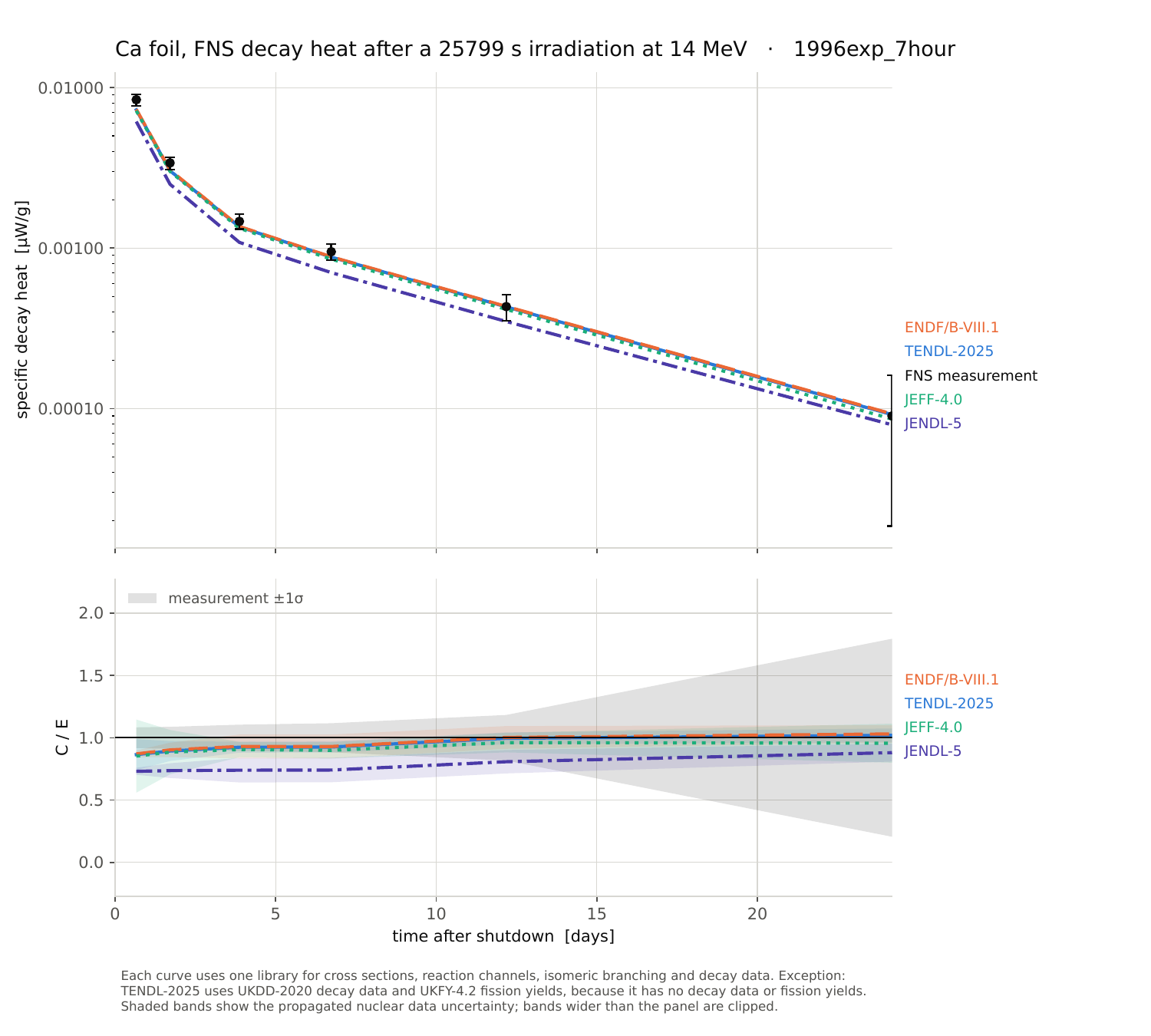}
\caption{Ca: 1996 seven-hour campaign, irradiation position 7. Mean $|C/E-1|$ by library: ENDF/B-VIII.1 7\%, TENDL-2025 7\%, JEFF-4.0 9\%, JENDL-5 23\%. }
\end{figure}
\clearpage

JEFF-4.0 stands out at 86 percent from the measurement, against 15 percent for the next furthest library and 12 percent for TENDL-2025. The median measurement uncertainty is 9 percent.

\subsubsection*{Differences in the nuclear data}
\begin{itemize}\setlength{\itemsep}{6pt}
\item ENDF/B-VIII.1 gives the total reaction rate of O17(n,d)→N16 as 1.096e-13, against 2.351e-14 for the median of the others, a factor of 5.\\[2pt]{\itshape This is the total rate of the channel over all its products, so the difference is in the cross section, not in the split between states.}
\item JEFF-4.0 gives the total reaction rate of Ca44(n,p)→K44 as 2.551e-13, against 1.245e-13 for the median of the others, a factor of 2.
\end{itemize}
\subsubsection*{Suggested measurements}
\begin{itemize}\setlength{\itemsep}{4pt}
\item Natural Ca has 6 isotopes, so the measured heat combines several parents. Enriched single-isotope targets would distinguish their contributions. Candidate parents are Ca44, O17. Enrichment alone does not isolate a single reaction channel.
\item The sample is a compound that also contains O. Their activation adds to the measured heat; a metal or purer sample would remove it.
\item The reaction rates are averaged over the measured spectrum. Activation measurements at several well-defined neutron energies, including below 14 MeV, would show which energy range the differences come from.
\end{itemize}
\clearpage
\subsection{Cd: Cadmium}\label{foil:Cd}
\begin{figure}[H]
\centering
\includegraphics[width=\textwidth,height=0.82\textheight,keepaspectratio]{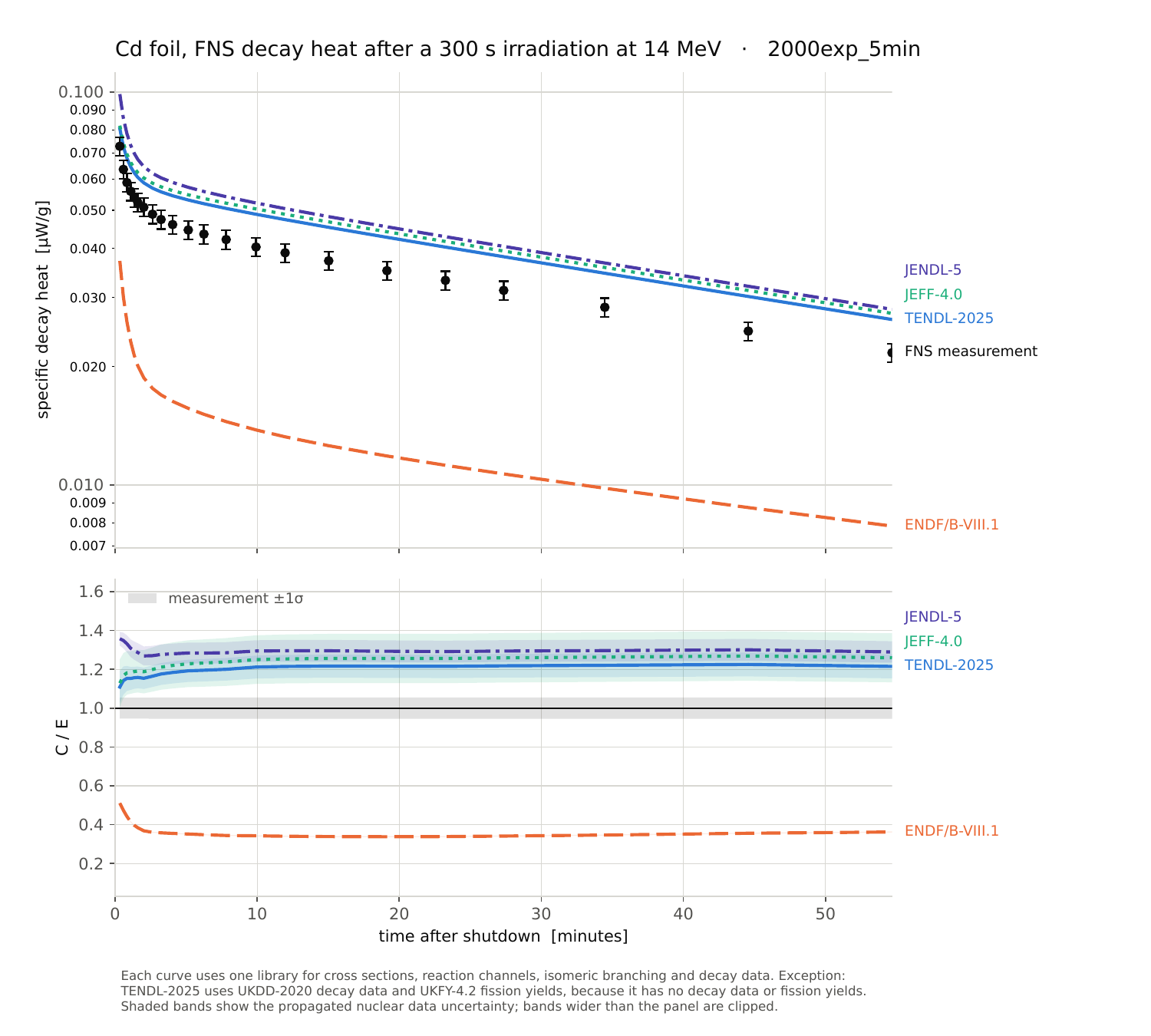}
\caption{Cd: decay heat against the measurement, with C/E beneath. Mean $|C/E-1|$ by library: TENDL-2025 19\%, JEFF-4.0 22\%, JENDL-5 30\%, ENDF/B-VIII.1 63\%.}
\end{figure}
\clearpage

\begin{figure}[H]
\centering
\includegraphics[width=\textwidth,height=0.82\textheight,keepaspectratio]{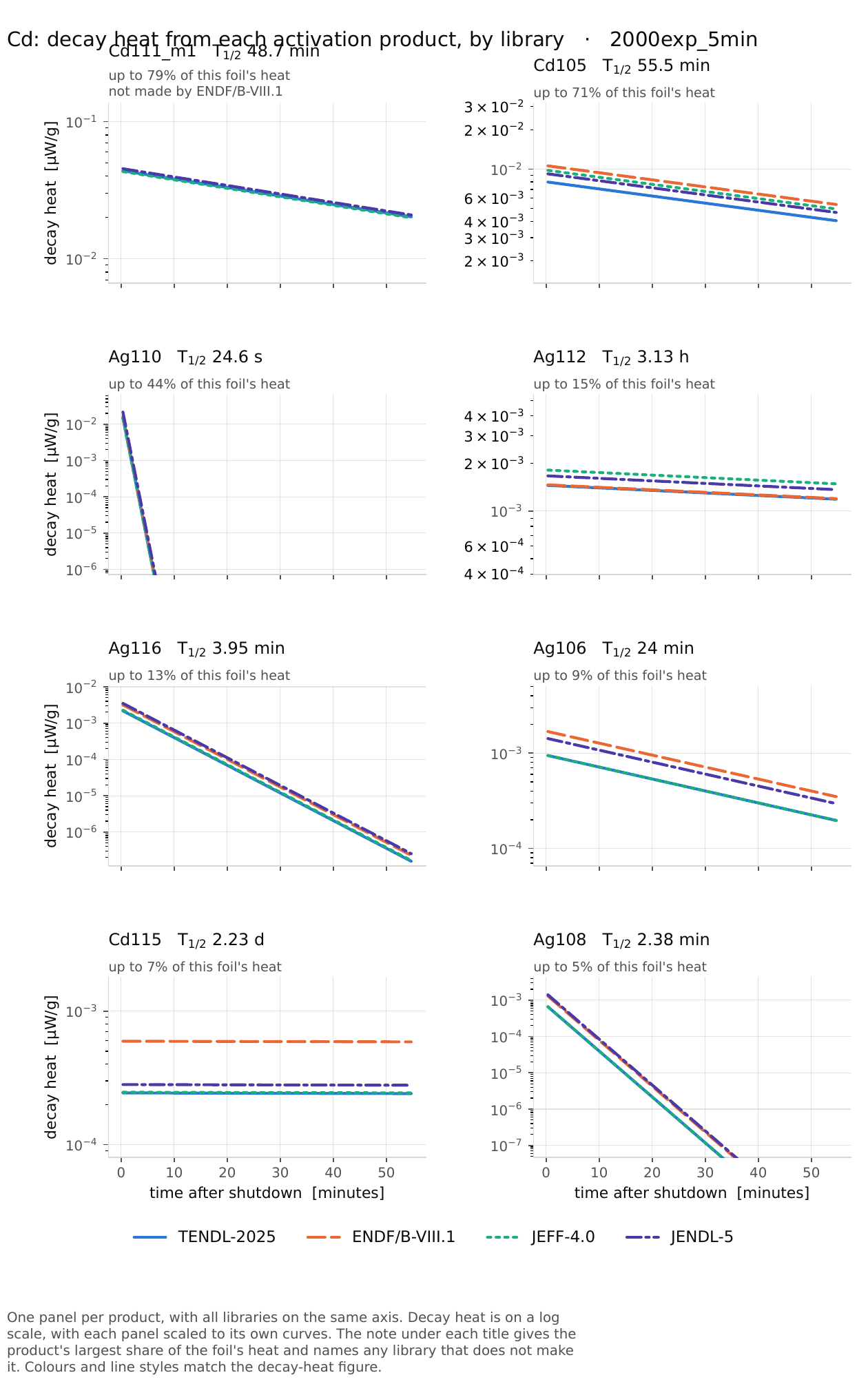}
\caption{Cd: share of the decay heat from each activation product during cooling.}
\end{figure}
\clearpage

\begin{figure}[H]
\centering
\includegraphics[width=\textwidth,height=0.82\textheight,keepaspectratio]{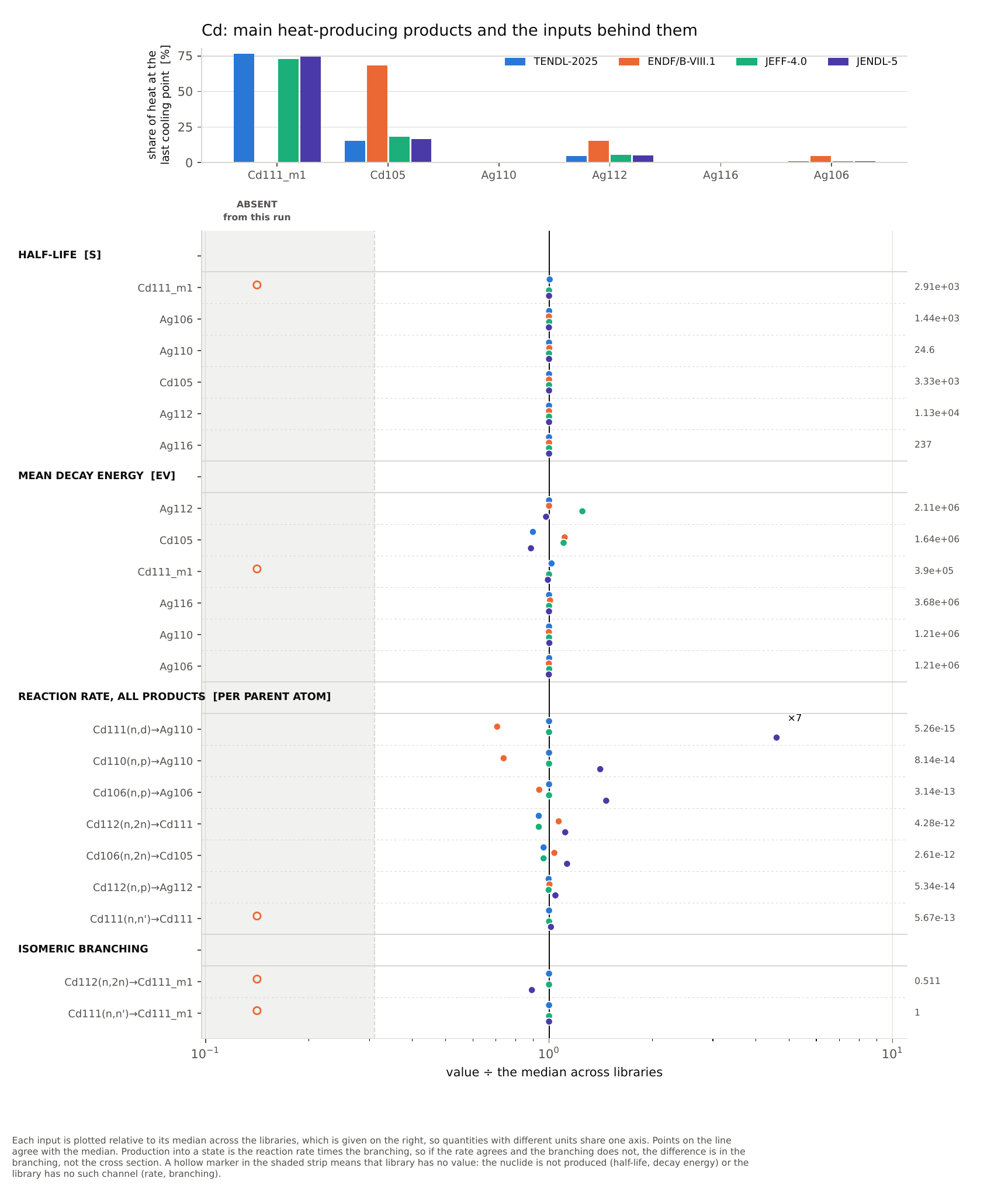}
\caption{Cd: inputs for the main heat-producing products, each relative to its median across the available libraries.}
\end{figure}
\clearpage

\begin{figure}[H]
\centering
\includegraphics[width=\textwidth,height=0.82\textheight,keepaspectratio]{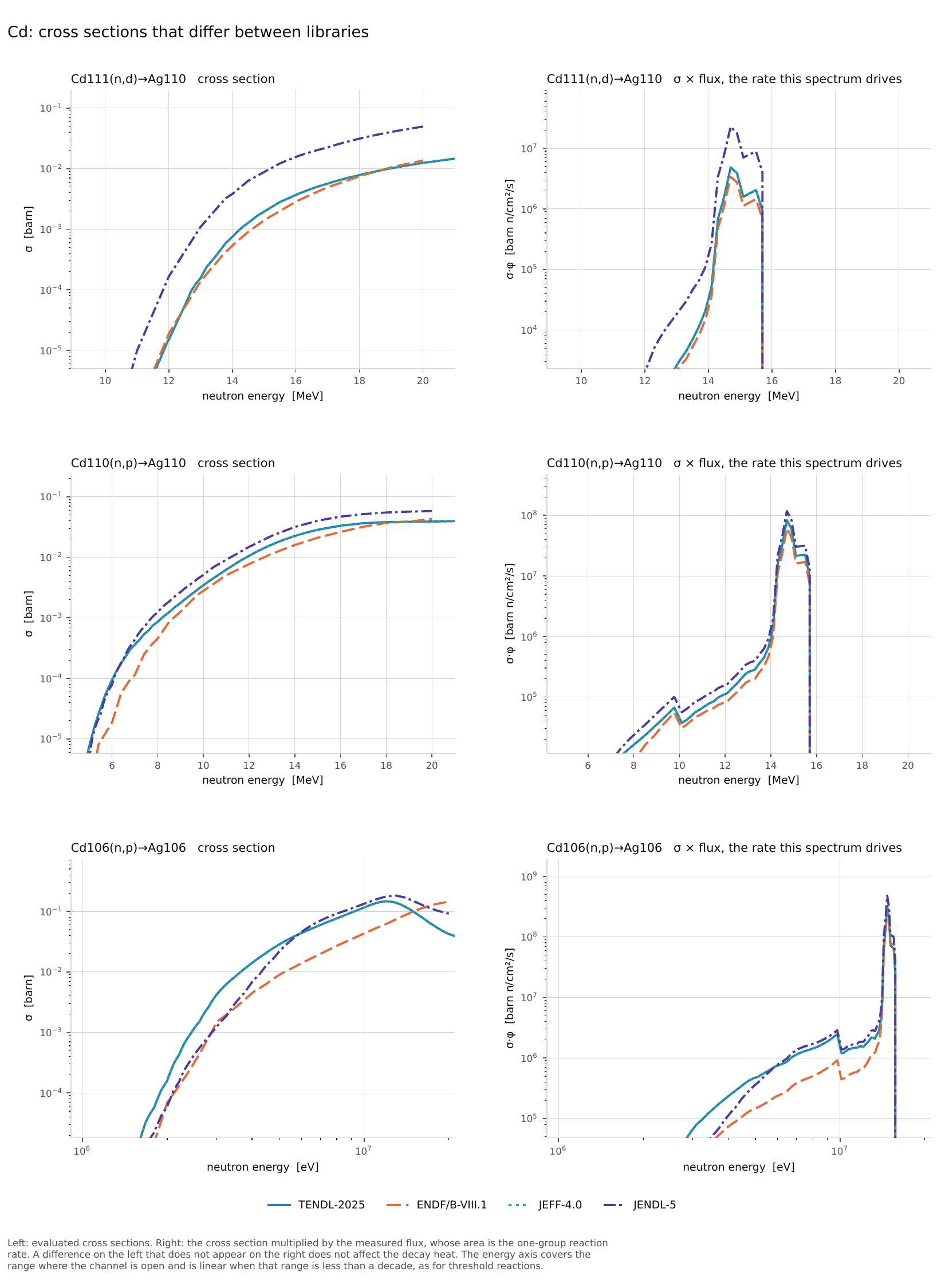}
\caption{Cd: cross sections of the channels whose spectrum-averaged rates differ, with the neutron spectrum and their product.}
\end{figure}
\clearpage

The libraries are between 19 and 63 percent from the measurement, which has a median uncertainty of 5 percent. TENDL-2025 is closest and ENDF/B-VIII.1 furthest.

\subsubsection*{Differences in the nuclear data}
\begin{itemize}\setlength{\itemsep}{6pt}
\item JENDL-5 gives the total reaction rate of Cd111(n,d)→Ag110 as 2.416e-14, against 5.258e-15 for the median of the others, a factor of 5.\\[2pt]{\itshape This is the total rate of the channel over all its products, so the difference is in the cross section, not in the split between states.}
\item ENDF/B-VIII.1 gives no heat from Cd111\_m1; the other libraries do.\\[2pt]{\itshape The heat breakdown alone does not show why. Possible causes are a missing reaction channel, a different state assignment, different decay data or a reporting threshold.}
\end{itemize}
\subsubsection*{Suggested measurements}
\begin{itemize}\setlength{\itemsep}{4pt}
\item Natural Cd has 8 isotopes, so the measured heat combines several parents. Enriched single-isotope targets would distinguish their contributions. The candidate parent is Cd111. Enrichment alone does not isolate a single reaction channel.
\item This foil makes Cd115 (half-life 2.23 days), which gives up to 7 percent of the calculated heat. A five-minute irradiation and 55 minute measurement only see the start of this contribution. There is no seven-hour FNS measurement of this foil. A longer irradiation followed for days to months would be needed to test this product's production and decay.
\item The reaction rates are averaged over the measured spectrum. Activation measurements at several well-defined neutron energies, including below 14 MeV, would show which energy range the differences come from.
\end{itemize}
\clearpage
\subsection{Ce: Cerium}\label{foil:Ce}
\begin{figure}[H]
\centering
\includegraphics[width=\textwidth,height=0.82\textheight,keepaspectratio]{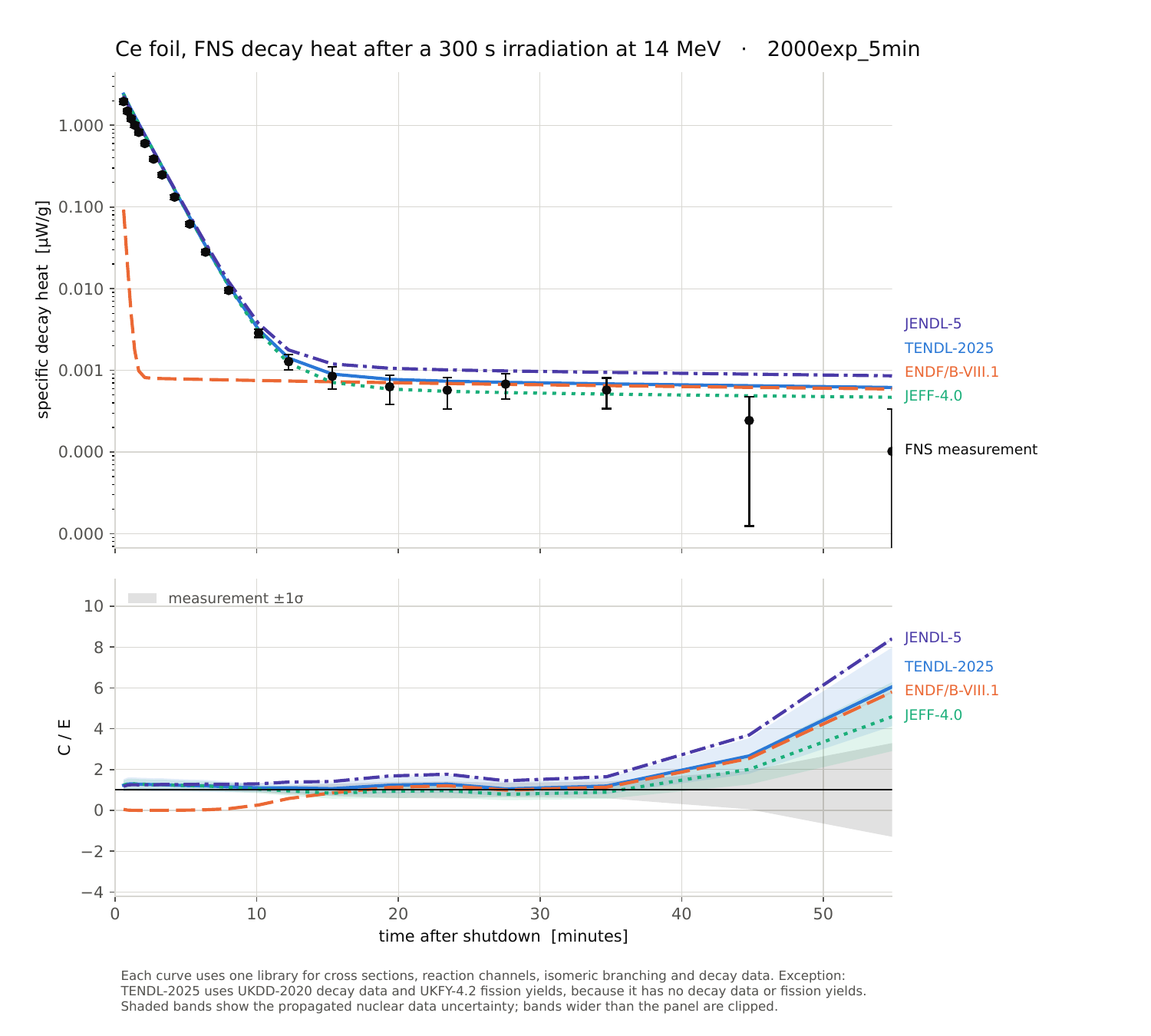}
\caption{Ce: decay heat against the measurement, with C/E beneath. Mean $|C/E-1|$ by library: JEFF-4.0 39\%, TENDL-2025 51\%, JENDL-5 80\%, ENDF/B-VIII.1 95\%.}
\end{figure}
\clearpage

\begin{figure}[H]
\centering
\includegraphics[width=\textwidth,height=0.82\textheight,keepaspectratio]{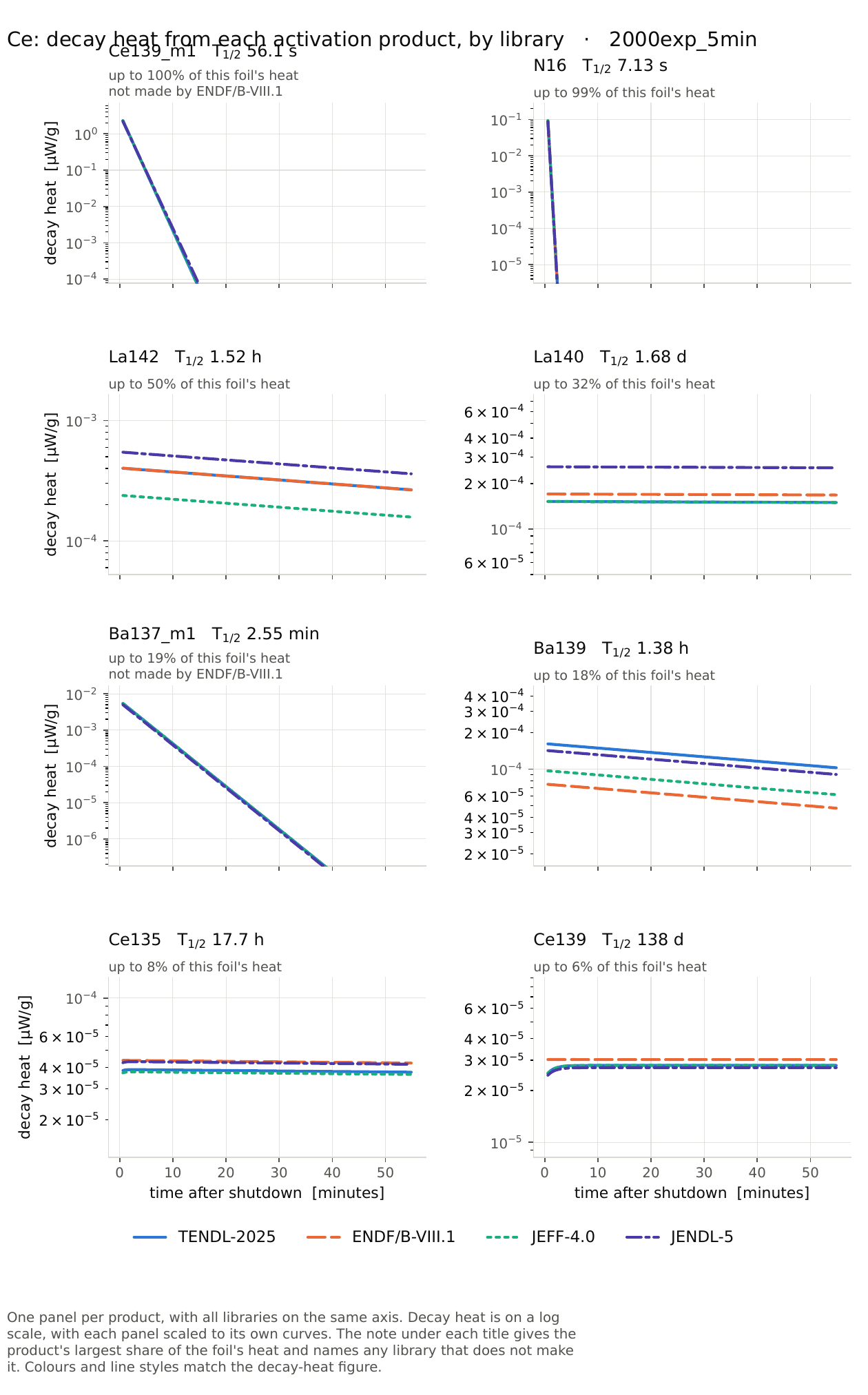}
\caption{Ce: share of the decay heat from each activation product during cooling.}
\end{figure}
\clearpage

\begin{figure}[H]
\centering
\includegraphics[width=\textwidth,height=0.82\textheight,keepaspectratio]{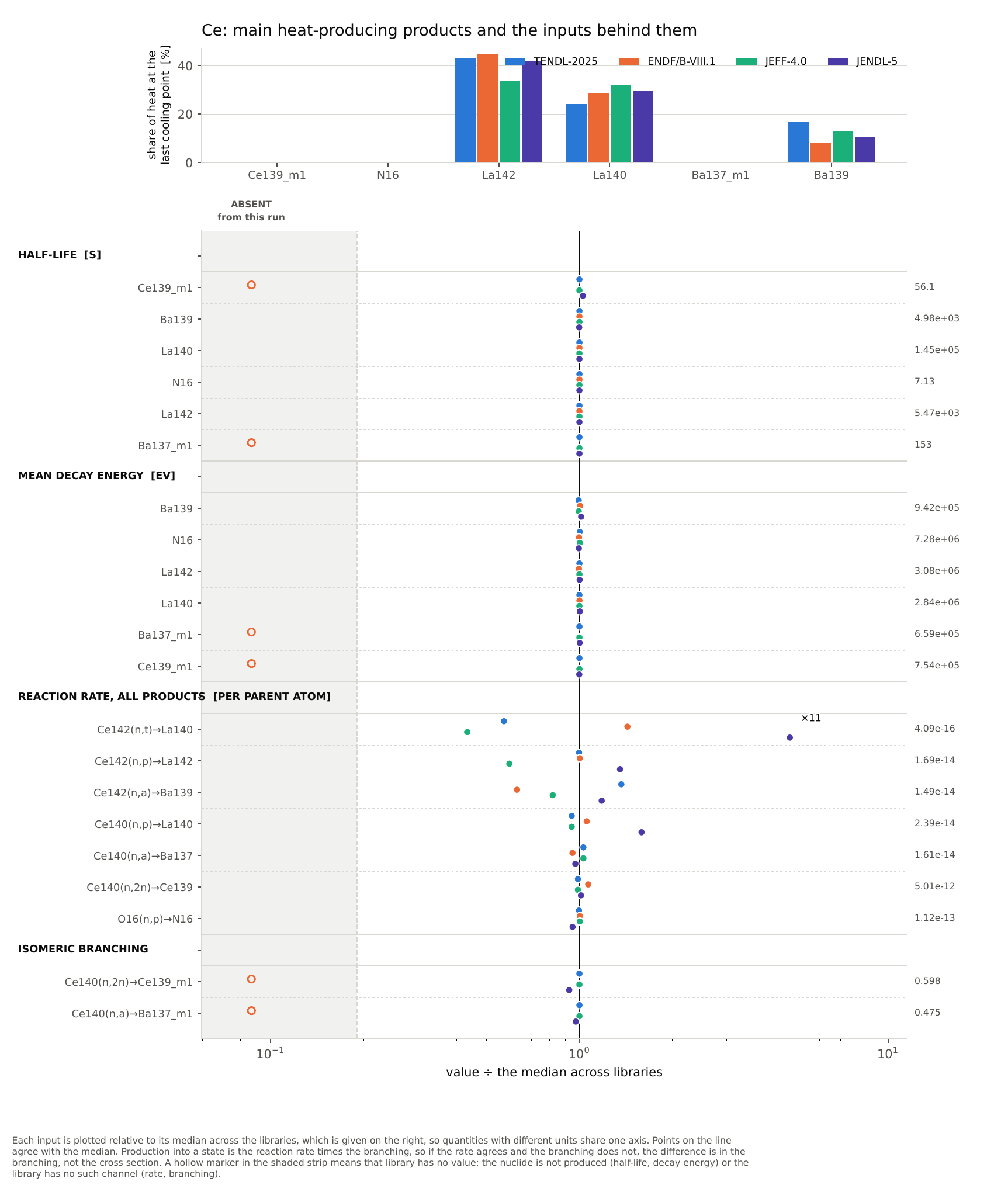}
\caption{Ce: inputs for the main heat-producing products, each relative to its median across the available libraries.}
\end{figure}
\clearpage

\begin{figure}[H]
\centering
\includegraphics[width=\textwidth,height=0.82\textheight,keepaspectratio]{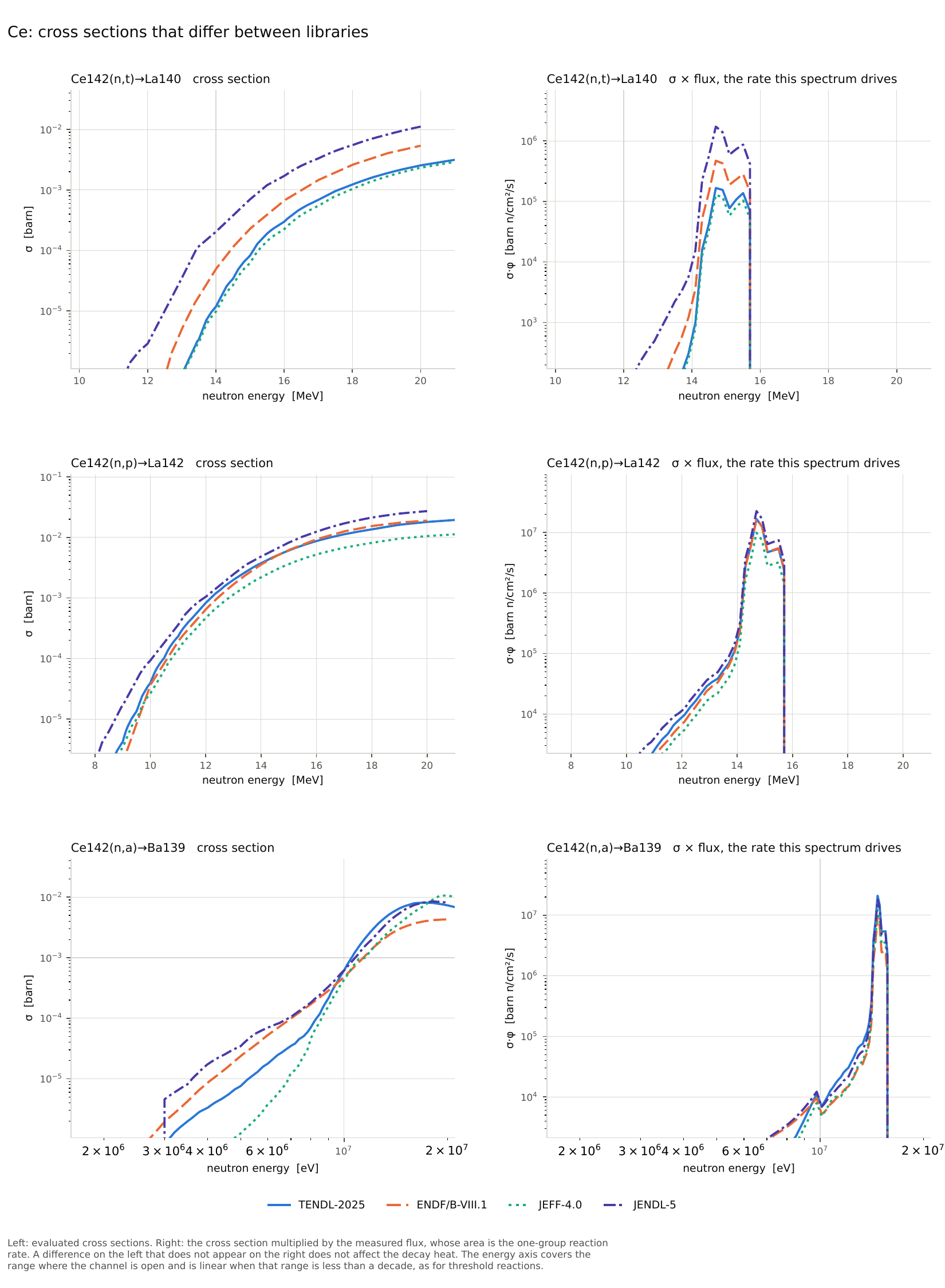}
\caption{Ce: cross sections of the channels whose spectrum-averaged rates differ, with the neutron spectrum and their product.}
\end{figure}
\clearpage

The libraries are between 39 and 95 percent from the measurement, which has a median uncertainty of 7 percent. JEFF-4.0 is closest and ENDF/B-VIII.1 furthest.

\subsubsection*{Differences in the nuclear data}
\begin{itemize}\setlength{\itemsep}{6pt}
\item JENDL-5 gives the total reaction rate of Ce142(n,t)→La140 as 1.964e-15, against 2.328e-16 for the median of the others, a factor of 8.\\[2pt]{\itshape This is the total rate of the channel over all its products, so the difference is in the cross section, not in the split between states.}
\item ENDF/B-VIII.1 gives no heat from Ce139\_m1; the other libraries do.\\[2pt]{\itshape The heat breakdown alone does not show why. Possible causes are a missing reaction channel, a different state assignment, different decay data or a reporting threshold.}
\item ENDF/B-VIII.1 gives no heat from Ba137\_m1; the other libraries do.
\end{itemize}
\subsubsection*{Suggested measurements}
\begin{itemize}\setlength{\itemsep}{4pt}
\item Natural Ce has 4 isotopes, so the measured heat combines several parents. Enriched single-isotope targets would distinguish their contributions. The candidate parent is Ce142. Enrichment alone does not isolate a single reaction channel.
\item The sample is a compound that also contains O. Their activation adds to the measured heat; a metal or purer sample would remove it.
\item This foil makes La140 (half-life 1.68 days), which gives up to 32 percent of the calculated heat. A five-minute irradiation and 55 minute measurement only see the start of this contribution. There is no seven-hour FNS measurement of this foil. A longer irradiation followed for days to months would be needed to test this product's production and decay.
\item The reaction rates are averaged over the measured spectrum. Activation measurements at several well-defined neutron energies, including below 14 MeV, would show which energy range the differences come from.
\end{itemize}
\clearpage
\subsection{Cl: Chlorine}\label{foil:Cl}
\begin{figure}[H]
\centering
\includegraphics[width=\textwidth,height=0.82\textheight,keepaspectratio]{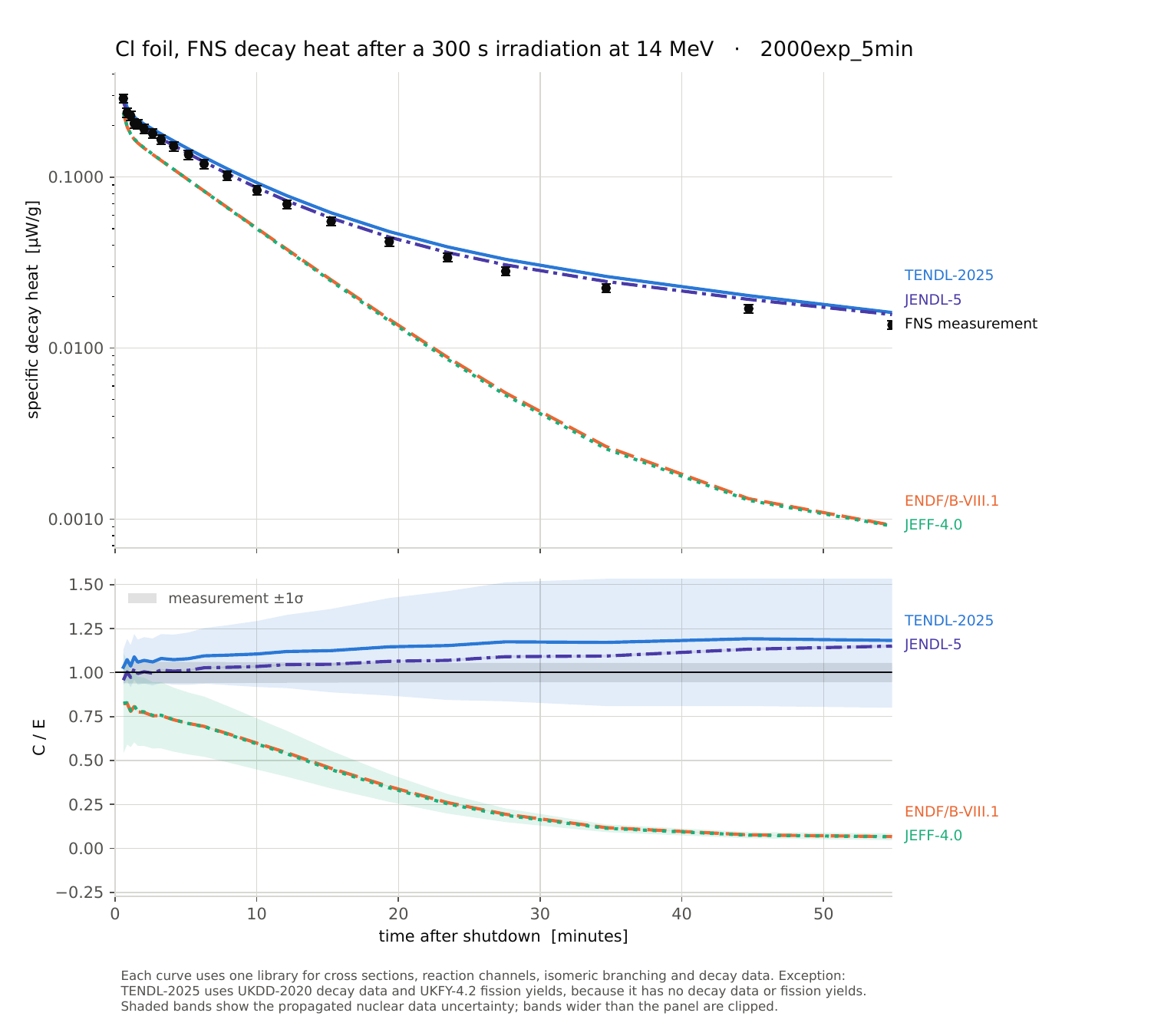}
\caption{Cl: decay heat against the measurement, with C/E beneath. Mean $|C/E-1|$ by library: JENDL-5 4\%, TENDL-2025 11\%, ENDF/B-VIII.1 44\%, JEFF-4.0 44\%.}
\end{figure}
\clearpage

\begin{figure}[H]
\centering
\includegraphics[width=\textwidth,height=0.82\textheight,keepaspectratio]{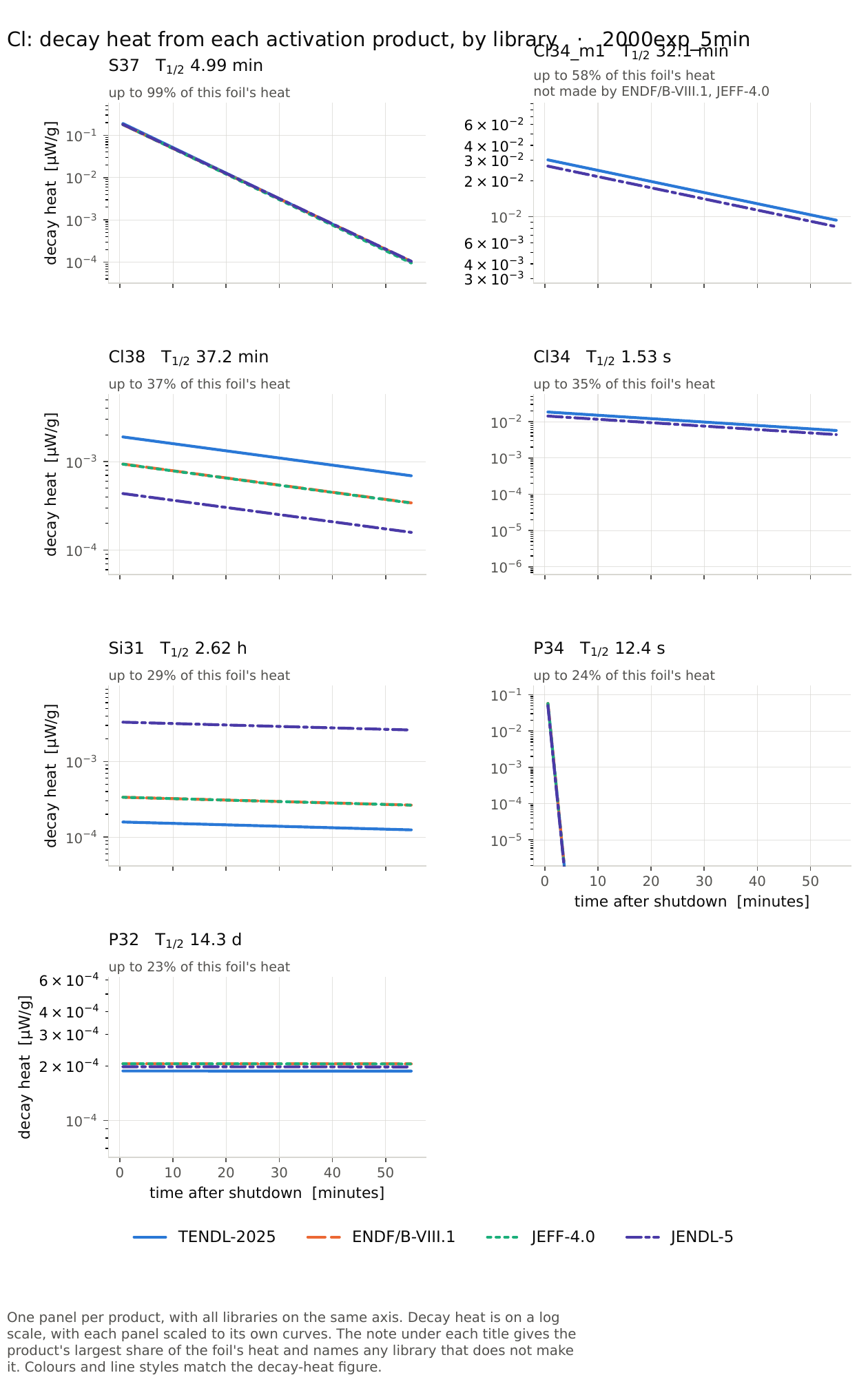}
\caption{Cl: share of the decay heat from each activation product during cooling.}
\end{figure}
\clearpage

\begin{figure}[H]
\centering
\includegraphics[width=\textwidth,height=0.82\textheight,keepaspectratio]{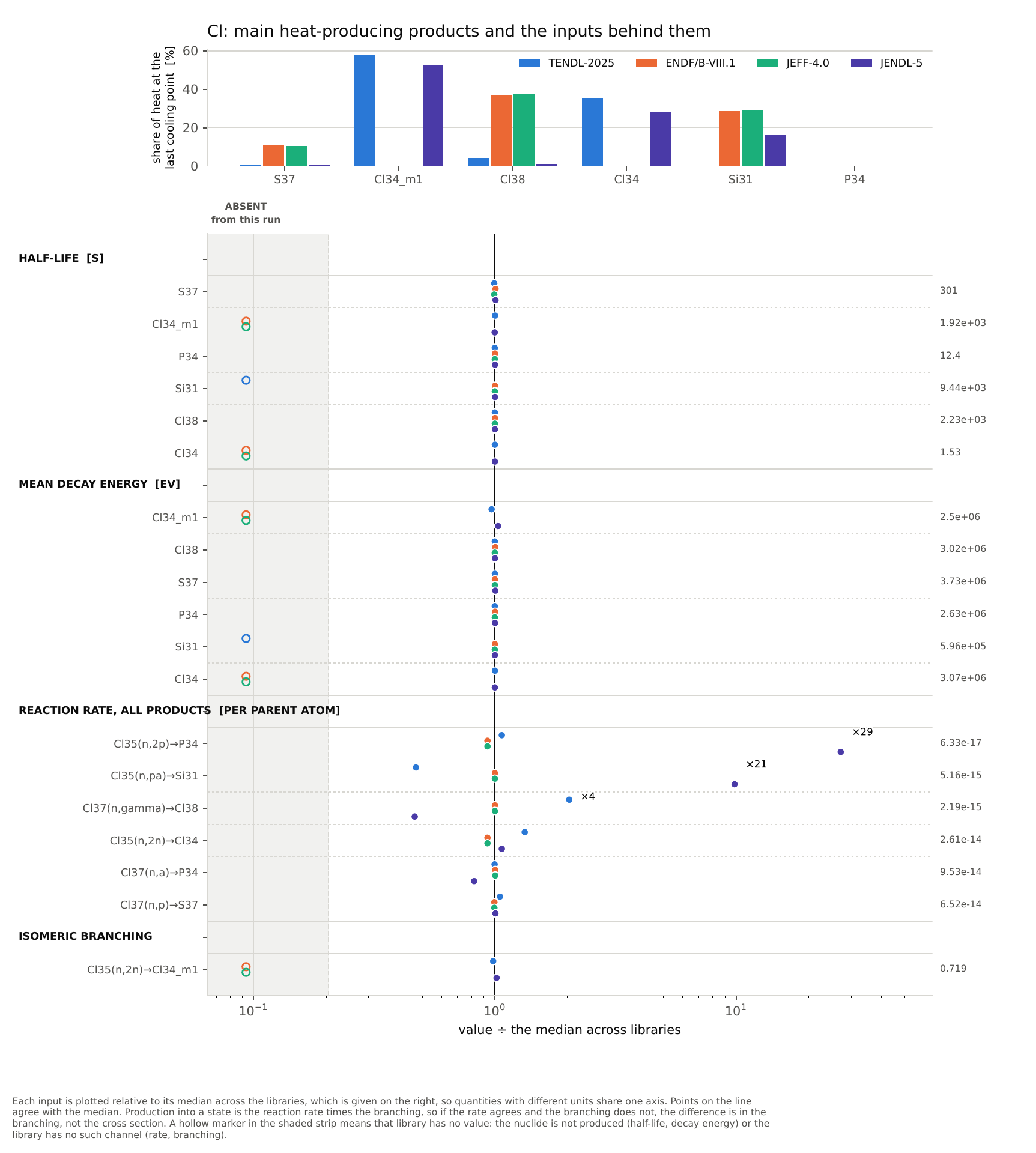}
\caption{Cl: inputs for the main heat-producing products, each relative to its median across the available libraries.}
\end{figure}
\clearpage

\begin{figure}[H]
\centering
\includegraphics[width=\textwidth,height=0.82\textheight,keepaspectratio]{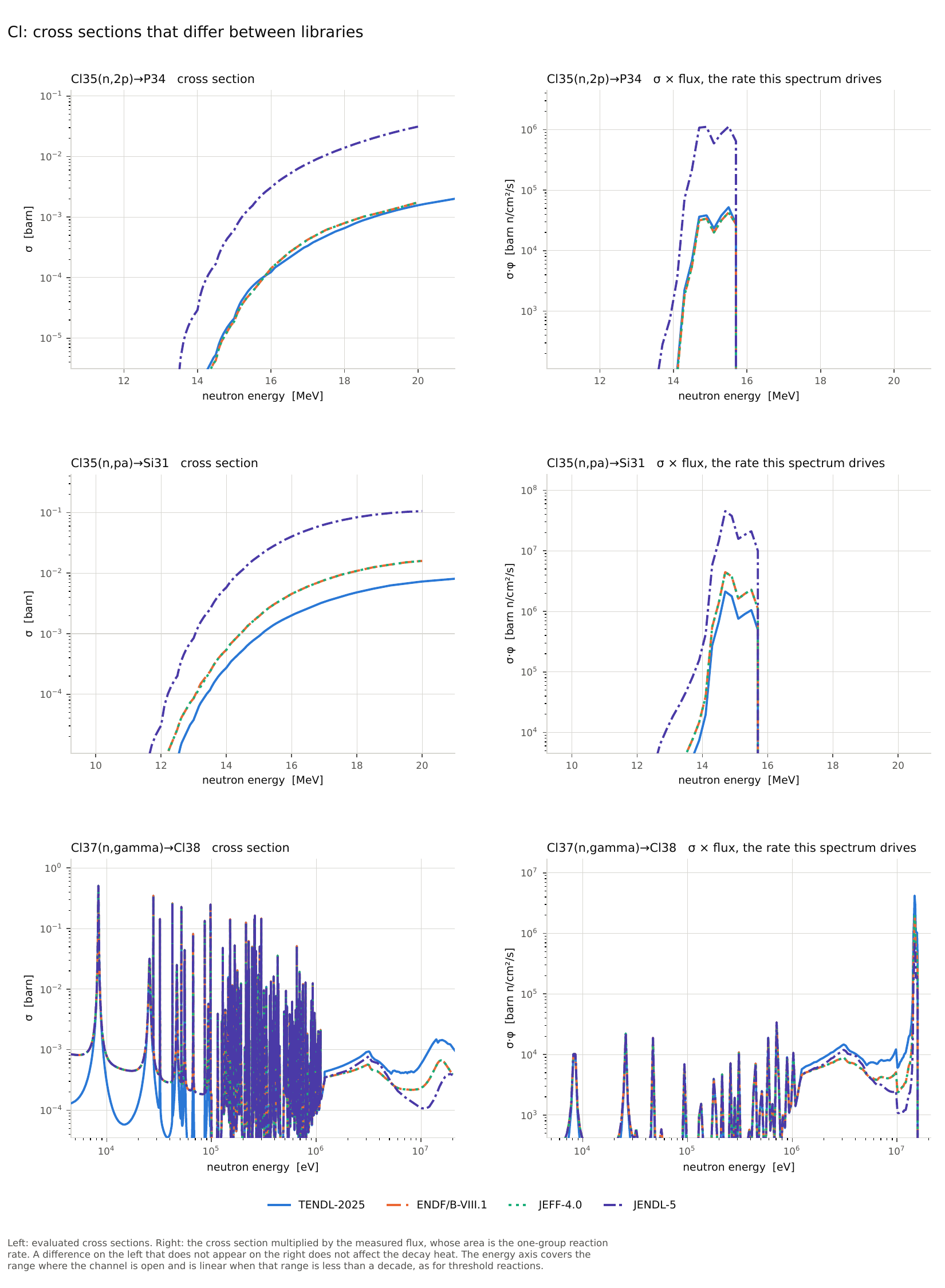}
\caption{Cl: cross sections of the channels whose spectrum-averaged rates differ, with the neutron spectrum and their product.}
\end{figure}
\clearpage

The libraries are between 4 and 44 percent from the measurement, which has a median uncertainty of 6 percent. JENDL-5 is closest and JEFF-4.0 furthest.

\subsubsection*{Differences in the nuclear data}
\begin{itemize}\setlength{\itemsep}{6pt}
\item JENDL-5 gives the total reaction rate of Cl35(n,2p)→P34 as 1.718e-15, against 5.897e-17 for the median of the others, a factor of 29.\\[2pt]{\itshape This is the total rate of the channel over all its products, so the difference is in the cross section, not in the split between states.}
\item JENDL-5 gives the total reaction rate of Cl35(n,pa)→Si31 as 5.084e-14, against 5.161e-15 for the median of the others, a factor of 10.
\item JENDL-5 gives the total reaction rate of Cl37(n,gamma)→Cl38 as 1.019e-15, against 2.192e-15 for the median of the others, a factor of 2.
\item ENDF/B-VIII.1, JEFF-4.0 give no heat from Cl34\_m1; the other libraries do.\\[2pt]{\itshape The heat breakdown alone does not show why. Possible causes are a missing reaction channel, a different state assignment, different decay data or a reporting threshold.}
\end{itemize}
\subsubsection*{Suggested measurements}
\begin{itemize}\setlength{\itemsep}{4pt}
\item Natural Cl has 2 isotopes, so the measured heat combines several parents. Enriched single-isotope targets would distinguish their contributions. Candidate parents are Cl35, Cl37. Enrichment alone does not isolate a single reaction channel.
\item The sample is a compound that also contains C, H. Their activation adds to the measured heat; a metal or purer sample would remove it.
\item This foil makes P32 (half-life 14.3 days), which gives up to 23 percent of the calculated heat. A five-minute irradiation and 55 minute measurement only see the start of this contribution. There is no seven-hour FNS measurement of this foil. A longer irradiation followed for days to months would be needed to test this product's production and decay.
\item The reaction rates are averaged over the measured spectrum. Activation measurements at several well-defined neutron energies, including below 14 MeV, would show which energy range the differences come from.
\end{itemize}
\clearpage
\subsection{Co: Cobalt}\label{foil:Co}
\begin{figure}[H]
\centering
\includegraphics[width=\textwidth,height=0.82\textheight,keepaspectratio]{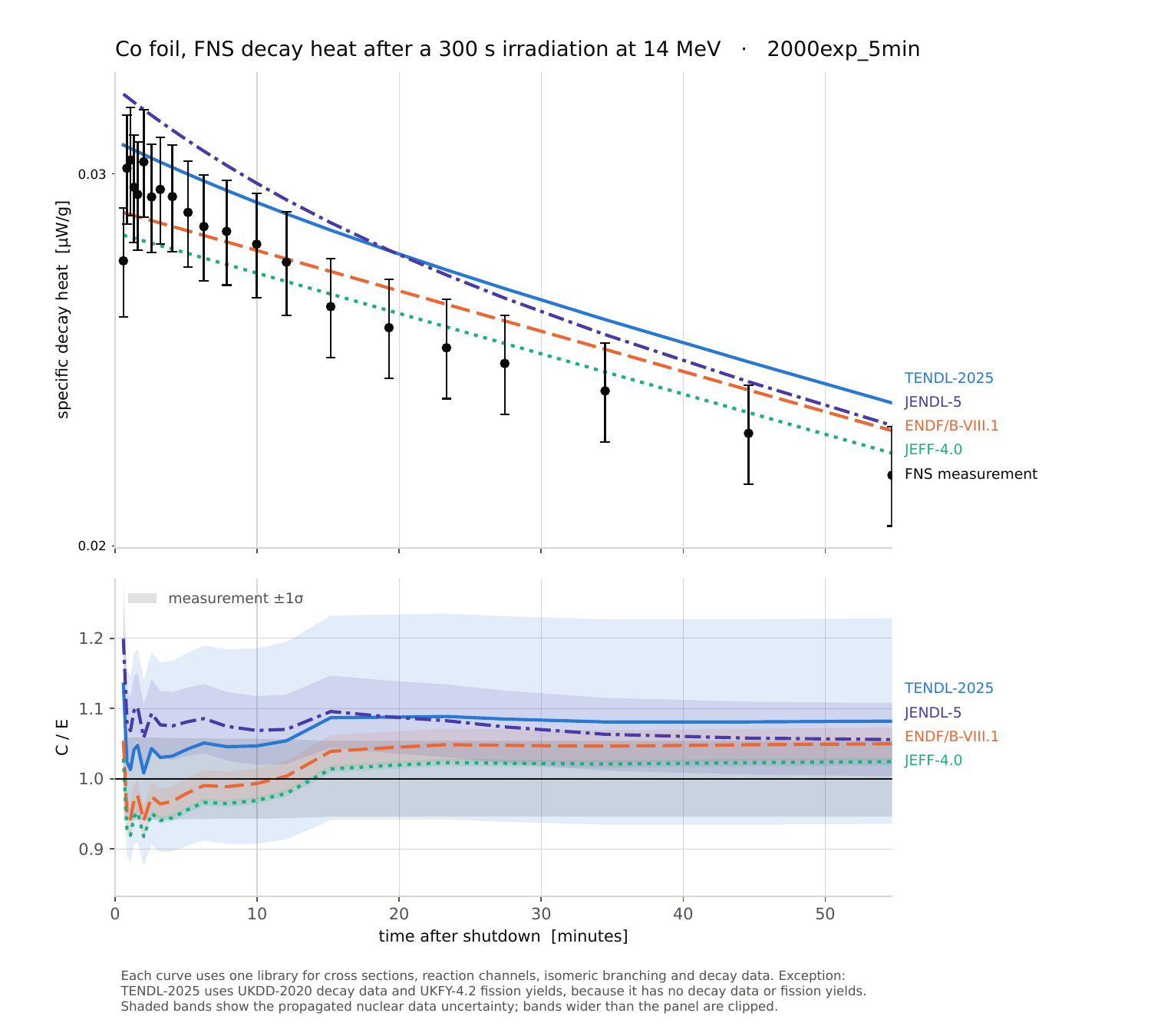}
\caption{Co: decay heat against the measurement, with C/E beneath. Mean $|C/E-1|$ by library: ENDF/B-VIII.1 4\%, JEFF-4.0 4\%, TENDL-2025 6\%, JENDL-5 8\%.}
\end{figure}
\clearpage

\begin{figure}[H]
\centering
\includegraphics[width=\textwidth,height=0.82\textheight,keepaspectratio]{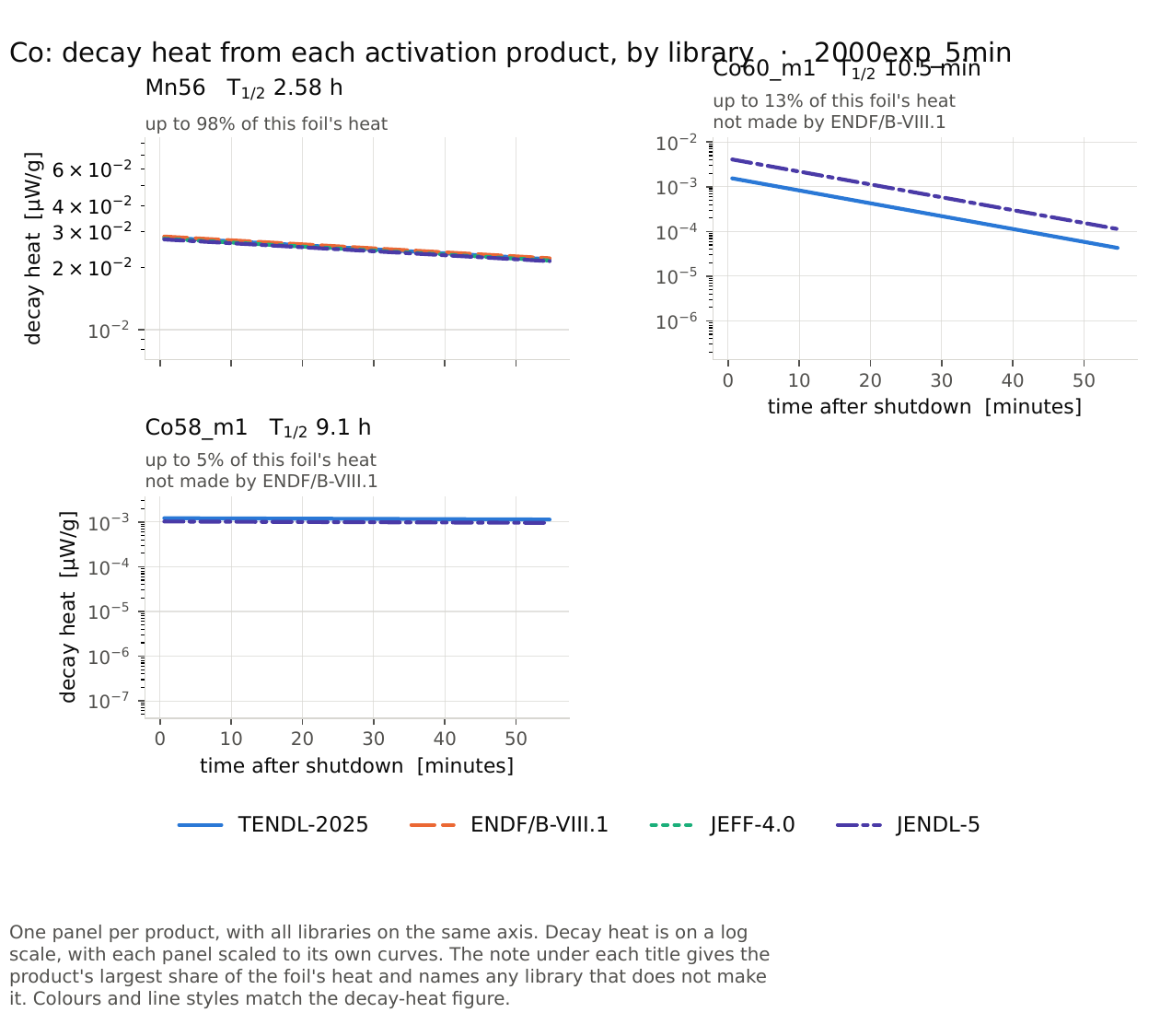}
\caption{Co: share of the decay heat from each activation product during cooling.}
\end{figure}
\clearpage

\begin{figure}[H]
\centering
\includegraphics[width=\textwidth,height=0.82\textheight,keepaspectratio]{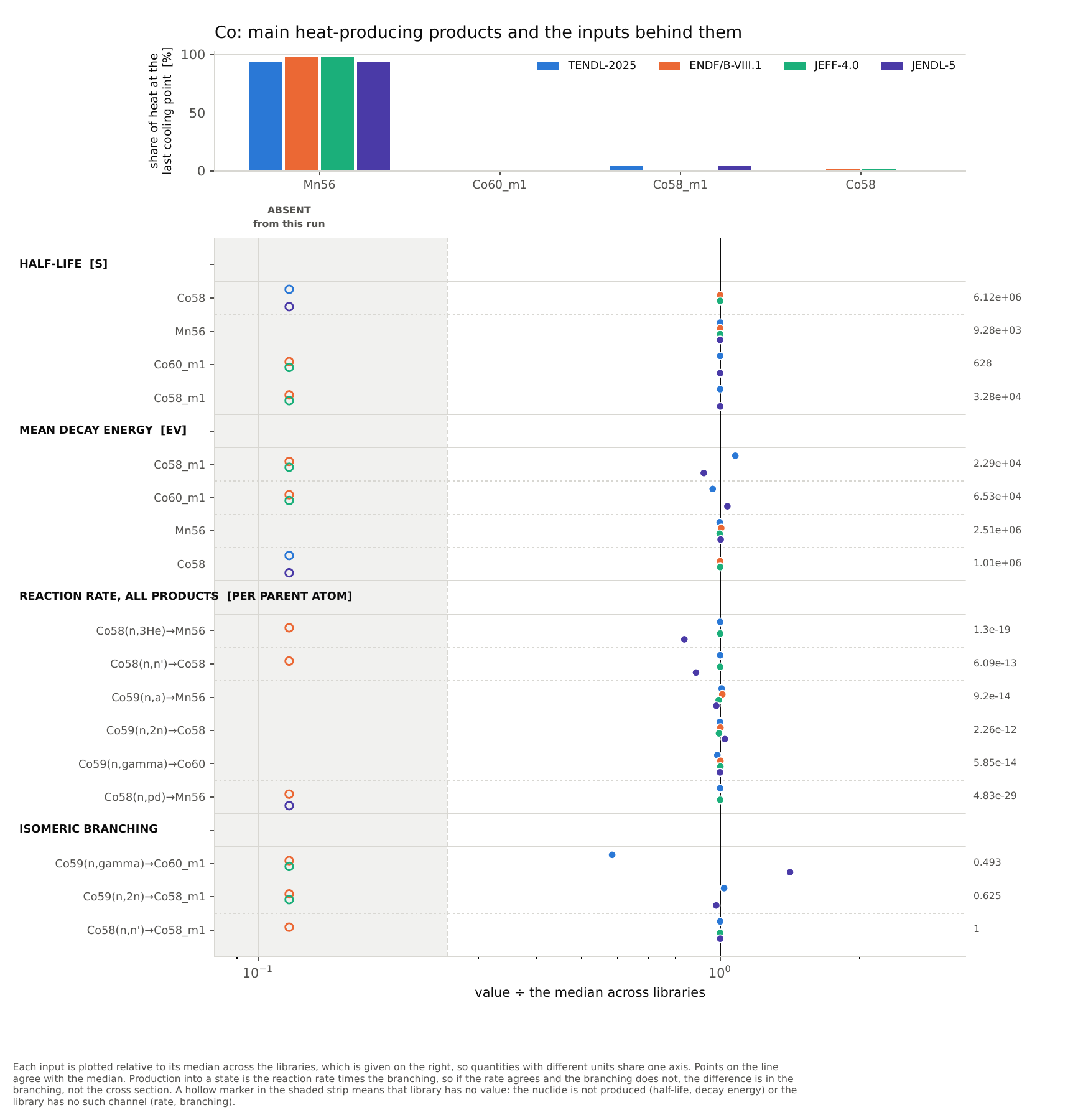}
\caption{Co: inputs for the main heat-producing products, each relative to its median across the available libraries.}
\end{figure}
\clearpage

\begin{figure}[H]
\centering
\includegraphics[width=\textwidth,height=0.82\textheight,keepaspectratio]{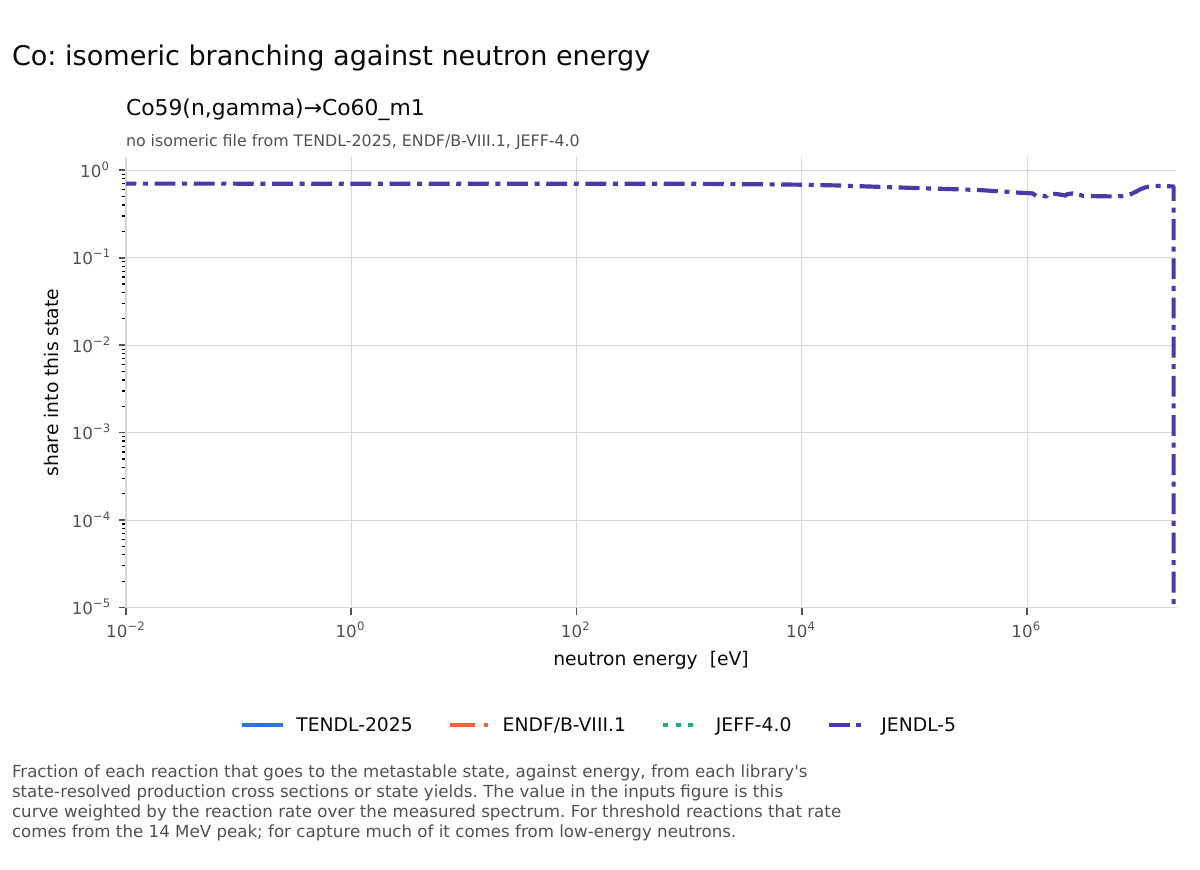}
\caption{Co: isomeric branching ratio against neutron energy for the channels whose spectrum-averaged ratio differs between libraries.}
\end{figure}
\clearpage

\subsubsection*{1996 seven-hour campaign}
The 1996 campaign irradiated this foil for 7.57 hours at position 7 and measured it out to 402.95 days. In the five-minute calculation Co58 (half-life 70.9 days) gives up to 2 percent of the heat, and the longer irradiation and cooling test its production and decay data. ENDF/B-VIII.1 is closest. This campaign uses a different spectrum, so it is not included in the main rankings.

\begin{figure}[H]
\centering
\includegraphics[width=\textwidth,height=0.82\textheight,keepaspectratio]{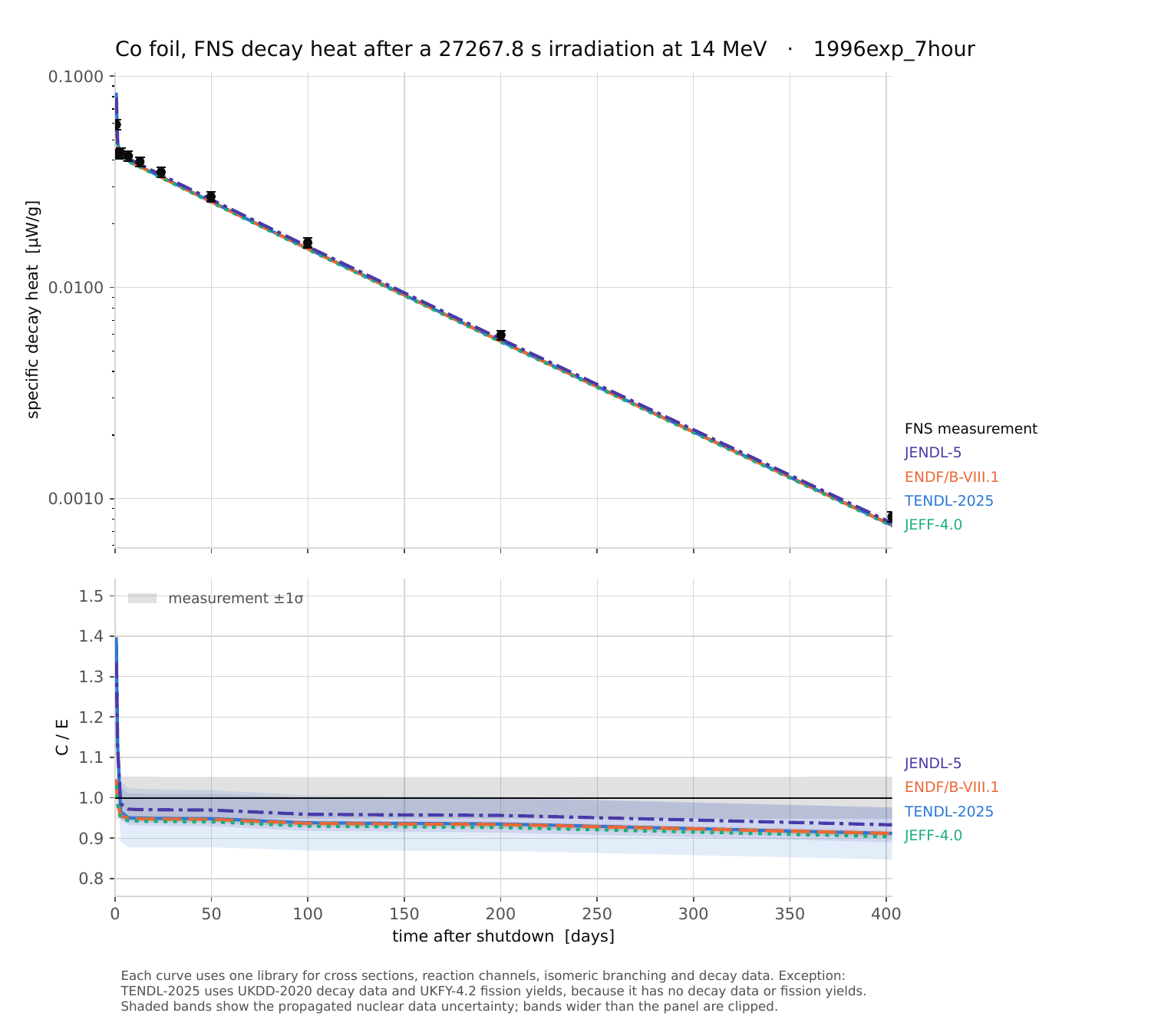}
\caption{Co: 1996 seven-hour campaign, irradiation position 7. Mean $|C/E-1|$ by library: ENDF/B-VIII.1 5\%, JEFF-4.0 6\%, JENDL-5 7\%, TENDL-2025 10\%. }
\end{figure}
\clearpage

Mean absolute C/E deviations range from 4 to 8 percent. The spread of 5 percentage points is smaller than the median measurement uncertainty of 6 percent, so the measurement cannot separate the libraries.

\subsubsection*{Differences in the nuclear data}
\begin{itemize}\setlength{\itemsep}{6pt}
\item Only JENDL-5 and TENDL-2025 have the isomeric branching of Co59(n,gamma)→Co60\_m1, and they differ by a factor of 2: 0.6983 against 0.2877.\\[2pt]{\itshape Isomeric branching sets how a channel's production is split between the ground and metastable states. It should be checked together with the total rate of the channel.}
\item ENDF/B-VIII.1 gives no heat from Co60\_m1; the other libraries do.\\[2pt]{\itshape The heat breakdown alone does not show why. Possible causes are a missing reaction channel, a different state assignment, different decay data or a reporting threshold.}
\item ENDF/B-VIII.1 gives no heat from Co58\_m1; the other libraries do.
\end{itemize}
\subsubsection*{Suggested measurements}
\begin{itemize}\setlength{\itemsep}{4pt}
\item Co has a single stable isotope, so the foil is already a single-isotope target and enrichment would not help.
\item Measuring the activity of each state separately, for example by gamma spectrometry where the states have distinct lines, would constrain the isomeric ratio. This needs corrections for feeding between states, detector efficiency and irradiation history.
\end{itemize}
\clearpage
\subsection{Cr: Chromium}\label{foil:Cr}
\begin{figure}[H]
\centering
\includegraphics[width=\textwidth,height=0.82\textheight,keepaspectratio]{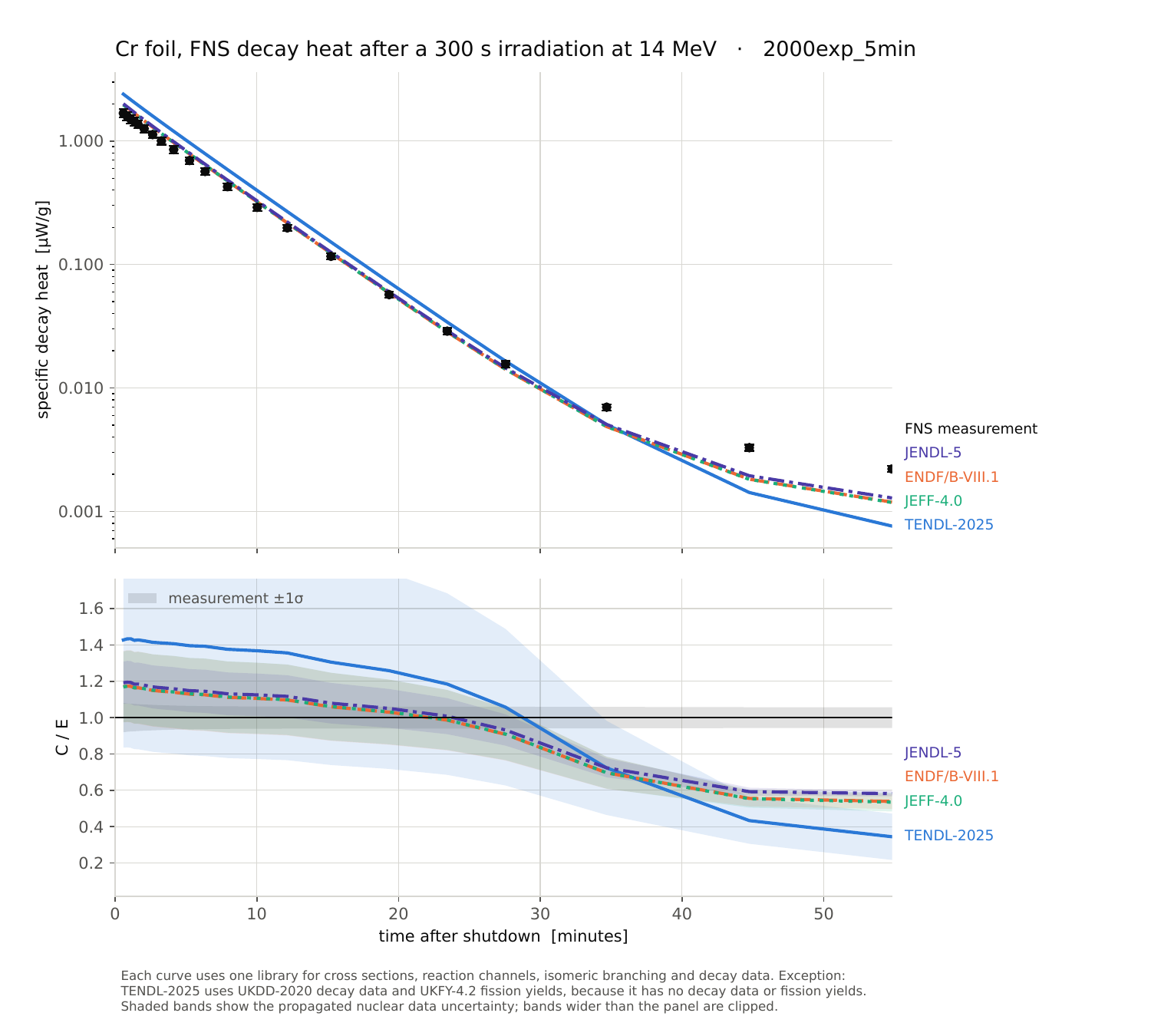}
\caption{Cr: decay heat against the measurement, with C/E beneath. Mean $|C/E-1|$ by library: ENDF/B-VIII.1 16\%, JEFF-4.0 16\%, JENDL-5 17\%, TENDL-2025 38\%.}
\end{figure}
\clearpage

\begin{figure}[H]
\centering
\includegraphics[width=\textwidth,height=0.82\textheight,keepaspectratio]{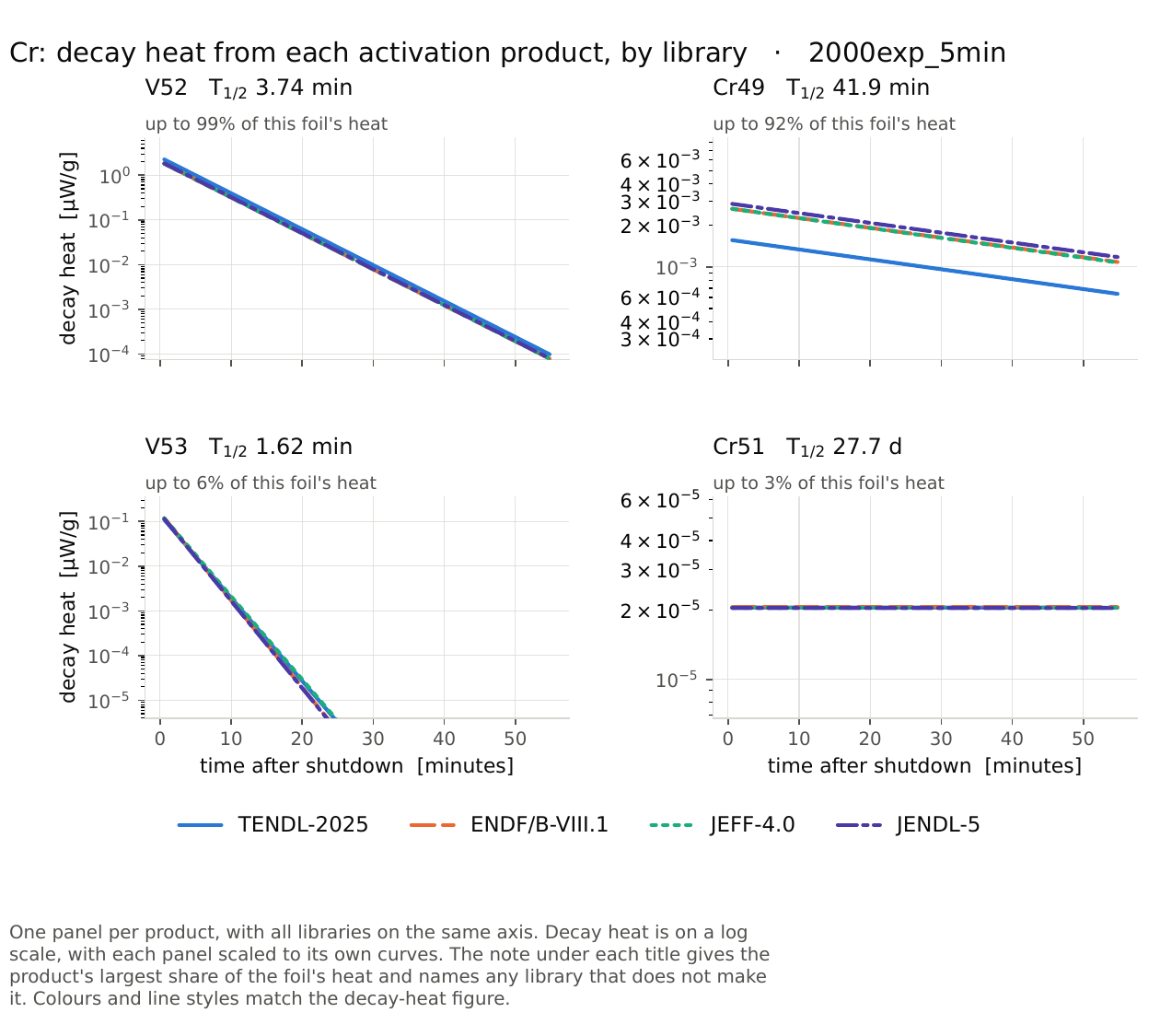}
\caption{Cr: share of the decay heat from each activation product during cooling.}
\end{figure}
\clearpage

\begin{figure}[H]
\centering
\includegraphics[width=\textwidth,height=0.82\textheight,keepaspectratio]{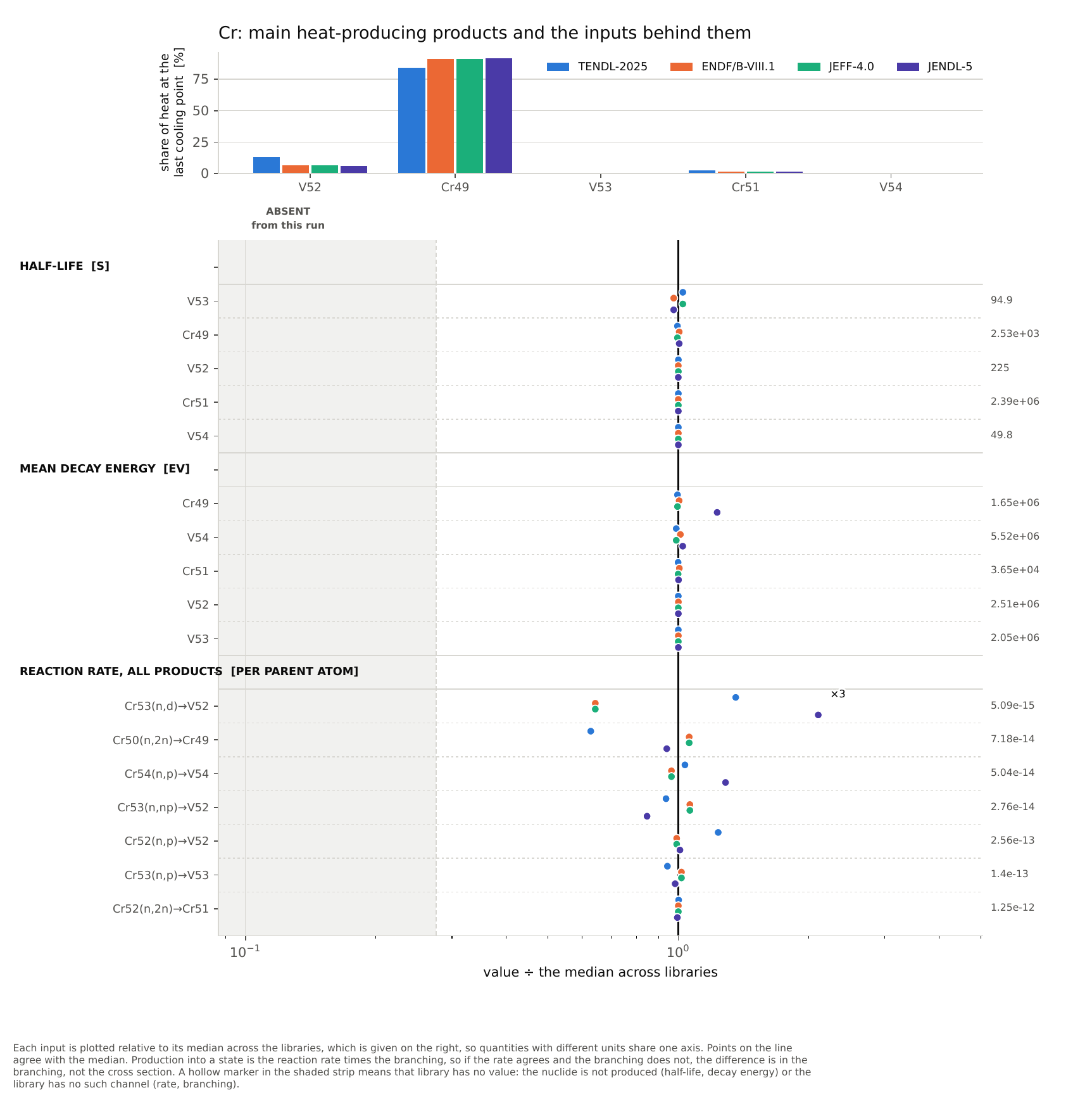}
\caption{Cr: inputs for the main heat-producing products, each relative to its median across the available libraries.}
\end{figure}
\clearpage

\begin{figure}[H]
\centering
\includegraphics[width=\textwidth,height=0.82\textheight,keepaspectratio]{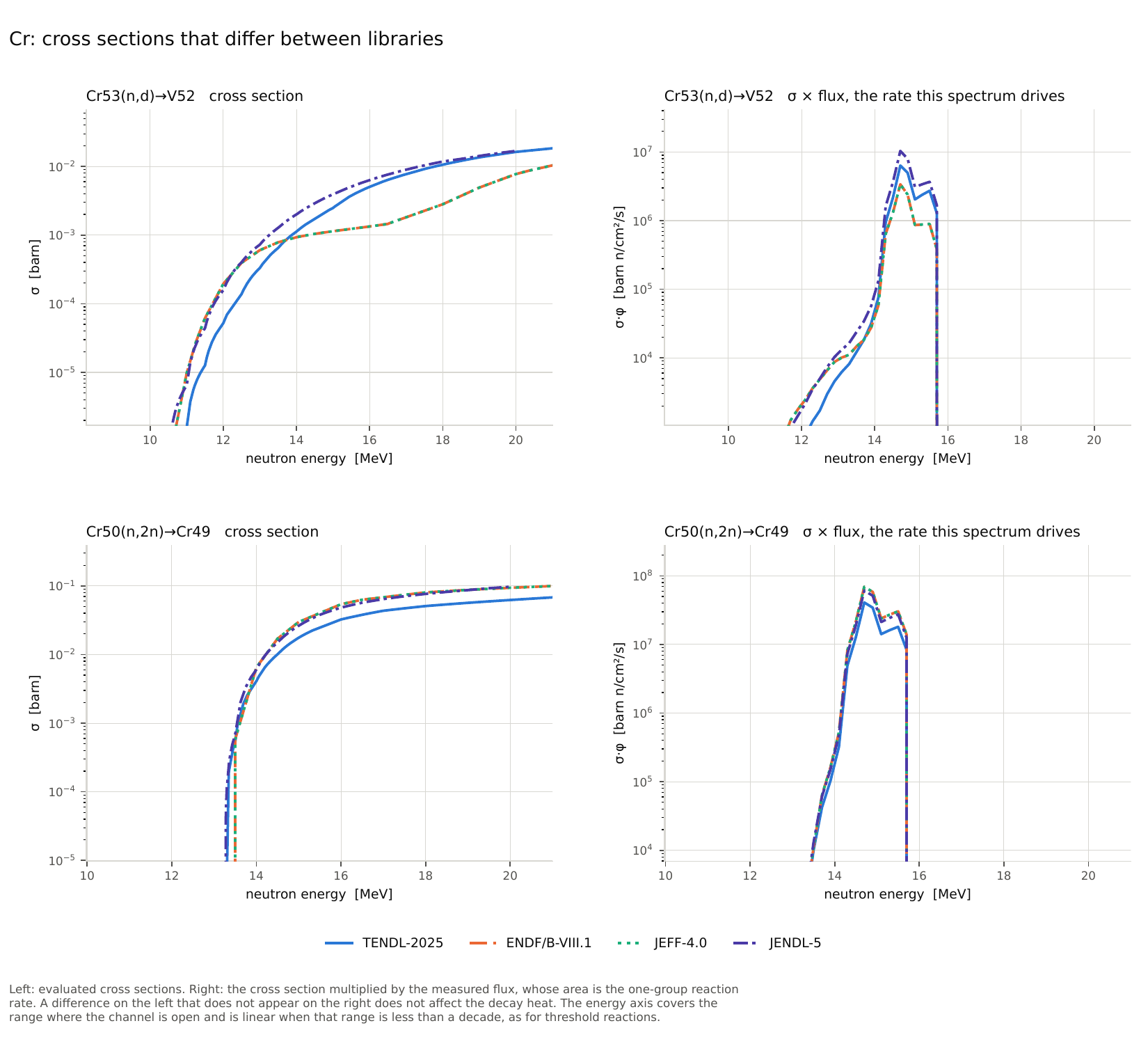}
\caption{Cr: cross sections of the channels whose spectrum-averaged rates differ, with the neutron spectrum and their product.}
\end{figure}
\clearpage

\subsubsection*{1996 seven-hour campaign}
The 1996 campaign irradiated this foil for 7.17 hours at position 7 and measured it out to 197.96 days. In the five-minute calculation Cr51 (half-life 27.7 days) gives up to 3 percent of the heat, and the longer irradiation and cooling test its production and decay data. JENDL-5 is closest. This campaign uses a different spectrum, so it is not included in the main rankings.

\begin{figure}[H]
\centering
\includegraphics[width=\textwidth,height=0.82\textheight,keepaspectratio]{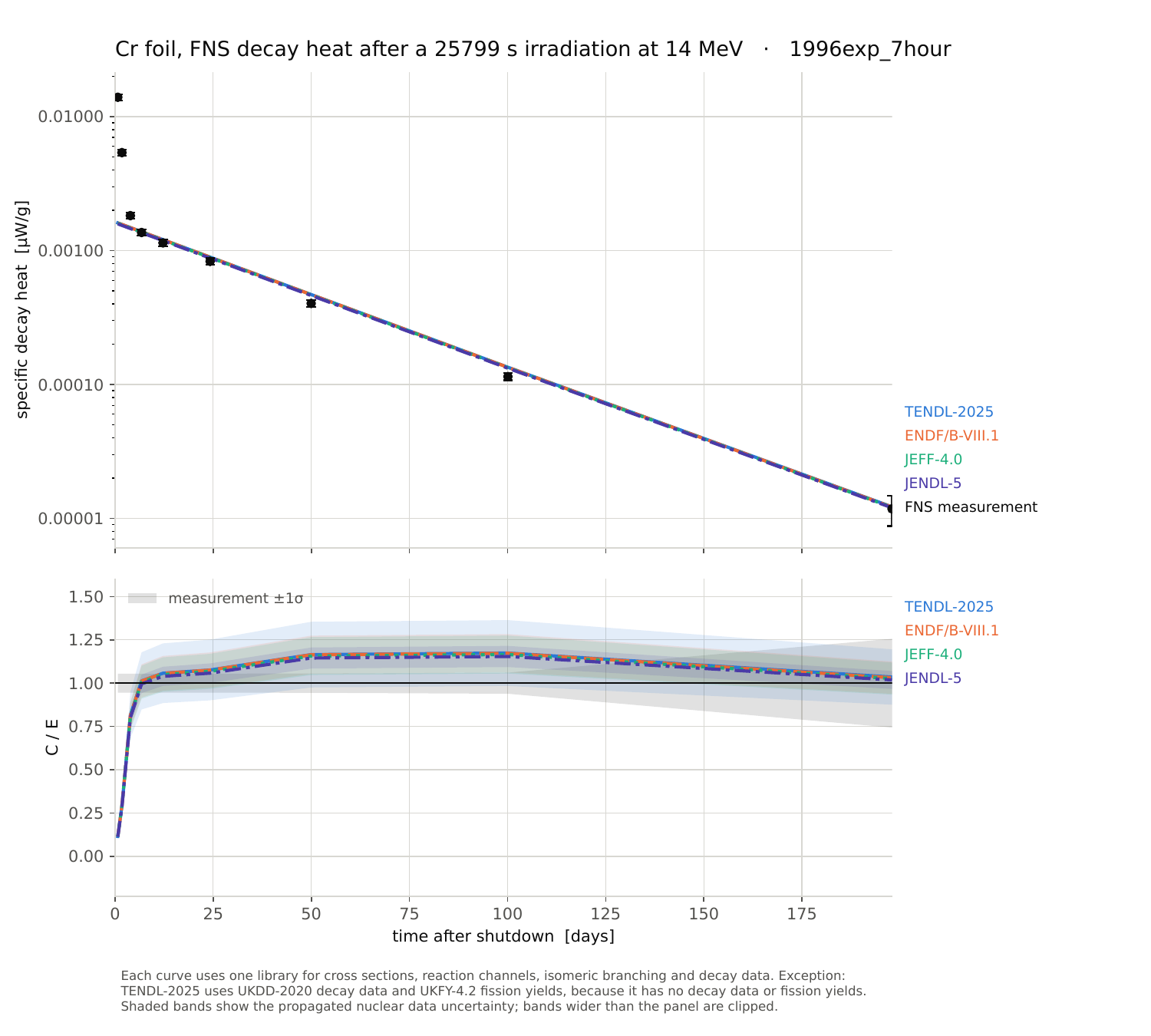}
\caption{Cr: 1996 seven-hour campaign, irradiation position 7. Mean $|C/E-1|$ by library: JENDL-5 25\%, JEFF-4.0 25\%, ENDF/B-VIII.1 25\%, TENDL-2025 26\%. }
\end{figure}
\clearpage

The libraries are between 16 and 38 percent from the measurement, which has a median uncertainty of 6 percent. ENDF/B-VIII.1 is closest and TENDL-2025 furthest.

\subsubsection*{Differences in the nuclear data}
\begin{itemize}\setlength{\itemsep}{6pt}
\item JENDL-5 gives the total reaction rate of Cr53(n,d)→V52 as 1.072e-14, against 3.274e-15 for the median of the others, a factor of 3.\\[2pt]{\itshape This is the total rate of the channel over all its products, so the difference is in the cross section, not in the split between states.}
\end{itemize}
\subsubsection*{Suggested measurements}
\begin{itemize}\setlength{\itemsep}{4pt}
\item Natural Cr has 4 isotopes, so the measured heat combines several parents. Enriched single-isotope targets would distinguish their contributions. The candidate parent is Cr53. Enrichment alone does not isolate a single reaction channel.
\item The reaction rates are averaged over the measured spectrum. Activation measurements at several well-defined neutron energies, including below 14 MeV, would show which energy range the differences come from.
\end{itemize}
\clearpage
\subsection{Cs: Cesium}\label{foil:Cs}
\begin{figure}[H]
\centering
\includegraphics[width=\textwidth,height=0.82\textheight,keepaspectratio]{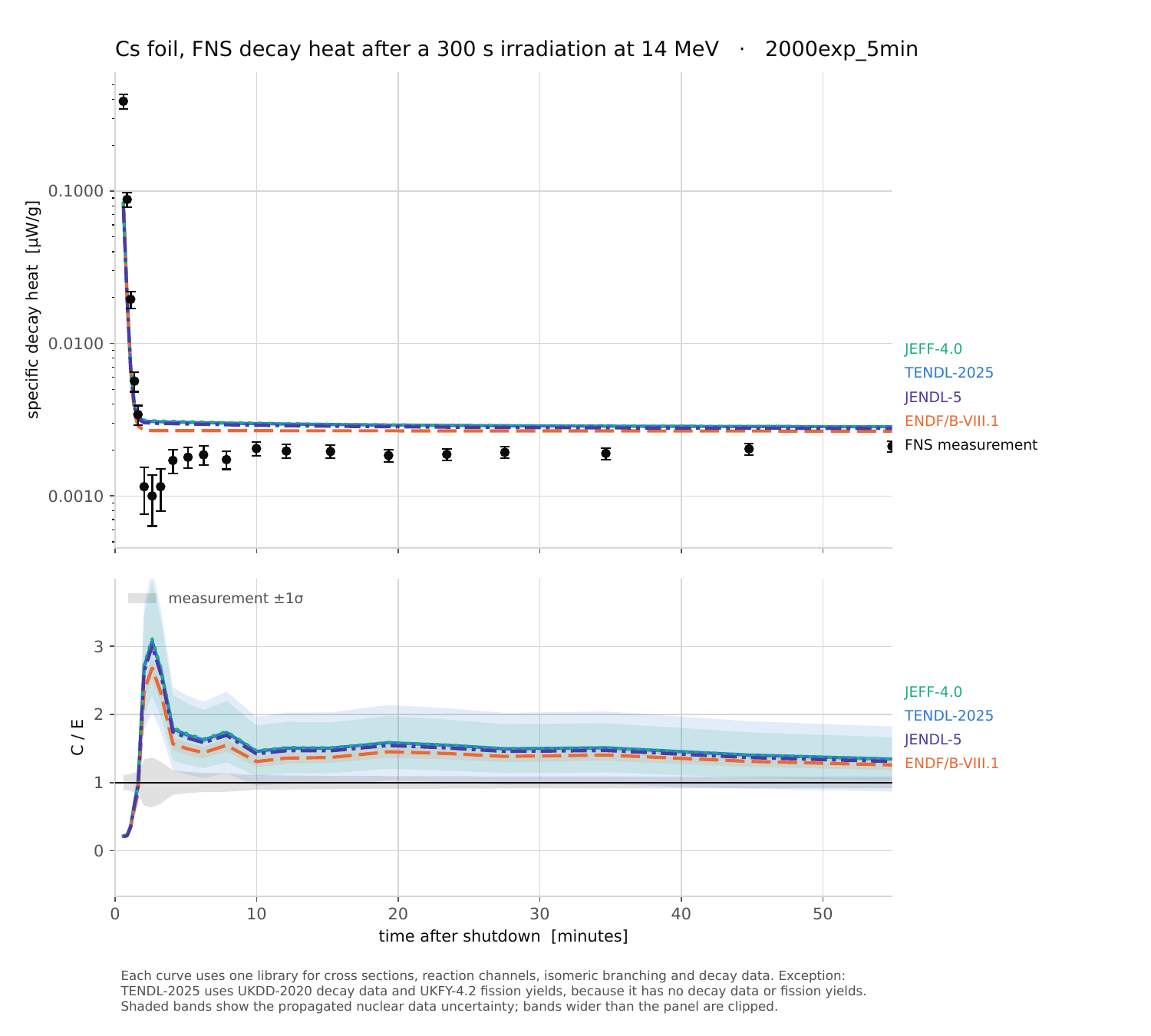}
\caption{Cs: decay heat against the measurement, with C/E beneath. Mean $|C/E-1|$ by library: ENDF/B-VIII.1 59\%, JENDL-5 69\%, TENDL-2025 72\%, JEFF-4.0 73\%.}
\end{figure}
\clearpage

\begin{figure}[H]
\centering
\includegraphics[width=\textwidth,height=0.82\textheight,keepaspectratio]{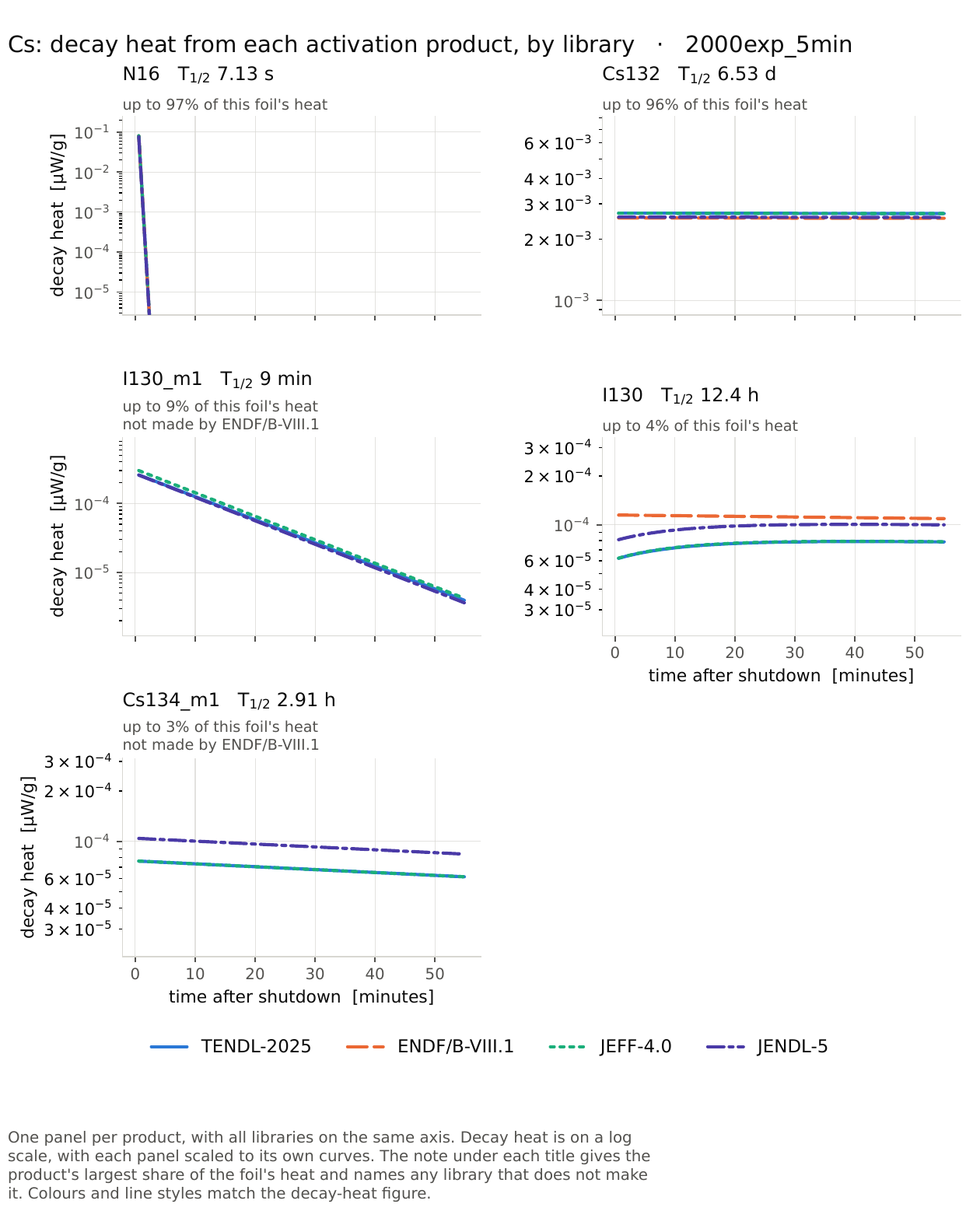}
\caption{Cs: share of the decay heat from each activation product during cooling.}
\end{figure}
\clearpage

\begin{figure}[H]
\centering
\includegraphics[width=\textwidth,height=0.82\textheight,keepaspectratio]{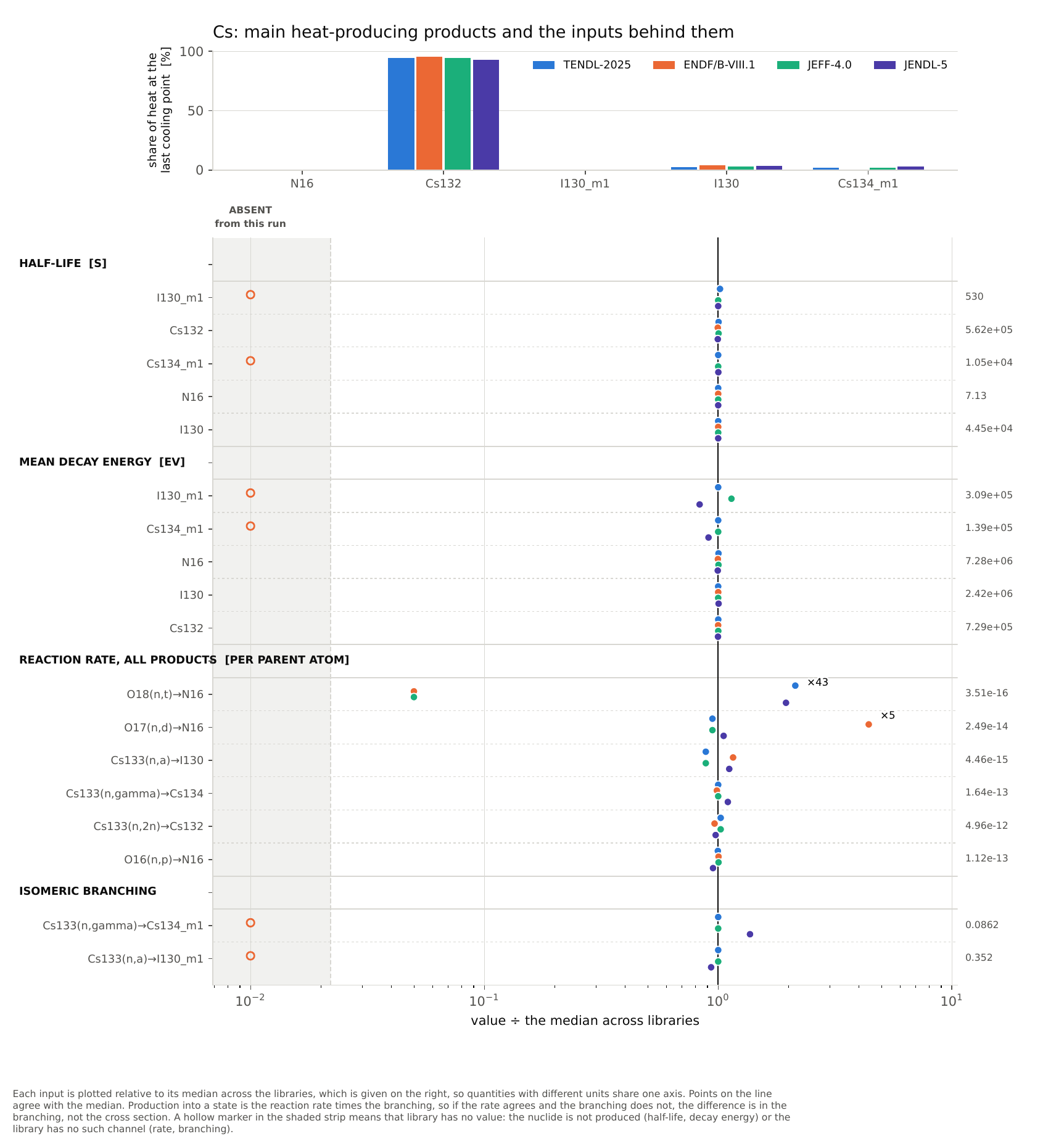}
\caption{Cs: inputs for the main heat-producing products, each relative to its median across the available libraries.}
\end{figure}
\clearpage

\begin{figure}[H]
\centering
\includegraphics[width=\textwidth,height=0.82\textheight,keepaspectratio]{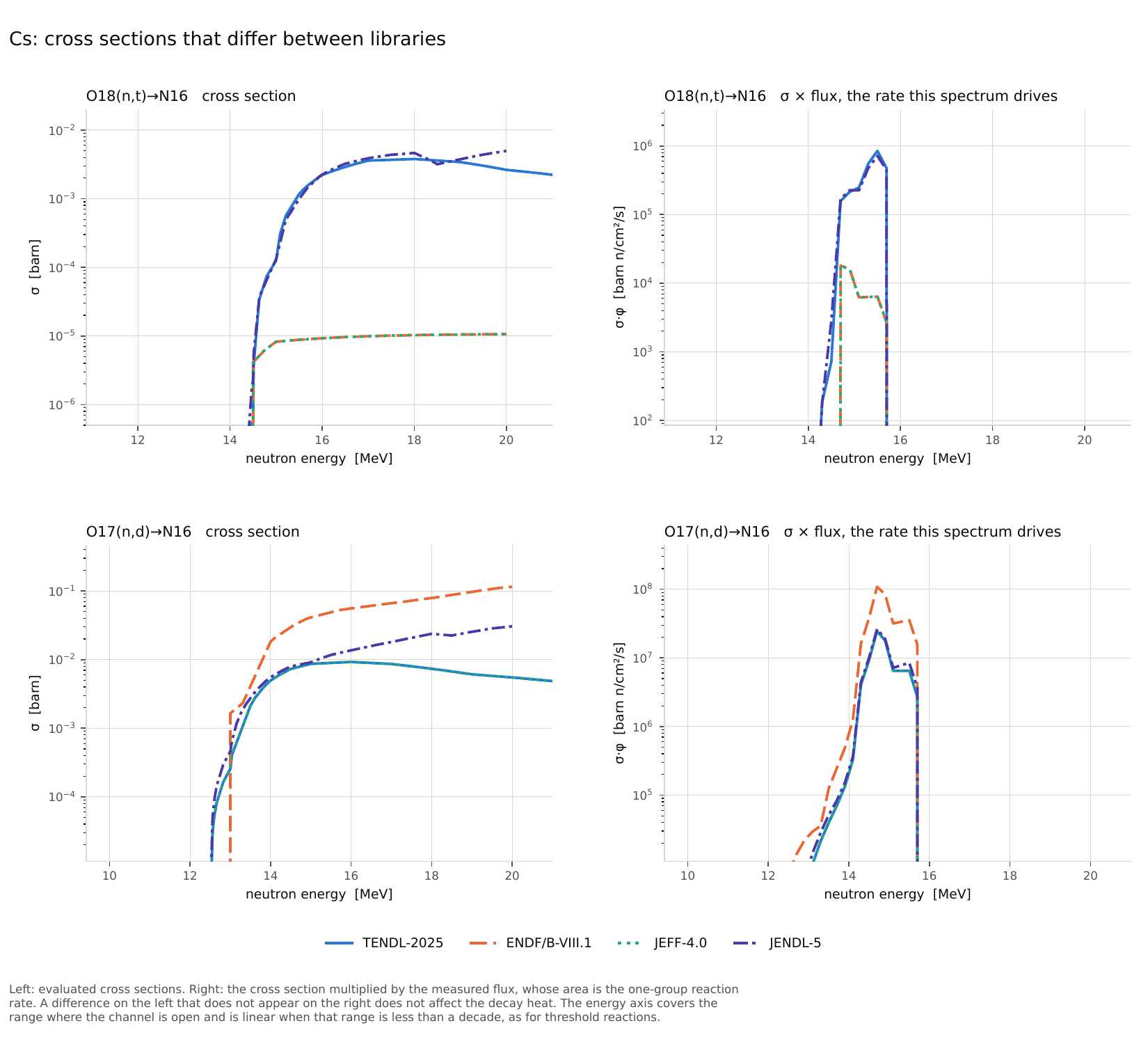}
\caption{Cs: cross sections of the channels whose spectrum-averaged rates differ, with the neutron spectrum and their product.}
\end{figure}
\clearpage

The libraries are between 59 and 73 percent from the measurement, which has a median uncertainty of 11 percent. ENDF/B-VIII.1 is closest and JEFF-4.0 furthest.

\subsubsection*{Differences in the nuclear data}
\begin{itemize}\setlength{\itemsep}{6pt}
\item ENDF/B-VIII.1 gives the total reaction rate of O18(n,t)→N16 as 1.751e-17, against 6.84e-16 for the median of the others, a factor of 39.\\[2pt]{\itshape This is the total rate of the channel over all its products, so the difference is in the cross section, not in the split between states.}
\item ENDF/B-VIII.1 gives the total reaction rate of O17(n,d)→N16 as 1.096e-13, against 2.351e-14 for the median of the others, a factor of 5.
\item ENDF/B-VIII.1 gives no heat from I130\_m1; the other libraries do.\\[2pt]{\itshape The heat breakdown alone does not show why. Possible causes are a missing reaction channel, a different state assignment, different decay data or a reporting threshold.}
\item ENDF/B-VIII.1 gives no heat from Cs134\_m1; the other libraries do.
\end{itemize}
\subsubsection*{Suggested measurements}
\begin{itemize}\setlength{\itemsep}{4pt}
\item Cs has a single stable isotope, so the foil is already a single-isotope target and enrichment would not help.
\item The sample is a compound that also contains C, O. Their activation adds to the measured heat; a metal or purer sample would remove it.
\item This foil makes Cs132 (half-life 6.53 days), which gives up to 96 percent of the calculated heat. A five-minute irradiation and 55 minute measurement only see the start of this contribution. There is no seven-hour FNS measurement of this foil. A longer irradiation followed for days to months would be needed to test this product's production and decay.
\item The reaction rates are averaged over the measured spectrum. Activation measurements at several well-defined neutron energies, including below 14 MeV, would show which energy range the differences come from.
\end{itemize}
\clearpage
\subsection{Cu: Copper}\label{foil:Cu}
\begin{figure}[H]
\centering
\includegraphics[width=\textwidth,height=0.82\textheight,keepaspectratio]{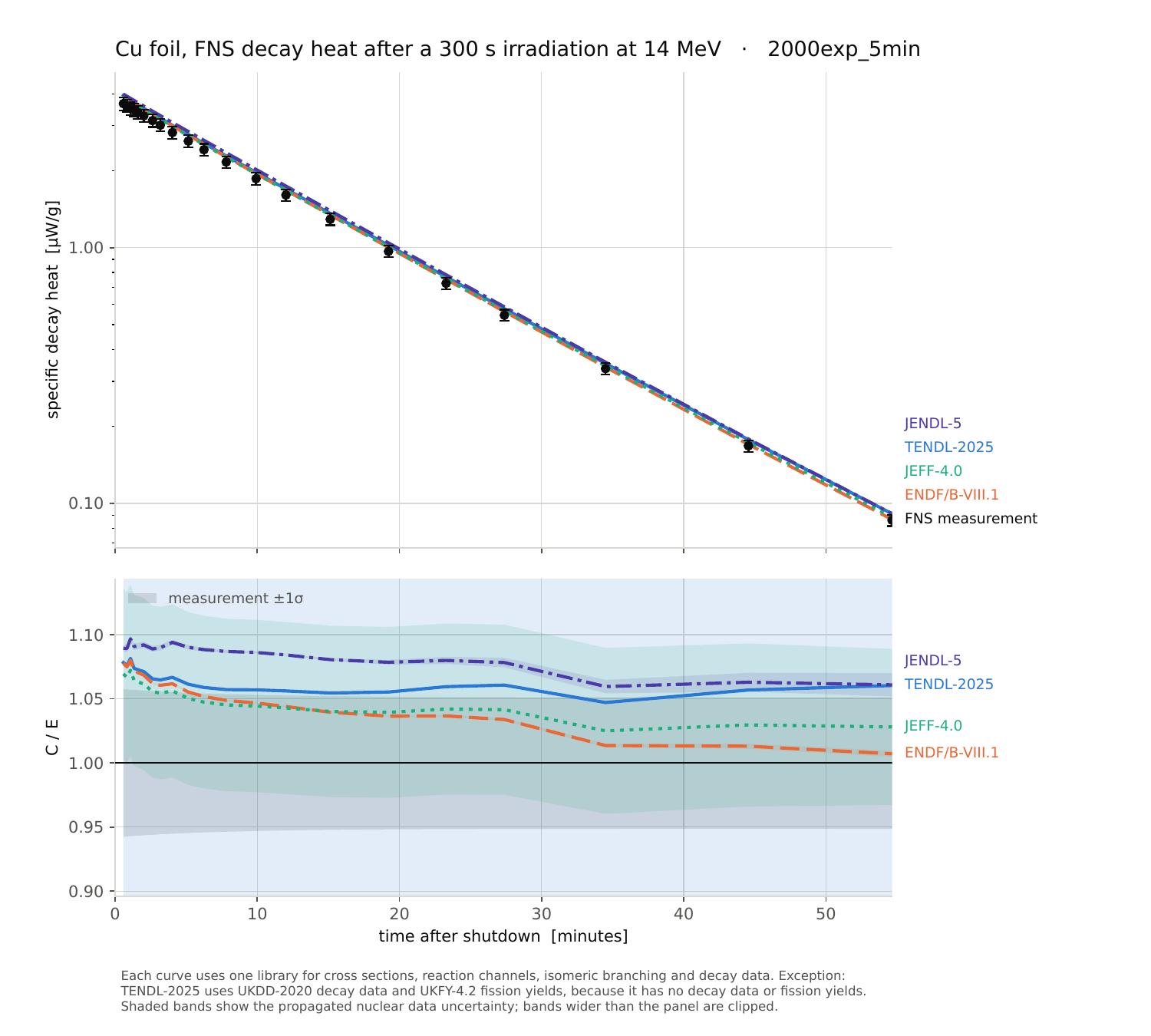}
\caption{Cu: decay heat against the measurement, with C/E beneath. Mean $|C/E-1|$ by library: JEFF-4.0 5\%, ENDF/B-VIII.1 5\%, TENDL-2025 6\%, JENDL-5 8\%.}
\end{figure}
\clearpage

\begin{figure}[H]
\centering
\includegraphics[width=\textwidth,height=0.82\textheight,keepaspectratio]{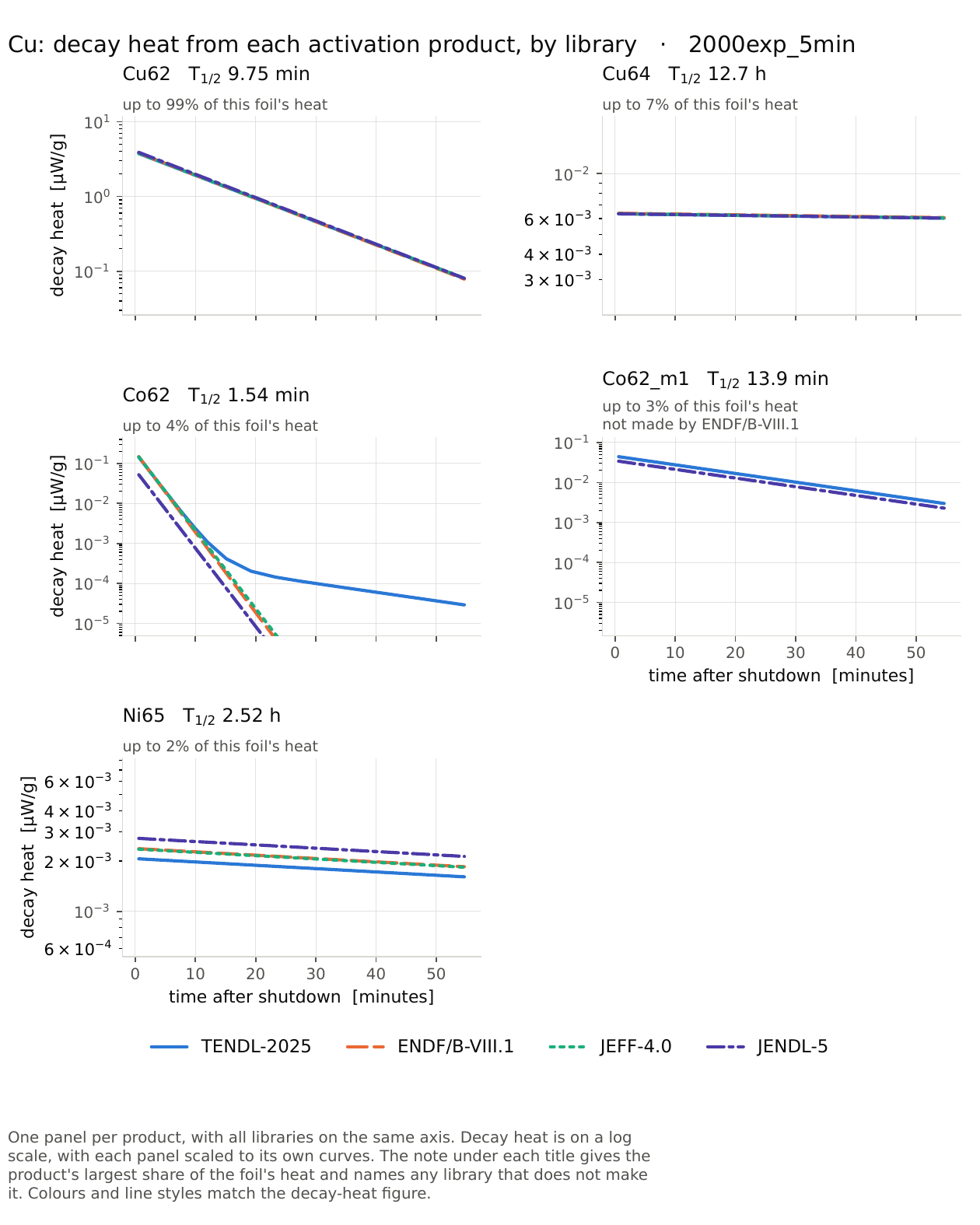}
\caption{Cu: share of the decay heat from each activation product during cooling.}
\end{figure}
\clearpage

\begin{figure}[H]
\centering
\includegraphics[width=\textwidth,height=0.82\textheight,keepaspectratio]{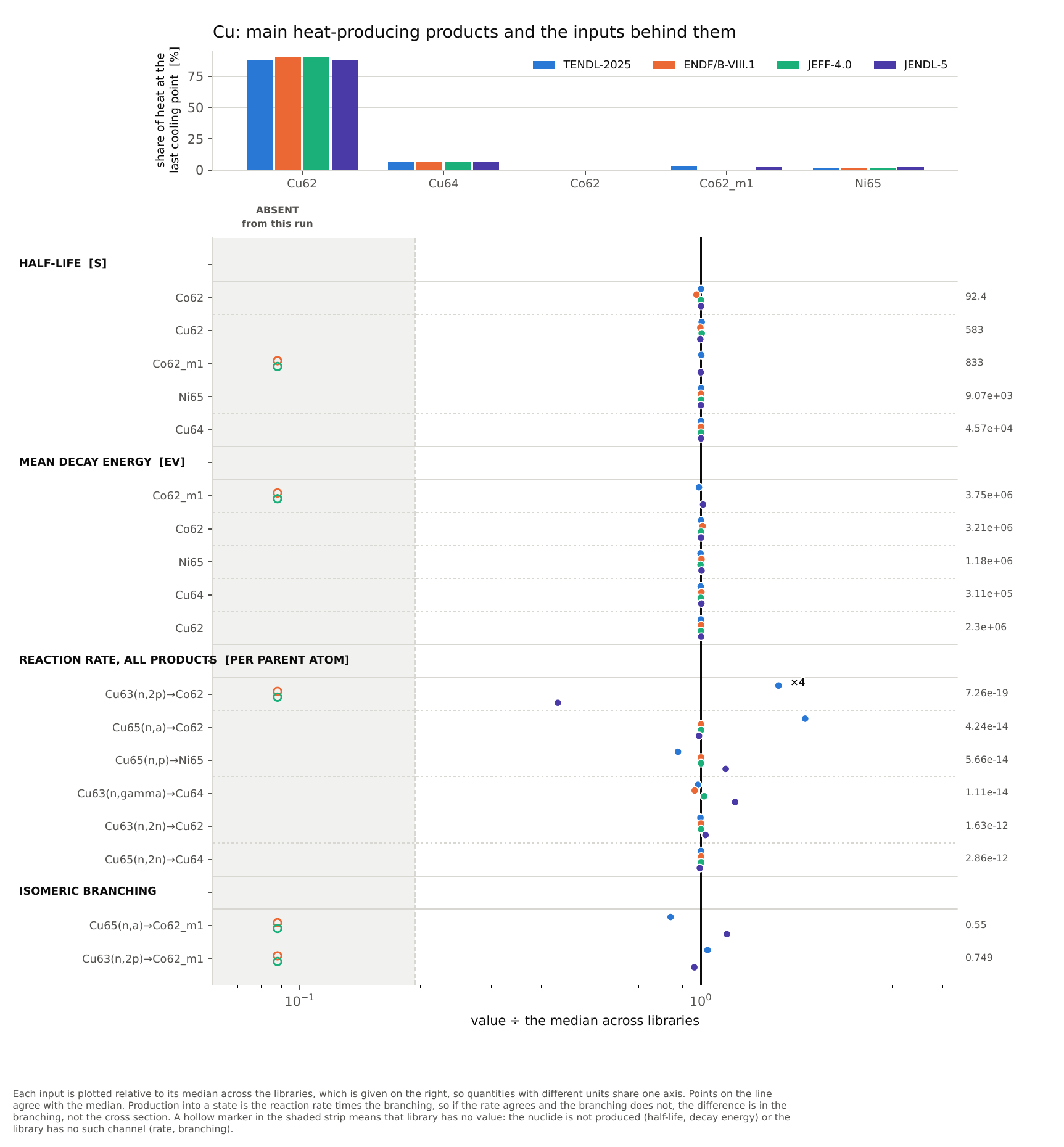}
\caption{Cu: inputs for the main heat-producing products, each relative to its median across the available libraries.}
\end{figure}
\clearpage

\begin{figure}[H]
\centering
\includegraphics[width=\textwidth,height=0.82\textheight,keepaspectratio]{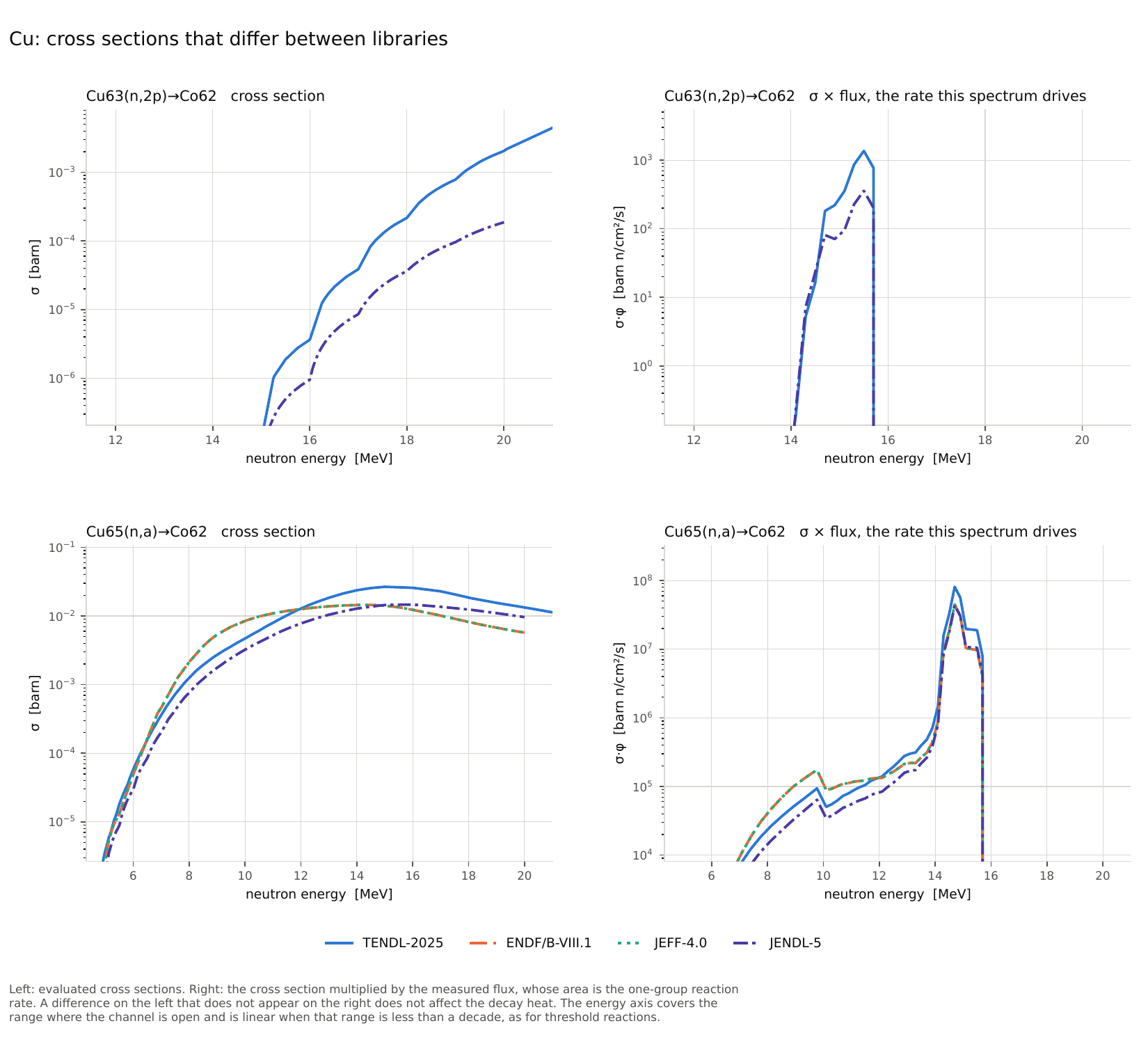}
\caption{Cu: cross sections of the channels whose spectrum-averaged rates differ, with the neutron spectrum and their product.}
\end{figure}
\clearpage

Mean absolute C/E deviations range from 5 to 8 percent. The spread of 3 percentage points is smaller than the median measurement uncertainty of 5 percent, so the measurement cannot separate the libraries.

\subsubsection*{Differences in the nuclear data}
\begin{itemize}\setlength{\itemsep}{6pt}
\item Only TENDL-2025 and JENDL-5 have the total reaction rate of Cu63(n,2p)→Co62, and they differ by a factor of 4: 1.133e-18 against 3.191e-19.\\[2pt]{\itshape This is the total rate of the channel over all its products, so the difference is in the cross section, not in the split between states.}
\item ENDF/B-VIII.1 gives no heat from Co62\_m1; the other libraries do.\\[2pt]{\itshape The heat breakdown alone does not show why. Possible causes are a missing reaction channel, a different state assignment, different decay data or a reporting threshold.}
\end{itemize}
\subsubsection*{Suggested measurements}
\begin{itemize}\setlength{\itemsep}{4pt}
\item Natural Cu has 2 isotopes, so the measured heat combines several parents. Enriched single-isotope targets would distinguish their contributions. The candidate parent is Cu63. Enrichment alone does not isolate a single reaction channel.
\item The reaction rates are averaged over the measured spectrum. Activation measurements at several well-defined neutron energies, including below 14 MeV, would show which energy range the differences come from.
\end{itemize}
\clearpage
\subsection{Dy: Dysprosium}\label{foil:Dy}
\begin{figure}[H]
\centering
\includegraphics[width=\textwidth,height=0.82\textheight,keepaspectratio]{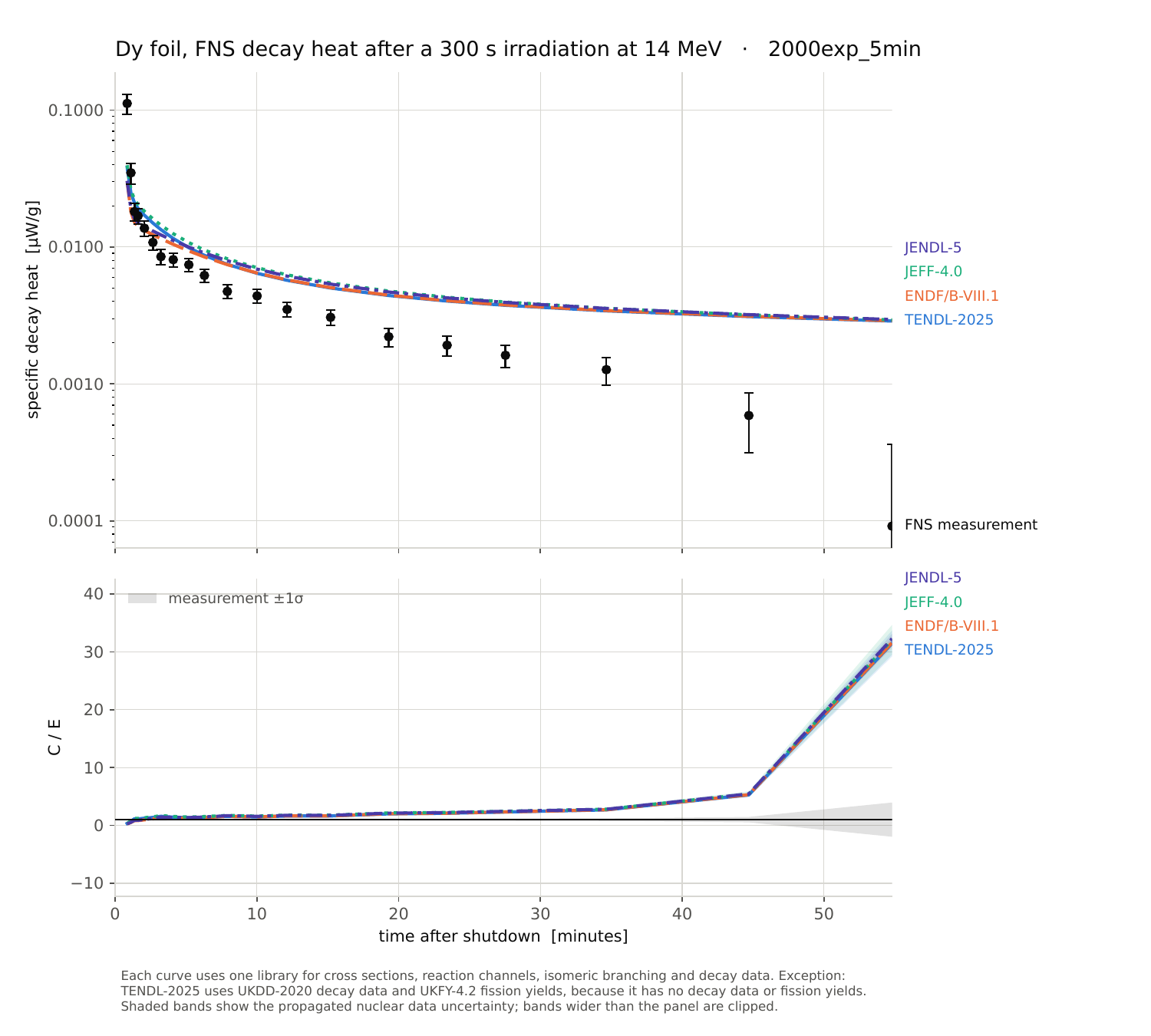}
\caption{Dy: decay heat against the measurement, with C/E beneath. Mean $|C/E-1|$ by library: ENDF/B-VIII.1 227\%, TENDL-2025 229\%, JENDL-5 236\%, JEFF-4.0 242\%.}
\end{figure}
\clearpage

\begin{figure}[H]
\centering
\includegraphics[width=\textwidth,height=0.82\textheight,keepaspectratio]{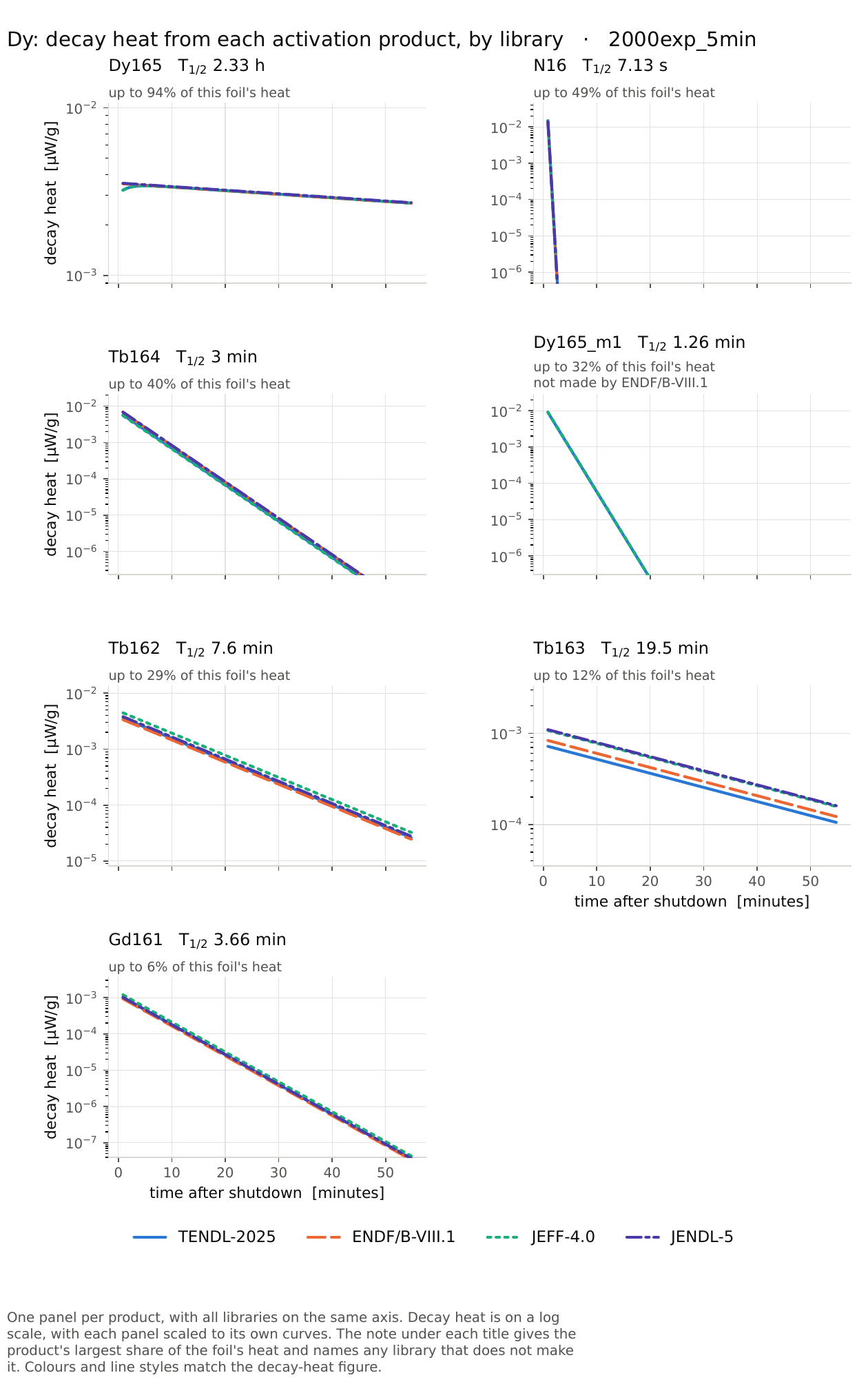}
\caption{Dy: share of the decay heat from each activation product during cooling.}
\end{figure}
\clearpage

\begin{figure}[H]
\centering
\includegraphics[width=\textwidth,height=0.82\textheight,keepaspectratio]{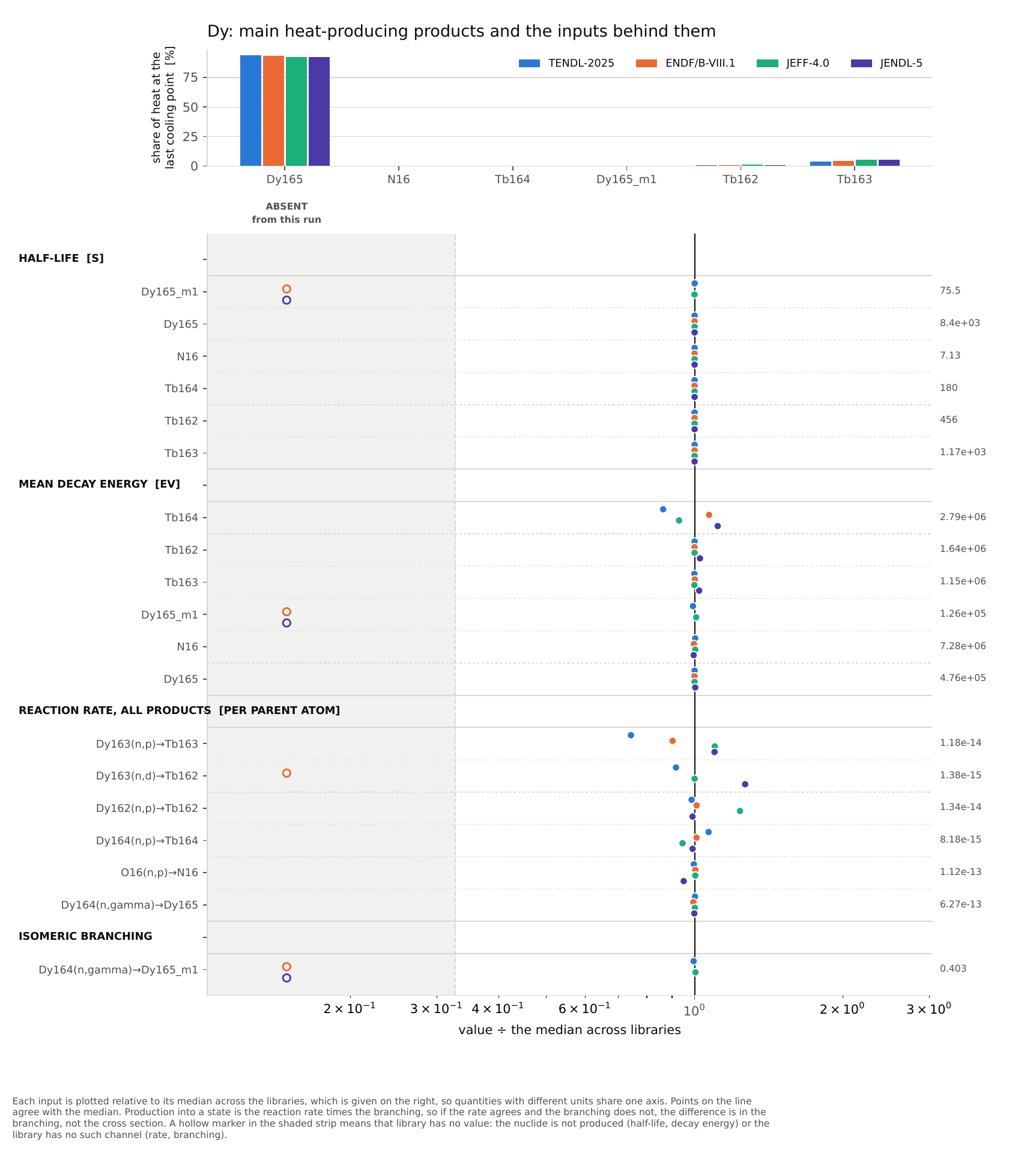}
\caption{Dy: inputs for the main heat-producing products, each relative to its median across the available libraries.}
\end{figure}
\clearpage

The libraries are between 227 and 242 percent from the measurement, which has a median uncertainty of 13 percent. ENDF/B-VIII.1 is closest and JEFF-4.0 furthest.

\subsubsection*{Differences in the nuclear data}
\begin{itemize}\setlength{\itemsep}{6pt}
\item ENDF/B-VIII.1 gives no heat from Dy165\_m1; the other libraries do.\\[2pt]{\itshape The heat breakdown alone does not show why. Possible causes are a missing reaction channel, a different state assignment, different decay data or a reporting threshold.}
\end{itemize}
\subsubsection*{Suggested measurements}
\begin{itemize}\setlength{\itemsep}{4pt}
\item Natural Dy has 7 isotopes, so the measured heat combines several parents. Enriched single-isotope targets would distinguish their contributions. Enrichment alone does not isolate a single reaction channel.
\item The sample is a compound that also contains O. Their activation adds to the measured heat; a metal or purer sample would remove it.
\end{itemize}
\clearpage
\subsection{Er: Erbium}\label{foil:Er}
\begin{figure}[H]
\centering
\includegraphics[width=\textwidth,height=0.82\textheight,keepaspectratio]{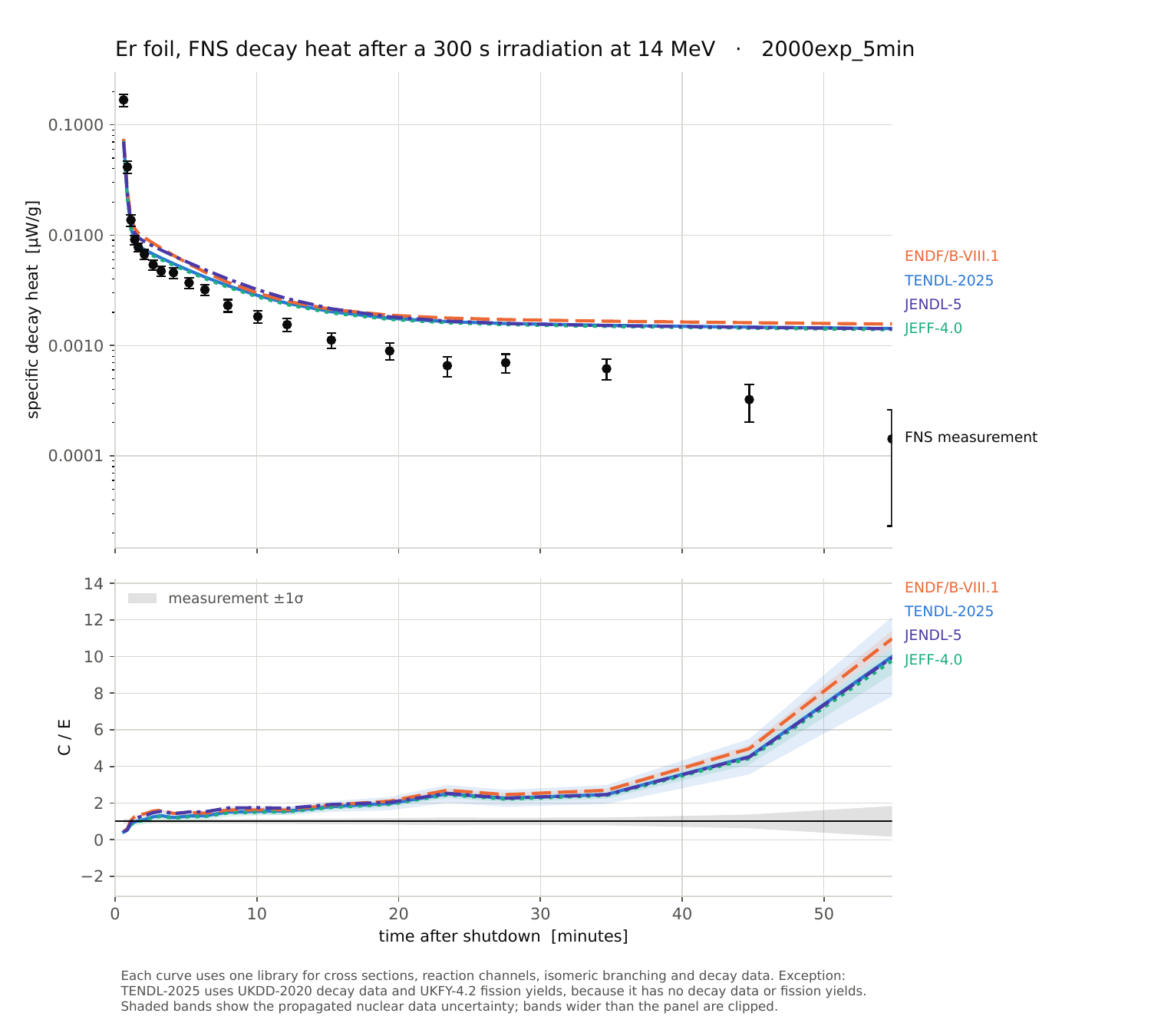}
\caption{Er: decay heat against the measurement, with C/E beneath. Mean $|C/E-1|$ by library: JEFF-4.0 105\%, TENDL-2025 109\%, JENDL-5 120\%, ENDF/B-VIII.1 130\%.}
\end{figure}
\clearpage

\begin{figure}[H]
\centering
\includegraphics[width=\textwidth,height=0.82\textheight,keepaspectratio]{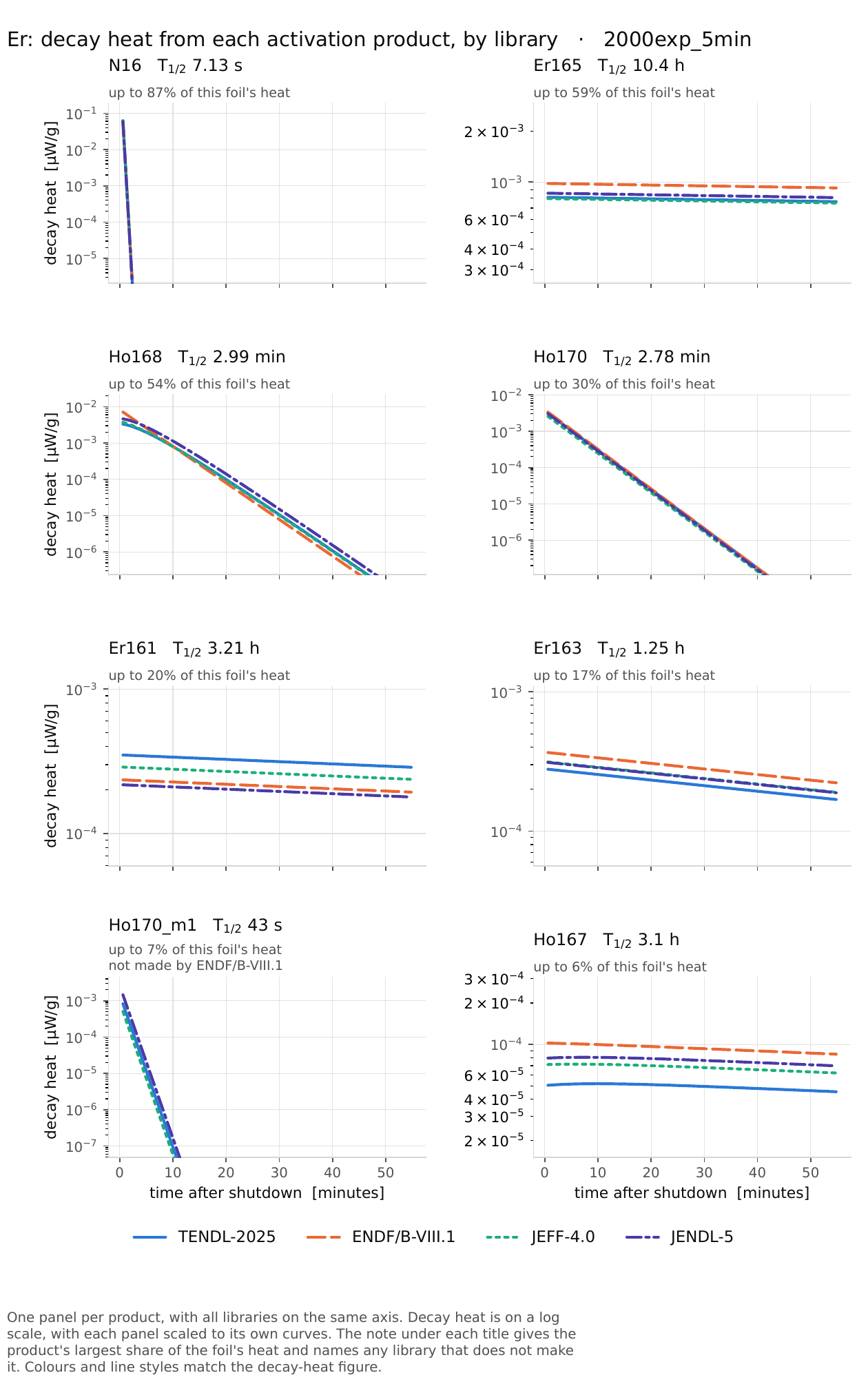}
\caption{Er: share of the decay heat from each activation product during cooling.}
\end{figure}
\clearpage

\begin{figure}[H]
\centering
\includegraphics[width=\textwidth,height=0.82\textheight,keepaspectratio]{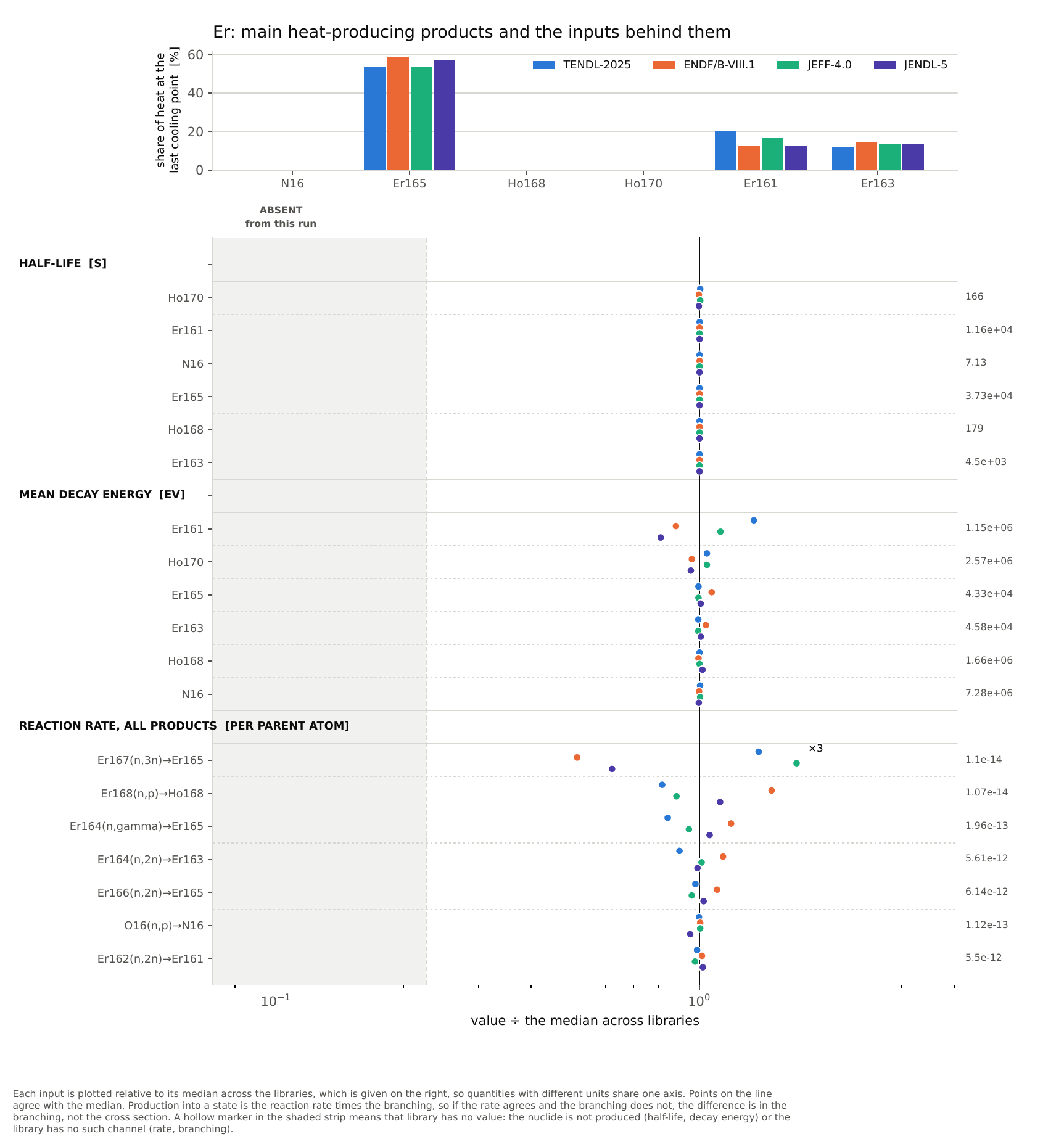}
\caption{Er: inputs for the main heat-producing products, each relative to its median across the available libraries.}
\end{figure}
\clearpage

\begin{figure}[H]
\centering
\includegraphics[width=\textwidth,height=0.82\textheight,keepaspectratio]{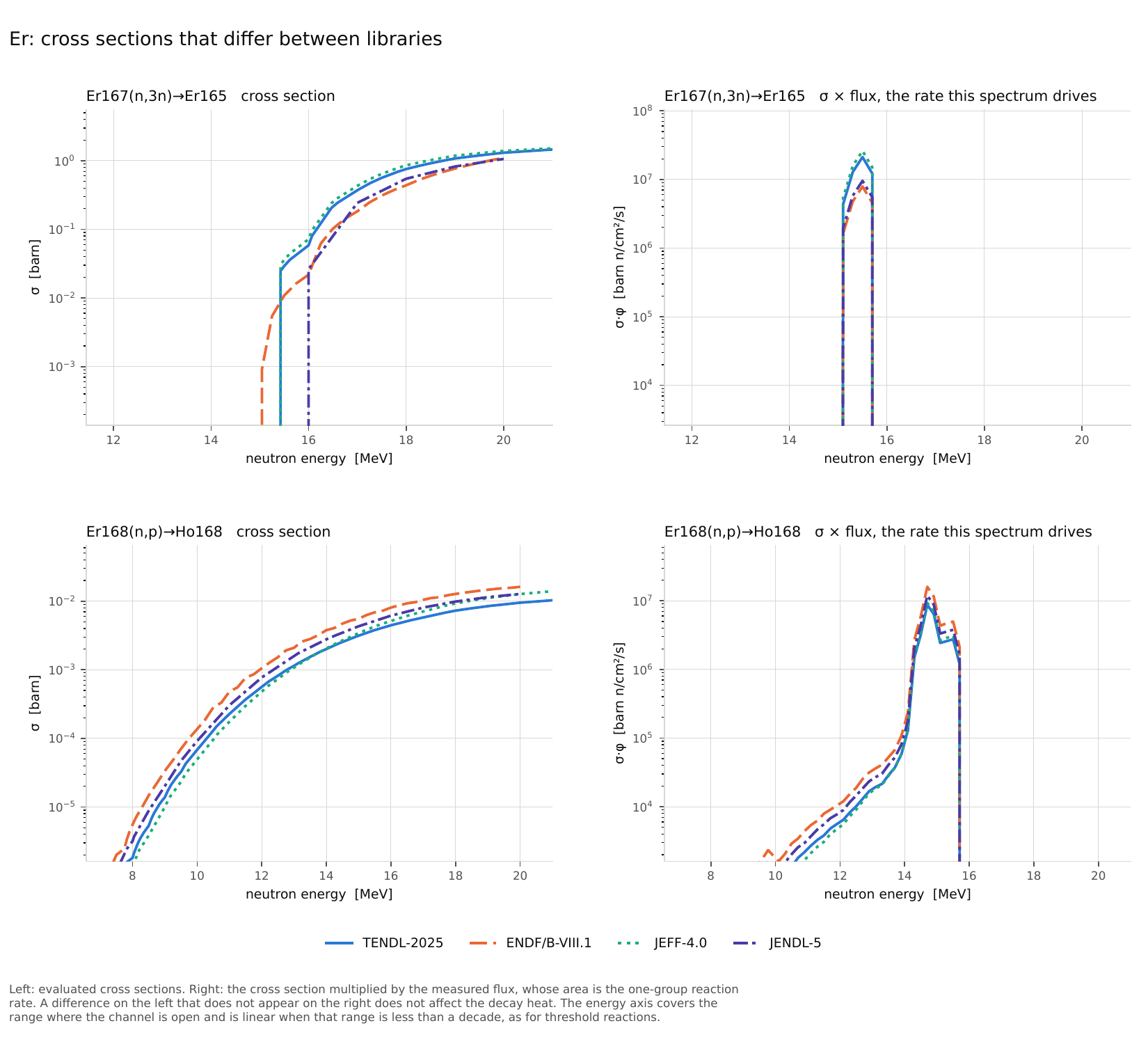}
\caption{Er: cross sections of the channels whose spectrum-averaged rates differ, with the neutron spectrum and their product.}
\end{figure}
\clearpage

The libraries are between 105 and 130 percent from the measurement, which has a median uncertainty of 13 percent. JEFF-4.0 is closest and ENDF/B-VIII.1 furthest.

\subsubsection*{Differences in the nuclear data}
\begin{itemize}\setlength{\itemsep}{6pt}
\item ENDF/B-VIII.1 gives the total reaction rate of Er167(n,3n)→Er165 as 5.655e-15, against 1.517e-14 for the median of the others, a factor of 3.\\[2pt]{\itshape This is the total rate of the channel over all its products, so the difference is in the cross section, not in the split between states.}
\end{itemize}
\subsubsection*{Suggested measurements}
\begin{itemize}\setlength{\itemsep}{4pt}
\item Natural Er has 6 isotopes, so the measured heat combines several parents. Enriched single-isotope targets would distinguish their contributions. The candidate parent is Er167. Enrichment alone does not isolate a single reaction channel.
\item The sample is a compound that also contains O. Their activation adds to the measured heat; a metal or purer sample would remove it.
\item This foil makes Er169 (half-life 9.38 days), which gives up to 3 percent of the calculated heat. A five-minute irradiation and 55 minute measurement only see the start of this contribution. There is no seven-hour FNS measurement of this foil. A longer irradiation followed for days to months would be needed to test this product's production and decay.
\item The reaction rates are averaged over the measured spectrum. Activation measurements at several well-defined neutron energies, including below 14 MeV, would show which energy range the differences come from.
\end{itemize}
\clearpage
\subsection{Eu: Europium}\label{foil:Eu}
\begin{figure}[H]
\centering
\includegraphics[width=\textwidth,height=0.82\textheight,keepaspectratio]{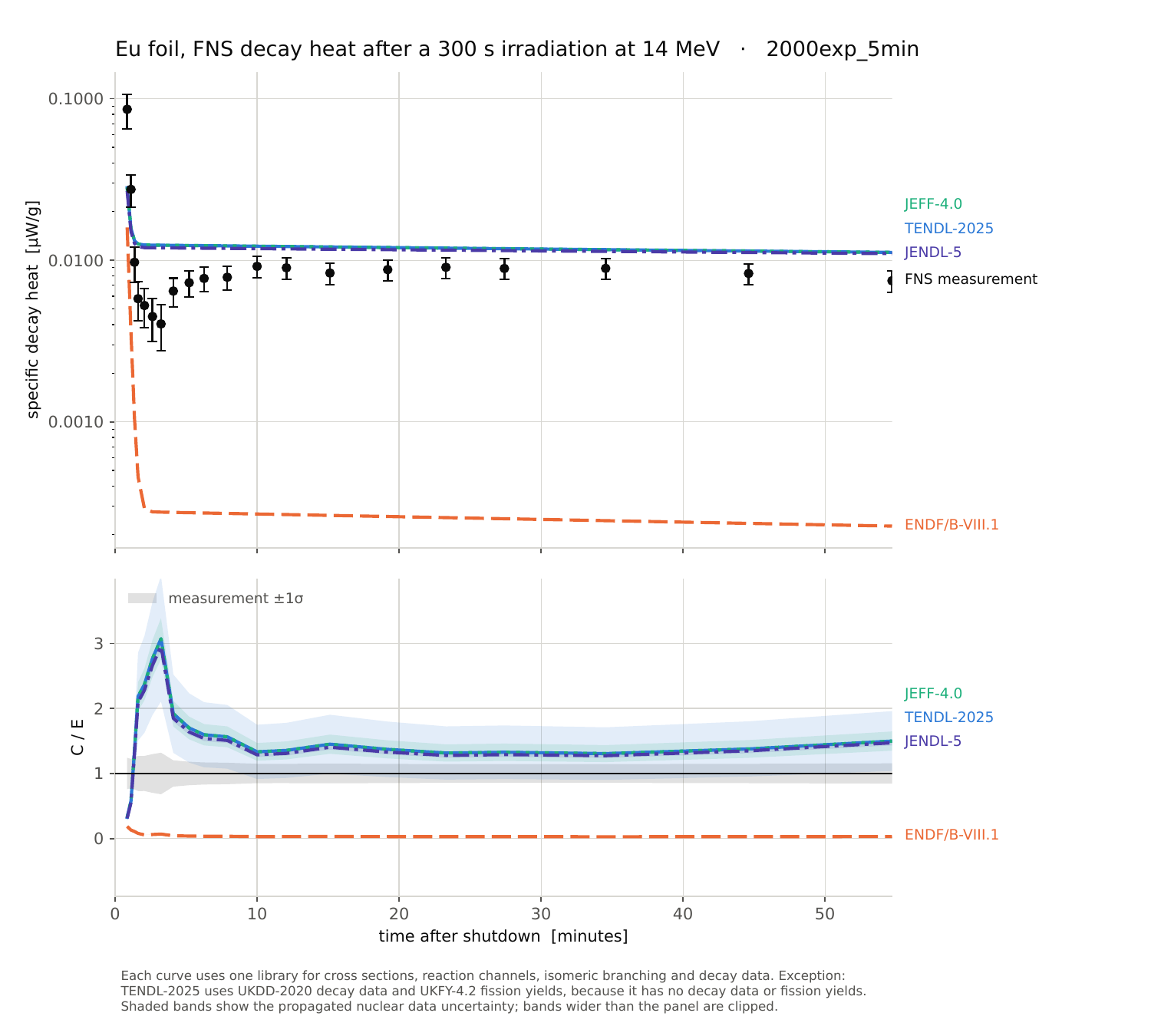}
\caption{Eu: decay heat against the measurement, with C/E beneath. Mean $|C/E-1|$ by library: JENDL-5 65\%, JEFF-4.0 70\%, TENDL-2025 70\%, ENDF/B-VIII.1 94\%.}
\end{figure}
\clearpage

\begin{figure}[H]
\centering
\includegraphics[width=\textwidth,height=0.82\textheight,keepaspectratio]{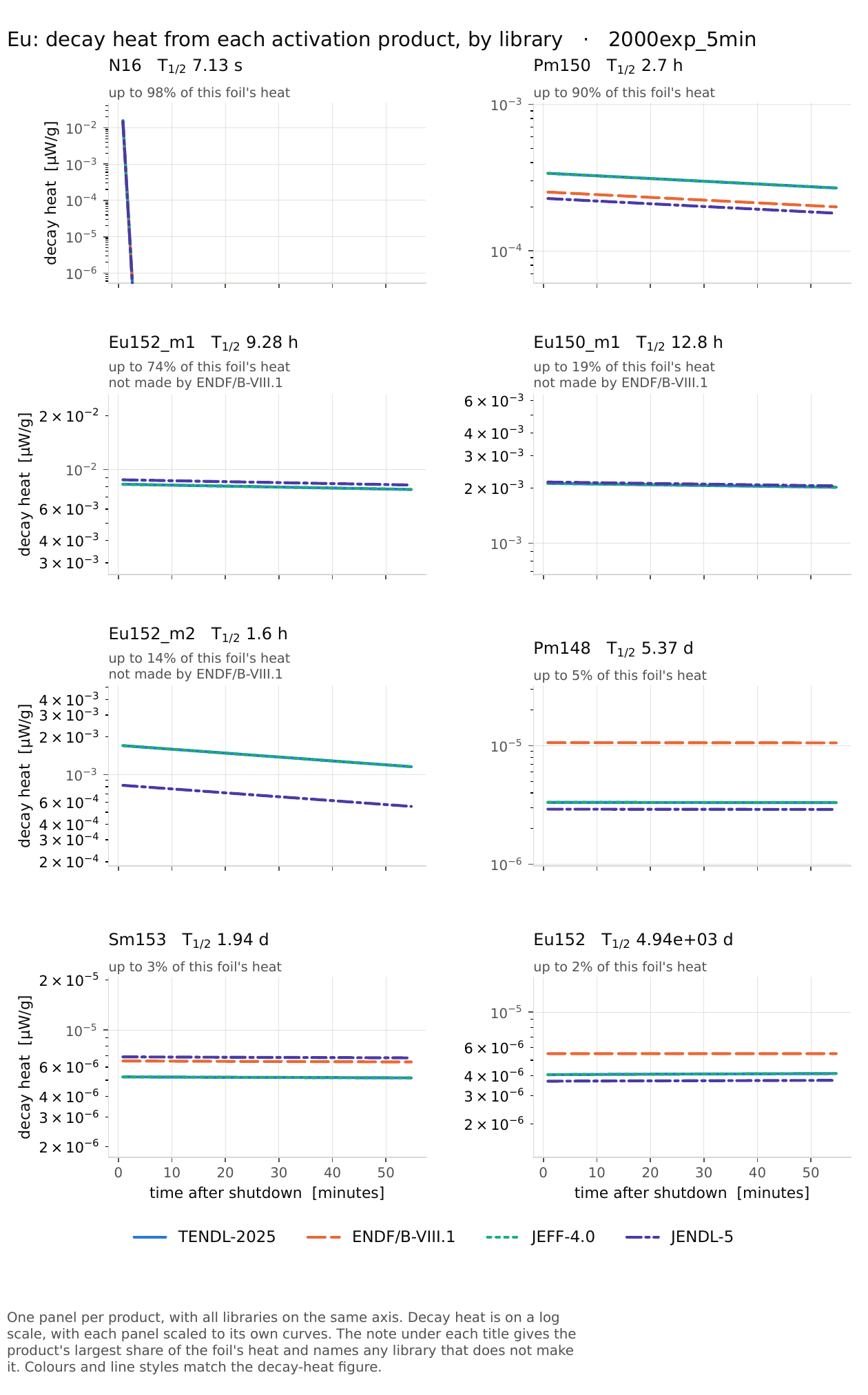}
\caption{Eu: share of the decay heat from each activation product during cooling.}
\end{figure}
\clearpage

\begin{figure}[H]
\centering
\includegraphics[width=\textwidth,height=0.82\textheight,keepaspectratio]{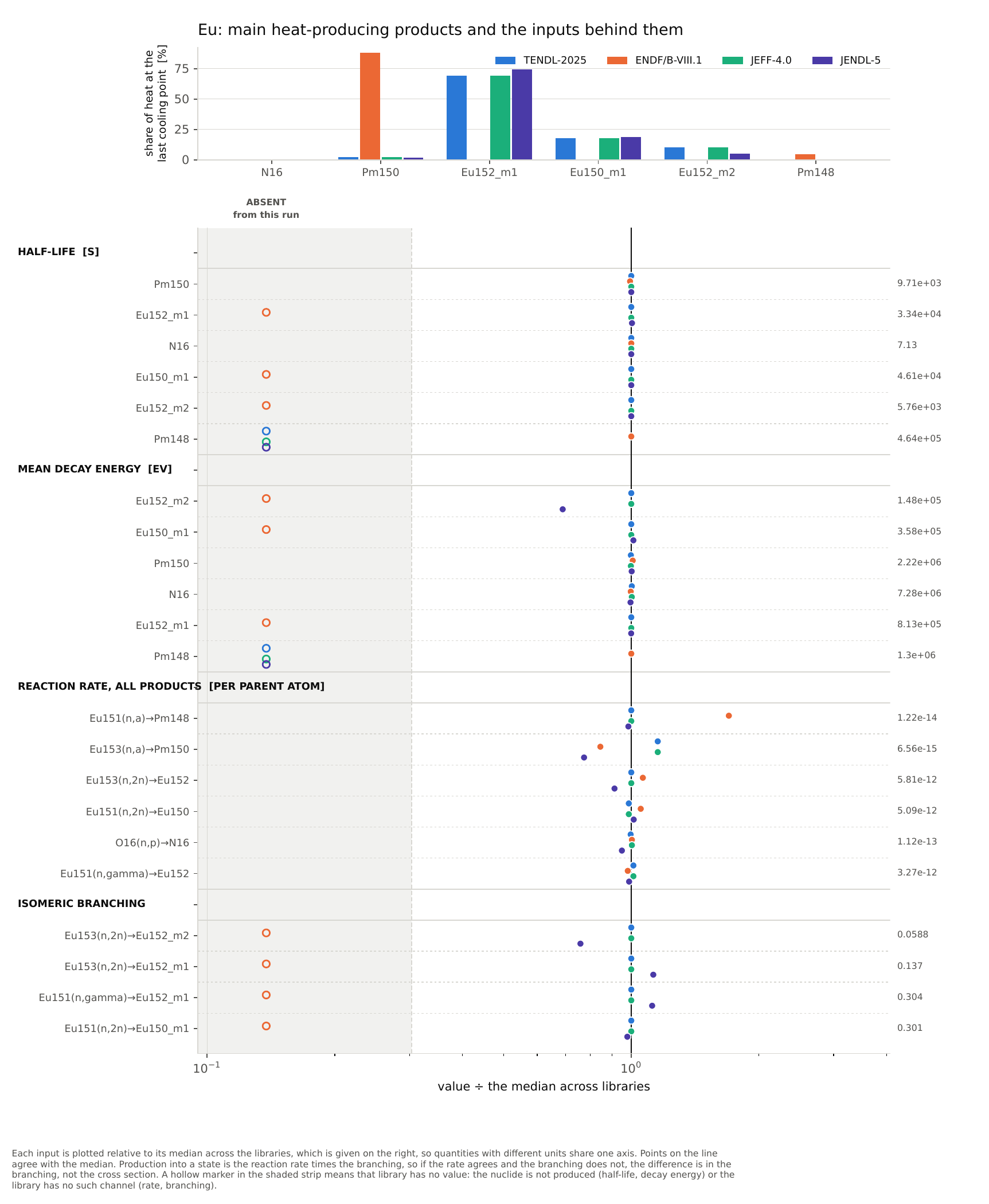}
\caption{Eu: inputs for the main heat-producing products, each relative to its median across the available libraries.}
\end{figure}
\clearpage

\begin{figure}[H]
\centering
\includegraphics[width=\textwidth,height=0.82\textheight,keepaspectratio]{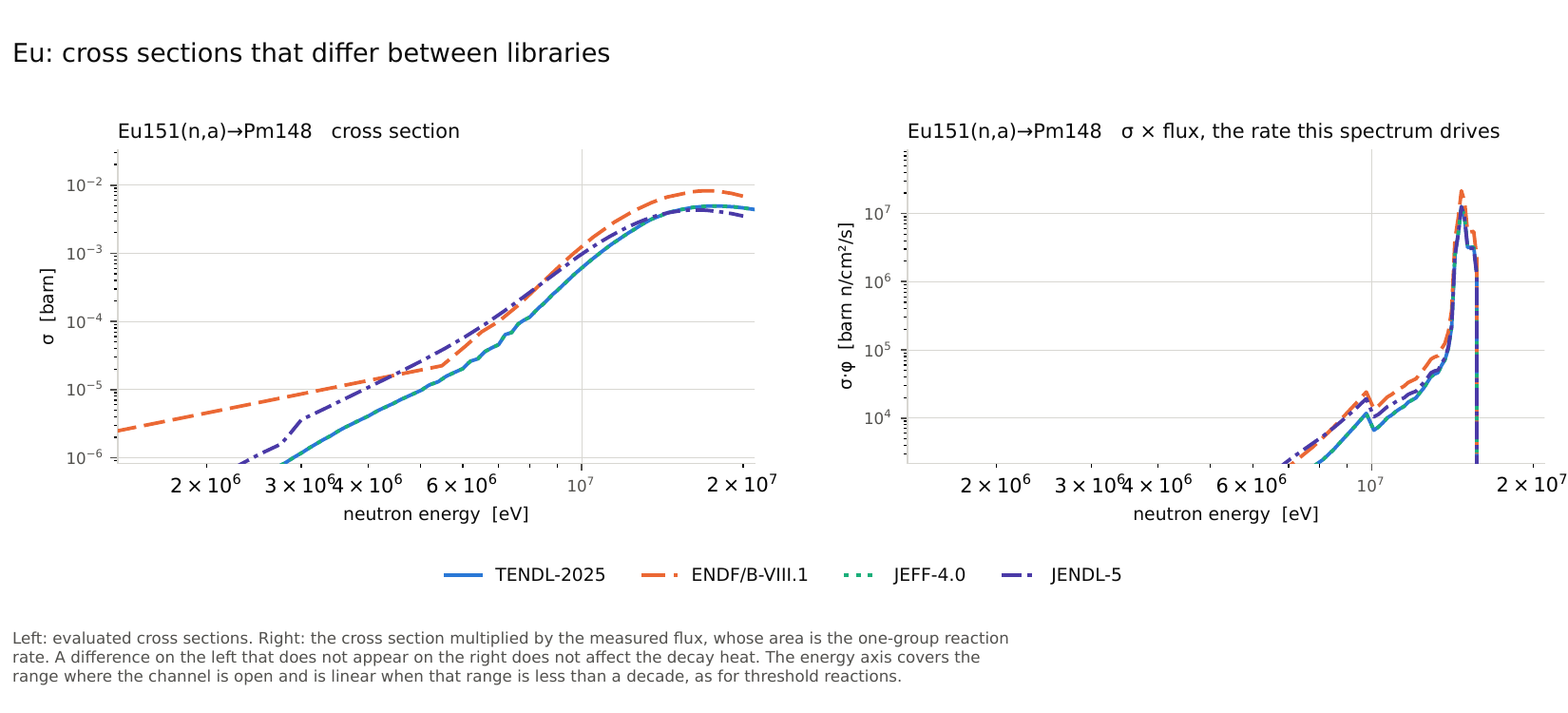}
\caption{Eu: cross sections of the channels whose spectrum-averaged rates differ, with the neutron spectrum and their product.}
\end{figure}
\clearpage

The libraries are between 65 and 94 percent from the measurement, which has a median uncertainty of 17 percent. JENDL-5 is closest and ENDF/B-VIII.1 furthest.

\subsubsection*{Differences in the nuclear data}
\begin{itemize}\setlength{\itemsep}{6pt}
\item ENDF/B-VIII.1 gives no heat from Eu152\_m1; the other libraries do.\\[2pt]{\itshape The heat breakdown alone does not show why. Possible causes are a missing reaction channel, a different state assignment, different decay data or a reporting threshold.}
\item ENDF/B-VIII.1 gives no heat from Eu150\_m1; the other libraries do.
\item ENDF/B-VIII.1 gives no heat from Eu152\_m2; the other libraries do.
\end{itemize}
\subsubsection*{Suggested measurements}
\begin{itemize}\setlength{\itemsep}{4pt}
\item Natural Eu has 2 isotopes, so the measured heat combines several parents. Enriched single-isotope targets would distinguish their contributions. Enrichment alone does not isolate a single reaction channel.
\item The sample is a compound that also contains O. Their activation adds to the measured heat; a metal or purer sample would remove it.
\item This foil makes Pm148 (half-life 5.37 days), which gives up to 5 percent of the calculated heat. A five-minute irradiation and 55 minute measurement only see the start of this contribution. There is no seven-hour FNS measurement of this foil. A longer irradiation followed for days to months would be needed to test this product's production and decay.
\end{itemize}
\clearpage
\subsection{F: Fluorine}\label{foil:F}
\begin{figure}[H]
\centering
\includegraphics[width=\textwidth,height=0.82\textheight,keepaspectratio]{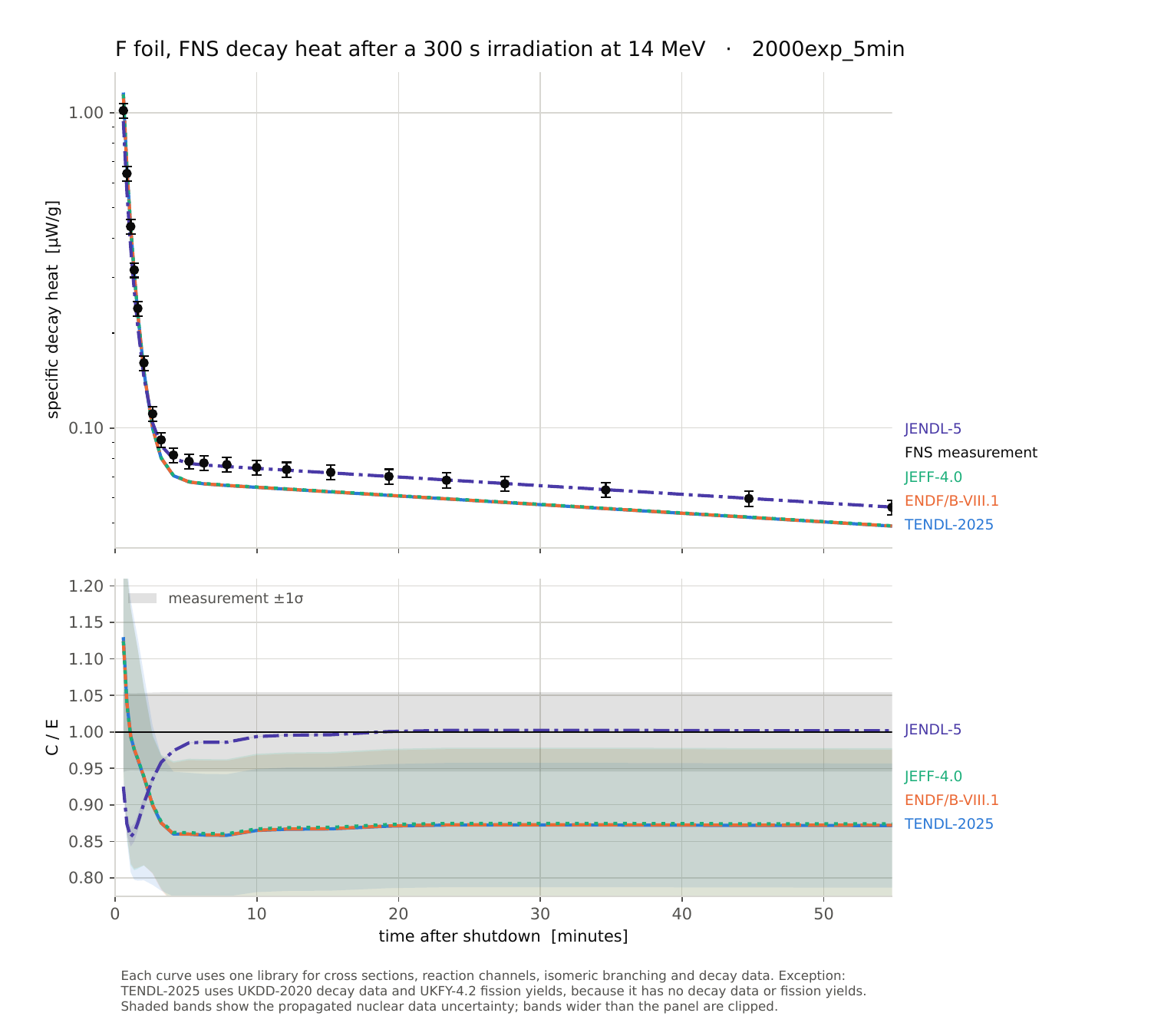}
\caption{F: decay heat against the measurement, with C/E beneath. Mean $|C/E-1|$ by library: JENDL-5 4\%, JEFF-4.0 11\%, ENDF/B-VIII.1 11\%, TENDL-2025 11\%.}
\end{figure}
\clearpage

\begin{figure}[H]
\centering
\includegraphics[width=\textwidth,height=0.82\textheight,keepaspectratio]{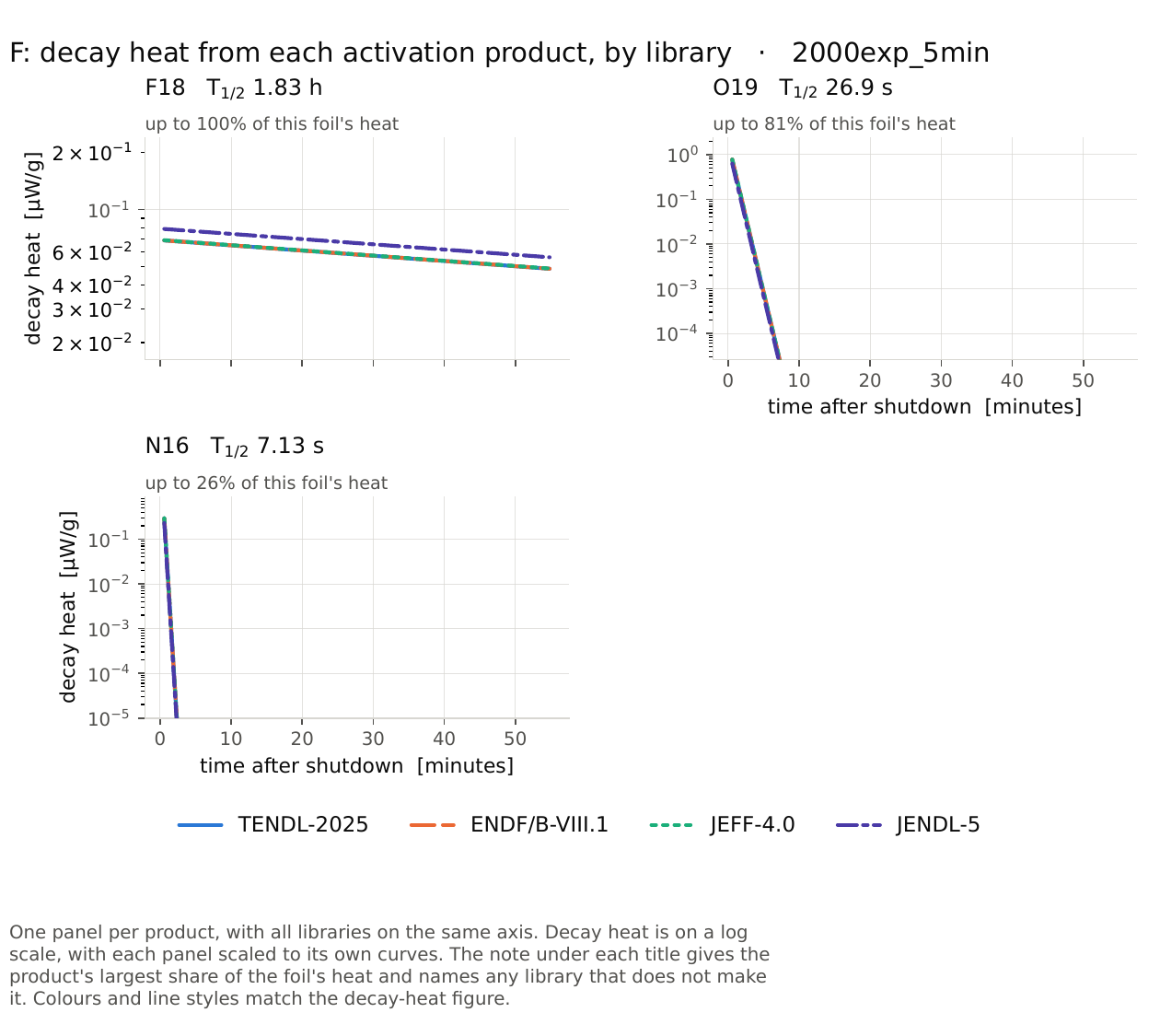}
\caption{F: share of the decay heat from each activation product during cooling.}
\end{figure}
\clearpage

\begin{figure}[H]
\centering
\includegraphics[width=\textwidth,height=0.82\textheight,keepaspectratio]{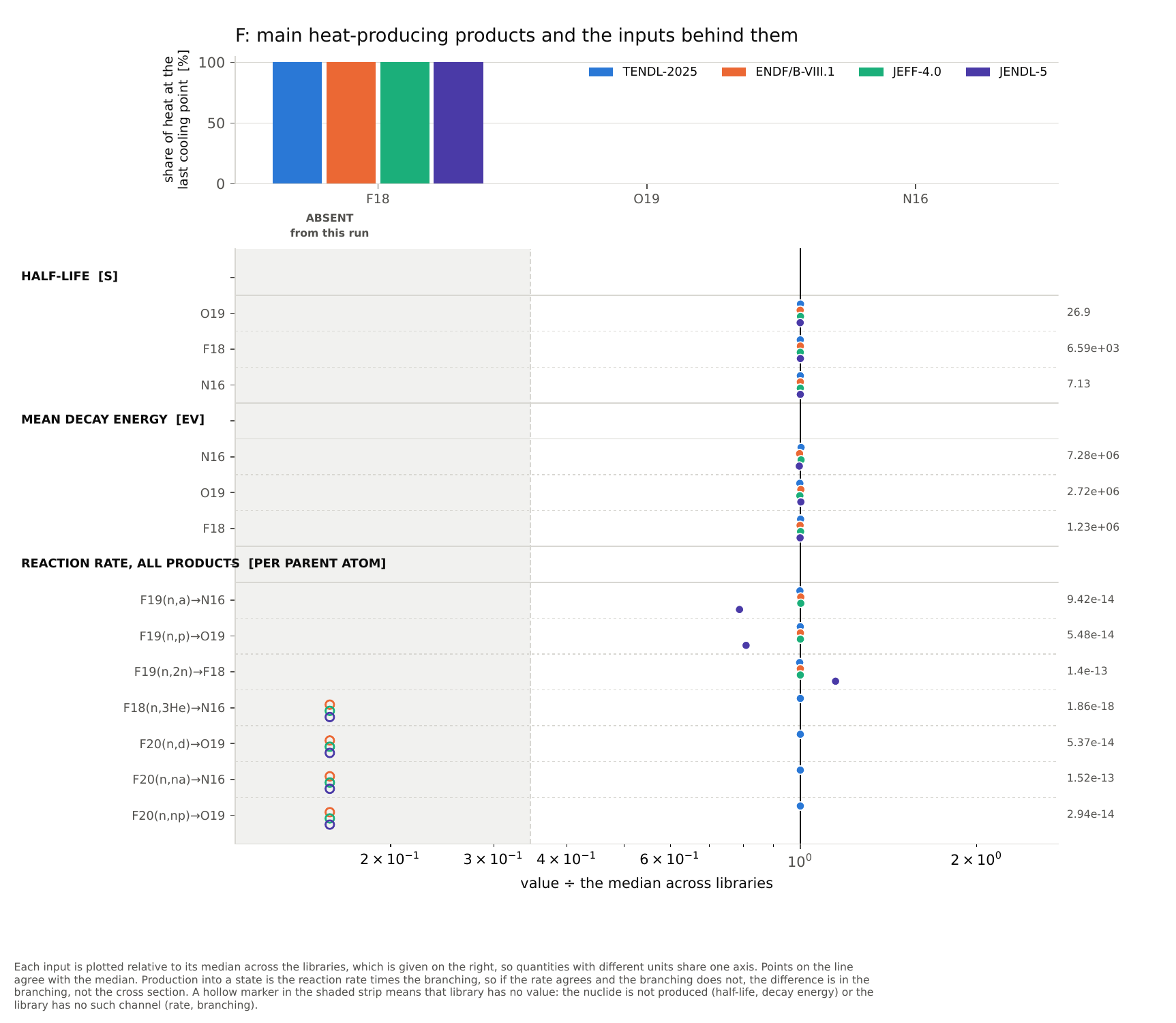}
\caption{F: inputs for the main heat-producing products, each relative to its median across the available libraries.}
\end{figure}
\clearpage

The libraries are between 4 and 11 percent from the measurement, which has a median uncertainty of 5 percent. JENDL-5 is closest and TENDL-2025 furthest.

\subsubsection*{Suggested measurements}
\begin{itemize}\setlength{\itemsep}{4pt}
\item No input differs enough between the libraries to be listed. An error shared by all of them, or a smaller difference, is still possible.
\end{itemize}
\clearpage
\subsection{Fe: Iron}\label{foil:Fe}
\begin{figure}[H]
\centering
\includegraphics[width=\textwidth,height=0.82\textheight,keepaspectratio]{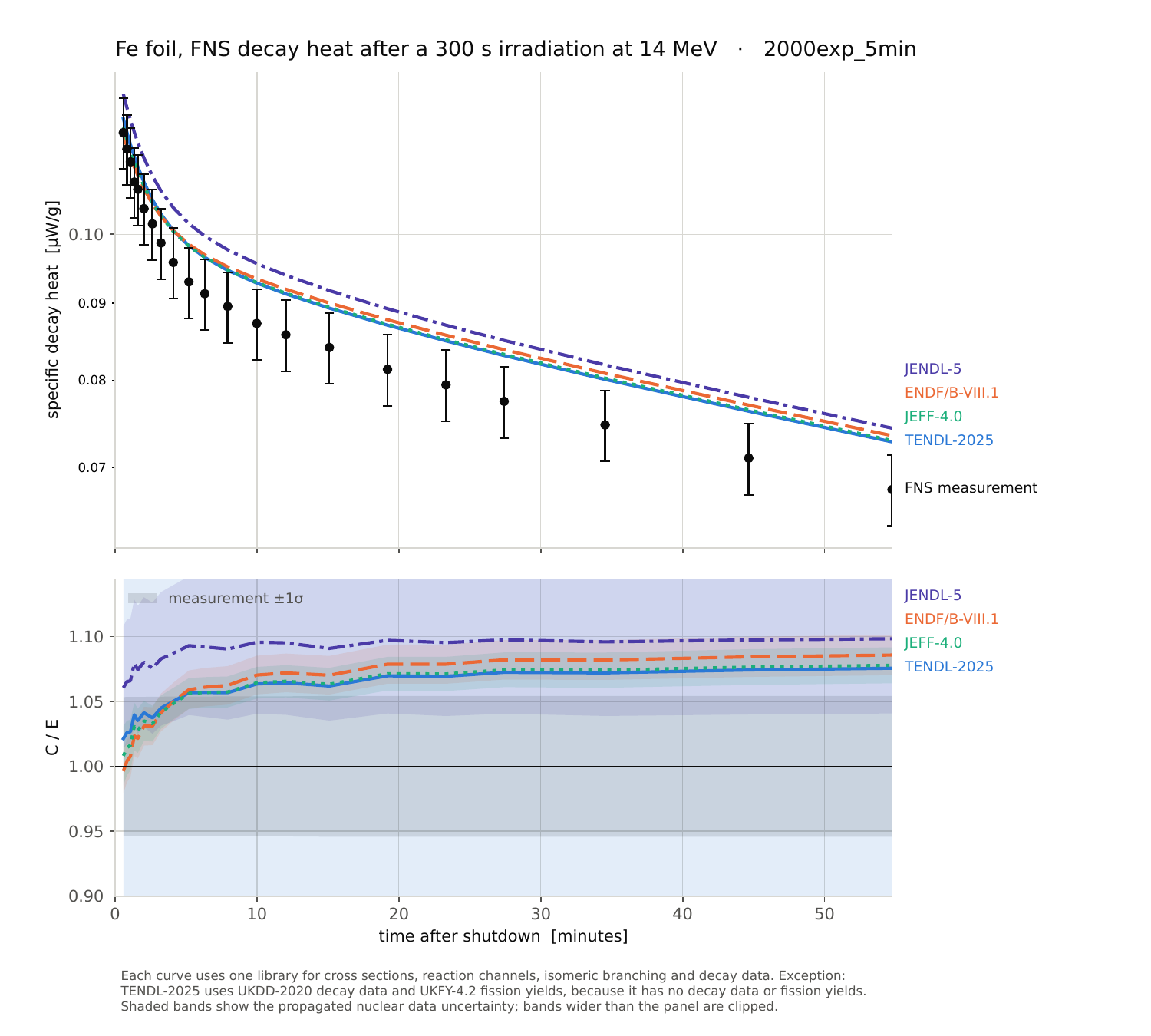}
\caption{Fe: decay heat against the measurement, with C/E beneath. Mean $|C/E-1|$ by library: JEFF-4.0 5\%, ENDF/B-VIII.1 5\%, TENDL-2025 5\%, JENDL-5 9\%.}
\end{figure}
\clearpage

\begin{figure}[H]
\centering
\includegraphics[width=\textwidth,height=0.82\textheight,keepaspectratio]{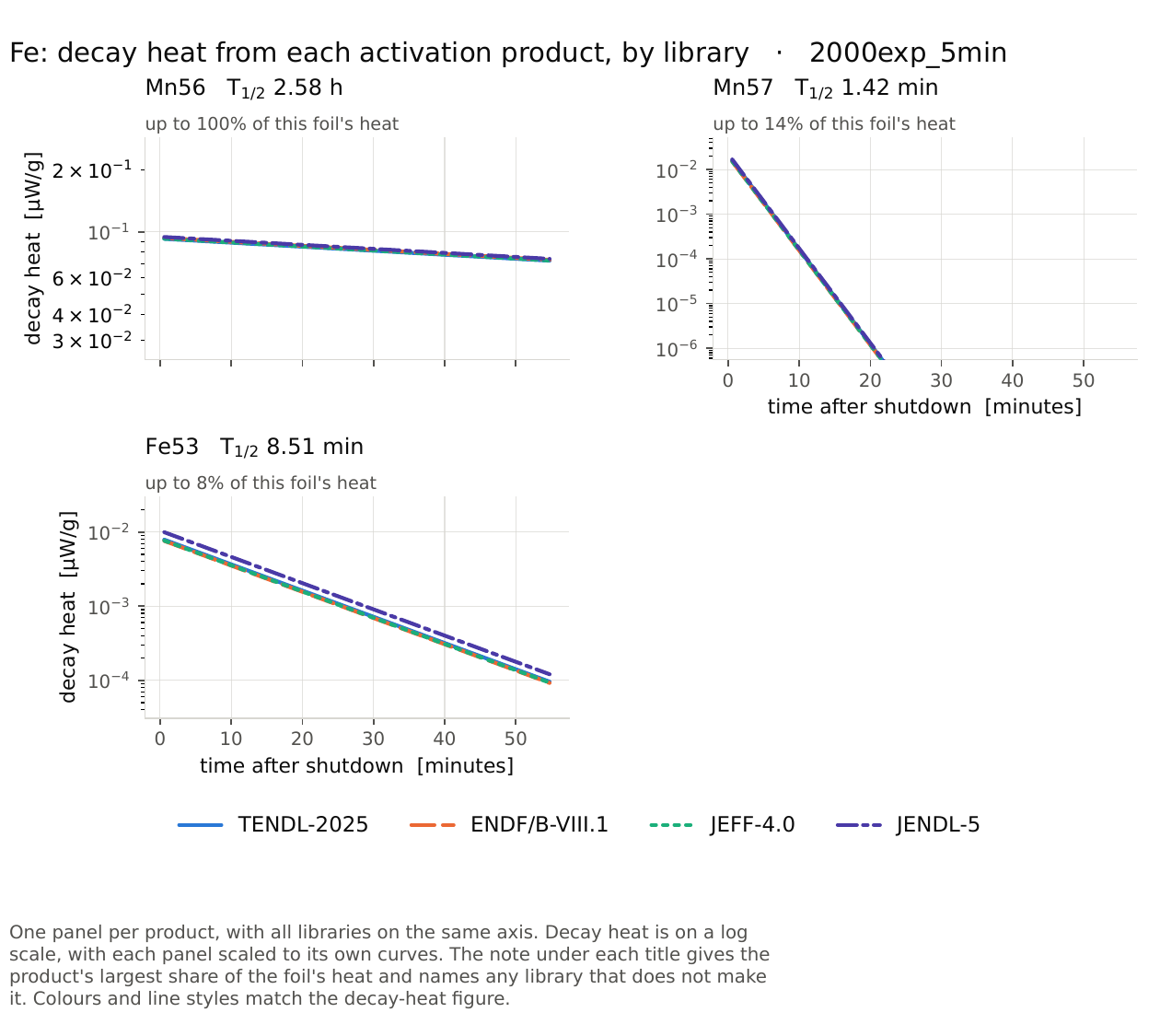}
\caption{Fe: share of the decay heat from each activation product during cooling.}
\end{figure}
\clearpage

\begin{figure}[H]
\centering
\includegraphics[width=\textwidth,height=0.82\textheight,keepaspectratio]{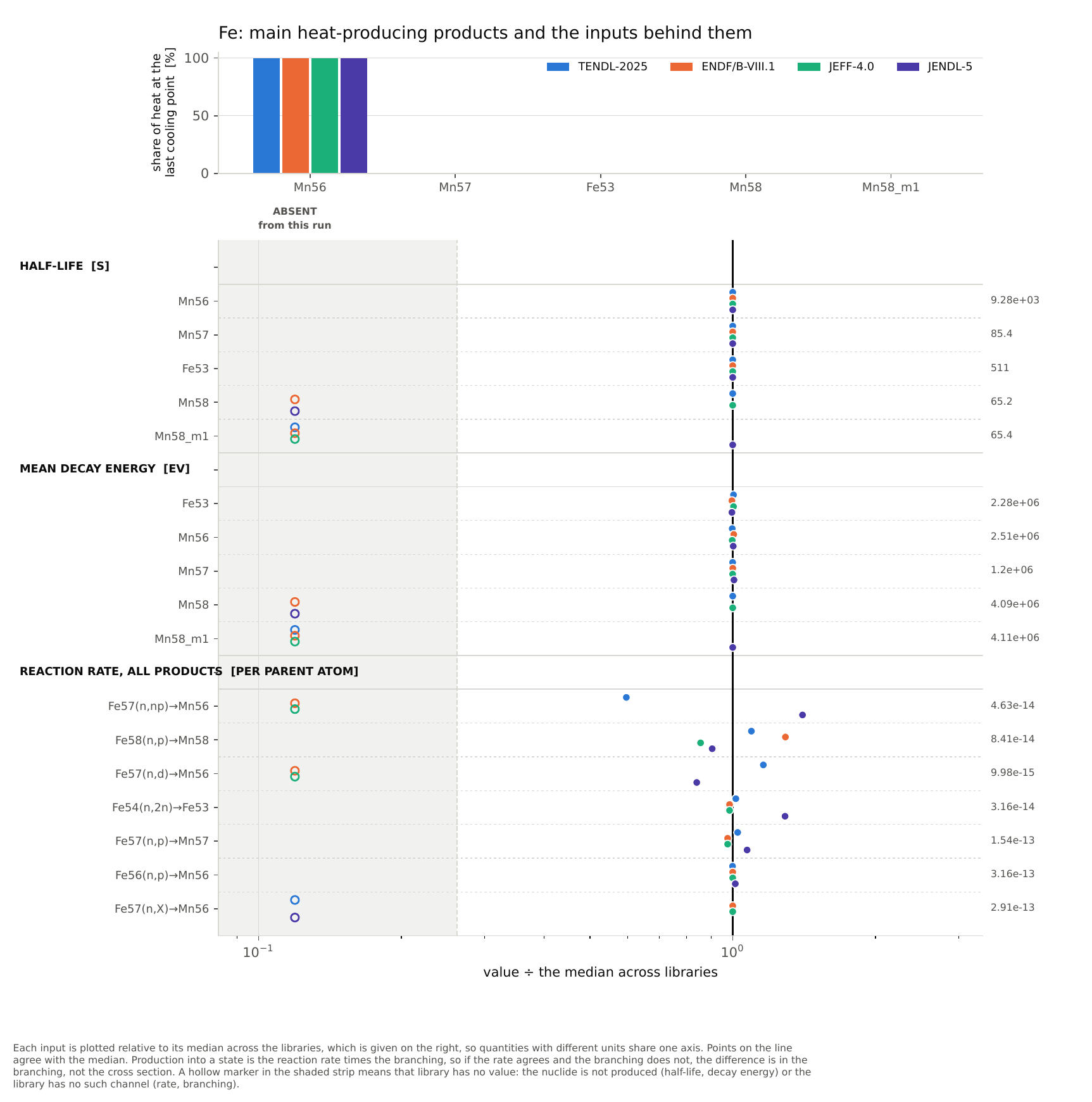}
\caption{Fe: inputs for the main heat-producing products, each relative to its median across the available libraries.}
\end{figure}
\clearpage

\begin{figure}[H]
\centering
\includegraphics[width=\textwidth,height=0.82\textheight,keepaspectratio]{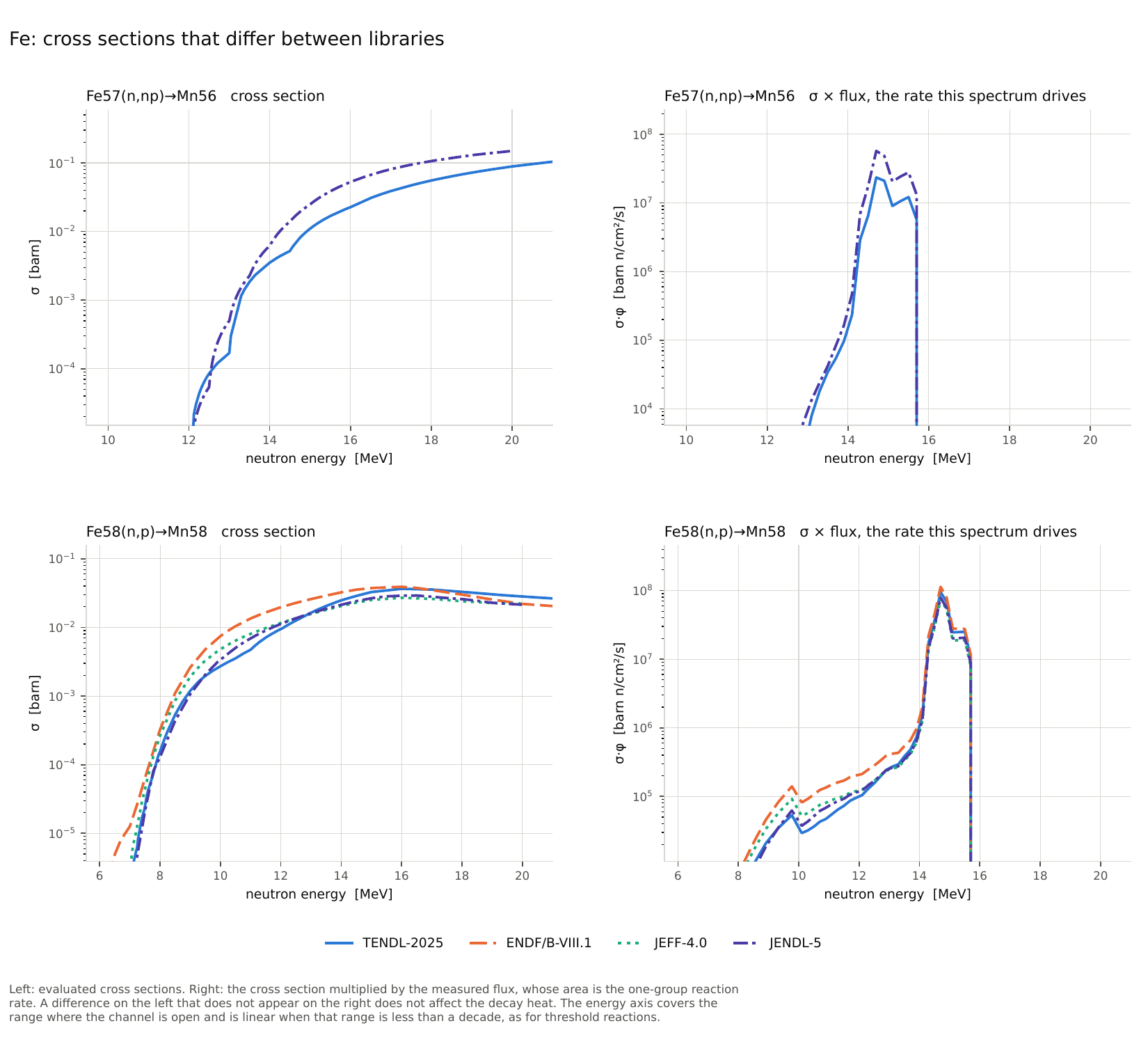}
\caption{Fe: cross sections of the channels whose spectrum-averaged rates differ, with the neutron spectrum and their product.}
\end{figure}
\clearpage

Mean absolute C/E deviations range from 5 to 9 percent. The spread of 4 percentage points is smaller than the median measurement uncertainty of 5 percent, so the measurement cannot separate the libraries.

\subsubsection*{Differences in the nuclear data}
\begin{itemize}\setlength{\itemsep}{6pt}
\item Only JENDL-5 and TENDL-2025 have the total reaction rate of Fe57(n,np)→Mn56, and they differ by a factor of 2: 6.503e-14 against 2.764e-14.\\[2pt]{\itshape This is the total rate of the channel over all its products, so the difference is in the cross section, not in the split between states.}
\end{itemize}
\subsubsection*{Suggested measurements}
\begin{itemize}\setlength{\itemsep}{4pt}
\item Natural Fe has 4 isotopes, so the measured heat combines several parents. Enriched single-isotope targets would distinguish their contributions. The candidate parent is Fe57. Enrichment alone does not isolate a single reaction channel.
\item The reaction rates are averaged over the measured spectrum. Activation measurements at several well-defined neutron energies, including below 14 MeV, would show which energy range the differences come from.
\end{itemize}
\clearpage
\subsection{Ga: Gallium}\label{foil:Ga}
\begin{figure}[H]
\centering
\includegraphics[width=\textwidth,height=0.82\textheight,keepaspectratio]{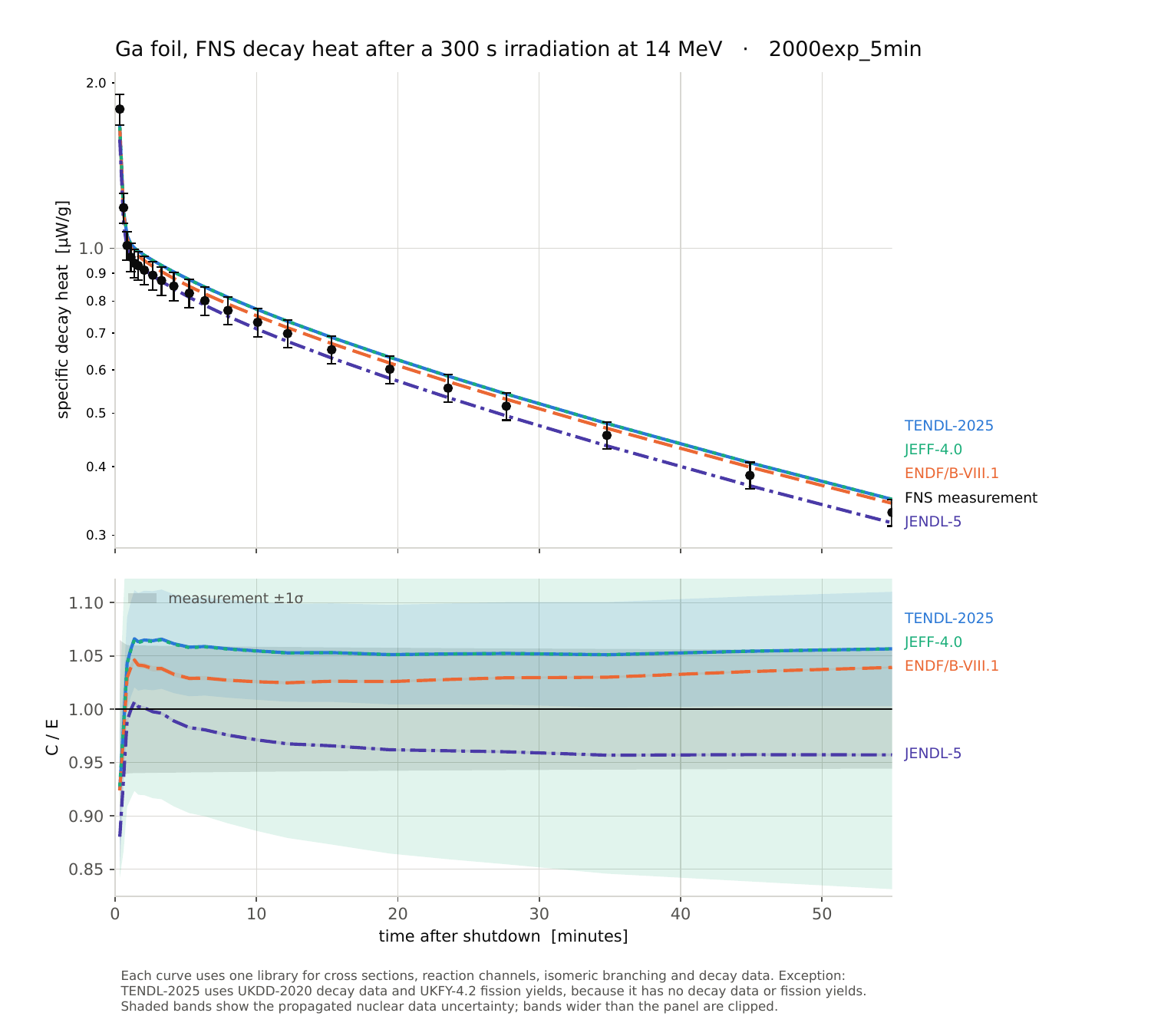}
\caption{Ga: decay heat against the measurement, with C/E beneath. Mean $|C/E-1|$ by library: JENDL-5 3\%, ENDF/B-VIII.1 3\%, JEFF-4.0 6\%, TENDL-2025 6\%.}
\end{figure}
\clearpage

\begin{figure}[H]
\centering
\includegraphics[width=\textwidth,height=0.82\textheight,keepaspectratio]{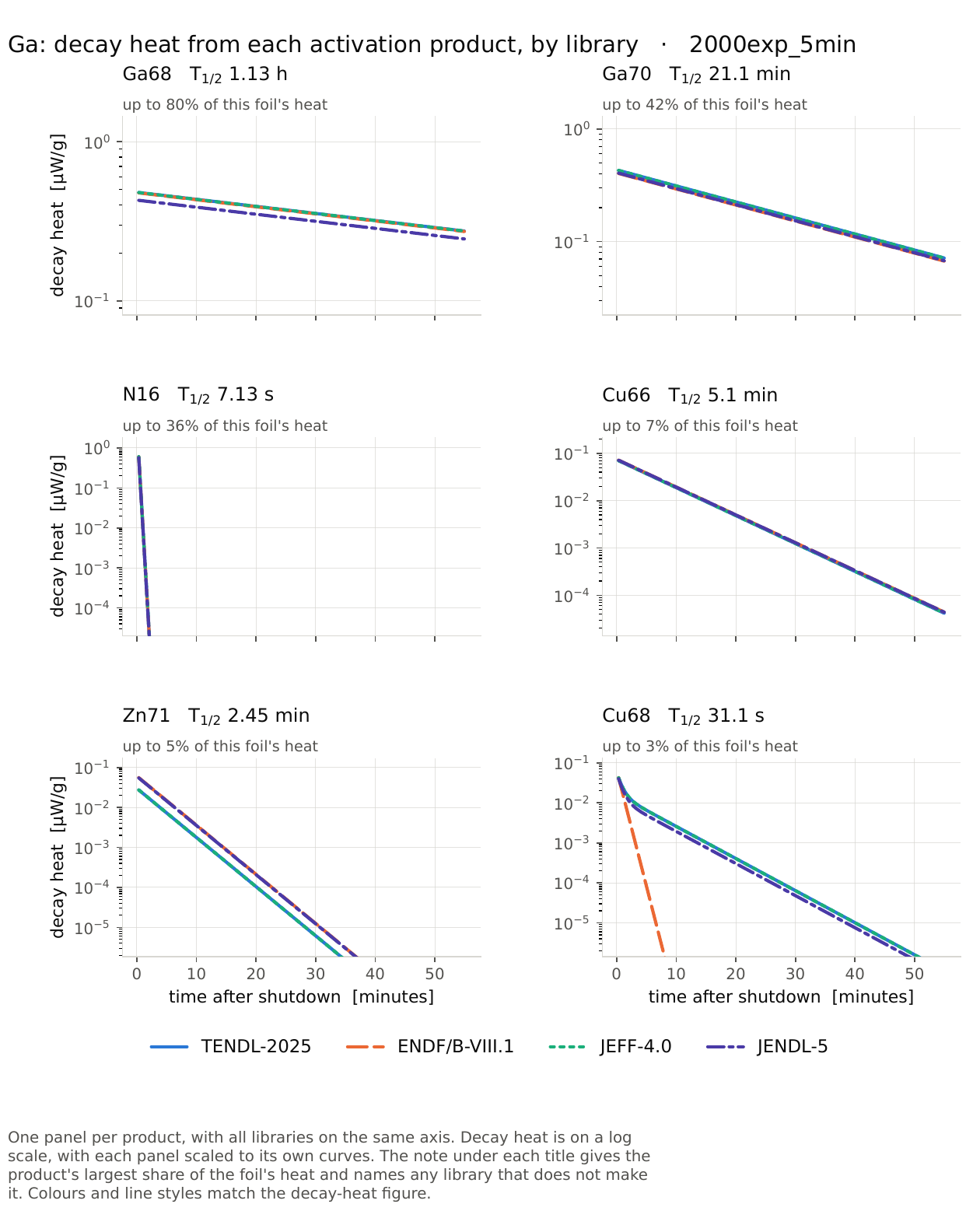}
\caption{Ga: share of the decay heat from each activation product during cooling.}
\end{figure}
\clearpage

\begin{figure}[H]
\centering
\includegraphics[width=\textwidth,height=0.82\textheight,keepaspectratio]{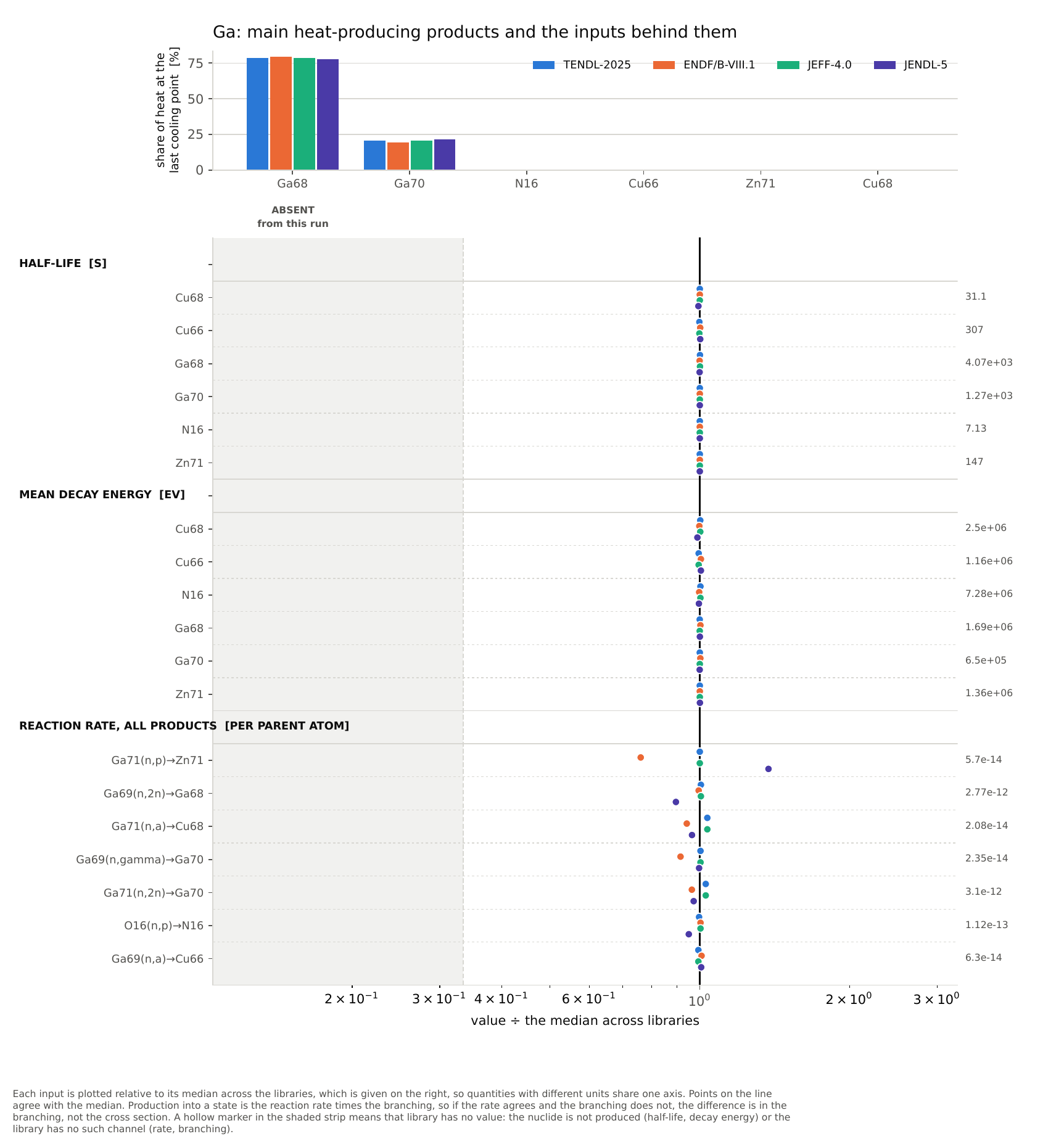}
\caption{Ga: inputs for the main heat-producing products, each relative to its median across the available libraries.}
\end{figure}
\clearpage

\begin{figure}[H]
\centering
\includegraphics[width=\textwidth,height=0.82\textheight,keepaspectratio]{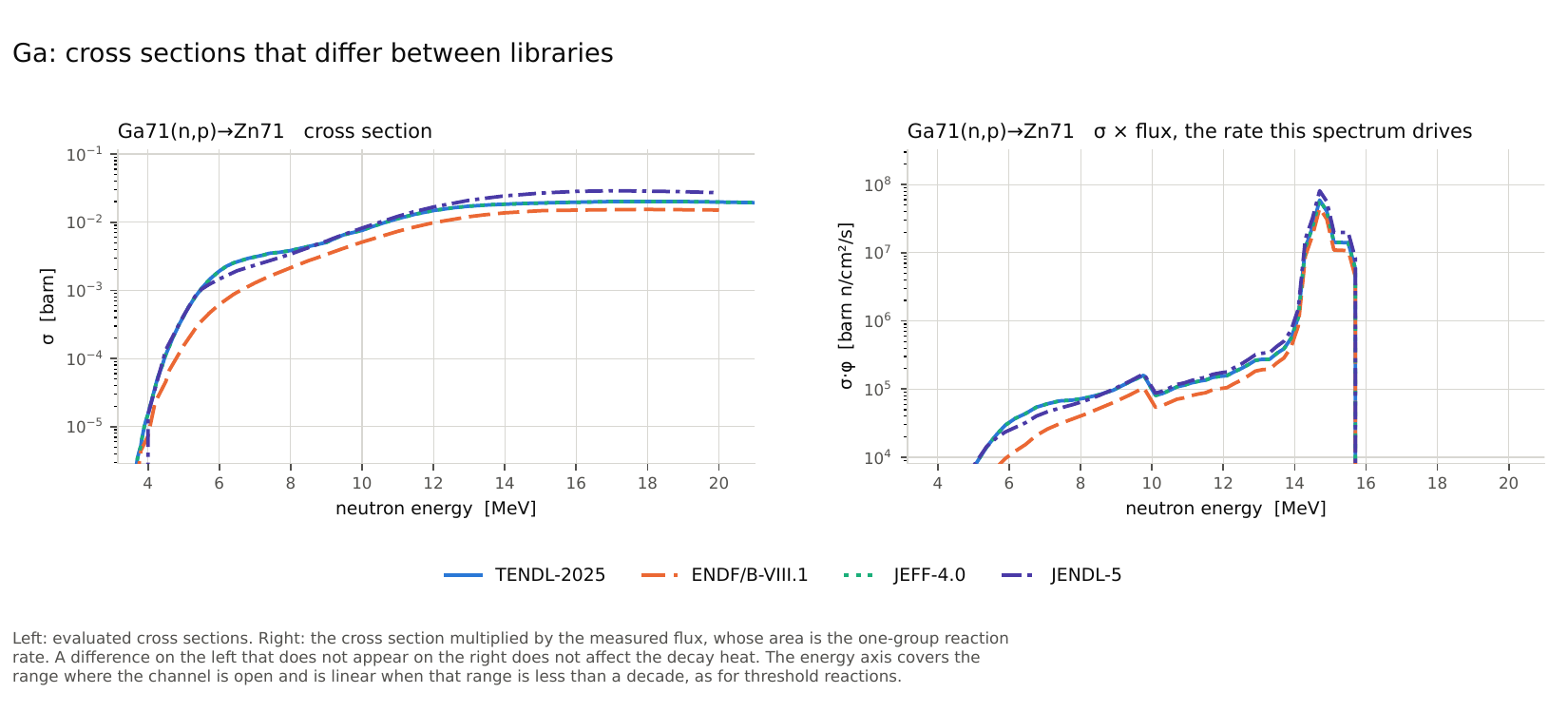}
\caption{Ga: cross sections of the channels whose spectrum-averaged rates differ, with the neutron spectrum and their product.}
\end{figure}
\clearpage

Mean absolute C/E deviations range from 3 to 6 percent. The spread of 3 percentage points is smaller than the median measurement uncertainty of 6 percent, so the measurement cannot separate the libraries.

\subsubsection*{Suggested measurements}
\begin{itemize}\setlength{\itemsep}{4pt}
\item No input differs enough between the libraries to be listed. An error shared by all of them, or a smaller difference, is still possible.
\end{itemize}
\clearpage
\subsection{Gd: Gadolinium}\label{foil:Gd}
\begin{figure}[H]
\centering
\includegraphics[width=\textwidth,height=0.82\textheight,keepaspectratio]{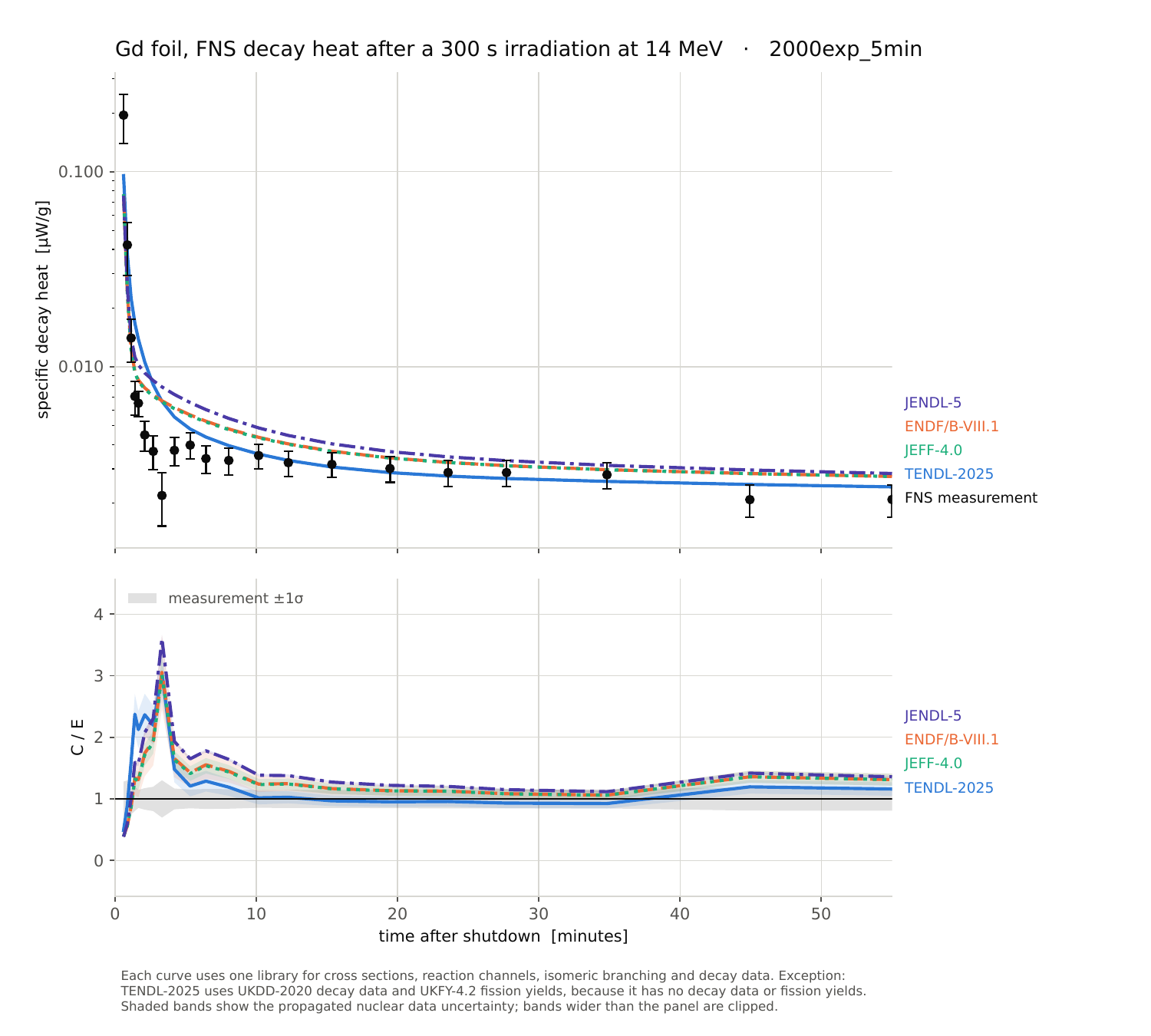}
\caption{Gd: decay heat against the measurement, with C/E beneath. Mean $|C/E-1|$ by library: JEFF-4.0 44\%, ENDF/B-VIII.1 45\%, TENDL-2025 48\%, JENDL-5 60\%.}
\end{figure}
\clearpage

\begin{figure}[H]
\centering
\includegraphics[width=\textwidth,height=0.82\textheight,keepaspectratio]{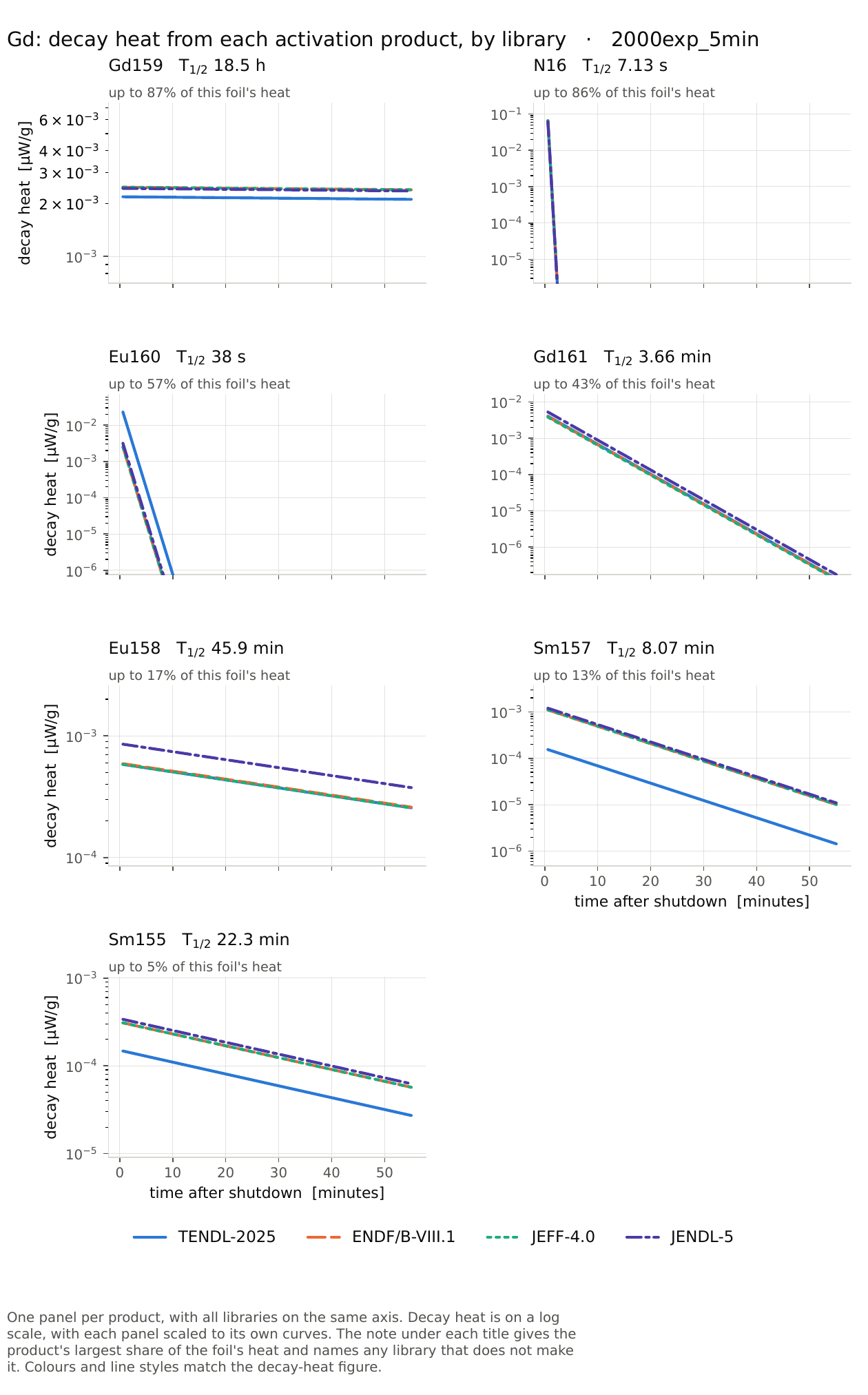}
\caption{Gd: share of the decay heat from each activation product during cooling.}
\end{figure}
\clearpage

\begin{figure}[H]
\centering
\includegraphics[width=\textwidth,height=0.82\textheight,keepaspectratio]{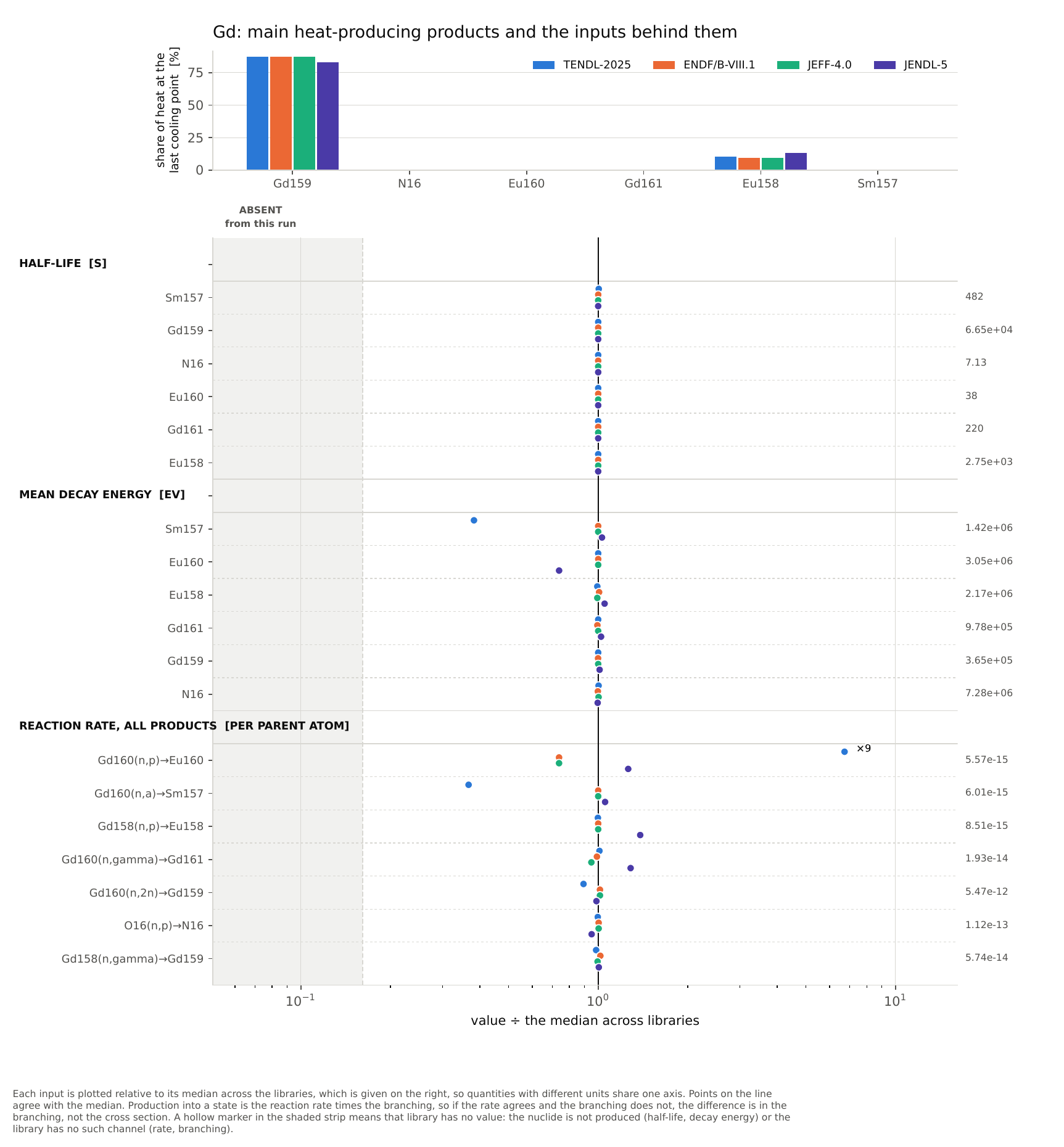}
\caption{Gd: inputs for the main heat-producing products, each relative to its median across the available libraries.}
\end{figure}
\clearpage

\begin{figure}[H]
\centering
\includegraphics[width=\textwidth,height=0.82\textheight,keepaspectratio]{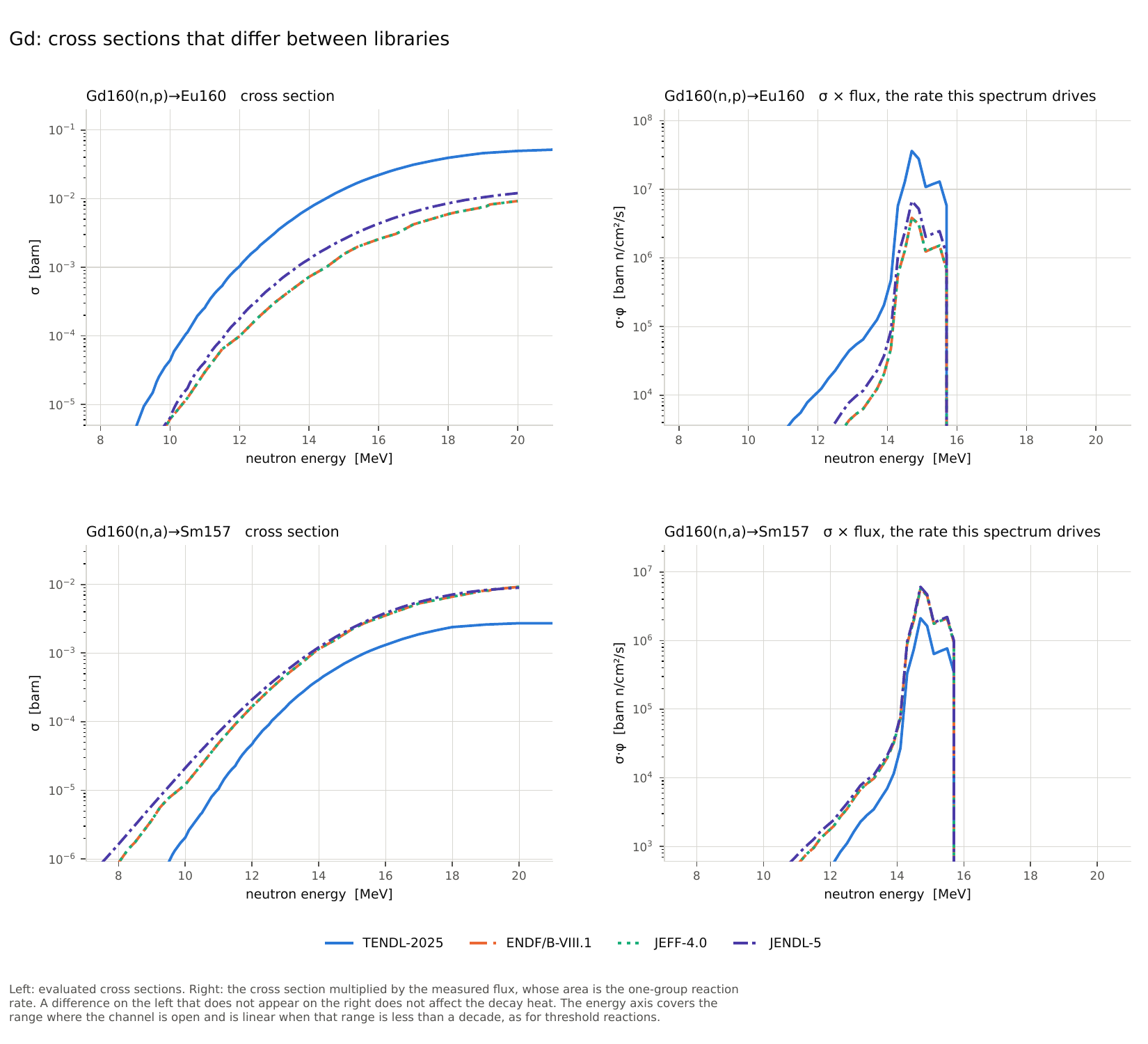}
\caption{Gd: cross sections of the channels whose spectrum-averaged rates differ, with the neutron spectrum and their product.}
\end{figure}
\clearpage

The libraries are between 44 and 60 percent from the measurement, which has a median uncertainty of 16 percent. JEFF-4.0 is closest and JENDL-5 furthest.

\subsubsection*{Differences in the nuclear data}
\begin{itemize}\setlength{\itemsep}{6pt}
\item TENDL-2025 gives the total reaction rate of Gd160(n,p)→Eu160 as 3.757e-14, against 4.117e-15 for the median of the others, a factor of 9.\\[2pt]{\itshape This is the total rate of the channel over all its products, so the difference is in the cross section, not in the split between states.}
\item TENDL-2025 gives the total reaction rate of Gd160(n,a)→Sm157 as 2.201e-15, against 6.007e-15 for the median of the others, a factor of 3.
\item TENDL-2025 gives the mean decay energy of Sm157 as 5.44e+05 eV, against 1.424e+06 eV for the median of the others, a factor of 3.\\[2pt]{\itshape For a given activity the heat is proportional to the mean decay energy, so this difference goes straight into the heat. Compare the decay schemes and how the emitted energy is counted.}
\end{itemize}
\subsubsection*{Suggested measurements}
\begin{itemize}\setlength{\itemsep}{4pt}
\item Natural Gd has 7 isotopes, so the measured heat combines several parents. Enriched single-isotope targets would distinguish their contributions. The candidate parent is Gd160. Enrichment alone does not isolate a single reaction channel.
\item The sample is a compound that also contains O. Their activation adds to the measured heat; a metal or purer sample would remove it.
\item The reaction rates are averaged over the measured spectrum. Activation measurements at several well-defined neutron energies, including below 14 MeV, would show which energy range the differences come from.
\end{itemize}
\clearpage
\subsection{Ge: Germanium}\label{foil:Ge}
\begin{figure}[H]
\centering
\includegraphics[width=\textwidth,height=0.82\textheight,keepaspectratio]{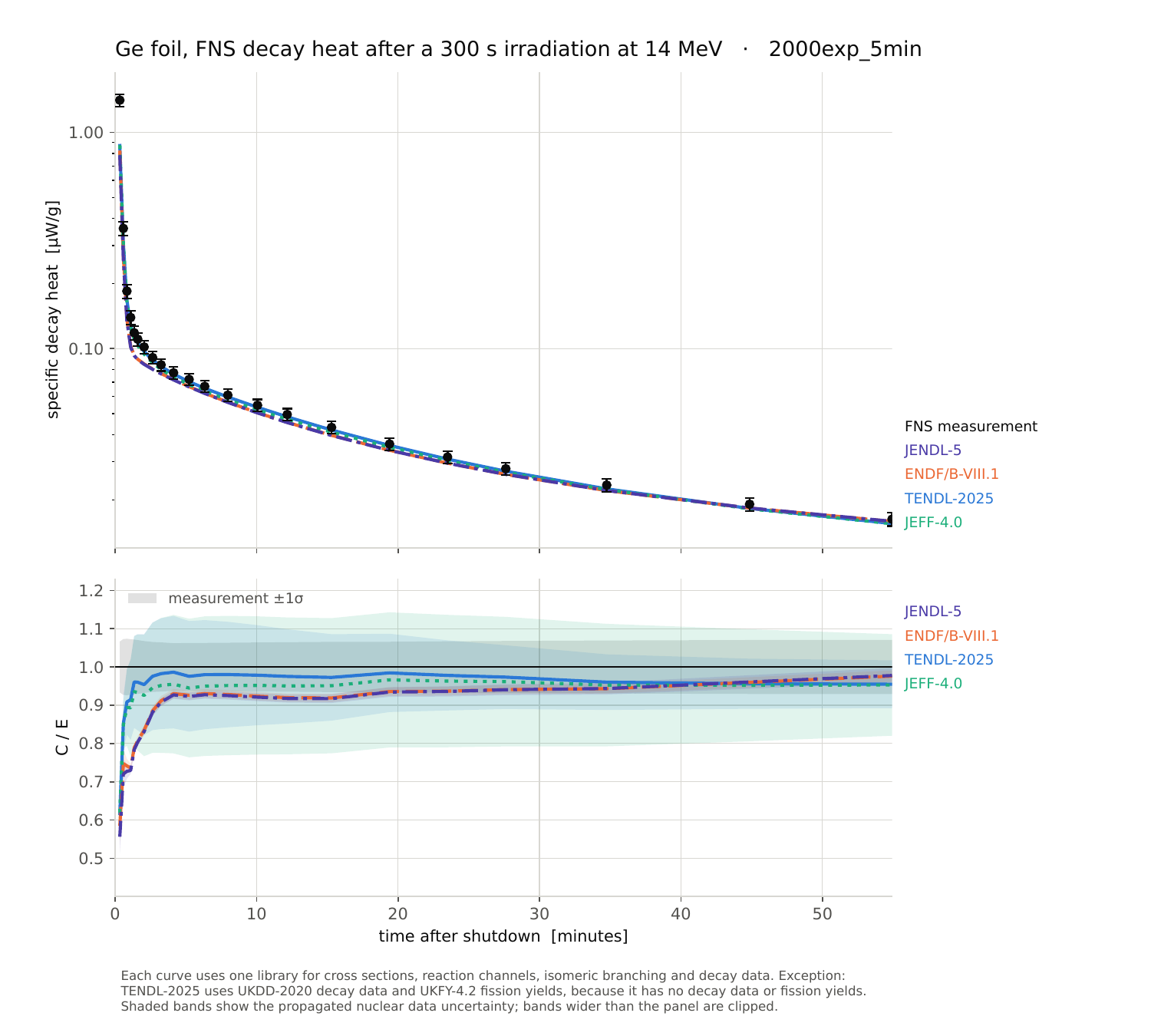}
\caption{Ge: decay heat against the measurement, with C/E beneath. Mean $|C/E-1|$ by library: TENDL-2025 6\%, JEFF-4.0 8\%, ENDF/B-VIII.1 13\%, JENDL-5 13\%.}
\end{figure}
\clearpage

\begin{figure}[H]
\centering
\includegraphics[width=\textwidth,height=0.82\textheight,keepaspectratio]{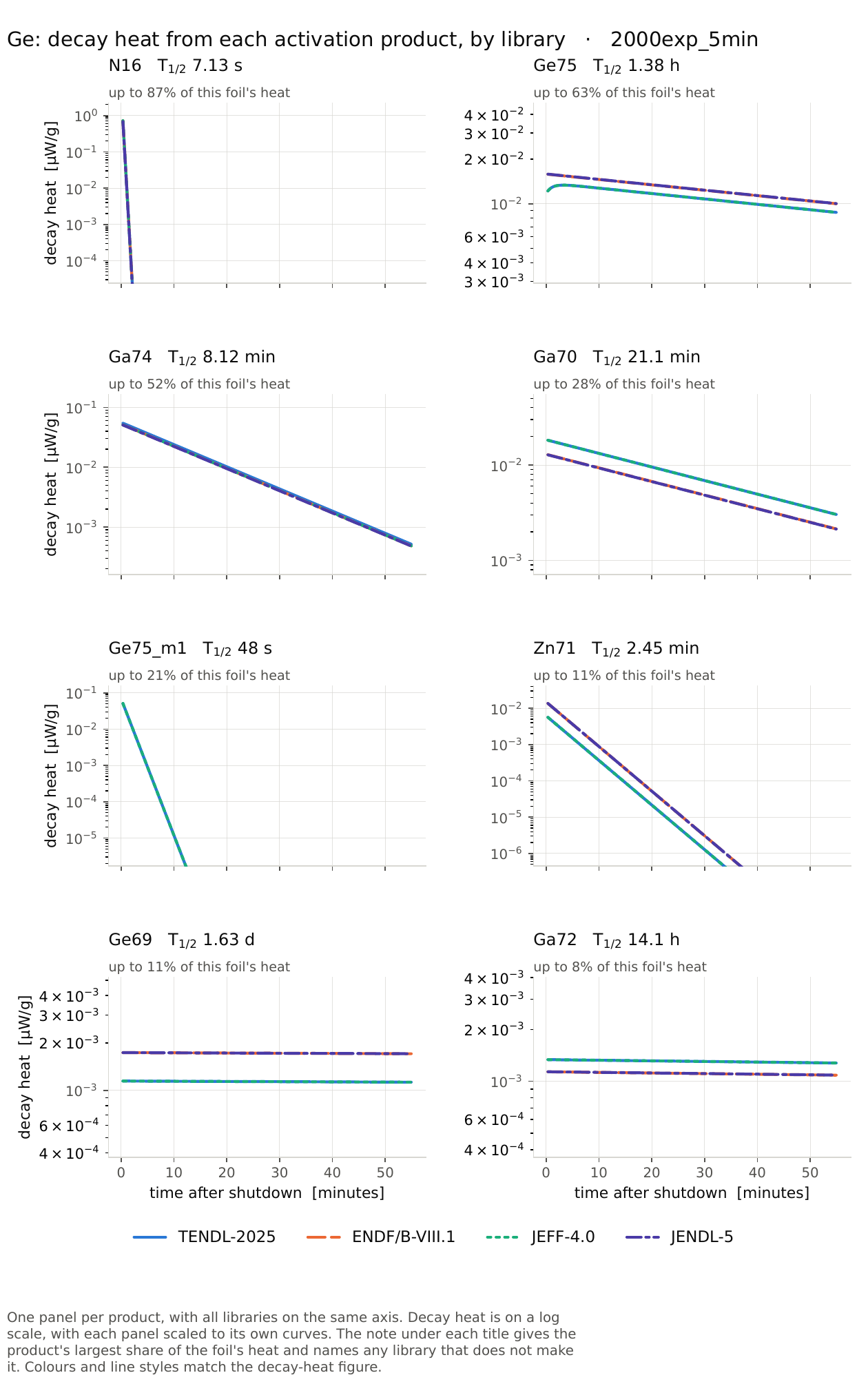}
\caption{Ge: share of the decay heat from each activation product during cooling.}
\end{figure}
\clearpage

\begin{figure}[H]
\centering
\includegraphics[width=\textwidth,height=0.82\textheight,keepaspectratio]{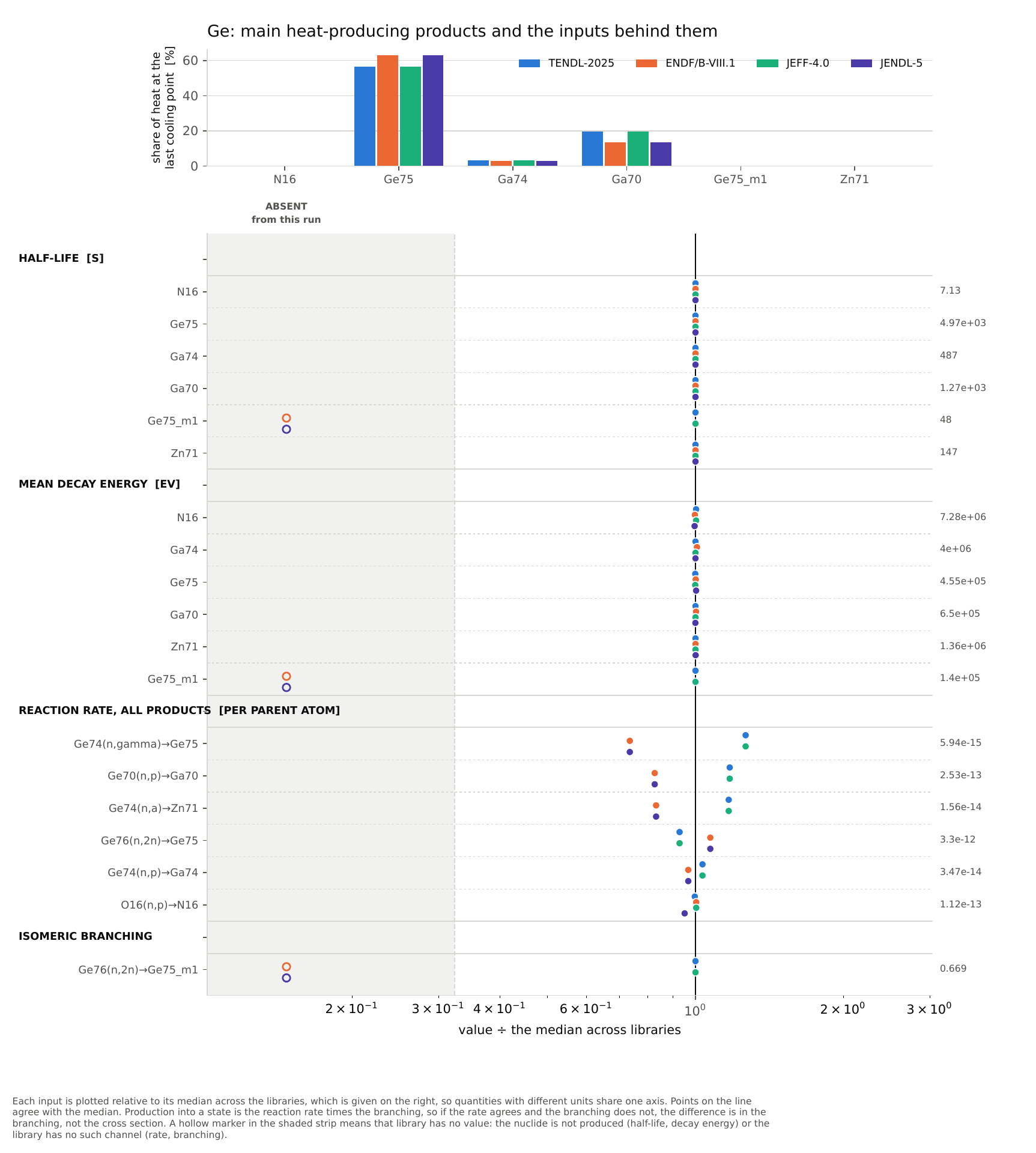}
\caption{Ge: inputs for the main heat-producing products, each relative to its median across the available libraries.}
\end{figure}
\clearpage

\begin{figure}[H]
\centering
\includegraphics[width=\textwidth,height=0.82\textheight,keepaspectratio]{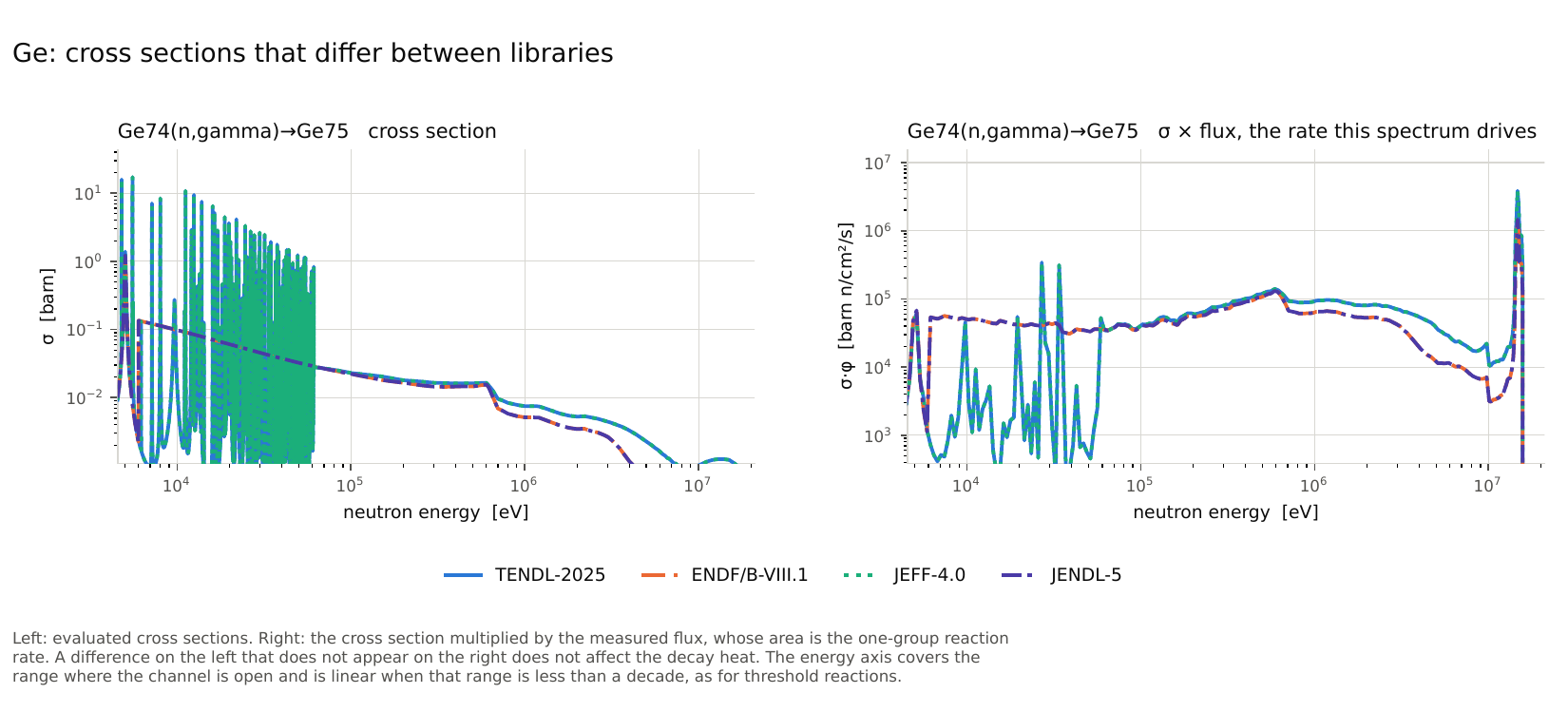}
\caption{Ge: cross sections of the channels whose spectrum-averaged rates differ, with the neutron spectrum and their product.}
\end{figure}
\clearpage

The libraries are between 6 and 13 percent from the measurement, which has a median uncertainty of 7 percent. TENDL-2025 is closest and JENDL-5 furthest.

\subsubsection*{Suggested measurements}
\begin{itemize}\setlength{\itemsep}{4pt}
\item This foil makes Ge69 (half-life 1.63 days), which gives up to 11 percent of the calculated heat. A five-minute irradiation and 55 minute measurement only see the start of this contribution. There is no seven-hour FNS measurement of this foil. A longer irradiation followed for days to months would be needed to test this product's production and decay.
\item No input differs enough between the libraries to be listed. An error shared by all of them, or a smaller difference, is still possible.
\end{itemize}
\clearpage
\subsection{Hf: Hafnium}\label{foil:Hf}
\begin{figure}[H]
\centering
\includegraphics[width=\textwidth,height=0.82\textheight,keepaspectratio]{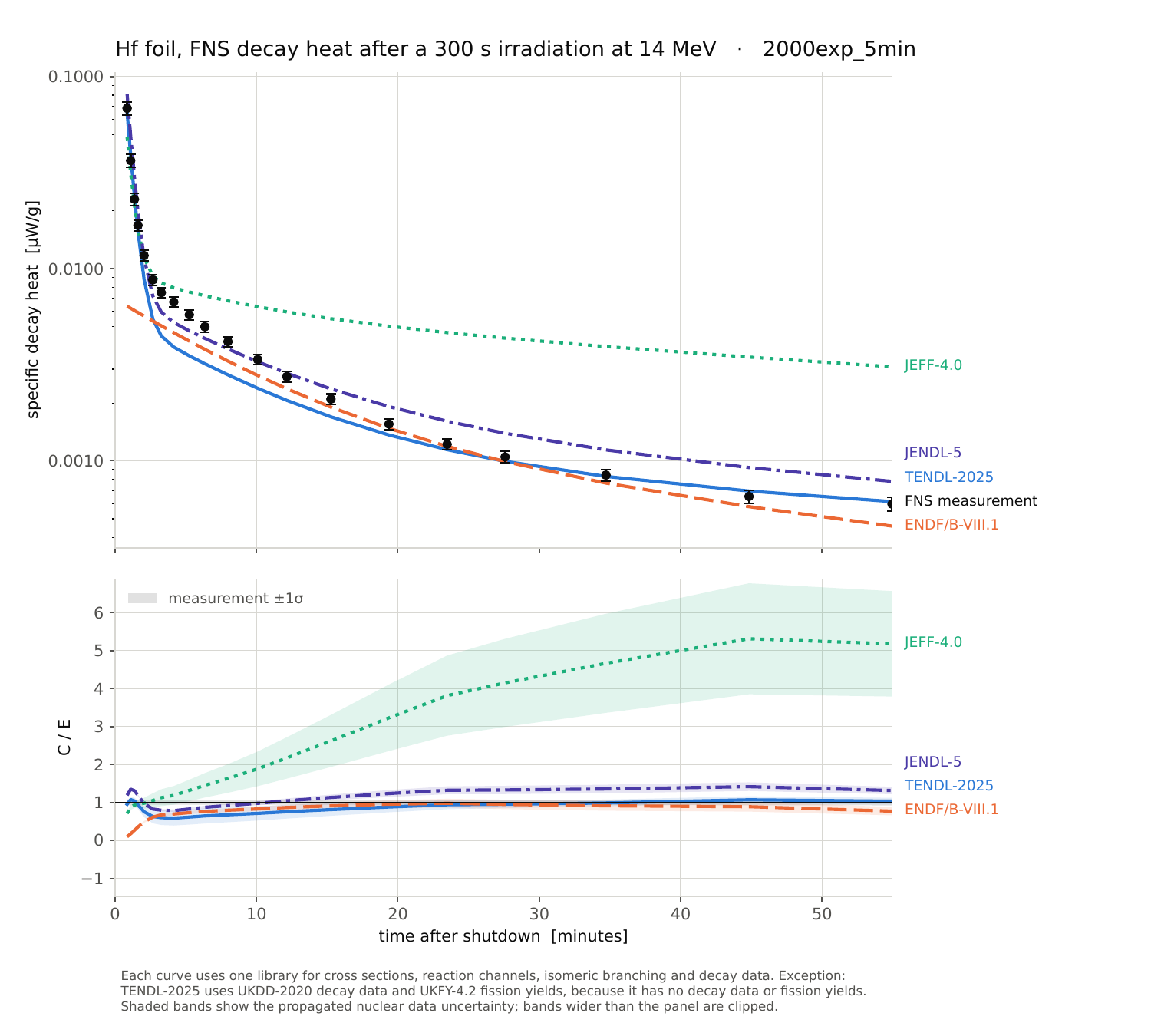}
\caption{Hf: decay heat against the measurement, with C/E beneath. Mean $|C/E-1|$ by library: TENDL-2025 19\%, JENDL-5 21\%, ENDF/B-VIII.1 32\%, JEFF-4.0 132\%.}
\end{figure}
\clearpage

\begin{figure}[H]
\centering
\includegraphics[width=\textwidth,height=0.82\textheight,keepaspectratio]{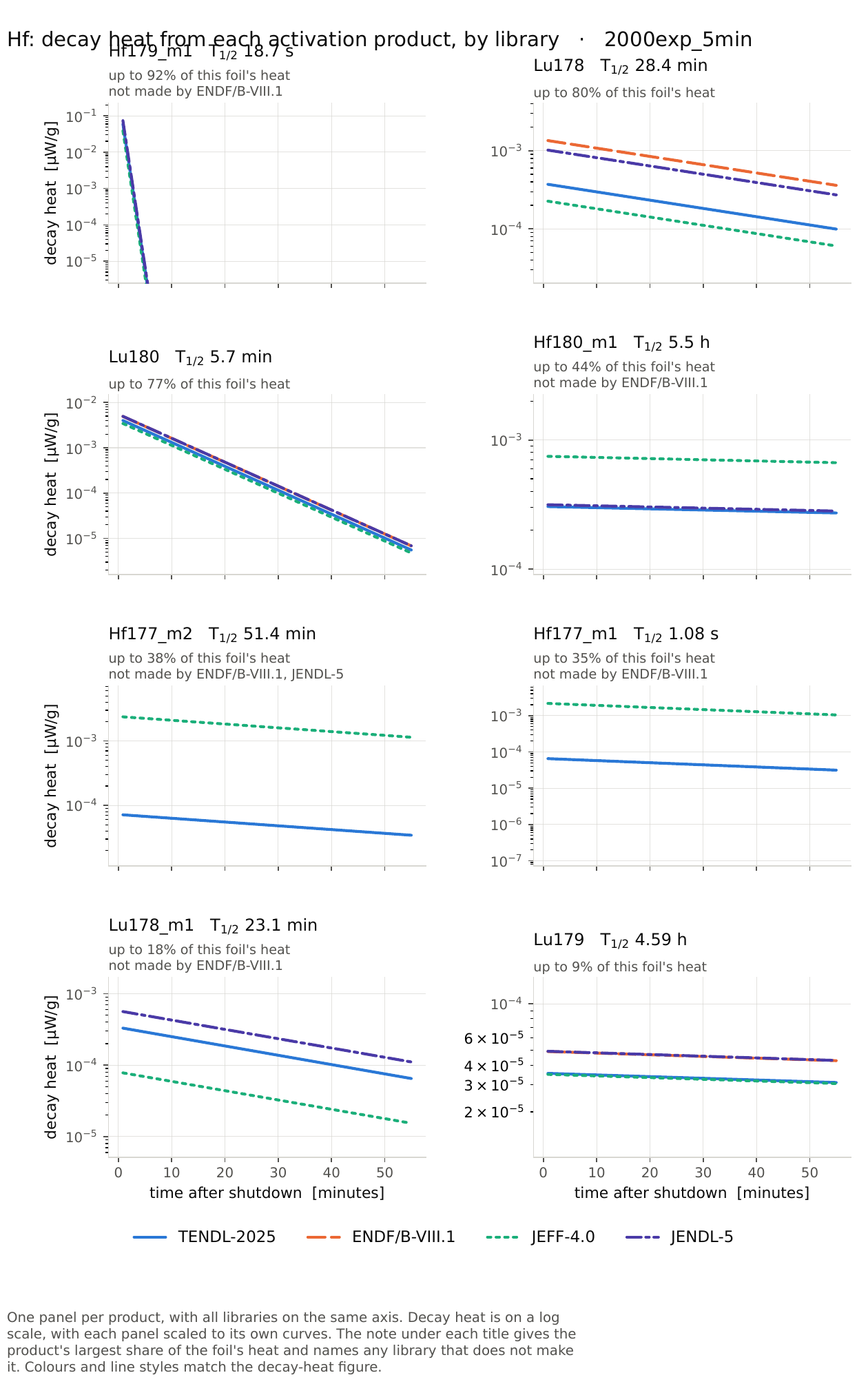}
\caption{Hf: share of the decay heat from each activation product during cooling.}
\end{figure}
\clearpage

\begin{figure}[H]
\centering
\includegraphics[width=\textwidth,height=0.82\textheight,keepaspectratio]{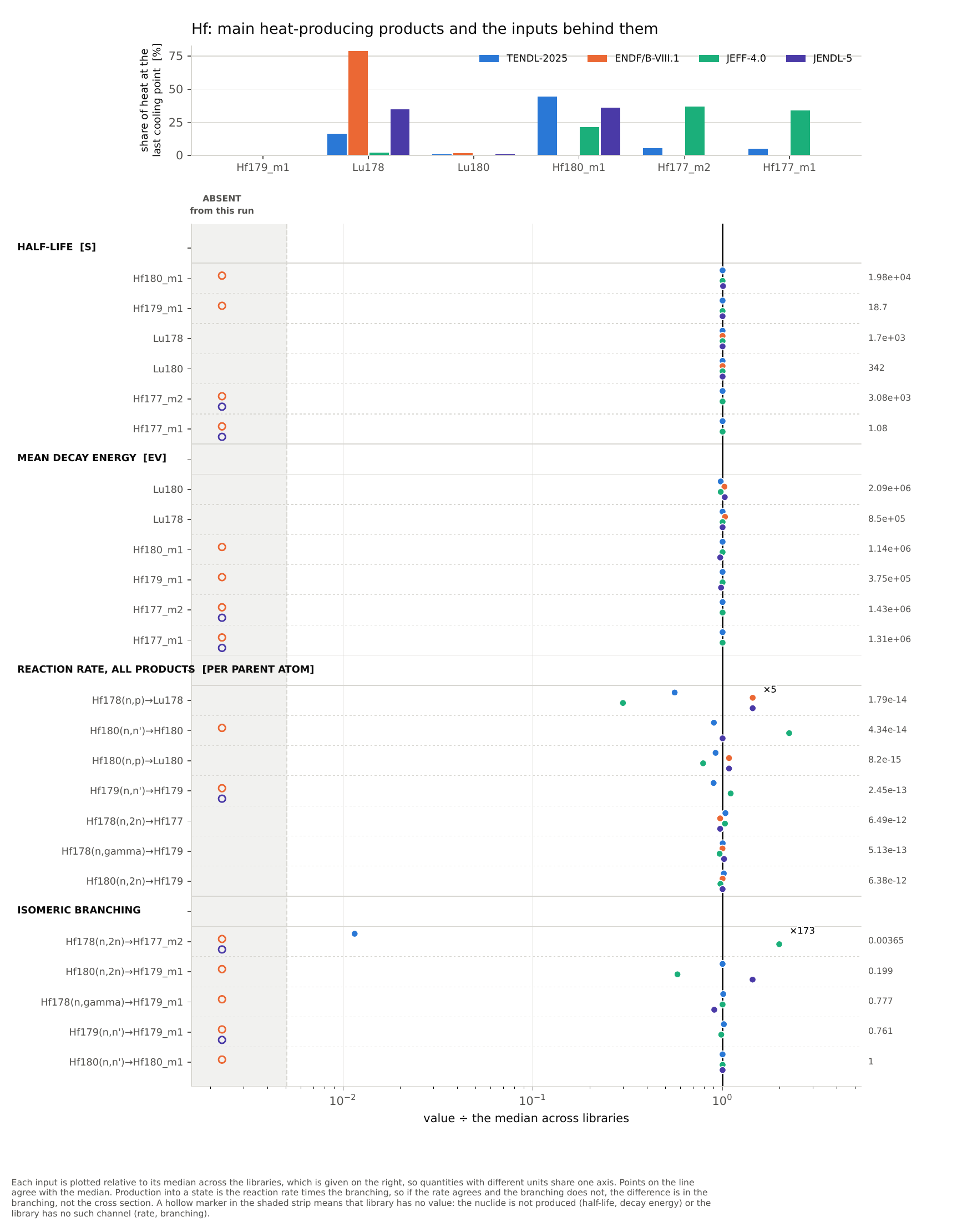}
\caption{Hf: inputs for the main heat-producing products, each relative to its median across the available libraries.}
\end{figure}
\clearpage

\begin{figure}[H]
\centering
\includegraphics[width=\textwidth,height=0.82\textheight,keepaspectratio]{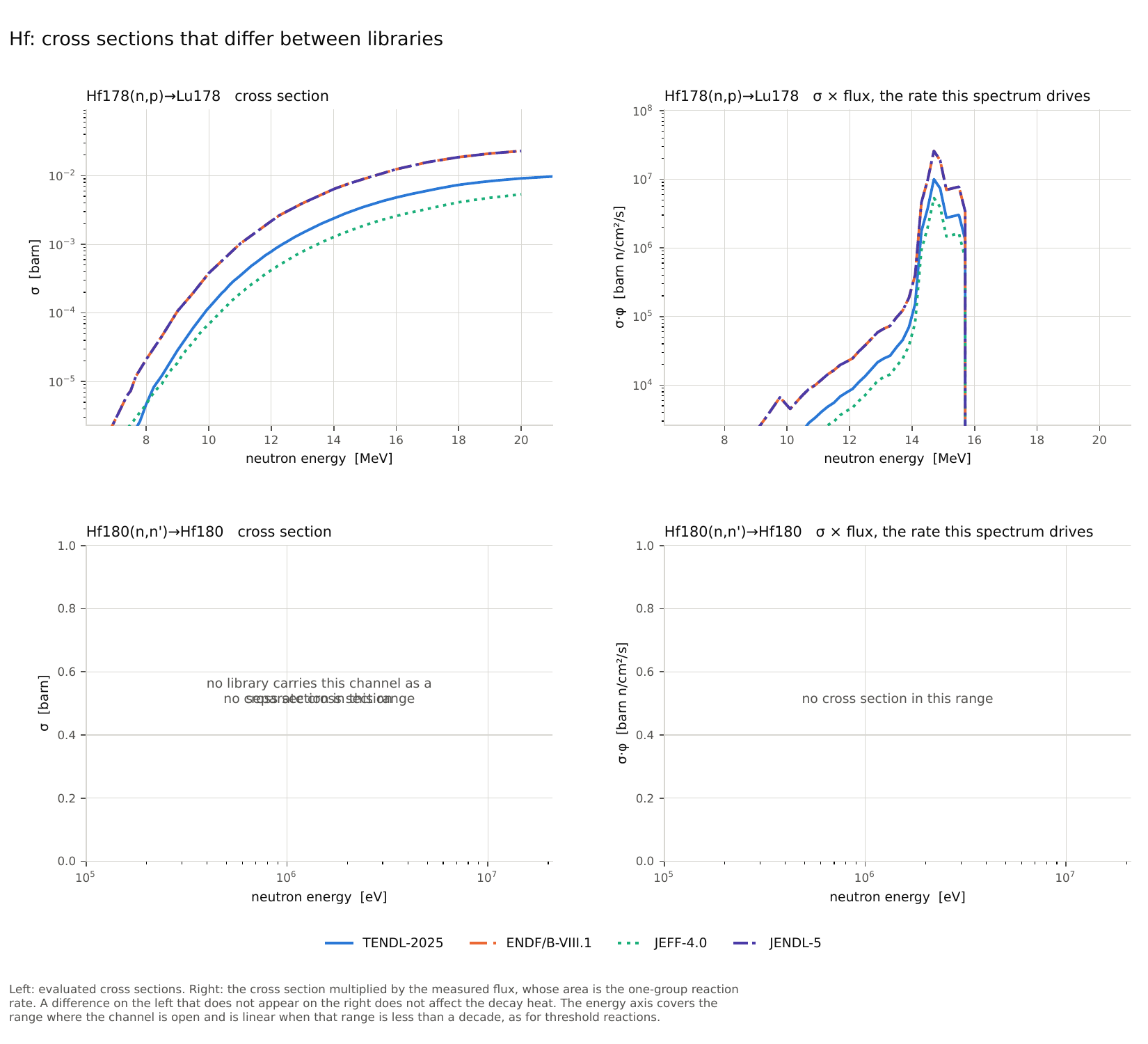}
\caption{Hf: cross sections of the channels whose spectrum-averaged rates differ, with the neutron spectrum and their product.}
\end{figure}
\clearpage

\begin{figure}[H]
\centering
\includegraphics[width=\textwidth,height=0.82\textheight,keepaspectratio]{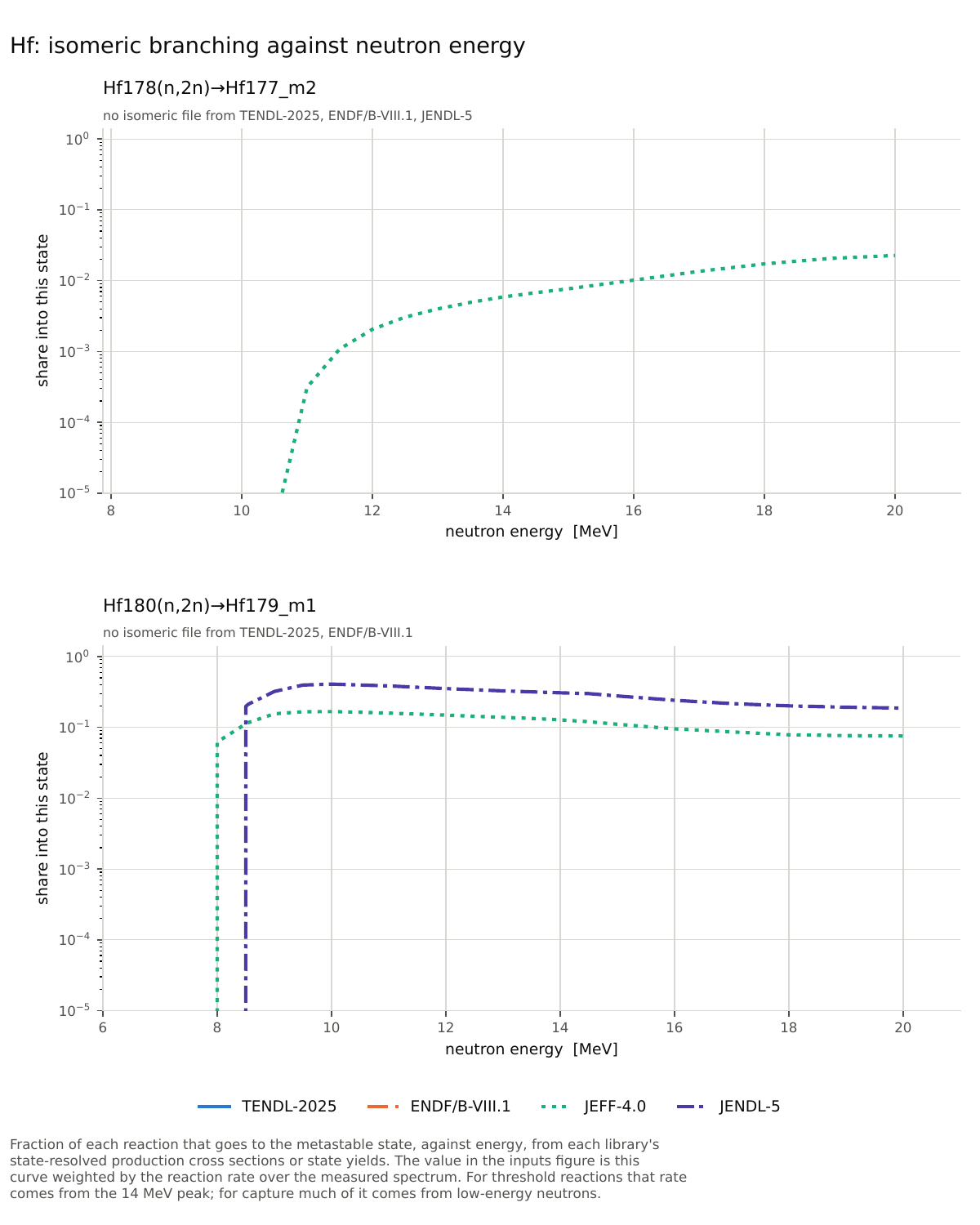}
\caption{Hf: isomeric branching ratio against neutron energy for the channels whose spectrum-averaged ratio differs between libraries.}
\end{figure}
\clearpage

JEFF-4.0 stands out at 132 percent from the measurement, against 32 percent for the next furthest library and 19 percent for TENDL-2025. The median measurement uncertainty is 6 percent.

\subsubsection*{Differences in the nuclear data}
\begin{itemize}\setlength{\itemsep}{6pt}
\item Only JEFF-4.0 and TENDL-2025 have the isomeric branching of Hf178(n,2n)→Hf177\_m2, and they differ by a factor of 173: 0.007267 against 4.204e-05.\\[2pt]{\itshape Isomeric branching sets how a channel's production is split between the ground and metastable states. It should be checked together with the total rate of the channel.}
\item JEFF-4.0 gives the total reaction rate of Hf178(n,p)→Lu178 as 5.332e-15, against 2.574e-14 for the median of the others, a factor of 5.\\[2pt]{\itshape This is the total rate of the channel over all its products, so the difference is in the cross section, not in the split between states.}
\item JEFF-4.0 gives the total reaction rate of Hf180(n,n')→Hf180 as 9.729e-14, against 4.12e-14 for the median of the others, a factor of 2.
\item JEFF-4.0 gives the isomeric branching of Hf180(n,2n)→Hf179\_m1 as 0.1152, against 0.243 for the median of the others, a factor of 2.
\item ENDF/B-VIII.1 gives no heat from Hf179\_m1; the other libraries do.\\[2pt]{\itshape The heat breakdown alone does not show why. Possible causes are a missing reaction channel, a different state assignment, different decay data or a reporting threshold.}
\item ENDF/B-VIII.1 gives no heat from Hf180\_m1; the other libraries do.
\item ENDF/B-VIII.1, JENDL-5 give no heat from Hf177\_m2; the other libraries do.
\end{itemize}
\subsubsection*{Suggested measurements}
\begin{itemize}\setlength{\itemsep}{4pt}
\item Natural Hf has 6 isotopes, so the measured heat combines several parents. Enriched single-isotope targets would distinguish their contributions. Candidate parents are Hf178, Hf180. Enrichment alone does not isolate a single reaction channel.
\item This foil makes Hf175 (half-life 70 days), which gives up to 2 percent of the calculated heat. A five-minute irradiation and 55 minute measurement only see the start of this contribution. There is no seven-hour FNS measurement of this foil. A longer irradiation followed for days to months would be needed to test this product's production and decay.
\item The reaction rates are averaged over the measured spectrum. Activation measurements at several well-defined neutron energies, including below 14 MeV, would show which energy range the differences come from.
\item Measuring the activity of each state separately, for example by gamma spectrometry where the states have distinct lines, would constrain the isomeric ratio. This needs corrections for feeding between states, detector efficiency and irradiation history.
\end{itemize}
\clearpage
\subsection{Hg: Mercury}\label{foil:Hg}
\begin{figure}[H]
\centering
\includegraphics[width=\textwidth,height=0.82\textheight,keepaspectratio]{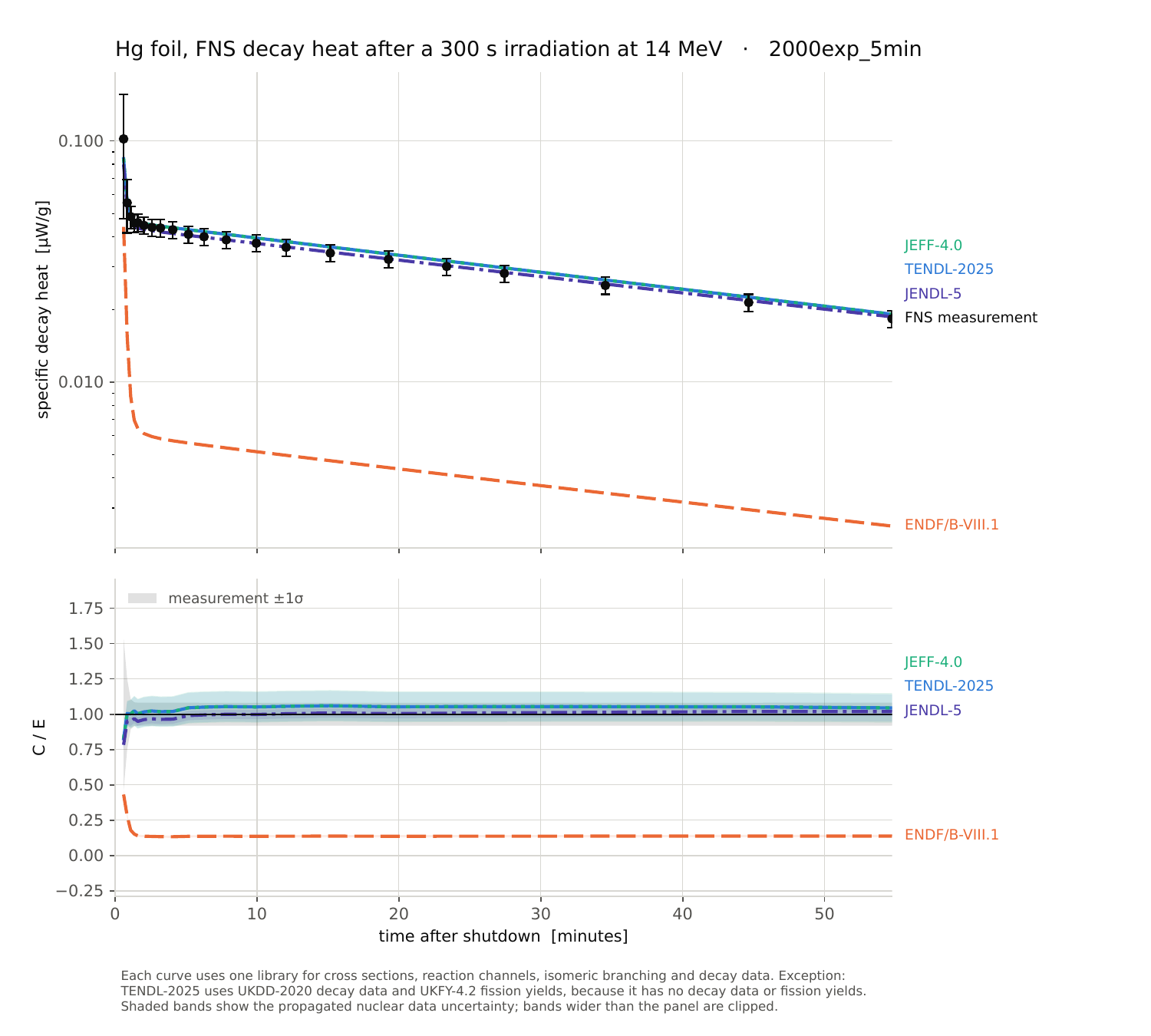}
\caption{Hg: decay heat against the measurement, with C/E beneath. Mean $|C/E-1|$ by library: JENDL-5 3\%, JEFF-4.0 4\%, TENDL-2025 4\%, ENDF/B-VIII.1 84\%.}
\end{figure}
\clearpage

\begin{figure}[H]
\centering
\includegraphics[width=\textwidth,height=0.82\textheight,keepaspectratio]{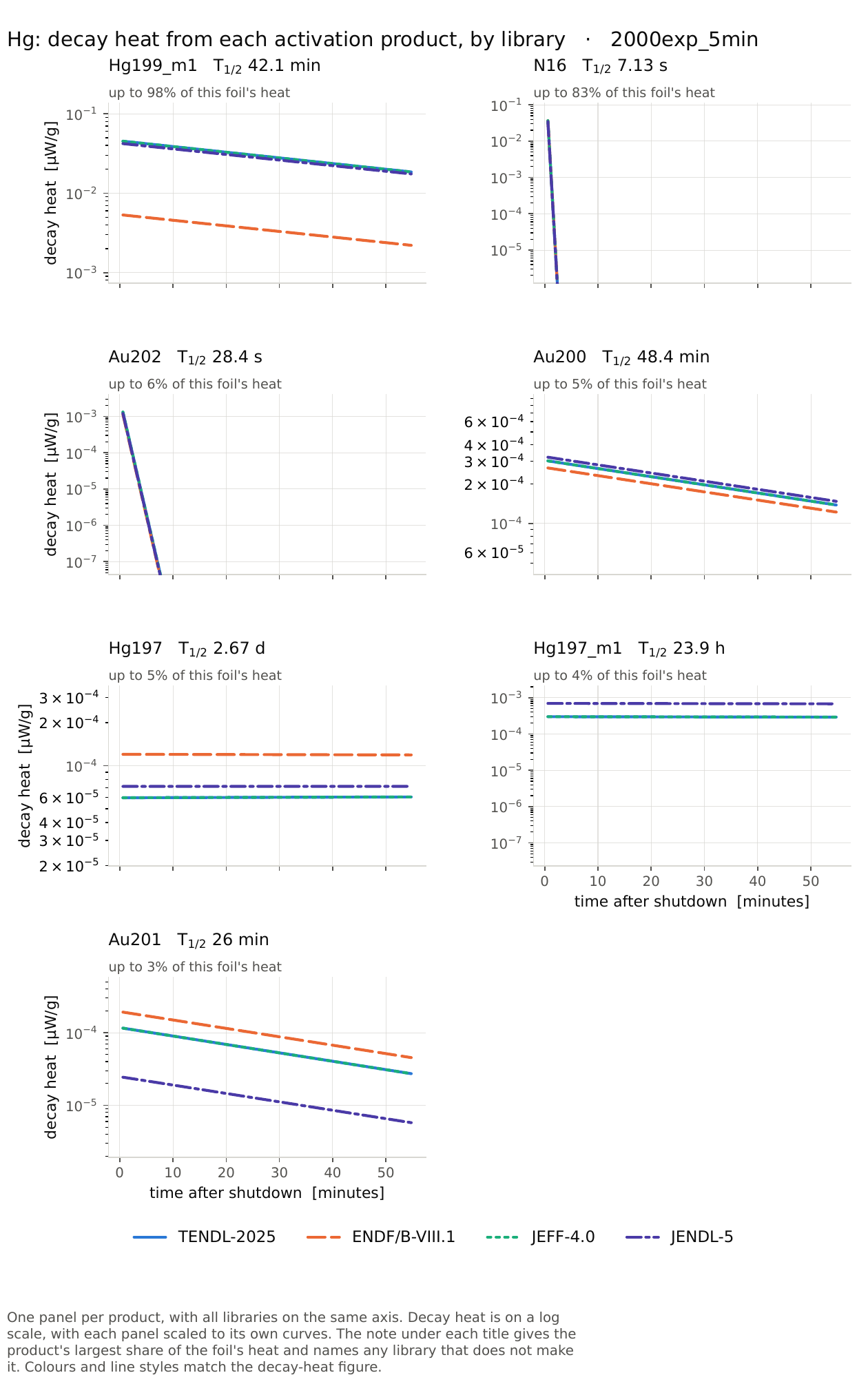}
\caption{Hg: share of the decay heat from each activation product during cooling.}
\end{figure}
\clearpage

\begin{figure}[H]
\centering
\includegraphics[width=\textwidth,height=0.82\textheight,keepaspectratio]{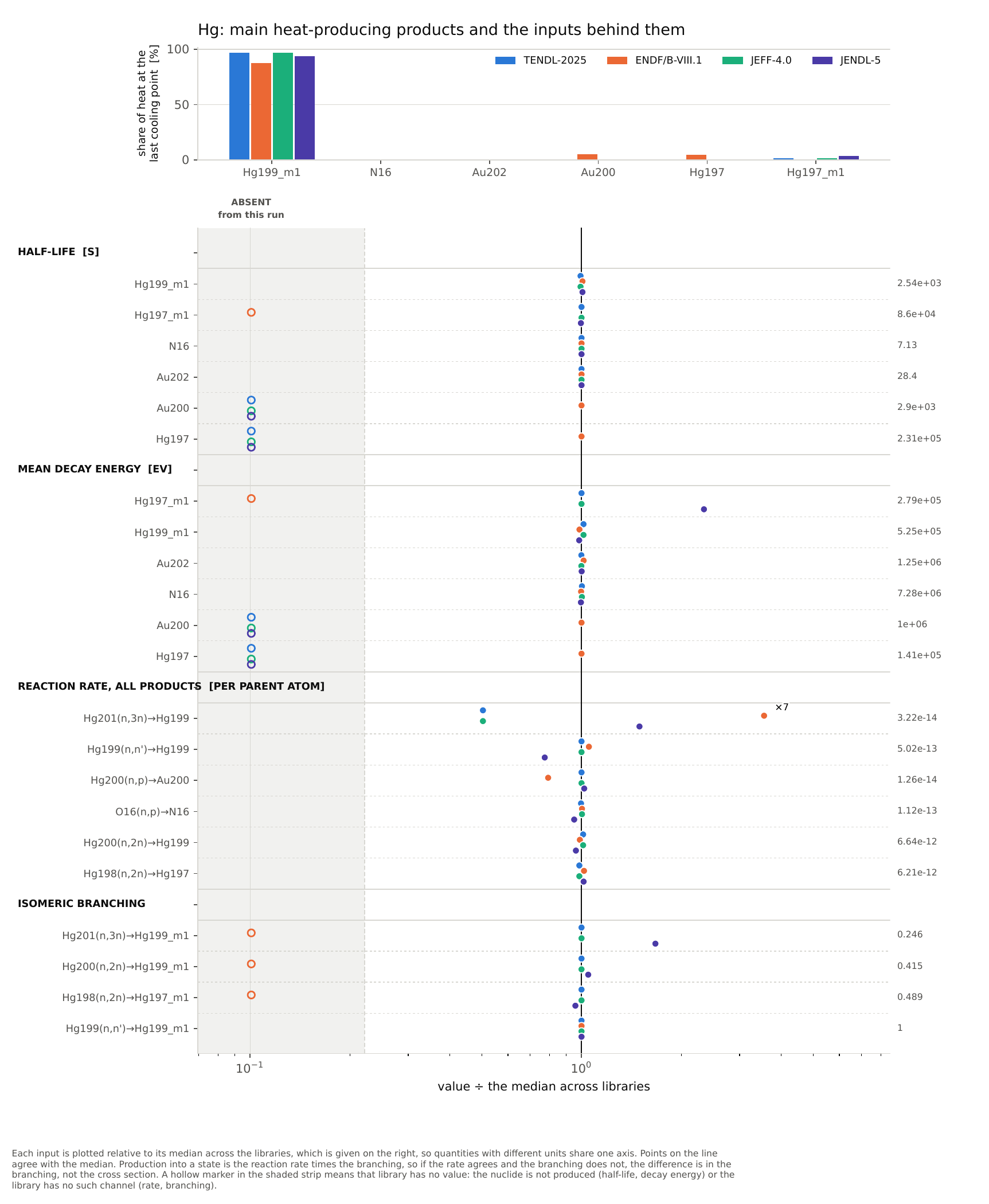}
\caption{Hg: inputs for the main heat-producing products, each relative to its median across the available libraries.}
\end{figure}
\clearpage

\begin{figure}[H]
\centering
\includegraphics[width=\textwidth,height=0.82\textheight,keepaspectratio]{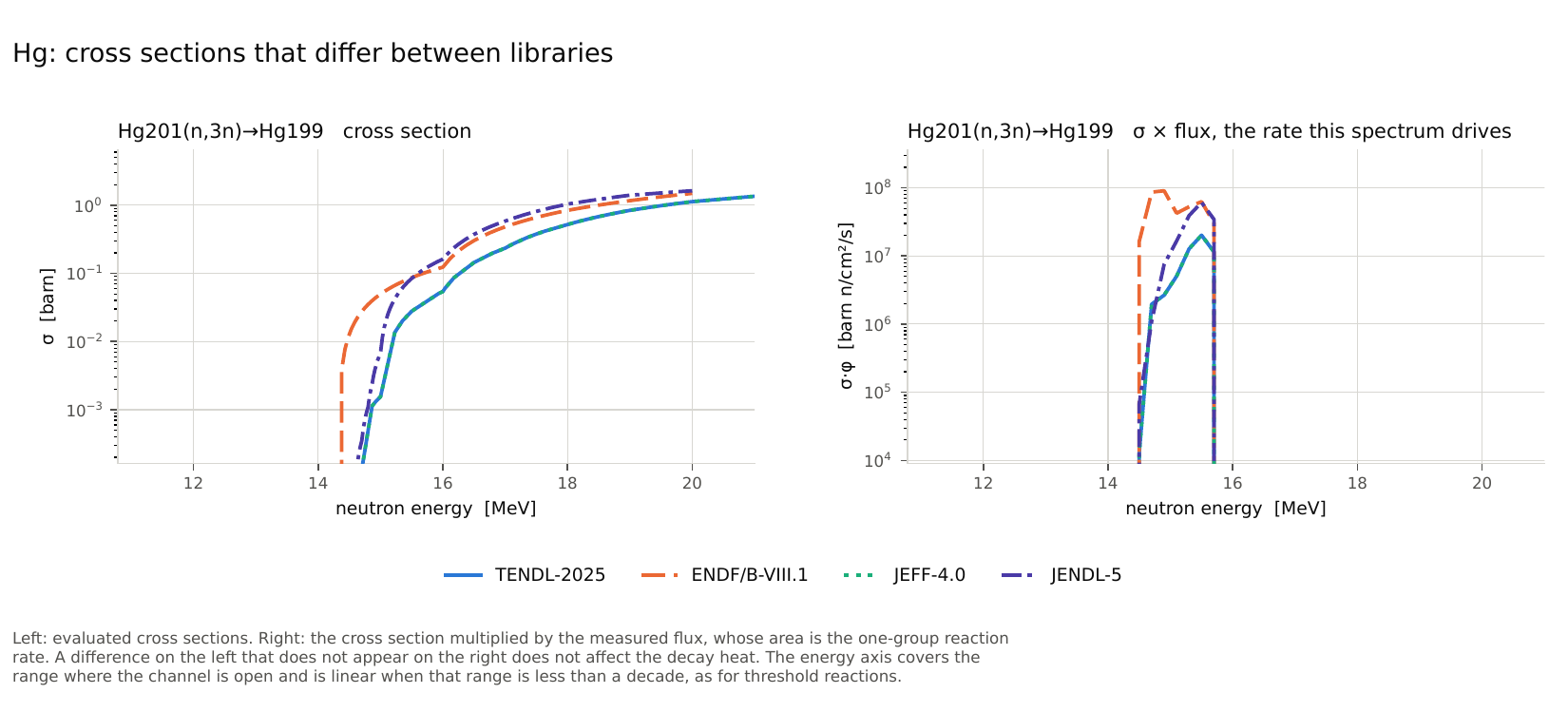}
\caption{Hg: cross sections of the channels whose spectrum-averaged rates differ, with the neutron spectrum and their product.}
\end{figure}
\clearpage

\begin{figure}[H]
\centering
\includegraphics[width=\textwidth,height=0.82\textheight,keepaspectratio]{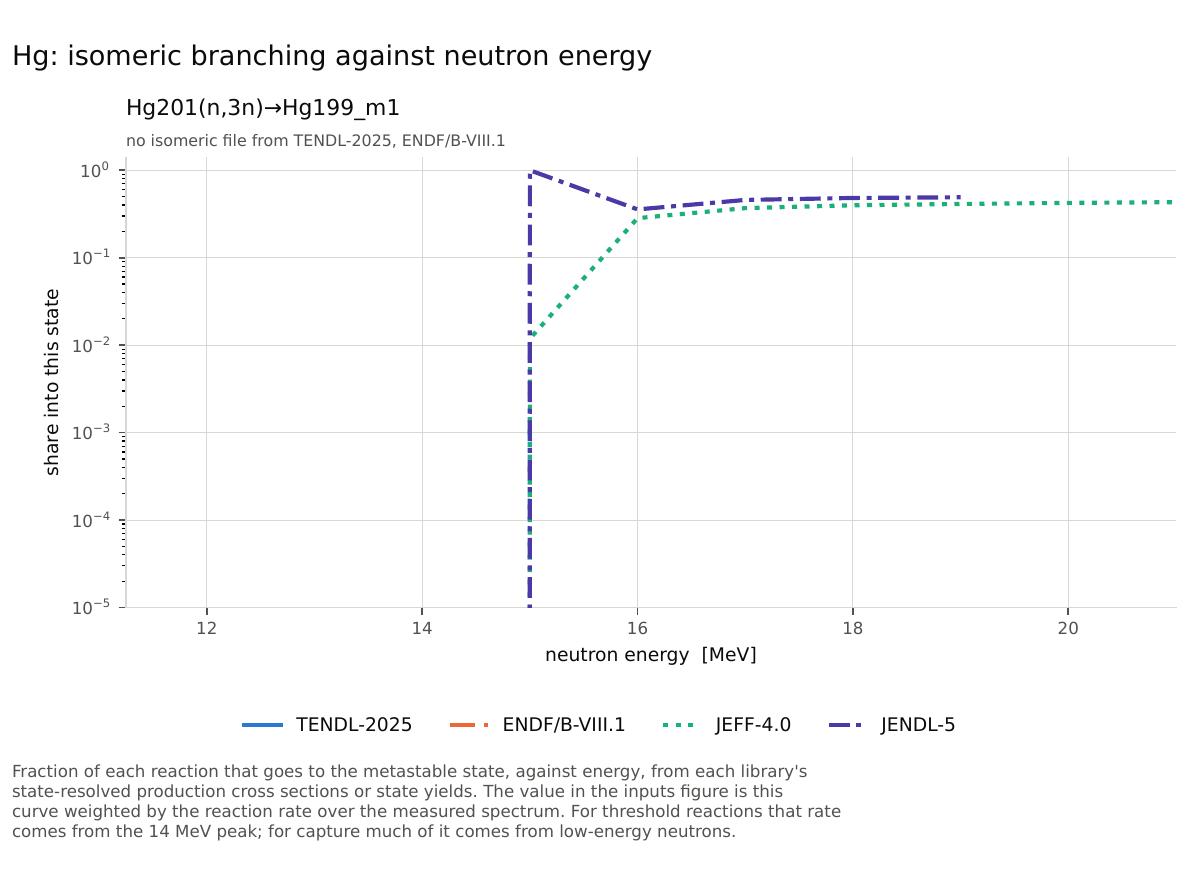}
\caption{Hg: isomeric branching ratio against neutron energy for the channels whose spectrum-averaged ratio differs between libraries.}
\end{figure}
\clearpage

ENDF/B-VIII.1 stands out at 84 percent from the measurement, against 4 percent for the next furthest library and 3 percent for JENDL-5. The median measurement uncertainty is 8 percent.

\subsubsection*{Differences in the nuclear data}
\begin{itemize}\setlength{\itemsep}{6pt}
\item ENDF/B-VIII.1 gives the total reaction rate of Hg201(n,3n)→Hg199 as 1.145e-13, against 1.623e-14 for the median of the others, a factor of 7.\\[2pt]{\itshape This is the total rate of the channel over all its products, so the difference is in the cross section, not in the split between states.}
\item JENDL-5 gives the mean decay energy of Hg197\_m1 as 6.522e+05 eV, against 2.785e+05 eV for the median of the others, a factor of 2.\\[2pt]{\itshape For a given activity the heat is proportional to the mean decay energy, so this difference goes straight into the heat. Compare the decay schemes and how the emitted energy is counted.}
\end{itemize}
\subsubsection*{Suggested measurements}
\begin{itemize}\setlength{\itemsep}{4pt}
\item Natural Hg has 7 isotopes, so the measured heat combines several parents. Enriched single-isotope targets would distinguish their contributions. The candidate parent is Hg201. Enrichment alone does not isolate a single reaction channel.
\item The sample is a compound that also contains O. Their activation adds to the measured heat; a metal or purer sample would remove it.
\item This foil makes Hg197 (half-life 2.67 days), which gives up to 5 percent of the calculated heat. A five-minute irradiation and 55 minute measurement only see the start of this contribution. There is no seven-hour FNS measurement of this foil. A longer irradiation followed for days to months would be needed to test this product's production and decay.
\item The reaction rates are averaged over the measured spectrum. Activation measurements at several well-defined neutron energies, including below 14 MeV, would show which energy range the differences come from.
\end{itemize}
\clearpage
\subsection{Ho: Holmium}\label{foil:Ho}
\begin{figure}[H]
\centering
\includegraphics[width=\textwidth,height=0.82\textheight,keepaspectratio]{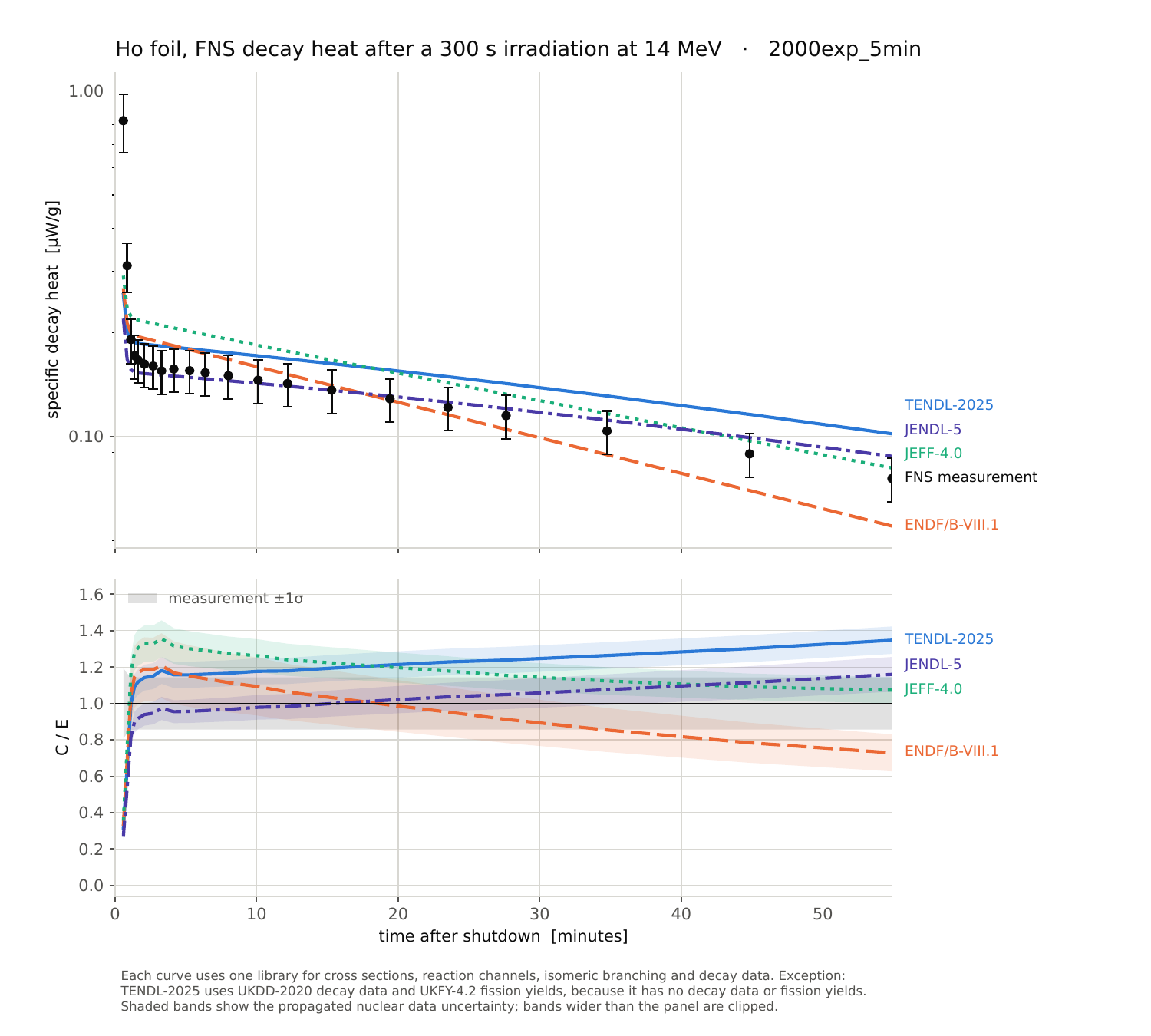}
\caption{Ho: decay heat against the measurement, with C/E beneath. Mean $|C/E-1|$ by library: JENDL-5 11\%, ENDF/B-VIII.1 17\%, TENDL-2025 21\%, JEFF-4.0 26\%.}
\end{figure}
\clearpage

\begin{figure}[H]
\centering
\includegraphics[width=\textwidth,height=0.82\textheight,keepaspectratio]{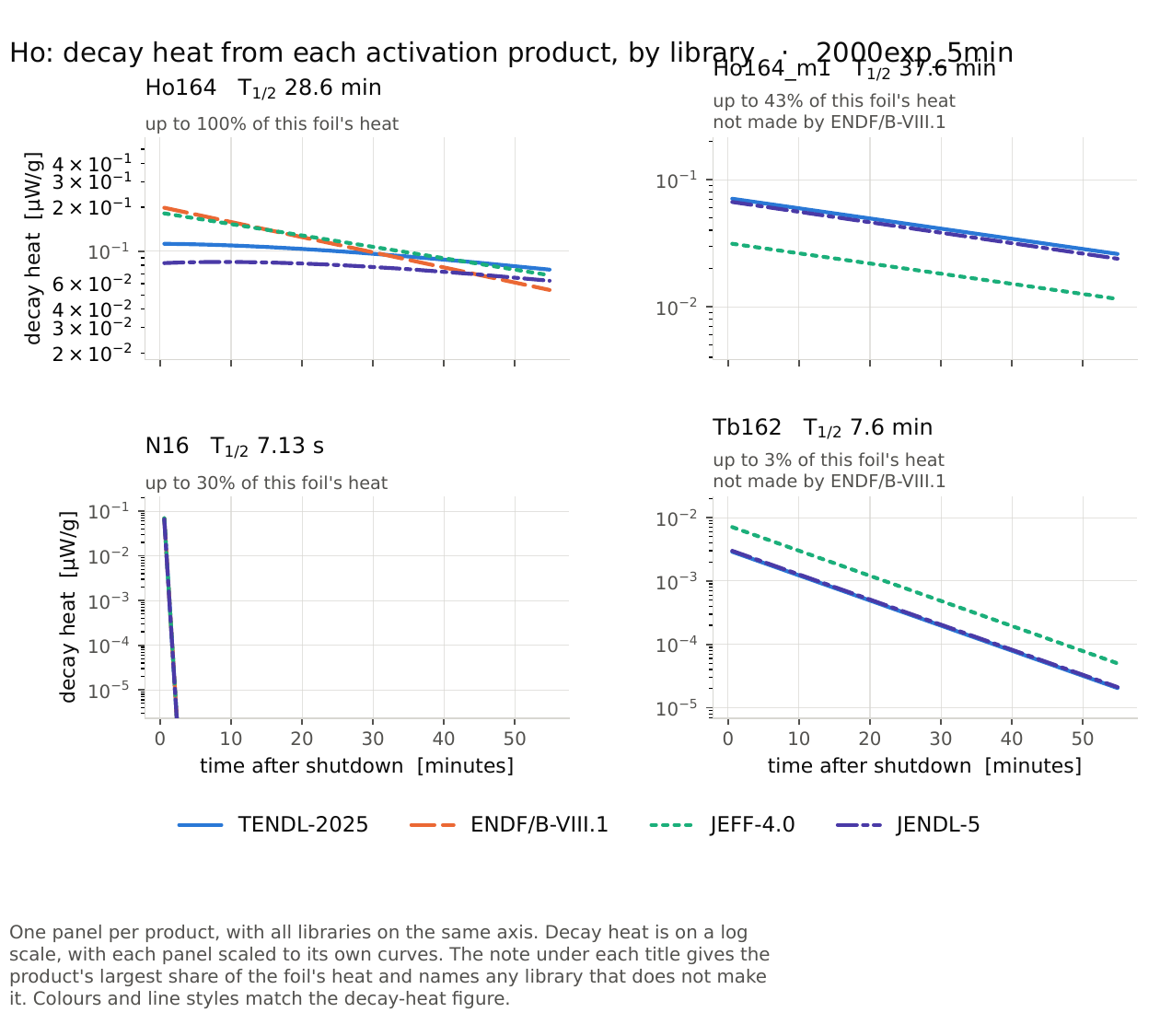}
\caption{Ho: share of the decay heat from each activation product during cooling.}
\end{figure}
\clearpage

\begin{figure}[H]
\centering
\includegraphics[width=\textwidth,height=0.82\textheight,keepaspectratio]{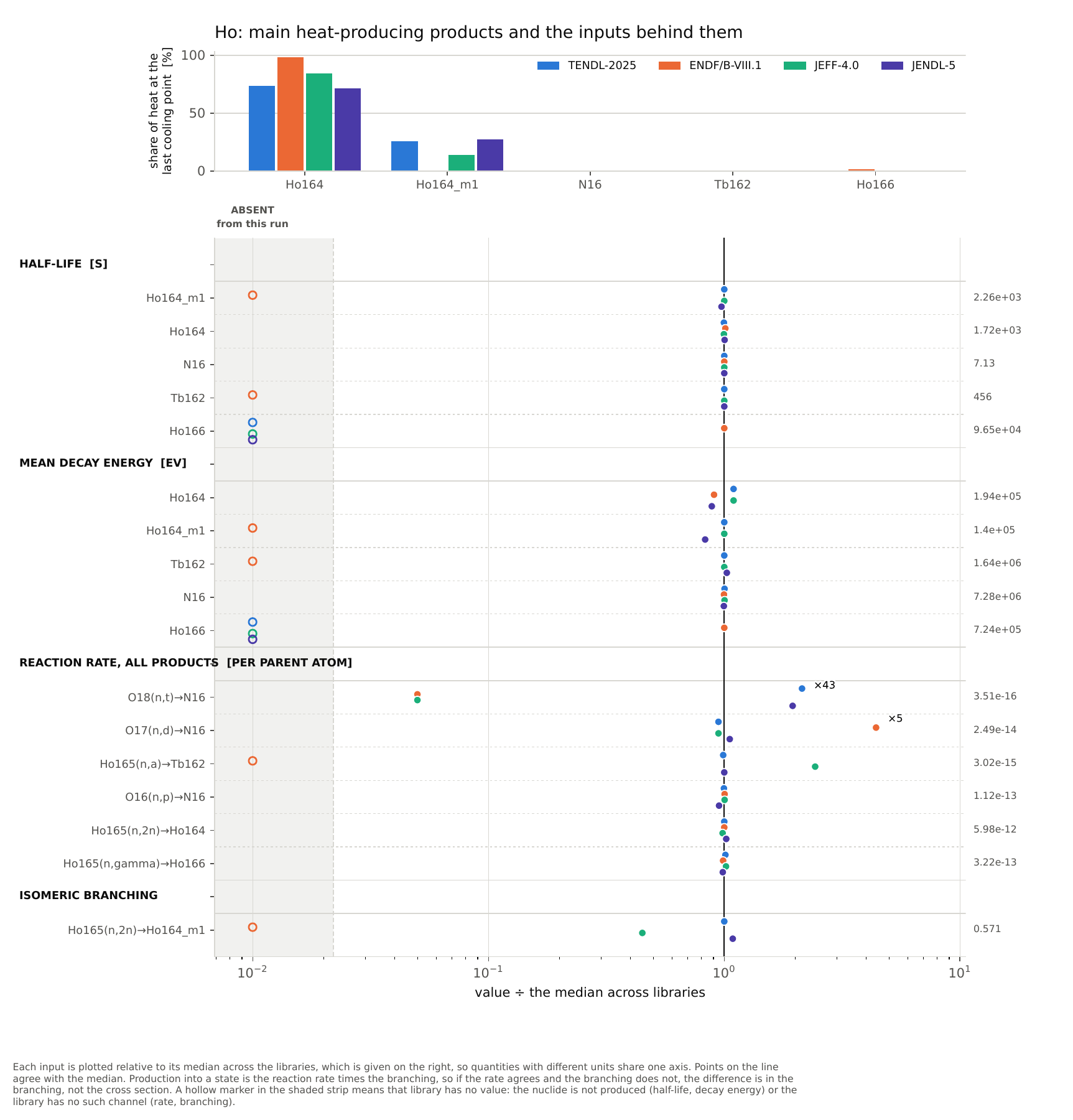}
\caption{Ho: inputs for the main heat-producing products, each relative to its median across the available libraries.}
\end{figure}
\clearpage

\begin{figure}[H]
\centering
\includegraphics[width=\textwidth,height=0.82\textheight,keepaspectratio]{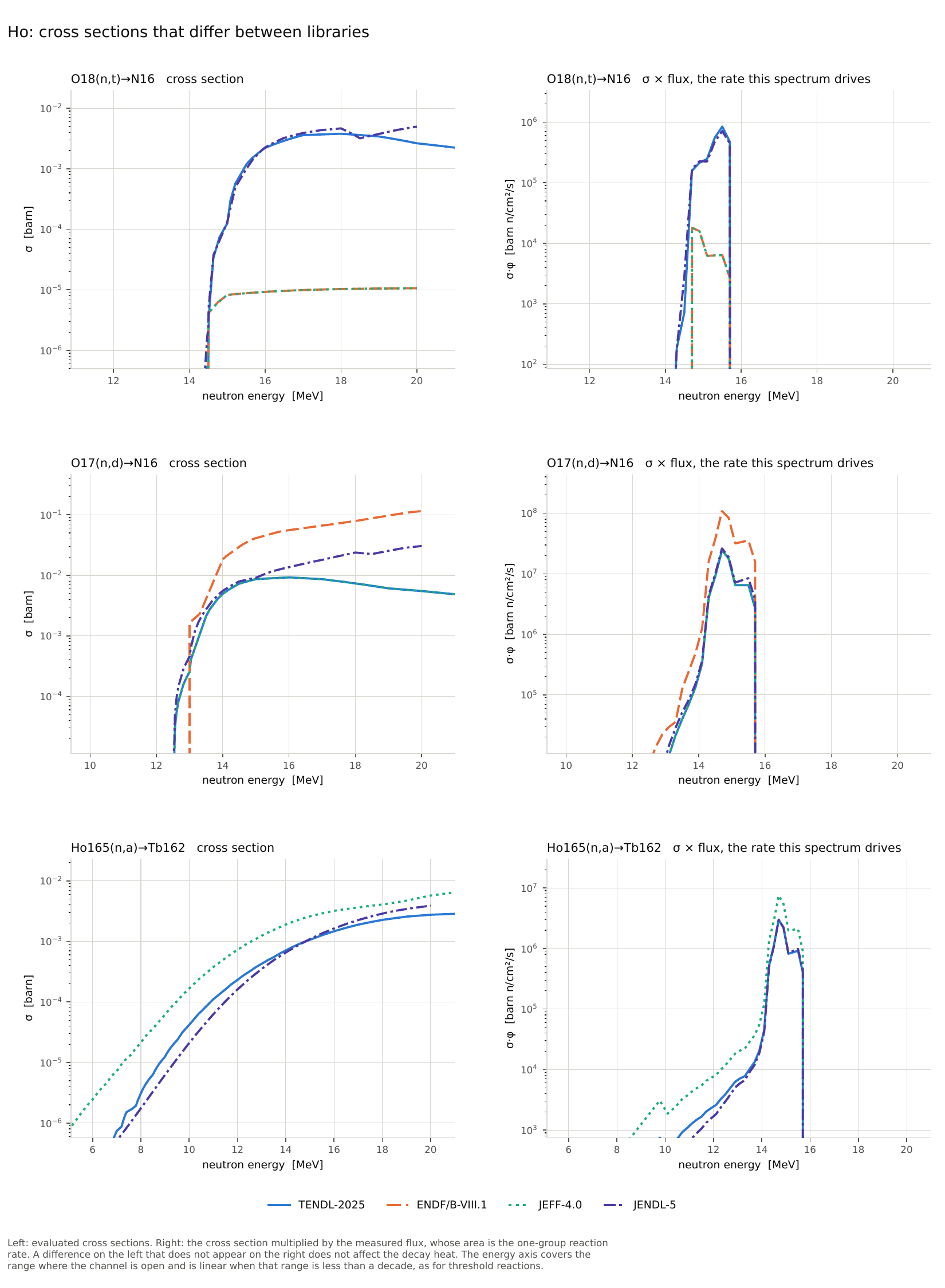}
\caption{Ho: cross sections of the channels whose spectrum-averaged rates differ, with the neutron spectrum and their product.}
\end{figure}
\clearpage

\begin{figure}[H]
\centering
\includegraphics[width=\textwidth,height=0.82\textheight,keepaspectratio]{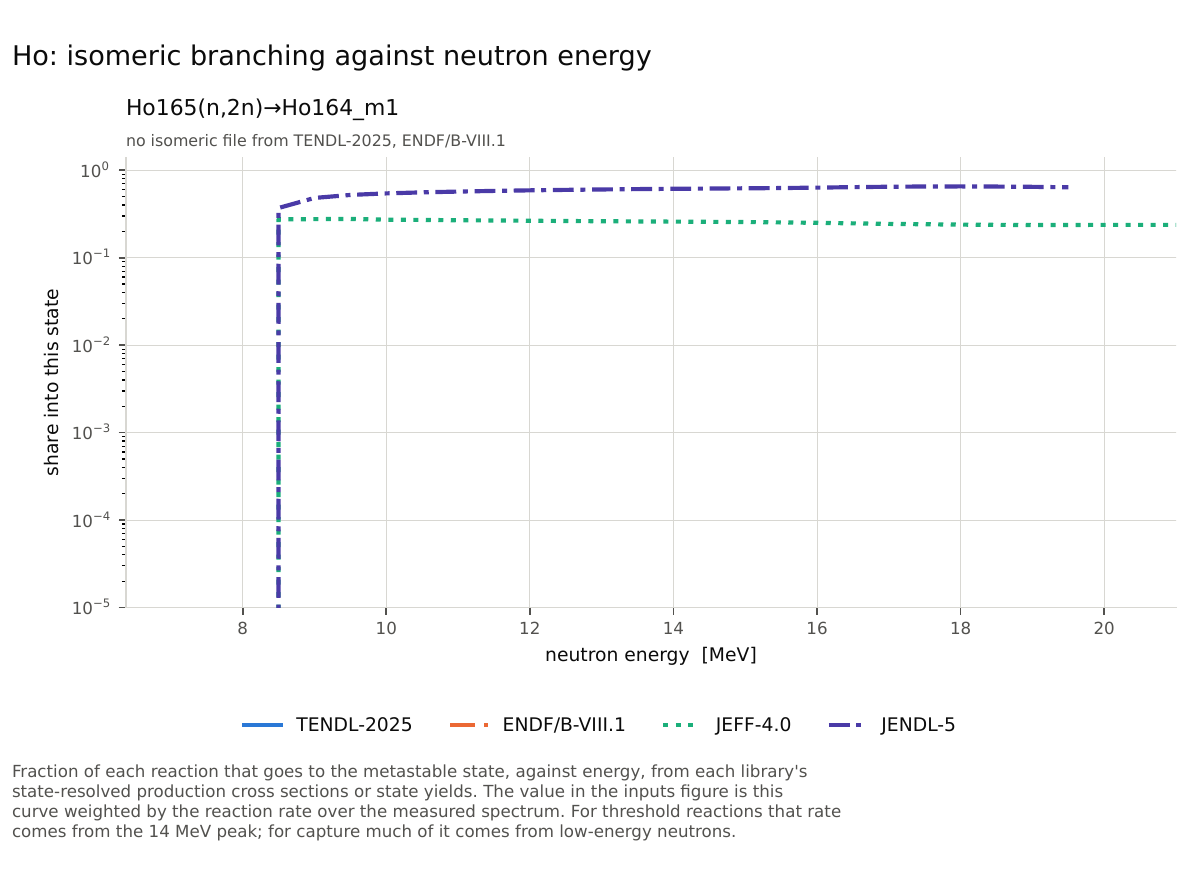}
\caption{Ho: isomeric branching ratio against neutron energy for the channels whose spectrum-averaged ratio differs between libraries.}
\end{figure}
\clearpage

The libraries are between 11 and 26 percent from the measurement, which has a median uncertainty of 14 percent. JENDL-5 is closest and JEFF-4.0 furthest.

\subsubsection*{Differences in the nuclear data}
\begin{itemize}\setlength{\itemsep}{6pt}
\item ENDF/B-VIII.1 gives the total reaction rate of O18(n,t)→N16 as 1.751e-17, against 6.84e-16 for the median of the others, a factor of 39.\\[2pt]{\itshape This is the total rate of the channel over all its products, so the difference is in the cross section, not in the split between states.}
\item ENDF/B-VIII.1 gives the total reaction rate of O17(n,d)→N16 as 1.096e-13, against 2.351e-14 for the median of the others, a factor of 5.
\item JEFF-4.0 gives the total reaction rate of Ho165(n,a)→Tb162 as 7.34e-15, against 3.005e-15 for the median of the others, a factor of 2.
\item JEFF-4.0 gives the isomeric branching of Ho165(n,2n)→Ho164\_m1 as 0.2564, against 0.5954 for the median of the others, a factor of 2.\\[2pt]{\itshape Isomeric branching sets how a channel's production is split between the ground and metastable states. It should be checked together with the total rate of the channel.}
\item ENDF/B-VIII.1 gives no heat from Ho164\_m1; the other libraries do.\\[2pt]{\itshape The heat breakdown alone does not show why. Possible causes are a missing reaction channel, a different state assignment, different decay data or a reporting threshold.}
\item ENDF/B-VIII.1 gives no heat from Tb162; the other libraries do.
\end{itemize}
\subsubsection*{Suggested measurements}
\begin{itemize}\setlength{\itemsep}{4pt}
\item Ho has a single stable isotope, so the foil is already a single-isotope target and enrichment would not help.
\item The sample is a compound that also contains O. Their activation adds to the measured heat; a metal or purer sample would remove it.
\item This foil makes Ho166 (half-life 1.12 days), which gives up to 1 percent of the calculated heat. A five-minute irradiation and 55 minute measurement only see the start of this contribution. There is no seven-hour FNS measurement of this foil. A longer irradiation followed for days to months would be needed to test this product's production and decay.
\item The reaction rates are averaged over the measured spectrum. Activation measurements at several well-defined neutron energies, including below 14 MeV, would show which energy range the differences come from.
\item Measuring the activity of each state separately, for example by gamma spectrometry where the states have distinct lines, would constrain the isomeric ratio. This needs corrections for feeding between states, detector efficiency and irradiation history.
\end{itemize}
\clearpage
\subsection{I: Iodine}\label{foil:I}
\begin{figure}[H]
\centering
\includegraphics[width=\textwidth,height=0.82\textheight,keepaspectratio]{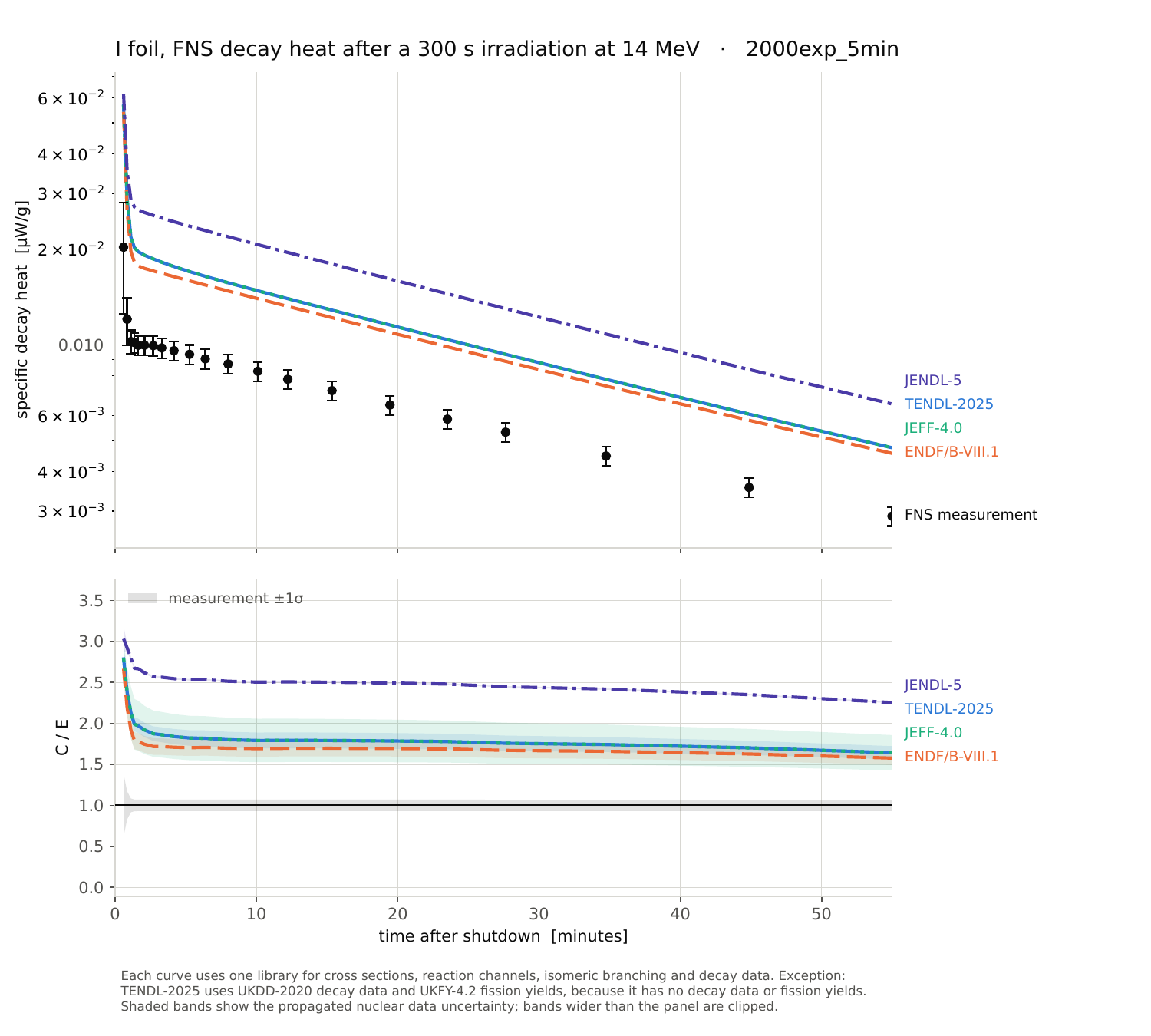}
\caption{I: decay heat against the measurement, with C/E beneath. Mean $|C/E-1|$ by library: ENDF/B-VIII.1 78\%, TENDL-2025 90\%, JEFF-4.0 91\%, JENDL-5 157\%.}
\end{figure}
\clearpage

\begin{figure}[H]
\centering
\includegraphics[width=\textwidth,height=0.82\textheight,keepaspectratio]{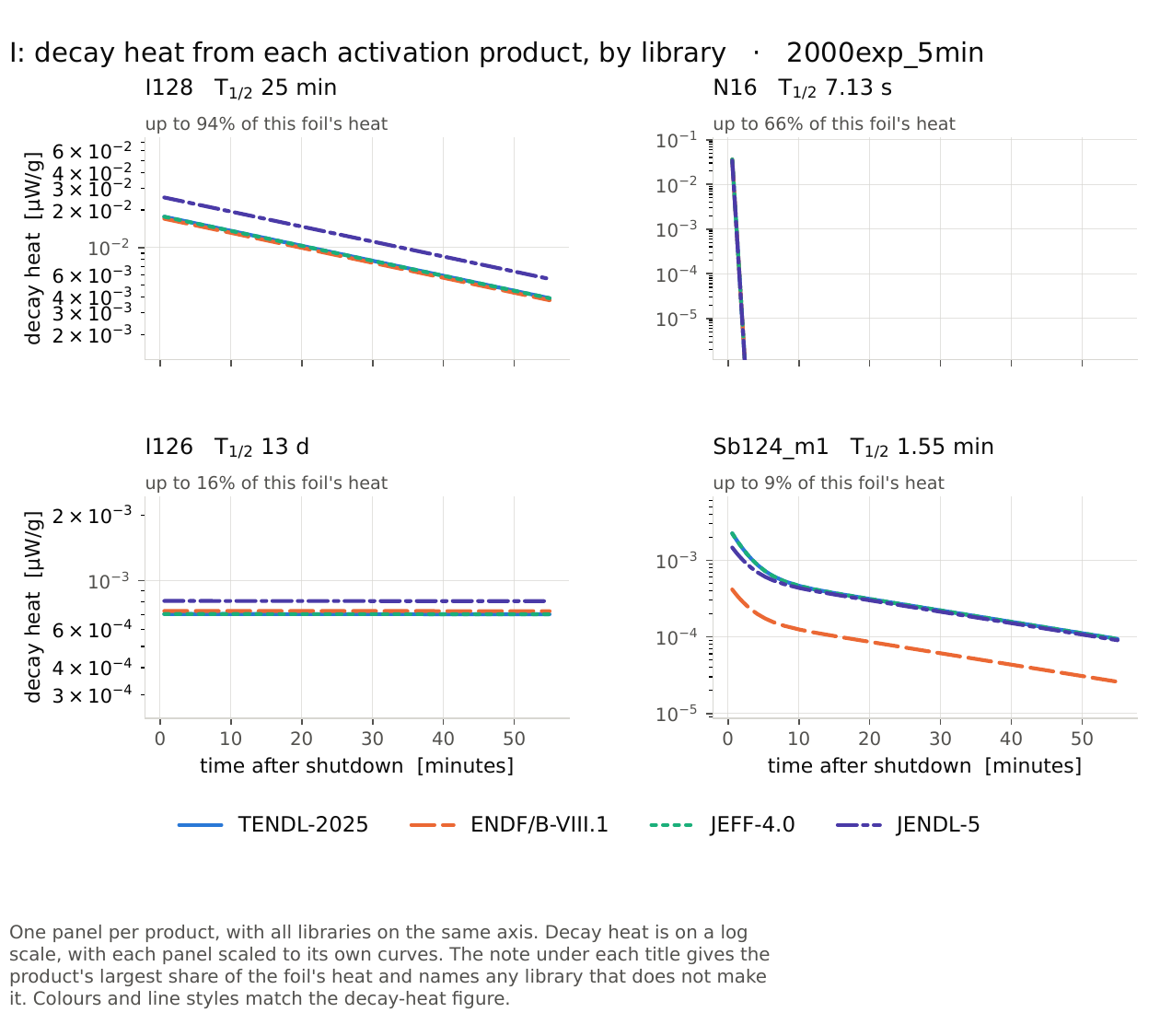}
\caption{I: share of the decay heat from each activation product during cooling.}
\end{figure}
\clearpage

\begin{figure}[H]
\centering
\includegraphics[width=\textwidth,height=0.82\textheight,keepaspectratio]{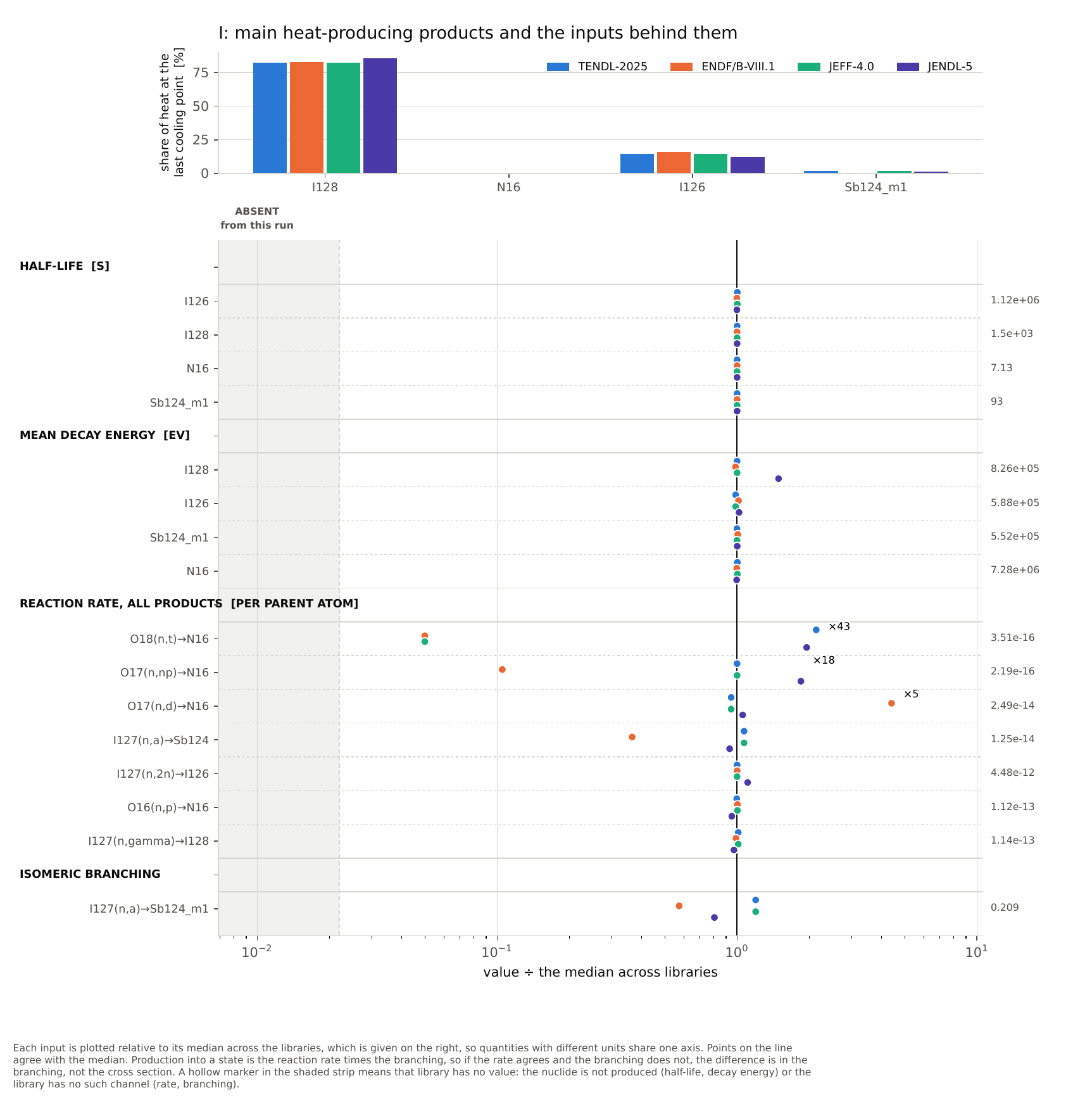}
\caption{I: inputs for the main heat-producing products, each relative to its median across the available libraries.}
\end{figure}
\clearpage

\begin{figure}[H]
\centering
\includegraphics[width=\textwidth,height=0.82\textheight,keepaspectratio]{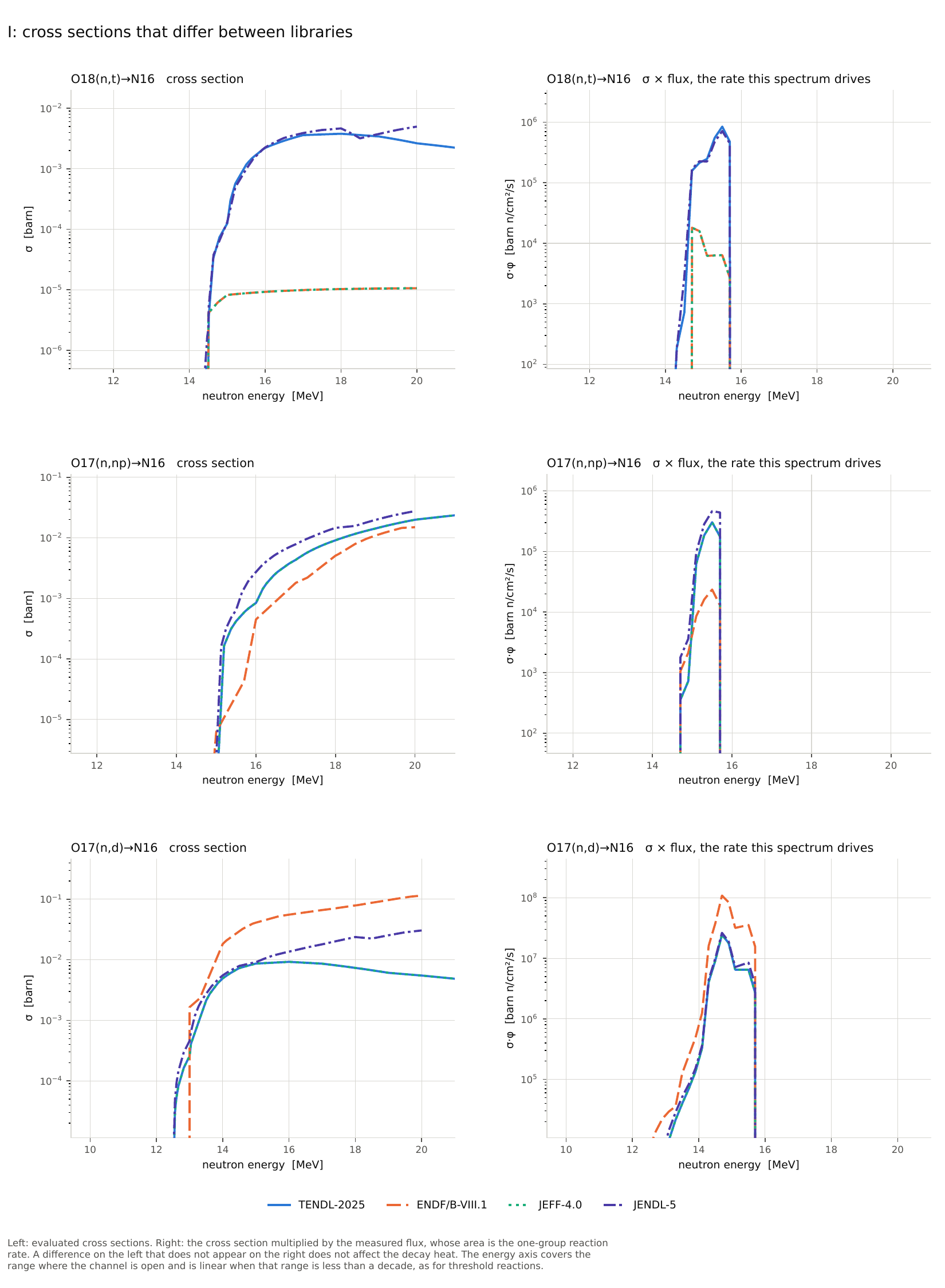}
\caption{I: cross sections of the channels whose spectrum-averaged rates differ, with the neutron spectrum and their product.}
\end{figure}
\clearpage

\begin{figure}[H]
\centering
\includegraphics[width=\textwidth,height=0.82\textheight,keepaspectratio]{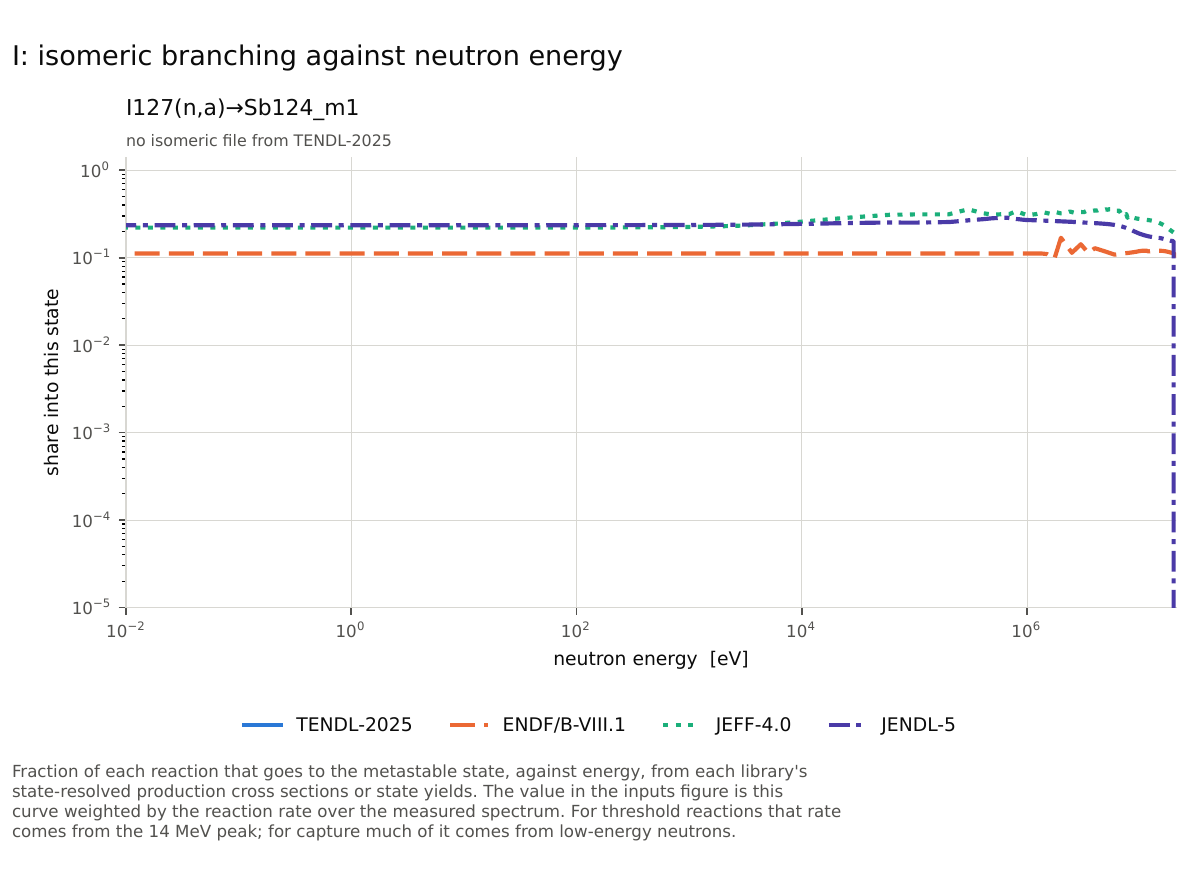}
\caption{I: isomeric branching ratio against neutron energy for the channels whose spectrum-averaged ratio differs between libraries.}
\end{figure}
\clearpage

The libraries are between 78 and 157 percent from the measurement, which has a median uncertainty of 7 percent. ENDF/B-VIII.1 is closest and JENDL-5 furthest.

\subsubsection*{Differences in the nuclear data}
\begin{itemize}\setlength{\itemsep}{6pt}
\item ENDF/B-VIII.1 gives the total reaction rate of O18(n,t)→N16 as 1.751e-17, against 6.84e-16 for the median of the others, a factor of 39.\\[2pt]{\itshape This is the total rate of the channel over all its products, so the difference is in the cross section, not in the split between states.}
\item ENDF/B-VIII.1 gives the total reaction rate of O17(n,np)→N16 as 2.295e-17, against 2.186e-16 for the median of the others, a factor of 10.
\item ENDF/B-VIII.1 gives the total reaction rate of O17(n,d)→N16 as 1.096e-13, against 2.351e-14 for the median of the others, a factor of 5.
\item ENDF/B-VIII.1 gives the total reaction rate of I127(n,a)→Sb124 as 4.567e-15, against 1.336e-14 for the median of the others, a factor of 3.
\end{itemize}
\subsubsection*{Suggested measurements}
\begin{itemize}\setlength{\itemsep}{4pt}
\item I has a single stable isotope, so the foil is already a single-isotope target and enrichment would not help.
\item The sample is a compound that also contains C, H, O. Their activation adds to the measured heat; a metal or purer sample would remove it.
\item This foil makes I126 (half-life 13 days), which gives up to 16 percent of the calculated heat. A five-minute irradiation and 55 minute measurement only see the start of this contribution. There is no seven-hour FNS measurement of this foil. A longer irradiation followed for days to months would be needed to test this product's production and decay.
\item The reaction rates are averaged over the measured spectrum. Activation measurements at several well-defined neutron energies, including below 14 MeV, would show which energy range the differences come from.
\end{itemize}
\clearpage
\subsection{In: Indium}\label{foil:In}
\begin{figure}[H]
\centering
\includegraphics[width=\textwidth,height=0.82\textheight,keepaspectratio]{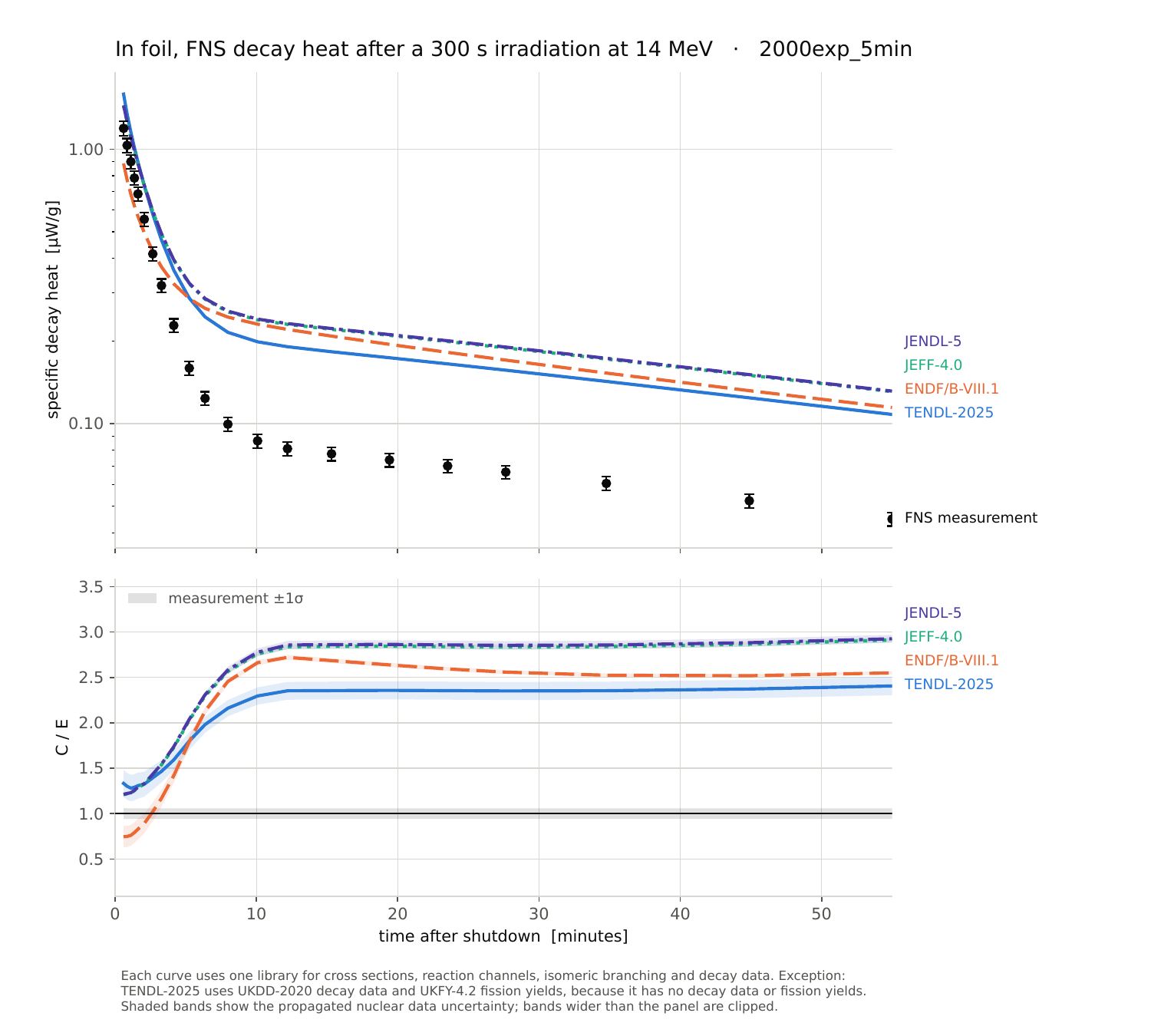}
\caption{In: decay heat against the measurement, with C/E beneath. Mean $|C/E-1|$ by library: TENDL-2025 88\%, ENDF/B-VIII.1 94\%, JEFF-4.0 113\%, JENDL-5 114\%.}
\end{figure}
\clearpage

\begin{figure}[H]
\centering
\includegraphics[width=\textwidth,height=0.82\textheight,keepaspectratio]{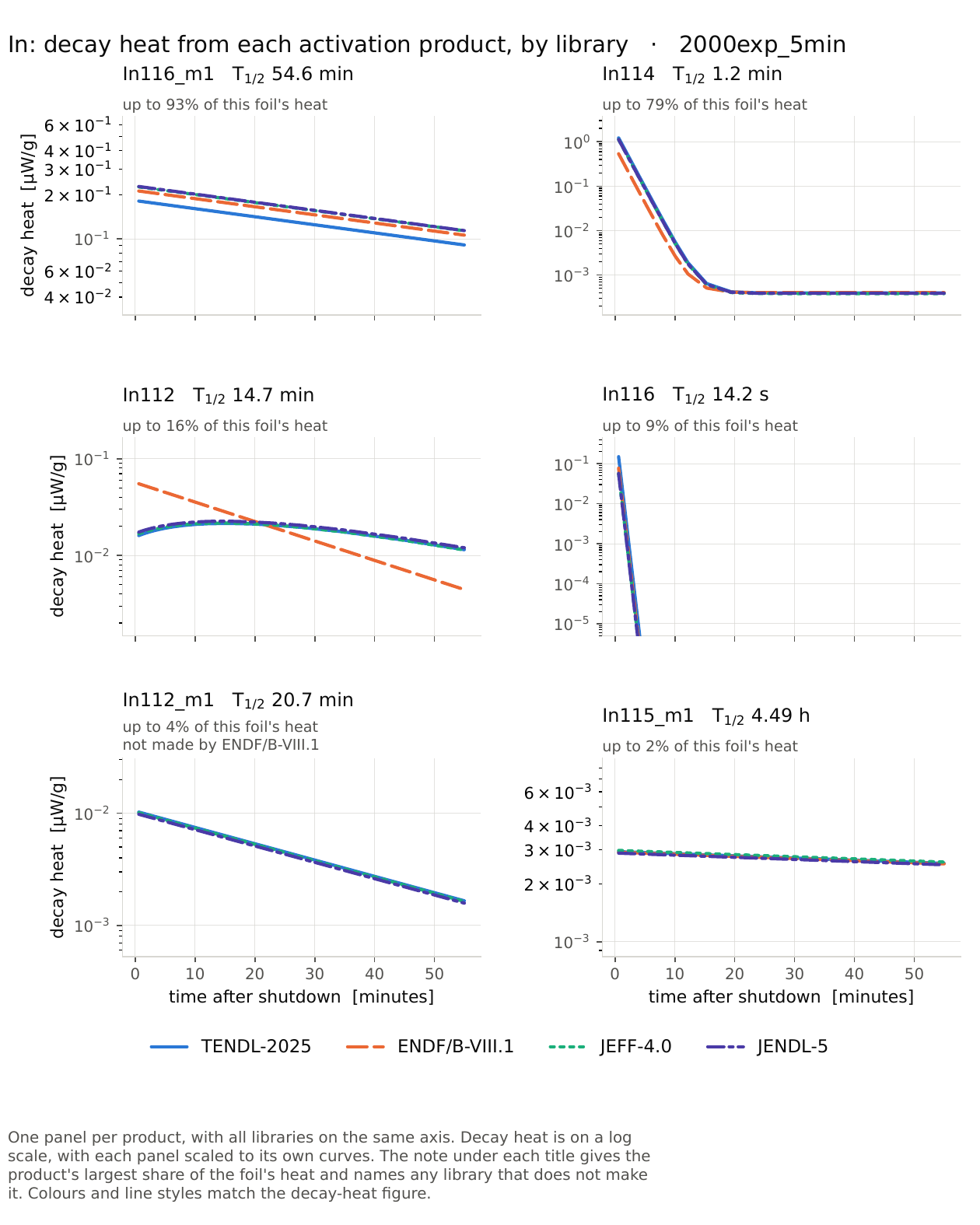}
\caption{In: share of the decay heat from each activation product during cooling.}
\end{figure}
\clearpage

\begin{figure}[H]
\centering
\includegraphics[width=\textwidth,height=0.82\textheight,keepaspectratio]{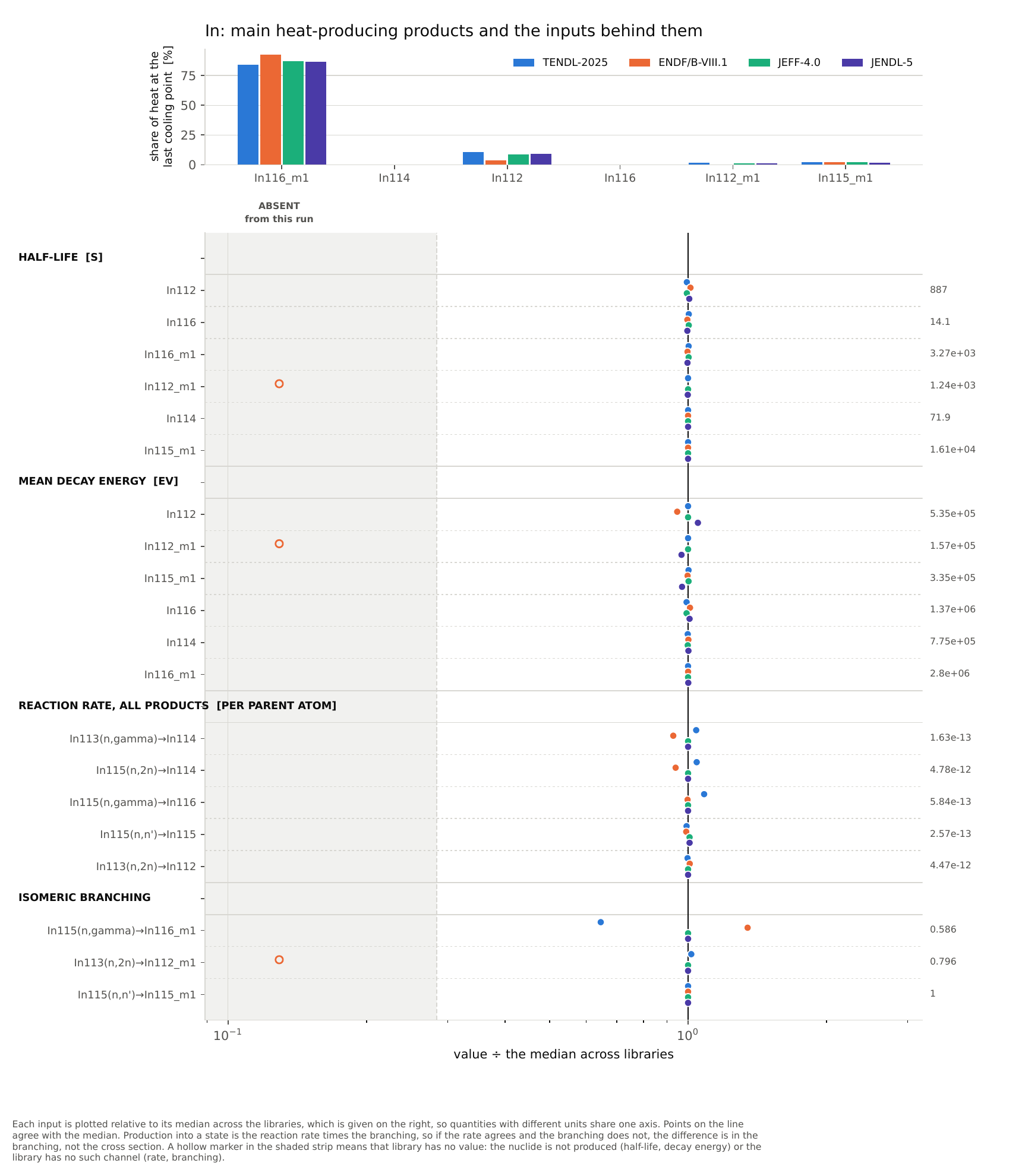}
\caption{In: inputs for the main heat-producing products, each relative to its median across the available libraries.}
\end{figure}
\clearpage

\begin{figure}[H]
\centering
\includegraphics[width=\textwidth,height=0.82\textheight,keepaspectratio]{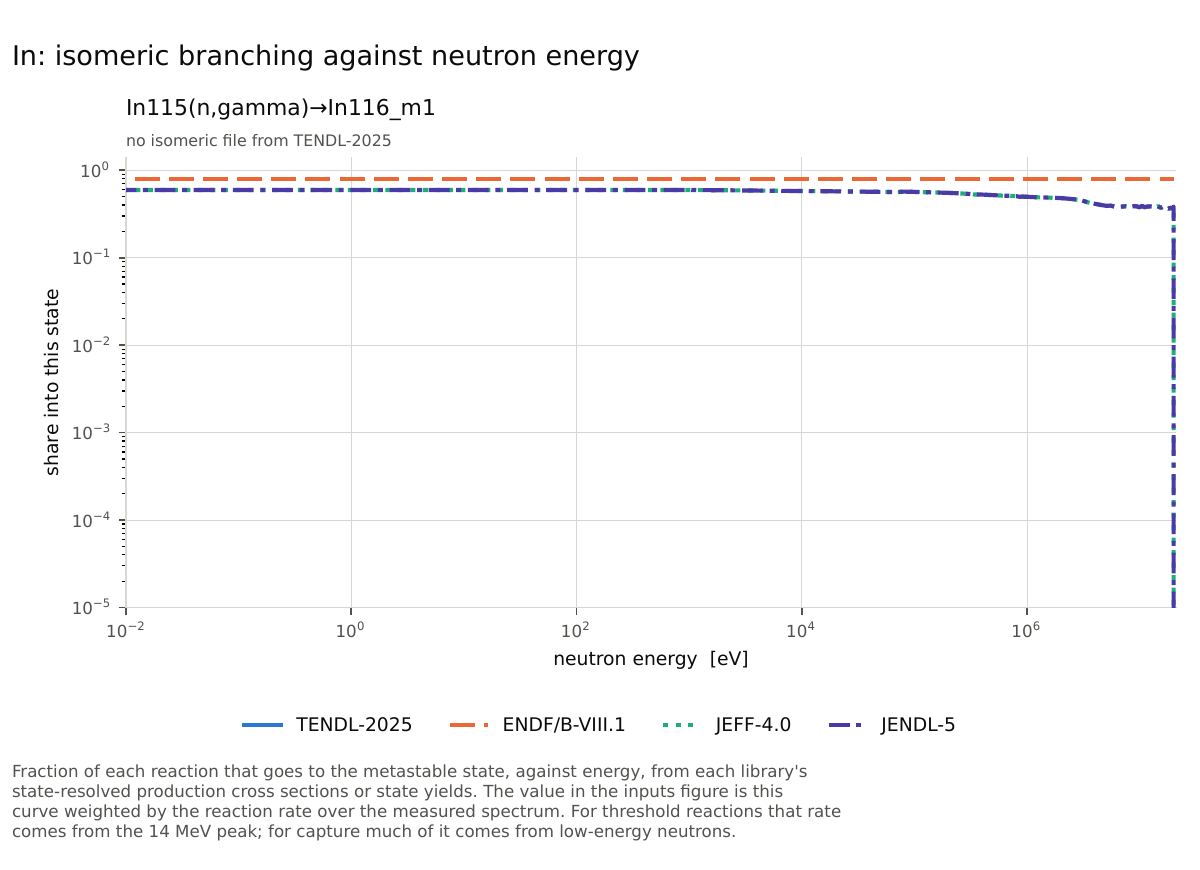}
\caption{In: isomeric branching ratio against neutron energy for the channels whose spectrum-averaged ratio differs between libraries.}
\end{figure}
\clearpage

The libraries are between 88 and 114 percent from the measurement, which has a median uncertainty of 6 percent. TENDL-2025 is closest and JENDL-5 furthest.

\subsubsection*{Differences in the nuclear data}
\begin{itemize}\setlength{\itemsep}{6pt}
\item ENDF/B-VIII.1 gives no heat from In112\_m1; the other libraries do.\\[2pt]{\itshape The heat breakdown alone does not show why. Possible causes are a missing reaction channel, a different state assignment, different decay data or a reporting threshold.}
\end{itemize}
\subsubsection*{Suggested measurements}
\begin{itemize}\setlength{\itemsep}{4pt}
\item Natural In has 2 isotopes, so the measured heat combines several parents. Enriched single-isotope targets would distinguish their contributions. Enrichment alone does not isolate a single reaction channel.
\end{itemize}
\clearpage
\subsection{Ir: Iridium}\label{foil:Ir}
\begin{figure}[H]
\centering
\includegraphics[width=\textwidth,height=0.82\textheight,keepaspectratio]{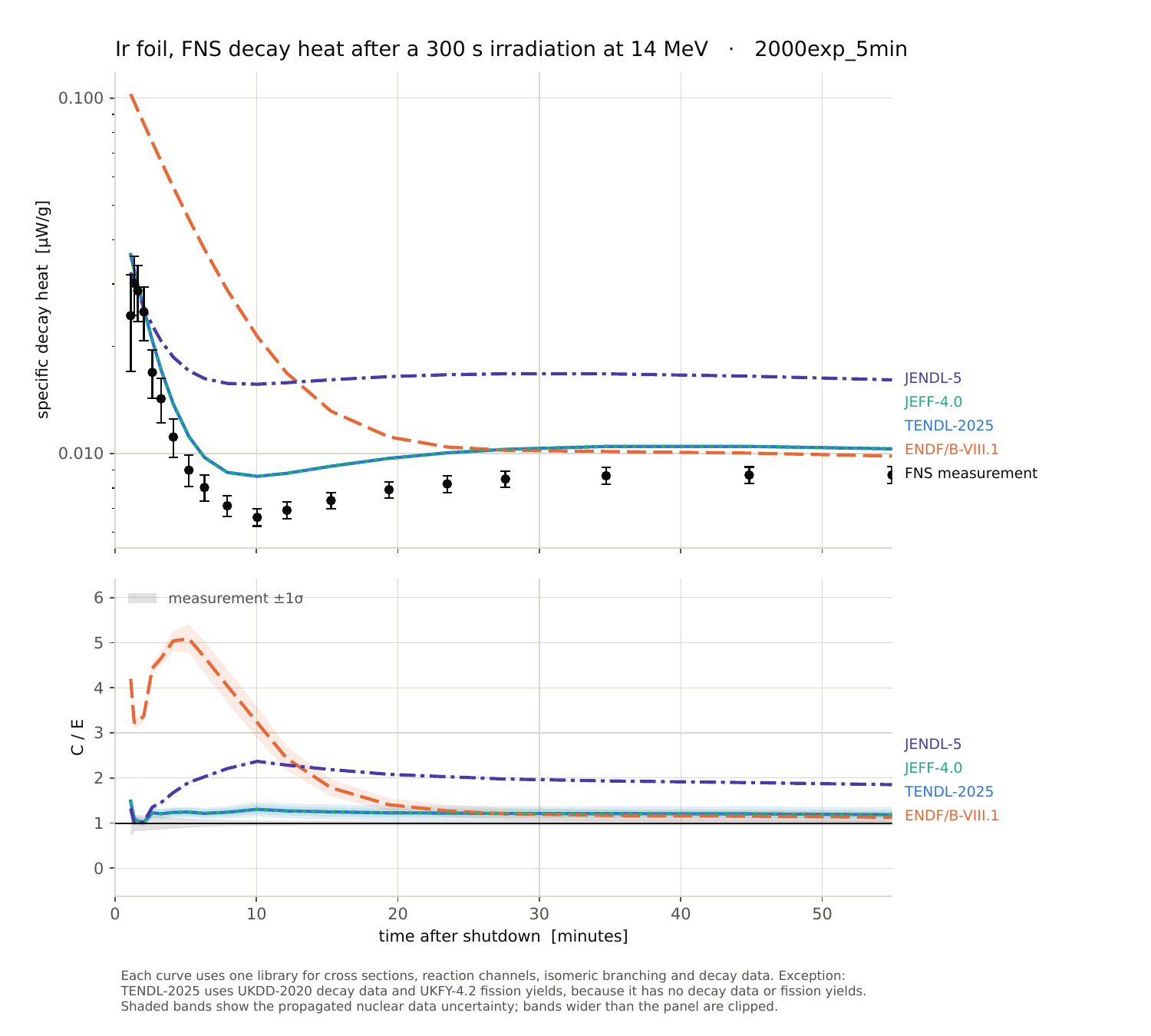}
\caption{Ir: decay heat against the measurement, with C/E beneath. Mean $|C/E-1|$ by library: TENDL-2025 22\%, JEFF-4.0 22\%, JENDL-5 77\%, ENDF/B-VIII.1 199\%.}
\end{figure}
\clearpage

\begin{figure}[H]
\centering
\includegraphics[width=\textwidth,height=0.82\textheight,keepaspectratio]{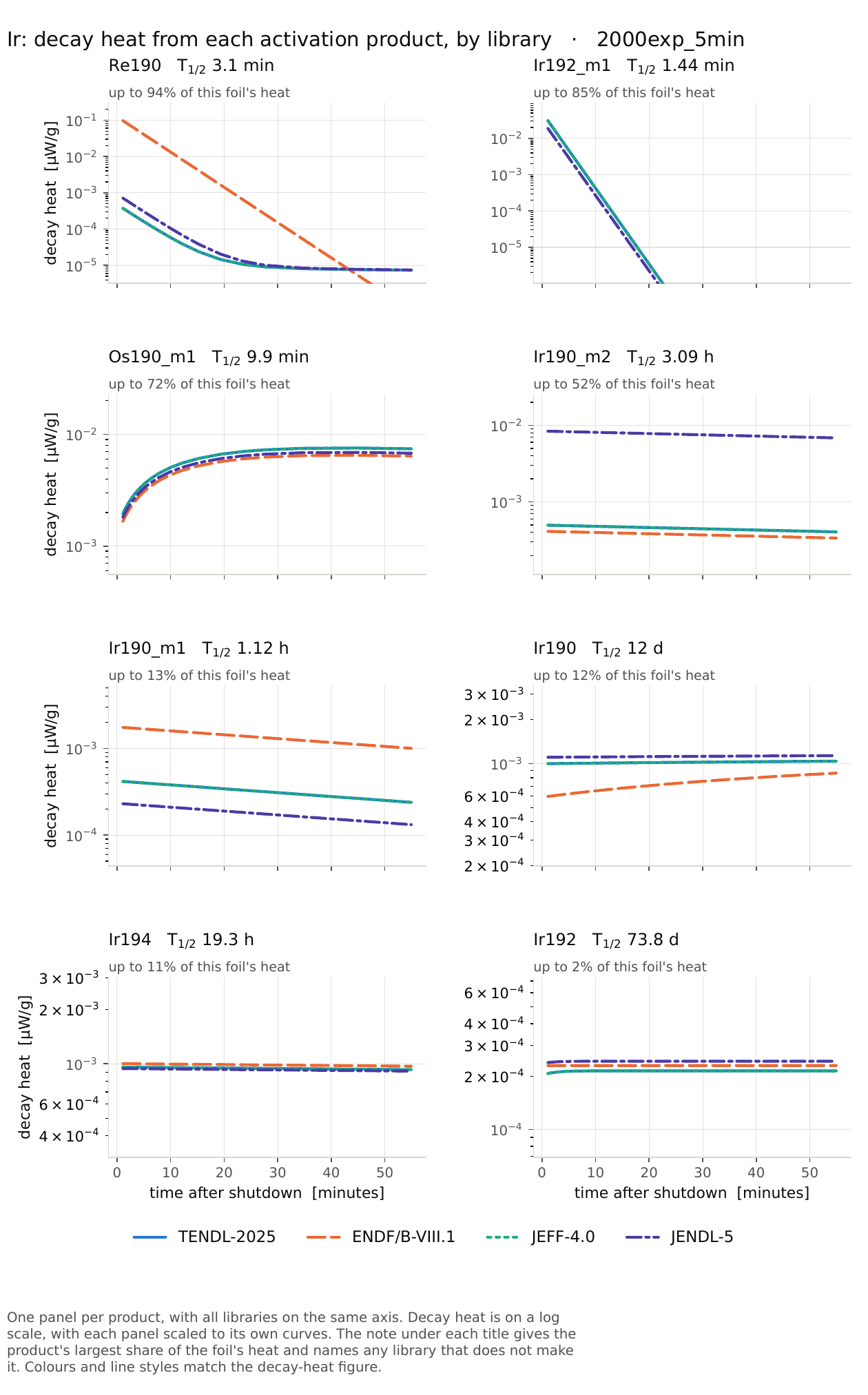}
\caption{Ir: share of the decay heat from each activation product during cooling.}
\end{figure}
\clearpage

\begin{figure}[H]
\centering
\includegraphics[width=\textwidth,height=0.82\textheight,keepaspectratio]{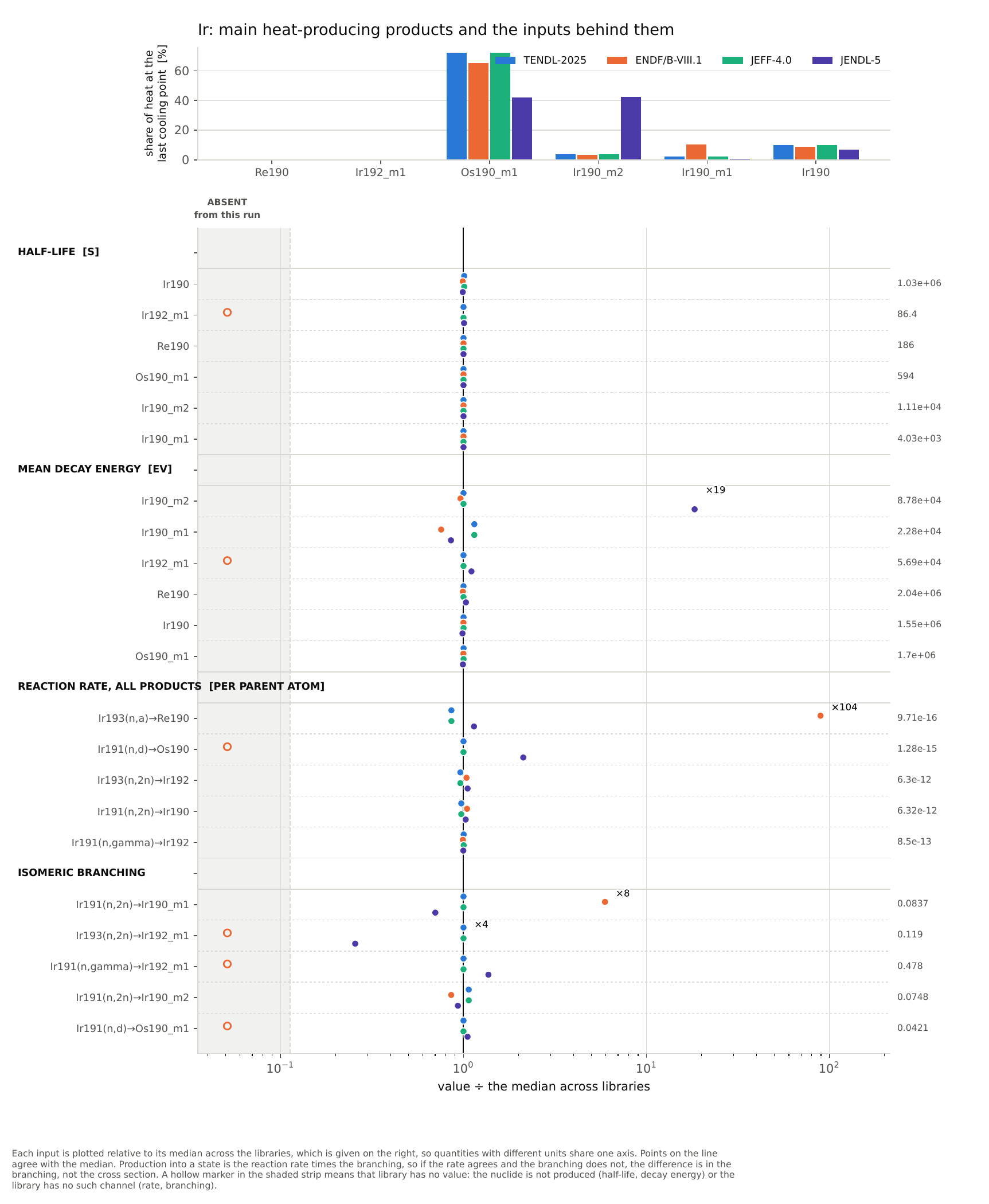}
\caption{Ir: inputs for the main heat-producing products, each relative to its median across the available libraries.}
\end{figure}
\clearpage

\begin{figure}[H]
\centering
\includegraphics[width=\textwidth,height=0.82\textheight,keepaspectratio]{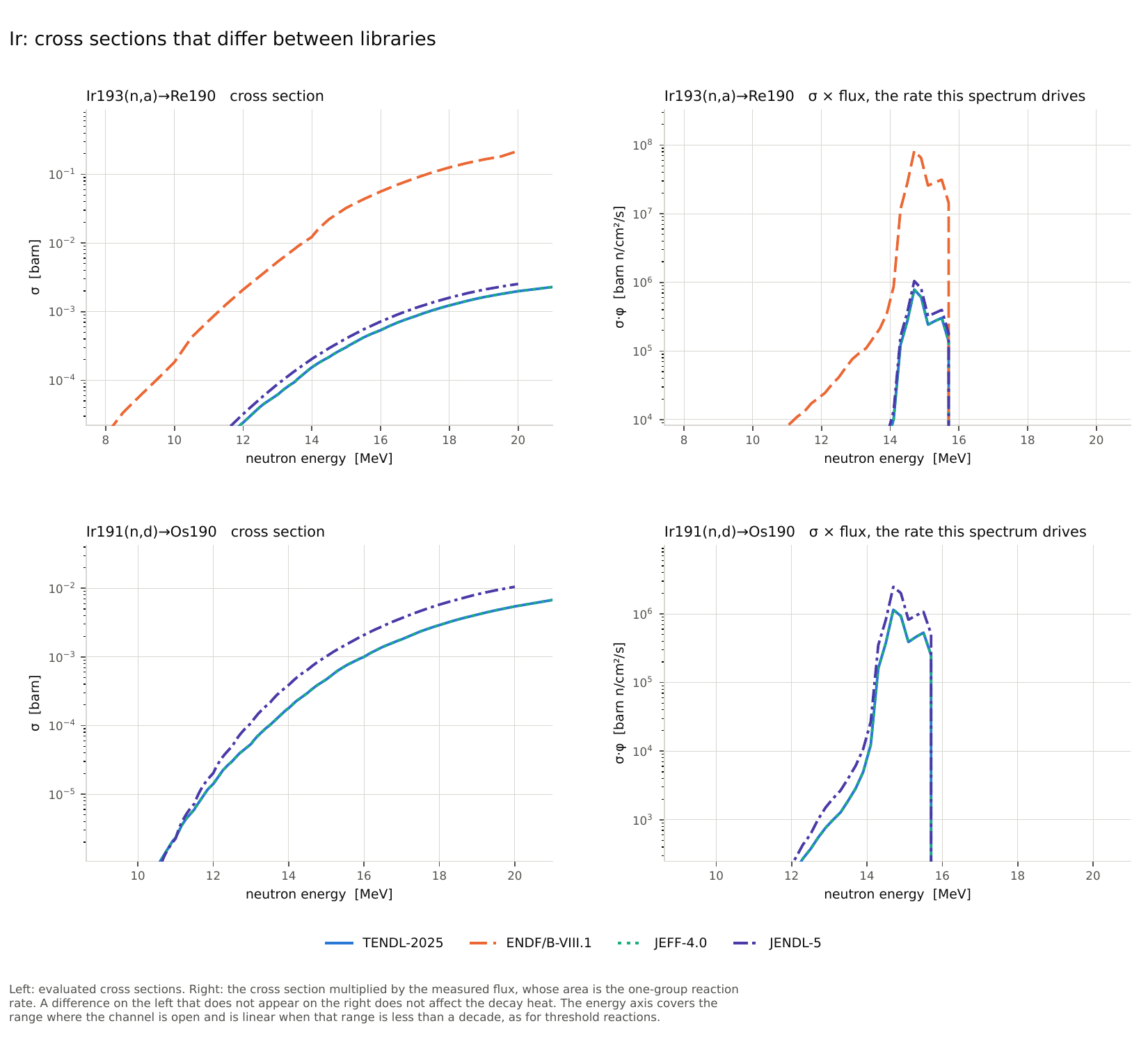}
\caption{Ir: cross sections of the channels whose spectrum-averaged rates differ, with the neutron spectrum and their product.}
\end{figure}
\clearpage

\begin{figure}[H]
\centering
\includegraphics[width=\textwidth,height=0.82\textheight,keepaspectratio]{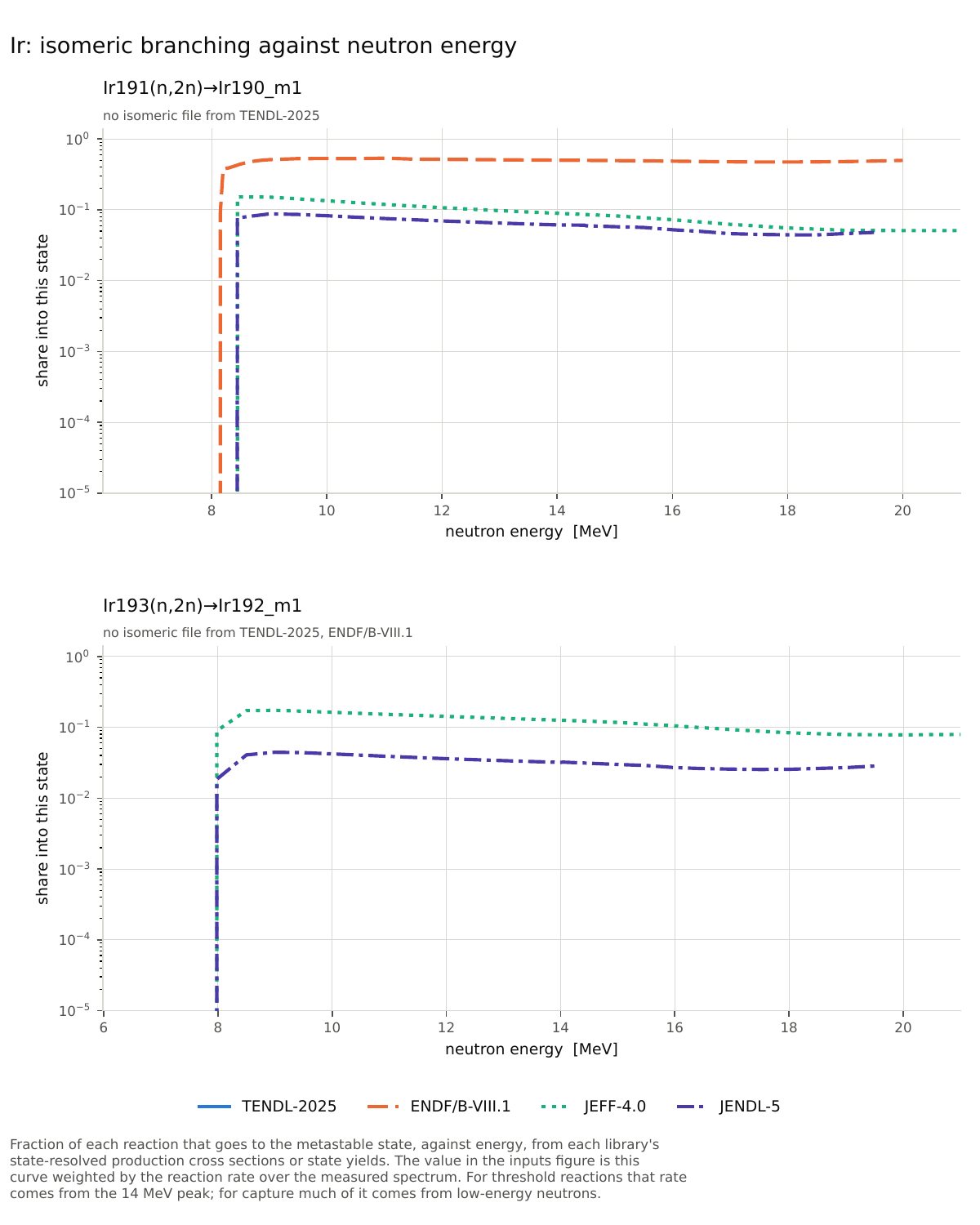}
\caption{Ir: isomeric branching ratio against neutron energy for the channels whose spectrum-averaged ratio differs between libraries.}
\end{figure}
\clearpage

ENDF/B-VIII.1 stands out at 199 percent from the measurement, against 77 percent for the next furthest library and 22 percent for TENDL-2025. The median measurement uncertainty is 7 percent.

\subsubsection*{Differences in the nuclear data}
\begin{itemize}\setlength{\itemsep}{6pt}
\item ENDF/B-VIII.1 gives the total reaction rate of Ir193(n,a)→Re190 as 8.681e-14, against 8.339e-16 for the median of the others, a factor of 104.\\[2pt]{\itshape This is the total rate of the channel over all its products, so the difference is in the cross section, not in the split between states.}
\item JENDL-5 gives the mean decay energy of Ir190\_m2 as 1.609e+06 eV, against 8.775e+04 eV for the median of the others, a factor of 18.\\[2pt]{\itshape For a given activity the heat is proportional to the mean decay energy, so this difference goes straight into the heat. Compare the decay schemes and how the emitted energy is counted.}
\item ENDF/B-VIII.1 gives the isomeric branching of Ir191(n,2n)→Ir190\_m1 as 0.4966, against 0.08371 for the median of the others, a factor of 6.\\[2pt]{\itshape Isomeric branching sets how a channel's production is split between the ground and metastable states. It should be checked together with the total rate of the channel.}
\item JENDL-5 gives the isomeric branching of Ir193(n,2n)→Ir192\_m1 as 0.03051, against 0.1192 for the median of the others, a factor of 4.
\end{itemize}
\subsubsection*{Suggested measurements}
\begin{itemize}\setlength{\itemsep}{4pt}
\item Natural Ir has 2 isotopes, so the measured heat combines several parents. Enriched single-isotope targets would distinguish their contributions. Candidate parents are Ir191, Ir193. Enrichment alone does not isolate a single reaction channel.
\item This foil makes Ir190 (half-life 12 days), which gives up to 12 percent of the calculated heat. A five-minute irradiation and 55 minute measurement only see the start of this contribution. There is no seven-hour FNS measurement of this foil. A longer irradiation followed for days to months would be needed to test this product's production and decay.
\item The reaction rates are averaged over the measured spectrum. Activation measurements at several well-defined neutron energies, including below 14 MeV, would show which energy range the differences come from.
\item Measuring the activity of each state separately, for example by gamma spectrometry where the states have distinct lines, would constrain the isomeric ratio. This needs corrections for feeding between states, detector efficiency and irradiation history.
\end{itemize}
\clearpage
\subsection{K: Potassium}\label{foil:K}
\begin{figure}[H]
\centering
\includegraphics[width=\textwidth,height=0.82\textheight,keepaspectratio]{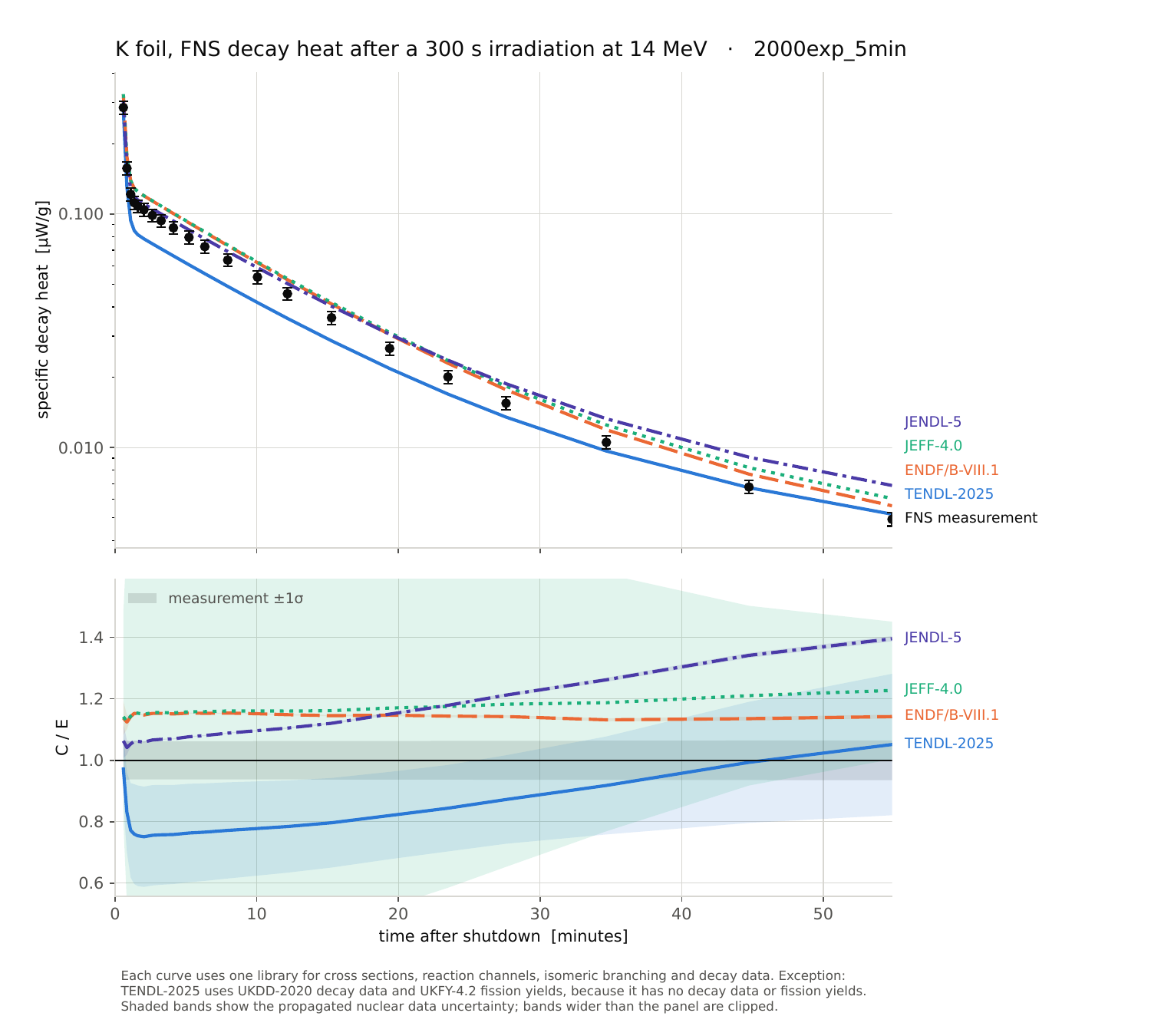}
\caption{K: decay heat against the measurement, with C/E beneath. Mean $|C/E-1|$ by library: JENDL-5 13\%, ENDF/B-VIII.1 15\%, JEFF-4.0 16\%, TENDL-2025 18\%.}
\end{figure}
\clearpage

\begin{figure}[H]
\centering
\includegraphics[width=\textwidth,height=0.82\textheight,keepaspectratio]{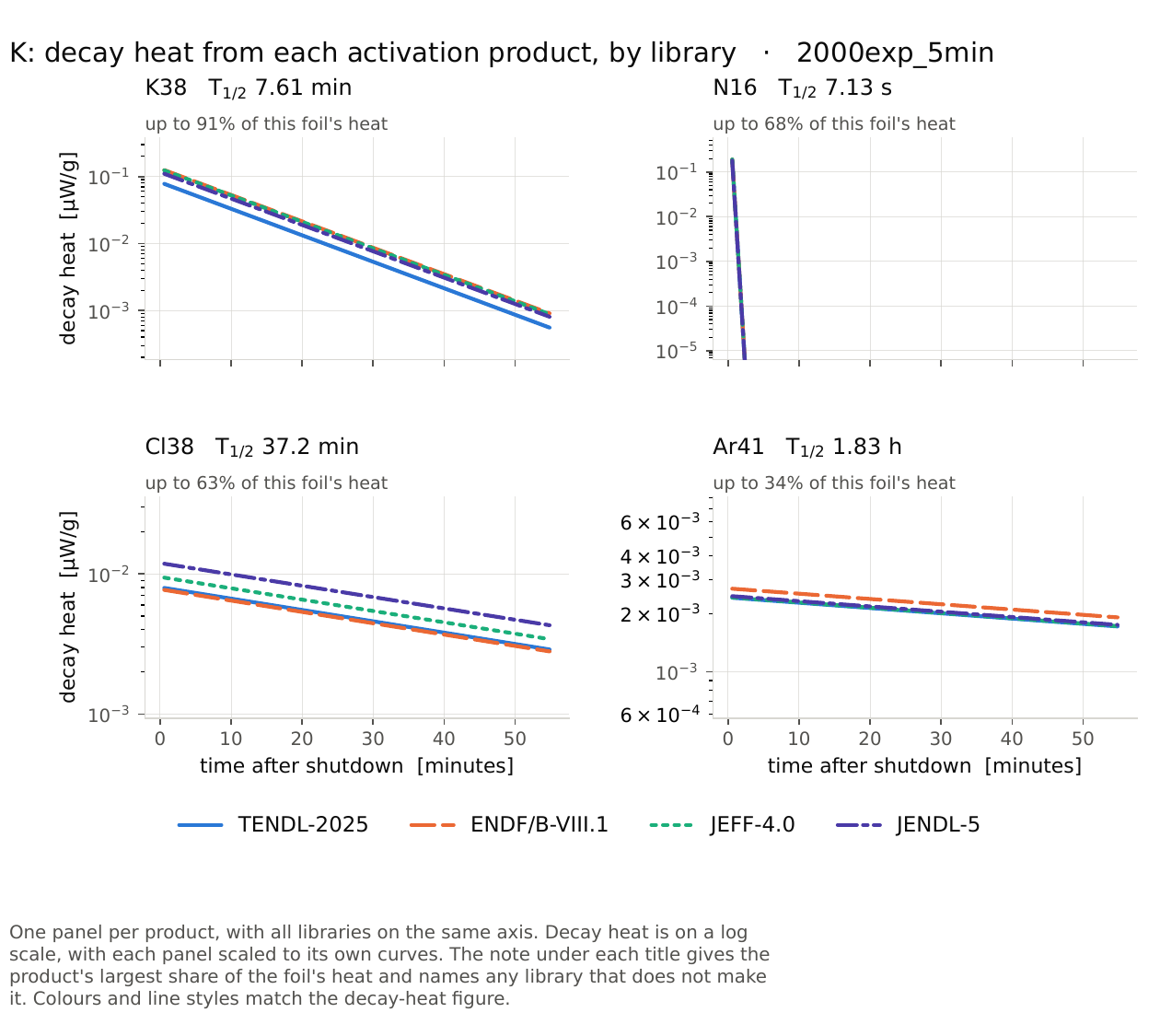}
\caption{K: share of the decay heat from each activation product during cooling.}
\end{figure}
\clearpage

\begin{figure}[H]
\centering
\includegraphics[width=\textwidth,height=0.82\textheight,keepaspectratio]{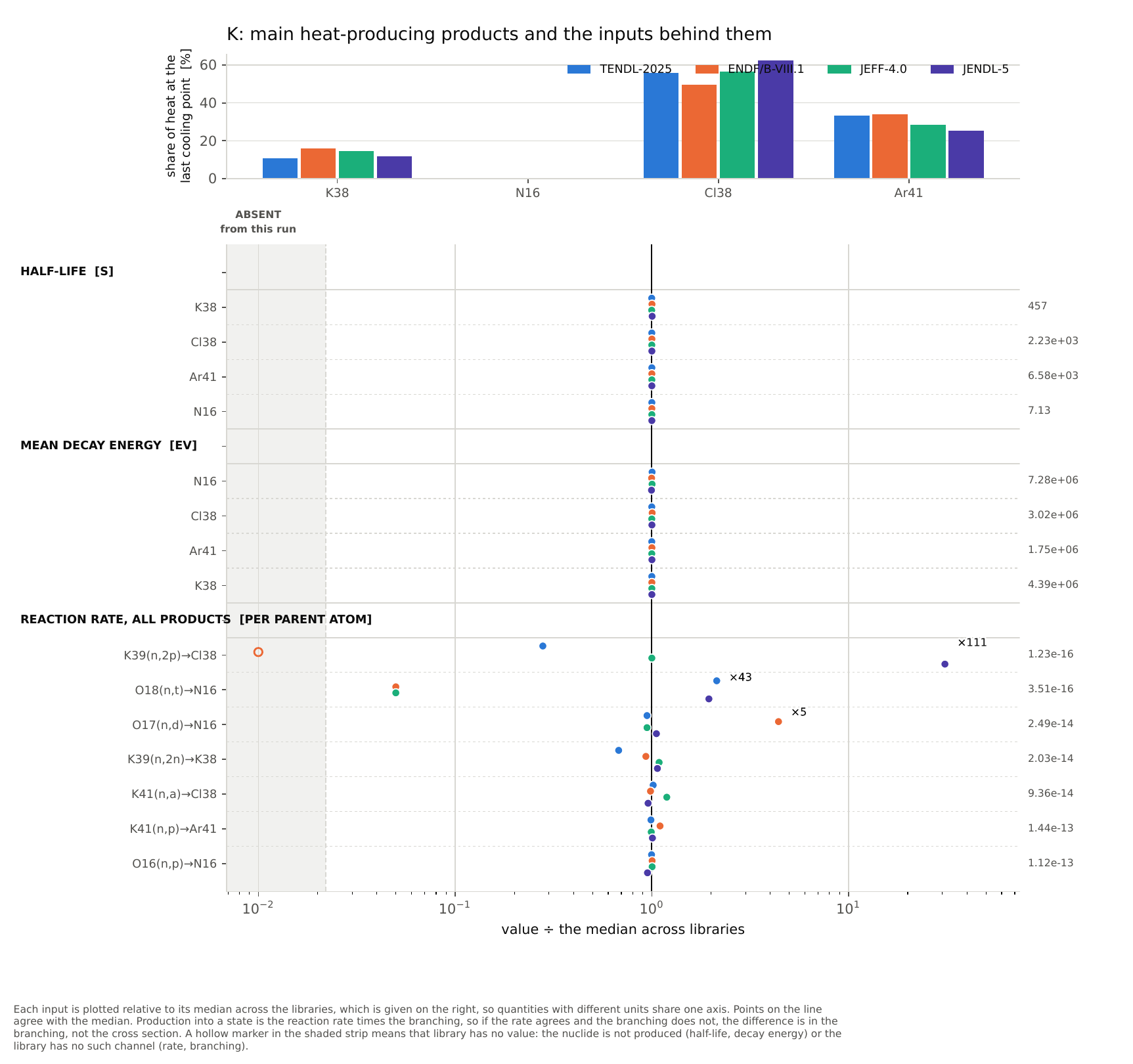}
\caption{K: inputs for the main heat-producing products, each relative to its median across the available libraries.}
\end{figure}
\clearpage

\begin{figure}[H]
\centering
\includegraphics[width=\textwidth,height=0.82\textheight,keepaspectratio]{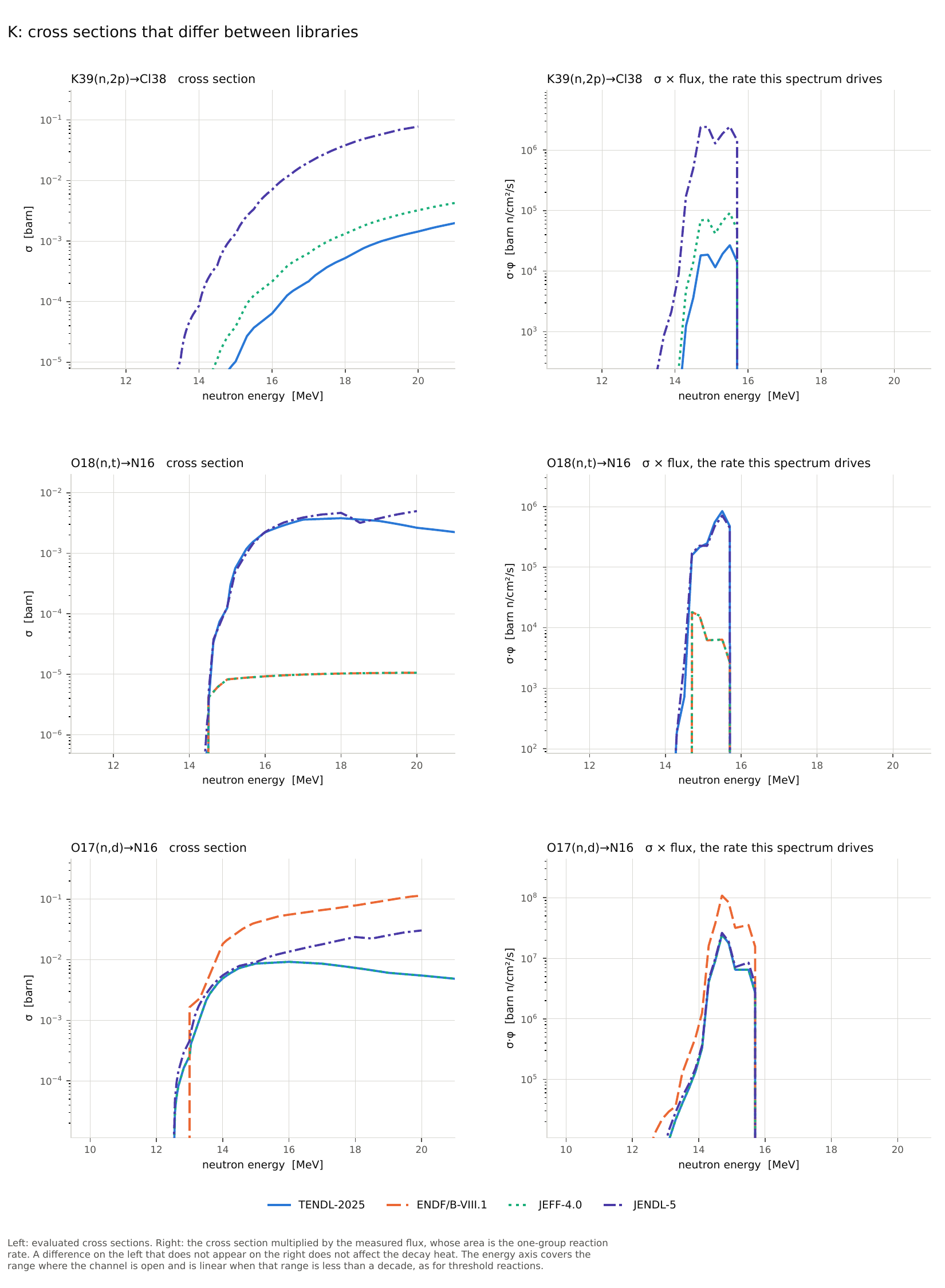}
\caption{K: cross sections of the channels whose spectrum-averaged rates differ, with the neutron spectrum and their product.}
\end{figure}
\clearpage

Mean absolute C/E deviations range from 13 to 18 percent. The spread of 6 percentage points is smaller than the median measurement uncertainty of 6 percent, so the measurement cannot separate the libraries.

\subsubsection*{Differences in the nuclear data}
\begin{itemize}\setlength{\itemsep}{6pt}
\item JENDL-5 gives the total reaction rate of K39(n,2p)→Cl38 as 3.788e-15, against 7.838e-17 for the median of the others, a factor of 48.\\[2pt]{\itshape This is the total rate of the channel over all its products, so the difference is in the cross section, not in the split between states.}
\item ENDF/B-VIII.1 gives the total reaction rate of O18(n,t)→N16 as 1.751e-17, against 6.84e-16 for the median of the others, a factor of 39.
\item ENDF/B-VIII.1 gives the total reaction rate of O17(n,d)→N16 as 1.096e-13, against 2.351e-14 for the median of the others, a factor of 5.
\end{itemize}
\subsubsection*{Suggested measurements}
\begin{itemize}\setlength{\itemsep}{4pt}
\item Natural K has 3 isotopes, so the measured heat combines several parents. Enriched single-isotope targets would distinguish their contributions. Candidate parents are K39, O17, O18. Enrichment alone does not isolate a single reaction channel.
\item The sample is a compound that also contains C, O. Their activation adds to the measured heat; a metal or purer sample would remove it.
\item The reaction rates are averaged over the measured spectrum. Activation measurements at several well-defined neutron energies, including below 14 MeV, would show which energy range the differences come from.
\end{itemize}
\clearpage
\subsection{La: Lanthanum}\label{foil:La}
\begin{figure}[H]
\centering
\includegraphics[width=\textwidth,height=0.82\textheight,keepaspectratio]{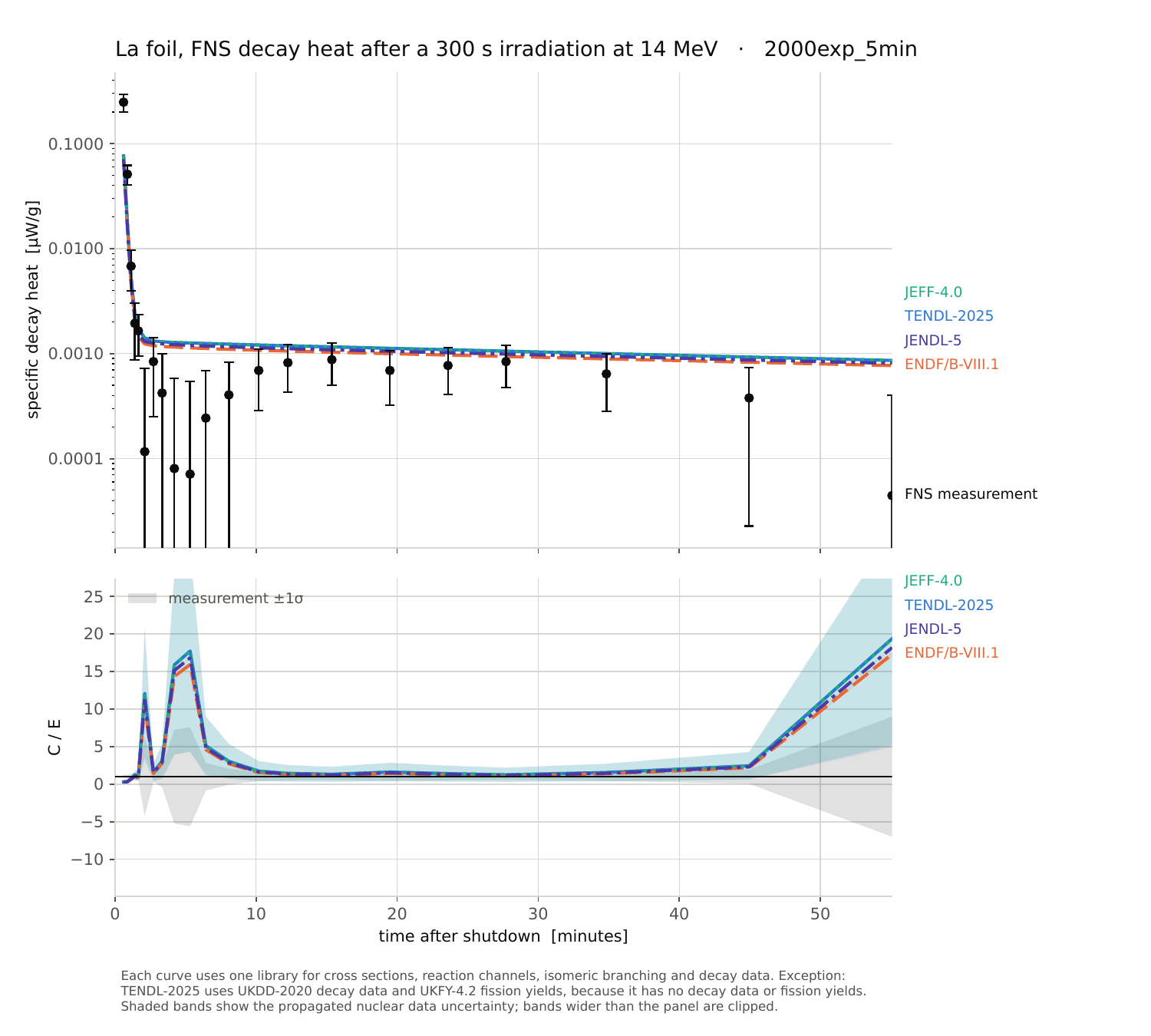}
\caption{La: decay heat against the measurement, with C/E beneath. Mean $|C/E-1|$ by library: ENDF/B-VIII.1 319\%, JENDL-5 340\%, JEFF-4.0 364\%, TENDL-2025 365\%.}
\end{figure}
\clearpage

\begin{figure}[H]
\centering
\includegraphics[width=\textwidth,height=0.82\textheight,keepaspectratio]{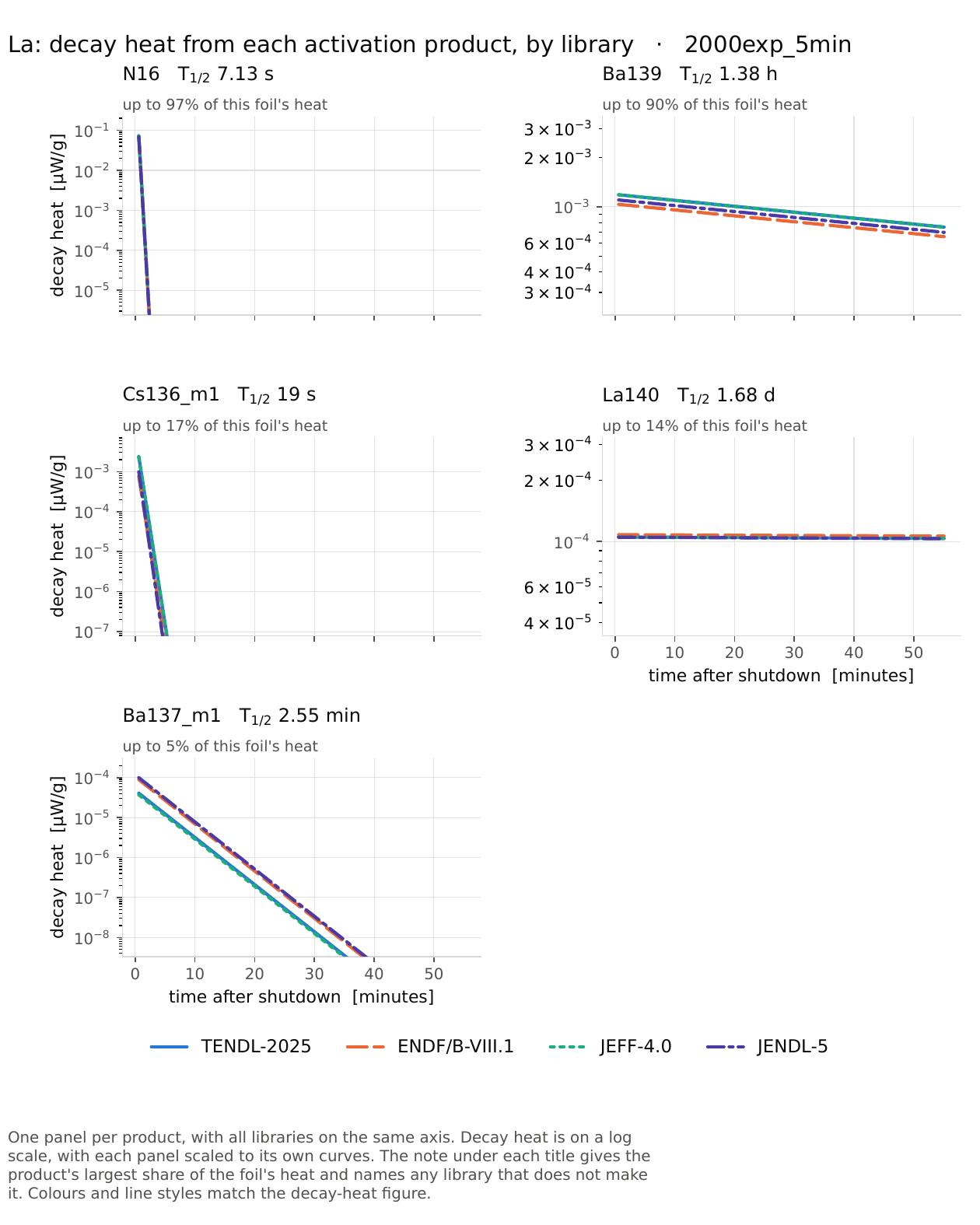}
\caption{La: share of the decay heat from each activation product during cooling.}
\end{figure}
\clearpage

\begin{figure}[H]
\centering
\includegraphics[width=\textwidth,height=0.82\textheight,keepaspectratio]{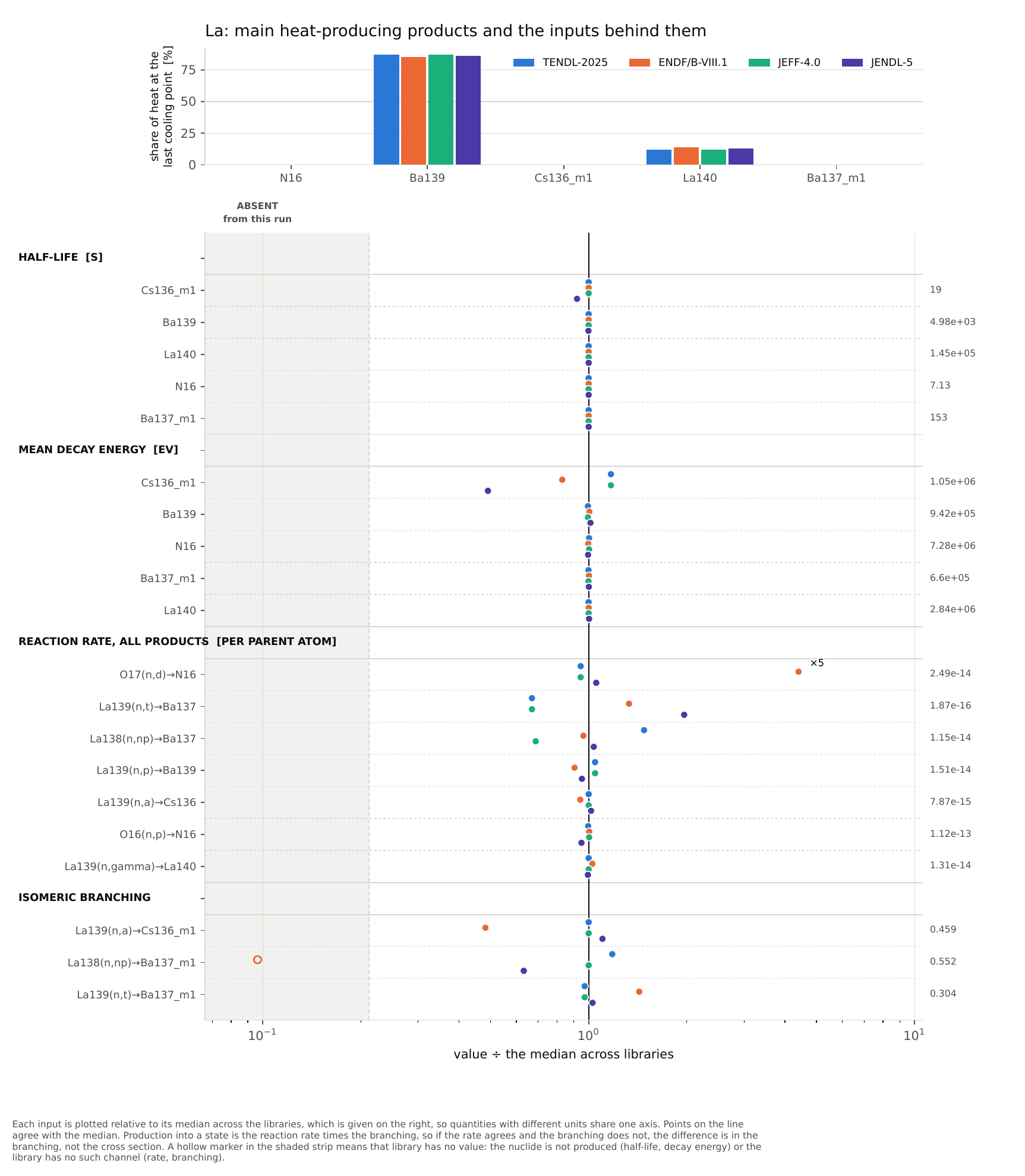}
\caption{La: inputs for the main heat-producing products, each relative to its median across the available libraries.}
\end{figure}
\clearpage

\begin{figure}[H]
\centering
\includegraphics[width=\textwidth,height=0.82\textheight,keepaspectratio]{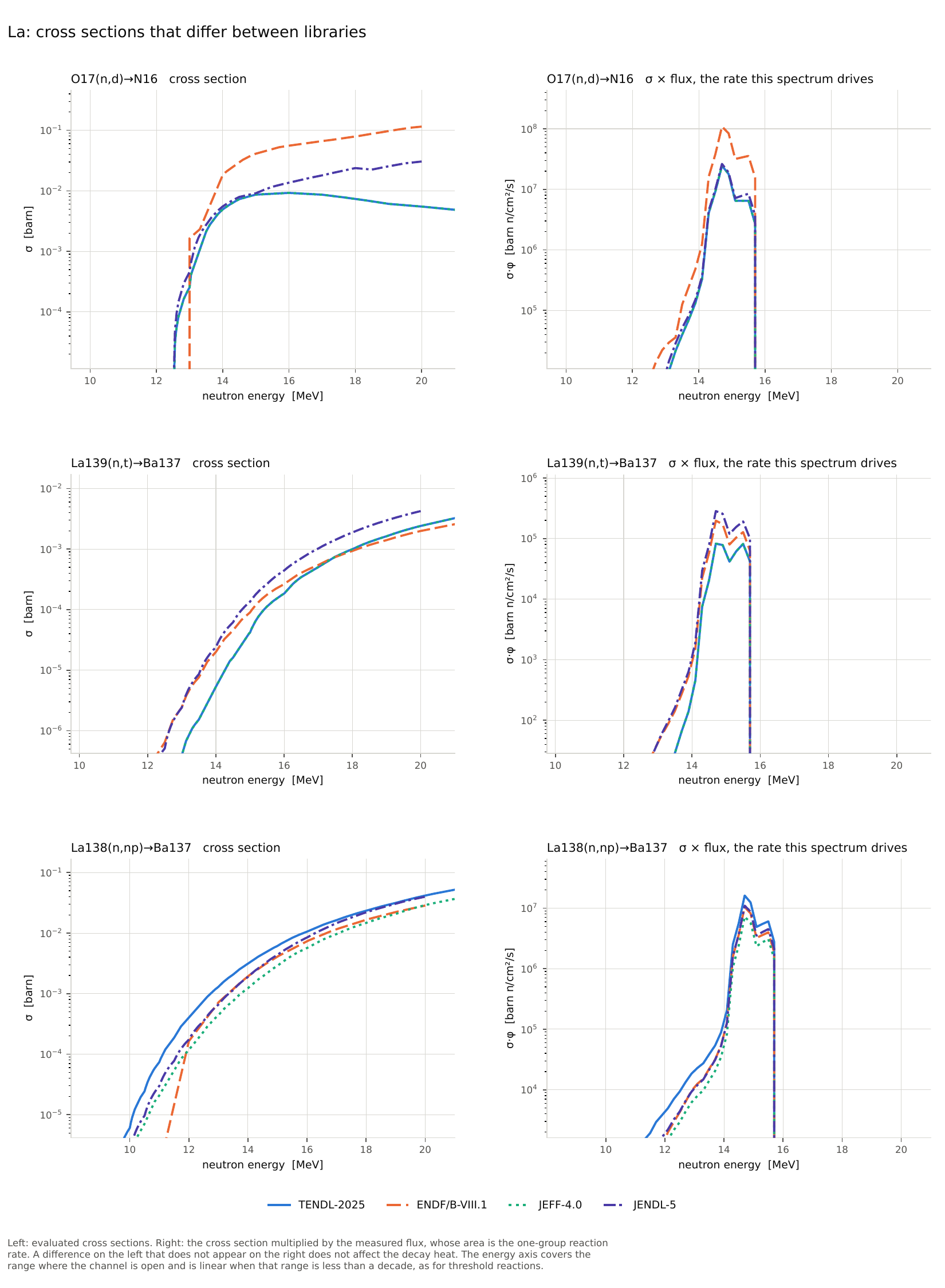}
\caption{La: cross sections of the channels whose spectrum-averaged rates differ, with the neutron spectrum and their product.}
\end{figure}
\clearpage

\begin{figure}[H]
\centering
\includegraphics[width=\textwidth,height=0.82\textheight,keepaspectratio]{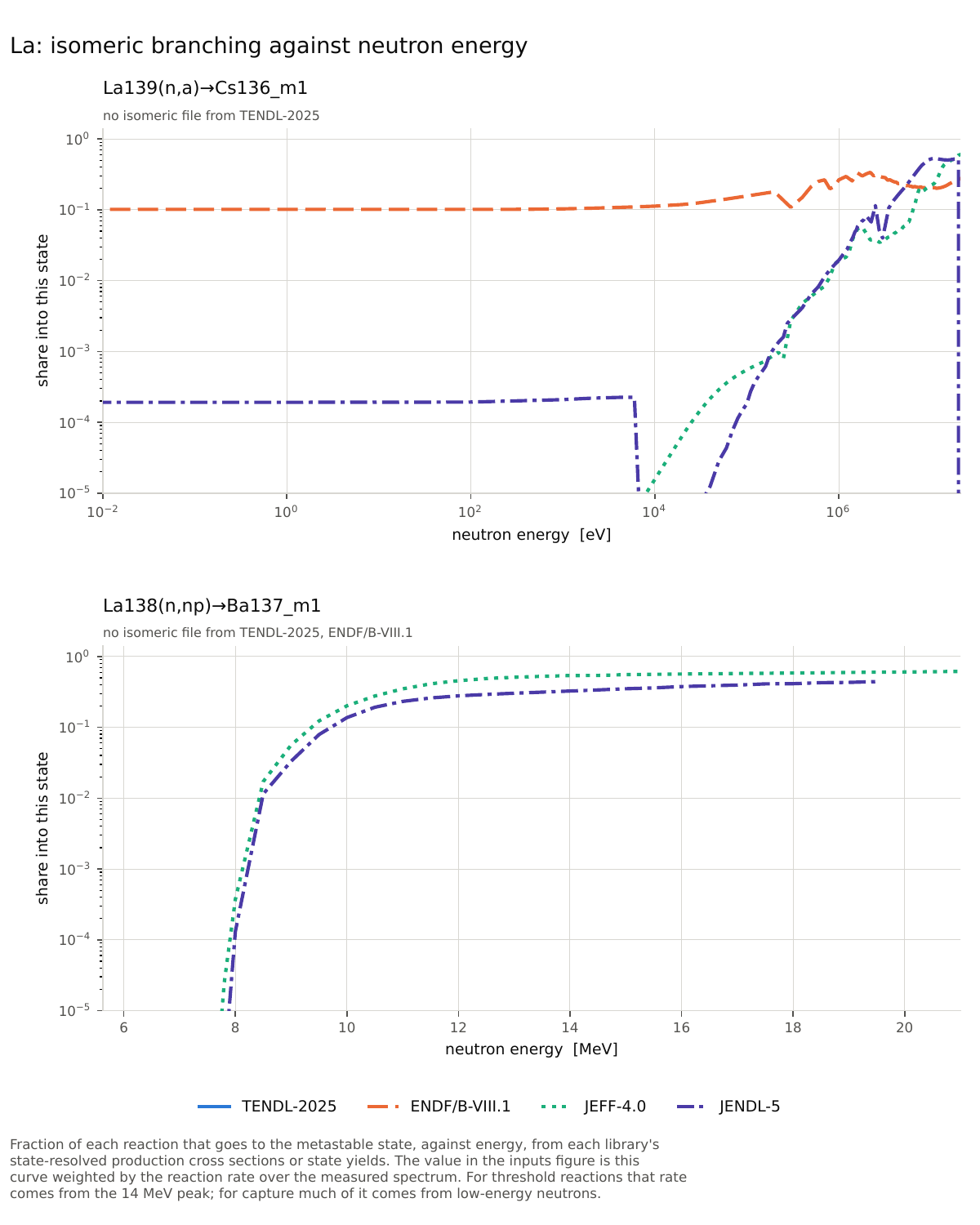}
\caption{La: isomeric branching ratio against neutron energy for the channels whose spectrum-averaged ratio differs between libraries.}
\end{figure}
\clearpage

Mean absolute C/E deviations range from 319 to 365 percent. The spread of 46 percentage points is smaller than the median measurement uncertainty of 56 percent, so the measurement cannot separate the libraries.

\subsubsection*{Differences in the nuclear data}
\begin{itemize}\setlength{\itemsep}{6pt}
\item ENDF/B-VIII.1 gives the total reaction rate of O17(n,d)→N16 as 1.096e-13, against 2.351e-14 for the median of the others, a factor of 5.\\[2pt]{\itshape This is the total rate of the channel over all its products, so the difference is in the cross section, not in the split between states.}
\item JENDL-5 gives the total reaction rate of La139(n,t)→Ba137 as 3.672e-16, against 1.253e-16 for the median of the others, a factor of 3.
\item JENDL-5 gives the mean decay energy of Cs136\_m1 as 5.173e+05 eV, against 1.233e+06 eV for the median of the others, a factor of 2.\\[2pt]{\itshape For a given activity the heat is proportional to the mean decay energy, so this difference goes straight into the heat. Compare the decay schemes and how the emitted energy is counted.}
\item ENDF/B-VIII.1 gives the isomeric branching of La139(n,a)→Cs136\_m1 as 0.2212, against 0.4587 for the median of the others, a factor of 2.\\[2pt]{\itshape Isomeric branching sets how a channel's production is split between the ground and metastable states. It should be checked together with the total rate of the channel.}
\end{itemize}
\subsubsection*{Suggested measurements}
\begin{itemize}\setlength{\itemsep}{4pt}
\item Natural La has 2 isotopes, so the measured heat combines several parents. Enriched single-isotope targets would distinguish their contributions. Candidate parents are La139, O17. Enrichment alone does not isolate a single reaction channel.
\item The sample is a compound that also contains O. Their activation adds to the measured heat; a metal or purer sample would remove it.
\item This foil makes La140 (half-life 1.68 days), which gives up to 14 percent of the calculated heat. A five-minute irradiation and 55 minute measurement only see the start of this contribution. There is no seven-hour FNS measurement of this foil. A longer irradiation followed for days to months would be needed to test this product's production and decay.
\item The reaction rates are averaged over the measured spectrum. Activation measurements at several well-defined neutron energies, including below 14 MeV, would show which energy range the differences come from.
\item Measuring the activity of each state separately, for example by gamma spectrometry where the states have distinct lines, would constrain the isomeric ratio. This needs corrections for feeding between states, detector efficiency and irradiation history.
\end{itemize}
\clearpage
\subsection{Lu: Lutetium}\label{foil:Lu}
\begin{figure}[H]
\centering
\includegraphics[width=\textwidth,height=0.82\textheight,keepaspectratio]{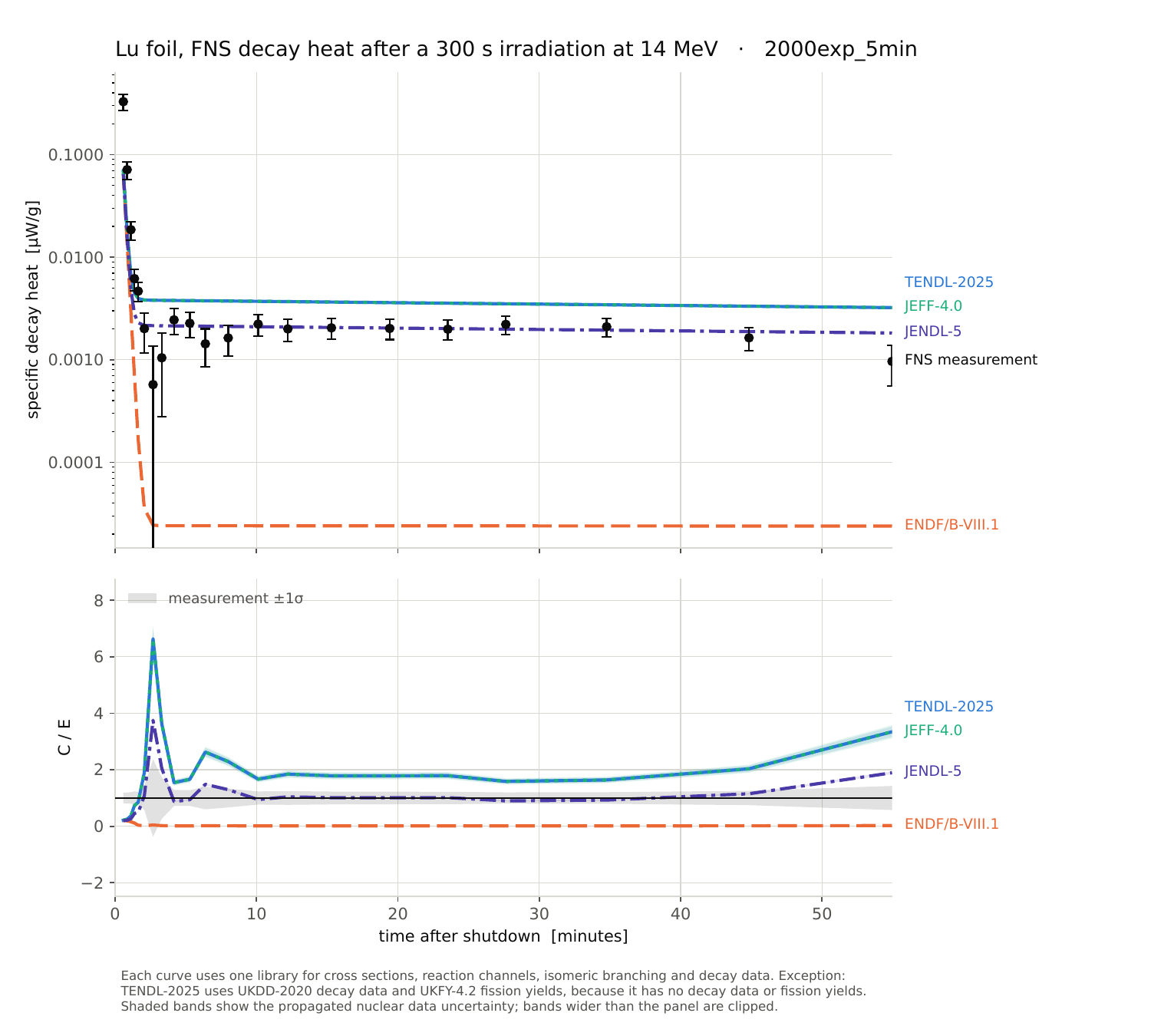}
\caption{Lu: decay heat against the measurement, with C/E beneath. Mean $|C/E-1|$ by library: JENDL-5 46\%, ENDF/B-VIII.1 95\%, JEFF-4.0 116\%, TENDL-2025 116\%.}
\end{figure}
\clearpage

\begin{figure}[H]
\centering
\includegraphics[width=\textwidth,height=0.82\textheight,keepaspectratio]{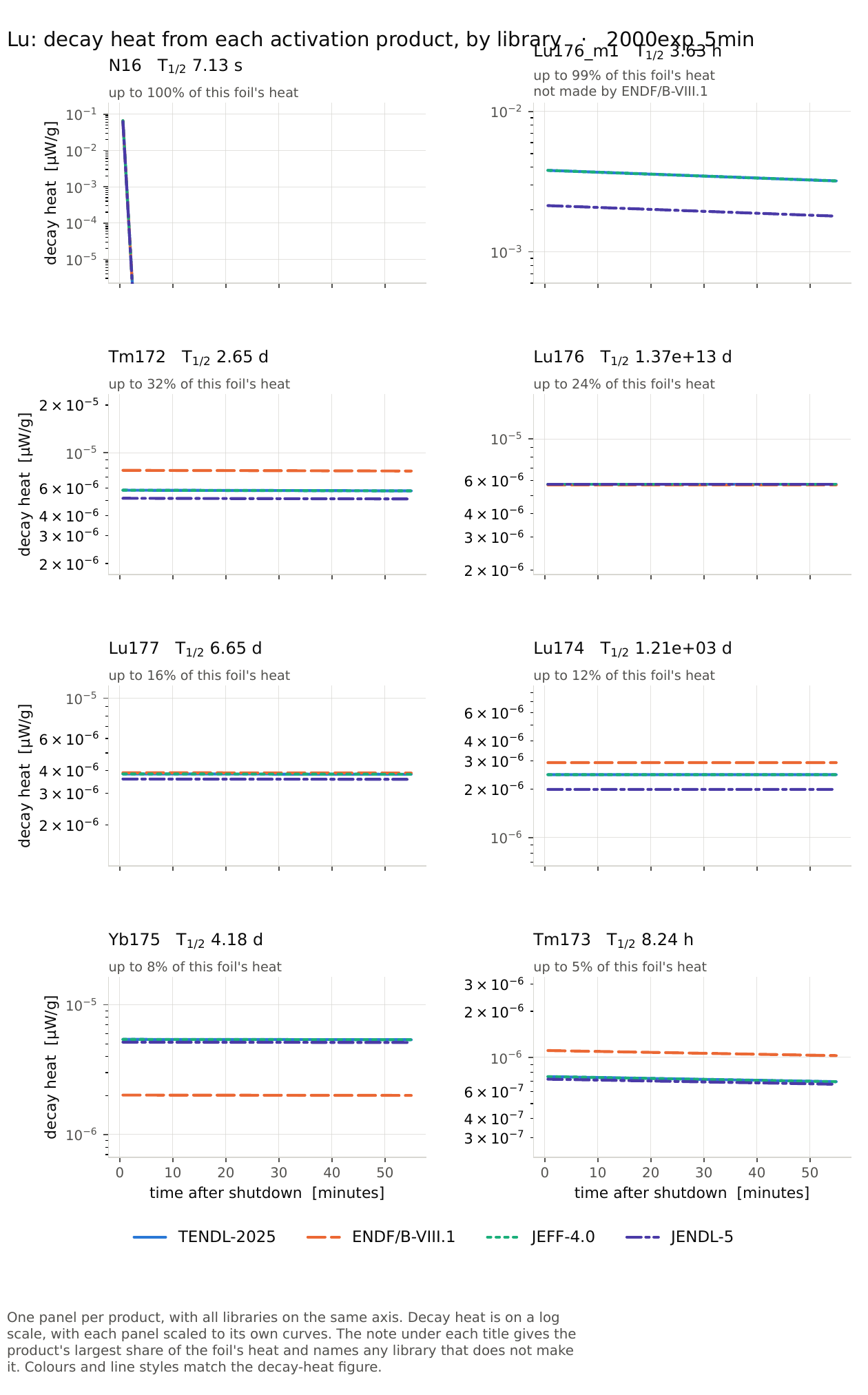}
\caption{Lu: share of the decay heat from each activation product during cooling.}
\end{figure}
\clearpage

\begin{figure}[H]
\centering
\includegraphics[width=\textwidth,height=0.82\textheight,keepaspectratio]{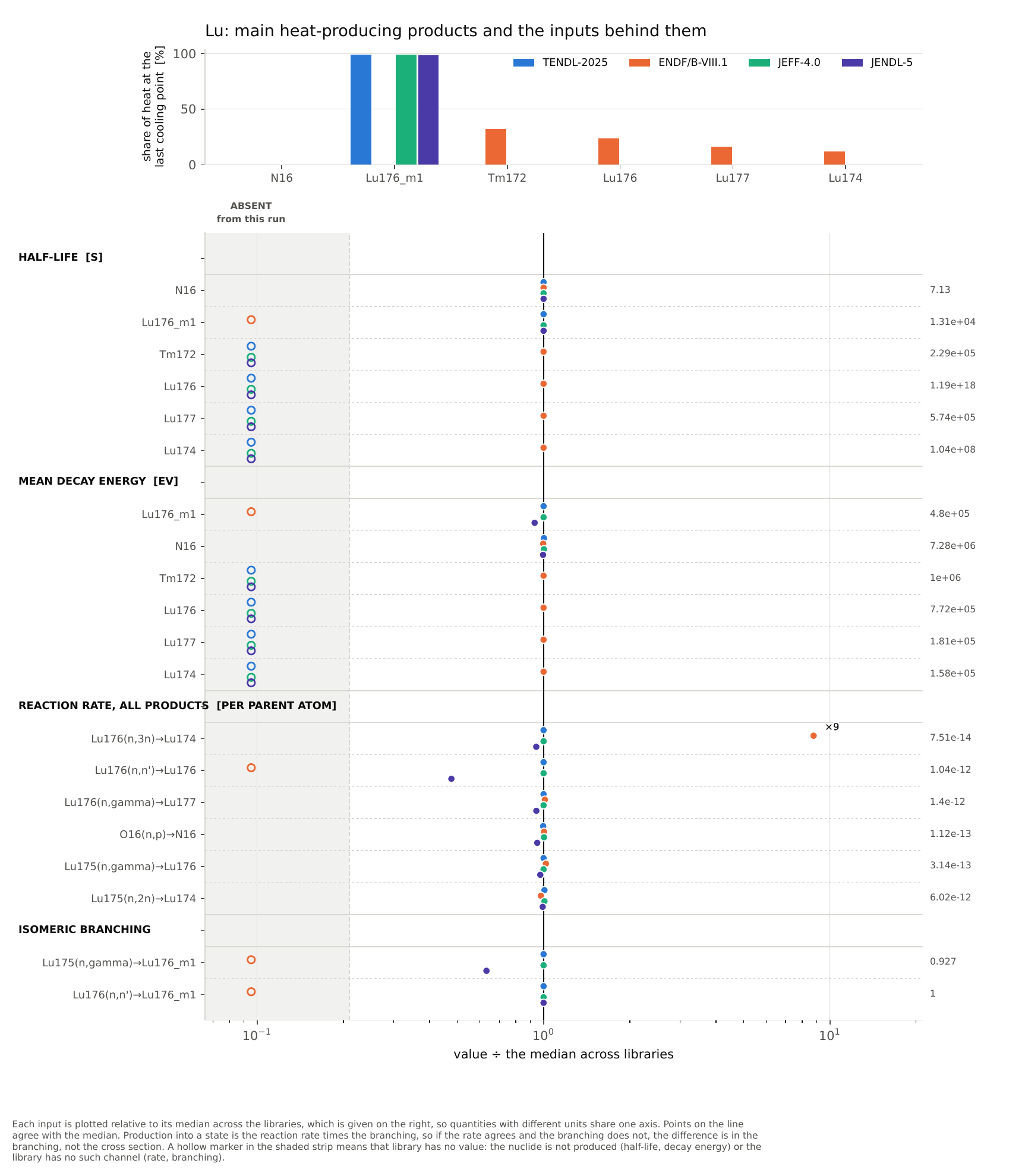}
\caption{Lu: inputs for the main heat-producing products, each relative to its median across the available libraries.}
\end{figure}
\clearpage

\begin{figure}[H]
\centering
\includegraphics[width=\textwidth,height=0.82\textheight,keepaspectratio]{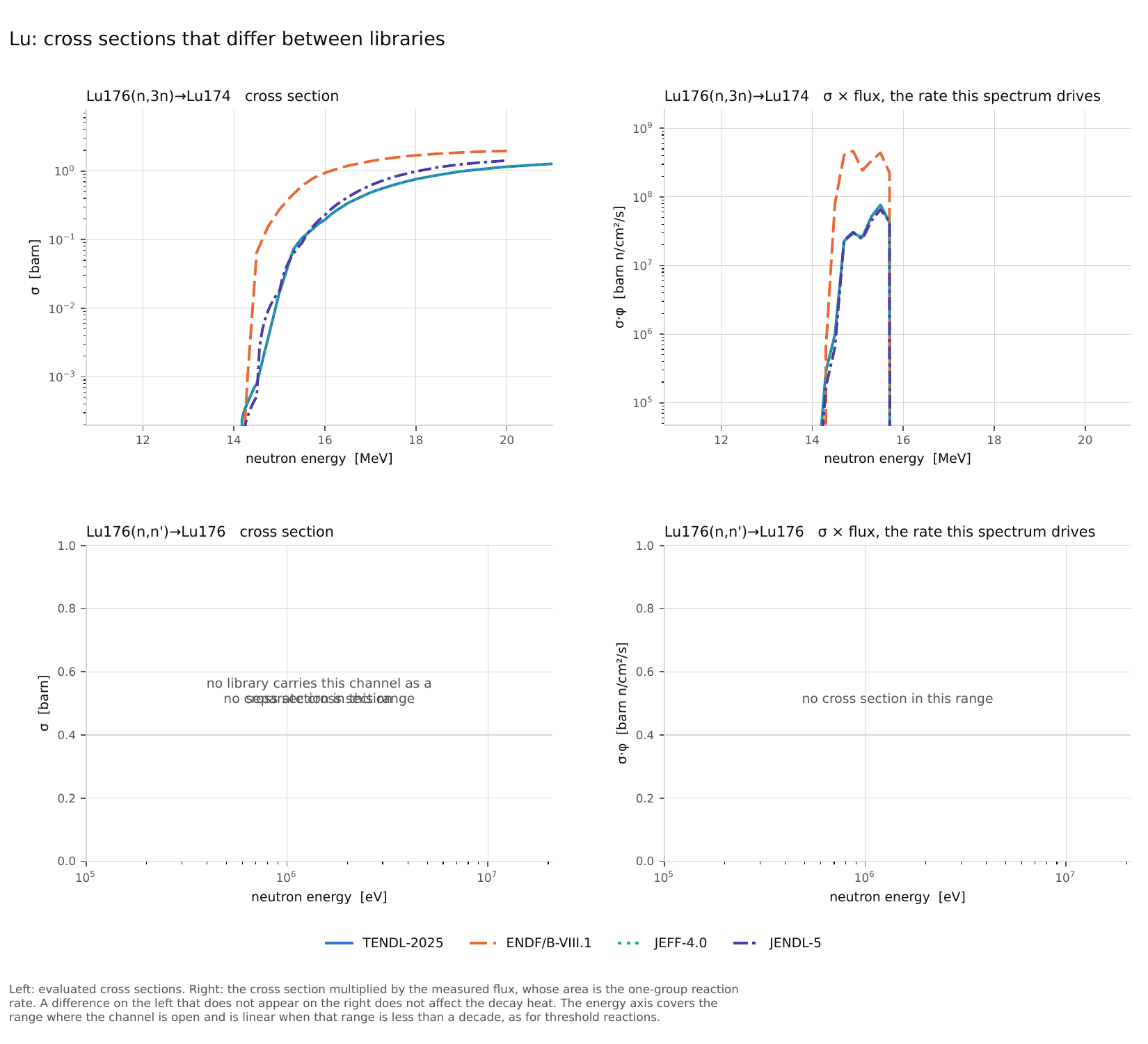}
\caption{Lu: cross sections of the channels whose spectrum-averaged rates differ, with the neutron spectrum and their product.}
\end{figure}
\clearpage

\begin{figure}[H]
\centering
\includegraphics[width=\textwidth,height=0.82\textheight,keepaspectratio]{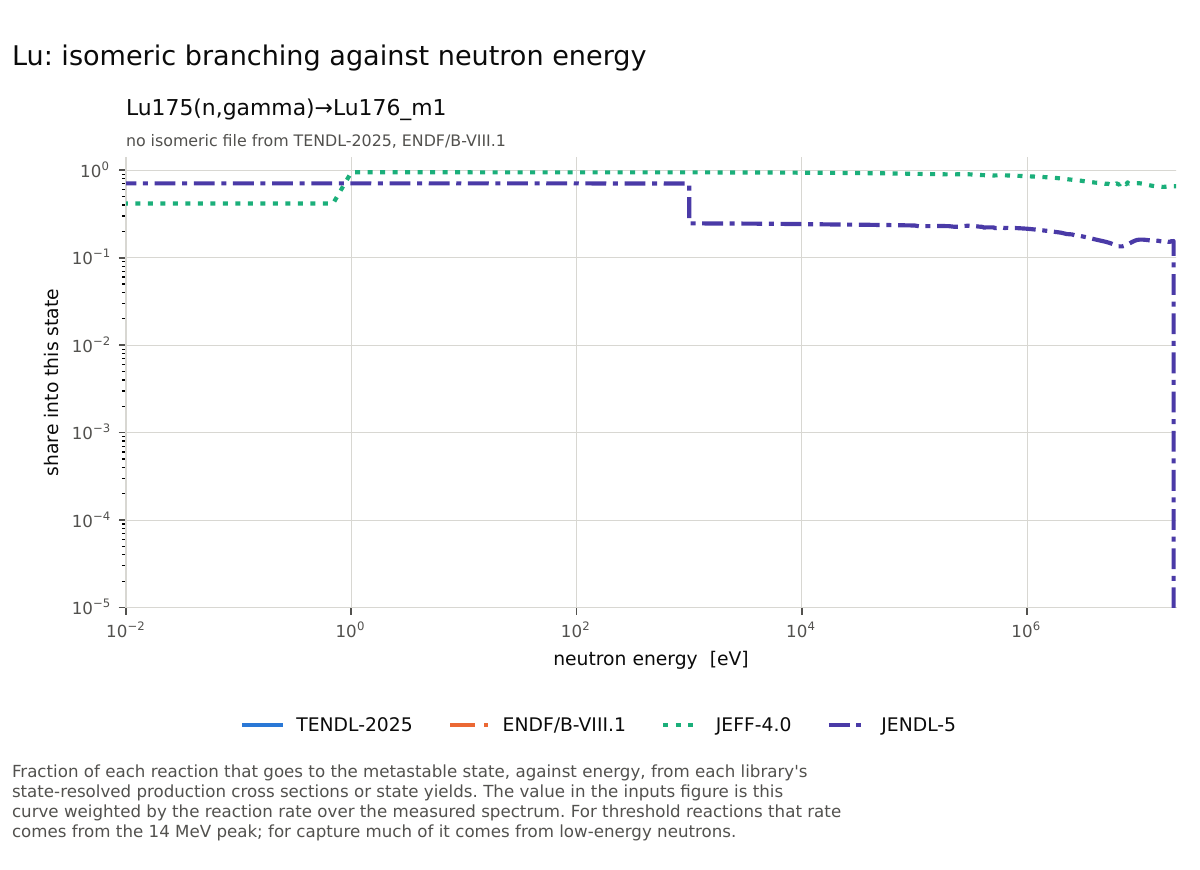}
\caption{Lu: isomeric branching ratio against neutron energy for the channels whose spectrum-averaged ratio differs between libraries.}
\end{figure}
\clearpage

The libraries are between 46 and 116 percent from the measurement, which has a median uncertainty of 24 percent. JENDL-5 is closest and TENDL-2025 furthest.

\subsubsection*{Differences in the nuclear data}
\begin{itemize}\setlength{\itemsep}{6pt}
\item ENDF/B-VIII.1 gives the total reaction rate of Lu176(n,3n)→Lu174 as 6.586e-13, against 7.51e-14 for the median of the others, a factor of 9.\\[2pt]{\itshape This is the total rate of the channel over all its products, so the difference is in the cross section, not in the split between states.}
\item JENDL-5 gives the total reaction rate of Lu176(n,n')→Lu176 as 4.973e-13, against 1.044e-12 for the median of the others, a factor of 2.
\item ENDF/B-VIII.1 gives no heat from Lu176\_m1; the other libraries do.\\[2pt]{\itshape The heat breakdown alone does not show why. Possible causes are a missing reaction channel, a different state assignment, different decay data or a reporting threshold.}
\end{itemize}
\subsubsection*{Suggested measurements}
\begin{itemize}\setlength{\itemsep}{4pt}
\item Natural Lu has 2 isotopes, so the measured heat combines several parents. Enriched single-isotope targets would distinguish their contributions. The candidate parent is Lu176. Enrichment alone does not isolate a single reaction channel.
\item The sample is a compound that also contains O. Their activation adds to the measured heat; a metal or purer sample would remove it.
\item This foil makes Tm172 (half-life 2.65 days), which gives up to 32 percent of the calculated heat. A five-minute irradiation and 55 minute measurement only see the start of this contribution. There is no seven-hour FNS measurement of this foil. A longer irradiation followed for days to months would be needed to test this product's production and decay.
\item The reaction rates are averaged over the measured spectrum. Activation measurements at several well-defined neutron energies, including below 14 MeV, would show which energy range the differences come from.
\end{itemize}
\clearpage
\subsection{Mg: Magnesium}\label{foil:Mg}
\begin{figure}[H]
\centering
\includegraphics[width=\textwidth,height=0.82\textheight,keepaspectratio]{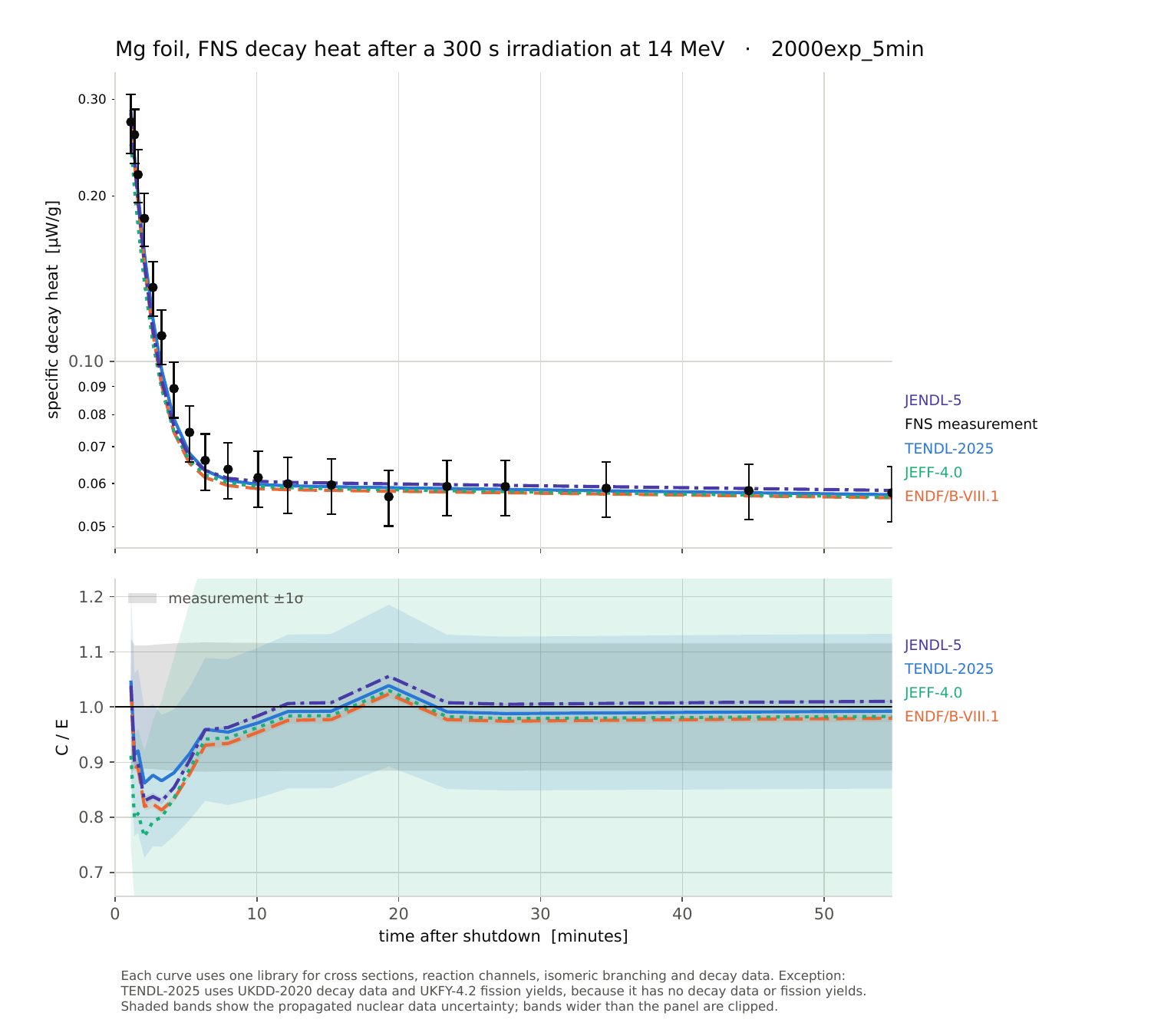}
\caption{Mg: decay heat against the measurement, with C/E beneath. Mean $|C/E-1|$ by library: TENDL-2025 5\%, JENDL-5 6\%, ENDF/B-VIII.1 8\%, JEFF-4.0 9\%.}
\end{figure}
\clearpage

\begin{figure}[H]
\centering
\includegraphics[width=\textwidth,height=0.82\textheight,keepaspectratio]{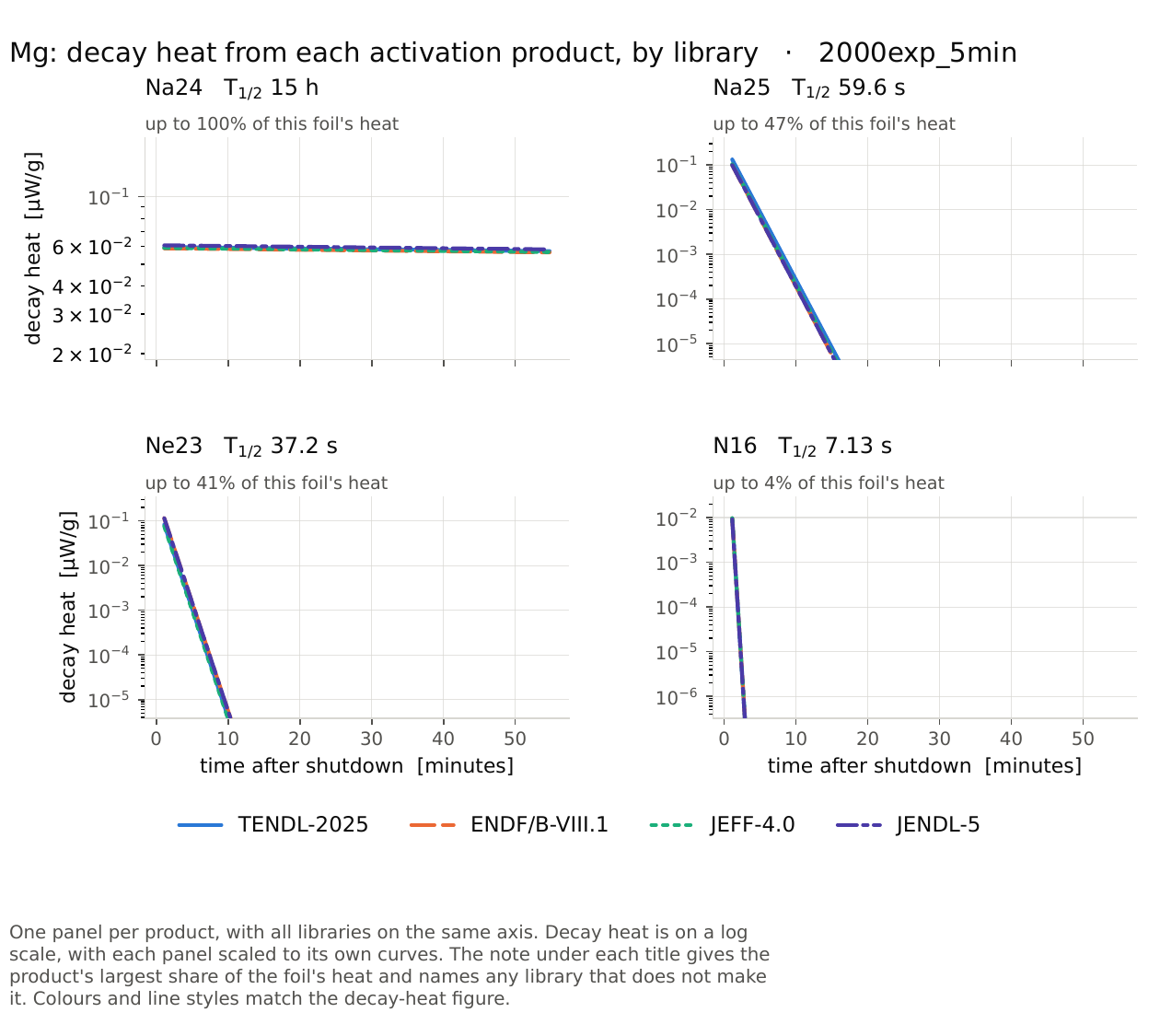}
\caption{Mg: share of the decay heat from each activation product during cooling.}
\end{figure}
\clearpage

\begin{figure}[H]
\centering
\includegraphics[width=\textwidth,height=0.82\textheight,keepaspectratio]{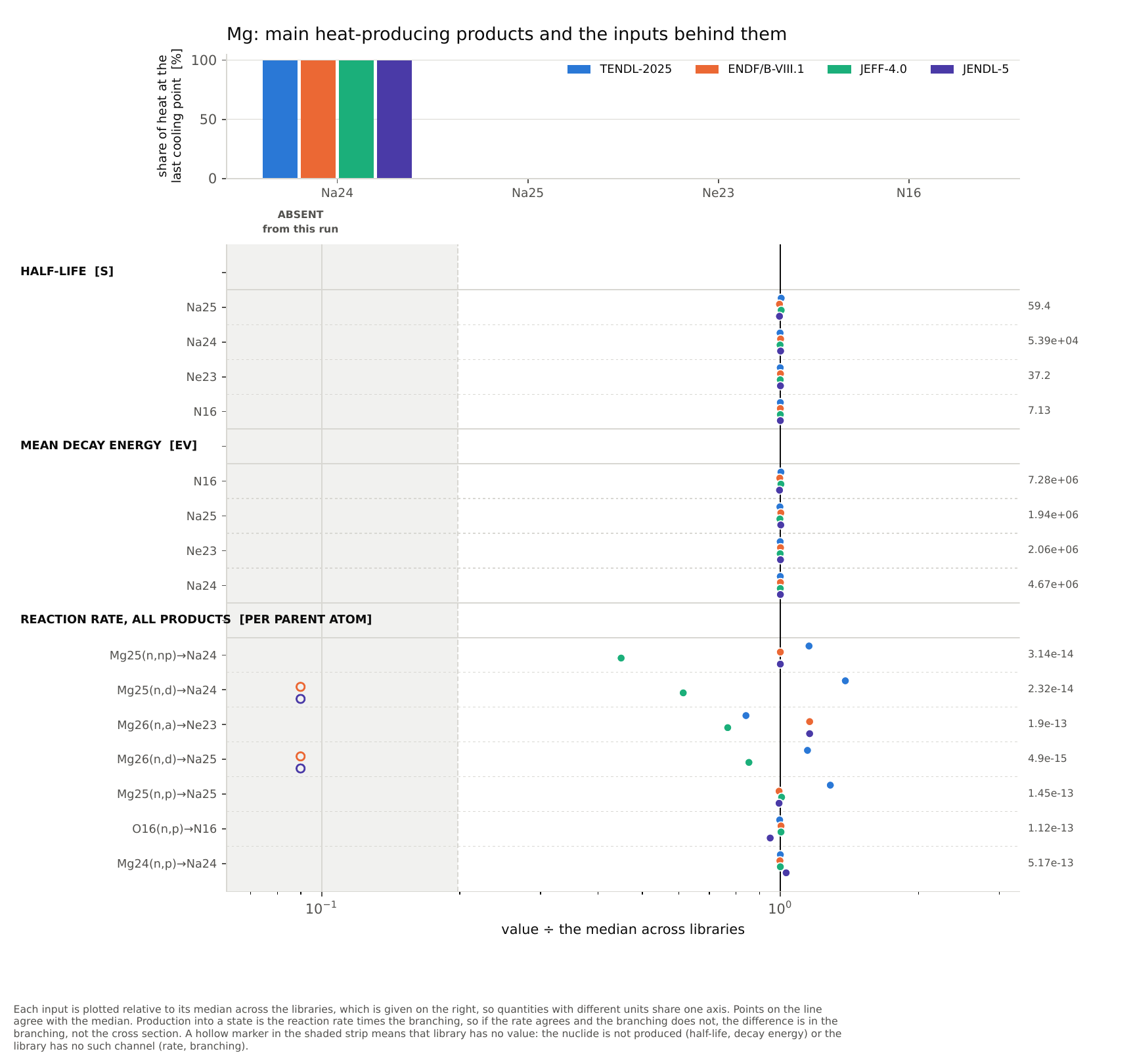}
\caption{Mg: inputs for the main heat-producing products, each relative to its median across the available libraries.}
\end{figure}
\clearpage

\begin{figure}[H]
\centering
\includegraphics[width=\textwidth,height=0.82\textheight,keepaspectratio]{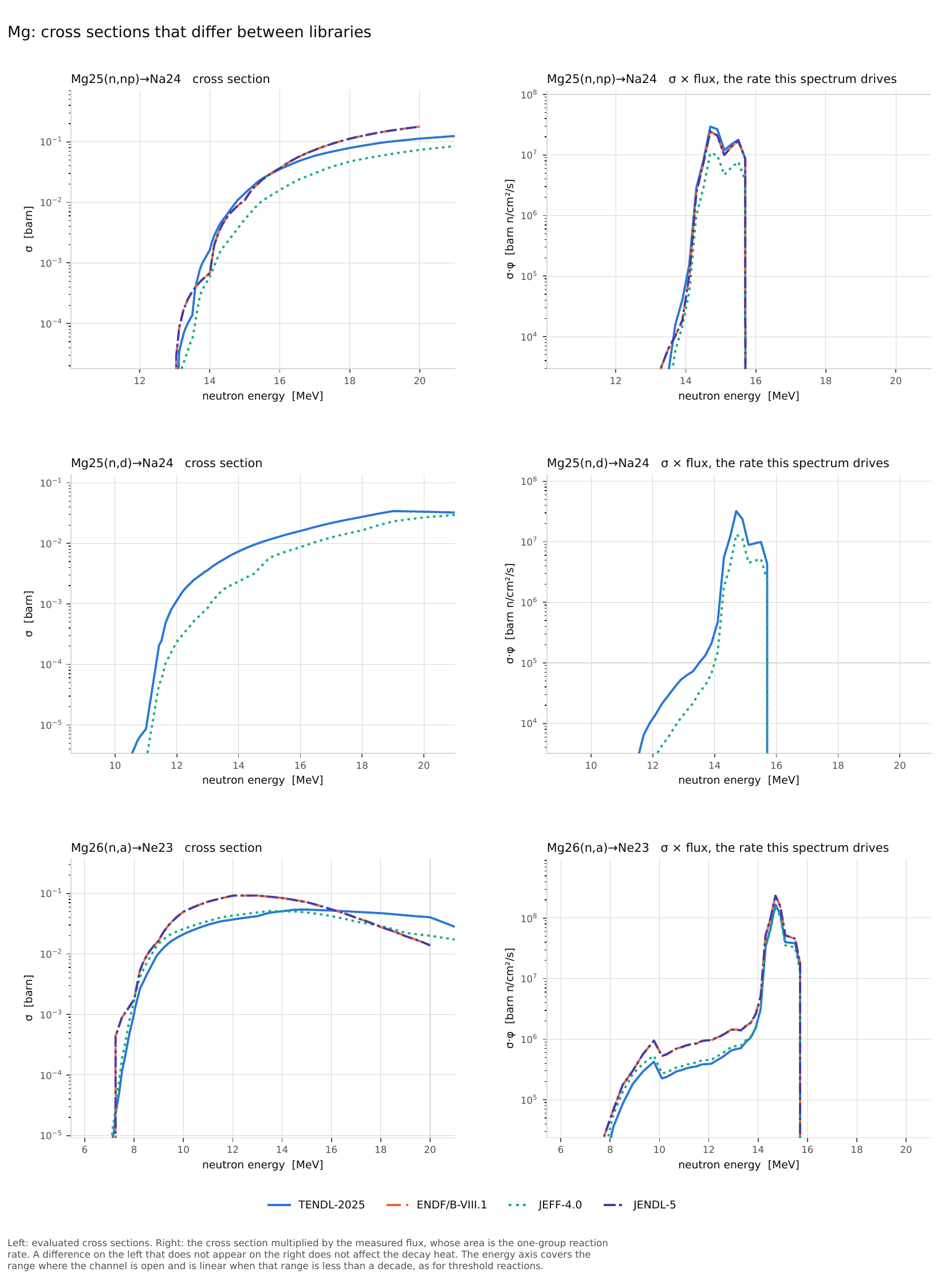}
\caption{Mg: cross sections of the channels whose spectrum-averaged rates differ, with the neutron spectrum and their product.}
\end{figure}
\clearpage

Mean absolute C/E deviations range from 5 to 9 percent. The spread of 4 percentage points is smaller than the median measurement uncertainty of 12 percent, so the measurement cannot separate the libraries.

\subsubsection*{Differences in the nuclear data}
\begin{itemize}\setlength{\itemsep}{6pt}
\item Only TENDL-2025 and JEFF-4.0 have the total reaction rate of Mg25(n,d)→Na24, and they differ by a factor of 2: 3.214e-14 against 1.424e-14.\\[2pt]{\itshape This is the total rate of the channel over all its products, so the difference is in the cross section, not in the split between states.}
\item JEFF-4.0 gives the total reaction rate of Mg25(n,np)→Na24 as 1.41e-14, against 3.135e-14 for the median of the others, a factor of 2.
\end{itemize}
\subsubsection*{Suggested measurements}
\begin{itemize}\setlength{\itemsep}{4pt}
\item Natural Mg has 3 isotopes, so the measured heat combines several parents. Enriched single-isotope targets would distinguish their contributions. The candidate parent is Mg25. Enrichment alone does not isolate a single reaction channel.
\item The sample is a compound that also contains O. Their activation adds to the measured heat; a metal or purer sample would remove it.
\item The reaction rates are averaged over the measured spectrum. Activation measurements at several well-defined neutron energies, including below 14 MeV, would show which energy range the differences come from.
\end{itemize}
\clearpage
\subsection{Mn: Manganese}\label{foil:Mn}
\begin{figure}[H]
\centering
\includegraphics[width=\textwidth,height=0.82\textheight,keepaspectratio]{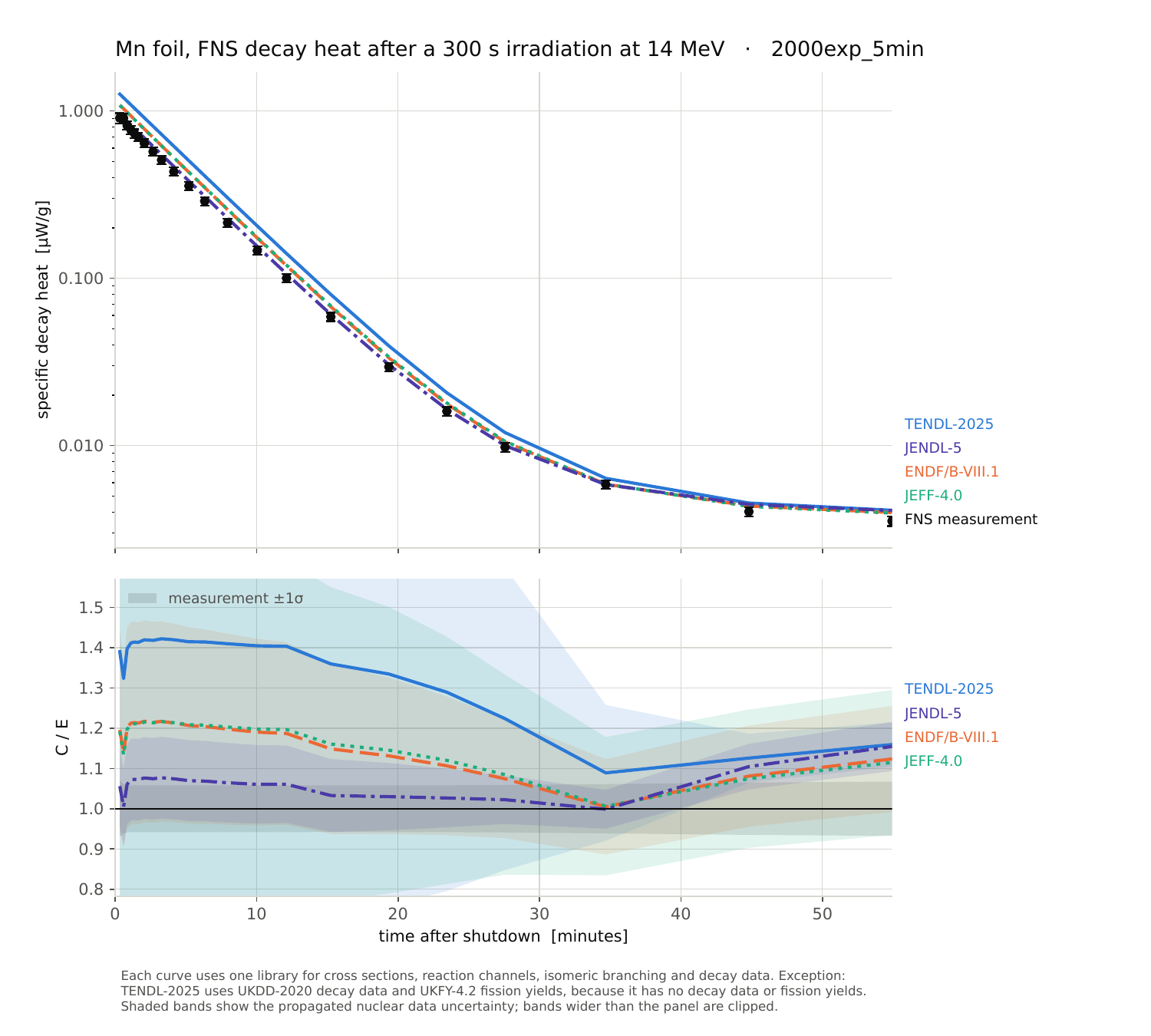}
\caption{Mn: decay heat against the measurement, with C/E beneath. Mean $|C/E-1|$ by library: JENDL-5 6\%, ENDF/B-VIII.1 17\%, JEFF-4.0 17\%, TENDL-2025 35\%.}
\end{figure}
\clearpage

\begin{figure}[H]
\centering
\includegraphics[width=\textwidth,height=0.82\textheight,keepaspectratio]{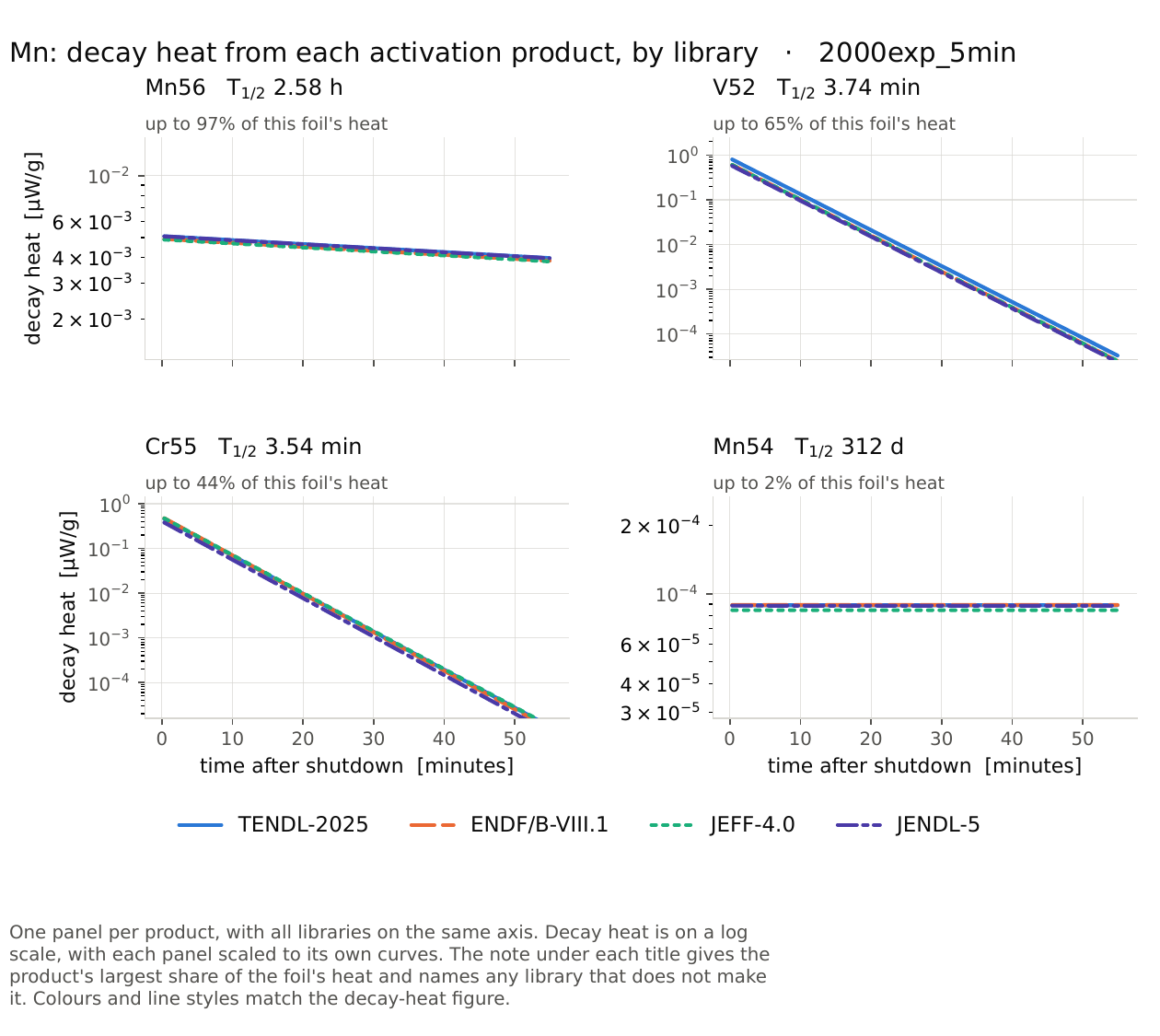}
\caption{Mn: share of the decay heat from each activation product during cooling.}
\end{figure}
\clearpage

\begin{figure}[H]
\centering
\includegraphics[width=\textwidth,height=0.82\textheight,keepaspectratio]{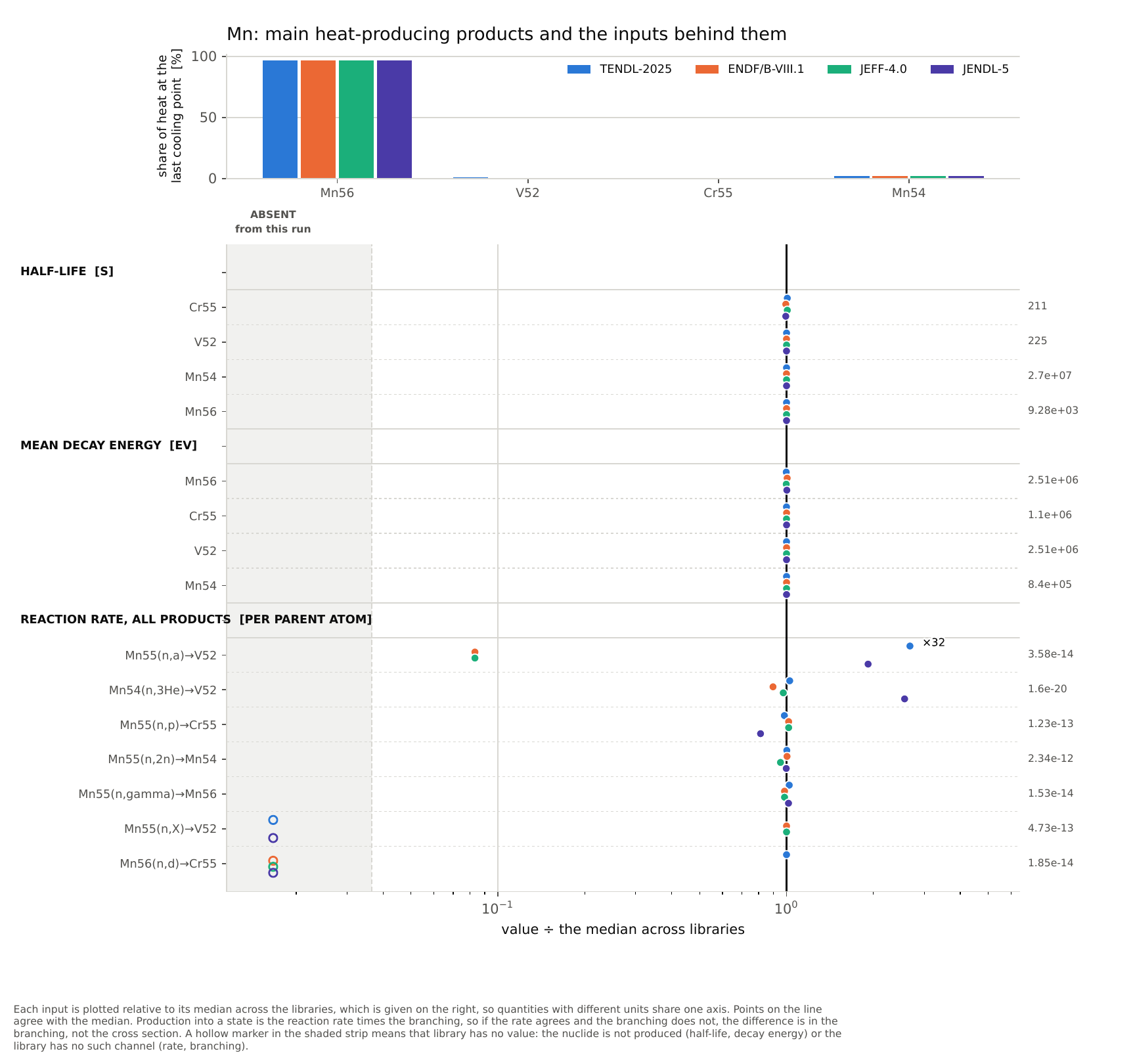}
\caption{Mn: inputs for the main heat-producing products, each relative to its median across the available libraries.}
\end{figure}
\clearpage

\begin{figure}[H]
\centering
\includegraphics[width=\textwidth,height=0.82\textheight,keepaspectratio]{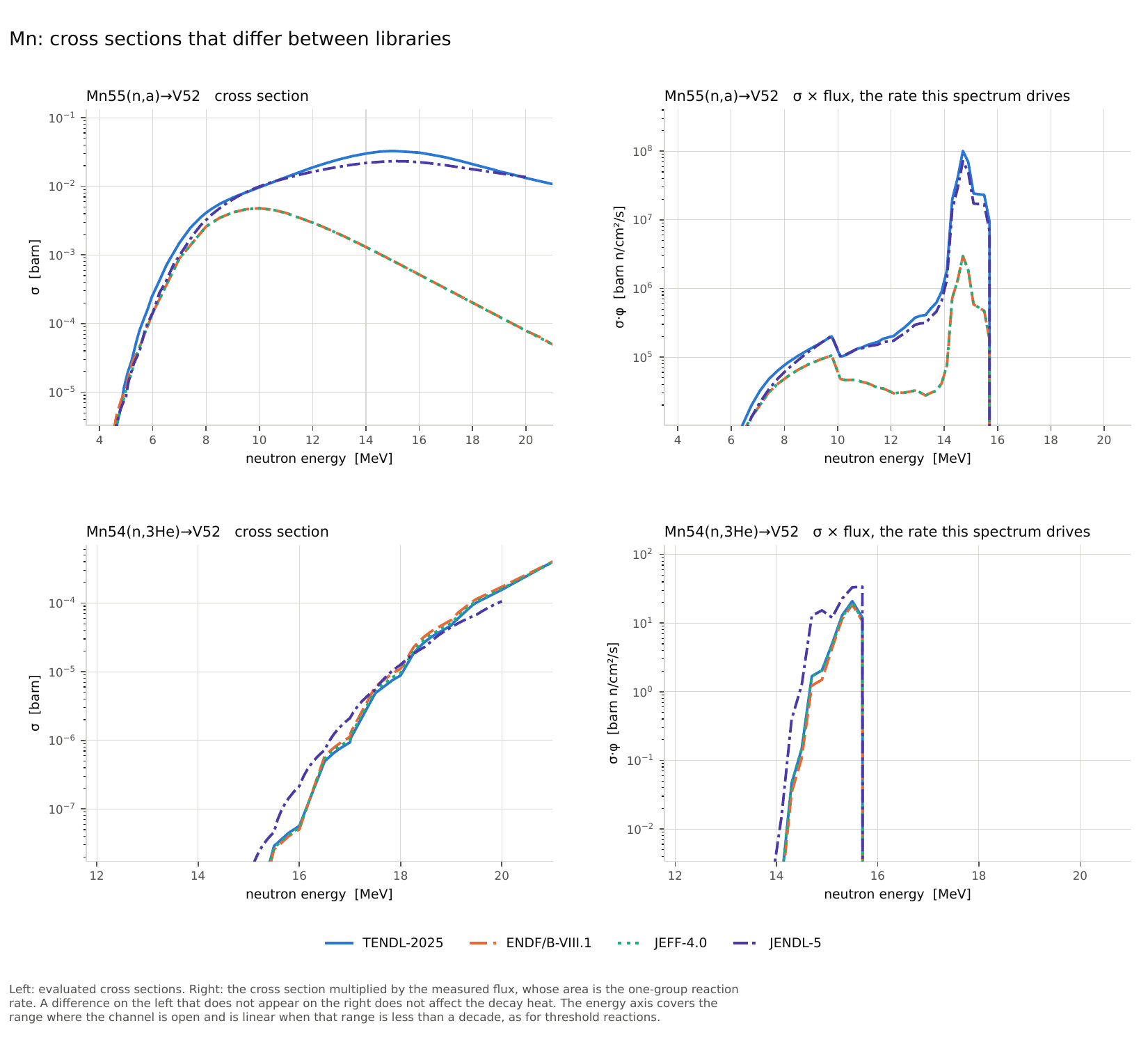}
\caption{Mn: cross sections of the channels whose spectrum-averaged rates differ, with the neutron spectrum and their product.}
\end{figure}
\clearpage

\subsubsection*{1996 seven-hour campaign}
The 1996 campaign irradiated this foil for 7.17 hours at position 7 and measured it out to 402.18 days. In the five-minute calculation Mn54 (half-life 312 days) gives up to 2 percent of the heat, and the longer irradiation and cooling test its production and decay data. JENDL-5 is closest. This campaign uses a different spectrum, so it is not included in the main rankings.

\begin{figure}[H]
\centering
\includegraphics[width=\textwidth,height=0.82\textheight,keepaspectratio]{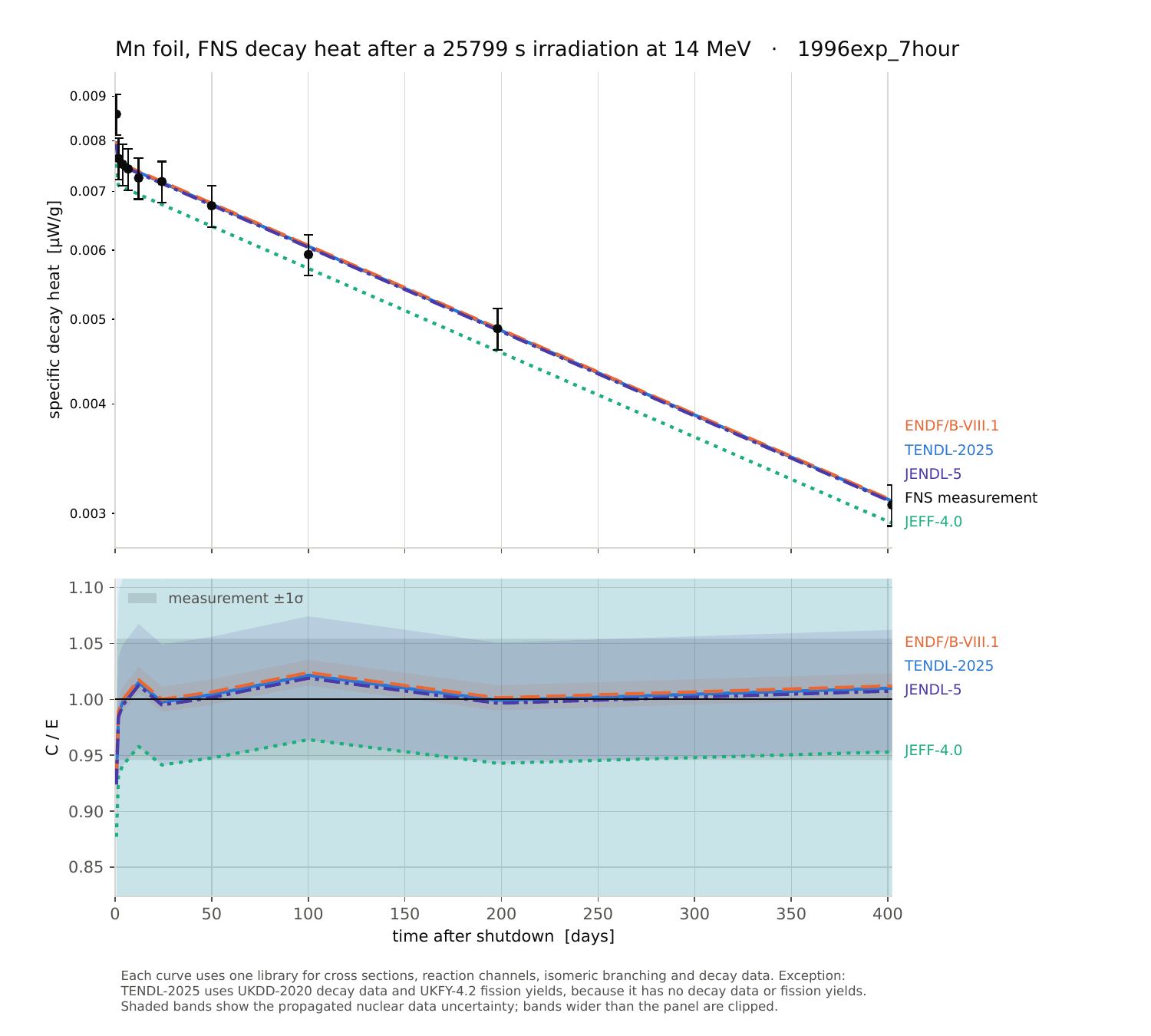}
\caption{Mn: 1996 seven-hour campaign, irradiation position 7. Mean $|C/E-1|$ by library: JENDL-5 1\%, TENDL-2025 1\%, ENDF/B-VIII.1 1\%, JEFF-4.0 6\%. }
\end{figure}
\clearpage

The libraries are between 6 and 35 percent from the measurement, which has a median uncertainty of 6 percent. JENDL-5 is closest and TENDL-2025 furthest.

\subsubsection*{Differences in the nuclear data}
\begin{itemize}\setlength{\itemsep}{6pt}
\item JEFF-4.0 gives the total reaction rate of Mn55(n,a)→V52 as 2.98e-15, against 6.86e-14 for the median of the others, a factor of 23.\\[2pt]{\itshape This is the total rate of the channel over all its products, so the difference is in the cross section, not in the split between states.}
\item JENDL-5 gives the total reaction rate of Mn54(n,3He)→V52 as 4.108e-20, against 1.561e-20 for the median of the others, a factor of 3.
\end{itemize}
\subsubsection*{Suggested measurements}
\begin{itemize}\setlength{\itemsep}{4pt}
\item Mn has a single stable isotope, so the foil is already a single-isotope target and enrichment would not help.
\item The reaction rates are averaged over the measured spectrum. Activation measurements at several well-defined neutron energies, including below 14 MeV, would show which energy range the differences come from.
\end{itemize}
\clearpage
\subsection{Mo: Molybdenum}\label{foil:Mo}
\begin{figure}[H]
\centering
\includegraphics[width=\textwidth,height=0.82\textheight,keepaspectratio]{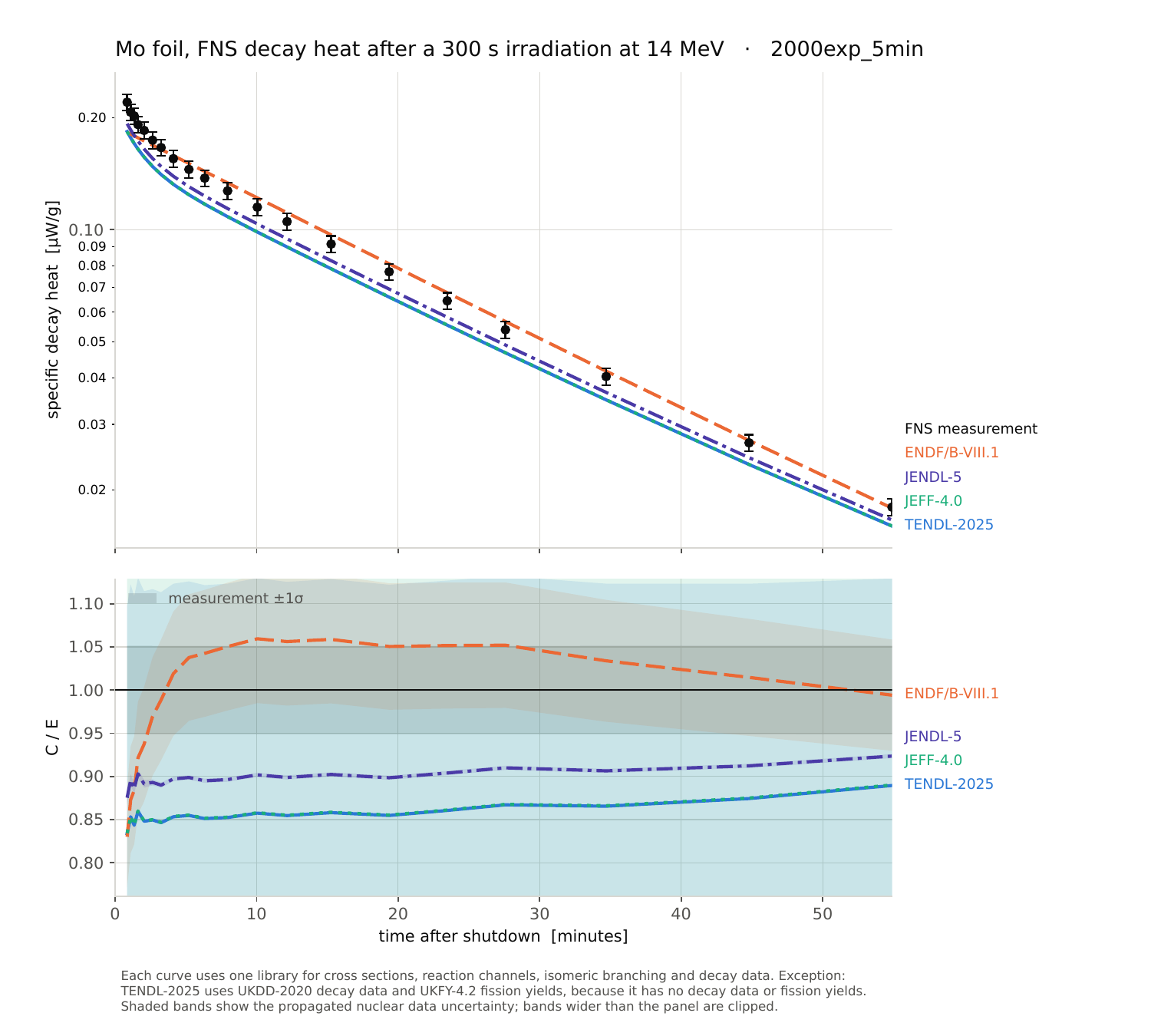}
\caption{Mo: decay heat against the measurement, with C/E beneath. Mean $|C/E-1|$ by library: ENDF/B-VIII.1 6\%, JENDL-5 10\%, JEFF-4.0 14\%, TENDL-2025 14\%.}
\end{figure}
\clearpage

\begin{figure}[H]
\centering
\includegraphics[width=\textwidth,height=0.82\textheight,keepaspectratio]{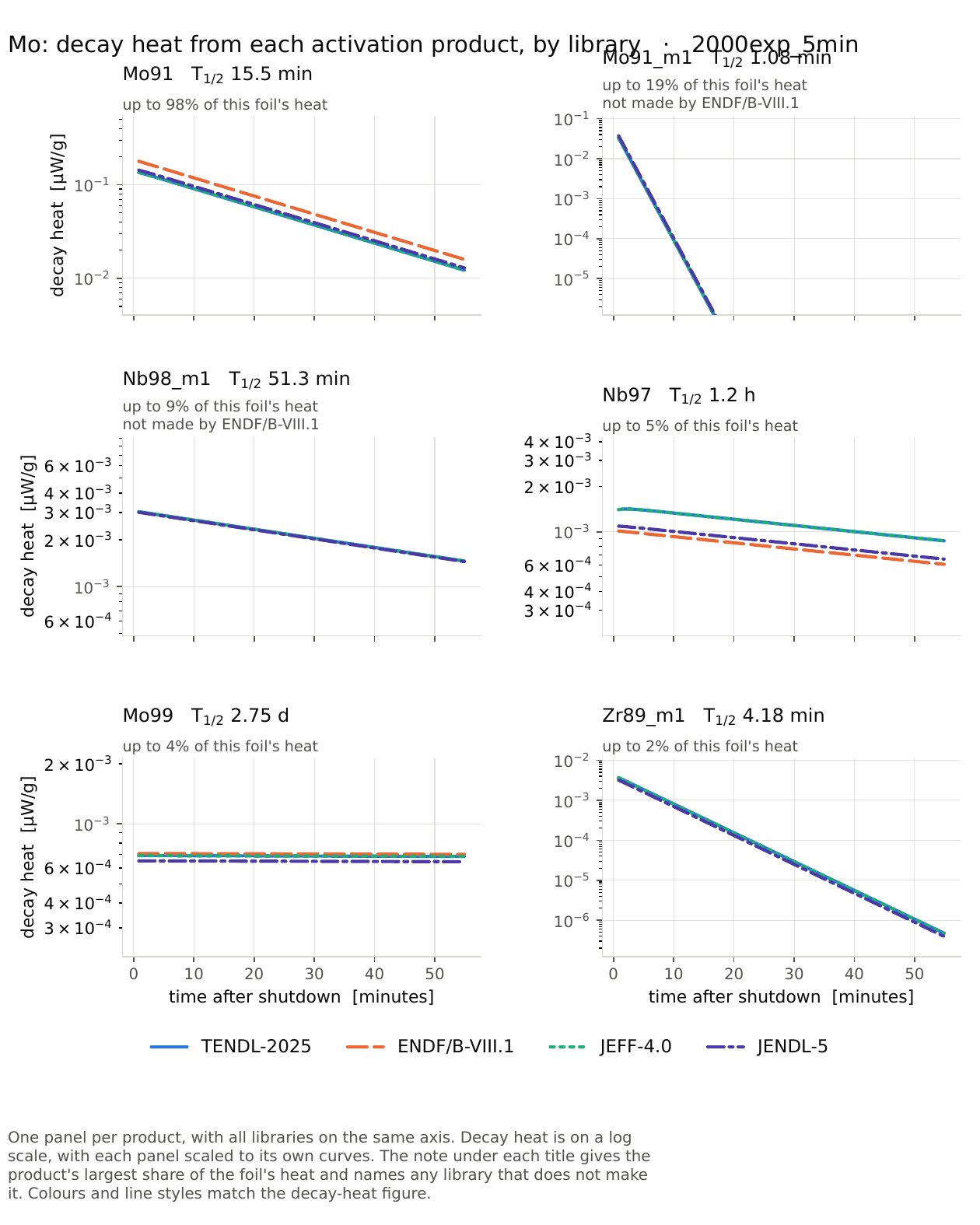}
\caption{Mo: share of the decay heat from each activation product during cooling.}
\end{figure}
\clearpage

\begin{figure}[H]
\centering
\includegraphics[width=\textwidth,height=0.82\textheight,keepaspectratio]{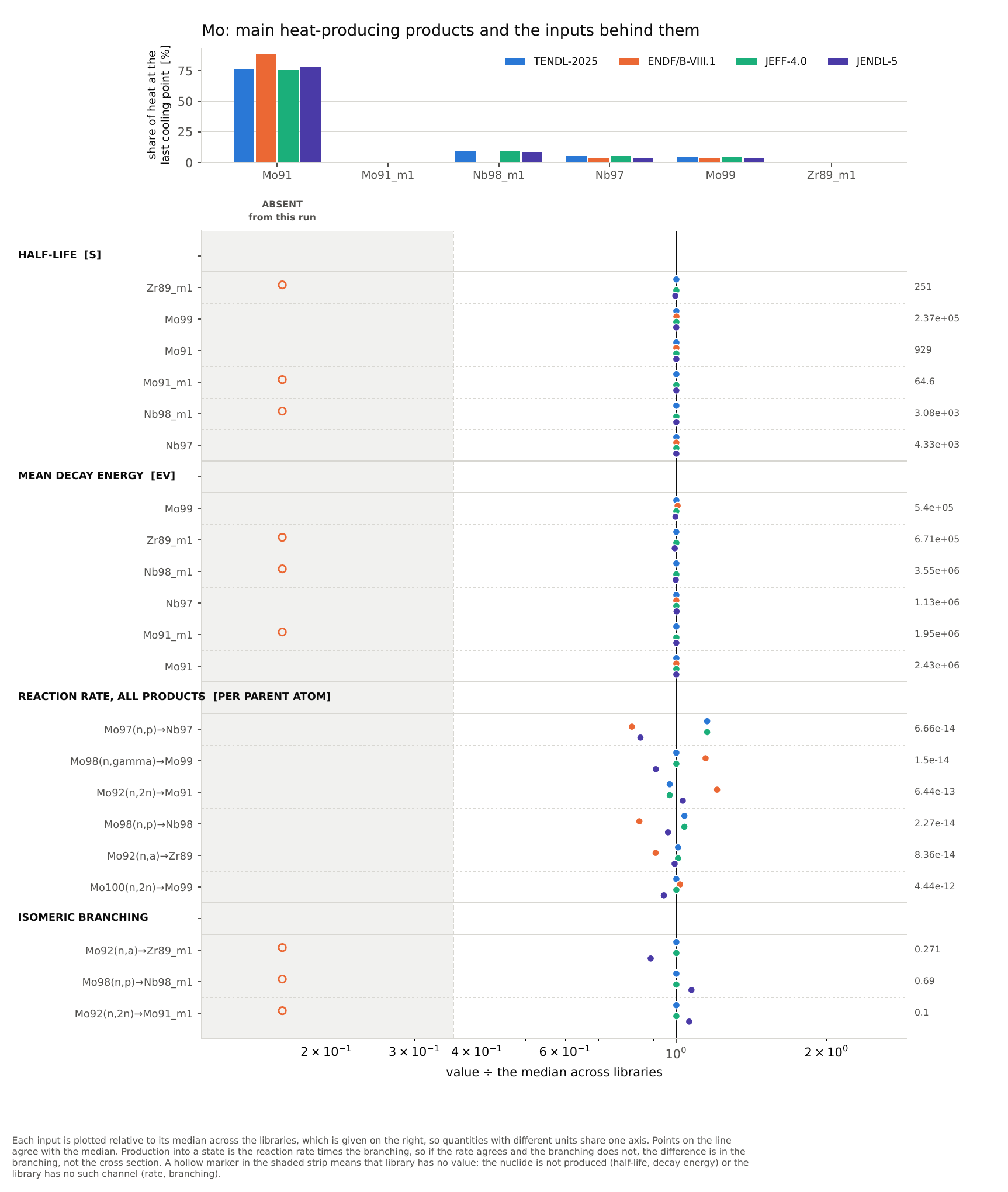}
\caption{Mo: inputs for the main heat-producing products, each relative to its median across the available libraries.}
\end{figure}
\clearpage

\subsubsection*{1996 seven-hour campaign}
The 1996 campaign irradiated this foil for 7.57 hours at position 7 and measured it out to 200.14 days. In the five-minute calculation Mo99 (half-life 2.75 days) gives up to 4 percent of the heat, and the longer irradiation and cooling test its production and decay data. JEFF-4.0 is closest. This campaign uses a different spectrum, so it is not included in the main rankings.

\begin{figure}[H]
\centering
\includegraphics[width=\textwidth,height=0.82\textheight,keepaspectratio]{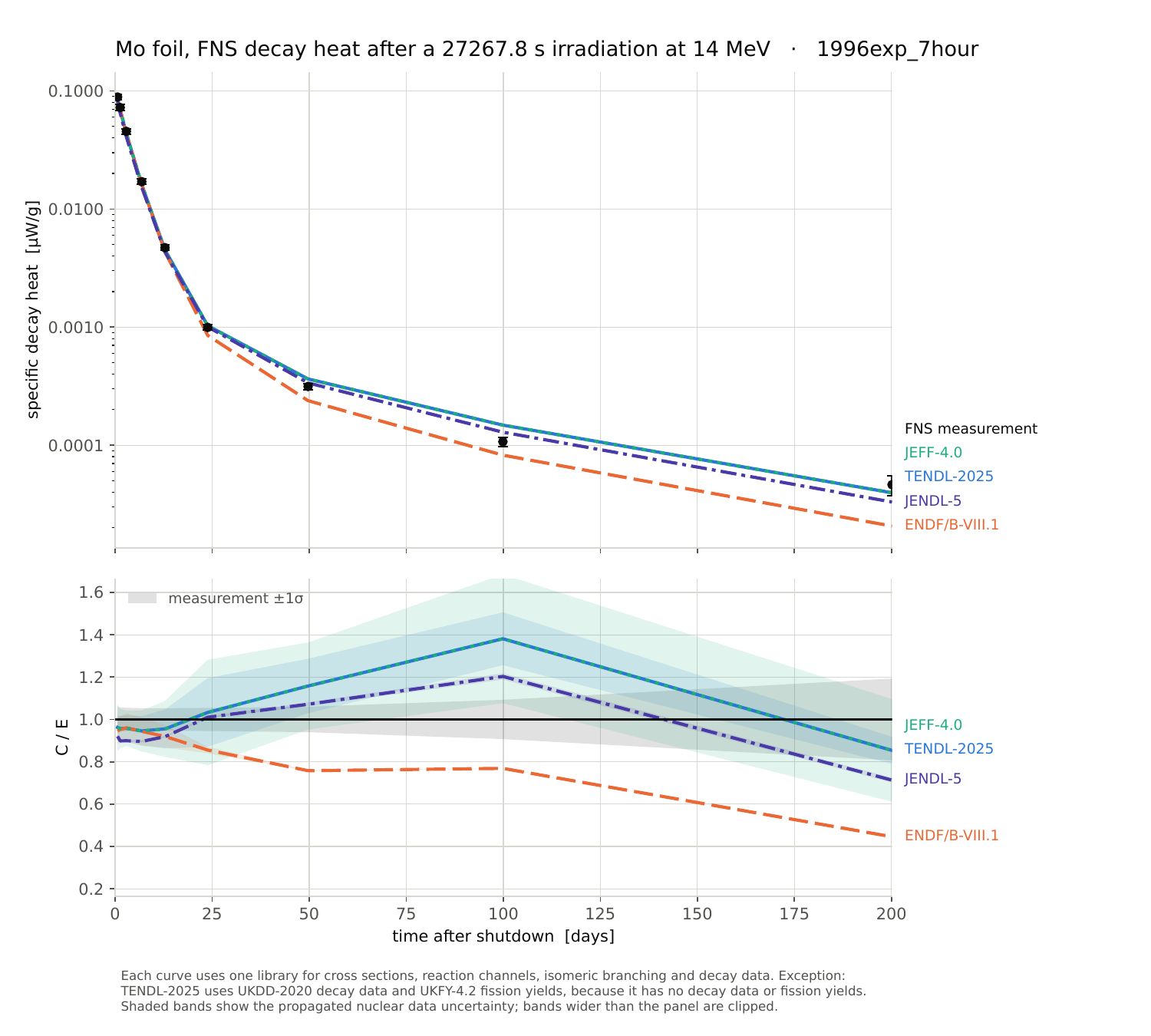}
\caption{Mo: 1996 seven-hour campaign, irradiation position 7. Mean $|C/E-1|$ by library: JEFF-4.0 10\%, TENDL-2025 10\%, JENDL-5 12\%, ENDF/B-VIII.1 16\%. }
\end{figure}
\clearpage

The libraries are between 6 and 14 percent from the measurement, which has a median uncertainty of 5 percent. ENDF/B-VIII.1 is closest and TENDL-2025 furthest.

\subsubsection*{Differences in the nuclear data}
\begin{itemize}\setlength{\itemsep}{6pt}
\item ENDF/B-VIII.1 gives no heat from Mo91\_m1; the other libraries do.\\[2pt]{\itshape The heat breakdown alone does not show why. Possible causes are a missing reaction channel, a different state assignment, different decay data or a reporting threshold.}
\item ENDF/B-VIII.1 gives no heat from Nb98\_m1; the other libraries do.
\end{itemize}
\subsubsection*{Suggested measurements}
\begin{itemize}\setlength{\itemsep}{4pt}
\item Natural Mo has 7 isotopes, so the measured heat combines several parents. Enriched single-isotope targets would distinguish their contributions. Enrichment alone does not isolate a single reaction channel.
\end{itemize}
\clearpage
\subsection{Na: Sodium}\label{foil:Na}
\begin{figure}[H]
\centering
\includegraphics[width=\textwidth,height=0.82\textheight,keepaspectratio]{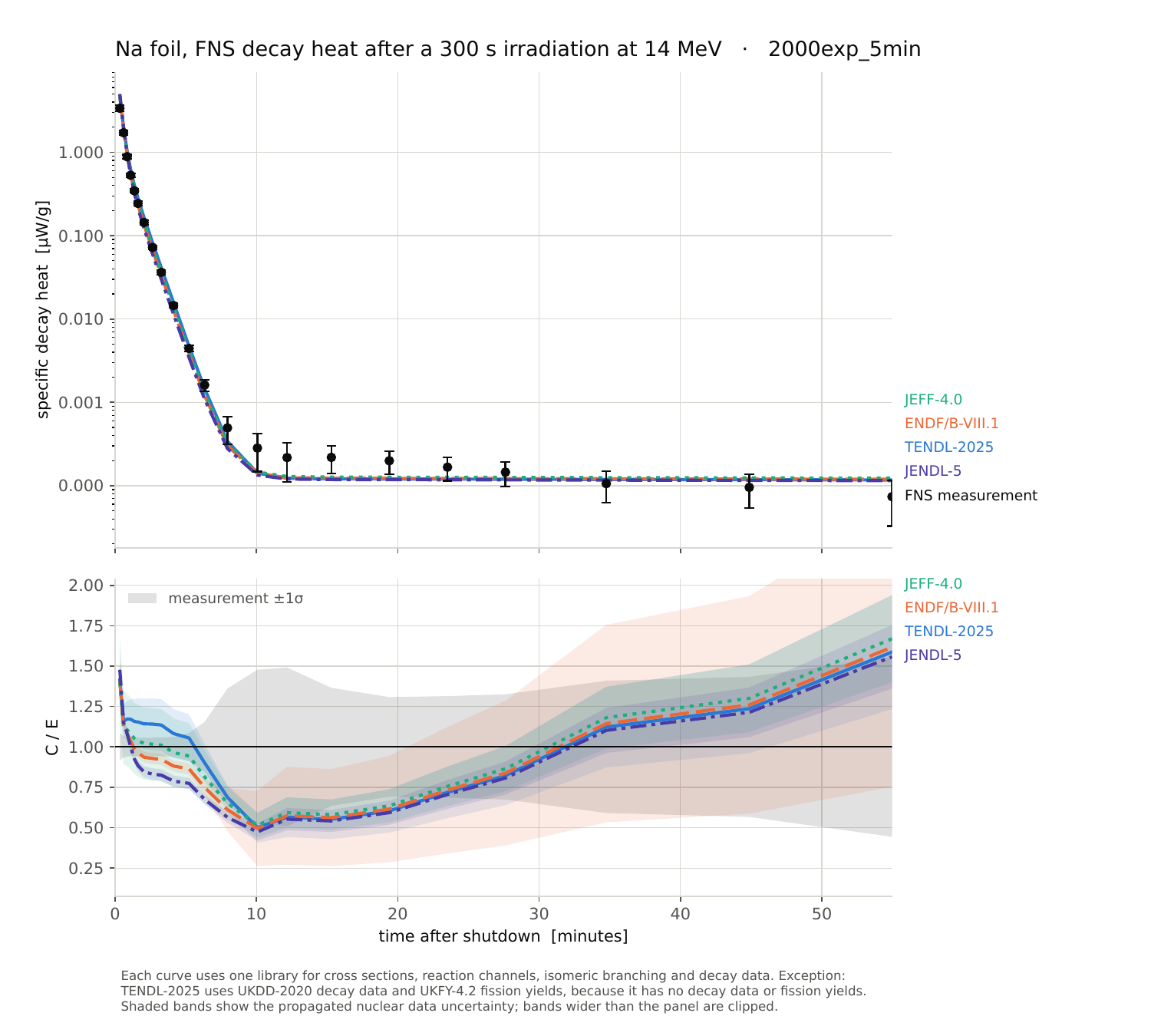}
\caption{Na: decay heat against the measurement, with C/E beneath. Mean $|C/E-1|$ by library: JEFF-4.0 21\%, ENDF/B-VIII.1 23\%, TENDL-2025 24\%, JENDL-5 26\%.}
\end{figure}
\clearpage

\begin{figure}[H]
\centering
\includegraphics[width=\textwidth,height=0.82\textheight,keepaspectratio]{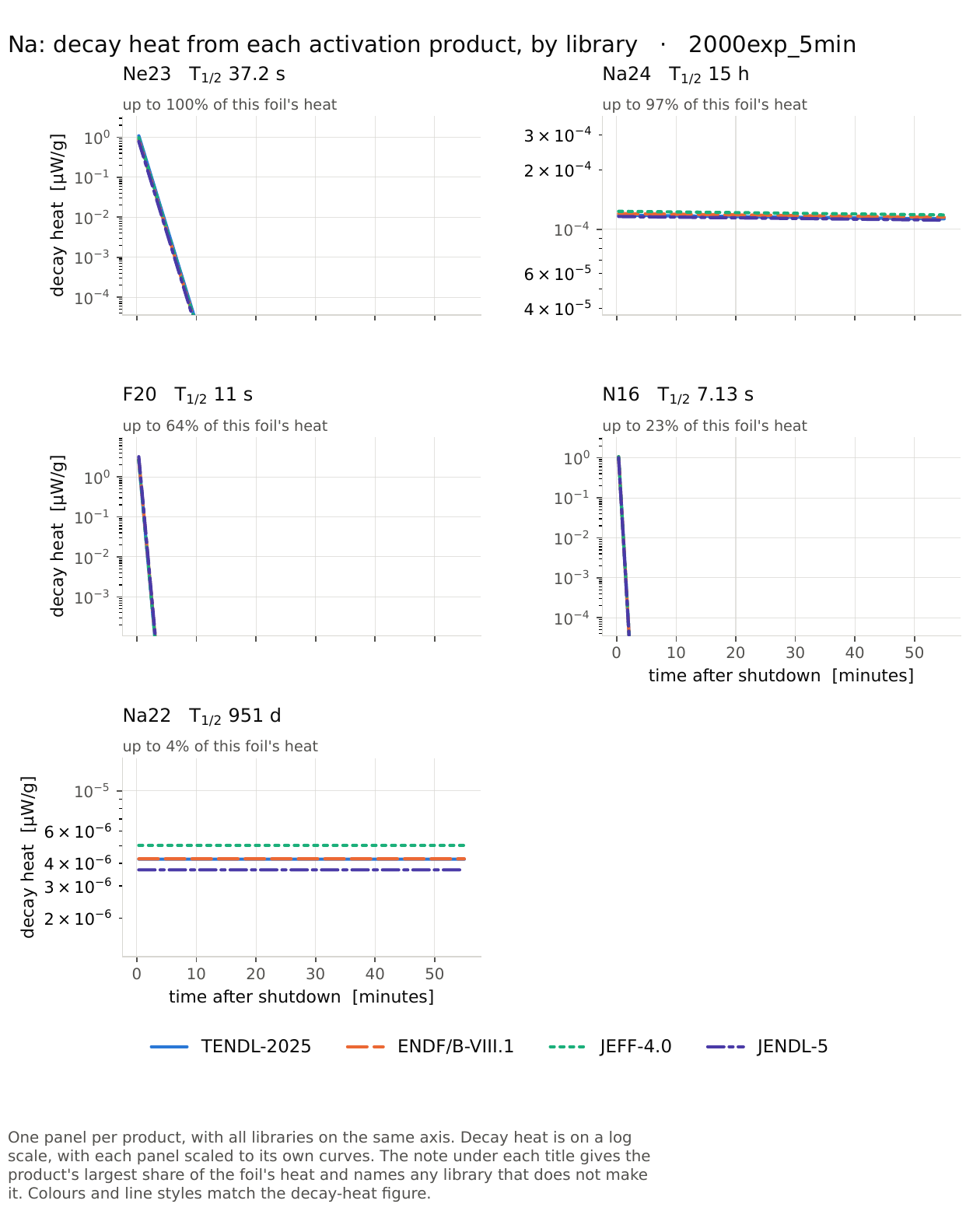}
\caption{Na: share of the decay heat from each activation product during cooling.}
\end{figure}
\clearpage

\begin{figure}[H]
\centering
\includegraphics[width=\textwidth,height=0.82\textheight,keepaspectratio]{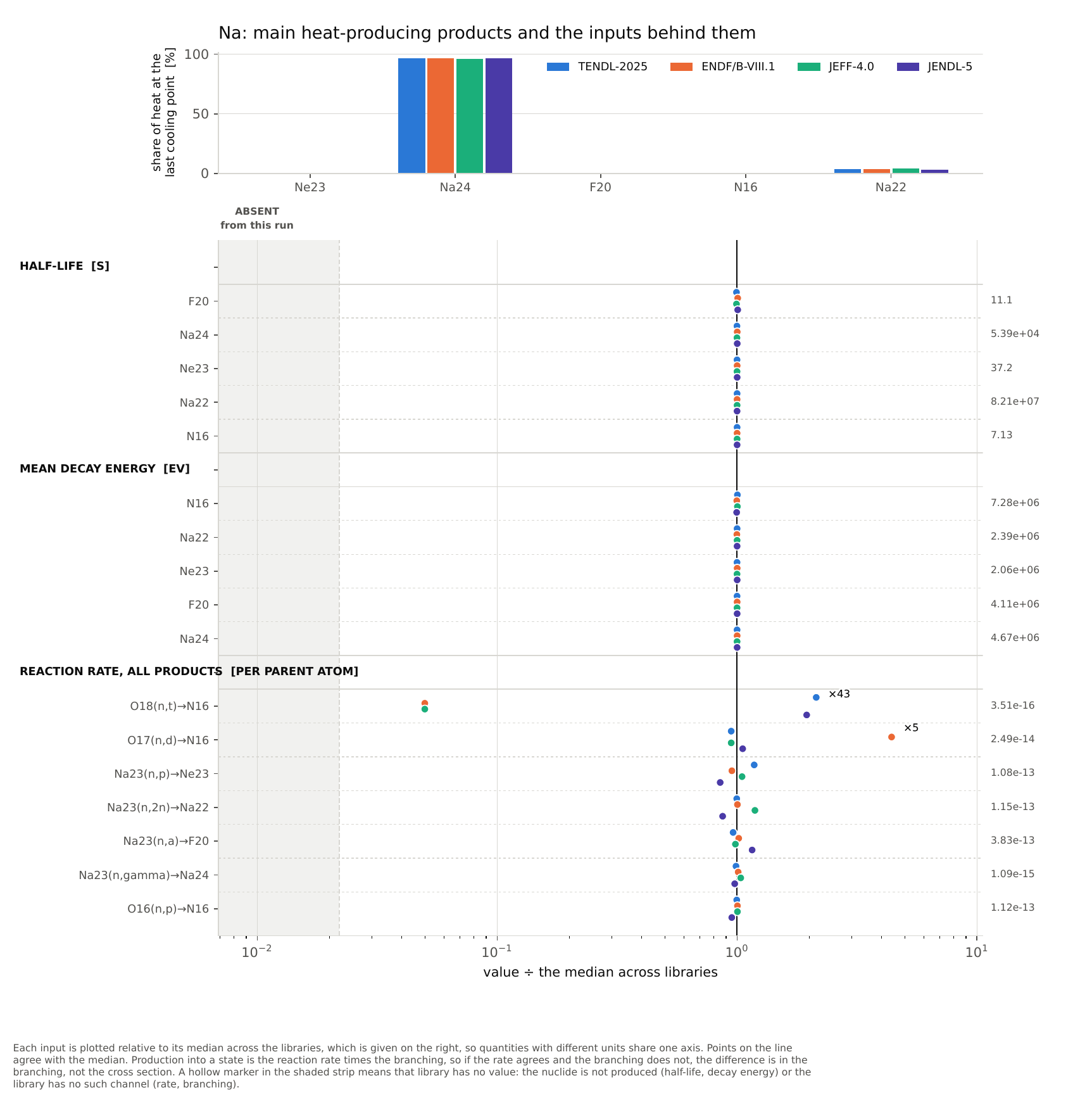}
\caption{Na: inputs for the main heat-producing products, each relative to its median across the available libraries.}
\end{figure}
\clearpage

\begin{figure}[H]
\centering
\includegraphics[width=\textwidth,height=0.82\textheight,keepaspectratio]{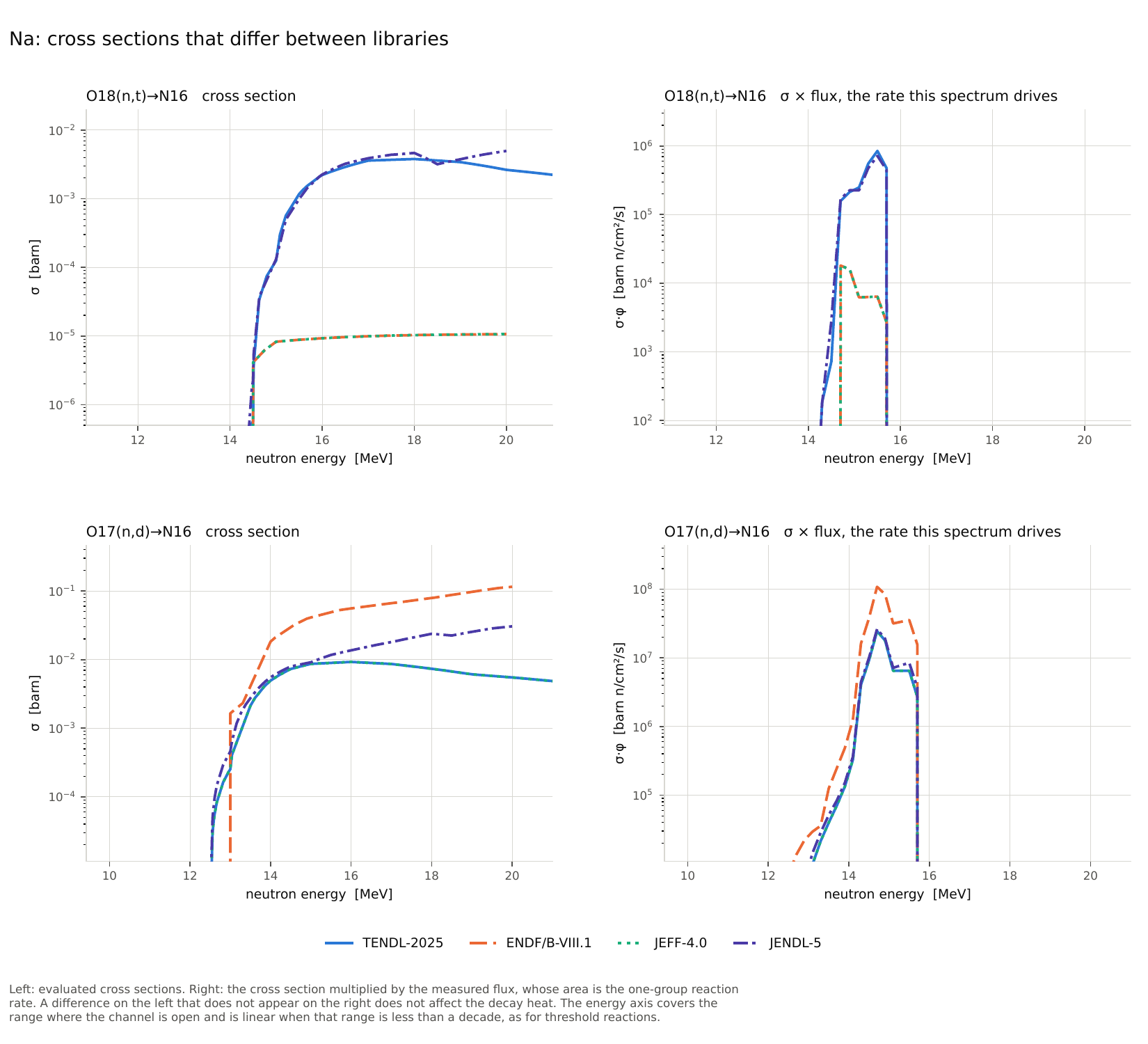}
\caption{Na: cross sections of the channels whose spectrum-averaged rates differ, with the neutron spectrum and their product.}
\end{figure}
\clearpage

\subsubsection*{1996 seven-hour campaign}
The 1996 campaign irradiated this foil for 7.17 hours at position 7 and measured it out to 402.16 days. In the five-minute calculation Na22 (half-life 951 days) gives up to 4 percent of the heat, and the longer irradiation and cooling test its production and decay data. JEFF-4.0 is closest. This campaign uses a different spectrum, so it is not included in the main rankings.

\begin{figure}[H]
\centering
\includegraphics[width=\textwidth,height=0.82\textheight,keepaspectratio]{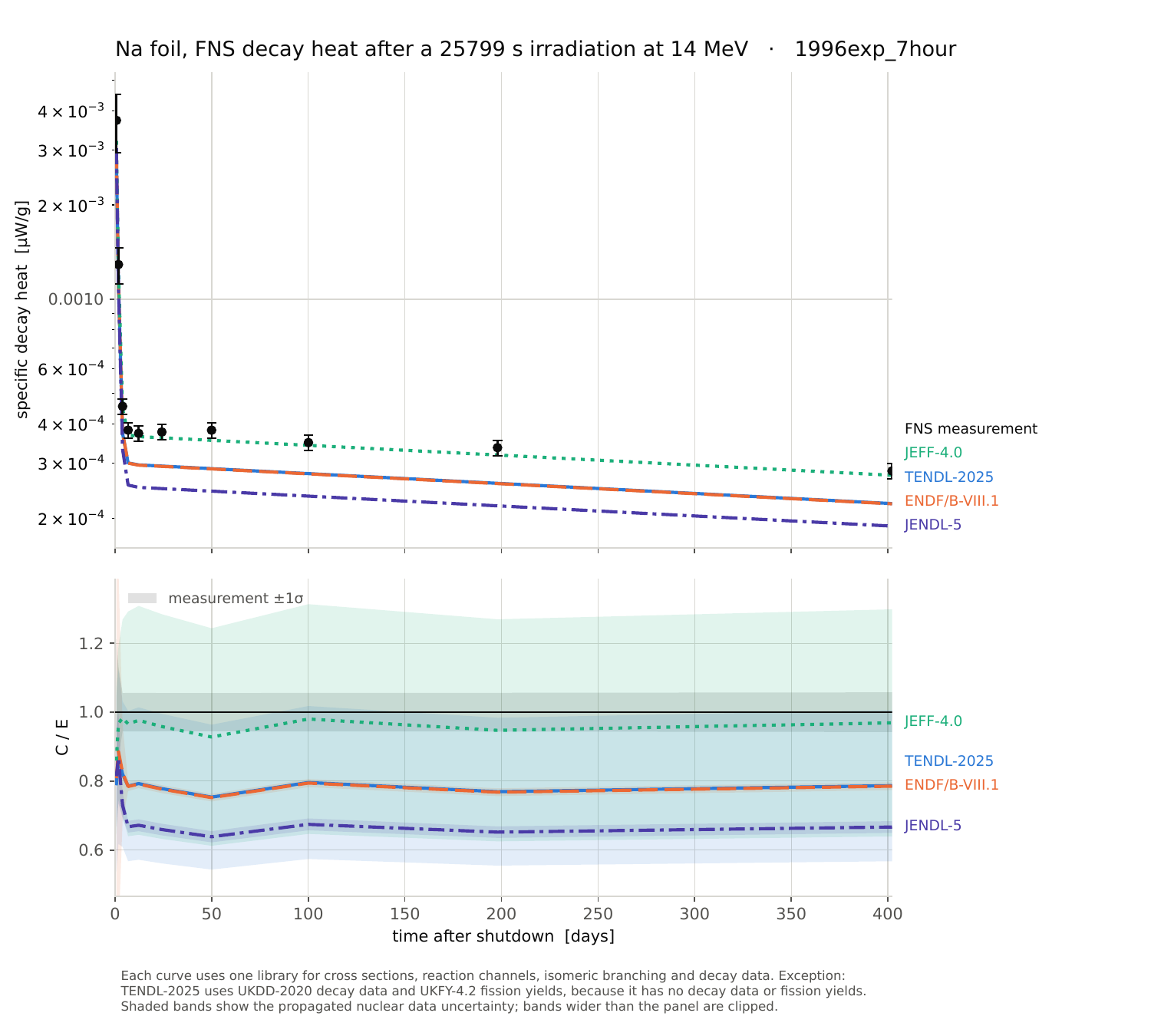}
\caption{Na: 1996 seven-hour campaign, irradiation position 7. Mean $|C/E-1|$ by library: JEFF-4.0 5\%, ENDF/B-VIII.1 20\%, TENDL-2025 21\%, JENDL-5 30\%. }
\end{figure}
\clearpage

Mean absolute C/E deviations range from 21 to 26 percent. The spread of 5 percentage points is smaller than the median measurement uncertainty of 12 percent, so the measurement cannot separate the libraries.

\subsubsection*{Differences in the nuclear data}
\begin{itemize}\setlength{\itemsep}{6pt}
\item ENDF/B-VIII.1 gives the total reaction rate of O18(n,t)→N16 as 1.751e-17, against 6.84e-16 for the median of the others, a factor of 39.\\[2pt]{\itshape This is the total rate of the channel over all its products, so the difference is in the cross section, not in the split between states.}
\item ENDF/B-VIII.1 gives the total reaction rate of O17(n,d)→N16 as 1.096e-13, against 2.351e-14 for the median of the others, a factor of 5.
\end{itemize}
\subsubsection*{Suggested measurements}
\begin{itemize}\setlength{\itemsep}{4pt}
\item Na has a single stable isotope, so the foil is already a single-isotope target and enrichment would not help.
\item The sample is a compound that also contains C, O. Their activation adds to the measured heat; a metal or purer sample would remove it.
\item The reaction rates are averaged over the measured spectrum. Activation measurements at several well-defined neutron energies, including below 14 MeV, would show which energy range the differences come from.
\end{itemize}
\clearpage
\subsection{Nb: Niobium}\label{foil:Nb}
\begin{figure}[H]
\centering
\includegraphics[width=\textwidth,height=0.82\textheight,keepaspectratio]{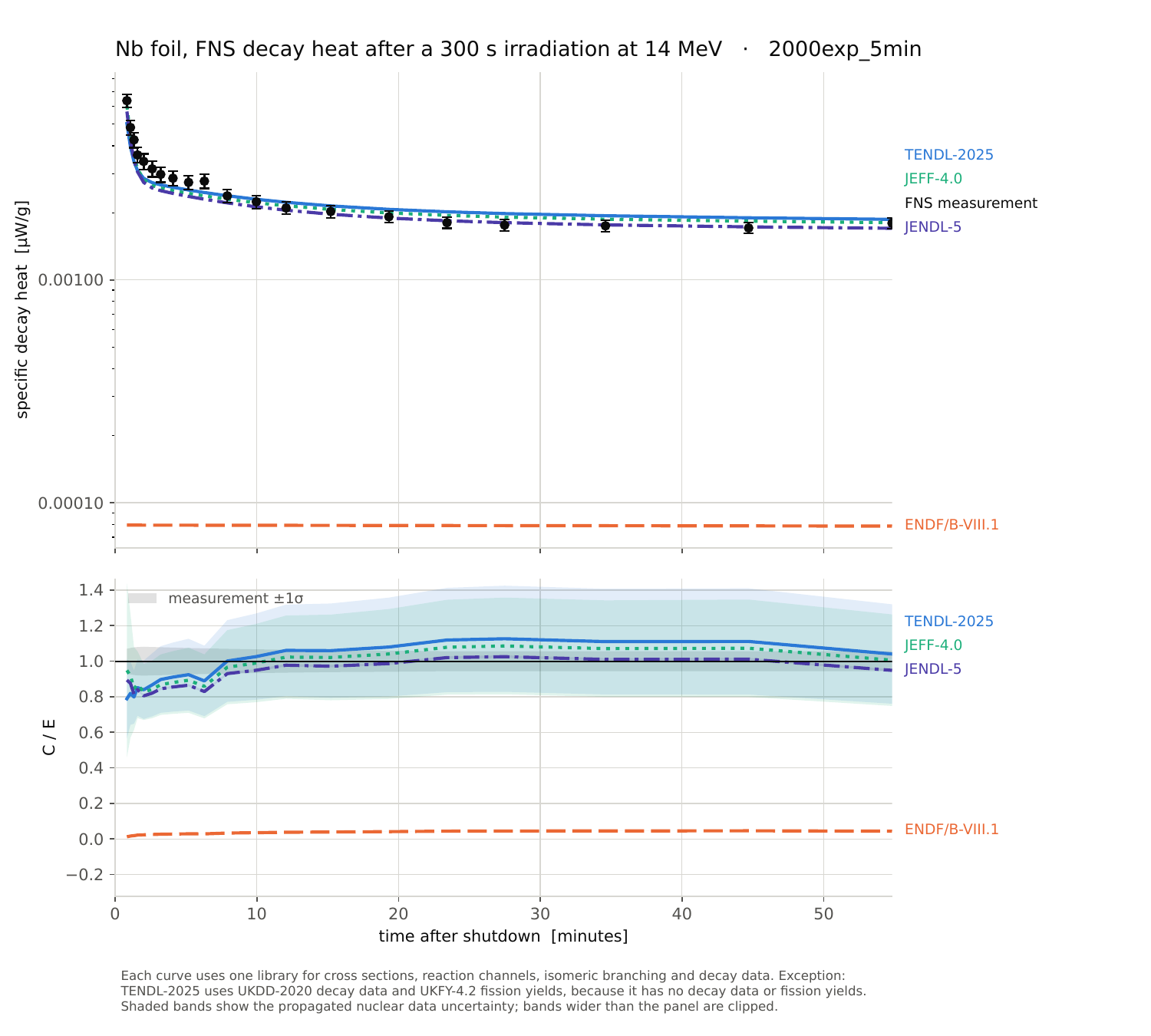}
\caption{Nb: decay heat against the measurement, with C/E beneath. Mean $|C/E-1|$ by library: JEFF-4.0 8\%, JENDL-5 9\%, TENDL-2025 11\%, ENDF/B-VIII.1 97\%.}
\end{figure}
\clearpage

\begin{figure}[H]
\centering
\includegraphics[width=\textwidth,height=0.82\textheight,keepaspectratio]{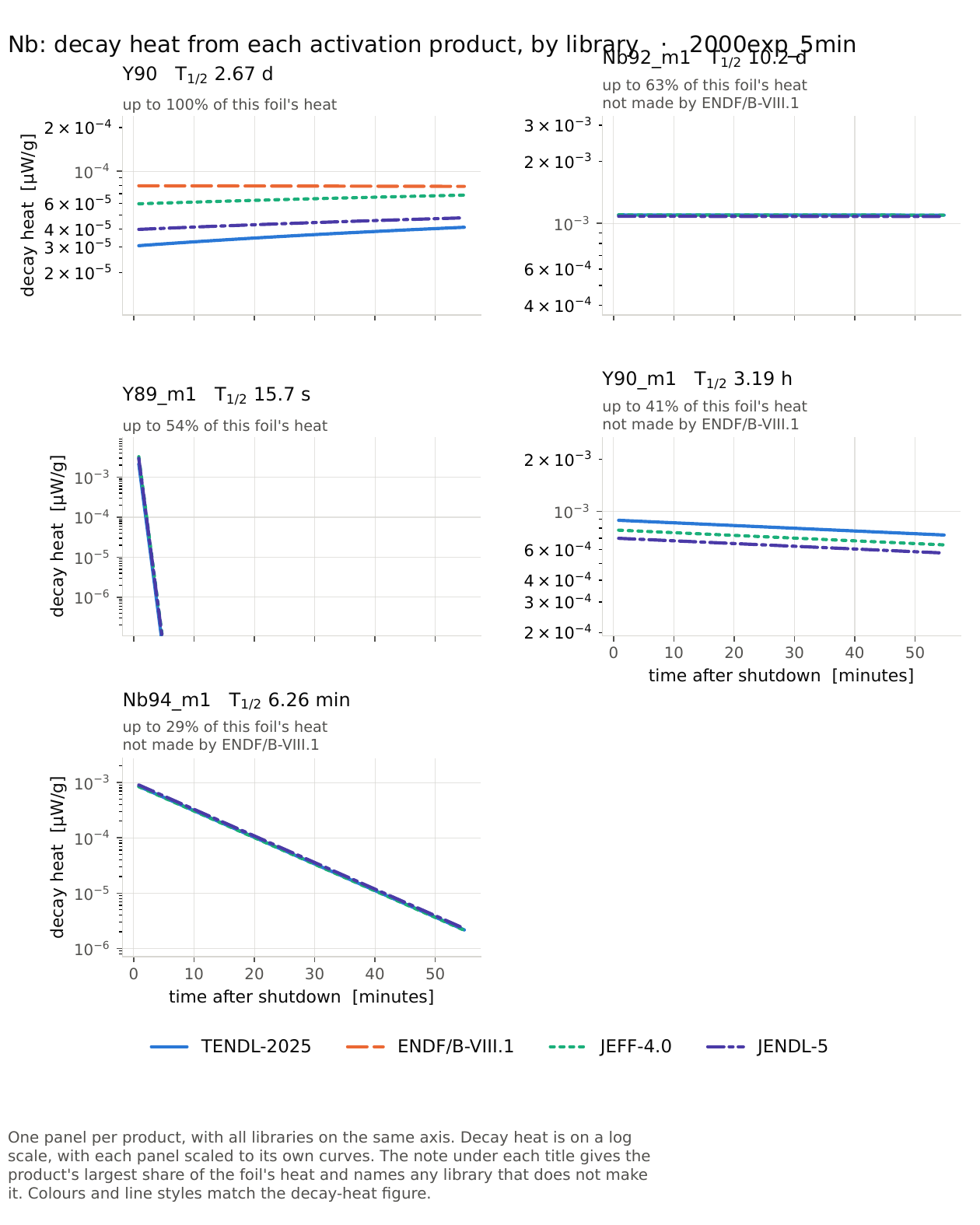}
\caption{Nb: share of the decay heat from each activation product during cooling.}
\end{figure}
\clearpage

\begin{figure}[H]
\centering
\includegraphics[width=\textwidth,height=0.82\textheight,keepaspectratio]{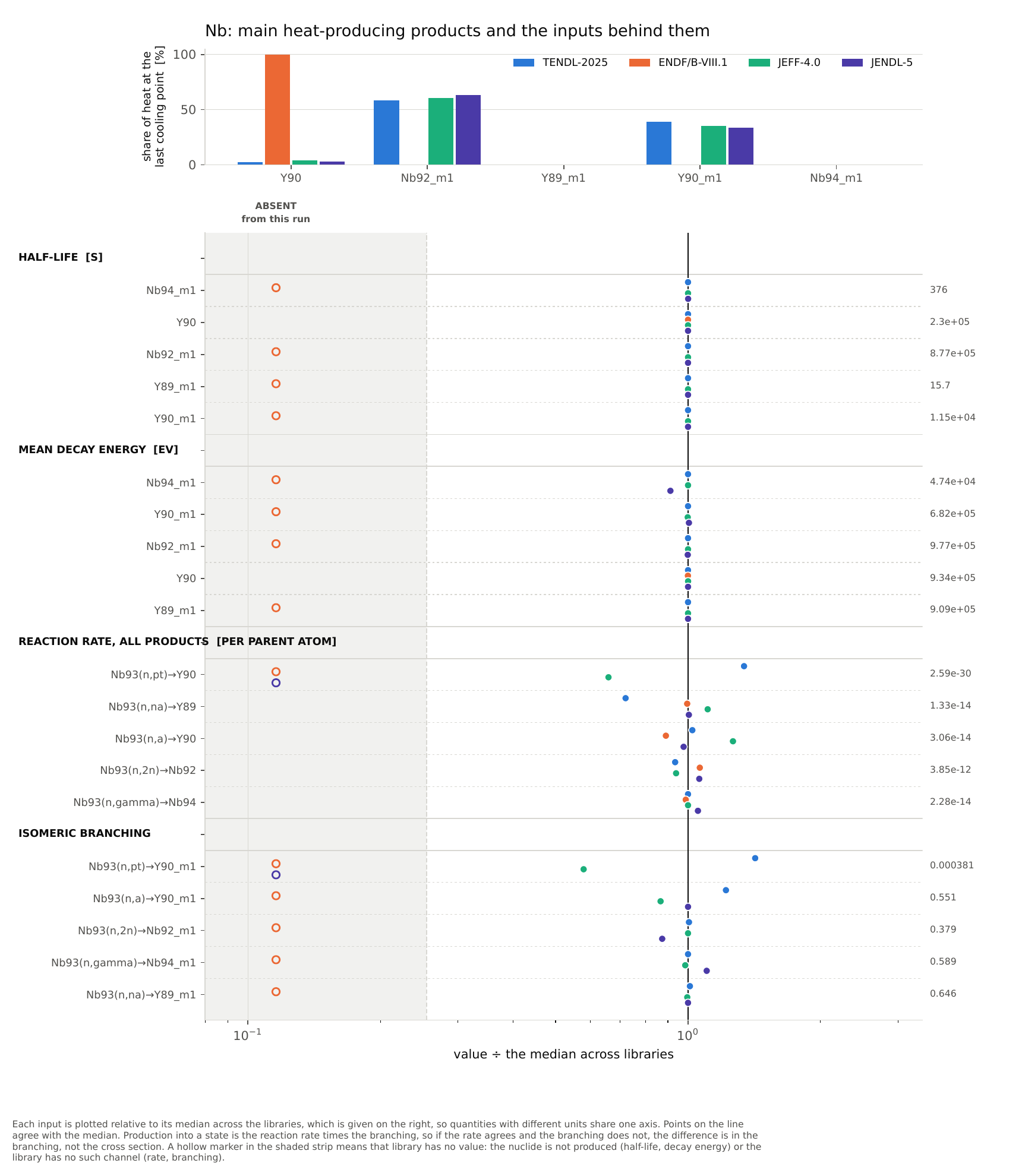}
\caption{Nb: inputs for the main heat-producing products, each relative to its median across the available libraries.}
\end{figure}
\clearpage

\begin{figure}[H]
\centering
\includegraphics[width=\textwidth,height=0.82\textheight,keepaspectratio]{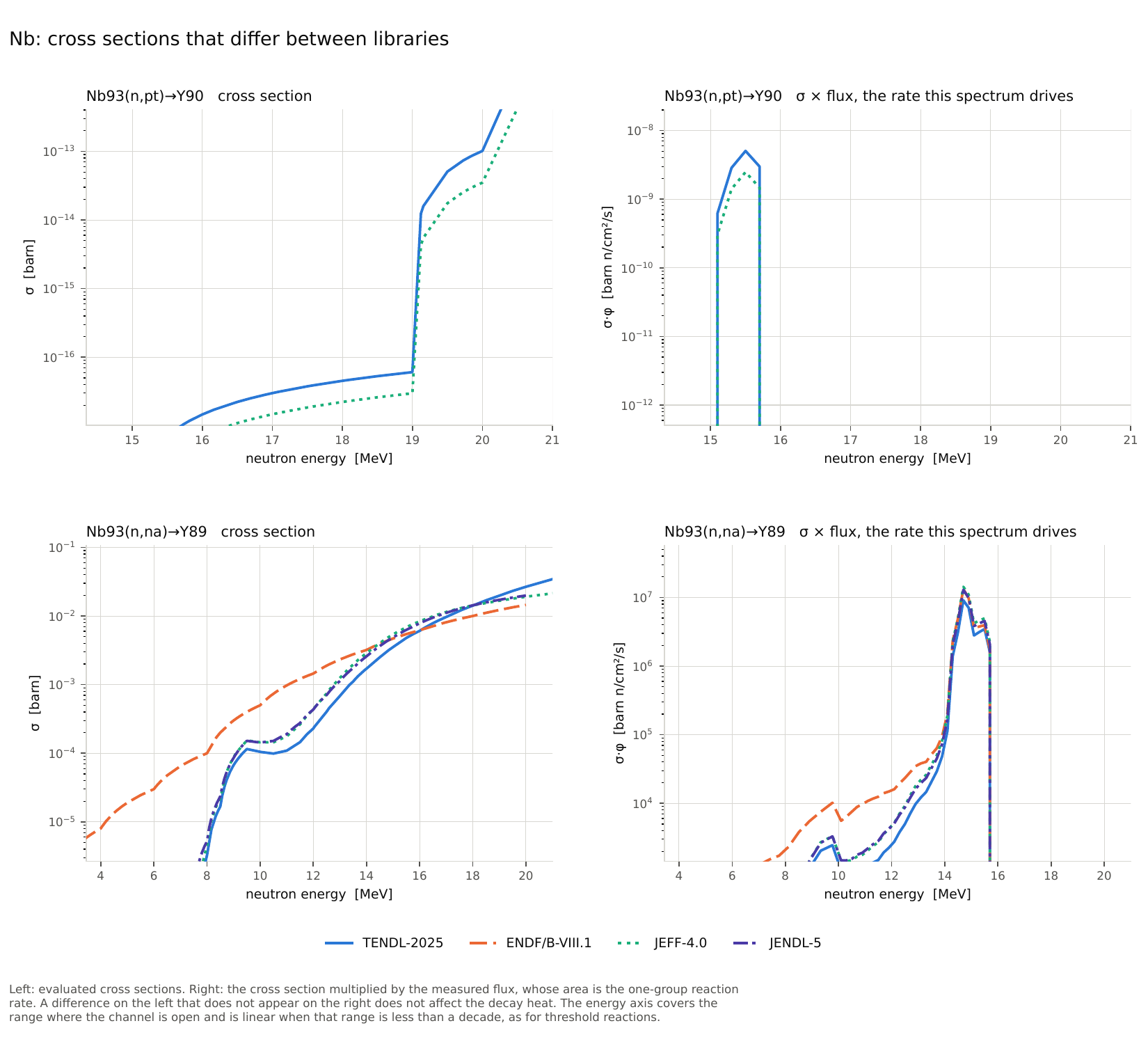}
\caption{Nb: cross sections of the channels whose spectrum-averaged rates differ, with the neutron spectrum and their product.}
\end{figure}
\clearpage

\begin{figure}[H]
\centering
\includegraphics[width=\textwidth,height=0.82\textheight,keepaspectratio]{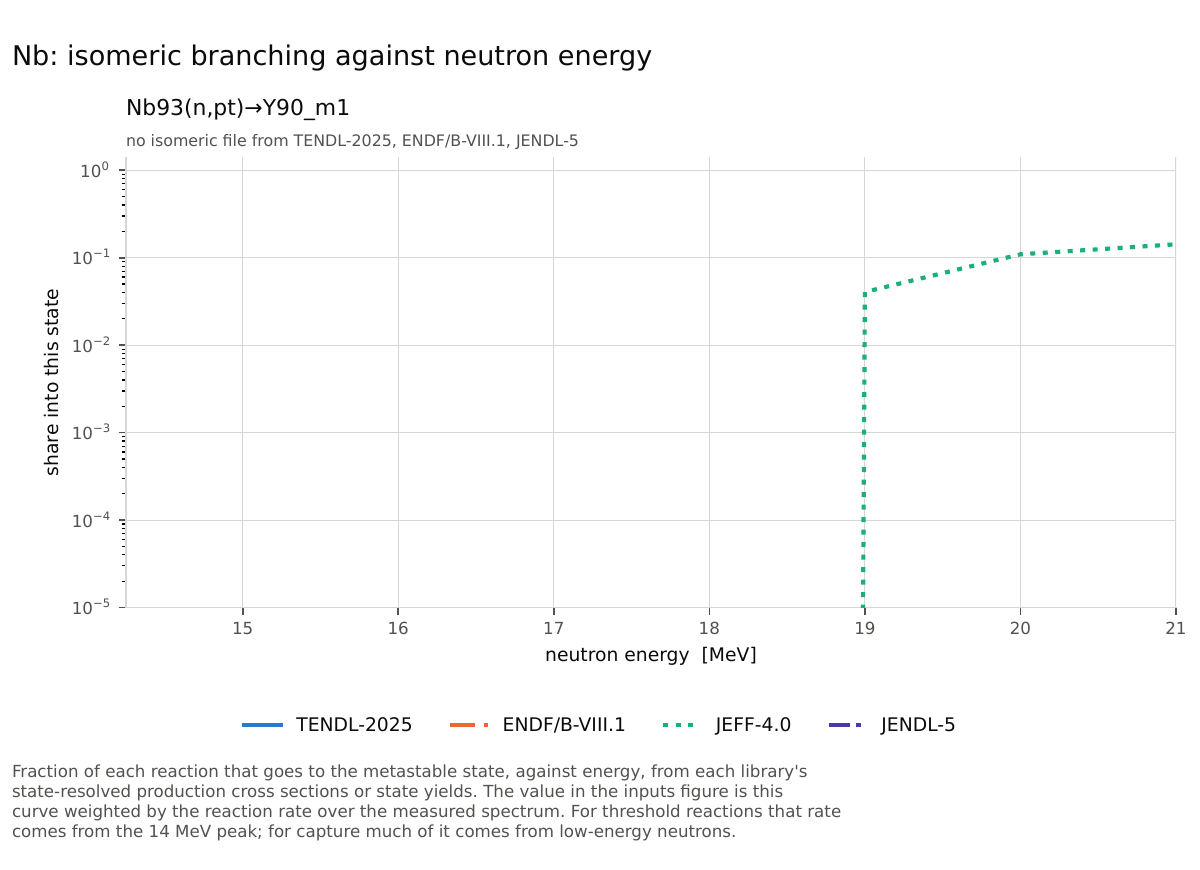}
\caption{Nb: isomeric branching ratio against neutron energy for the channels whose spectrum-averaged ratio differs between libraries.}
\end{figure}
\clearpage

\subsubsection*{1996 seven-hour campaign}
The 1996 campaign irradiated this foil for 7.57 hours at position 7 and measured it out to 99.89 days. In the five-minute calculation Y90 (half-life 2.67 days) gives up to 100 percent of the heat, and the longer irradiation and cooling test its production and decay data. JEFF-4.0 is closest. This campaign uses a different spectrum, so it is not included in the main rankings.

\begin{figure}[H]
\centering
\includegraphics[width=\textwidth,height=0.82\textheight,keepaspectratio]{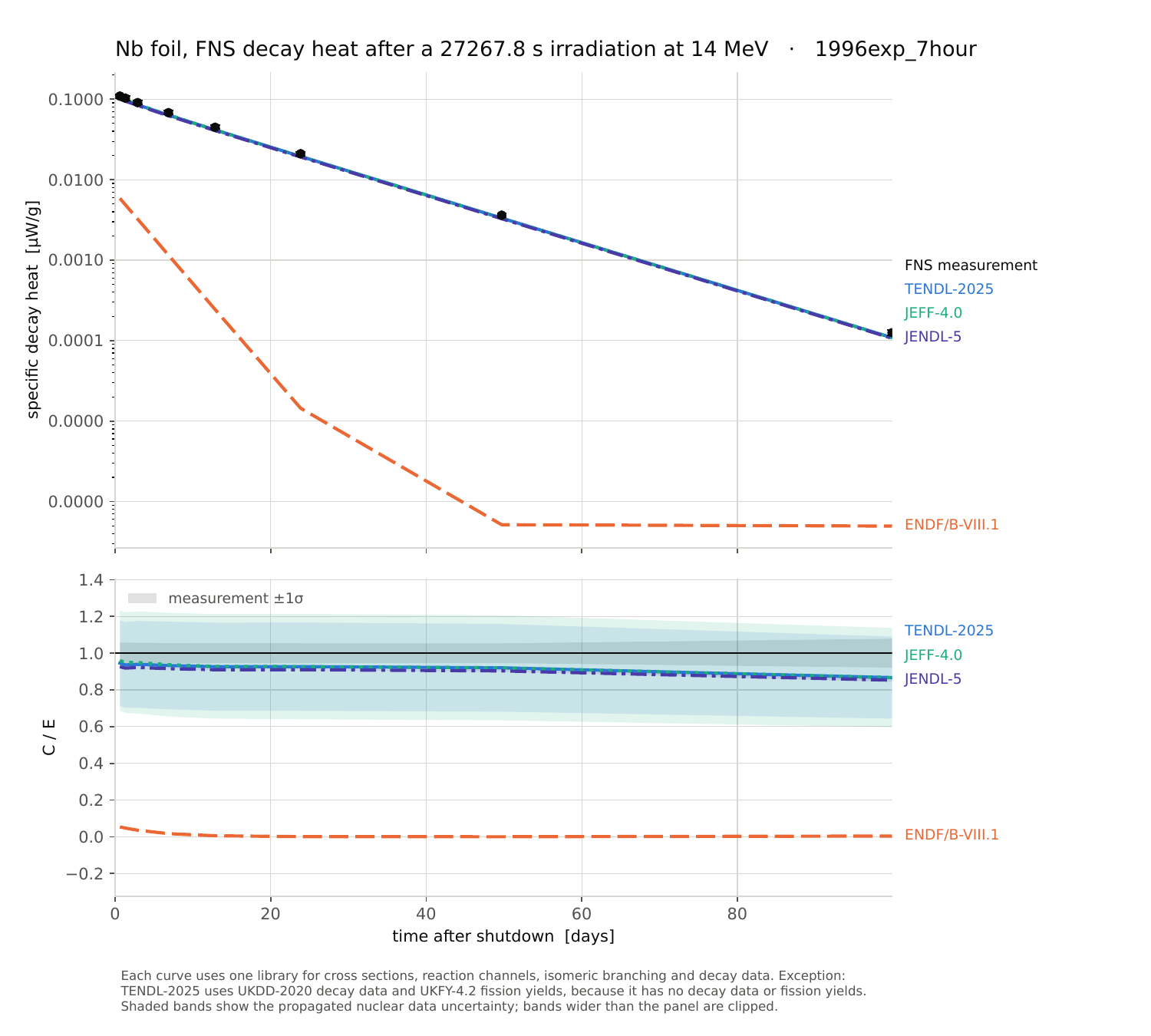}
\caption{Nb: 1996 seven-hour campaign, irradiation position 7. Mean $|C/E-1|$ by library: JEFF-4.0 7\%, TENDL-2025 8\%, JENDL-5 9\%, ENDF/B-VIII.1 98\%. }
\end{figure}
\clearpage

ENDF/B-VIII.1 stands out at 97 percent from the measurement, against 11 percent for the next furthest library and 8 percent for JEFF-4.0. The median measurement uncertainty is 7 percent.

\subsubsection*{Differences in the nuclear data}
\begin{itemize}\setlength{\itemsep}{6pt}
\item Only TENDL-2025 and JEFF-4.0 have the isomeric branching of Nb93(n,pt)→Y90\_m1, and they differ by a factor of 2: 0.0005419 against 0.0002208.\\[2pt]{\itshape Isomeric branching sets how a channel's production is split between the ground and metastable states. It should be checked together with the total rate of the channel.}
\item Only TENDL-2025 and JEFF-4.0 have the total reaction rate of Nb93(n,pt)→Y90, and they differ by a factor of 2: 3.477e-30 against 1.711e-30.\\[2pt]{\itshape This is the total rate of the channel over all its products, so the difference is in the cross section, not in the split between states.}
\item ENDF/B-VIII.1 gives no heat from Nb92\_m1; the other libraries do.\\[2pt]{\itshape The heat breakdown alone does not show why. Possible causes are a missing reaction channel, a different state assignment, different decay data or a reporting threshold.}
\item ENDF/B-VIII.1 gives no heat from Y90\_m1; the other libraries do.
\item ENDF/B-VIII.1 gives no heat from Nb94\_m1; the other libraries do.
\end{itemize}
\subsubsection*{Suggested measurements}
\begin{itemize}\setlength{\itemsep}{4pt}
\item Nb has a single stable isotope, so the foil is already a single-isotope target and enrichment would not help.
\item The reaction rates are averaged over the measured spectrum. Activation measurements at several well-defined neutron energies, including below 14 MeV, would show which energy range the differences come from.
\item Measuring the activity of each state separately, for example by gamma spectrometry where the states have distinct lines, would constrain the isomeric ratio. This needs corrections for feeding between states, detector efficiency and irradiation history.
\end{itemize}
\clearpage
\subsection{Nd: Neodymium}\label{foil:Nd}
\begin{figure}[H]
\centering
\includegraphics[width=\textwidth,height=0.82\textheight,keepaspectratio]{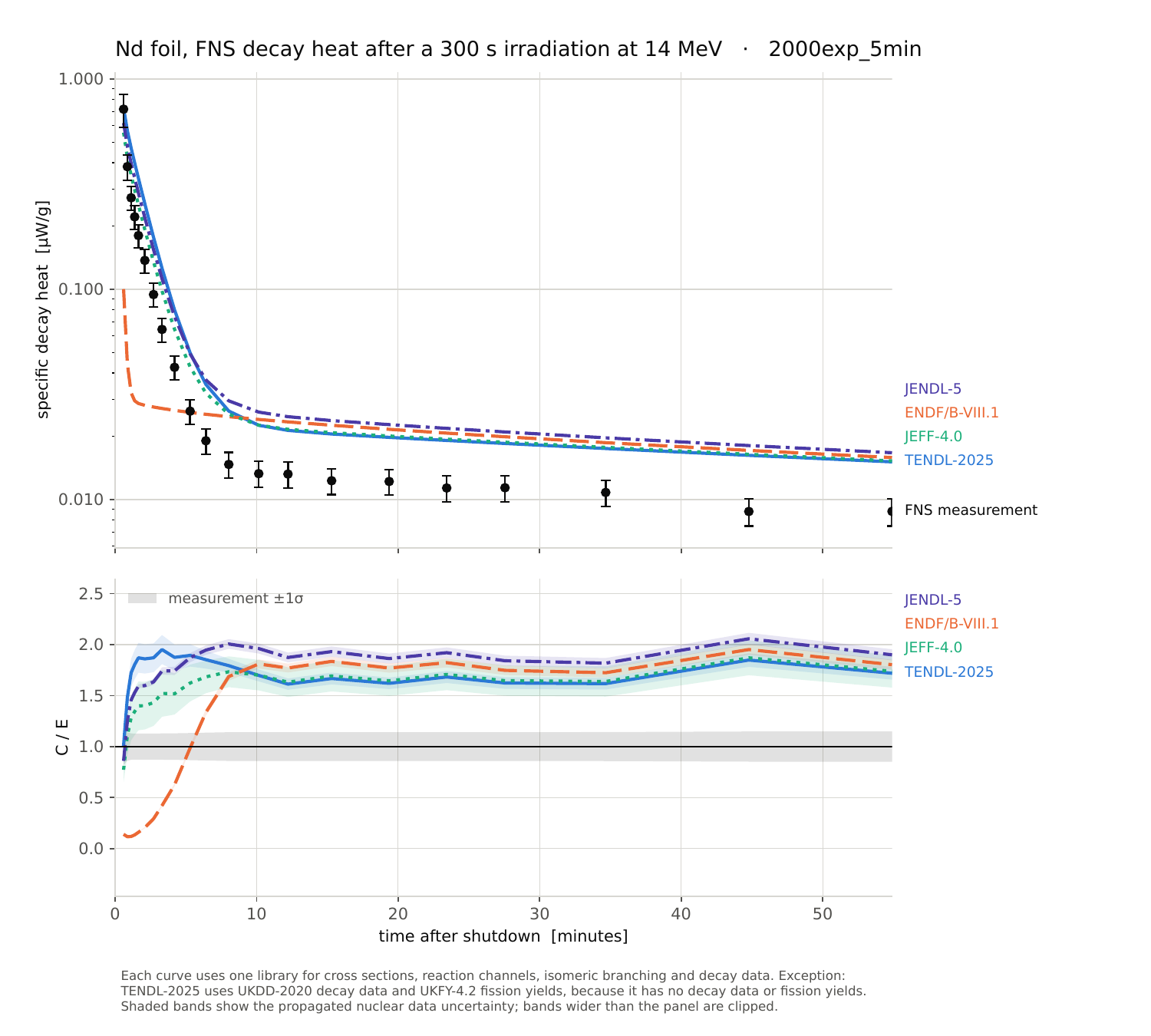}
\caption{Nd: decay heat against the measurement, with C/E beneath. Mean $|C/E-1|$ by library: JEFF-4.0 55\%, ENDF/B-VIII.1 72\%, TENDL-2025 72\%, JENDL-5 75\%.}
\end{figure}
\clearpage

\begin{figure}[H]
\centering
\includegraphics[width=\textwidth,height=0.82\textheight,keepaspectratio]{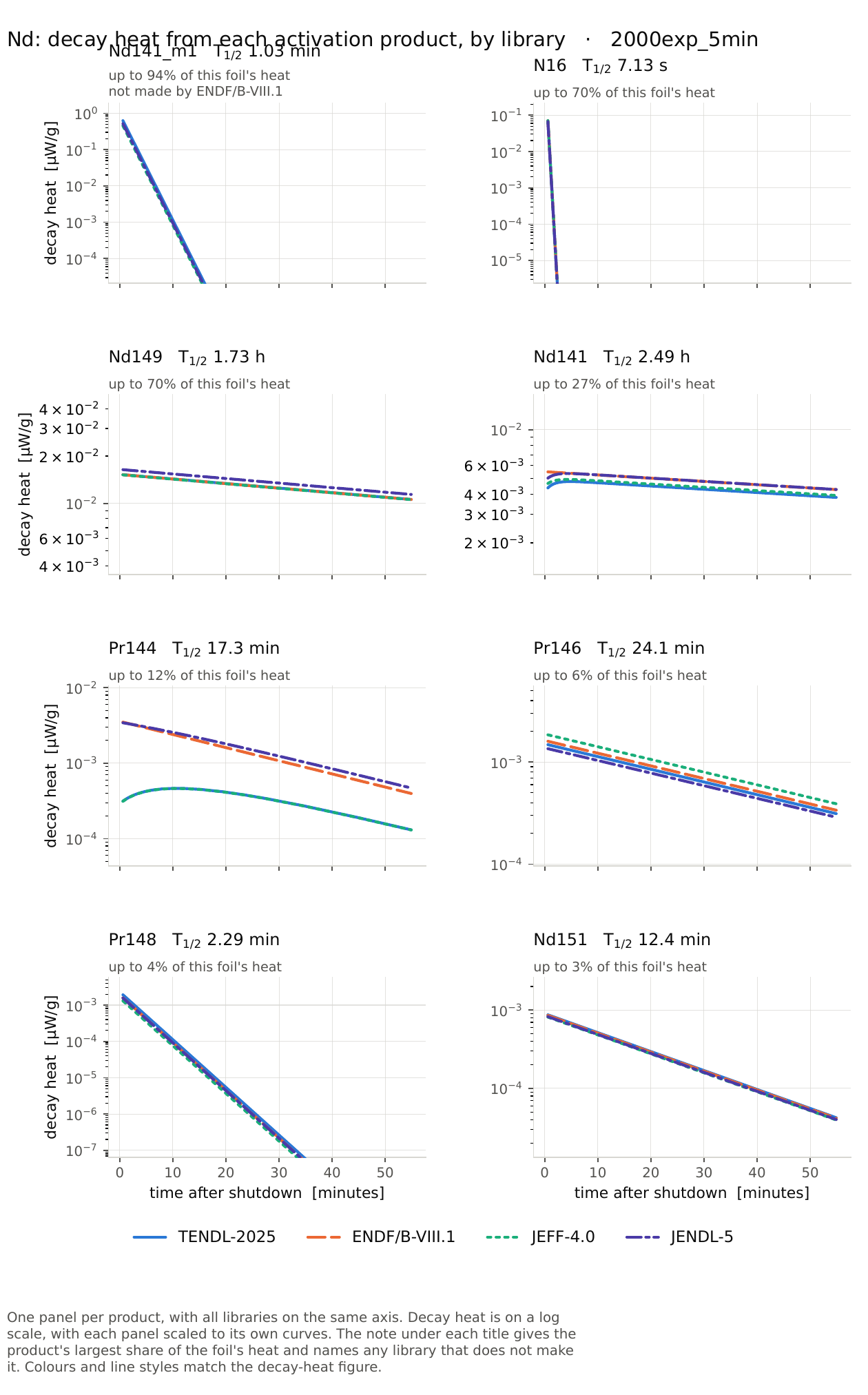}
\caption{Nd: share of the decay heat from each activation product during cooling.}
\end{figure}
\clearpage

\begin{figure}[H]
\centering
\includegraphics[width=\textwidth,height=0.82\textheight,keepaspectratio]{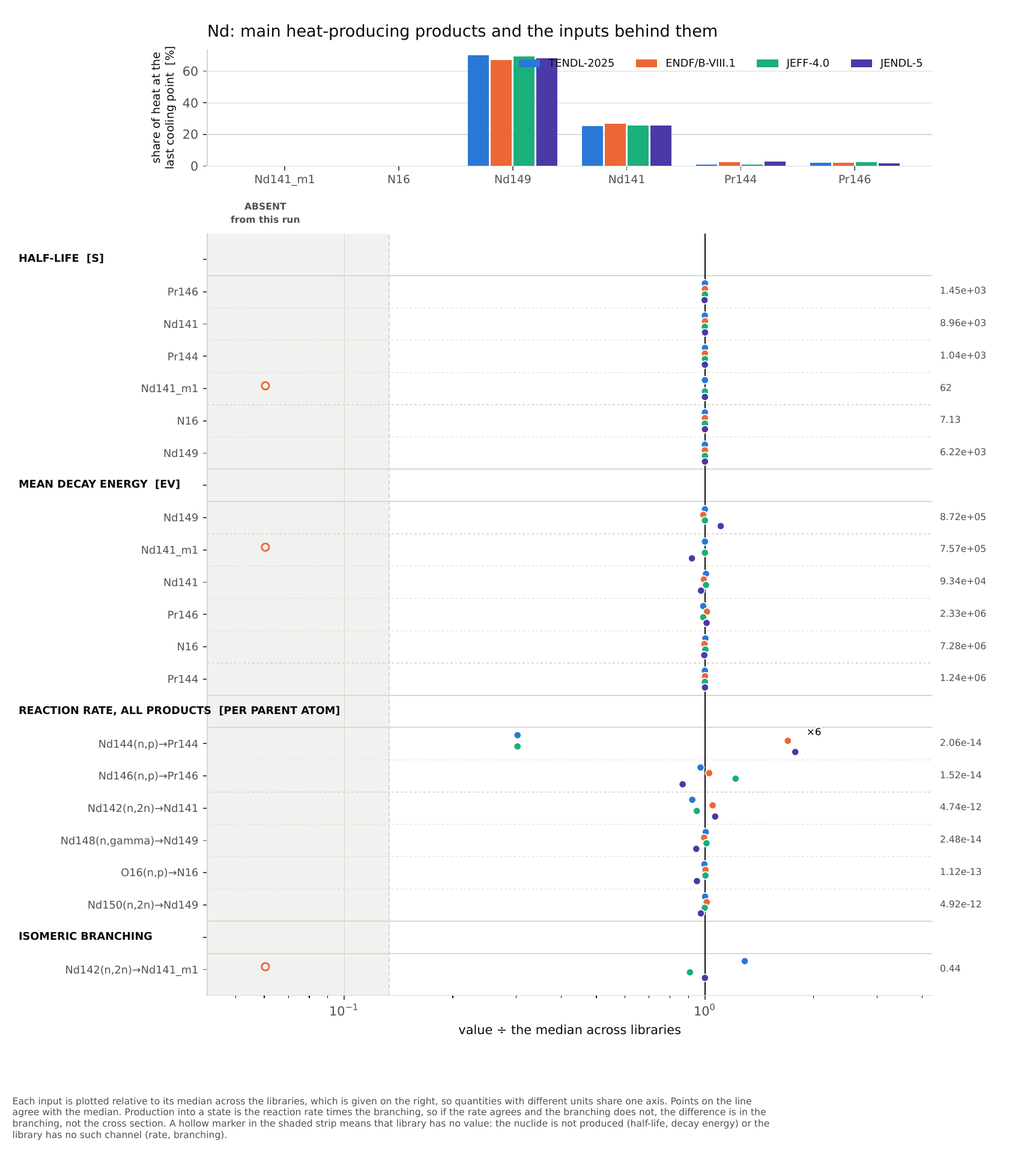}
\caption{Nd: inputs for the main heat-producing products, each relative to its median across the available libraries.}
\end{figure}
\clearpage

\begin{figure}[H]
\centering
\includegraphics[width=\textwidth,height=0.82\textheight,keepaspectratio]{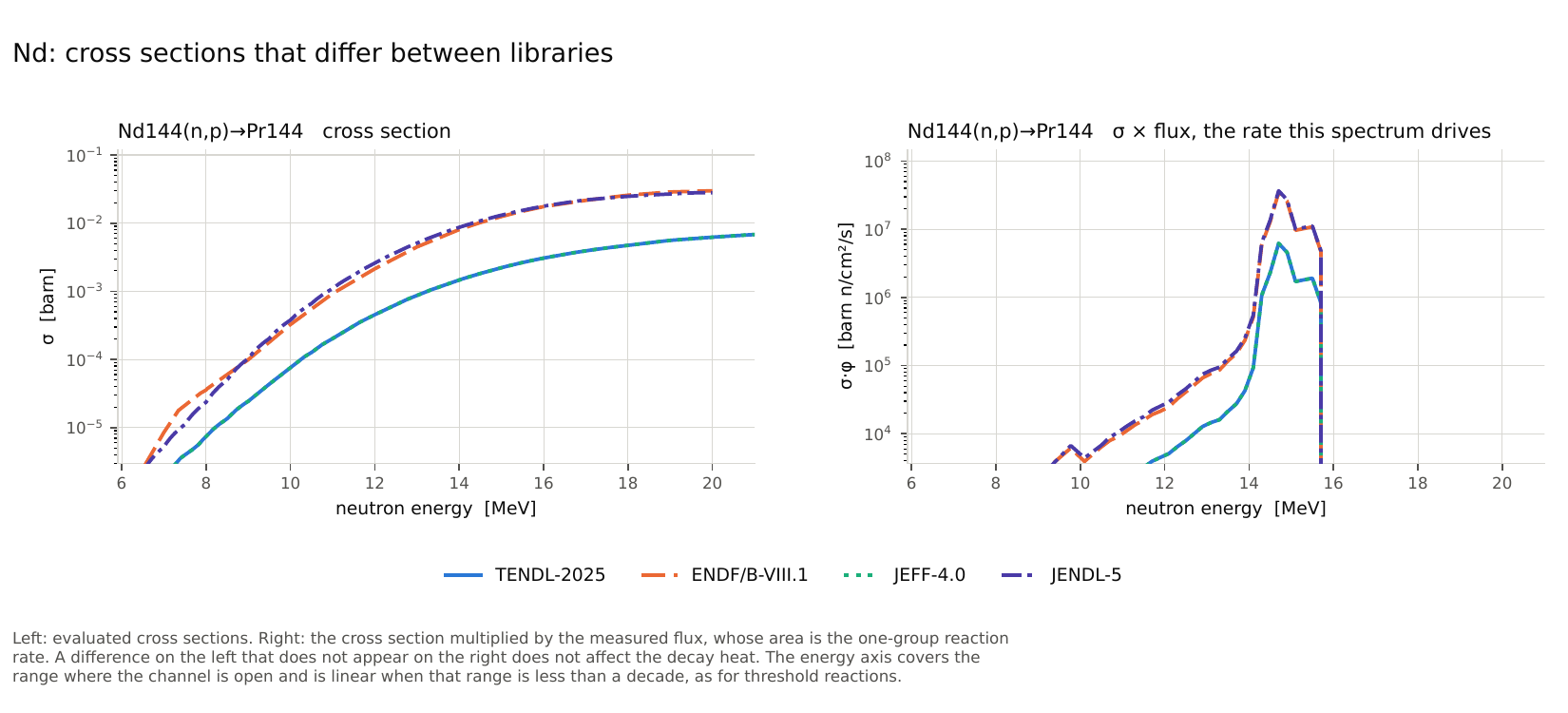}
\caption{Nd: cross sections of the channels whose spectrum-averaged rates differ, with the neutron spectrum and their product.}
\end{figure}
\clearpage

The libraries are between 55 and 75 percent from the measurement, which has a median uncertainty of 14 percent. JEFF-4.0 is closest and JENDL-5 furthest.

\subsubsection*{Differences in the nuclear data}
\begin{itemize}\setlength{\itemsep}{6pt}
\item JEFF-4.0 gives the total reaction rate of Nd144(n,p)→Pr144 as 6.217e-15, against 3.491e-14 for the median of the others, a factor of 6.\\[2pt]{\itshape This is the total rate of the channel over all its products, so the difference is in the cross section, not in the split between states.}
\item ENDF/B-VIII.1 gives no heat from Nd141\_m1; the other libraries do.\\[2pt]{\itshape The heat breakdown alone does not show why. Possible causes are a missing reaction channel, a different state assignment, different decay data or a reporting threshold.}
\end{itemize}
\subsubsection*{Suggested measurements}
\begin{itemize}\setlength{\itemsep}{4pt}
\item Natural Nd has 7 isotopes, so the measured heat combines several parents. Enriched single-isotope targets would distinguish their contributions. The candidate parent is Nd144. Enrichment alone does not isolate a single reaction channel.
\item The sample is a compound that also contains O. Their activation adds to the measured heat; a metal or purer sample would remove it.
\item The reaction rates are averaged over the measured spectrum. Activation measurements at several well-defined neutron energies, including below 14 MeV, would show which energy range the differences come from.
\end{itemize}
\clearpage
\subsection{Ni: Nickel}\label{foil:Ni}
\begin{figure}[H]
\centering
\includegraphics[width=\textwidth,height=0.82\textheight,keepaspectratio]{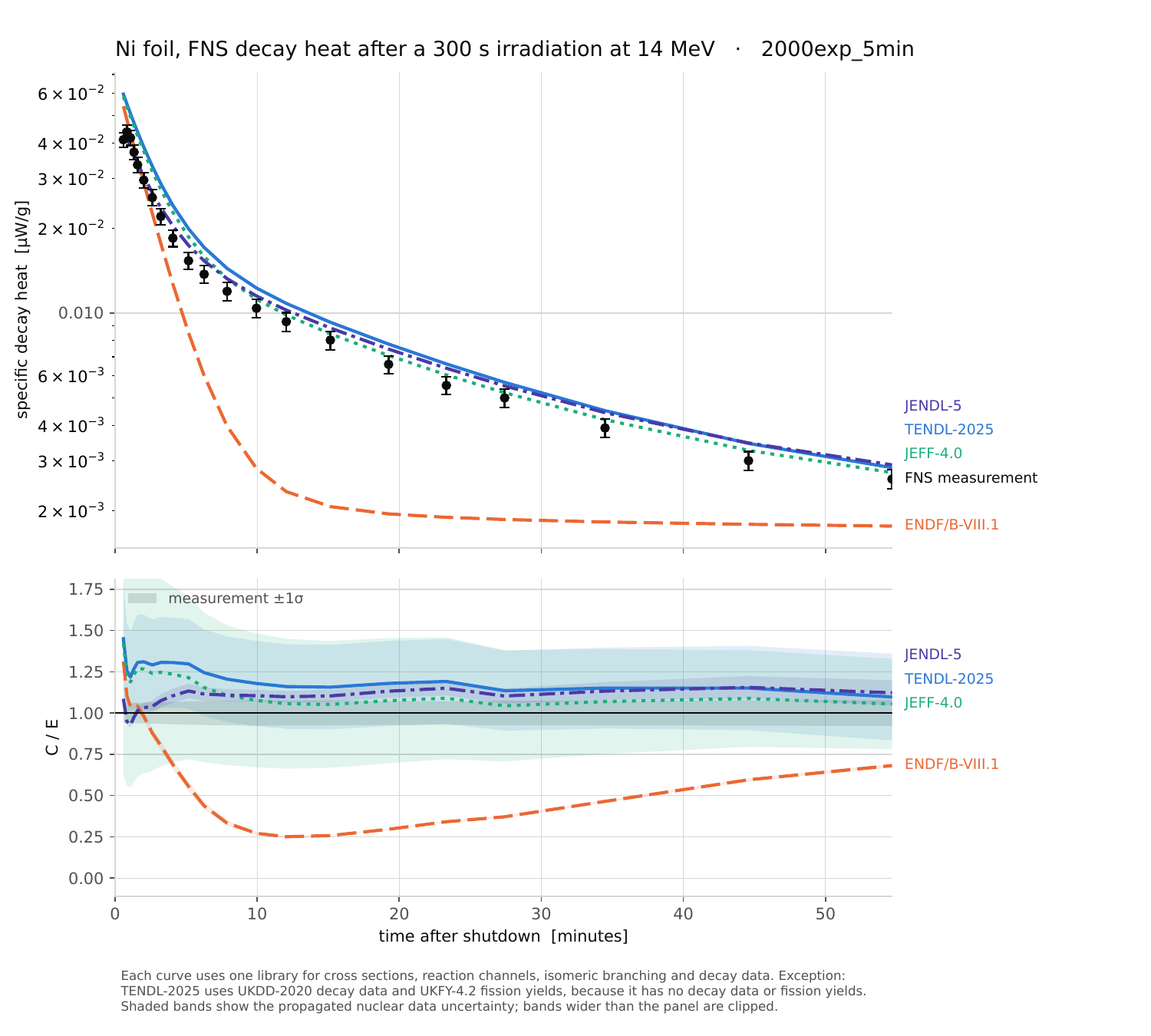}
\caption{Ni: decay heat against the measurement, with C/E beneath. Mean $|C/E-1|$ by library: JENDL-5 9\%, JEFF-4.0 16\%, TENDL-2025 23\%, ENDF/B-VIII.1 40\%.}
\end{figure}
\clearpage

\begin{figure}[H]
\centering
\includegraphics[width=\textwidth,height=0.82\textheight,keepaspectratio]{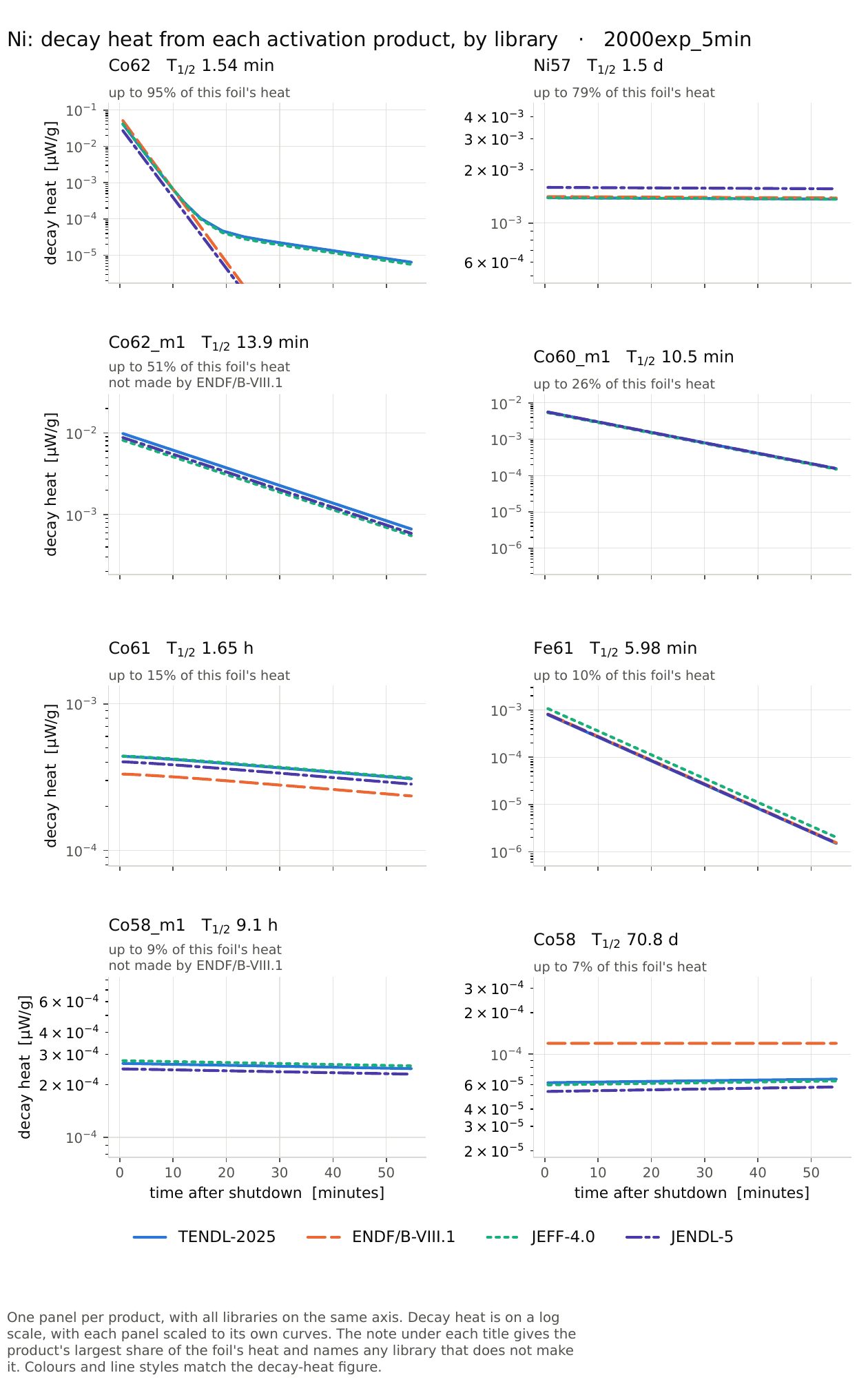}
\caption{Ni: share of the decay heat from each activation product during cooling.}
\end{figure}
\clearpage

\begin{figure}[H]
\centering
\includegraphics[width=\textwidth,height=0.82\textheight,keepaspectratio]{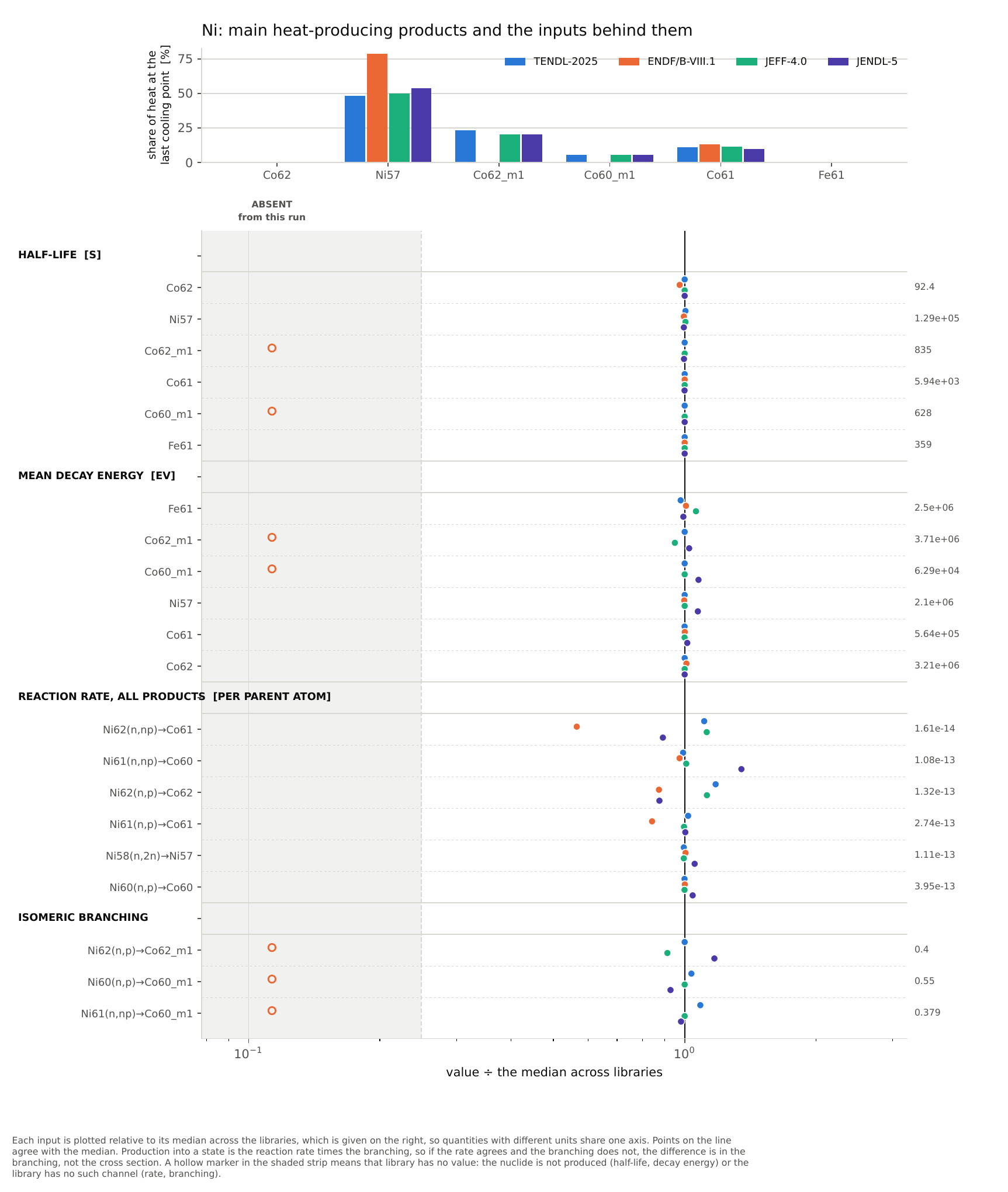}
\caption{Ni: inputs for the main heat-producing products, each relative to its median across the available libraries.}
\end{figure}
\clearpage

\begin{figure}[H]
\centering
\includegraphics[width=\textwidth,height=0.82\textheight,keepaspectratio]{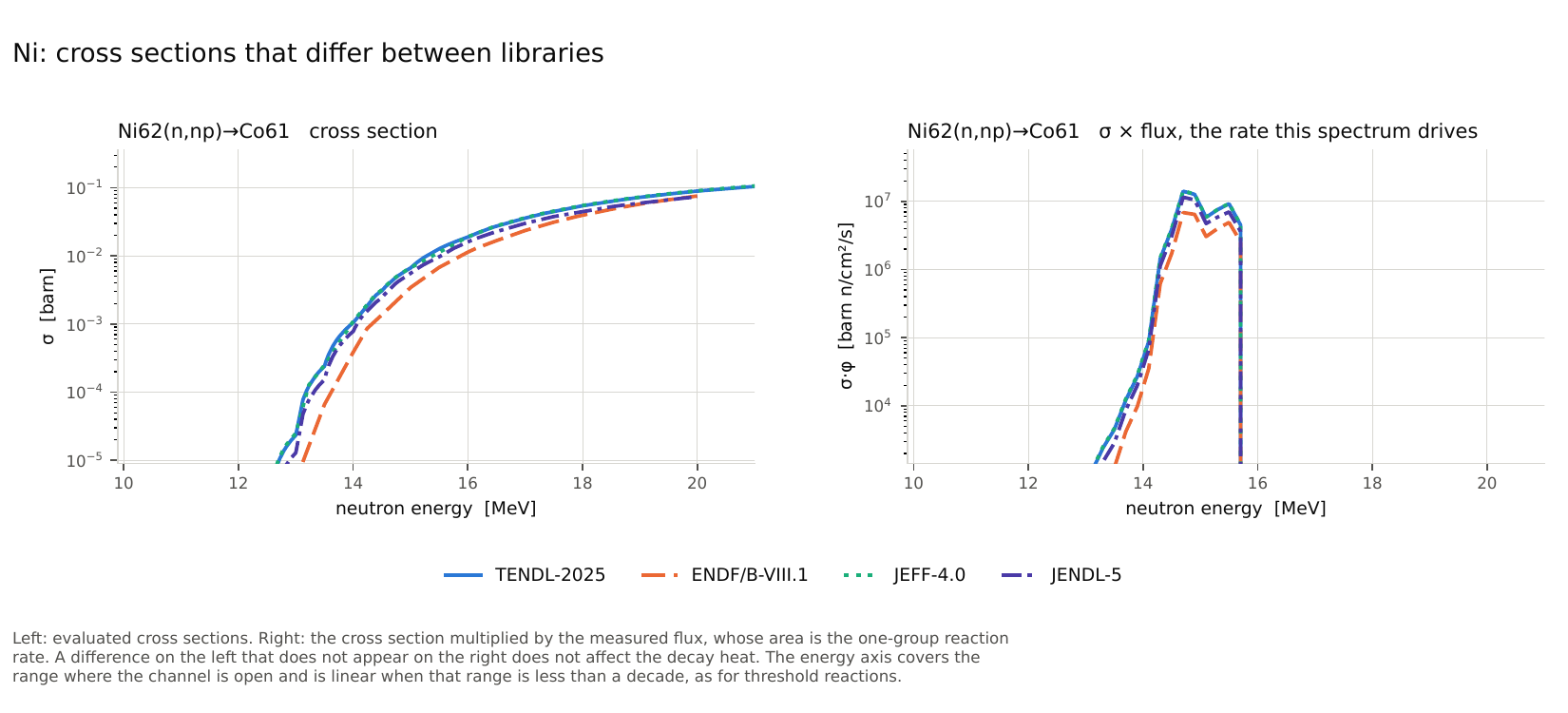}
\caption{Ni: cross sections of the channels whose spectrum-averaged rates differ, with the neutron spectrum and their product.}
\end{figure}
\clearpage

\subsubsection*{1996 seven-hour campaign}
The 1996 campaign irradiated this foil for 7.57 hours at position 7 and measured it out to 402.94 days. In the five-minute calculation Ni57 (half-life 1.5 days) gives up to 79 percent of the heat, and the longer irradiation and cooling test its production and decay data. ENDF/B-VIII.1 is closest. This campaign uses a different spectrum, so it is not included in the main rankings.

\begin{figure}[H]
\centering
\includegraphics[width=\textwidth,height=0.82\textheight,keepaspectratio]{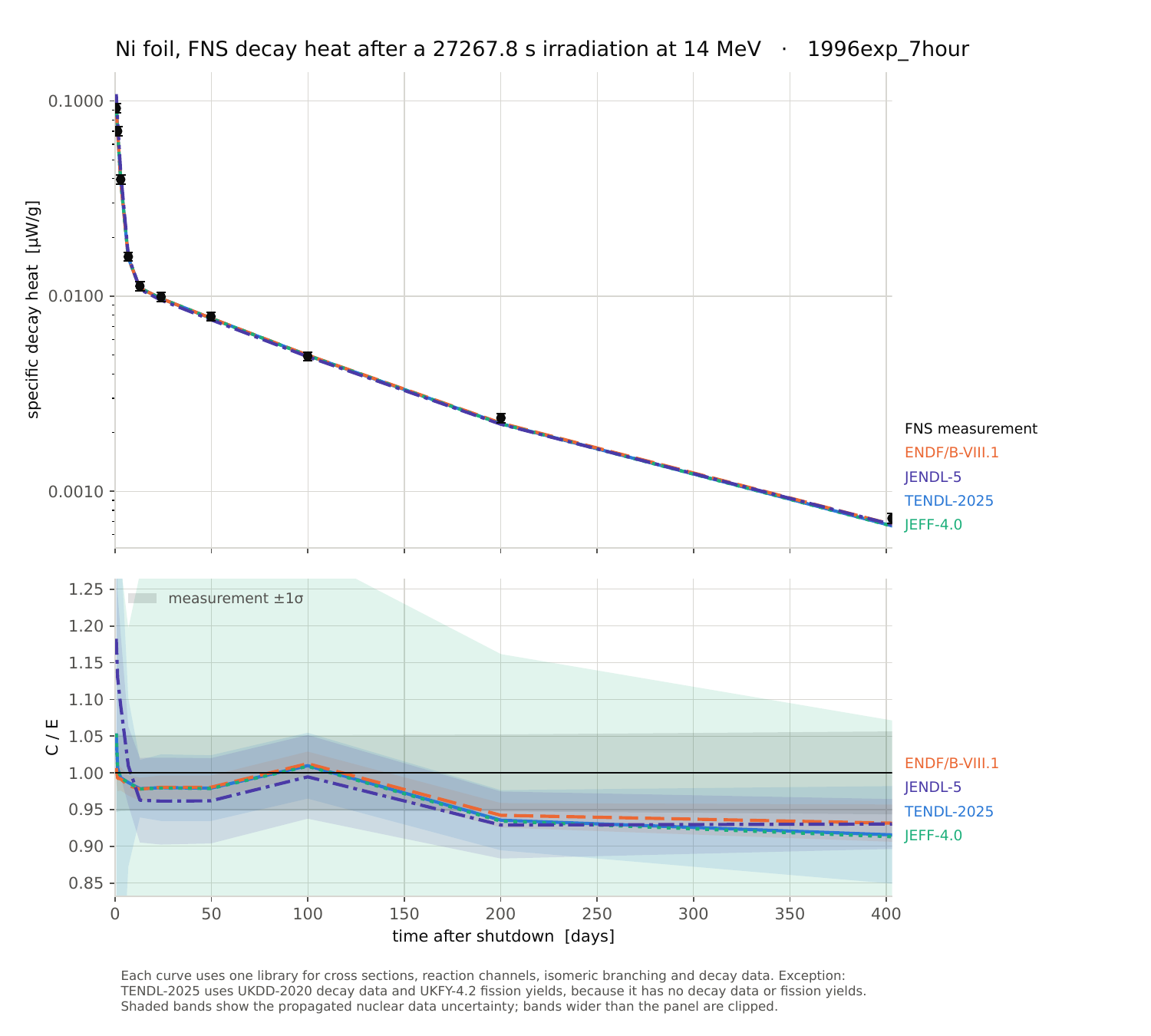}
\caption{Ni: 1996 seven-hour campaign, irradiation position 7. Mean $|C/E-1|$ by library: ENDF/B-VIII.1 2\%, TENDL-2025 3\%, JEFF-4.0 3\%, JENDL-5 7\%. }
\end{figure}
\clearpage

The libraries are between 9 and 40 percent from the measurement, which has a median uncertainty of 7 percent. JENDL-5 is closest and ENDF/B-VIII.1 furthest.

\subsubsection*{Differences in the nuclear data}
\begin{itemize}\setlength{\itemsep}{6pt}
\item ENDF/B-VIII.1 gives no heat from Co62\_m1; the other libraries do.\\[2pt]{\itshape The heat breakdown alone does not show why. Possible causes are a missing reaction channel, a different state assignment, different decay data or a reporting threshold.}
\end{itemize}
\subsubsection*{Suggested measurements}
\begin{itemize}\setlength{\itemsep}{4pt}
\item Natural Ni has 5 isotopes, so the measured heat combines several parents. Enriched single-isotope targets would distinguish their contributions. Enrichment alone does not isolate a single reaction channel.
\end{itemize}
\clearpage
\subsection{Os: Osmium}\label{foil:Os}
\begin{figure}[H]
\centering
\includegraphics[width=\textwidth,height=0.82\textheight,keepaspectratio]{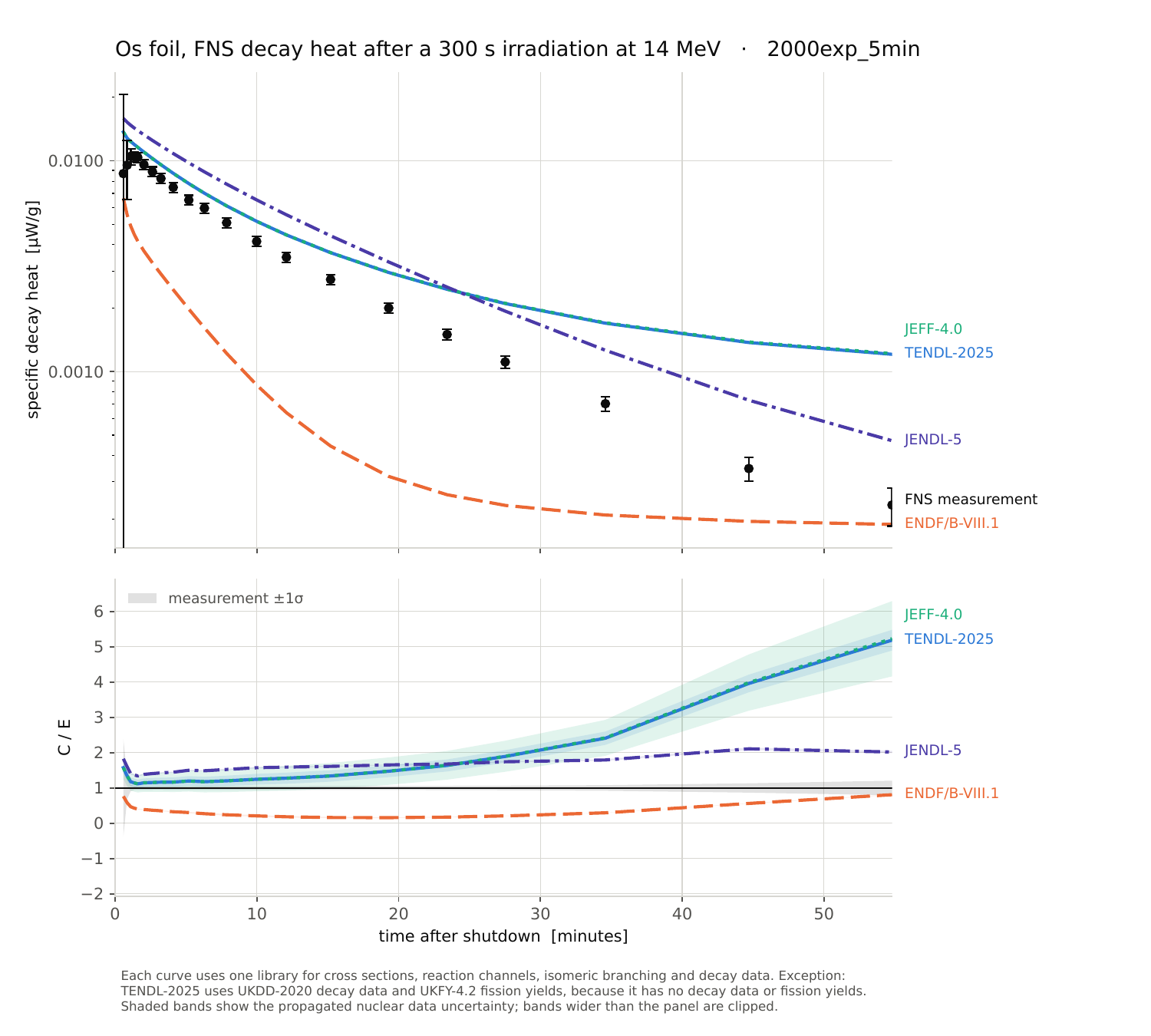}
\caption{Os: decay heat against the measurement, with C/E beneath. Mean $|C/E-1|$ by library: JENDL-5 59\%, ENDF/B-VIII.1 64\%, TENDL-2025 67\%, JEFF-4.0 67\%.}
\end{figure}
\clearpage

\begin{figure}[H]
\centering
\includegraphics[width=\textwidth,height=0.82\textheight,keepaspectratio]{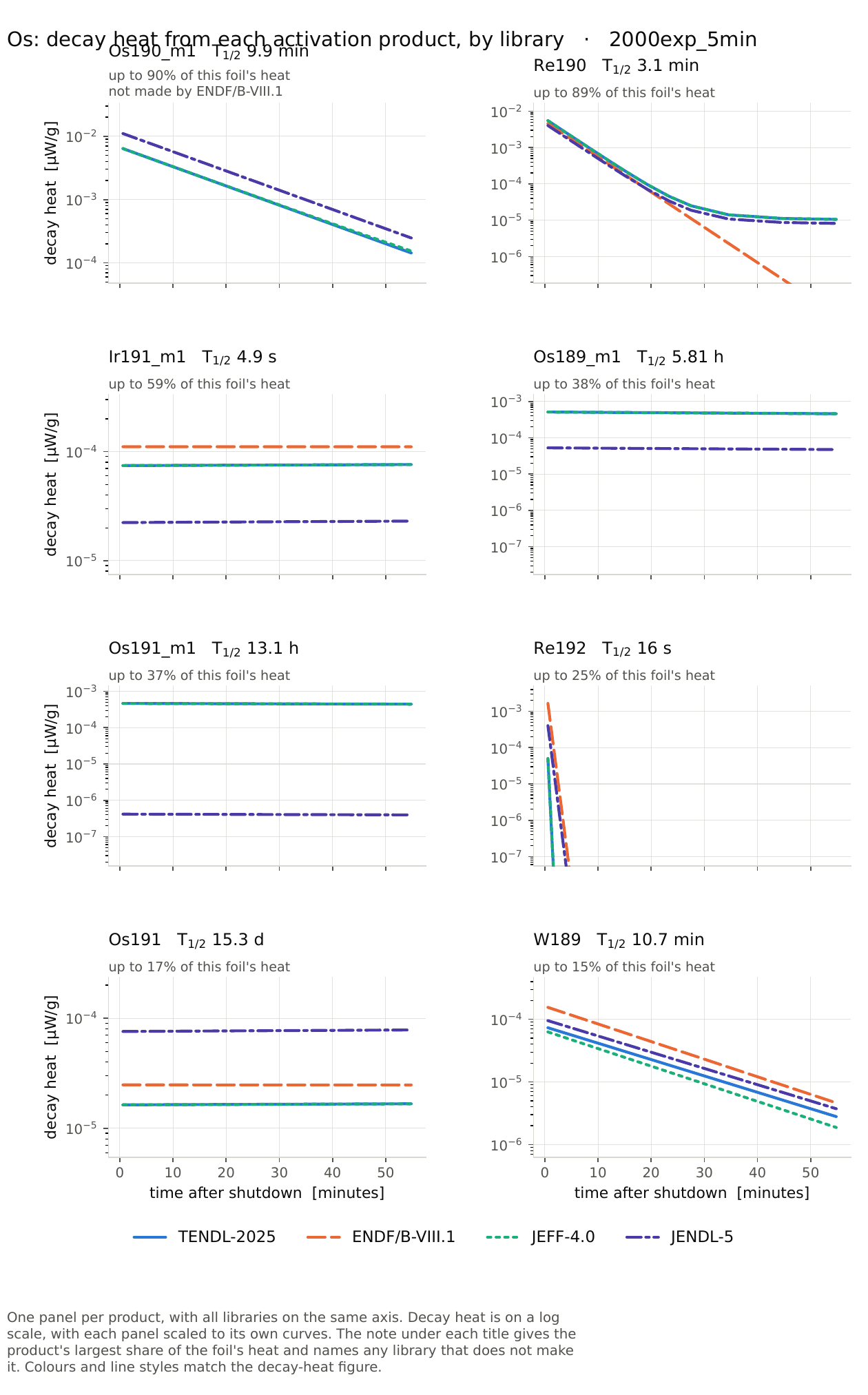}
\caption{Os: share of the decay heat from each activation product during cooling.}
\end{figure}
\clearpage

\begin{figure}[H]
\centering
\includegraphics[width=\textwidth,height=0.82\textheight,keepaspectratio]{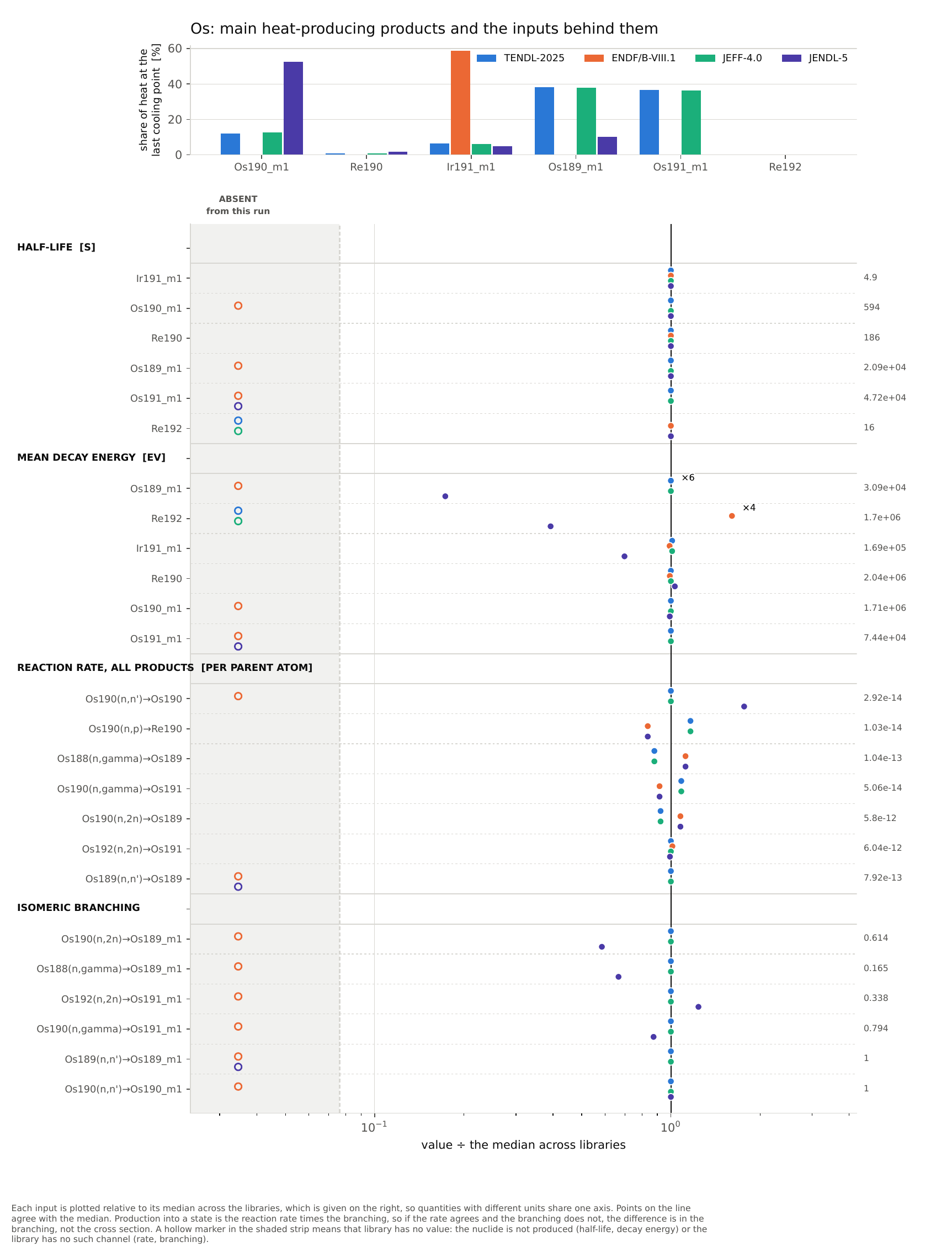}
\caption{Os: inputs for the main heat-producing products, each relative to its median across the available libraries.}
\end{figure}
\clearpage

\begin{figure}[H]
\centering
\includegraphics[width=\textwidth,height=0.82\textheight,keepaspectratio]{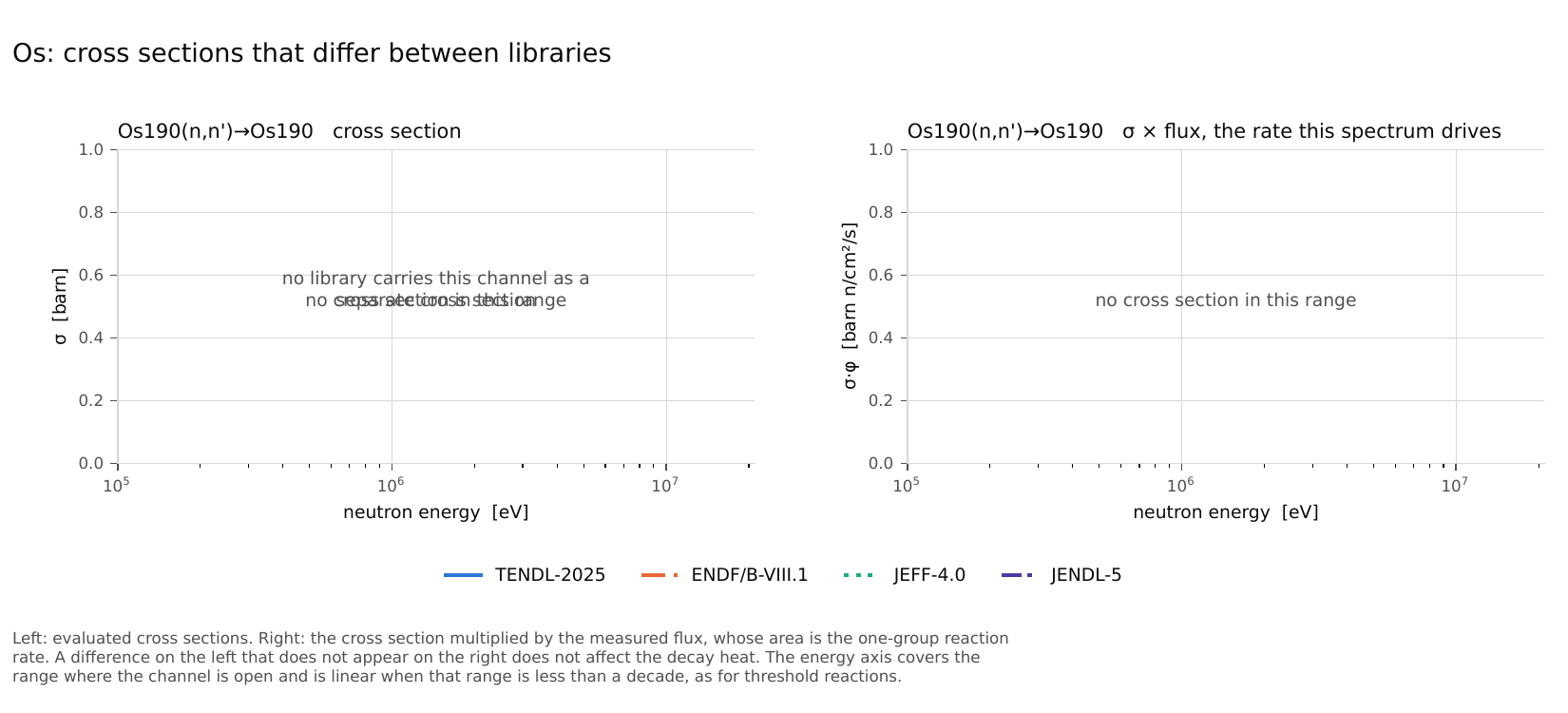}
\caption{Os: cross sections of the channels whose spectrum-averaged rates differ, with the neutron spectrum and their product.}
\end{figure}
\clearpage

\begin{figure}[H]
\centering
\includegraphics[width=\textwidth,height=0.82\textheight,keepaspectratio]{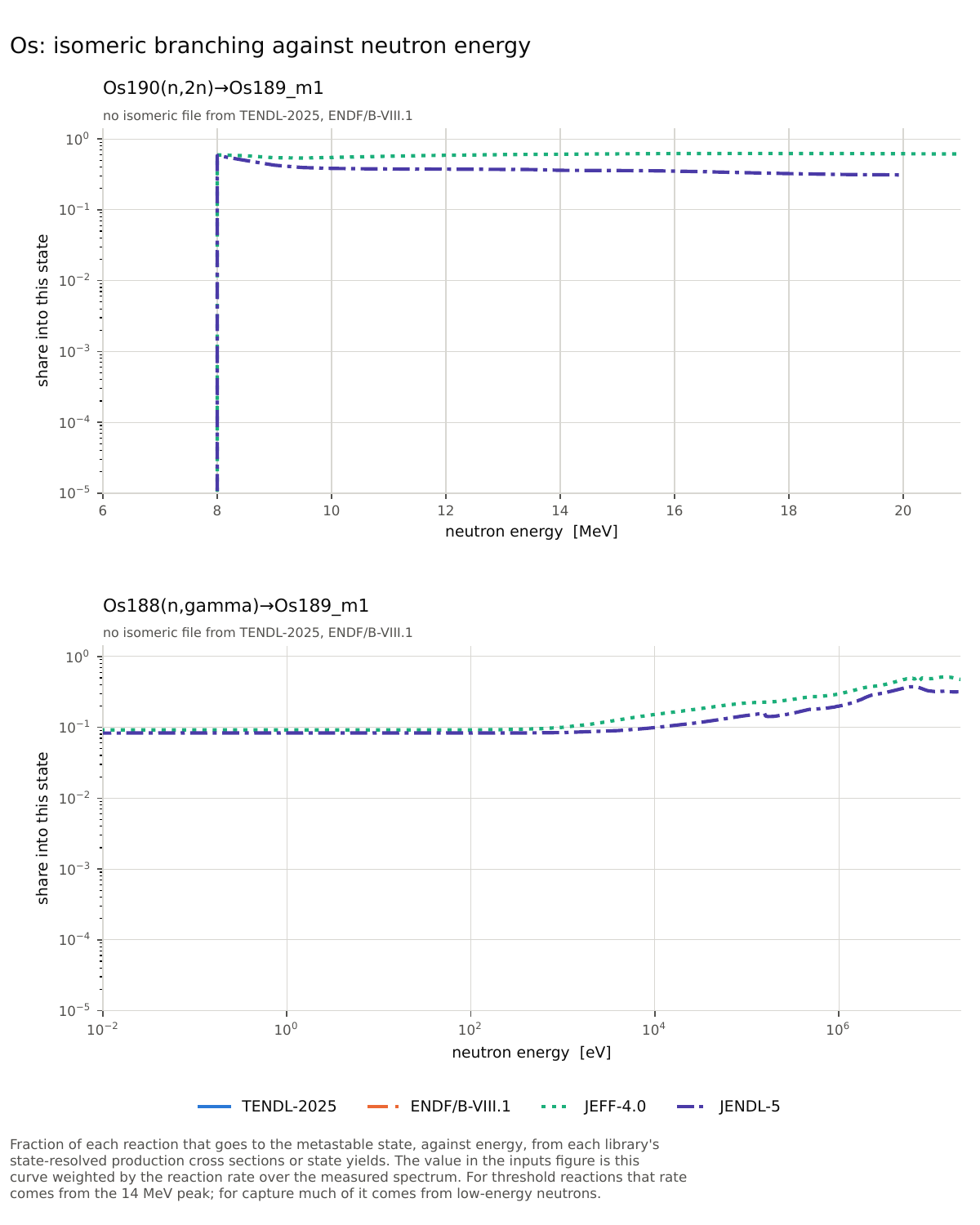}
\caption{Os: isomeric branching ratio against neutron energy for the channels whose spectrum-averaged ratio differs between libraries.}
\end{figure}
\clearpage

The libraries are between 59 and 67 percent from the measurement, which has a median uncertainty of 6 percent. JENDL-5 is closest and JEFF-4.0 furthest.

\subsubsection*{Differences in the nuclear data}
\begin{itemize}\setlength{\itemsep}{6pt}
\item JENDL-5 gives the mean decay energy of Os189\_m1 as 5351 eV, against 3.089e+04 eV for the median of the others, a factor of 6.\\[2pt]{\itshape For a given activity the heat is proportional to the mean decay energy, so this difference goes straight into the heat. Compare the decay schemes and how the emitted energy is counted.}
\item Only ENDF/B-VIII.1 and JENDL-5 have the mean decay energy of Re192, and they differ by a factor of 4: 2.74e+06 eV against 6.694e+05 eV.
\item ENDF/B-VIII.1 gives no heat from Os190\_m1; the other libraries do.\\[2pt]{\itshape The heat breakdown alone does not show why. Possible causes are a missing reaction channel, a different state assignment, different decay data or a reporting threshold.}
\end{itemize}
\subsubsection*{Suggested measurements}
\begin{itemize}\setlength{\itemsep}{4pt}
\item Natural Os has 7 isotopes, so the measured heat combines several parents. Enriched single-isotope targets would distinguish their contributions. Enrichment alone does not isolate a single reaction channel.
\item This foil makes Os191 (half-life 15.3 days), which gives up to 17 percent of the calculated heat. A five-minute irradiation and 55 minute measurement only see the start of this contribution. There is no seven-hour FNS measurement of this foil. A longer irradiation followed for days to months would be needed to test this product's production and decay.
\end{itemize}
\clearpage
\subsection{P: Phosphorus}\label{foil:P}
\begin{figure}[H]
\centering
\includegraphics[width=\textwidth,height=0.82\textheight,keepaspectratio]{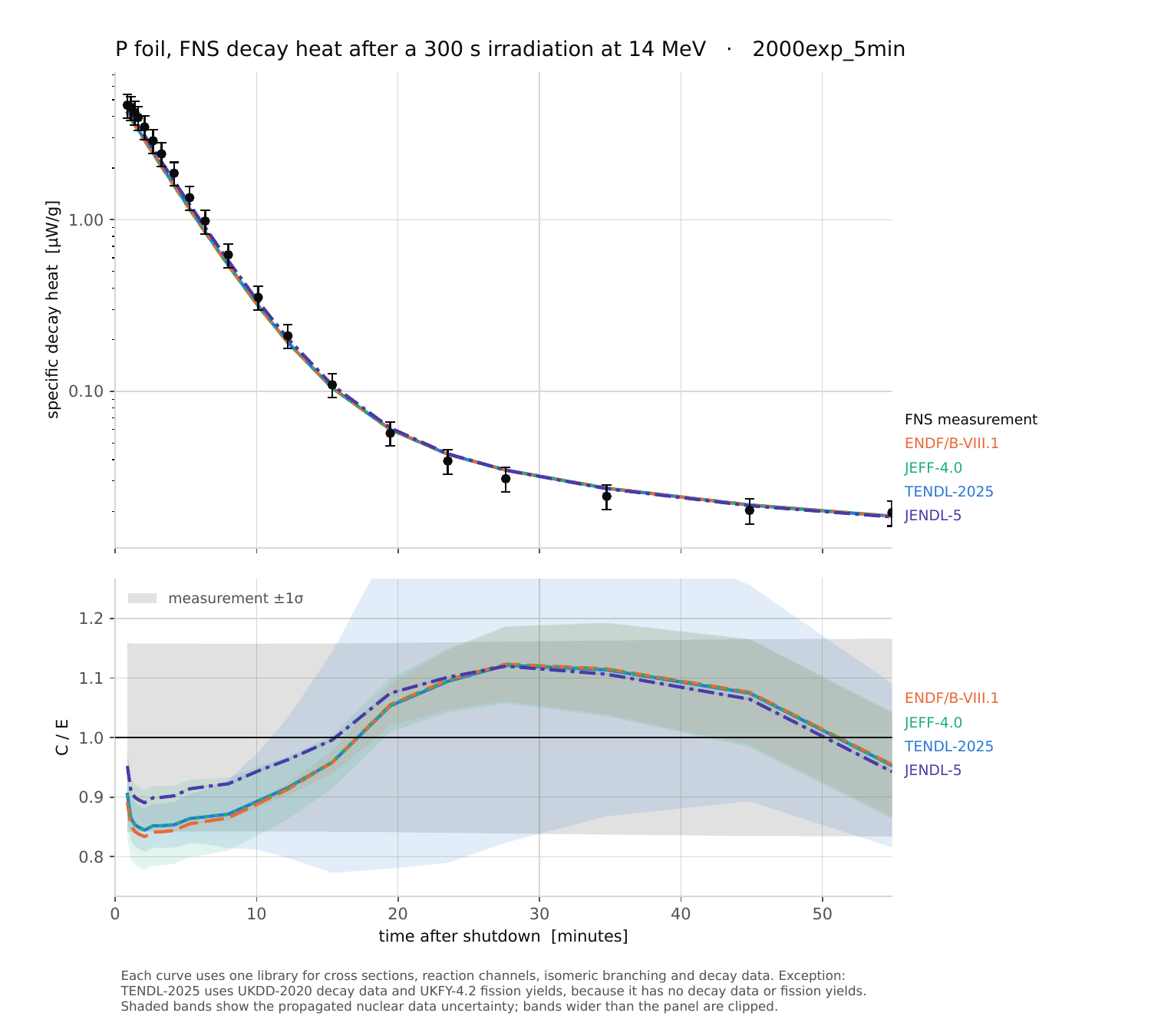}
\caption{P: decay heat against the measurement, with C/E beneath. Mean $|C/E-1|$ by library: JENDL-5 8\%, TENDL-2025 11\%, JEFF-4.0 11\%, ENDF/B-VIII.1 12\%.}
\end{figure}
\clearpage

\begin{figure}[H]
\centering
\includegraphics[width=\textwidth,height=0.82\textheight,keepaspectratio]{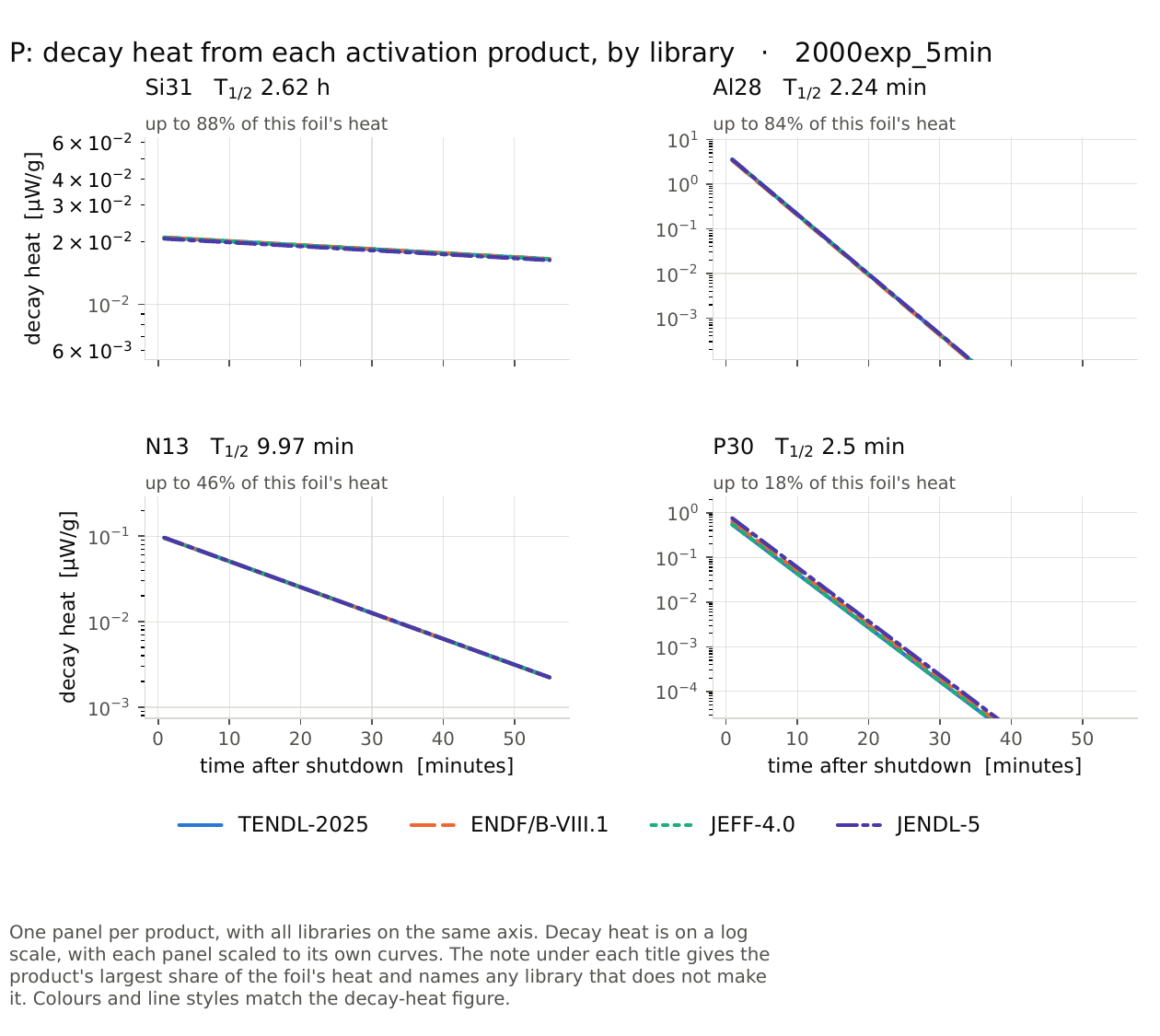}
\caption{P: share of the decay heat from each activation product during cooling.}
\end{figure}
\clearpage

\begin{figure}[H]
\centering
\includegraphics[width=\textwidth,height=0.82\textheight,keepaspectratio]{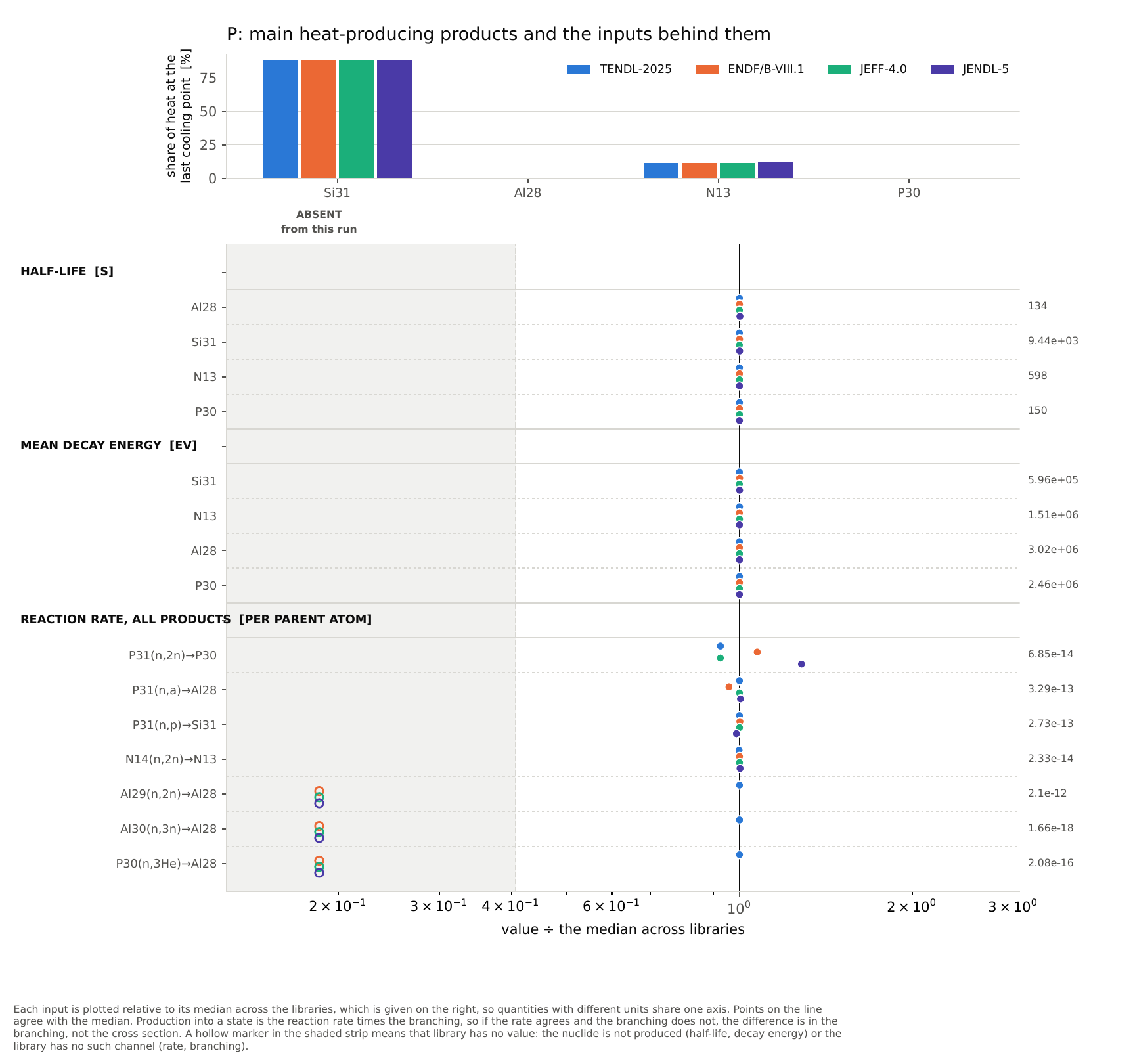}
\caption{P: inputs for the main heat-producing products, each relative to its median across the available libraries.}
\end{figure}
\clearpage

Mean absolute C/E deviations range from 8 to 12 percent. The spread of 4 percentage points is smaller than the median measurement uncertainty of 16 percent, so the measurement cannot separate the libraries.

\subsubsection*{Suggested measurements}
\begin{itemize}\setlength{\itemsep}{4pt}
\item No input differs enough between the libraries to be listed. An error shared by all of them, or a smaller difference, is still possible.
\end{itemize}
\clearpage
\subsection{Pb: Lead}\label{foil:Pb}
\begin{figure}[H]
\centering
\includegraphics[width=\textwidth,height=0.82\textheight,keepaspectratio]{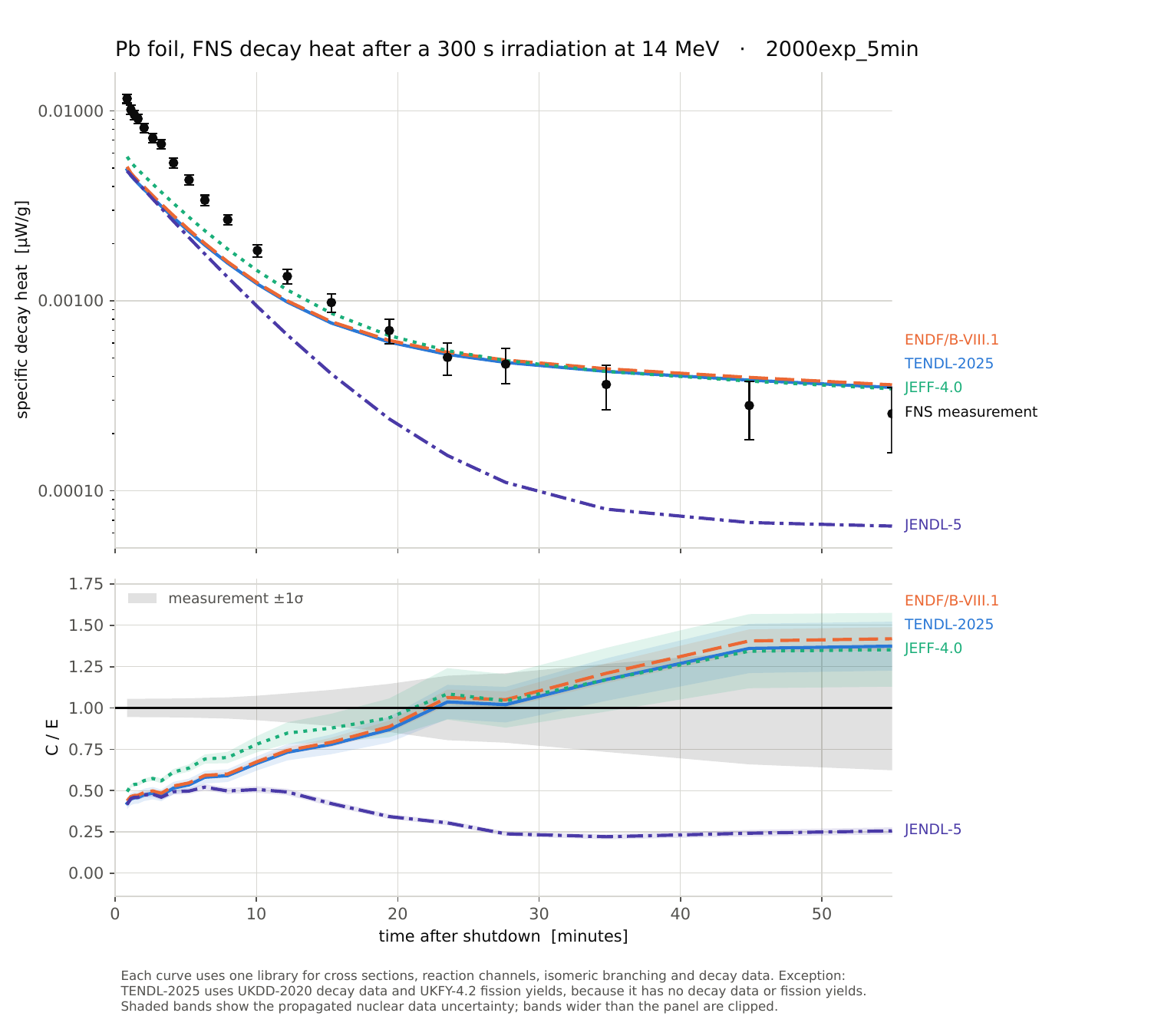}
\caption{Pb: decay heat against the measurement, with C/E beneath. Mean $|C/E-1|$ by library: JEFF-4.0 31\%, ENDF/B-VIII.1 37\%, TENDL-2025 37\%, JENDL-5 59\%.}
\end{figure}
\clearpage

\begin{figure}[H]
\centering
\includegraphics[width=\textwidth,height=0.82\textheight,keepaspectratio]{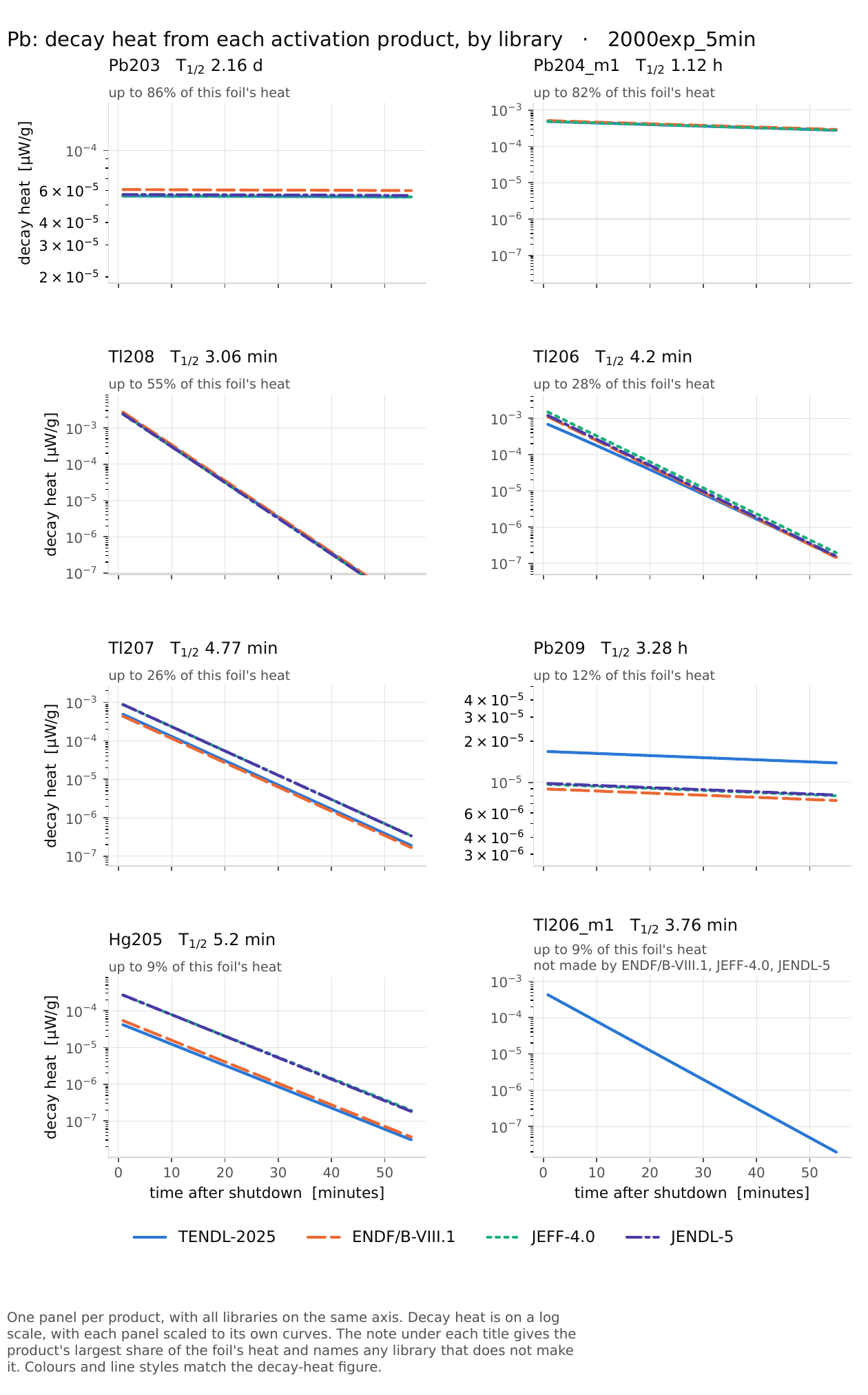}
\caption{Pb: share of the decay heat from each activation product during cooling.}
\end{figure}
\clearpage

\begin{figure}[H]
\centering
\includegraphics[width=\textwidth,height=0.82\textheight,keepaspectratio]{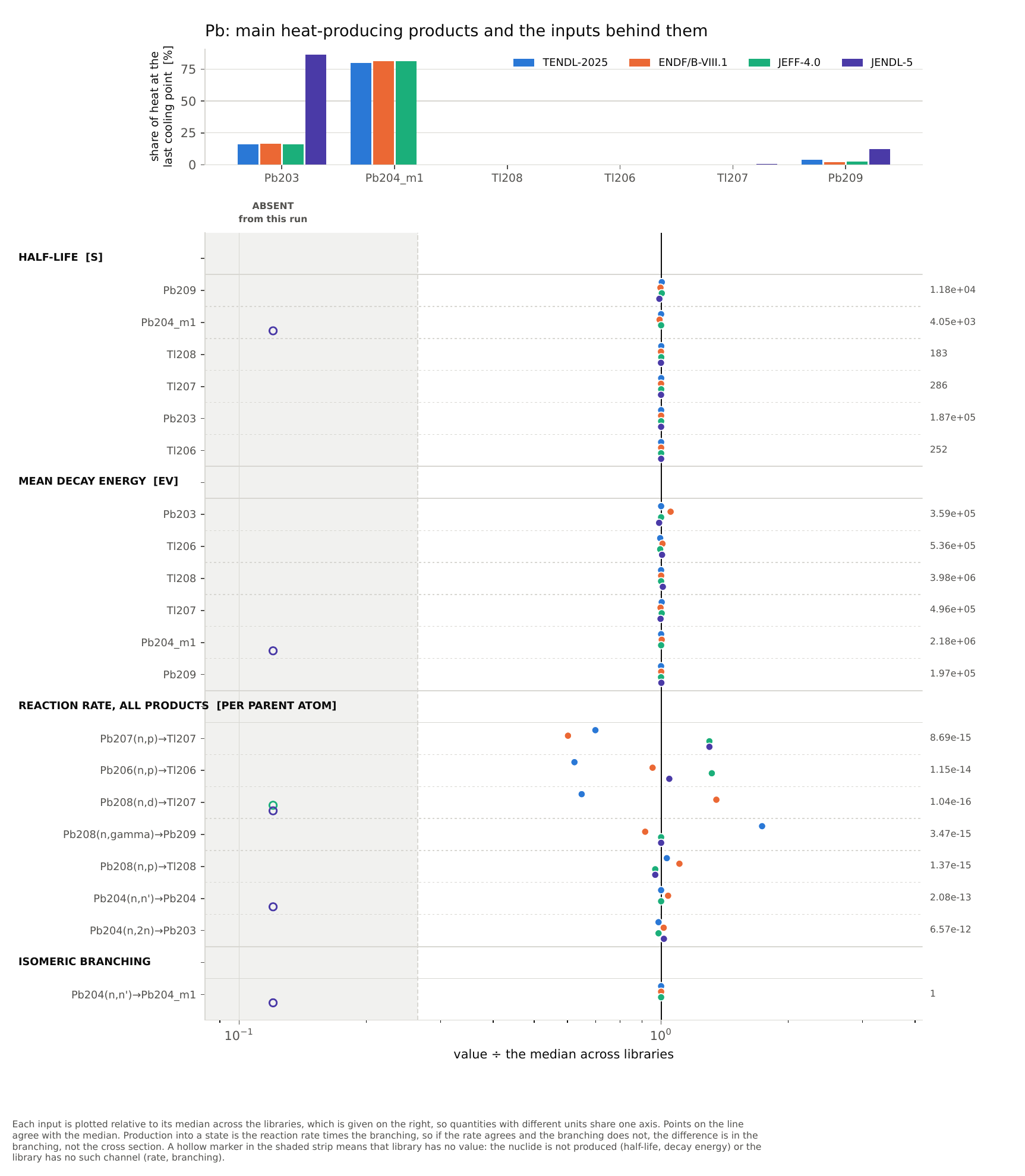}
\caption{Pb: inputs for the main heat-producing products, each relative to its median across the available libraries.}
\end{figure}
\clearpage

\begin{figure}[H]
\centering
\includegraphics[width=\textwidth,height=0.82\textheight,keepaspectratio]{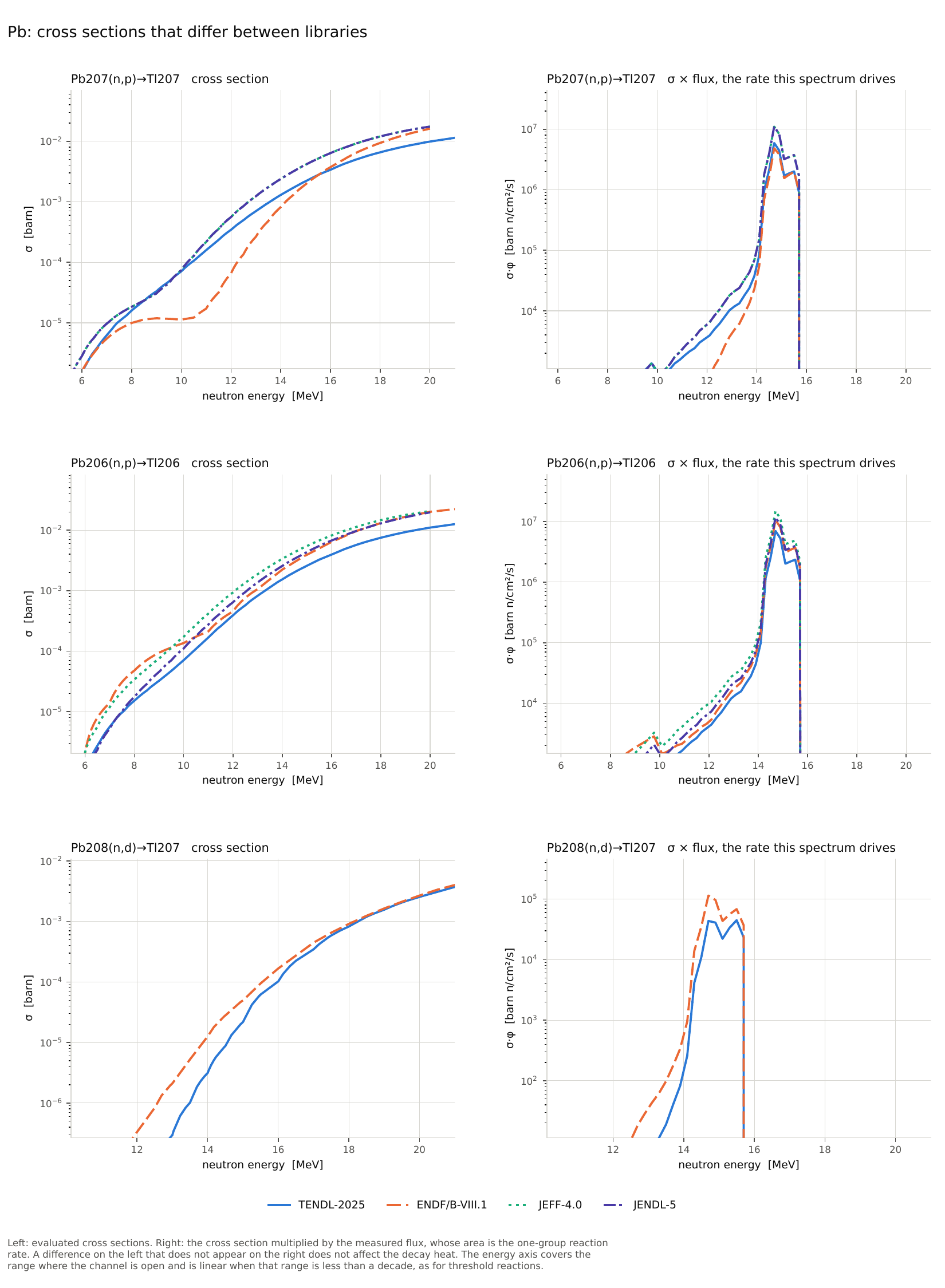}
\caption{Pb: cross sections of the channels whose spectrum-averaged rates differ, with the neutron spectrum and their product.}
\end{figure}
\clearpage

\subsubsection*{1996 seven-hour campaign}
The 1996 campaign irradiated this foil for 7.57 hours at position 7 and measured it out to 49.73 days. In the five-minute calculation Pb203 (half-life 2.16 days) gives up to 86 percent of the heat, and the longer irradiation and cooling test its production and decay data. ENDF/B-VIII.1 is closest. This campaign uses a different spectrum, so it is not included in the main rankings.

\begin{figure}[H]
\centering
\includegraphics[width=\textwidth,height=0.82\textheight,keepaspectratio]{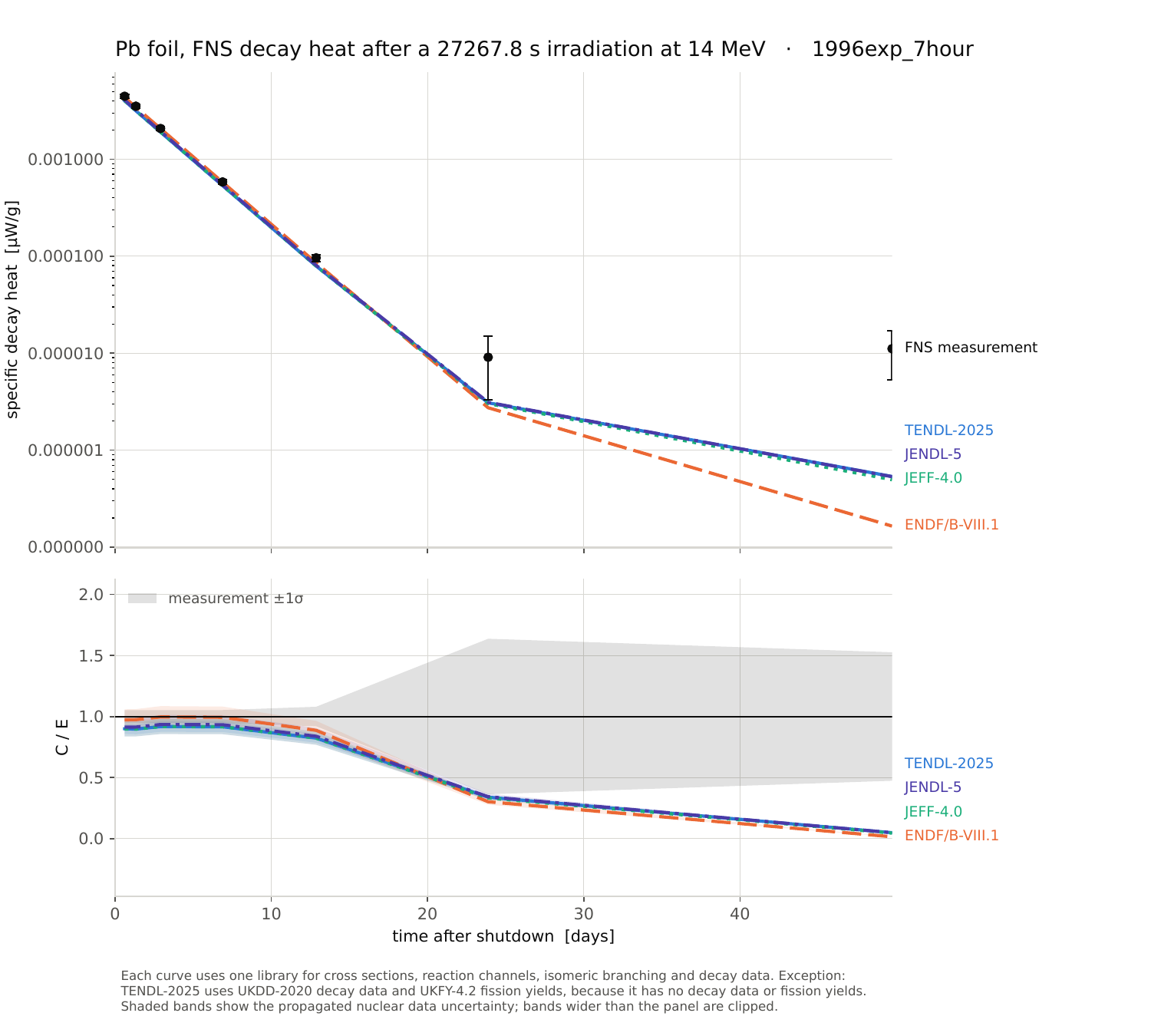}
\caption{Pb: 1996 seven-hour campaign, irradiation position 7. Mean $|C/E-1|$ by library: ENDF/B-VIII.1 27\%, JENDL-5 30\%, TENDL-2025 31\%, JEFF-4.0 31\%. }
\end{figure}
\clearpage

The libraries are between 31 and 59 percent from the measurement, which has a median uncertainty of 6 percent. JEFF-4.0 is closest and JENDL-5 furthest.

\subsubsection*{Differences in the nuclear data}
\begin{itemize}\setlength{\itemsep}{6pt}
\item ENDF/B-VIII.1 gives the total reaction rate of Pb207(n,p)→Tl207 as 5.228e-15, against 1.131e-14 for the median of the others, a factor of 2.\\[2pt]{\itshape This is the total rate of the channel over all its products, so the difference is in the cross section, not in the split between states.}
\item Only ENDF/B-VIII.1 and TENDL-2025 have the total reaction rate of Pb208(n,d)→Tl207, and they differ by a factor of 2: 1.404e-16 against 6.73e-17.
\end{itemize}
\subsubsection*{Suggested measurements}
\begin{itemize}\setlength{\itemsep}{4pt}
\item Natural Pb has 4 isotopes, so the measured heat combines several parents. Enriched single-isotope targets would distinguish their contributions. Candidate parents are Pb207, Pb208. Enrichment alone does not isolate a single reaction channel.
\item The reaction rates are averaged over the measured spectrum. Activation measurements at several well-defined neutron energies, including below 14 MeV, would show which energy range the differences come from.
\end{itemize}
\clearpage
\subsection{Pd: Palladium}\label{foil:Pd}
\begin{figure}[H]
\centering
\includegraphics[width=\textwidth,height=0.82\textheight,keepaspectratio]{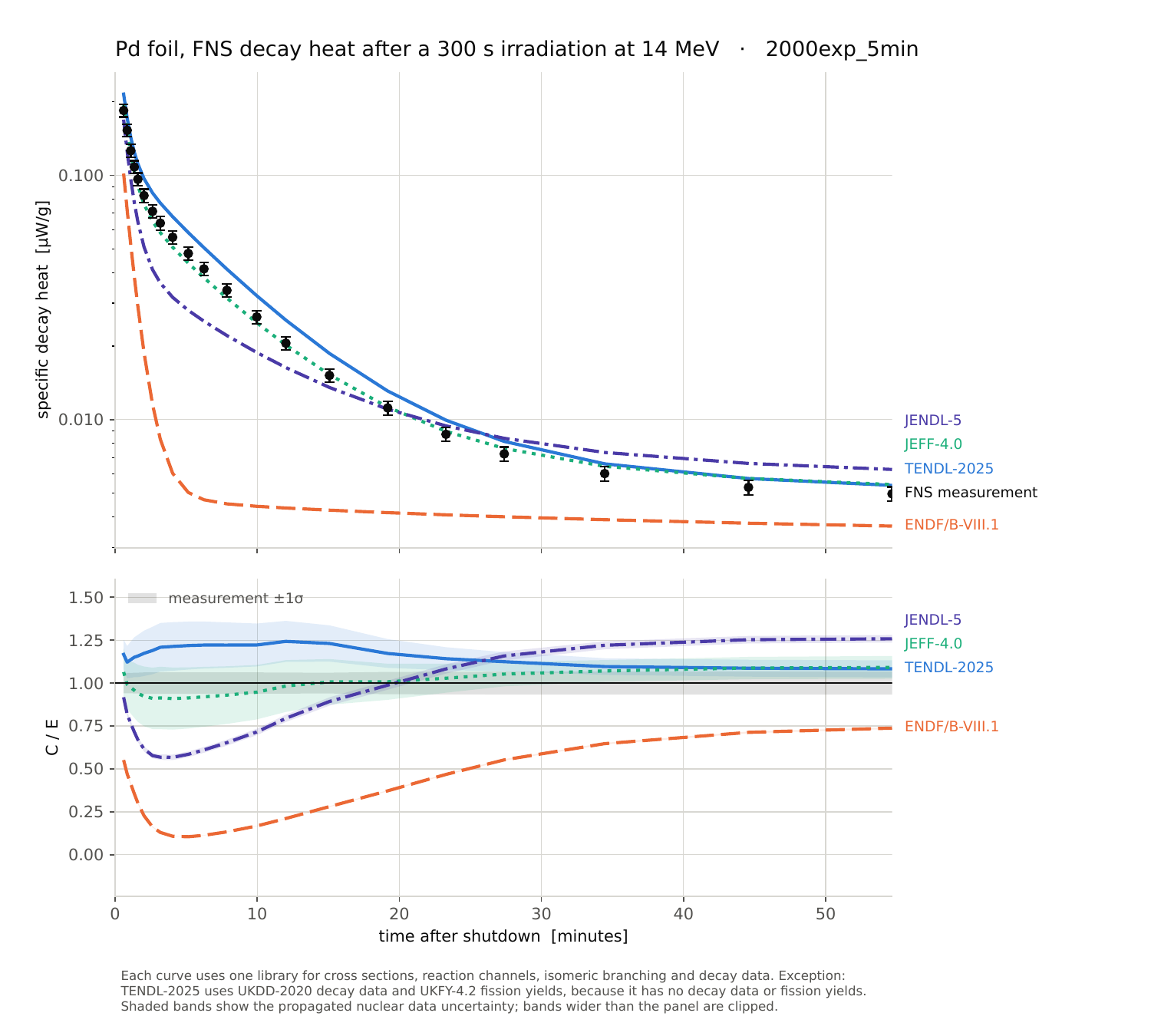}
\caption{Pd: decay heat against the measurement, with C/E beneath. Mean $|C/E-1|$ by library: JEFF-4.0 6\%, TENDL-2025 17\%, JENDL-5 26\%, ENDF/B-VIII.1 66\%.}
\end{figure}
\clearpage

\begin{figure}[H]
\centering
\includegraphics[width=\textwidth,height=0.82\textheight,keepaspectratio]{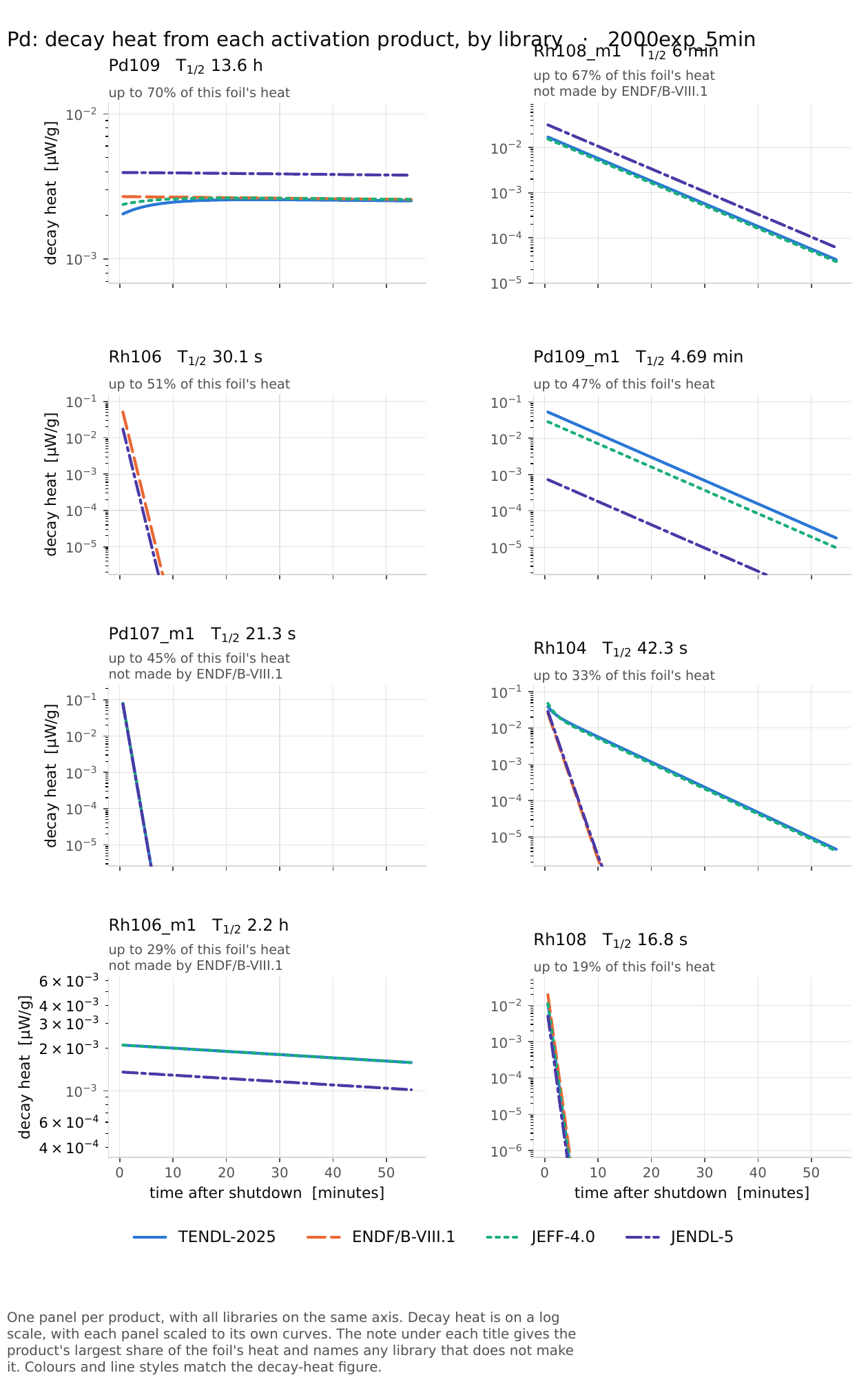}
\caption{Pd: share of the decay heat from each activation product during cooling.}
\end{figure}
\clearpage

\begin{figure}[H]
\centering
\includegraphics[width=\textwidth,height=0.82\textheight,keepaspectratio]{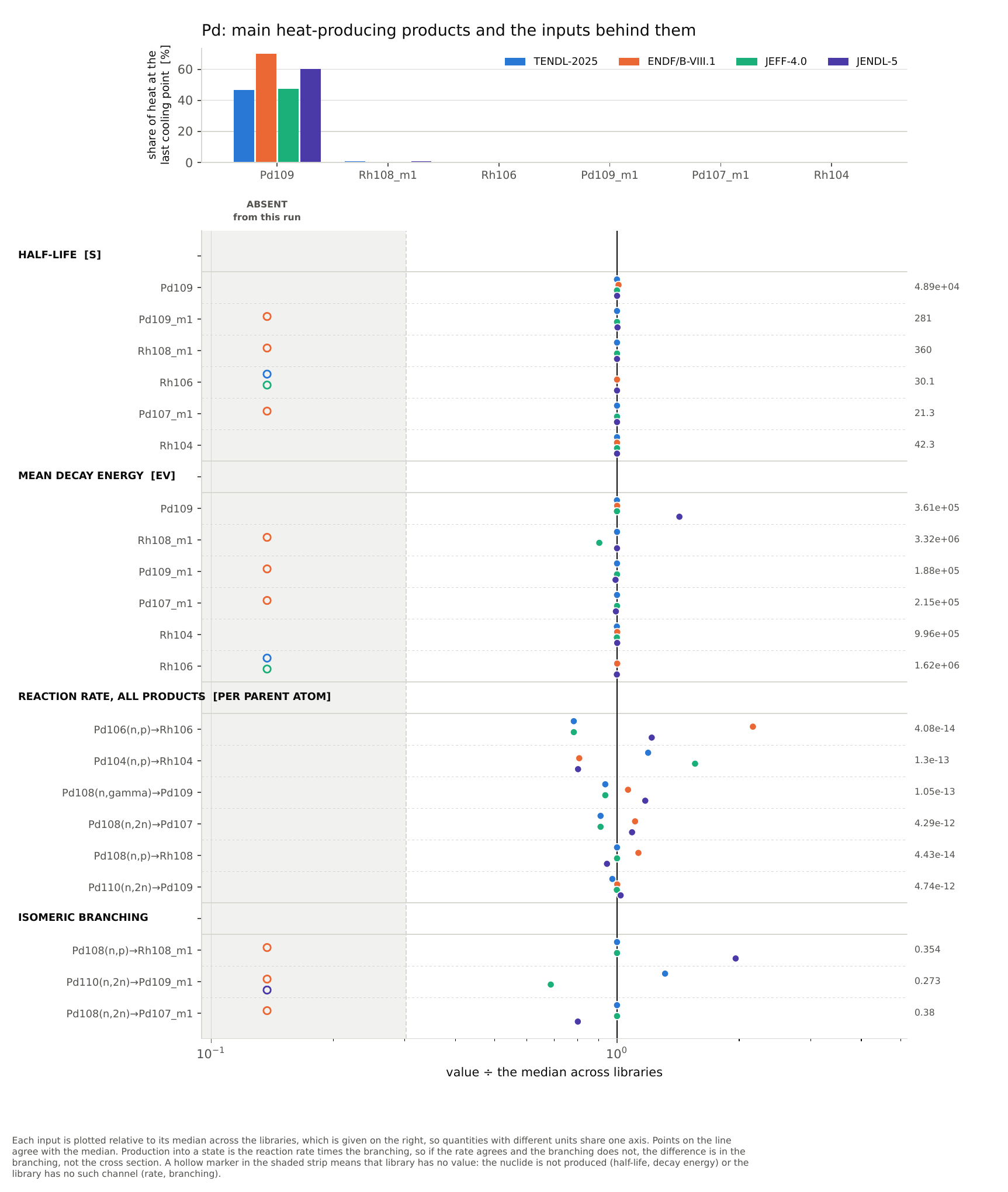}
\caption{Pd: inputs for the main heat-producing products, each relative to its median across the available libraries.}
\end{figure}
\clearpage

\begin{figure}[H]
\centering
\includegraphics[width=\textwidth,height=0.82\textheight,keepaspectratio]{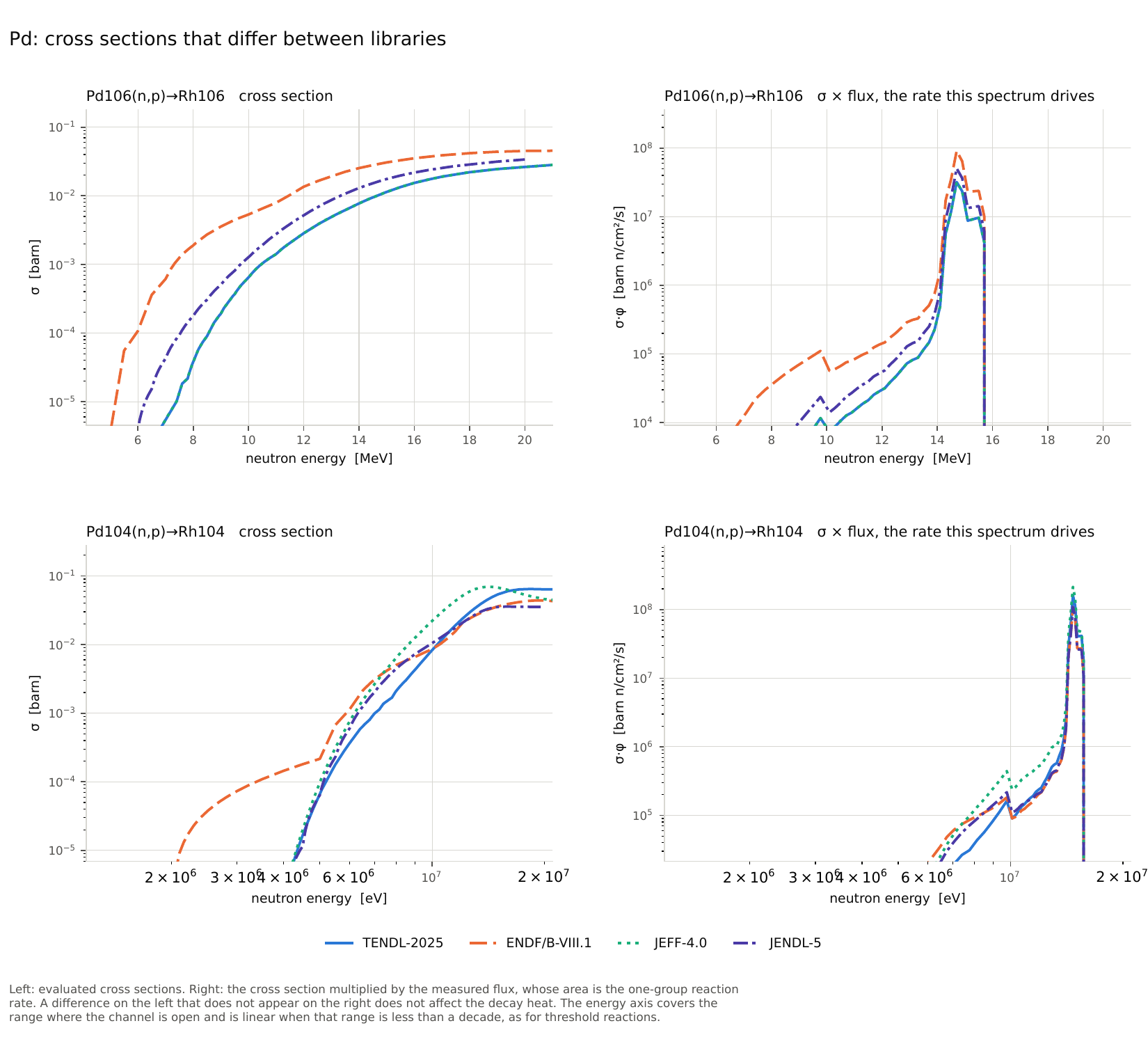}
\caption{Pd: cross sections of the channels whose spectrum-averaged rates differ, with the neutron spectrum and their product.}
\end{figure}
\clearpage

\begin{figure}[H]
\centering
\includegraphics[width=\textwidth,height=0.82\textheight,keepaspectratio]{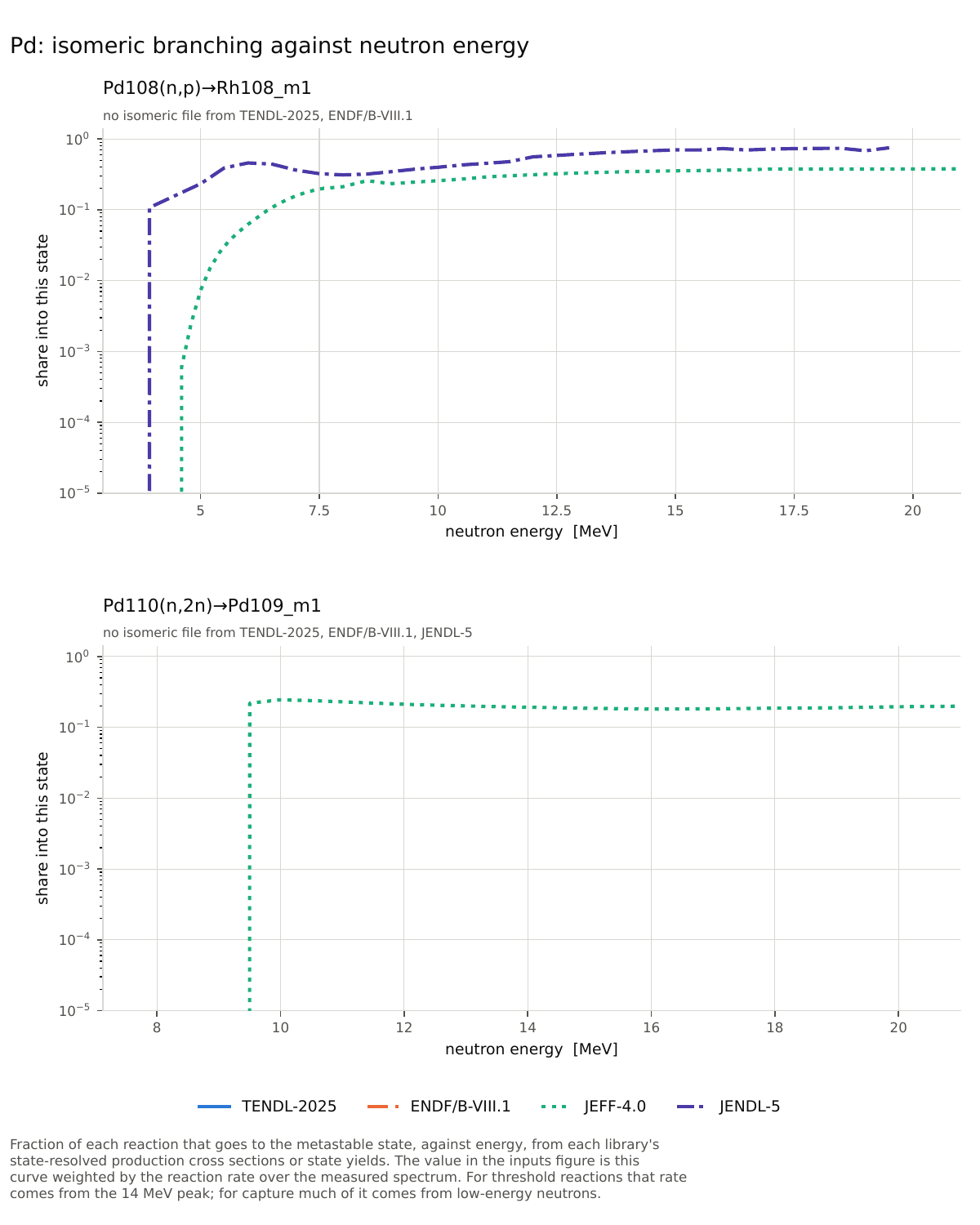}
\caption{Pd: isomeric branching ratio against neutron energy for the channels whose spectrum-averaged ratio differs between libraries.}
\end{figure}
\clearpage

The libraries are between 6 and 66 percent from the measurement, which has a median uncertainty of 6 percent. JEFF-4.0 is closest and ENDF/B-VIII.1 furthest.

\subsubsection*{Differences in the nuclear data}
\begin{itemize}\setlength{\itemsep}{6pt}
\item ENDF/B-VIII.1 gives the total reaction rate of Pd106(n,p)→Rh106 as 8.826e-14, against 3.195e-14 for the median of the others, a factor of 3.\\[2pt]{\itshape This is the total rate of the channel over all its products, so the difference is in the cross section, not in the split between states.}
\item ENDF/B-VIII.1 gives no heat from Rh108\_m1; the other libraries do.\\[2pt]{\itshape The heat breakdown alone does not show why. Possible causes are a missing reaction channel, a different state assignment, different decay data or a reporting threshold.}
\item ENDF/B-VIII.1 gives no heat from Pd107\_m1; the other libraries do.
\end{itemize}
\subsubsection*{Suggested measurements}
\begin{itemize}\setlength{\itemsep}{4pt}
\item Natural Pd has 6 isotopes, so the measured heat combines several parents. Enriched single-isotope targets would distinguish their contributions. The candidate parent is Pd106. Enrichment alone does not isolate a single reaction channel.
\item The reaction rates are averaged over the measured spectrum. Activation measurements at several well-defined neutron energies, including below 14 MeV, would show which energy range the differences come from.
\end{itemize}
\clearpage
\subsection{Pr: Praseodymium}\label{foil:Pr}
\begin{figure}[H]
\centering
\includegraphics[width=\textwidth,height=0.82\textheight,keepaspectratio]{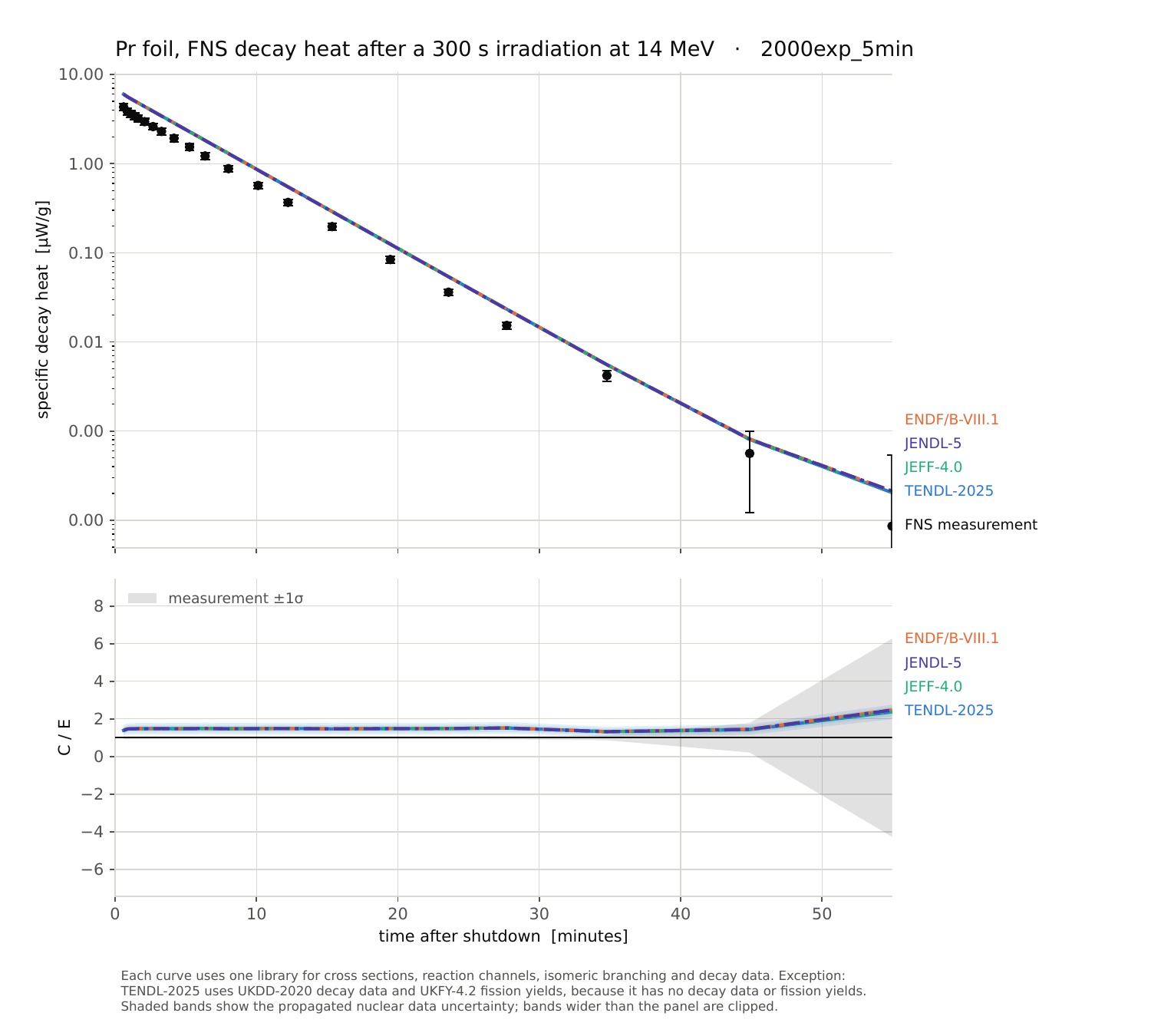}
\caption{Pr: decay heat against the measurement, with C/E beneath. Mean $|C/E-1|$ by library: TENDL-2025 51\%, JEFF-4.0 51\%, JENDL-5 52\%, ENDF/B-VIII.1 52\%.}
\end{figure}
\clearpage

\begin{figure}[H]
\centering
\includegraphics[width=\textwidth,height=0.82\textheight,keepaspectratio]{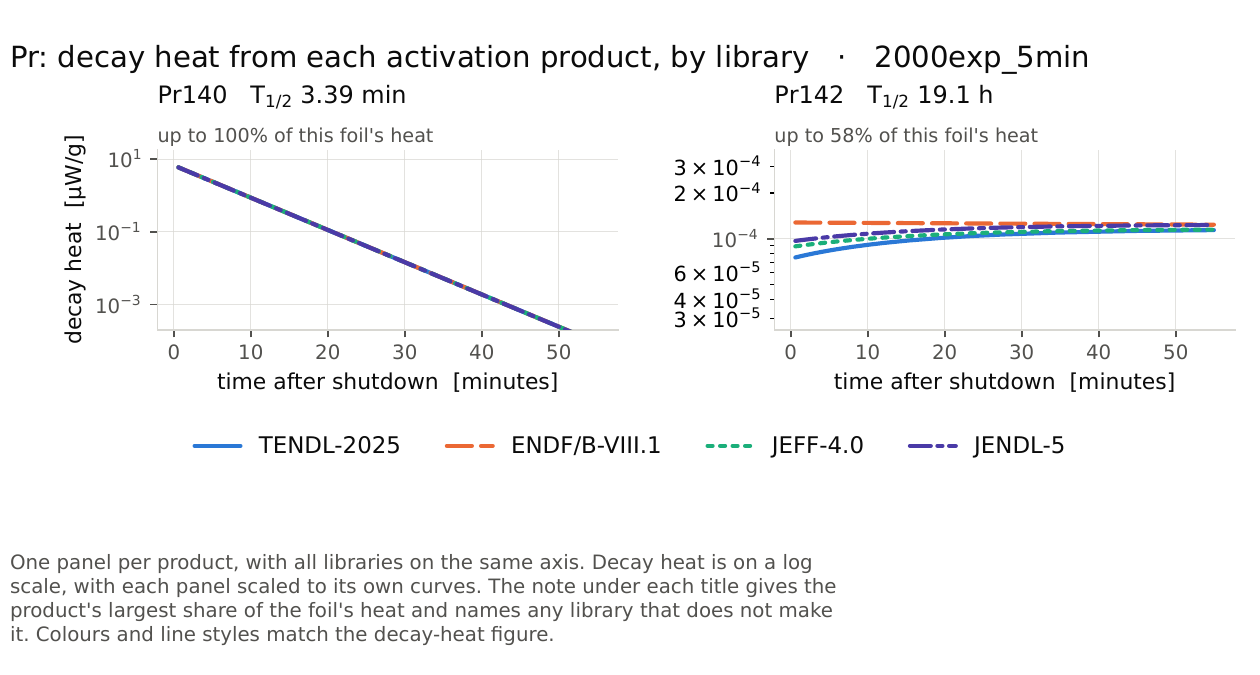}
\caption{Pr: share of the decay heat from each activation product during cooling.}
\end{figure}
\clearpage

\begin{figure}[H]
\centering
\includegraphics[width=\textwidth,height=0.82\textheight,keepaspectratio]{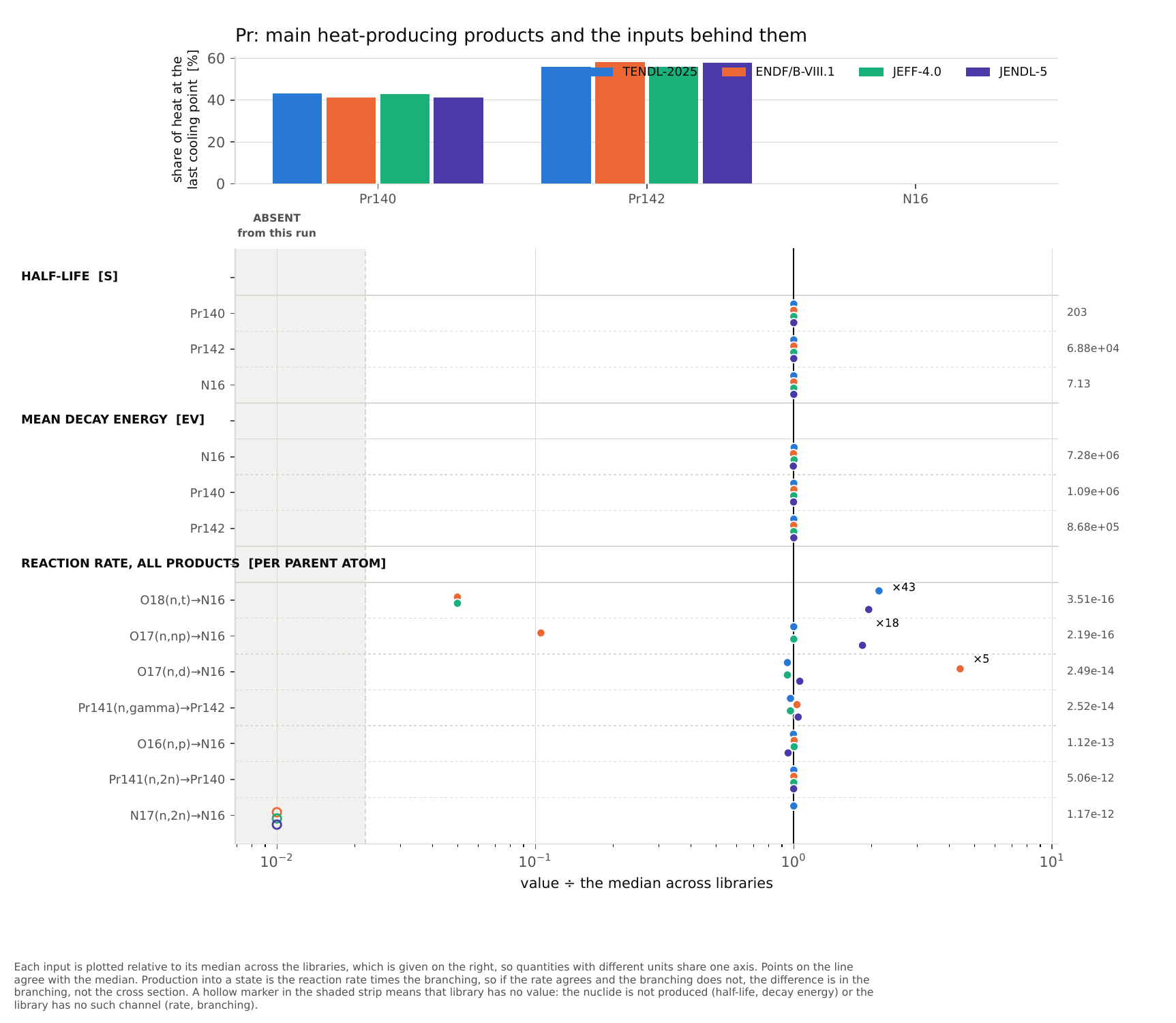}
\caption{Pr: inputs for the main heat-producing products, each relative to its median across the available libraries.}
\end{figure}
\clearpage

\begin{figure}[H]
\centering
\includegraphics[width=\textwidth,height=0.82\textheight,keepaspectratio]{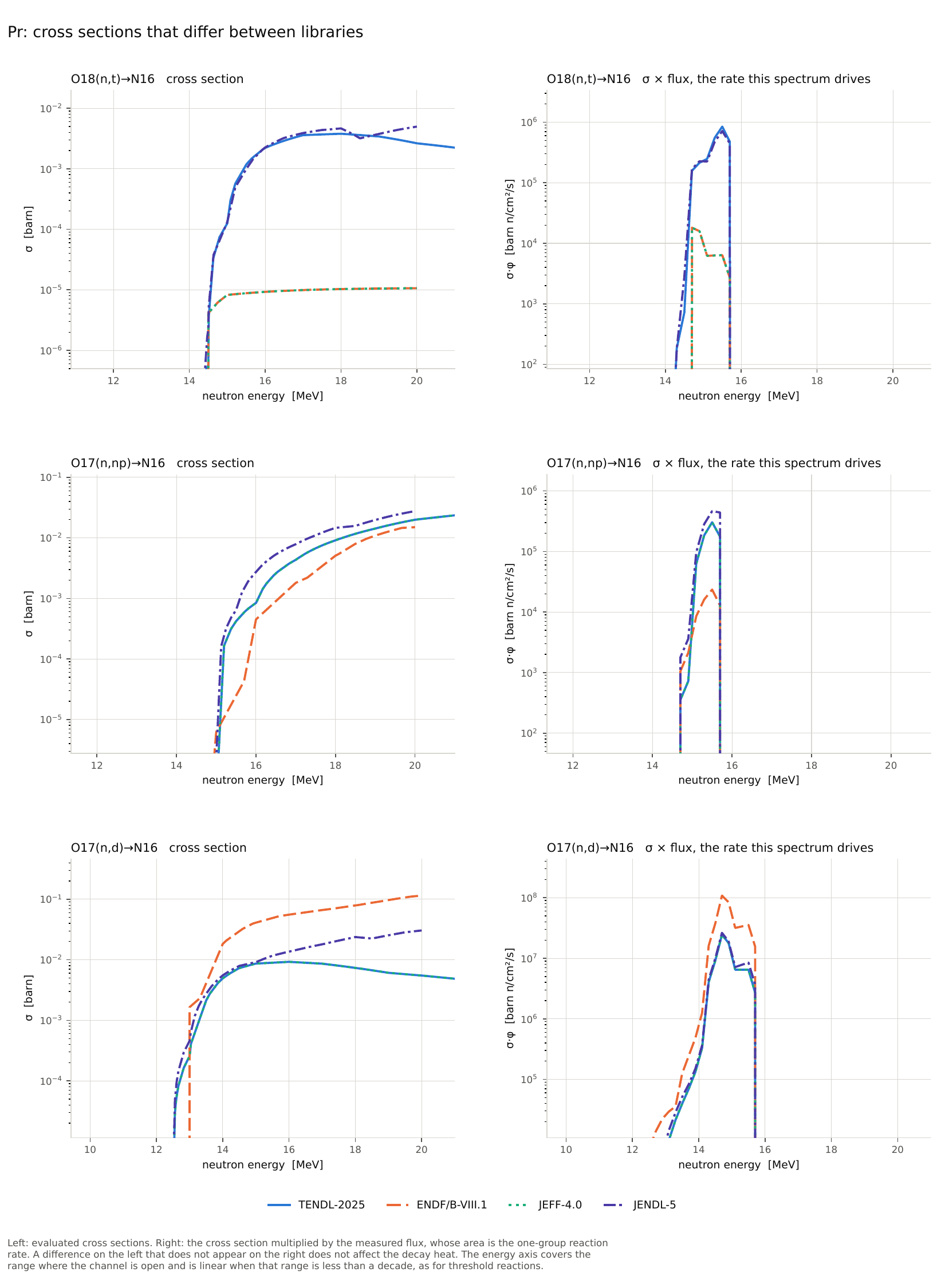}
\caption{Pr: cross sections of the channels whose spectrum-averaged rates differ, with the neutron spectrum and their product.}
\end{figure}
\clearpage

Mean absolute C/E deviations range from 51 to 52 percent. The spread of 1 percentage points is smaller than the median measurement uncertainty of 8 percent, so the measurement cannot separate the libraries.

\subsubsection*{Differences in the nuclear data}
\begin{itemize}\setlength{\itemsep}{6pt}
\item ENDF/B-VIII.1 gives the total reaction rate of O18(n,t)→N16 as 1.751e-17, against 6.84e-16 for the median of the others, a factor of 39.\\[2pt]{\itshape This is the total rate of the channel over all its products, so the difference is in the cross section, not in the split between states.}
\item ENDF/B-VIII.1 gives the total reaction rate of O17(n,np)→N16 as 2.295e-17, against 2.186e-16 for the median of the others, a factor of 10.
\item ENDF/B-VIII.1 gives the total reaction rate of O17(n,d)→N16 as 1.096e-13, against 2.351e-14 for the median of the others, a factor of 5.
\end{itemize}
\subsubsection*{Suggested measurements}
\begin{itemize}\setlength{\itemsep}{4pt}
\item Pr has a single stable isotope, so the foil is already a single-isotope target and enrichment would not help.
\item The sample is a compound that also contains O. Their activation adds to the measured heat; a metal or purer sample would remove it.
\item The reaction rates are averaged over the measured spectrum. Activation measurements at several well-defined neutron energies, including below 14 MeV, would show which energy range the differences come from.
\end{itemize}
\clearpage
\subsection{Pt: Platinum}\label{foil:Pt}
\begin{figure}[H]
\centering
\includegraphics[width=\textwidth,height=0.82\textheight,keepaspectratio]{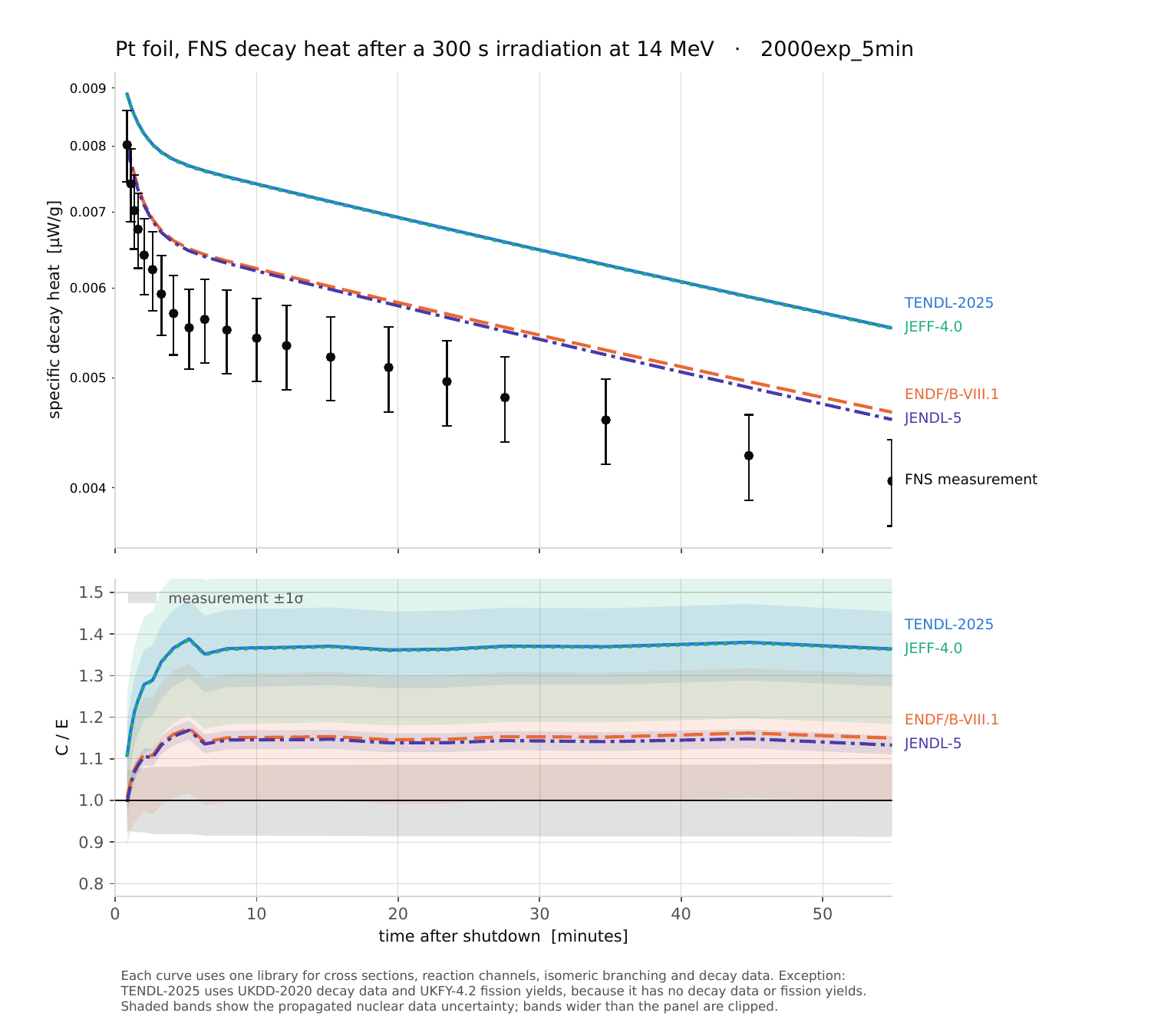}
\caption{Pt: decay heat against the measurement, with C/E beneath. Mean $|C/E-1|$ by library: JENDL-5 12\%, ENDF/B-VIII.1 13\%, JEFF-4.0 32\%, TENDL-2025 32\%.}
\end{figure}
\clearpage

\begin{figure}[H]
\centering
\includegraphics[width=\textwidth,height=0.82\textheight,keepaspectratio]{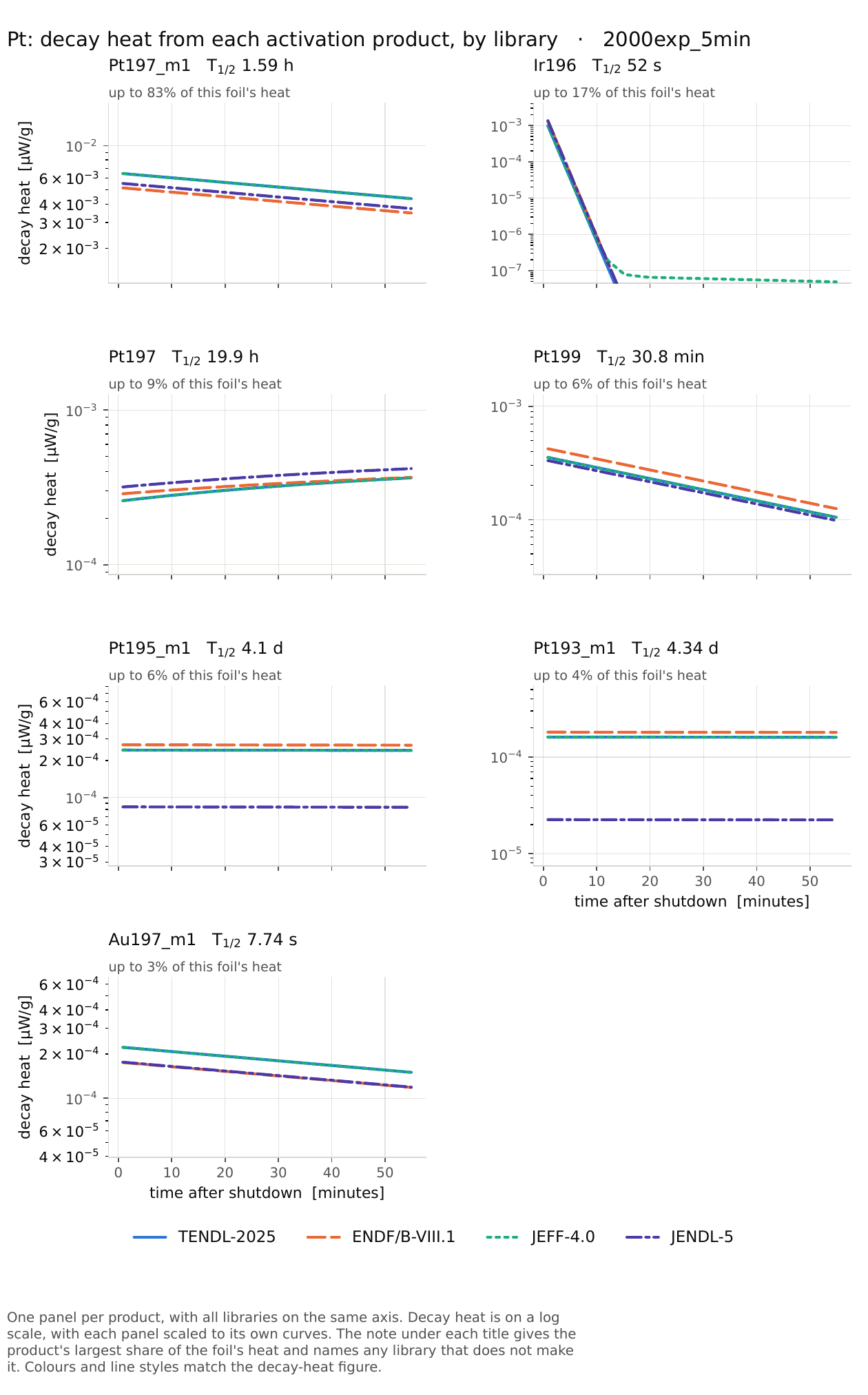}
\caption{Pt: share of the decay heat from each activation product during cooling.}
\end{figure}
\clearpage

\begin{figure}[H]
\centering
\includegraphics[width=\textwidth,height=0.82\textheight,keepaspectratio]{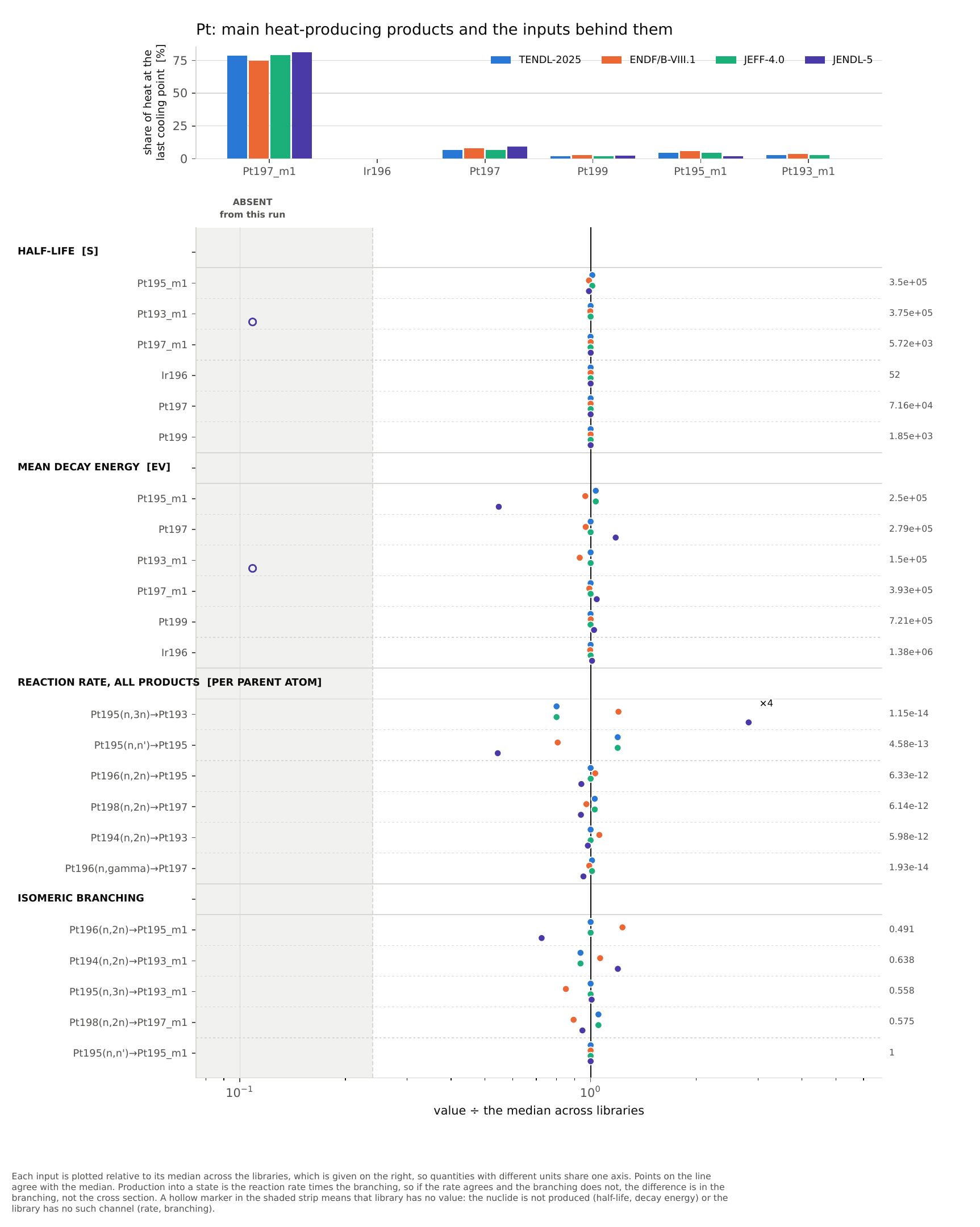}
\caption{Pt: inputs for the main heat-producing products, each relative to its median across the available libraries.}
\end{figure}
\clearpage

\begin{figure}[H]
\centering
\includegraphics[width=\textwidth,height=0.82\textheight,keepaspectratio]{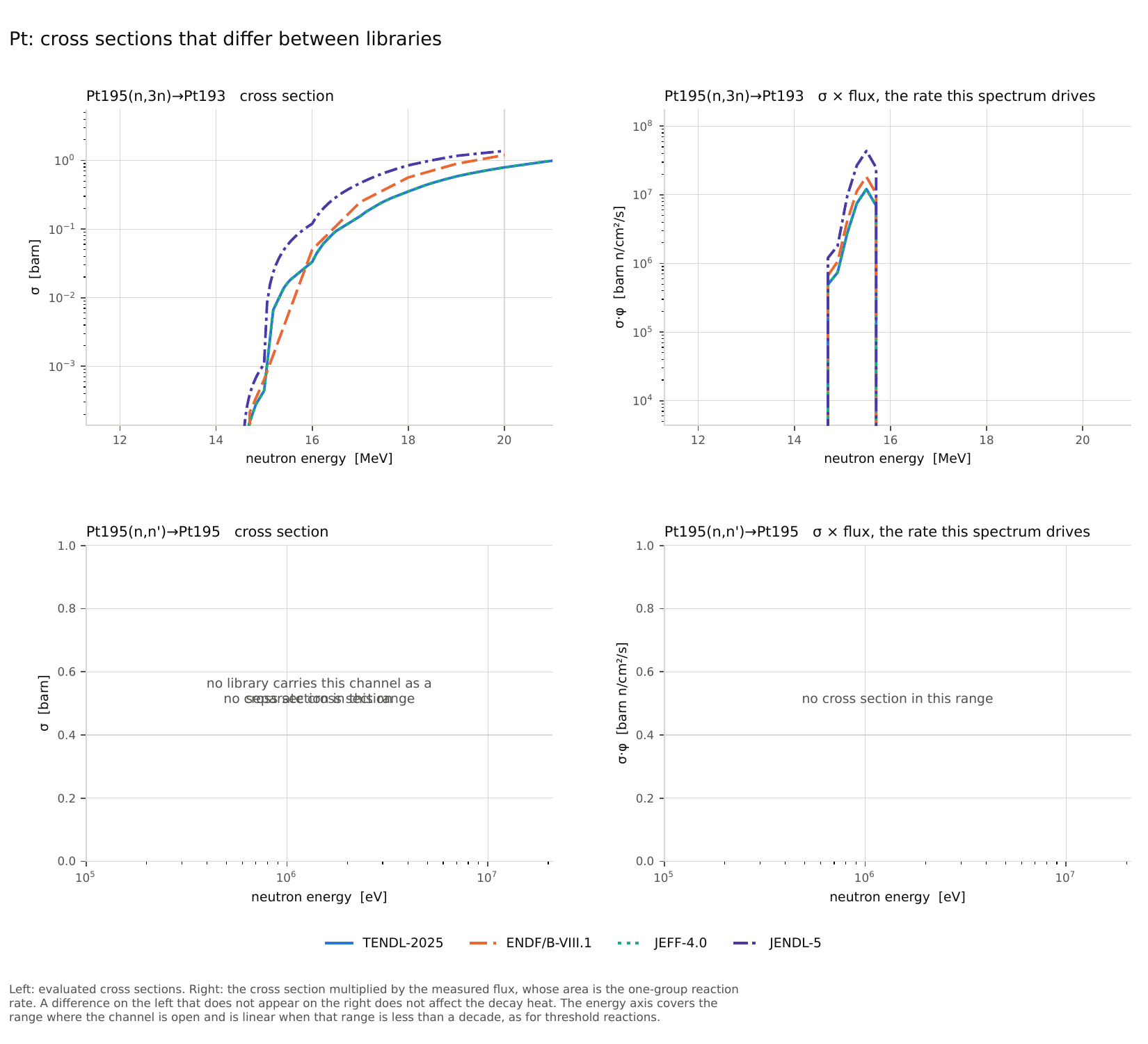}
\caption{Pt: cross sections of the channels whose spectrum-averaged rates differ, with the neutron spectrum and their product.}
\end{figure}
\clearpage

\begin{figure}[H]
\centering
\includegraphics[width=\textwidth,height=0.82\textheight,keepaspectratio]{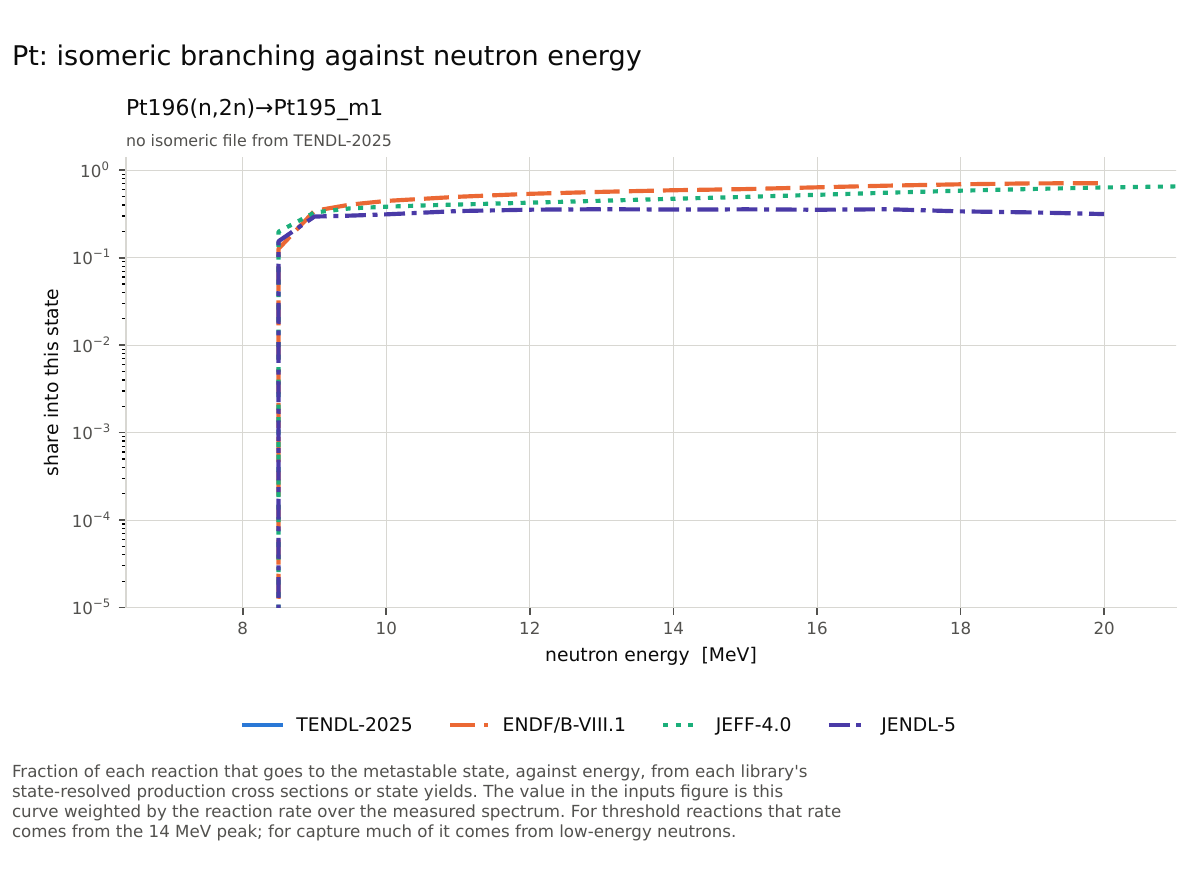}
\caption{Pt: isomeric branching ratio against neutron energy for the channels whose spectrum-averaged ratio differs between libraries.}
\end{figure}
\clearpage

The libraries are between 12 and 32 percent from the measurement, which has a median uncertainty of 8 percent. JENDL-5 is closest and TENDL-2025 furthest.

\subsubsection*{Differences in the nuclear data}
\begin{itemize}\setlength{\itemsep}{6pt}
\item JENDL-5 gives the total reaction rate of Pt195(n,3n)→Pt193 as 3.242e-14, against 9.188e-15 for the median of the others, a factor of 4.\\[2pt]{\itshape This is the total rate of the channel over all its products, so the difference is in the cross section, not in the split between states.}
\item JENDL-5 gives the total reaction rate of Pt195(n,n')→Pt195 as 2.49e-13, against 5.473e-13 for the median of the others, a factor of 2.
\end{itemize}
\subsubsection*{Suggested measurements}
\begin{itemize}\setlength{\itemsep}{4pt}
\item Natural Pt has 6 isotopes, so the measured heat combines several parents. Enriched single-isotope targets would distinguish their contributions. The candidate parent is Pt195. Enrichment alone does not isolate a single reaction channel.
\item This foil makes Pt195\textbackslash{}\_m1 (half-life 4.1 days), which gives up to 6 percent of the calculated heat. A five-minute irradiation and 55 minute measurement only see the start of this contribution. There is no seven-hour FNS measurement of this foil. A longer irradiation followed for days to months would be needed to test this product's production and decay.
\item The reaction rates are averaged over the measured spectrum. Activation measurements at several well-defined neutron energies, including below 14 MeV, would show which energy range the differences come from.
\end{itemize}
\clearpage
\subsection{Rb: Rubidium}\label{foil:Rb}
\begin{figure}[H]
\centering
\includegraphics[width=\textwidth,height=0.82\textheight,keepaspectratio]{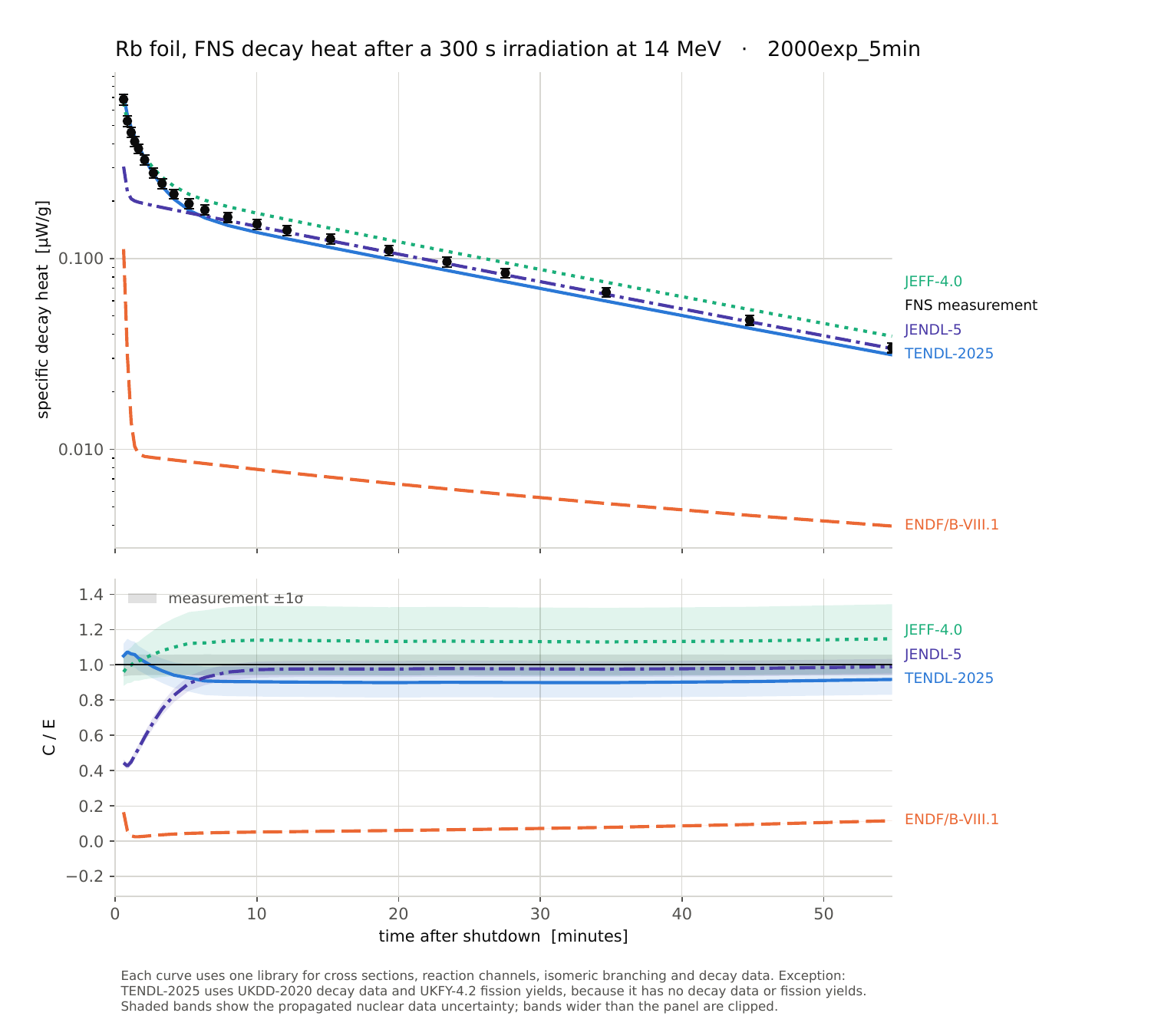}
\caption{Rb: decay heat against the measurement, with C/E beneath. Mean $|C/E-1|$ by library: TENDL-2025 7\%, JEFF-4.0 9\%, JENDL-5 20\%, ENDF/B-VIII.1 94\%.}
\end{figure}
\clearpage

\begin{figure}[H]
\centering
\includegraphics[width=\textwidth,height=0.82\textheight,keepaspectratio]{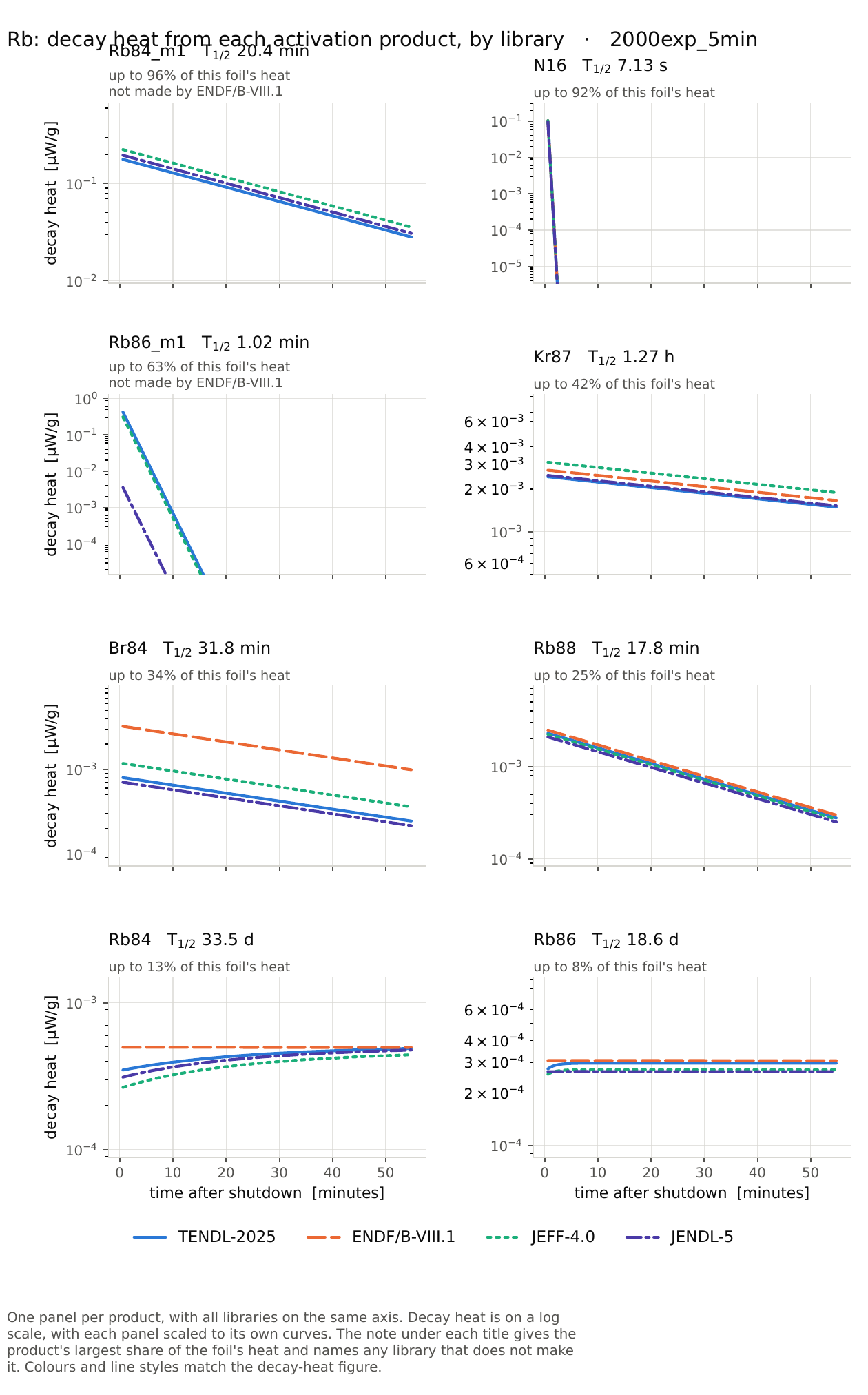}
\caption{Rb: share of the decay heat from each activation product during cooling.}
\end{figure}
\clearpage

\begin{figure}[H]
\centering
\includegraphics[width=\textwidth,height=0.82\textheight,keepaspectratio]{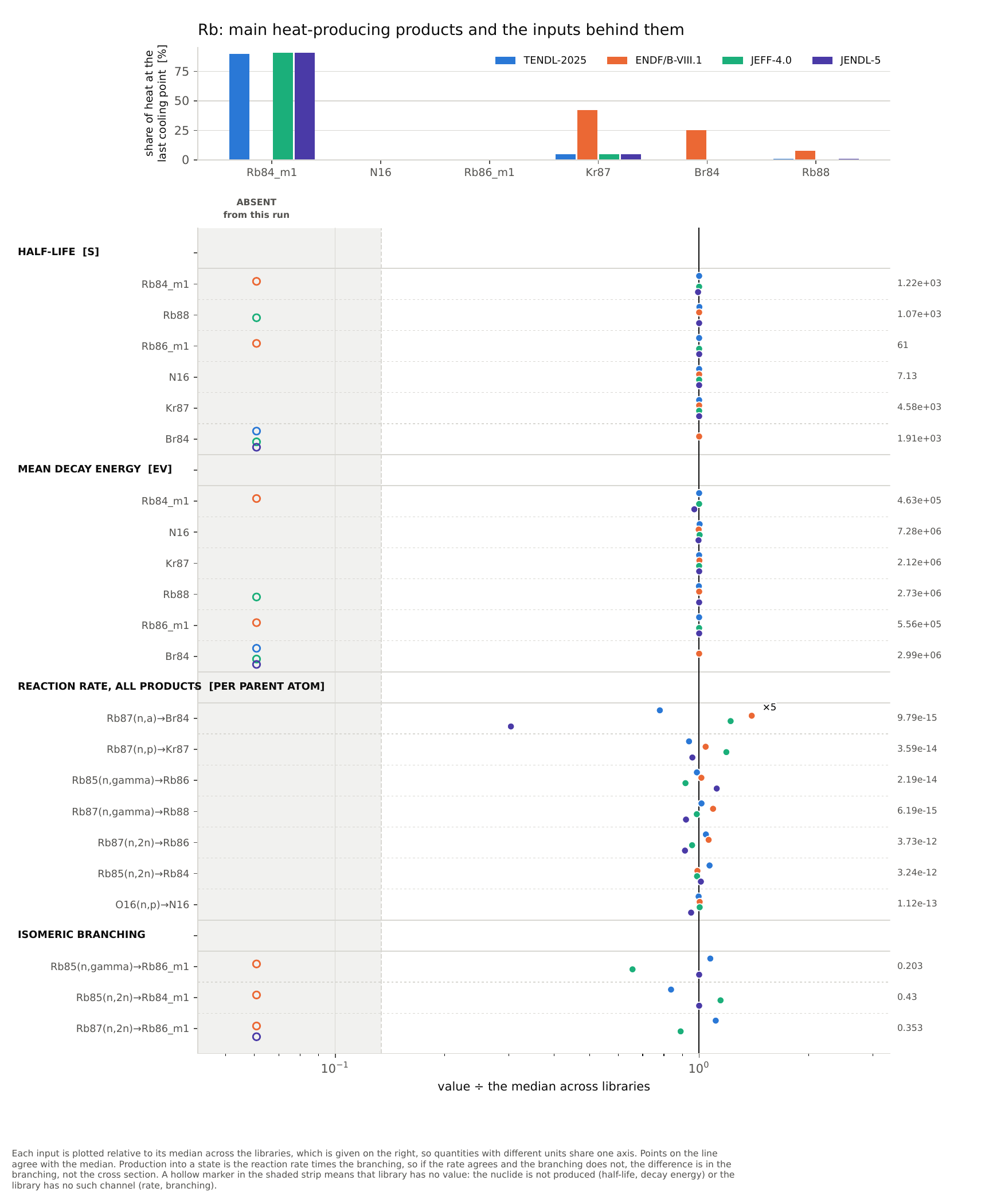}
\caption{Rb: inputs for the main heat-producing products, each relative to its median across the available libraries.}
\end{figure}
\clearpage

\begin{figure}[H]
\centering
\includegraphics[width=\textwidth,height=0.82\textheight,keepaspectratio]{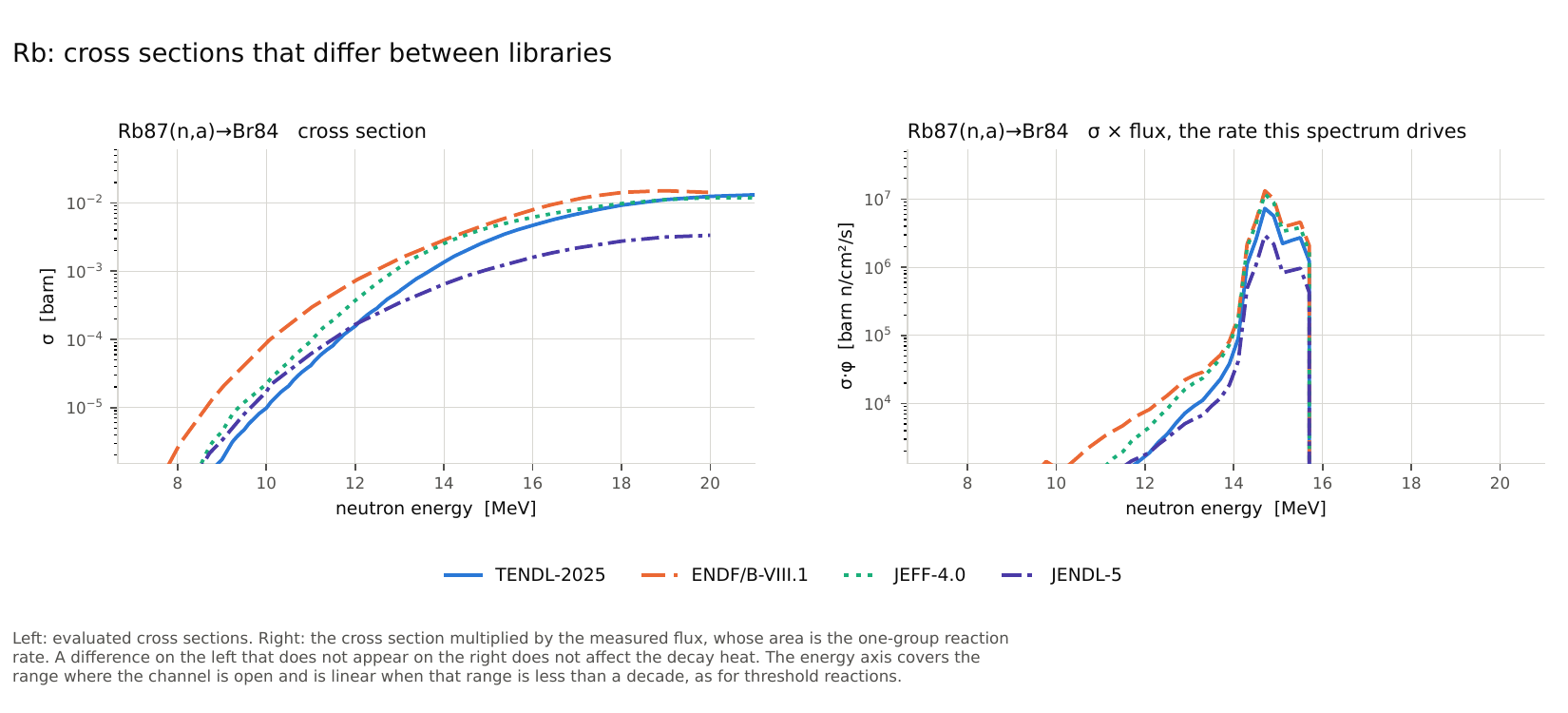}
\caption{Rb: cross sections of the channels whose spectrum-averaged rates differ, with the neutron spectrum and their product.}
\end{figure}
\clearpage

\begin{figure}[H]
\centering
\includegraphics[width=\textwidth,height=0.82\textheight,keepaspectratio]{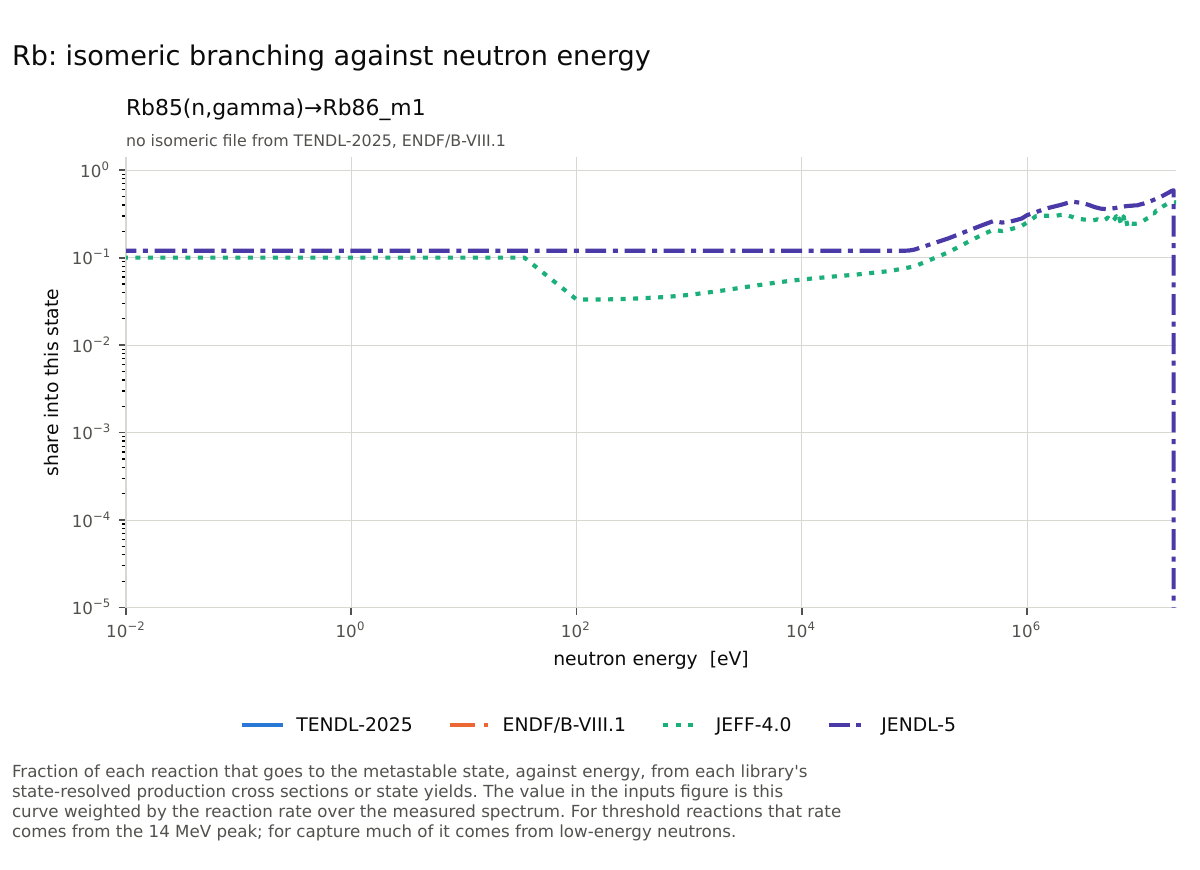}
\caption{Rb: isomeric branching ratio against neutron energy for the channels whose spectrum-averaged ratio differs between libraries.}
\end{figure}
\clearpage

ENDF/B-VIII.1 stands out at 94 percent from the measurement, against 20 percent for the next furthest library and 7 percent for TENDL-2025. The median measurement uncertainty is 6 percent.

\subsubsection*{Differences in the nuclear data}
\begin{itemize}\setlength{\itemsep}{6pt}
\item JENDL-5 gives the total reaction rate of Rb87(n,a)→Br84 as 2.976e-15, against 1.195e-14 for the median of the others, a factor of 4.\\[2pt]{\itshape This is the total rate of the channel over all its products, so the difference is in the cross section, not in the split between states.}
\item ENDF/B-VIII.1 gives no heat from Rb84\_m1; the other libraries do.\\[2pt]{\itshape The heat breakdown alone does not show why. Possible causes are a missing reaction channel, a different state assignment, different decay data or a reporting threshold.}
\item ENDF/B-VIII.1 gives no heat from Rb86\_m1; the other libraries do.
\end{itemize}
\subsubsection*{Suggested measurements}
\begin{itemize}\setlength{\itemsep}{4pt}
\item Natural Rb has 2 isotopes, so the measured heat combines several parents. Enriched single-isotope targets would distinguish their contributions. The candidate parent is Rb87. Enrichment alone does not isolate a single reaction channel.
\item The sample is a compound that also contains C, O. Their activation adds to the measured heat; a metal or purer sample would remove it.
\item This foil makes Rb84 (half-life 33.5 days), which gives up to 13 percent of the calculated heat. A five-minute irradiation and 55 minute measurement only see the start of this contribution. There is no seven-hour FNS measurement of this foil. A longer irradiation followed for days to months would be needed to test this product's production and decay.
\item The reaction rates are averaged over the measured spectrum. Activation measurements at several well-defined neutron energies, including below 14 MeV, would show which energy range the differences come from.
\end{itemize}
\clearpage
\subsection{Re: Rhenium}\label{foil:Re}
\begin{figure}[H]
\centering
\includegraphics[width=\textwidth,height=0.82\textheight,keepaspectratio]{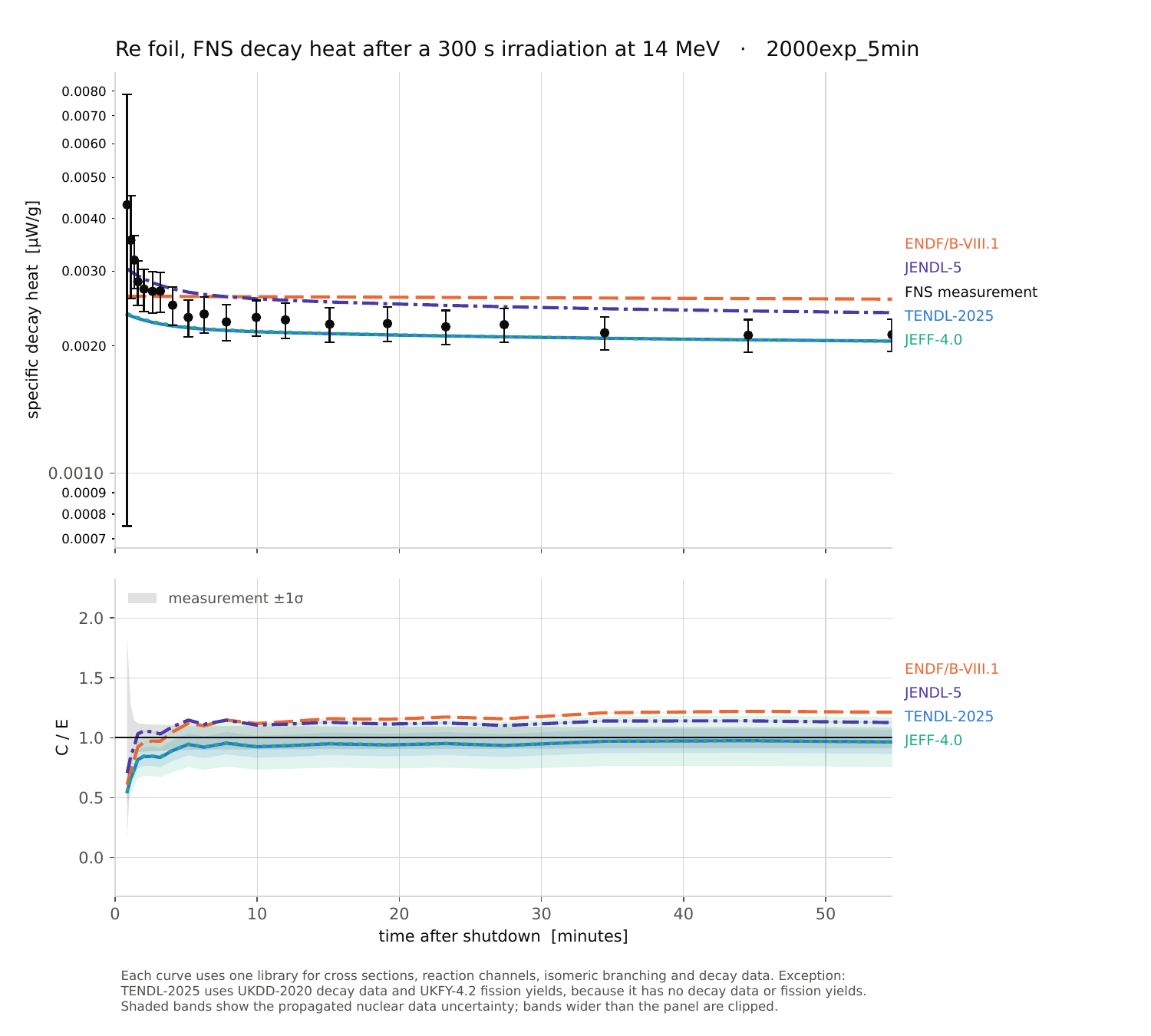}
\caption{Re: decay heat against the measurement, with C/E beneath. Mean $|C/E-1|$ by library: JENDL-5 11\%, JEFF-4.0 12\%, TENDL-2025 12\%, ENDF/B-VIII.1 15\%.}
\end{figure}
\clearpage

\begin{figure}[H]
\centering
\includegraphics[width=\textwidth,height=0.82\textheight,keepaspectratio]{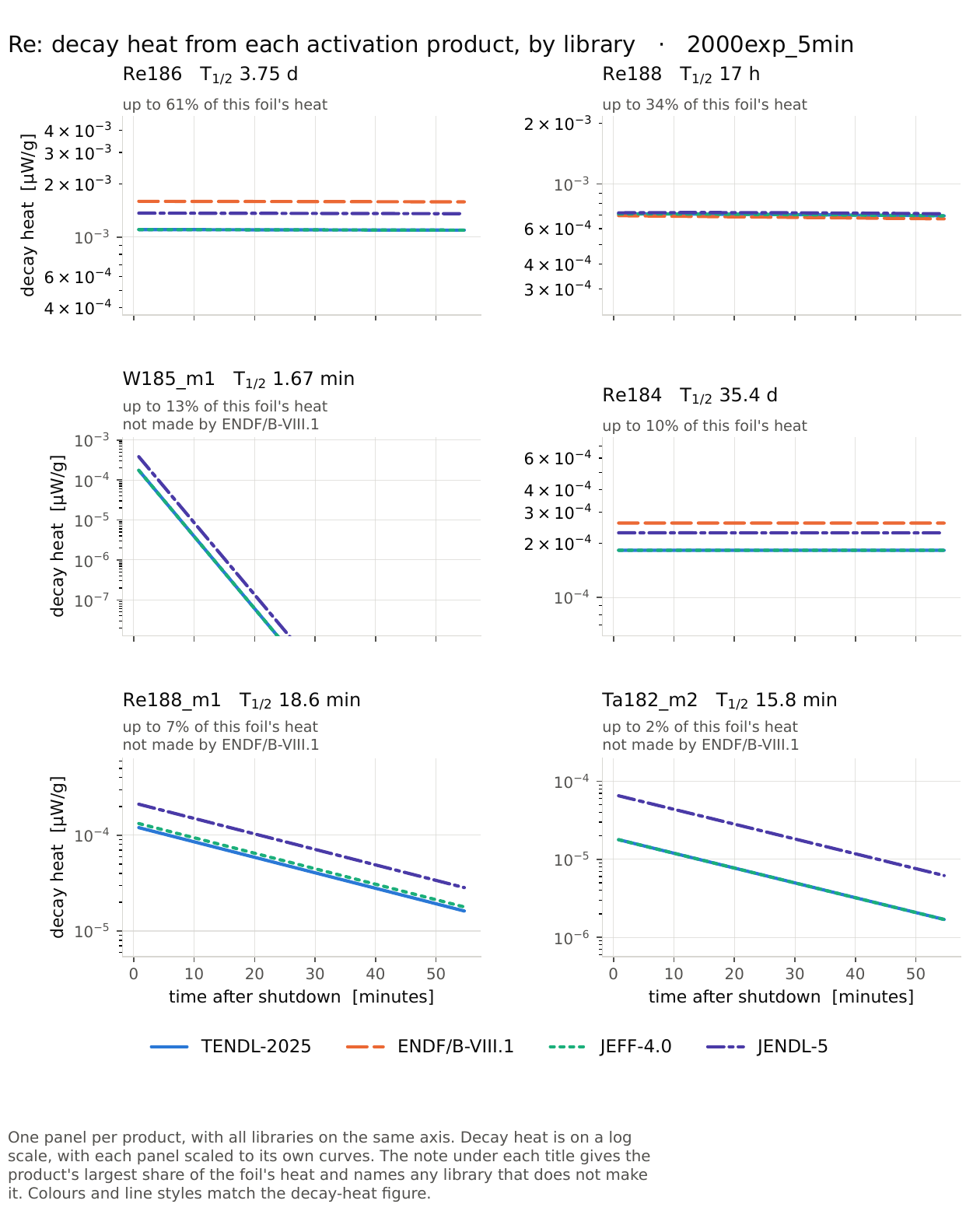}
\caption{Re: share of the decay heat from each activation product during cooling.}
\end{figure}
\clearpage

\begin{figure}[H]
\centering
\includegraphics[width=\textwidth,height=0.82\textheight,keepaspectratio]{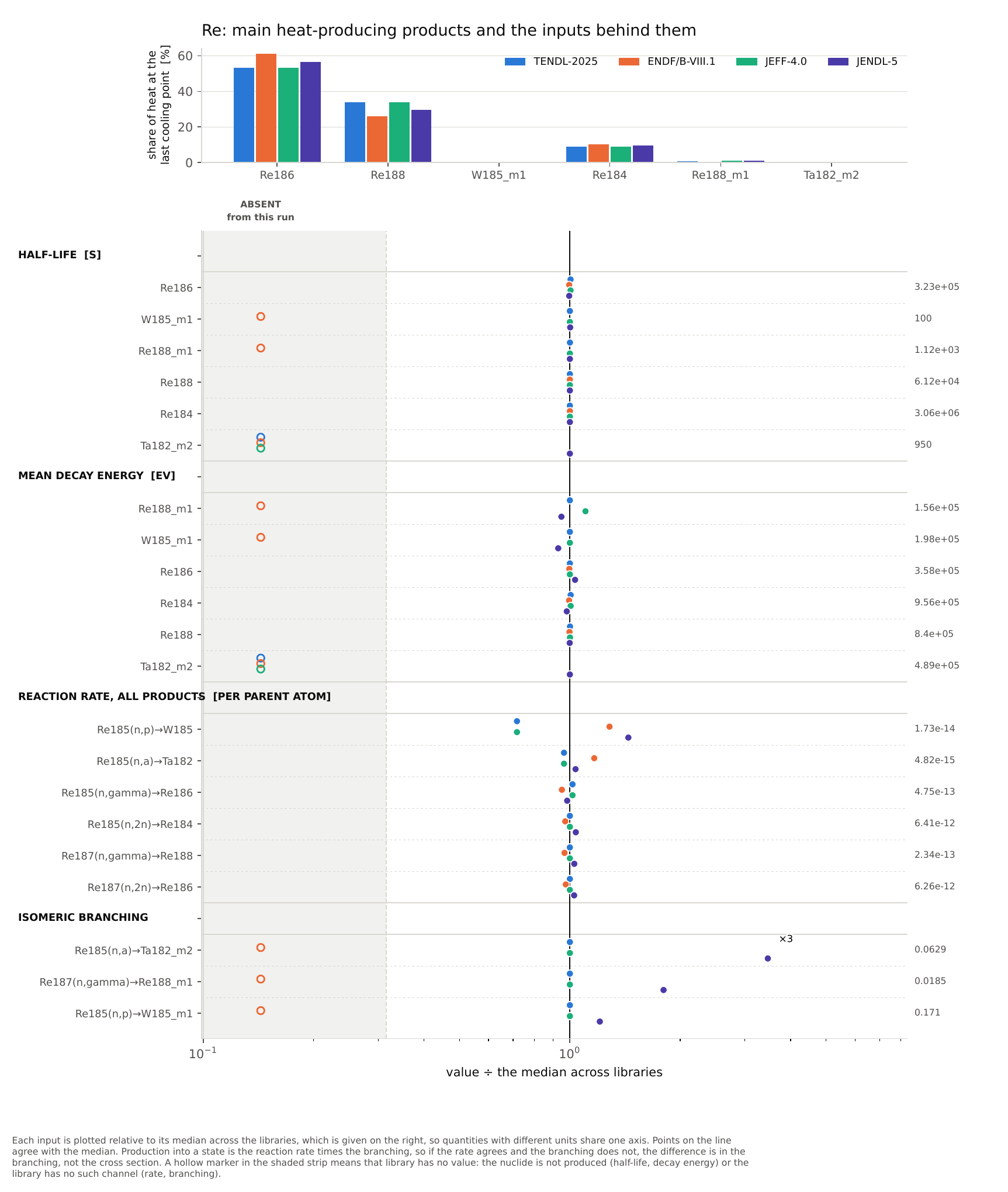}
\caption{Re: inputs for the main heat-producing products, each relative to its median across the available libraries.}
\end{figure}
\clearpage

\begin{figure}[H]
\centering
\includegraphics[width=\textwidth,height=0.82\textheight,keepaspectratio]{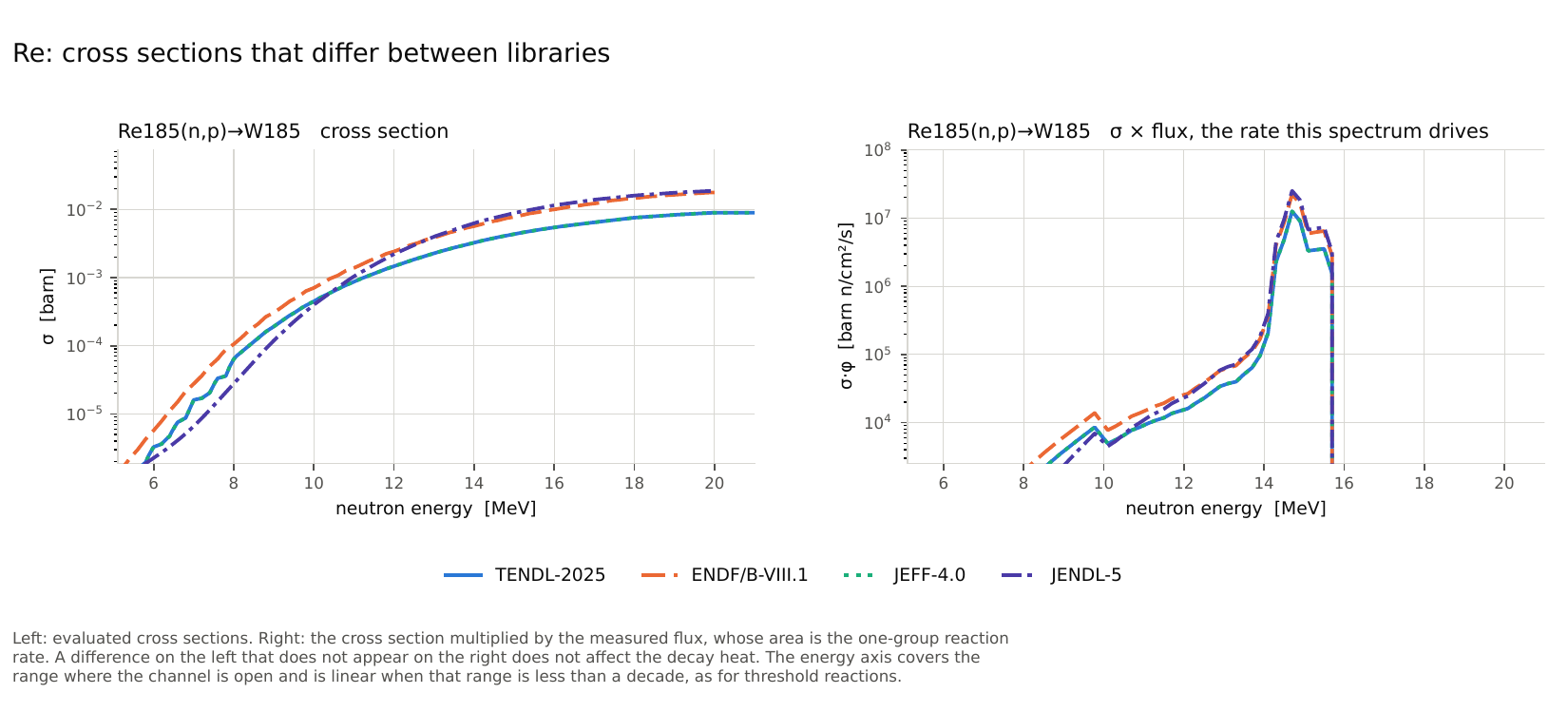}
\caption{Re: cross sections of the channels whose spectrum-averaged rates differ, with the neutron spectrum and their product.}
\end{figure}
\clearpage

\begin{figure}[H]
\centering
\includegraphics[width=\textwidth,height=0.82\textheight,keepaspectratio]{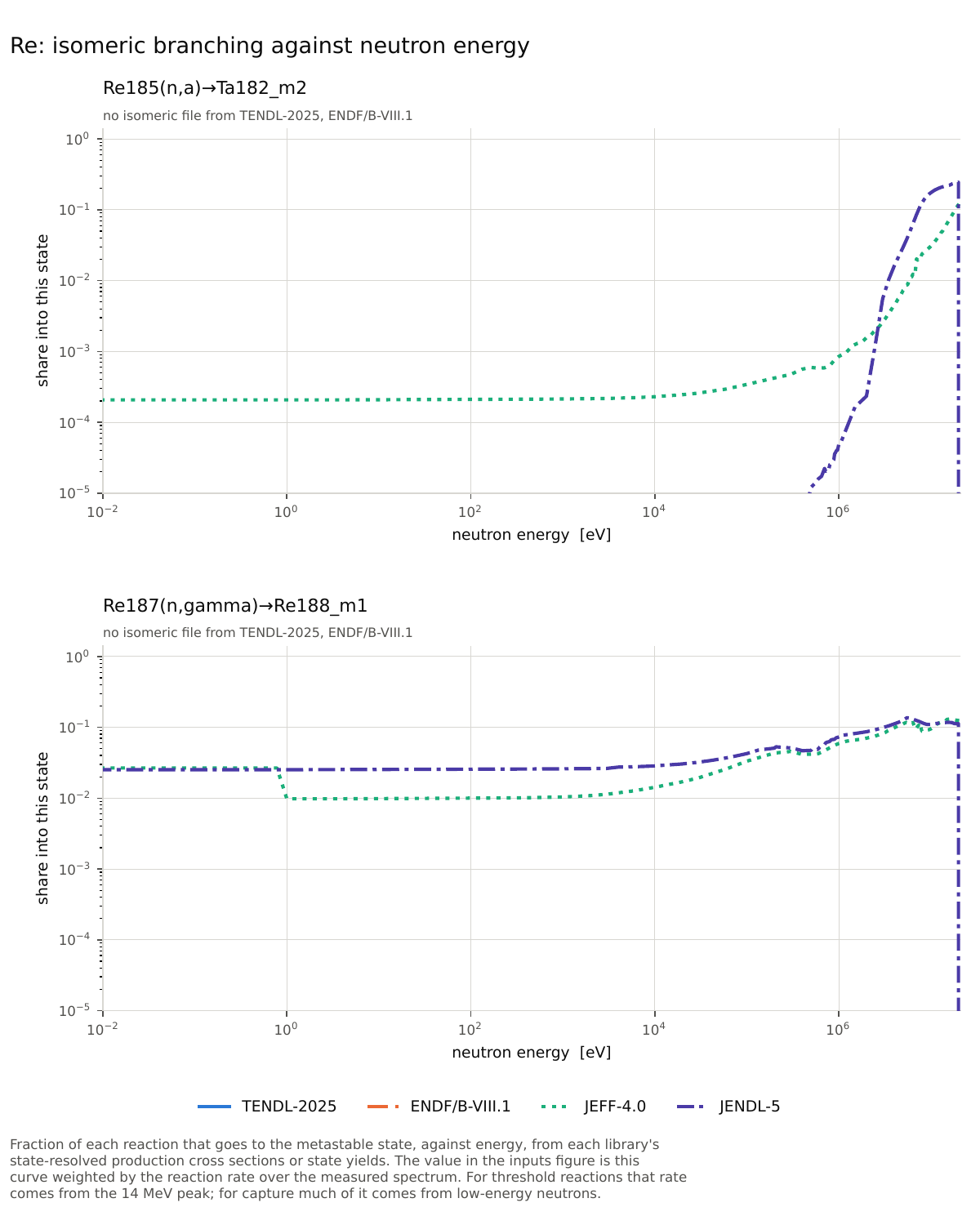}
\caption{Re: isomeric branching ratio against neutron energy for the channels whose spectrum-averaged ratio differs between libraries.}
\end{figure}
\clearpage

\subsubsection*{1996 seven-hour campaign}
The 1996 campaign irradiated this foil for 7.17 hours at position 7 and measured it out to 402.17 days. In the five-minute calculation Re186 (half-life 3.75 days) gives up to 61 percent of the heat, and the longer irradiation and cooling test its production and decay data. JENDL-5 is closest. This campaign uses a different spectrum, so it is not included in the main rankings.

\begin{figure}[H]
\centering
\includegraphics[width=\textwidth,height=0.82\textheight,keepaspectratio]{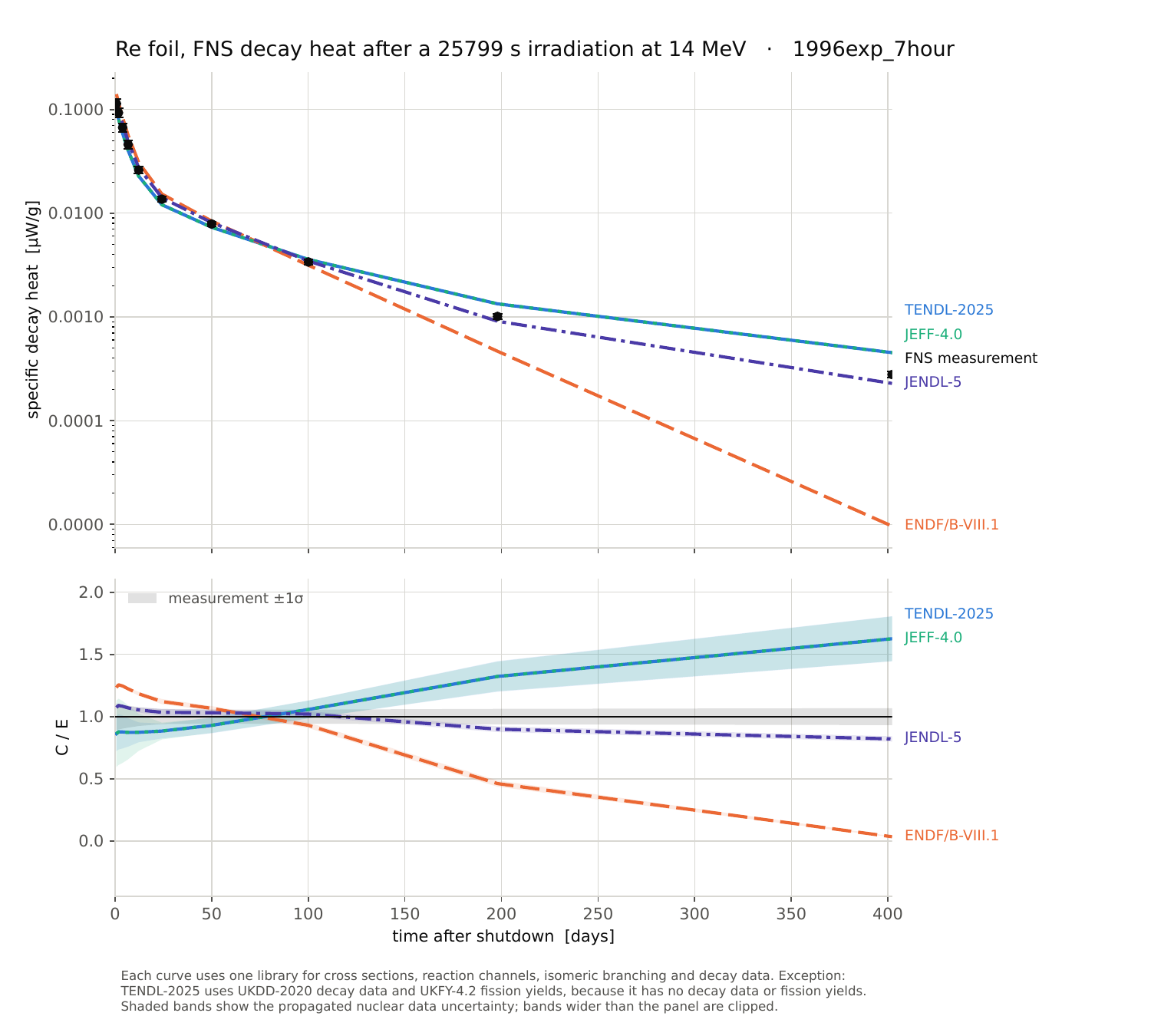}
\caption{Re: 1996 seven-hour campaign, irradiation position 7. Mean $|C/E-1|$ by library: JENDL-5 7\%, JEFF-4.0 18\%, TENDL-2025 18\%, ENDF/B-VIII.1 29\%. }
\end{figure}
\clearpage

Mean absolute C/E deviations range from 11 to 15 percent. The spread of 3 percentage points is smaller than the median measurement uncertainty of 10 percent, so the measurement cannot separate the libraries.

\subsubsection*{Differences in the nuclear data}
\begin{itemize}\setlength{\itemsep}{6pt}
\item JENDL-5 gives the isomeric branching of Re185(n,a)→Ta182\_m2 as 0.2181, against 0.06291 for the median of the others, a factor of 3.\\[2pt]{\itshape Isomeric branching sets how a channel's production is split between the ground and metastable states. It should be checked together with the total rate of the channel.}
\item JENDL-5 gives the total reaction rate of Re185(n,p)→W185 as 2.494e-14, against 1.239e-14 for the median of the others, a factor of 2.\\[2pt]{\itshape This is the total rate of the channel over all its products, so the difference is in the cross section, not in the split between states.}
\item ENDF/B-VIII.1 gives no heat from W185\_m1; the other libraries do.\\[2pt]{\itshape The heat breakdown alone does not show why. Possible causes are a missing reaction channel, a different state assignment, different decay data or a reporting threshold.}
\item ENDF/B-VIII.1 gives no heat from Re188\_m1; the other libraries do.
\item ENDF/B-VIII.1 gives no heat from Ta182\_m2; the other libraries do.
\end{itemize}
\subsubsection*{Suggested measurements}
\begin{itemize}\setlength{\itemsep}{4pt}
\item Natural Re has 2 isotopes, so the measured heat combines several parents. Enriched single-isotope targets would distinguish their contributions. The candidate parent is Re185. Enrichment alone does not isolate a single reaction channel.
\item The reaction rates are averaged over the measured spectrum. Activation measurements at several well-defined neutron energies, including below 14 MeV, would show which energy range the differences come from.
\item Measuring the activity of each state separately, for example by gamma spectrometry where the states have distinct lines, would constrain the isomeric ratio. This needs corrections for feeding between states, detector efficiency and irradiation history.
\end{itemize}
\clearpage
\subsection{Rh: Rhodium}\label{foil:Rh}
\begin{figure}[H]
\centering
\includegraphics[width=\textwidth,height=0.82\textheight,keepaspectratio]{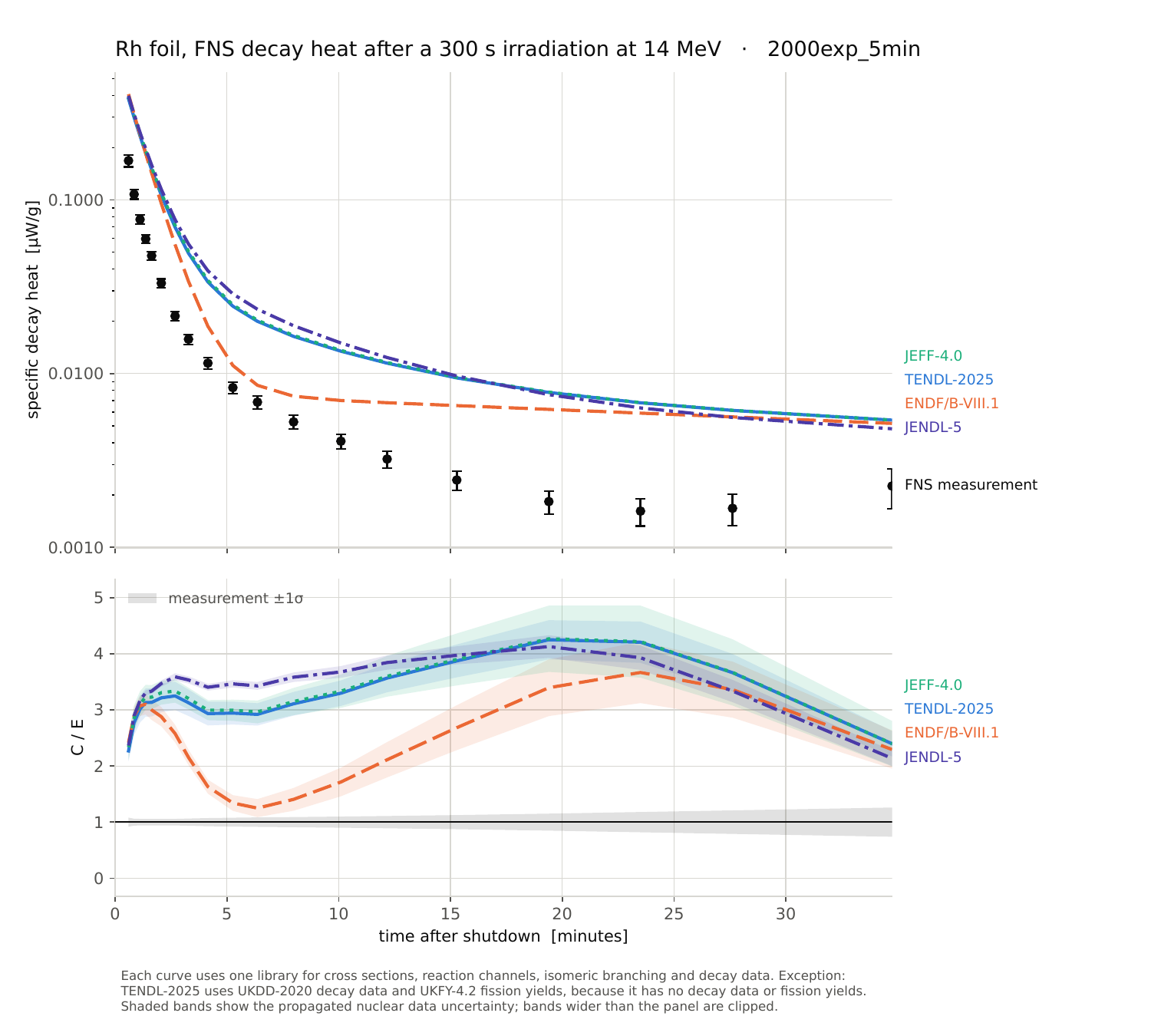}
\caption{Rh: decay heat against the measurement, with C/E beneath. Mean $|C/E-1|$ by library: ENDF/B-VIII.1 147\%, TENDL-2025 221\%, JEFF-4.0 227\%, JENDL-5 240\%.}
\end{figure}
\clearpage

\begin{figure}[H]
\centering
\includegraphics[width=\textwidth,height=0.82\textheight,keepaspectratio]{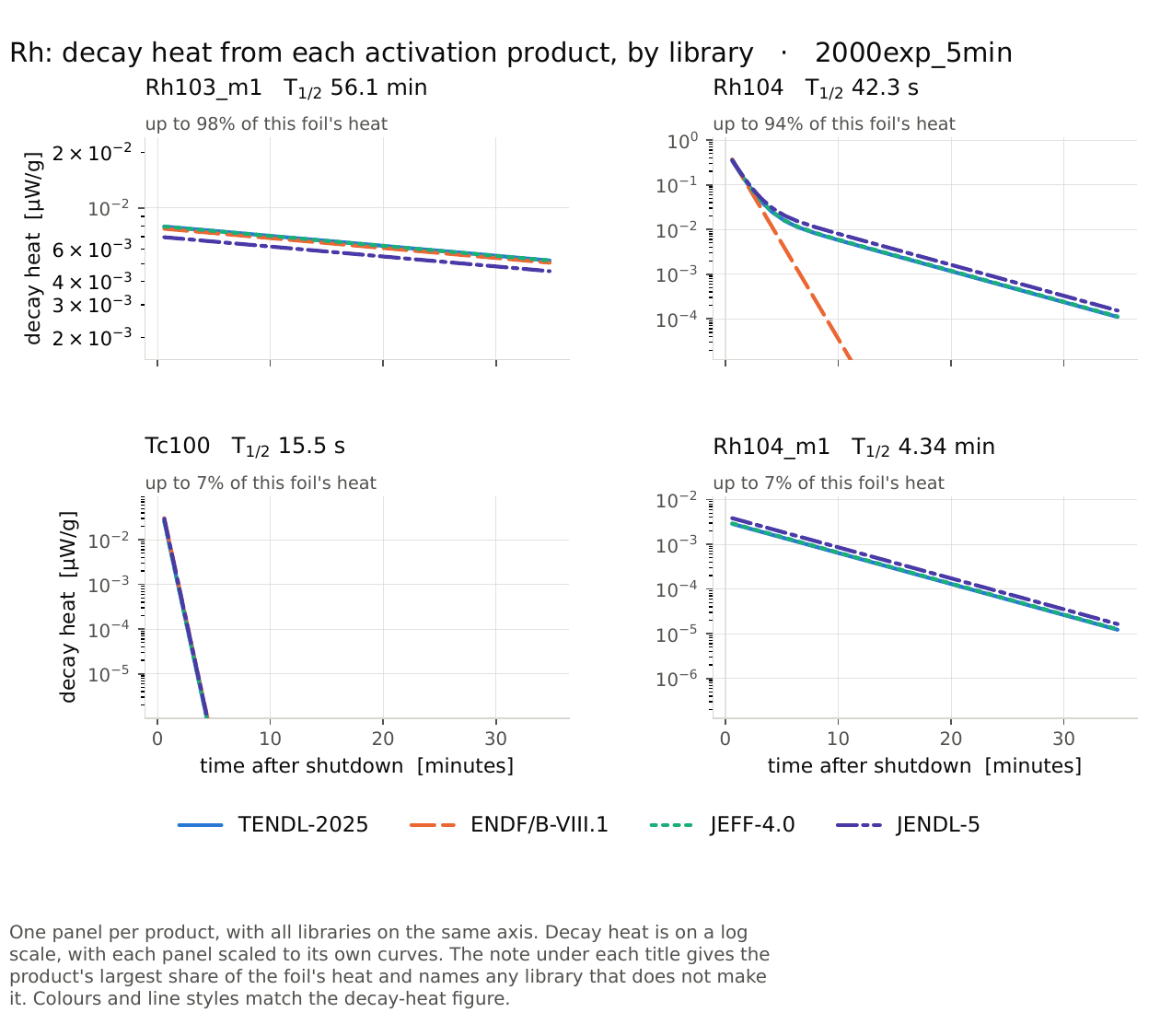}
\caption{Rh: share of the decay heat from each activation product during cooling.}
\end{figure}
\clearpage

\begin{figure}[H]
\centering
\includegraphics[width=\textwidth,height=0.82\textheight,keepaspectratio]{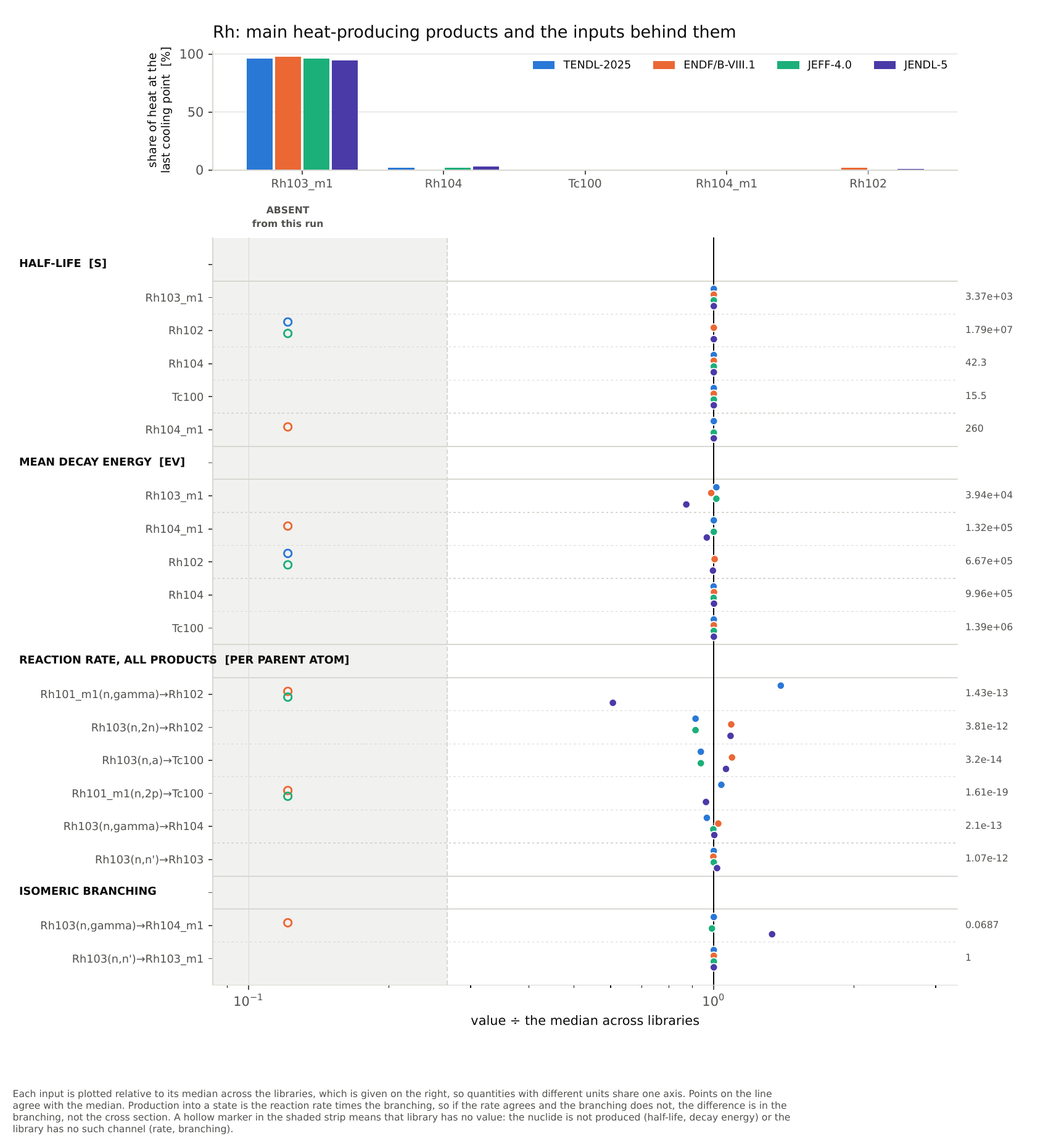}
\caption{Rh: inputs for the main heat-producing products, each relative to its median across the available libraries.}
\end{figure}
\clearpage

\begin{figure}[H]
\centering
\includegraphics[width=\textwidth,height=0.82\textheight,keepaspectratio]{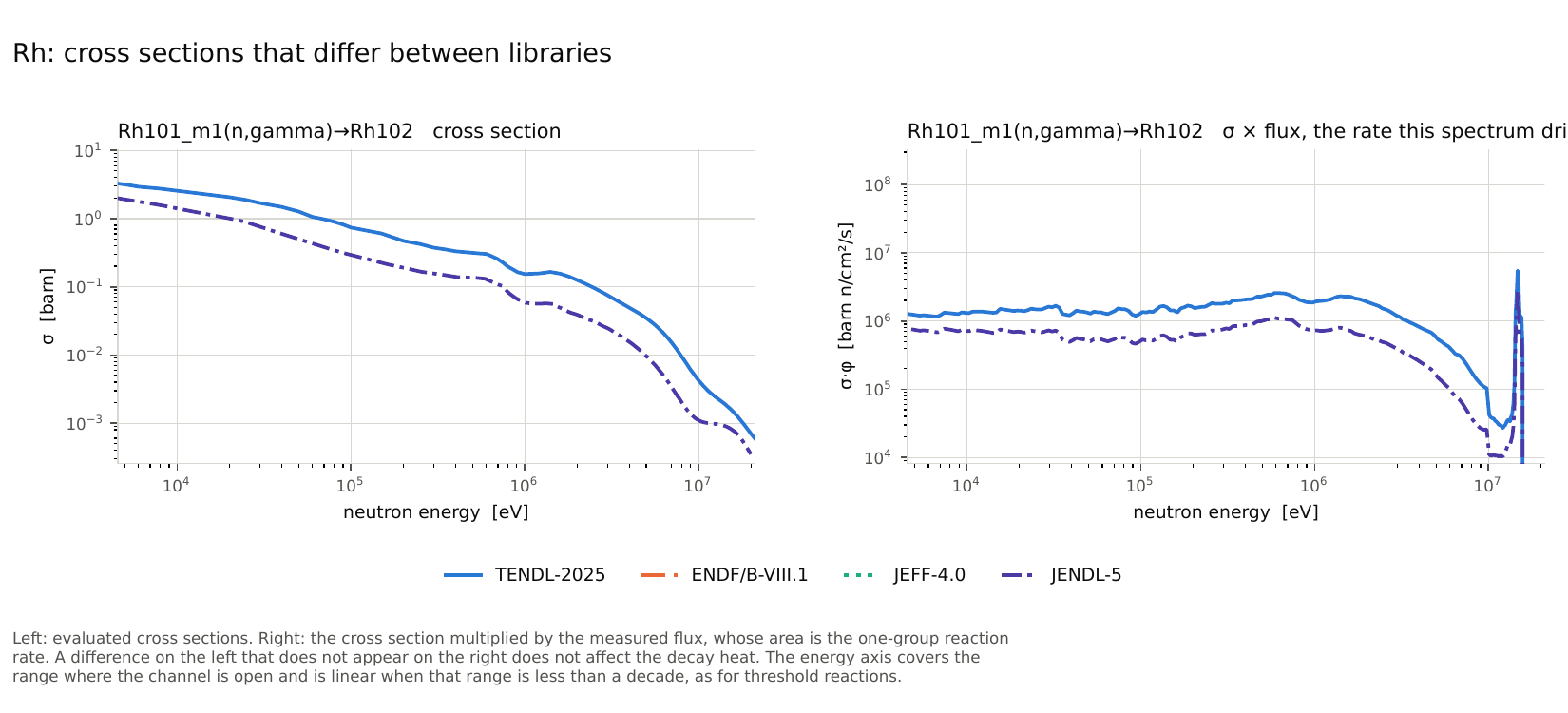}
\caption{Rh: cross sections of the channels whose spectrum-averaged rates differ, with the neutron spectrum and their product.}
\end{figure}
\clearpage

The libraries are between 147 and 240 percent from the measurement, which has a median uncertainty of 8 percent. ENDF/B-VIII.1 is closest and JENDL-5 furthest.

\subsubsection*{Differences in the nuclear data}
\begin{itemize}\setlength{\itemsep}{6pt}
\item Only TENDL-2025 and JENDL-5 have the total reaction rate of Rh101\_m1(n,gamma)→Rh102, and they differ by a factor of 2: 1.987e-13 against 8.653e-14.\\[2pt]{\itshape This is the total rate of the channel over all its products, so the difference is in the cross section, not in the split between states.}
\end{itemize}
\subsubsection*{Suggested measurements}
\begin{itemize}\setlength{\itemsep}{4pt}
\item Rh has a single stable isotope, so the foil is already a single-isotope target and enrichment would not help.
\item This foil makes Rh102 (half-life 207 days), which gives up to 2 percent of the calculated heat. A five-minute irradiation and 55 minute measurement only see the start of this contribution. There is no seven-hour FNS measurement of this foil. A longer irradiation followed for days to months would be needed to test this product's production and decay.
\item The reaction rates are averaged over the measured spectrum. Activation measurements at several well-defined neutron energies, including below 14 MeV, would show which energy range the differences come from.
\end{itemize}
\clearpage
\subsection{Ru: Ruthenium}\label{foil:Ru}
\begin{figure}[H]
\centering
\includegraphics[width=\textwidth,height=0.82\textheight,keepaspectratio]{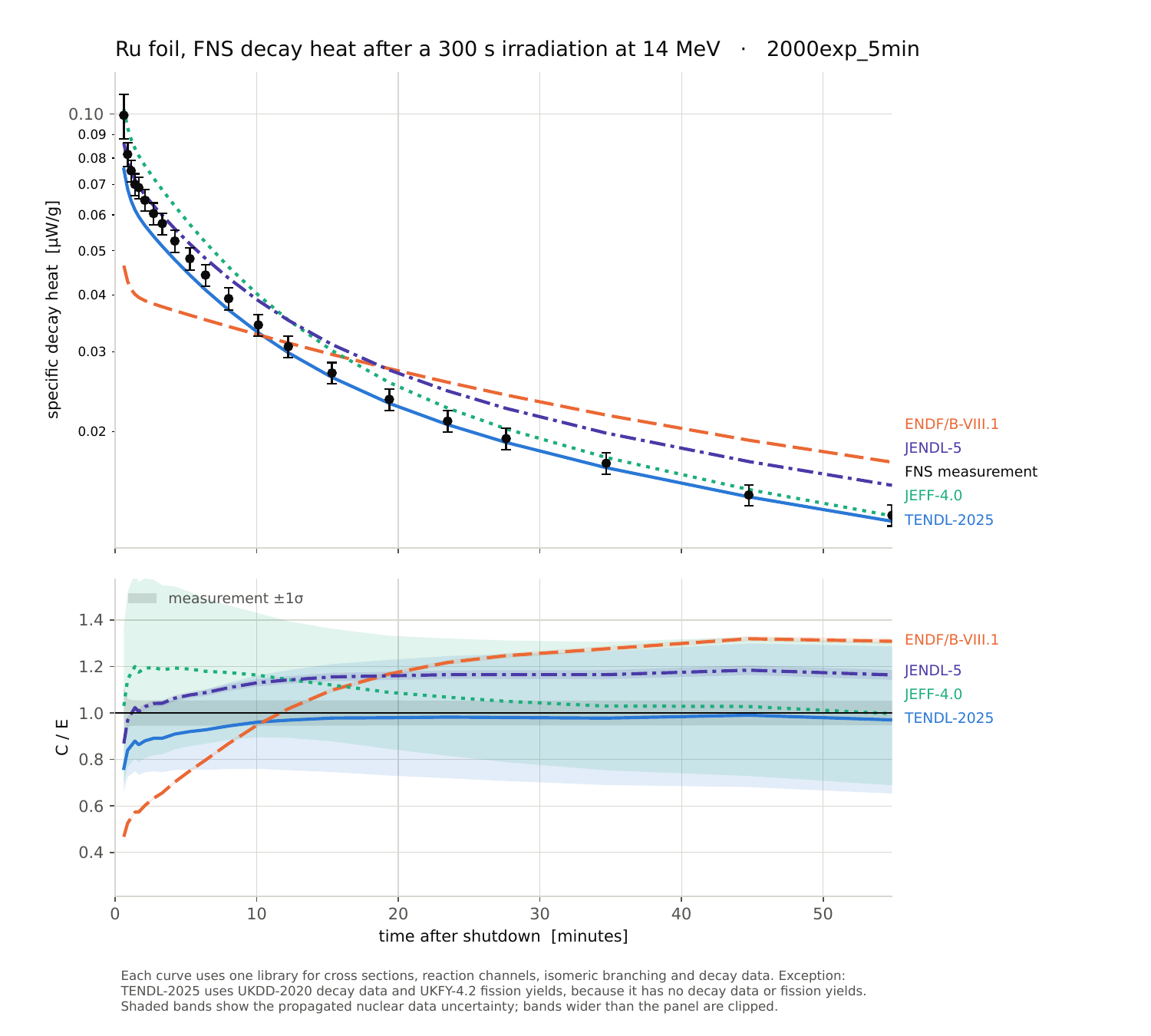}
\caption{Ru: decay heat against the measurement, with C/E beneath. Mean $|C/E-1|$ by library: TENDL-2025 8\%, JENDL-5 10\%, JEFF-4.0 13\%, ENDF/B-VIII.1 29\%.}
\end{figure}
\clearpage

\begin{figure}[H]
\centering
\includegraphics[width=\textwidth,height=0.82\textheight,keepaspectratio]{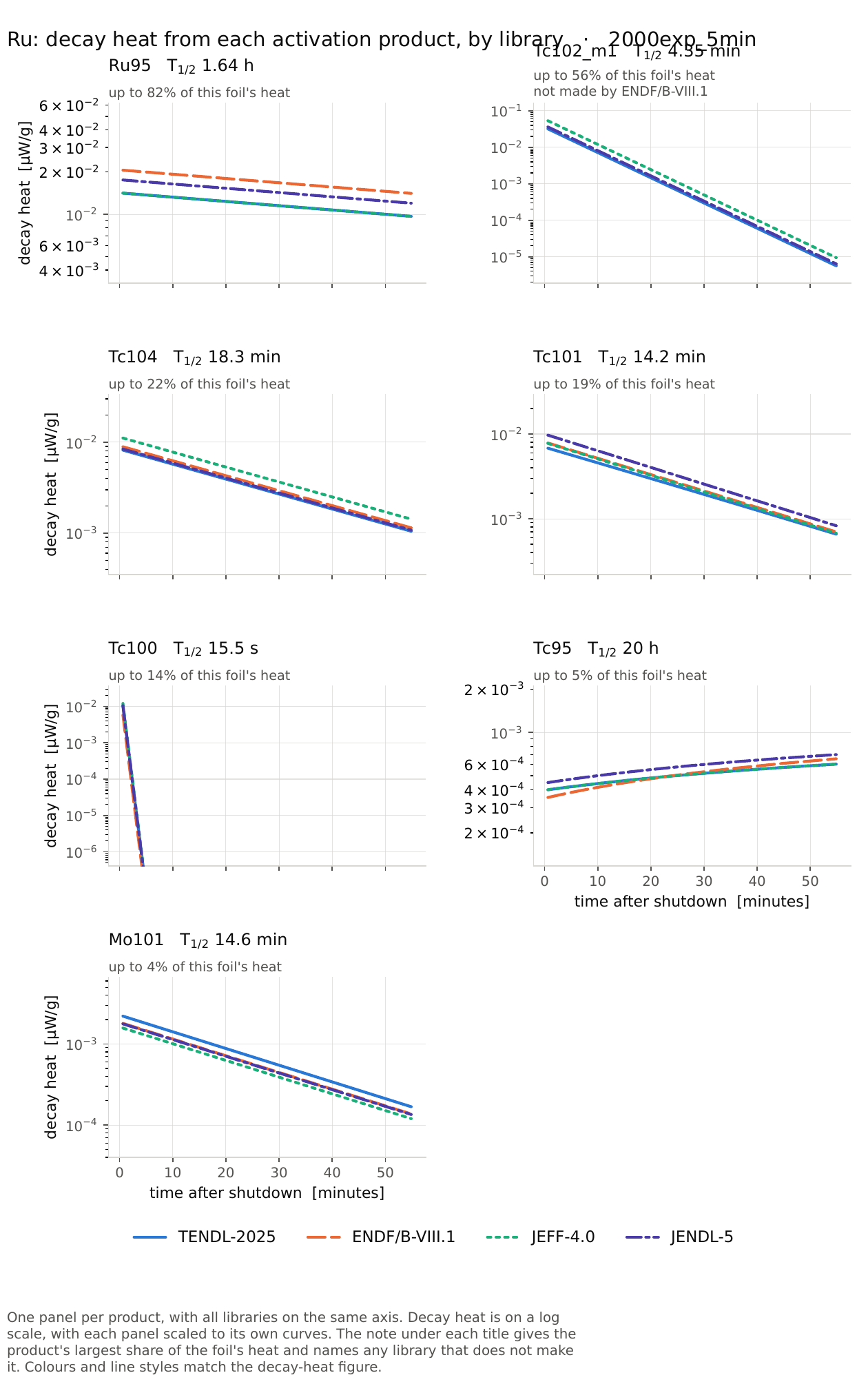}
\caption{Ru: share of the decay heat from each activation product during cooling.}
\end{figure}
\clearpage

\begin{figure}[H]
\centering
\includegraphics[width=\textwidth,height=0.82\textheight,keepaspectratio]{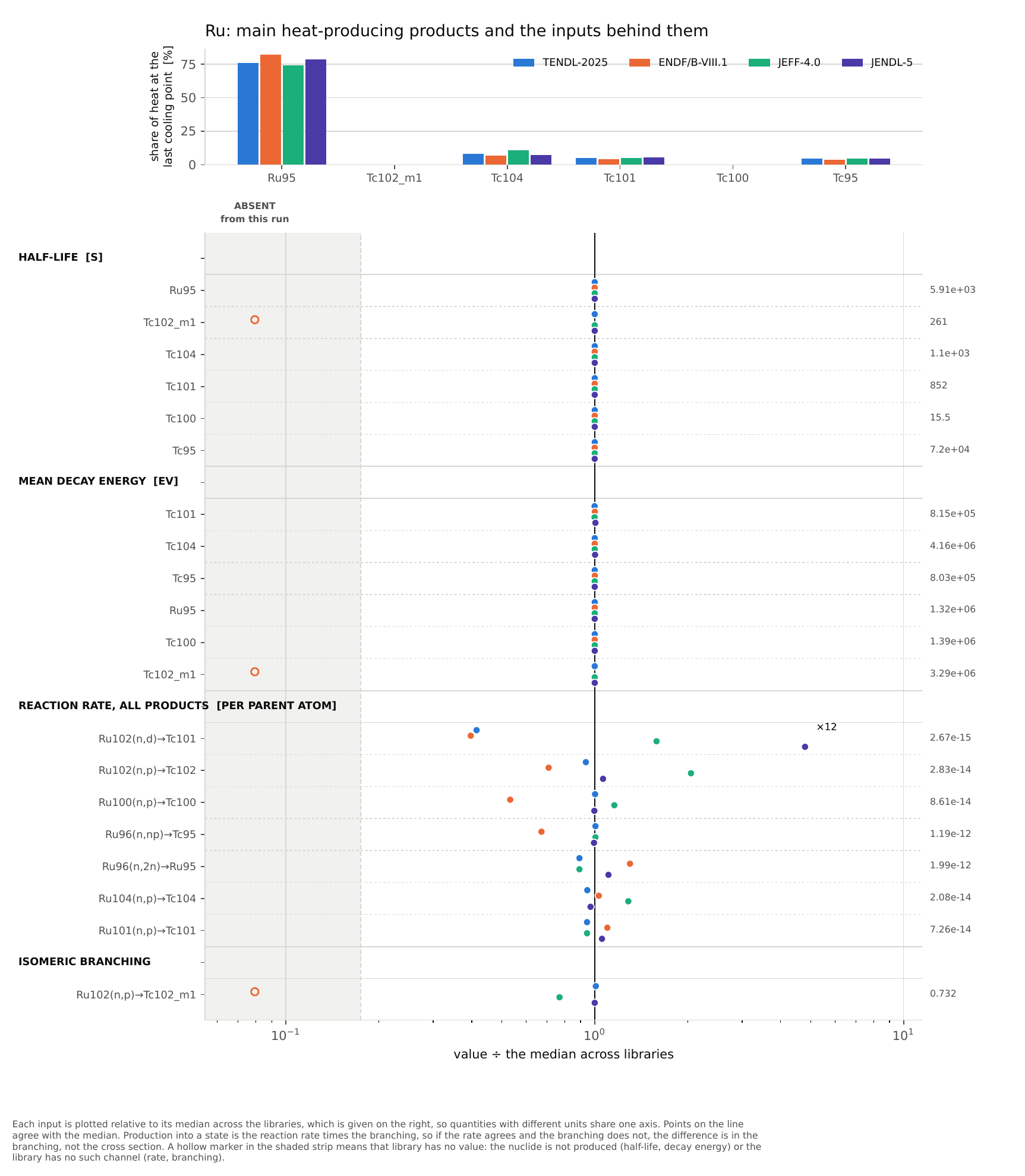}
\caption{Ru: inputs for the main heat-producing products, each relative to its median across the available libraries.}
\end{figure}
\clearpage

\begin{figure}[H]
\centering
\includegraphics[width=\textwidth,height=0.82\textheight,keepaspectratio]{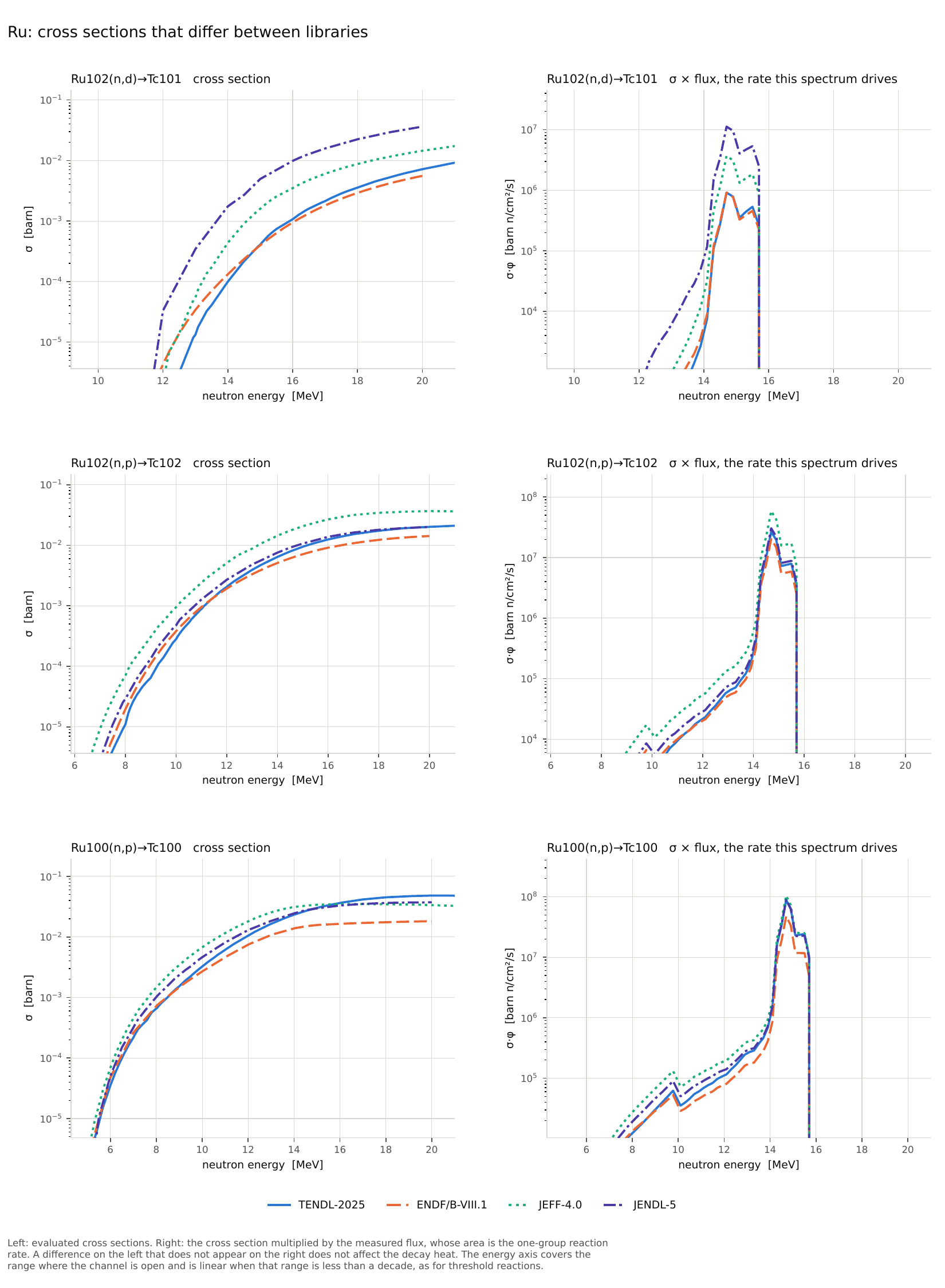}
\caption{Ru: cross sections of the channels whose spectrum-averaged rates differ, with the neutron spectrum and their product.}
\end{figure}
\clearpage

The libraries are between 8 and 29 percent from the measurement, which has a median uncertainty of 5 percent. TENDL-2025 is closest and ENDF/B-VIII.1 furthest.

\subsubsection*{Differences in the nuclear data}
\begin{itemize}\setlength{\itemsep}{6pt}
\item JENDL-5 gives the total reaction rate of Ru102(n,d)→Tc101 as 1.281e-14, against 1.107e-15 for the median of the others, a factor of 12.\\[2pt]{\itshape This is the total rate of the channel over all its products, so the difference is in the cross section, not in the split between states.}
\item JEFF-4.0 gives the total reaction rate of Ru102(n,p)→Tc102 as 5.795e-14, against 2.647e-14 for the median of the others, a factor of 2.
\item ENDF/B-VIII.1 gives no heat from Tc102\_m1; the other libraries do.\\[2pt]{\itshape The heat breakdown alone does not show why. Possible causes are a missing reaction channel, a different state assignment, different decay data or a reporting threshold.}
\end{itemize}
\subsubsection*{Suggested measurements}
\begin{itemize}\setlength{\itemsep}{4pt}
\item Natural Ru has 7 isotopes, so the measured heat combines several parents. Enriched single-isotope targets would distinguish their contributions. The candidate parent is Ru102. Enrichment alone does not isolate a single reaction channel.
\item The reaction rates are averaged over the measured spectrum. Activation measurements at several well-defined neutron energies, including below 14 MeV, would show which energy range the differences come from.
\end{itemize}
\clearpage
\subsection{S: Sulfur}\label{foil:S}
\begin{figure}[H]
\centering
\includegraphics[width=\textwidth,height=0.82\textheight,keepaspectratio]{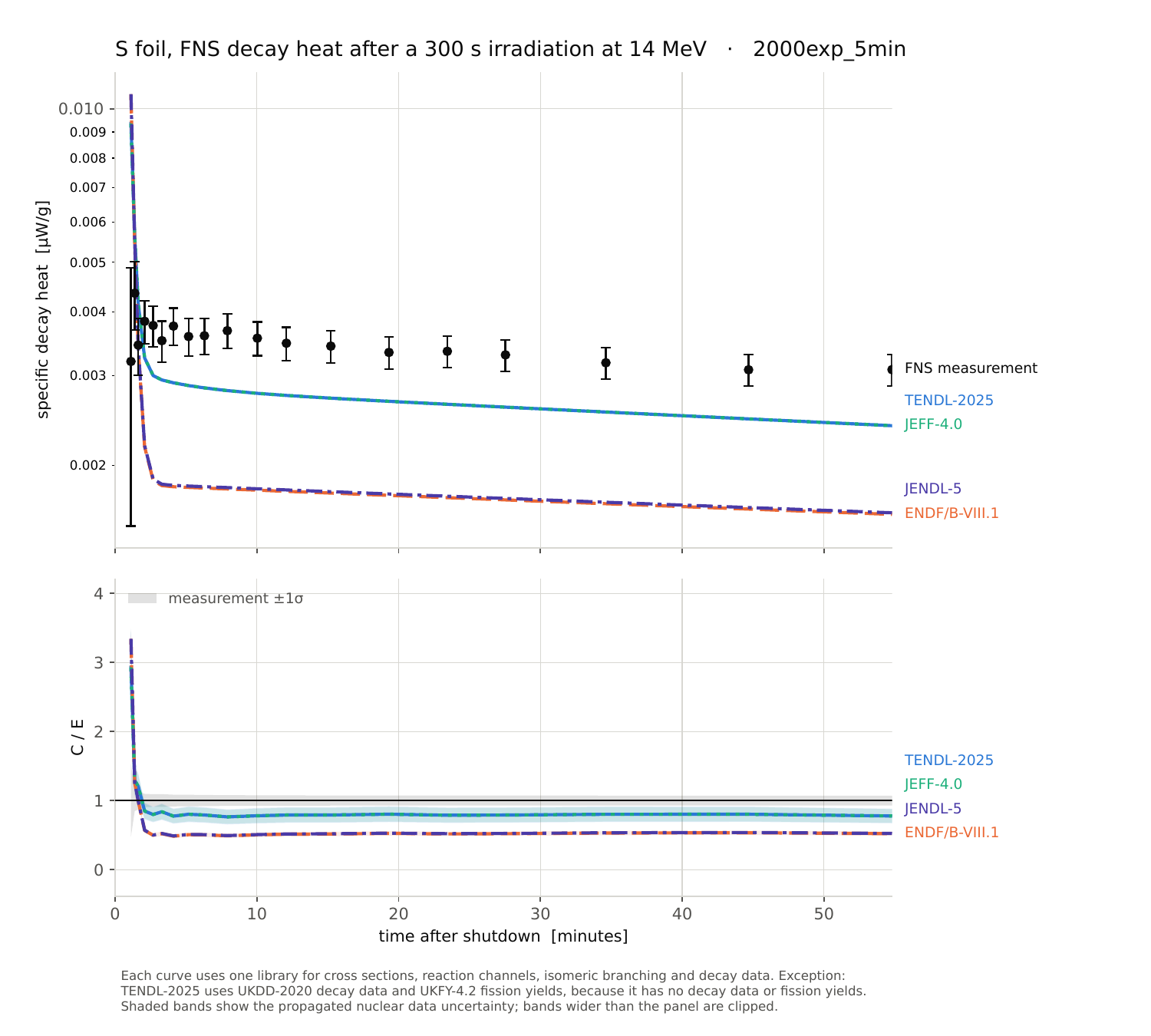}
\caption{S: decay heat against the measurement, with C/E beneath. Mean $|C/E-1|$ by library: JEFF-4.0 30\%, TENDL-2025 30\%, JENDL-5 54\%, ENDF/B-VIII.1 55\%.}
\end{figure}
\clearpage

\begin{figure}[H]
\centering
\includegraphics[width=\textwidth,height=0.82\textheight,keepaspectratio]{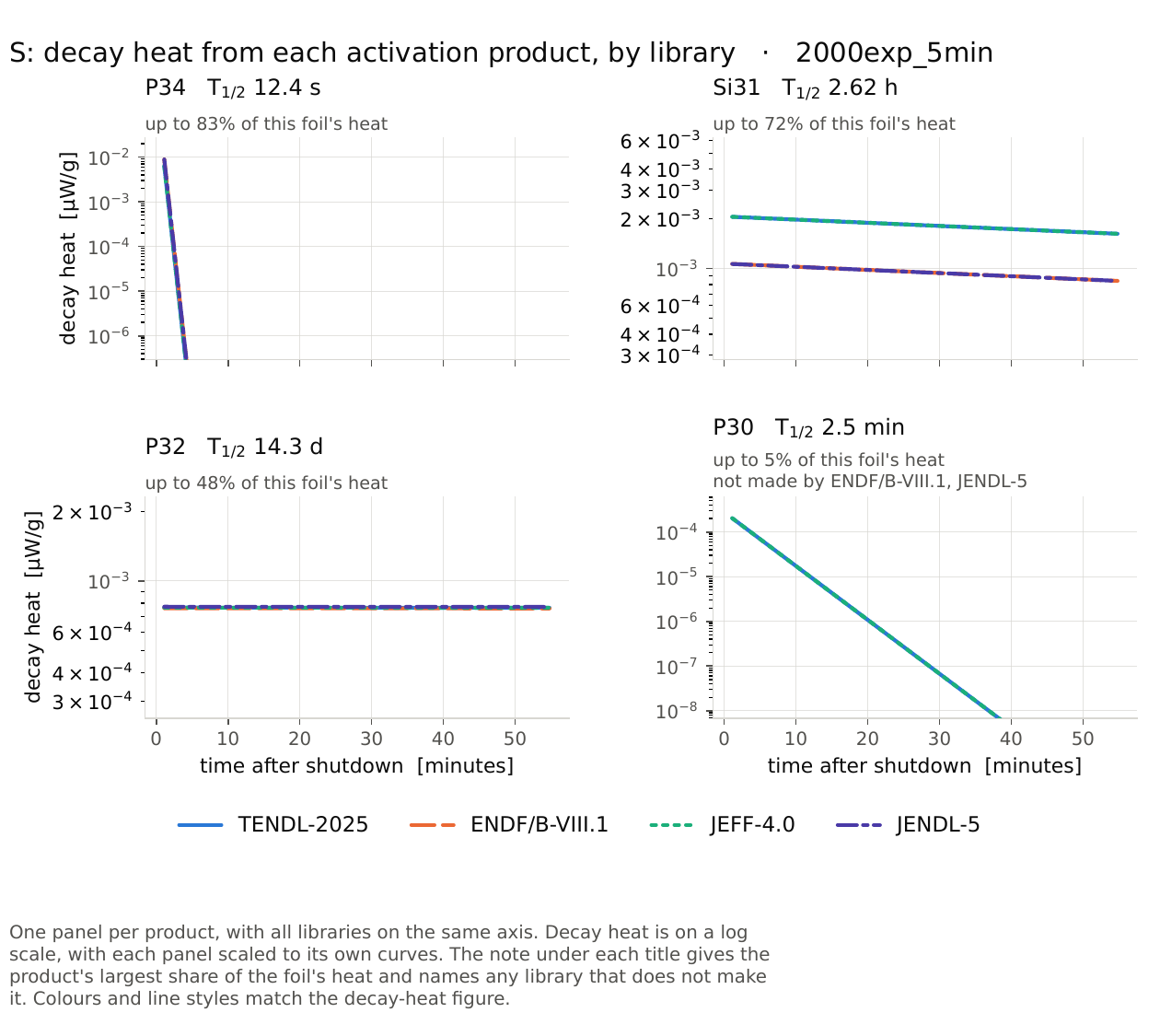}
\caption{S: share of the decay heat from each activation product during cooling.}
\end{figure}
\clearpage

\begin{figure}[H]
\centering
\includegraphics[width=\textwidth,height=0.82\textheight,keepaspectratio]{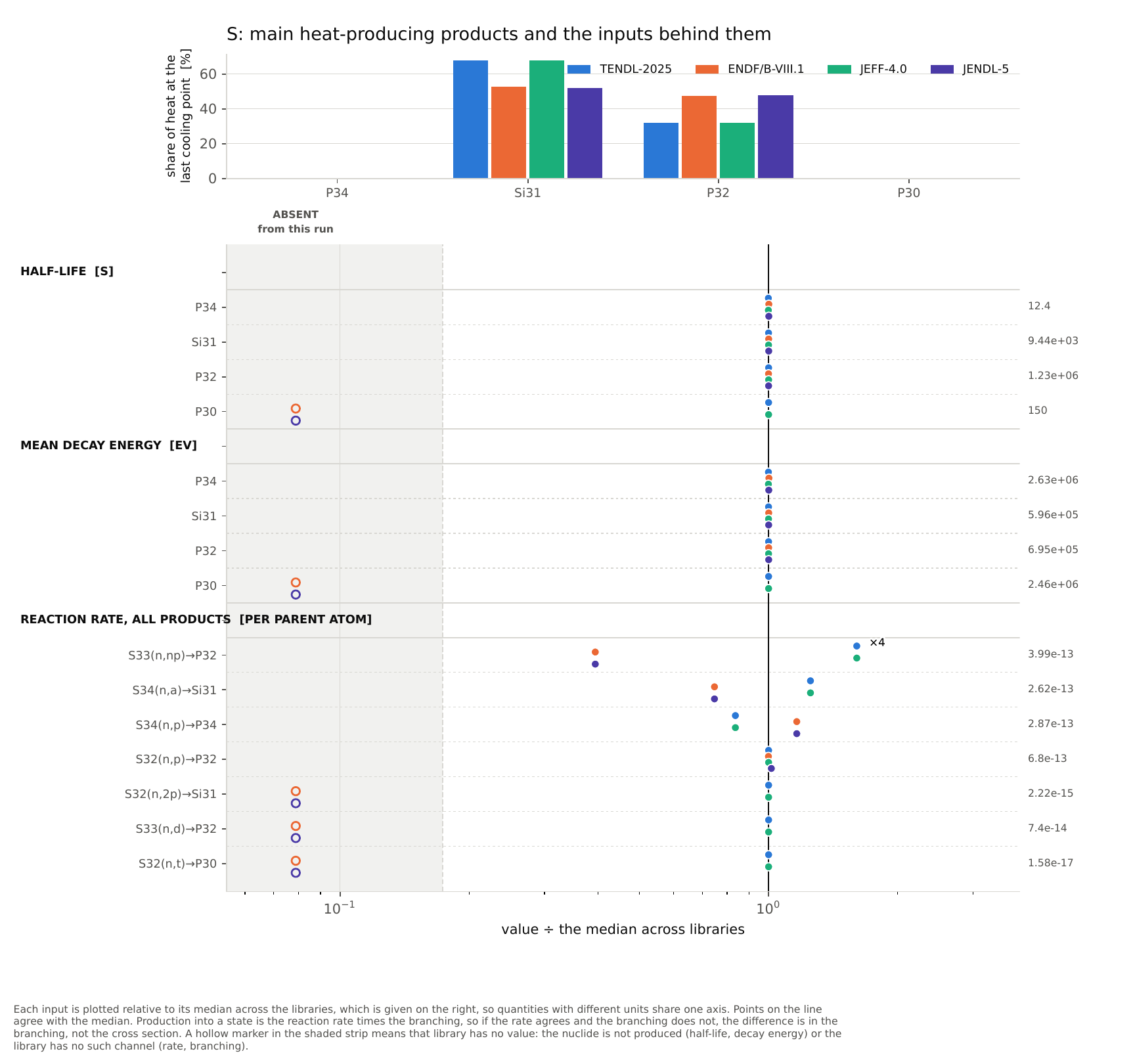}
\caption{S: inputs for the main heat-producing products, each relative to its median across the available libraries.}
\end{figure}
\clearpage

\begin{figure}[H]
\centering
\includegraphics[width=\textwidth,height=0.82\textheight,keepaspectratio]{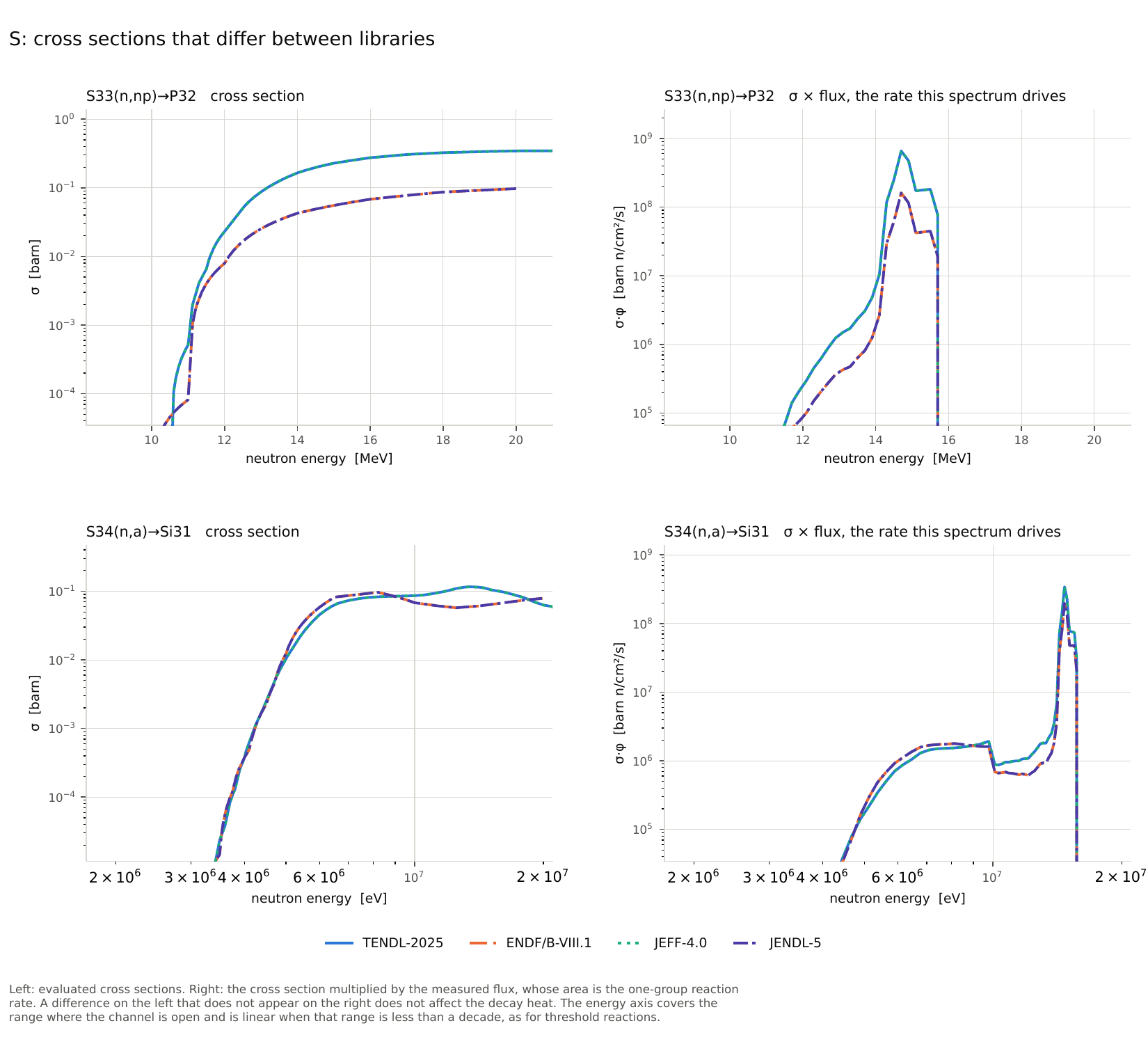}
\caption{S: cross sections of the channels whose spectrum-averaged rates differ, with the neutron spectrum and their product.}
\end{figure}
\clearpage

\subsubsection*{1996 seven-hour campaign}
The 1996 campaign irradiated this foil for 7.17 hours at position 7 and measured it out to 197.94 days. In the five-minute calculation P32 (half-life 14.3 days) gives up to 48 percent of the heat, and the longer irradiation and cooling test its production and decay data. JENDL-5 is closest. This campaign uses a different spectrum, so it is not included in the main rankings.

\begin{figure}[H]
\centering
\includegraphics[width=\textwidth,height=0.82\textheight,keepaspectratio]{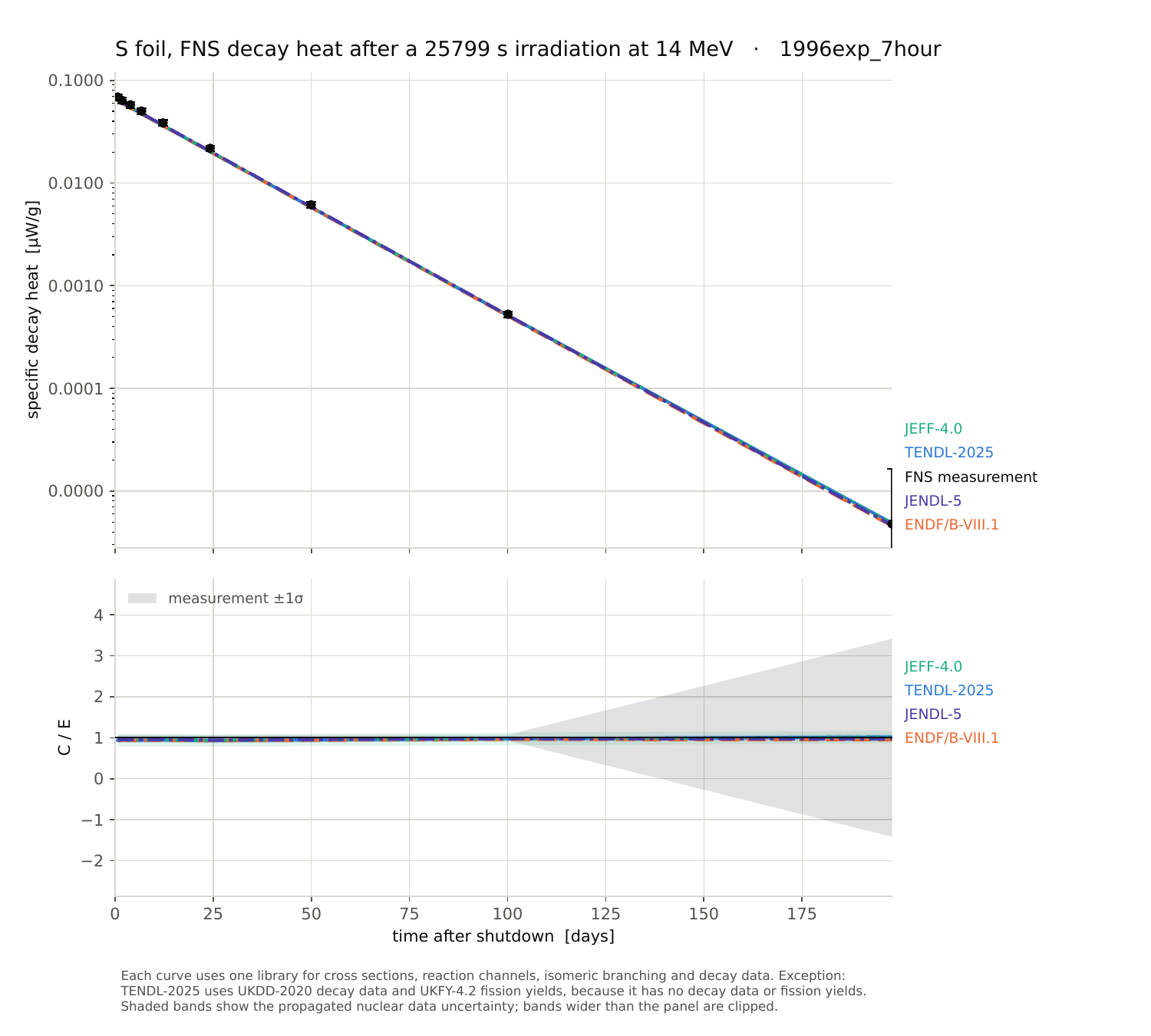}
\caption{S: 1996 seven-hour campaign, irradiation position 7. Mean $|C/E-1|$ by library: JENDL-5 5\%, JEFF-4.0 5\%, TENDL-2025 5\%, ENDF/B-VIII.1 6\%. }
\end{figure}
\clearpage

The libraries are between 30 and 55 percent from the measurement, which has a median uncertainty of 8 percent. TENDL-2025 is closest and ENDF/B-VIII.1 furthest.

\subsubsection*{Differences in the nuclear data}
\begin{itemize}\setlength{\itemsep}{6pt}
\item ENDF/B-VIII.1 gives the total reaction rate of S33(n,np)→P32 as 1.574e-13, against 6.413e-13 for the median of the others, a factor of 4.\\[2pt]{\itshape This is the total rate of the channel over all its products, so the difference is in the cross section, not in the split between states.}
\item ENDF/B-VIII.1, JENDL-5 give no heat from P30; the other libraries do.\\[2pt]{\itshape The heat breakdown alone does not show why. Possible causes are a missing reaction channel, a different state assignment, different decay data or a reporting threshold.}
\end{itemize}
\subsubsection*{Suggested measurements}
\begin{itemize}\setlength{\itemsep}{4pt}
\item Natural S has 4 isotopes, so the measured heat combines several parents. Enriched single-isotope targets would distinguish their contributions. The candidate parent is S33. Enrichment alone does not isolate a single reaction channel.
\item The reaction rates are averaged over the measured spectrum. Activation measurements at several well-defined neutron energies, including below 14 MeV, would show which energy range the differences come from.
\end{itemize}
\clearpage
\subsection{Sb: Antimony}\label{foil:Sb}
\begin{figure}[H]
\centering
\includegraphics[width=\textwidth,height=0.82\textheight,keepaspectratio]{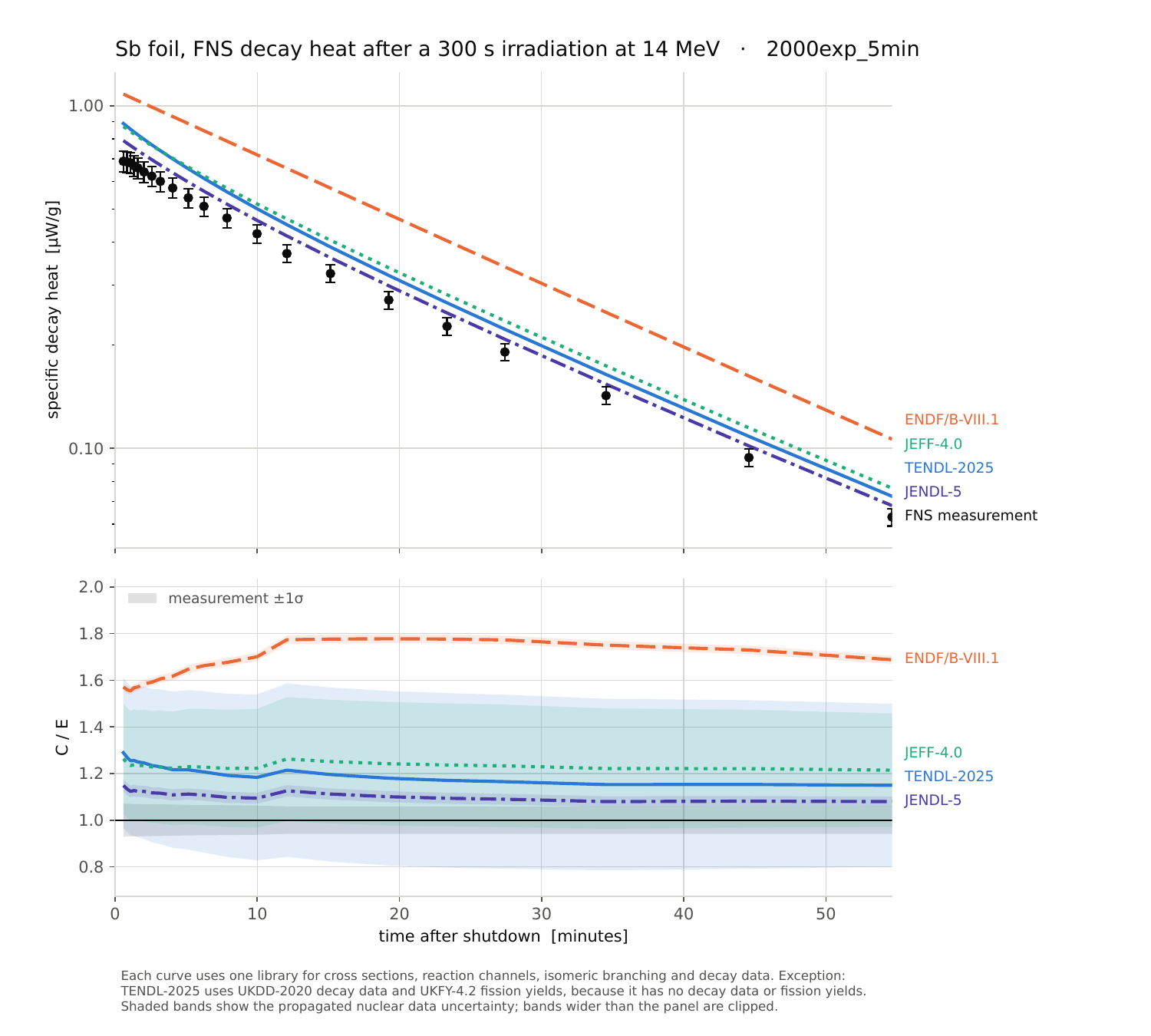}
\caption{Sb: decay heat against the measurement, with C/E beneath. Mean $|C/E-1|$ by library: JENDL-5 11\%, TENDL-2025 21\%, JEFF-4.0 23\%, ENDF/B-VIII.1 66\%.}
\end{figure}
\clearpage

\begin{figure}[H]
\centering
\includegraphics[width=\textwidth,height=0.82\textheight,keepaspectratio]{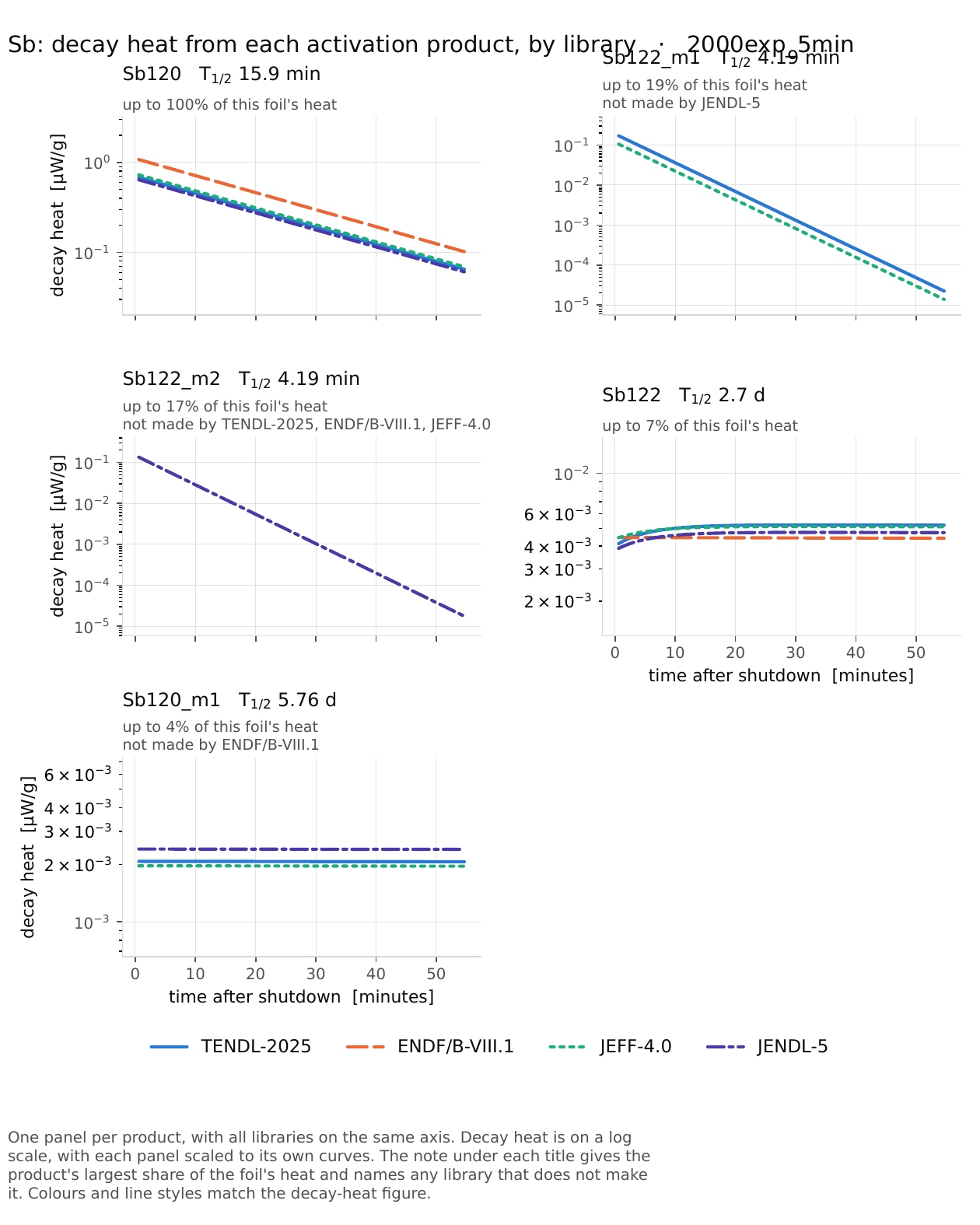}
\caption{Sb: share of the decay heat from each activation product during cooling.}
\end{figure}
\clearpage

\begin{figure}[H]
\centering
\includegraphics[width=\textwidth,height=0.82\textheight,keepaspectratio]{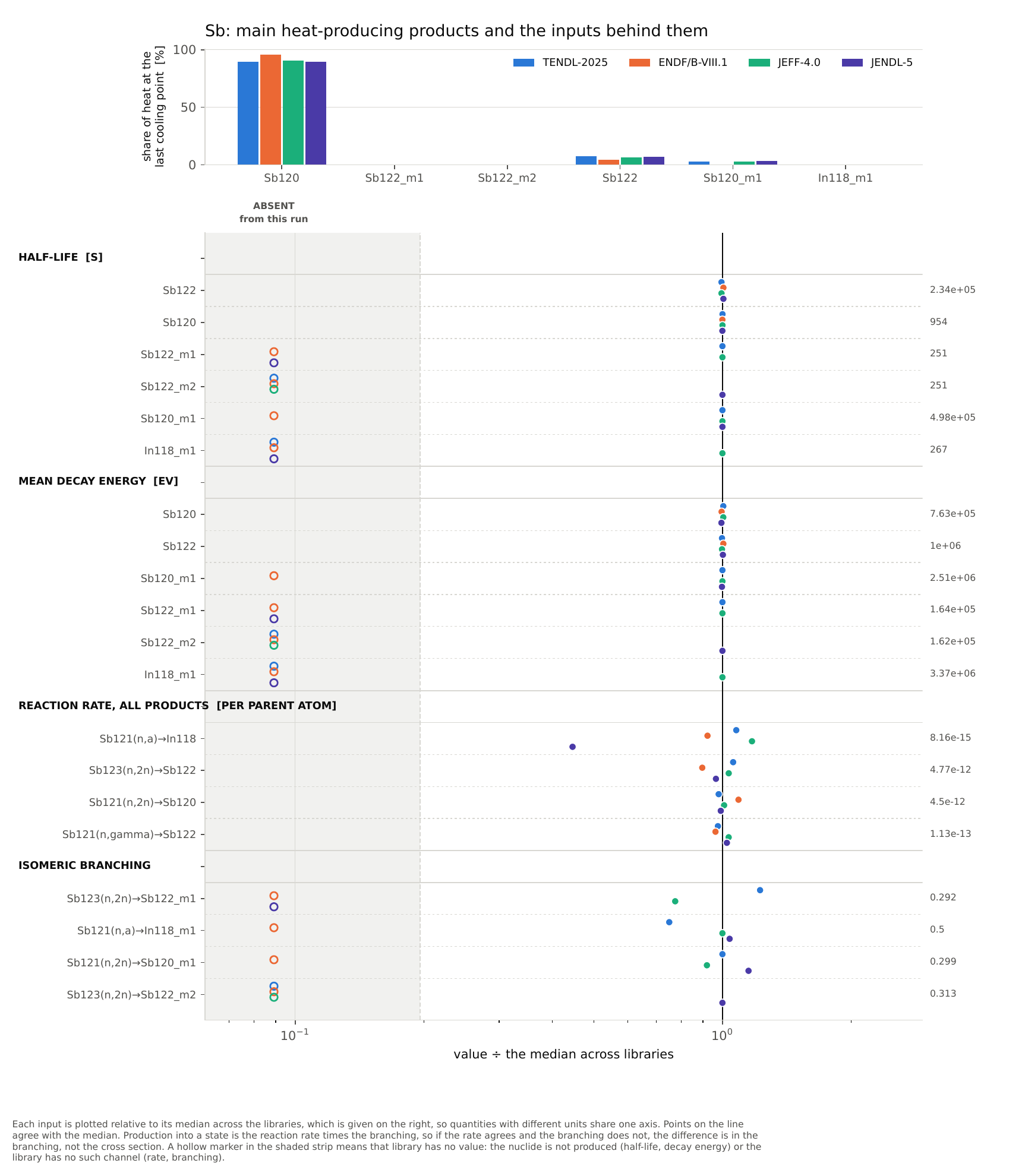}
\caption{Sb: inputs for the main heat-producing products, each relative to its median across the available libraries.}
\end{figure}
\clearpage

\begin{figure}[H]
\centering
\includegraphics[width=\textwidth,height=0.82\textheight,keepaspectratio]{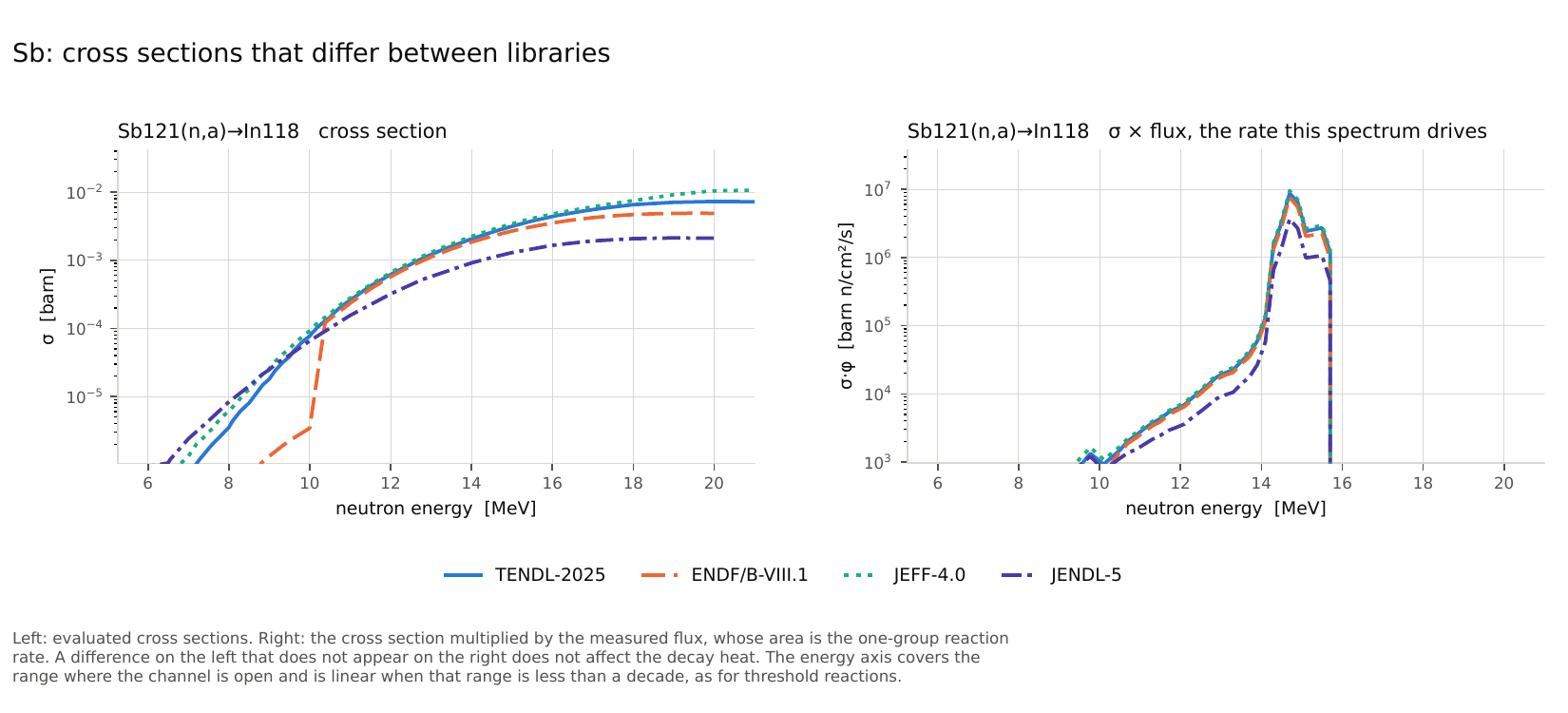}
\caption{Sb: cross sections of the channels whose spectrum-averaged rates differ, with the neutron spectrum and their product.}
\end{figure}
\clearpage

\begin{figure}[H]
\centering
\includegraphics[width=\textwidth,height=0.82\textheight,keepaspectratio]{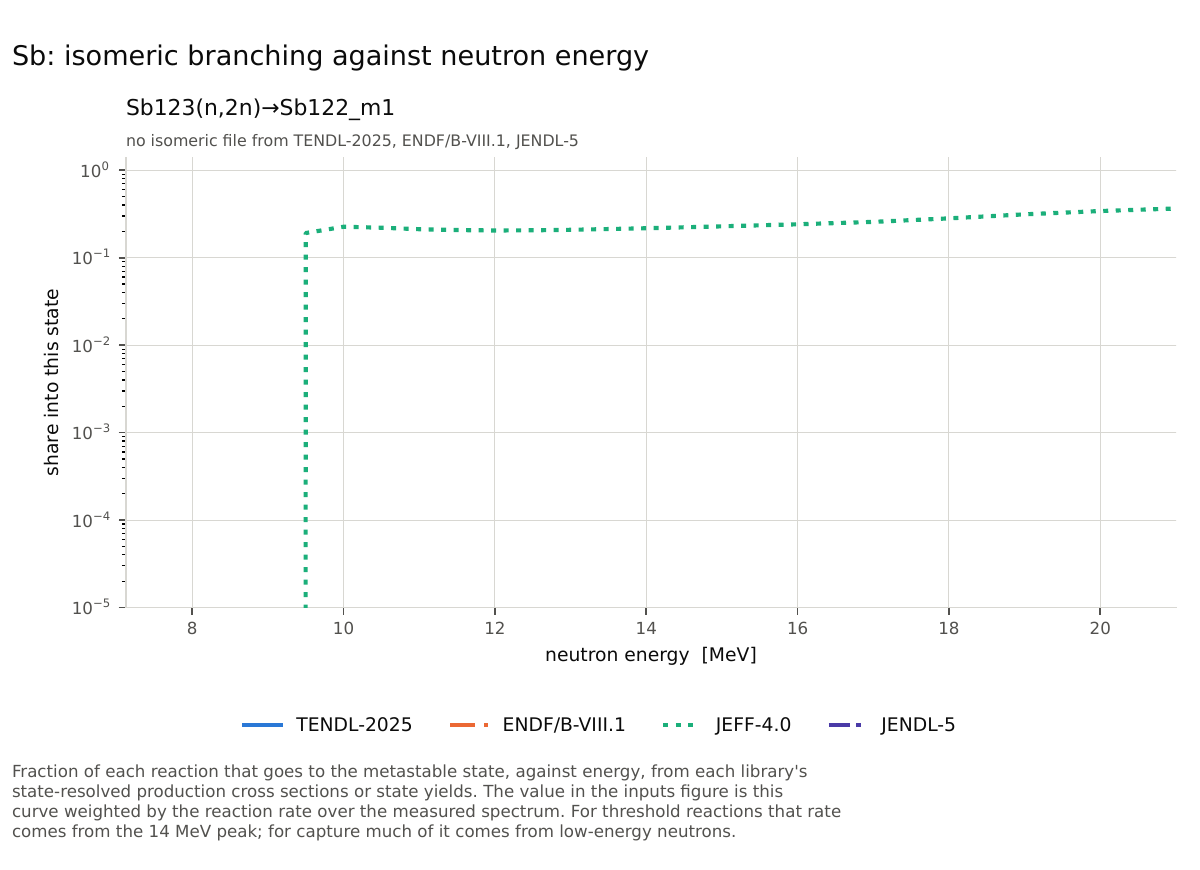}
\caption{Sb: isomeric branching ratio against neutron energy for the channels whose spectrum-averaged ratio differs between libraries.}
\end{figure}
\clearpage

ENDF/B-VIII.1 stands out at 66 percent from the measurement, against 23 percent for the next furthest library and 11 percent for JENDL-5. The median measurement uncertainty is 6 percent.

\subsubsection*{Differences in the nuclear data}
\begin{itemize}\setlength{\itemsep}{6pt}
\item JENDL-5 gives the total reaction rate of Sb121(n,a)→In118 as 3.638e-15, against 8.793e-15 for the median of the others, a factor of 2.\\[2pt]{\itshape This is the total rate of the channel over all its products, so the difference is in the cross section, not in the split between states.}
\item JENDL-5 gives no heat from Sb122\_m1; the other libraries do.\\[2pt]{\itshape The heat breakdown alone does not show why. Possible causes are a missing reaction channel, a different state assignment, different decay data or a reporting threshold.}
\item TENDL-2025, ENDF/B-VIII.1, JEFF-4.0 give no heat from Sb122\_m2; the other libraries do.
\item ENDF/B-VIII.1 gives no heat from Sb120\_m1; the other libraries do.
\end{itemize}
\subsubsection*{Suggested measurements}
\begin{itemize}\setlength{\itemsep}{4pt}
\item Natural Sb has 2 isotopes, so the measured heat combines several parents. Enriched single-isotope targets would distinguish their contributions. The candidate parent is Sb121. Enrichment alone does not isolate a single reaction channel.
\item This foil makes Sb122 (half-life 2.7 days), which gives up to 7 percent of the calculated heat. A five-minute irradiation and 55 minute measurement only see the start of this contribution. There is no seven-hour FNS measurement of this foil. A longer irradiation followed for days to months would be needed to test this product's production and decay.
\item The reaction rates are averaged over the measured spectrum. Activation measurements at several well-defined neutron energies, including below 14 MeV, would show which energy range the differences come from.
\end{itemize}
\clearpage
\subsection{Sc: Scandium}\label{foil:Sc}
\begin{figure}[H]
\centering
\includegraphics[width=\textwidth,height=0.82\textheight,keepaspectratio]{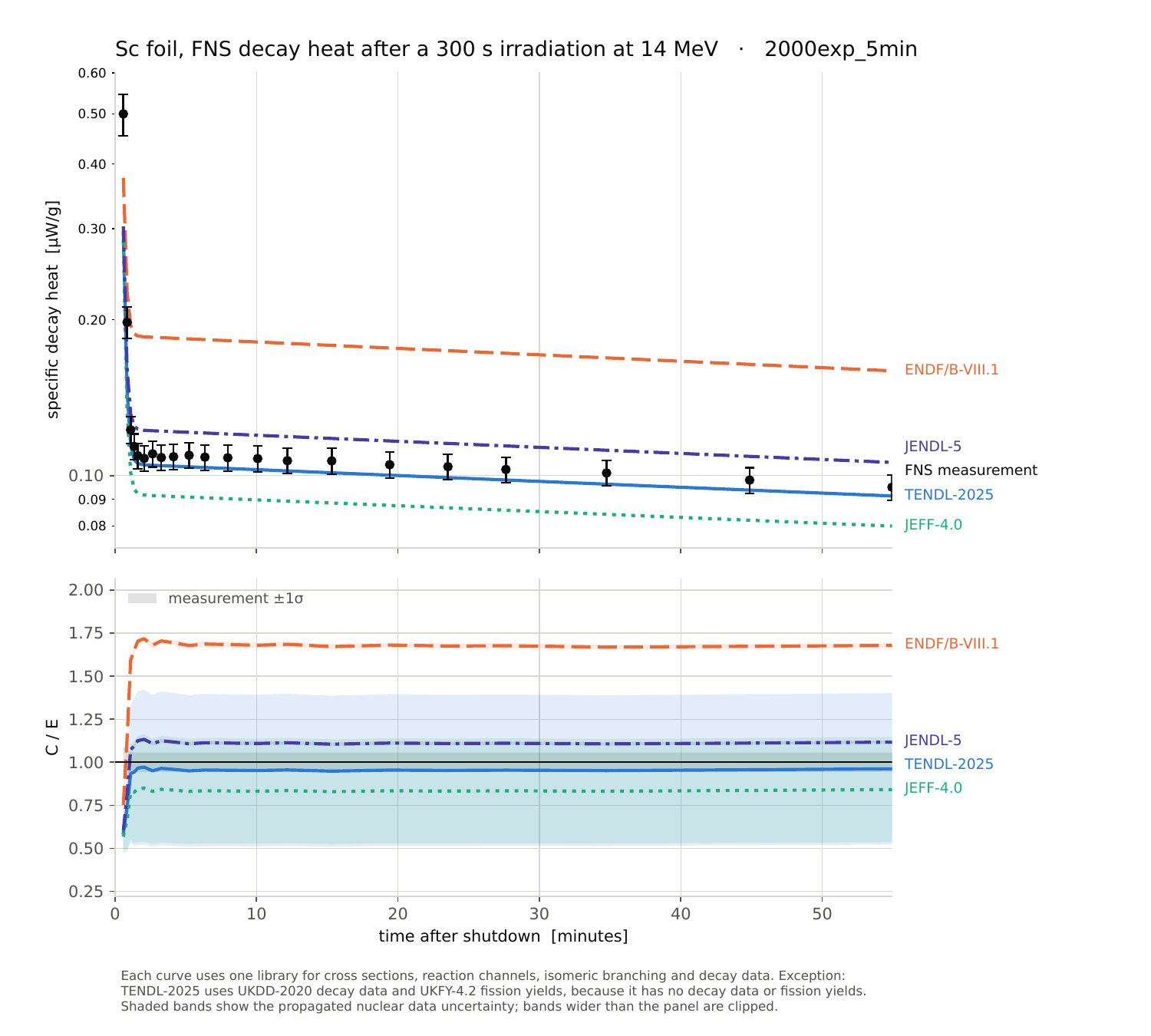}
\caption{Sc: decay heat against the measurement, with C/E beneath. Mean $|C/E-1|$ by library: TENDL-2025 7\%, JENDL-5 13\%, JEFF-4.0 19\%, ENDF/B-VIII.1 63\%.}
\end{figure}
\clearpage

\begin{figure}[H]
\centering
\includegraphics[width=\textwidth,height=0.82\textheight,keepaspectratio]{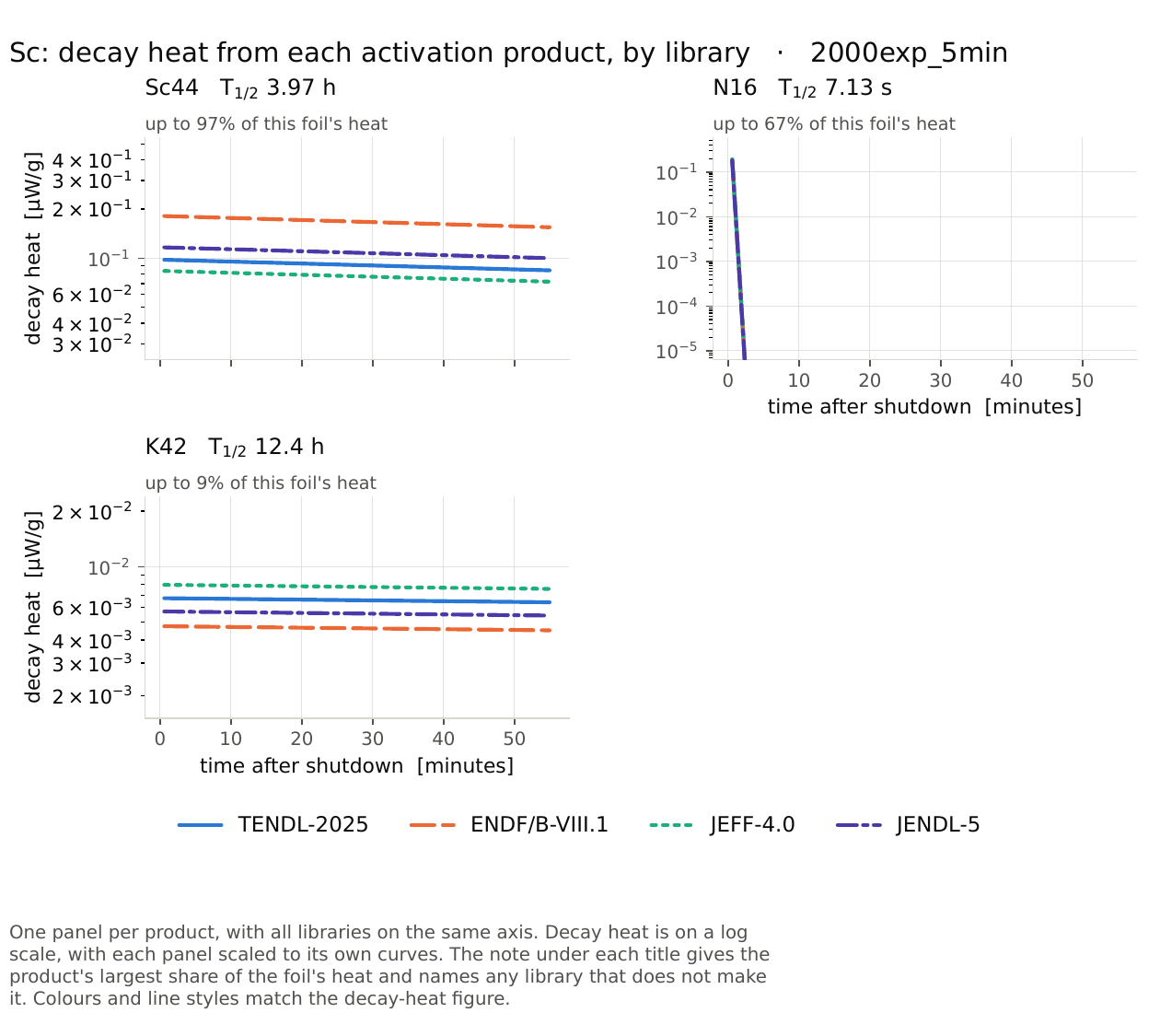}
\caption{Sc: share of the decay heat from each activation product during cooling.}
\end{figure}
\clearpage

\begin{figure}[H]
\centering
\includegraphics[width=\textwidth,height=0.82\textheight,keepaspectratio]{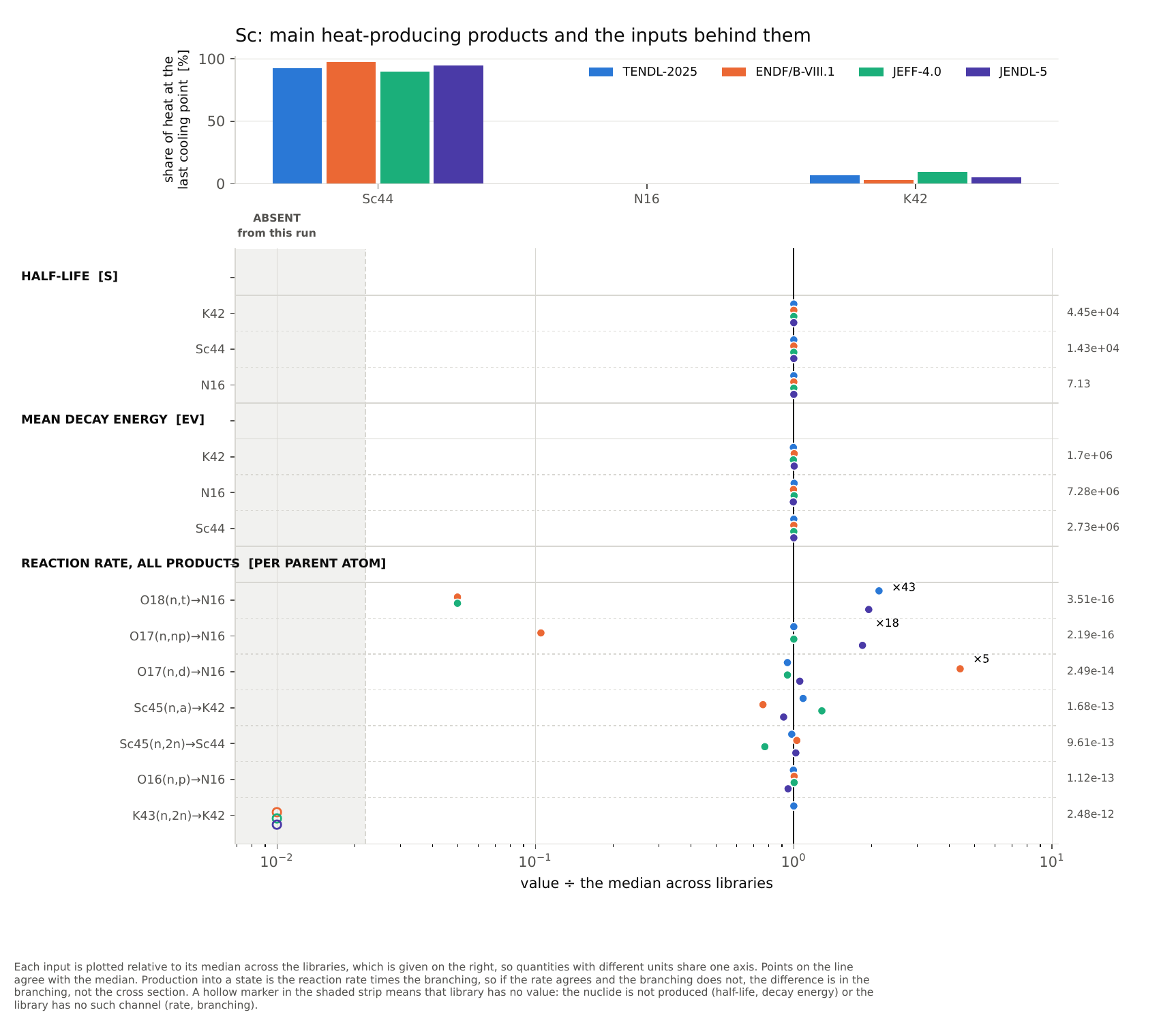}
\caption{Sc: inputs for the main heat-producing products, each relative to its median across the available libraries.}
\end{figure}
\clearpage

\begin{figure}[H]
\centering
\includegraphics[width=\textwidth,height=0.82\textheight,keepaspectratio]{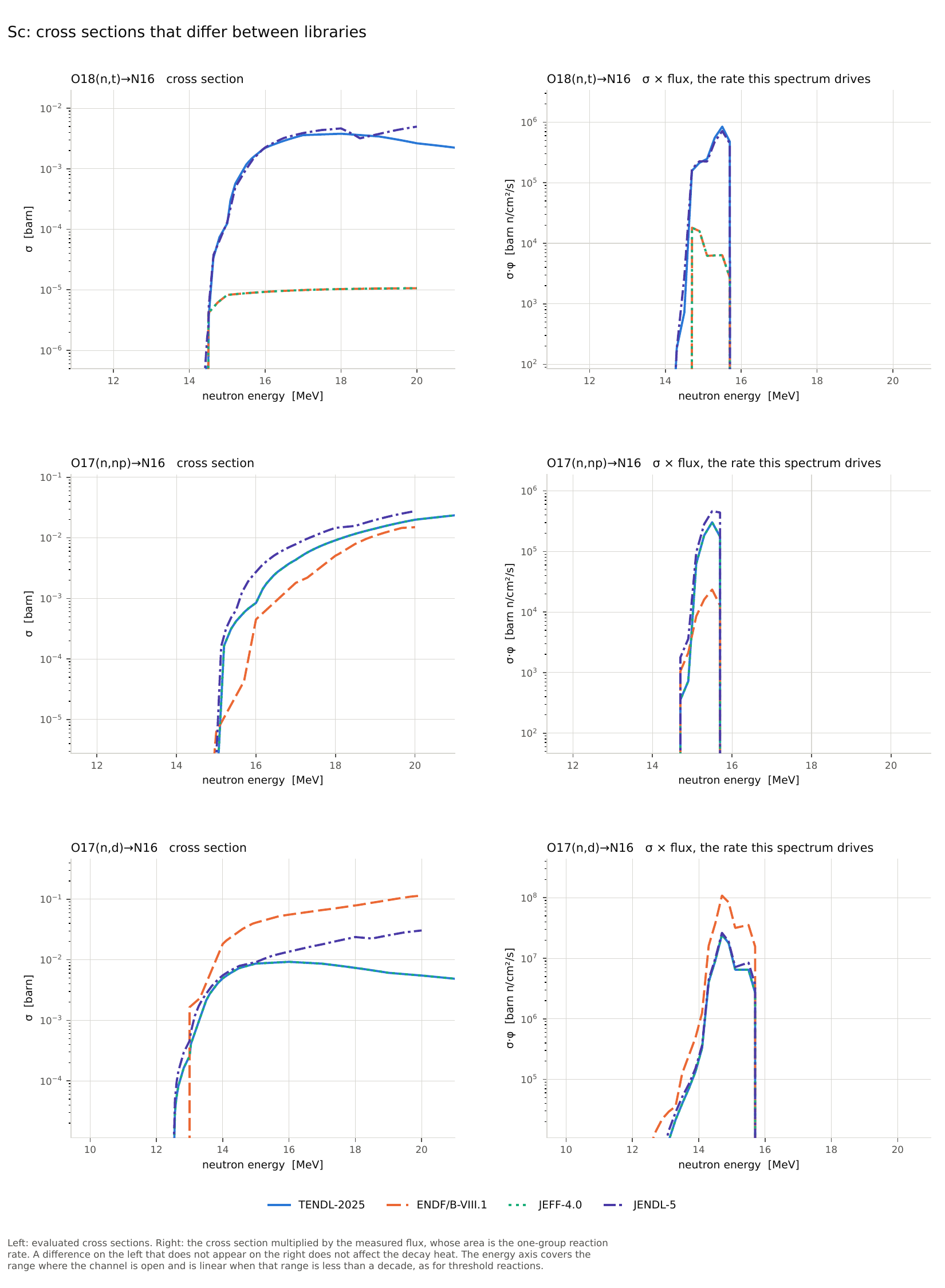}
\caption{Sc: cross sections of the channels whose spectrum-averaged rates differ, with the neutron spectrum and their product.}
\end{figure}
\clearpage

ENDF/B-VIII.1 stands out at 63 percent from the measurement, against 19 percent for the next furthest library and 7 percent for TENDL-2025. The median measurement uncertainty is 6 percent.

\subsubsection*{Differences in the nuclear data}
\begin{itemize}\setlength{\itemsep}{6pt}
\item ENDF/B-VIII.1 gives the total reaction rate of O18(n,t)→N16 as 1.751e-17, against 6.84e-16 for the median of the others, a factor of 39.\\[2pt]{\itshape This is the total rate of the channel over all its products, so the difference is in the cross section, not in the split between states.}
\item ENDF/B-VIII.1 gives the total reaction rate of O17(n,np)→N16 as 2.295e-17, against 2.186e-16 for the median of the others, a factor of 10.
\item ENDF/B-VIII.1 gives the total reaction rate of O17(n,d)→N16 as 1.096e-13, against 2.351e-14 for the median of the others, a factor of 5.
\end{itemize}
\subsubsection*{Suggested measurements}
\begin{itemize}\setlength{\itemsep}{4pt}
\item Sc has a single stable isotope, so the foil is already a single-isotope target and enrichment would not help.
\item The sample is a compound that also contains O. Their activation adds to the measured heat; a metal or purer sample would remove it.
\item The reaction rates are averaged over the measured spectrum. Activation measurements at several well-defined neutron energies, including below 14 MeV, would show which energy range the differences come from.
\end{itemize}
\clearpage
\subsection{Se: Selenium}\label{foil:Se}
\begin{figure}[H]
\centering
\includegraphics[width=\textwidth,height=0.82\textheight,keepaspectratio]{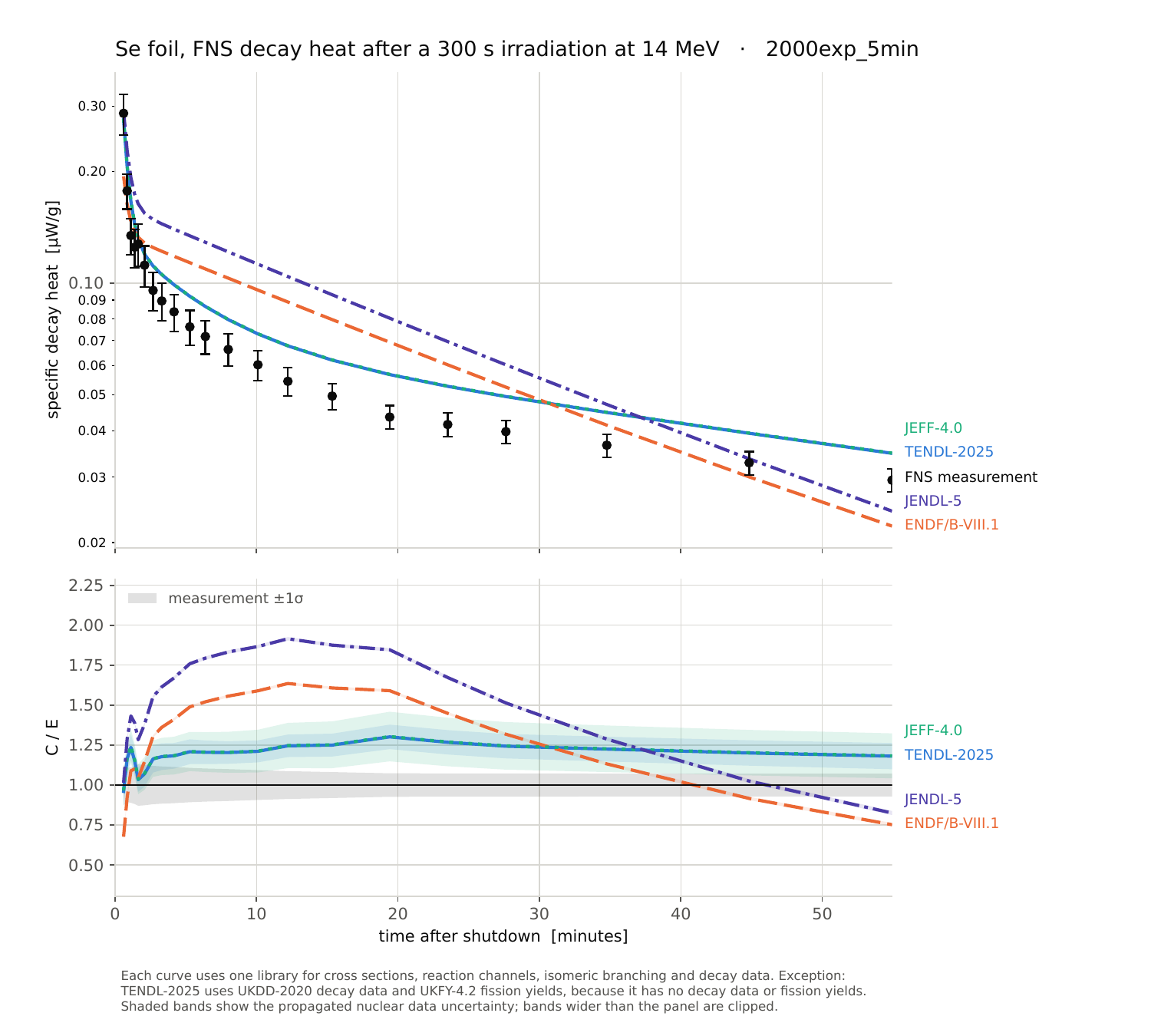}
\caption{Se: decay heat against the measurement, with C/E beneath. Mean $|C/E-1|$ by library: TENDL-2025 19\%, JEFF-4.0 19\%, ENDF/B-VIII.1 34\%, JENDL-5 53\%.}
\end{figure}
\clearpage

\begin{figure}[H]
\centering
\includegraphics[width=\textwidth,height=0.82\textheight,keepaspectratio]{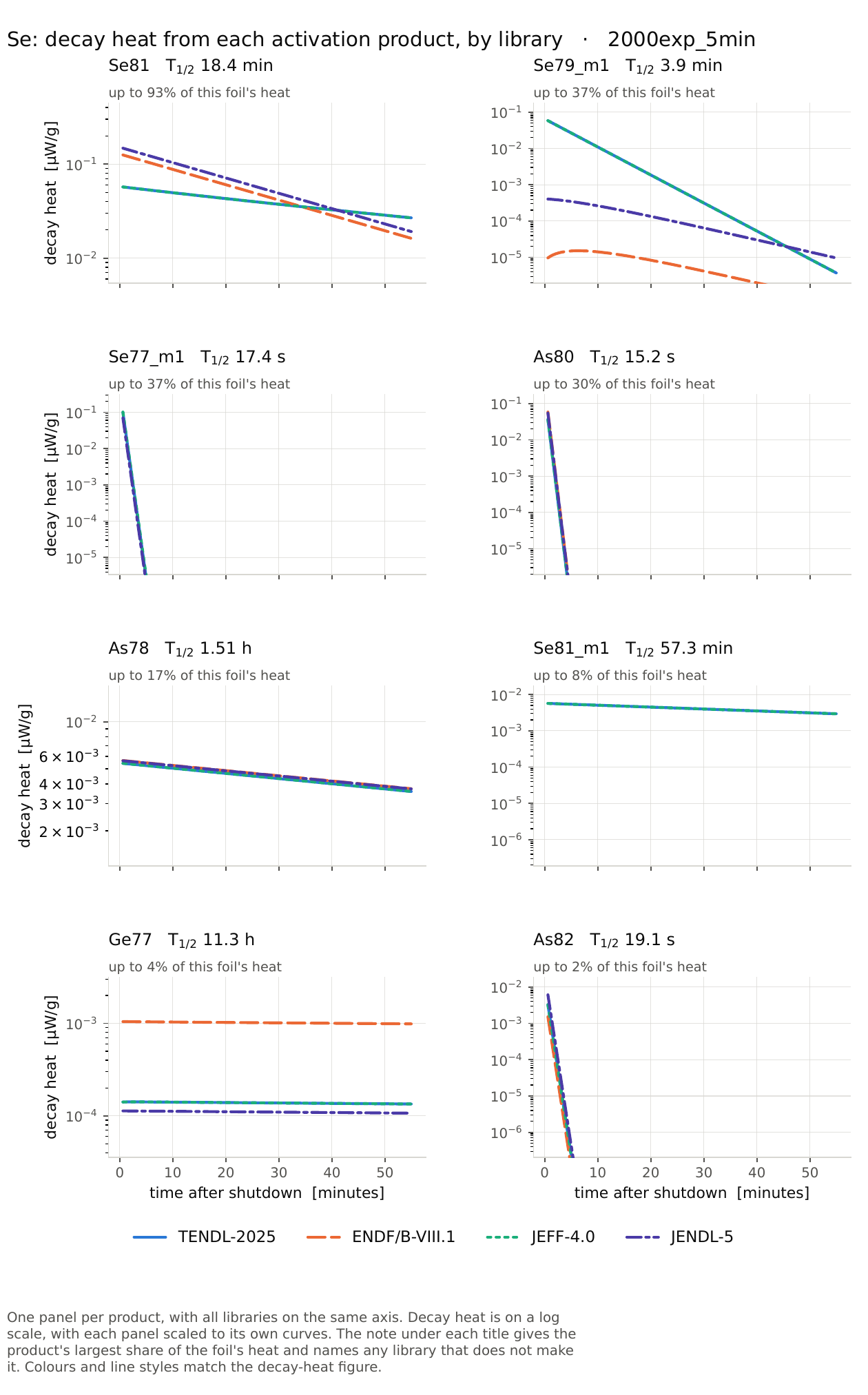}
\caption{Se: share of the decay heat from each activation product during cooling.}
\end{figure}
\clearpage

\begin{figure}[H]
\centering
\includegraphics[width=\textwidth,height=0.82\textheight,keepaspectratio]{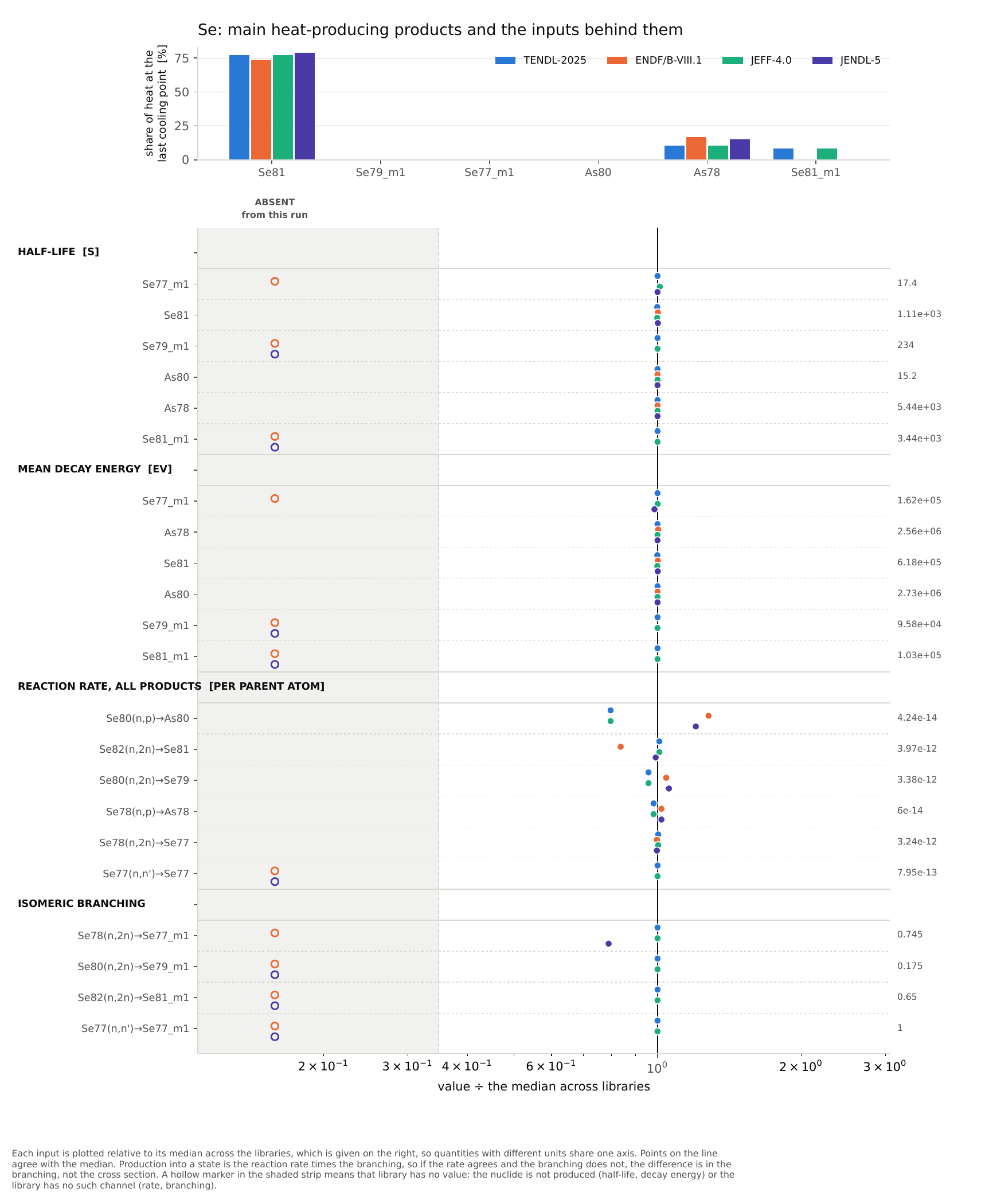}
\caption{Se: inputs for the main heat-producing products, each relative to its median across the available libraries.}
\end{figure}
\clearpage

\begin{figure}[H]
\centering
\includegraphics[width=\textwidth,height=0.82\textheight,keepaspectratio]{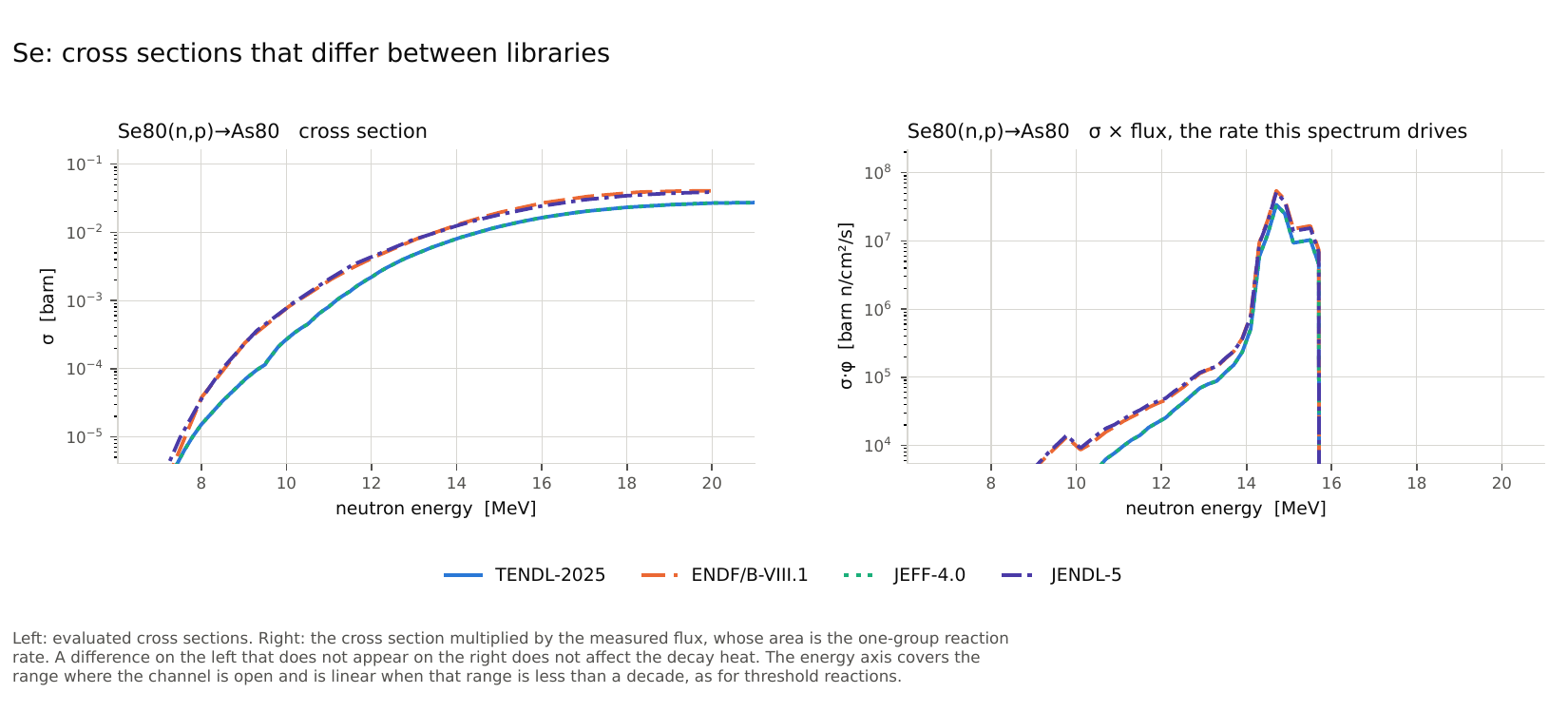}
\caption{Se: cross sections of the channels whose spectrum-averaged rates differ, with the neutron spectrum and their product.}
\end{figure}
\clearpage

The libraries are between 19 and 53 percent from the measurement, which has a median uncertainty of 10 percent. TENDL-2025 is closest and JENDL-5 furthest.

\subsubsection*{Suggested measurements}
\begin{itemize}\setlength{\itemsep}{4pt}
\item This foil makes As76 (half-life 1.09 days), which gives up to 1 percent of the calculated heat. A five-minute irradiation and 55 minute measurement only see the start of this contribution. There is no seven-hour FNS measurement of this foil. A longer irradiation followed for days to months would be needed to test this product's production and decay.
\item No input differs enough between the libraries to be listed. An error shared by all of them, or a smaller difference, is still possible.
\end{itemize}
\clearpage
\subsection{Si: Silicon}\label{foil:Si}
\begin{figure}[H]
\centering
\includegraphics[width=\textwidth,height=0.82\textheight,keepaspectratio]{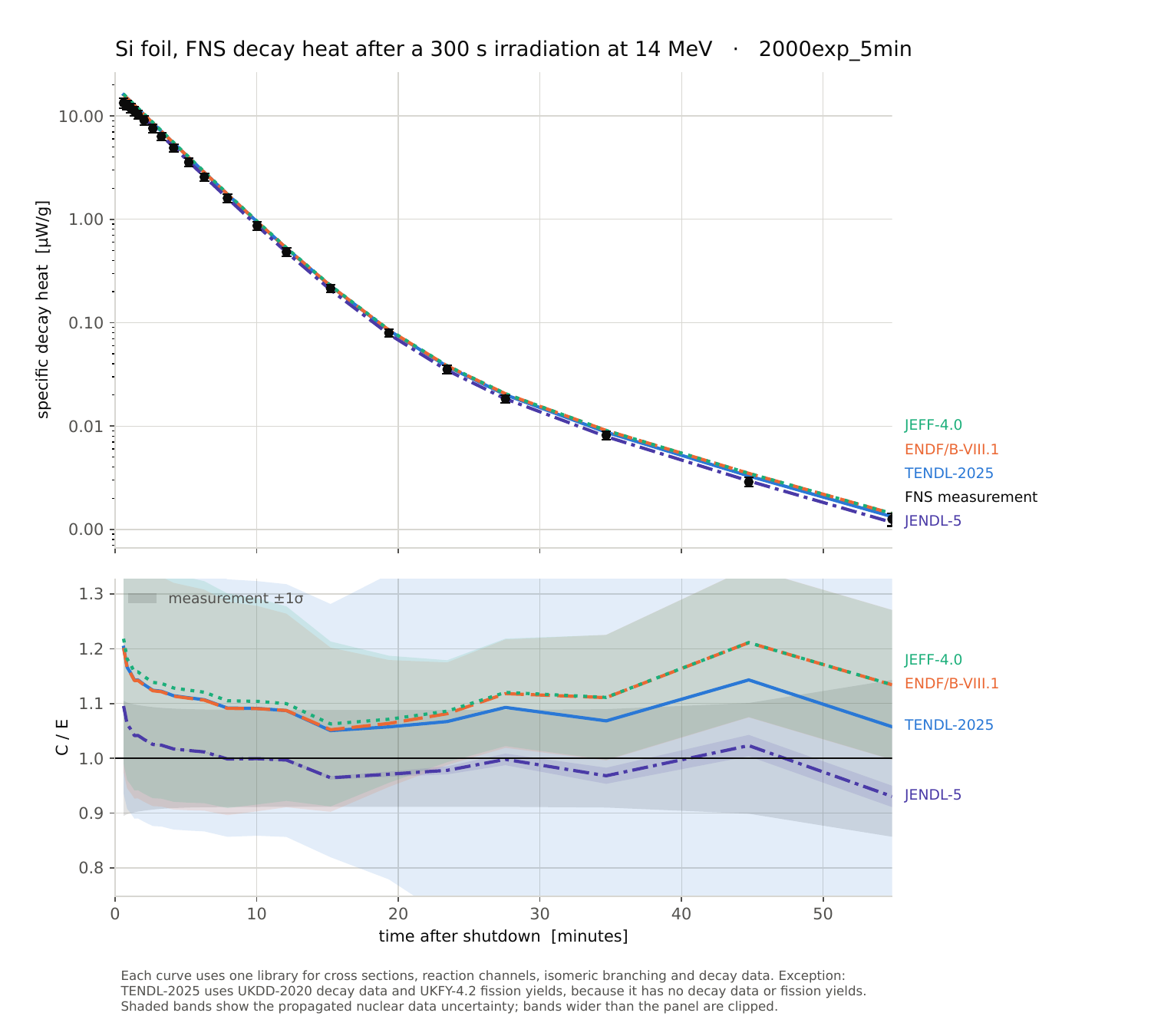}
\caption{Si: decay heat against the measurement, with C/E beneath. Mean $|C/E-1|$ by library: JENDL-5 3\%, TENDL-2025 11\%, ENDF/B-VIII.1 12\%, JEFF-4.0 13\%.}
\end{figure}
\clearpage

\begin{figure}[H]
\centering
\includegraphics[width=\textwidth,height=0.82\textheight,keepaspectratio]{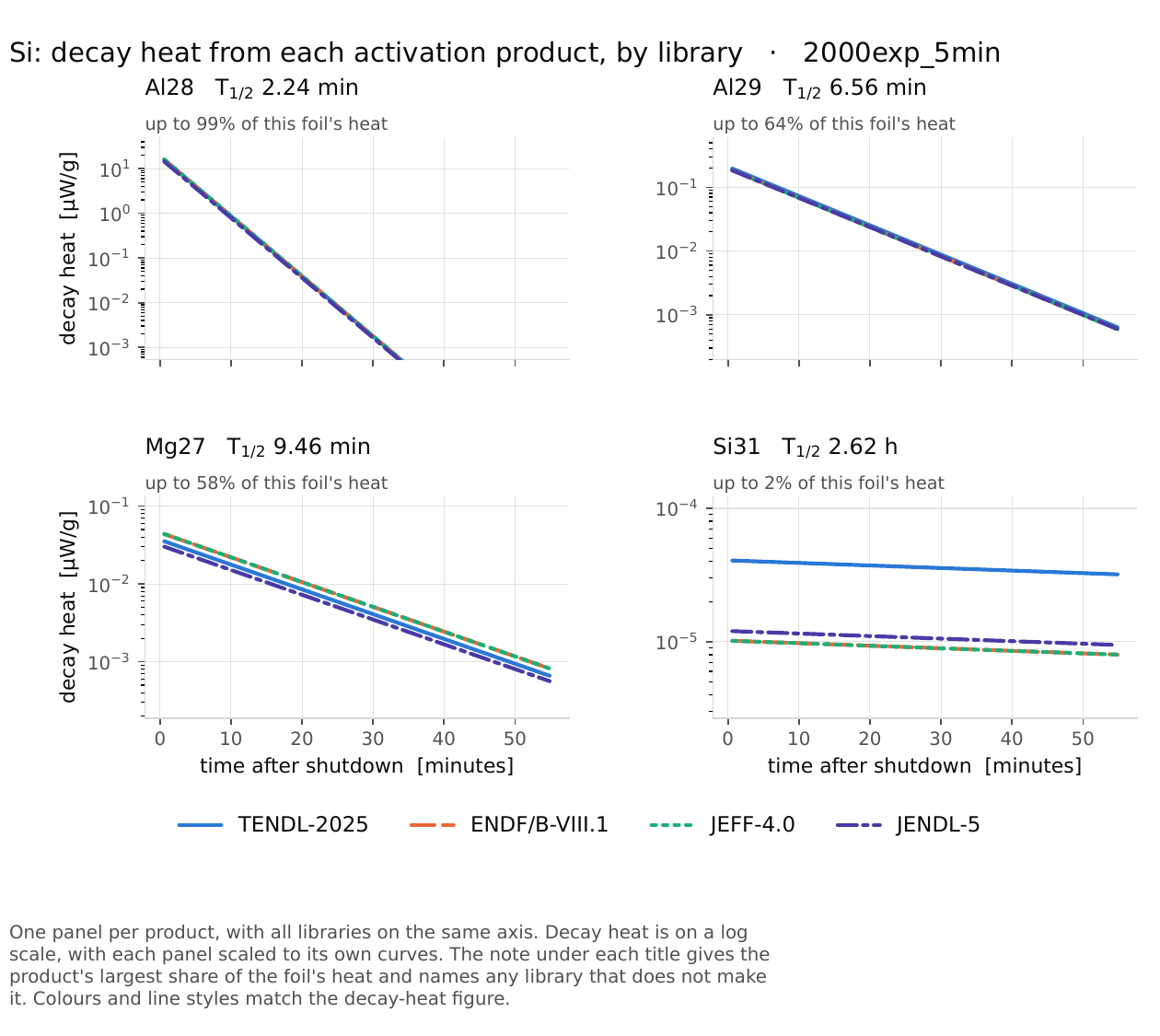}
\caption{Si: share of the decay heat from each activation product during cooling.}
\end{figure}
\clearpage

\begin{figure}[H]
\centering
\includegraphics[width=\textwidth,height=0.82\textheight,keepaspectratio]{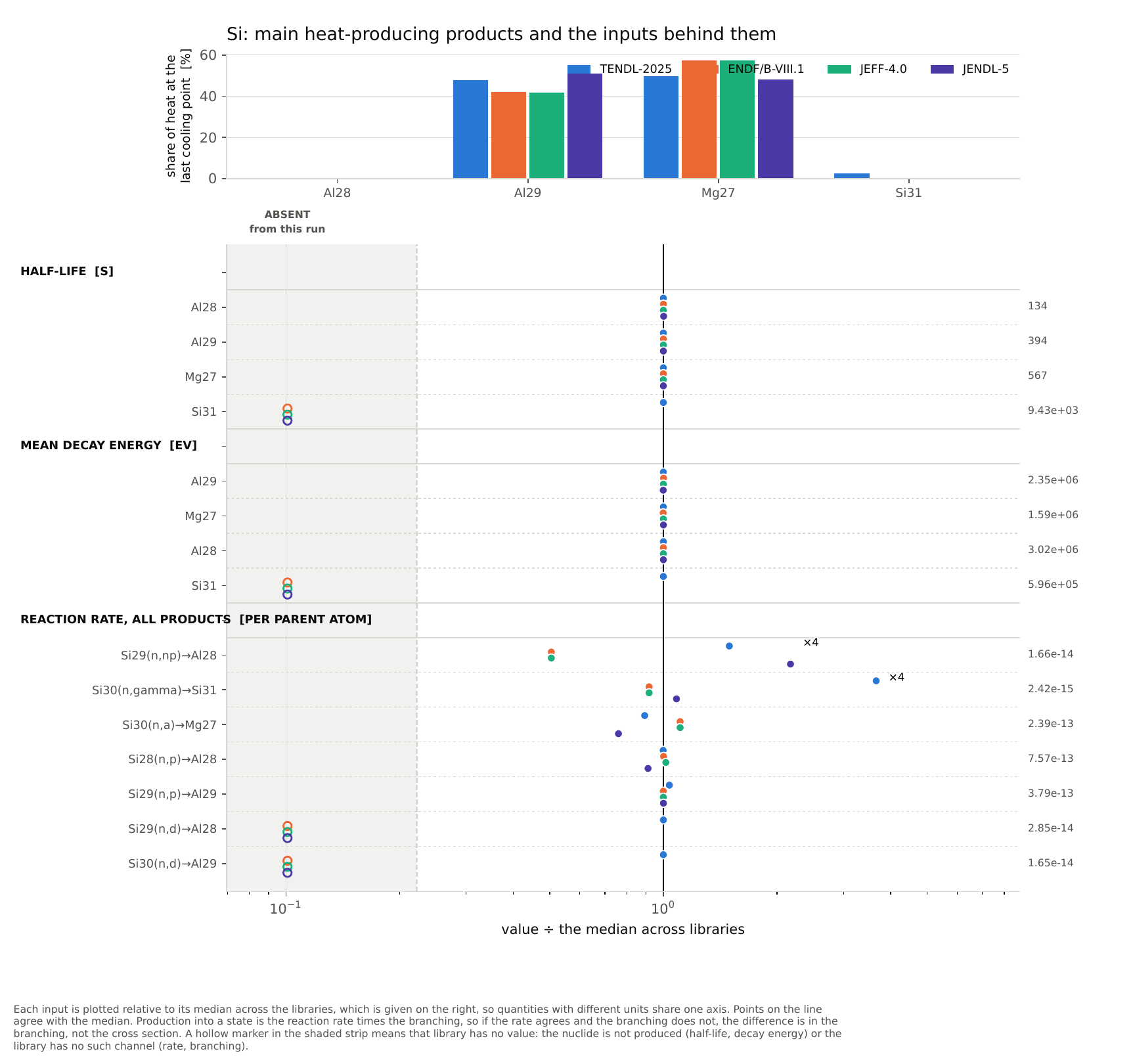}
\caption{Si: inputs for the main heat-producing products, each relative to its median across the available libraries.}
\end{figure}
\clearpage

\begin{figure}[H]
\centering
\includegraphics[width=\textwidth,height=0.82\textheight,keepaspectratio]{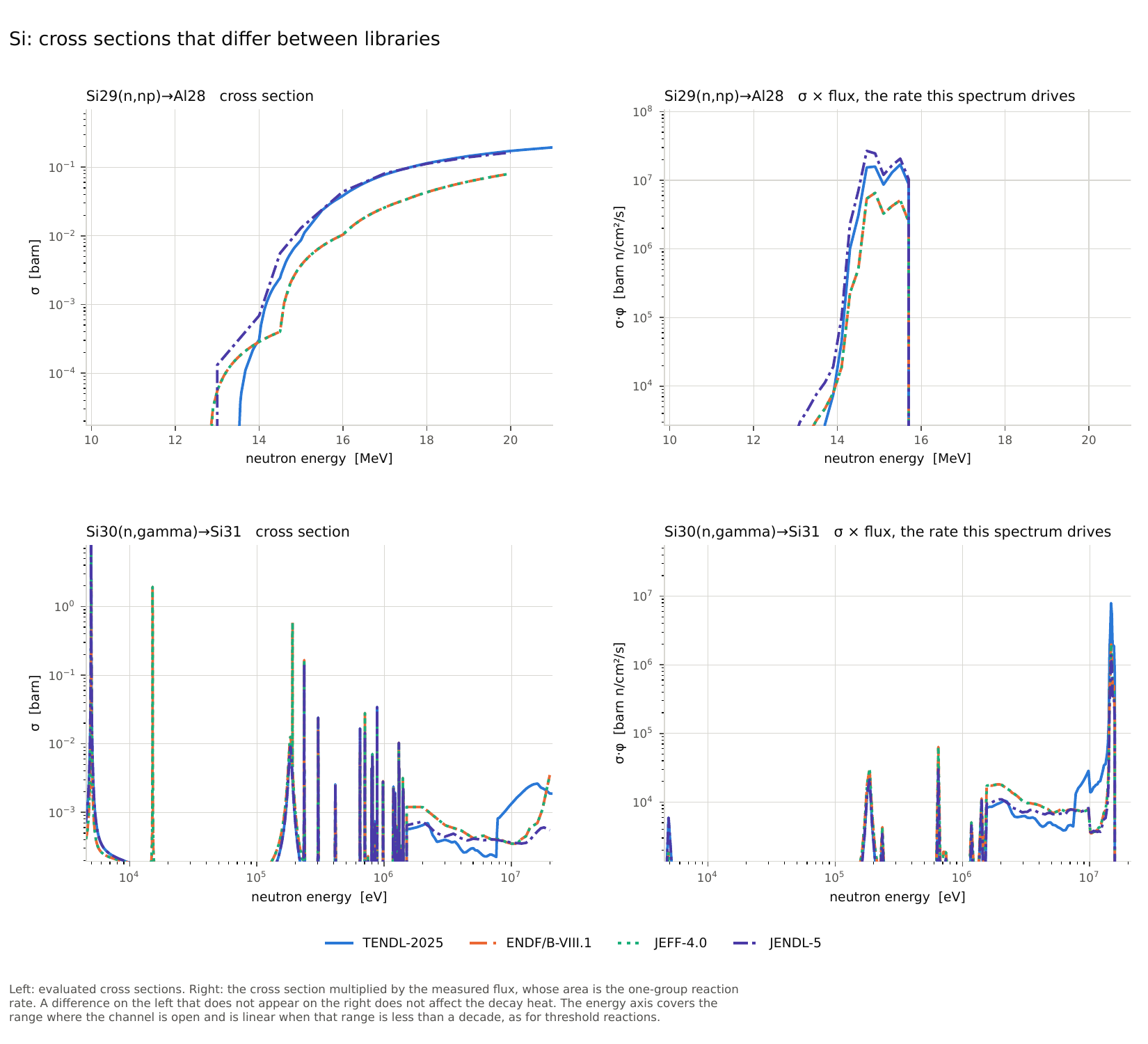}
\caption{Si: cross sections of the channels whose spectrum-averaged rates differ, with the neutron spectrum and their product.}
\end{figure}
\clearpage

The libraries are between 3 and 13 percent from the measurement, which has a median uncertainty of 9 percent. JENDL-5 is closest and JEFF-4.0 furthest.

\subsubsection*{Differences in the nuclear data}
\begin{itemize}\setlength{\itemsep}{6pt}
\item JENDL-5 gives the total reaction rate of Si29(n,np)→Al28 as 3.612e-14, against 8.393e-15 for the median of the others, a factor of 4.\\[2pt]{\itshape This is the total rate of the channel over all its products, so the difference is in the cross section, not in the split between states.}
\item TENDL-2025 gives the total reaction rate of Si30(n,gamma)→Si31 as 8.87e-15, against 2.219e-15 for the median of the others, a factor of 4.
\end{itemize}
\subsubsection*{Suggested measurements}
\begin{itemize}\setlength{\itemsep}{4pt}
\item Natural Si has 3 isotopes, so the measured heat combines several parents. Enriched single-isotope targets would distinguish their contributions. Candidate parents are Si29, Si30. Enrichment alone does not isolate a single reaction channel.
\item The reaction rates are averaged over the measured spectrum. Activation measurements at several well-defined neutron energies, including below 14 MeV, would show which energy range the differences come from.
\end{itemize}
\clearpage
\subsection{Sm: Samarium}\label{foil:Sm}
\begin{figure}[H]
\centering
\includegraphics[width=\textwidth,height=0.82\textheight,keepaspectratio]{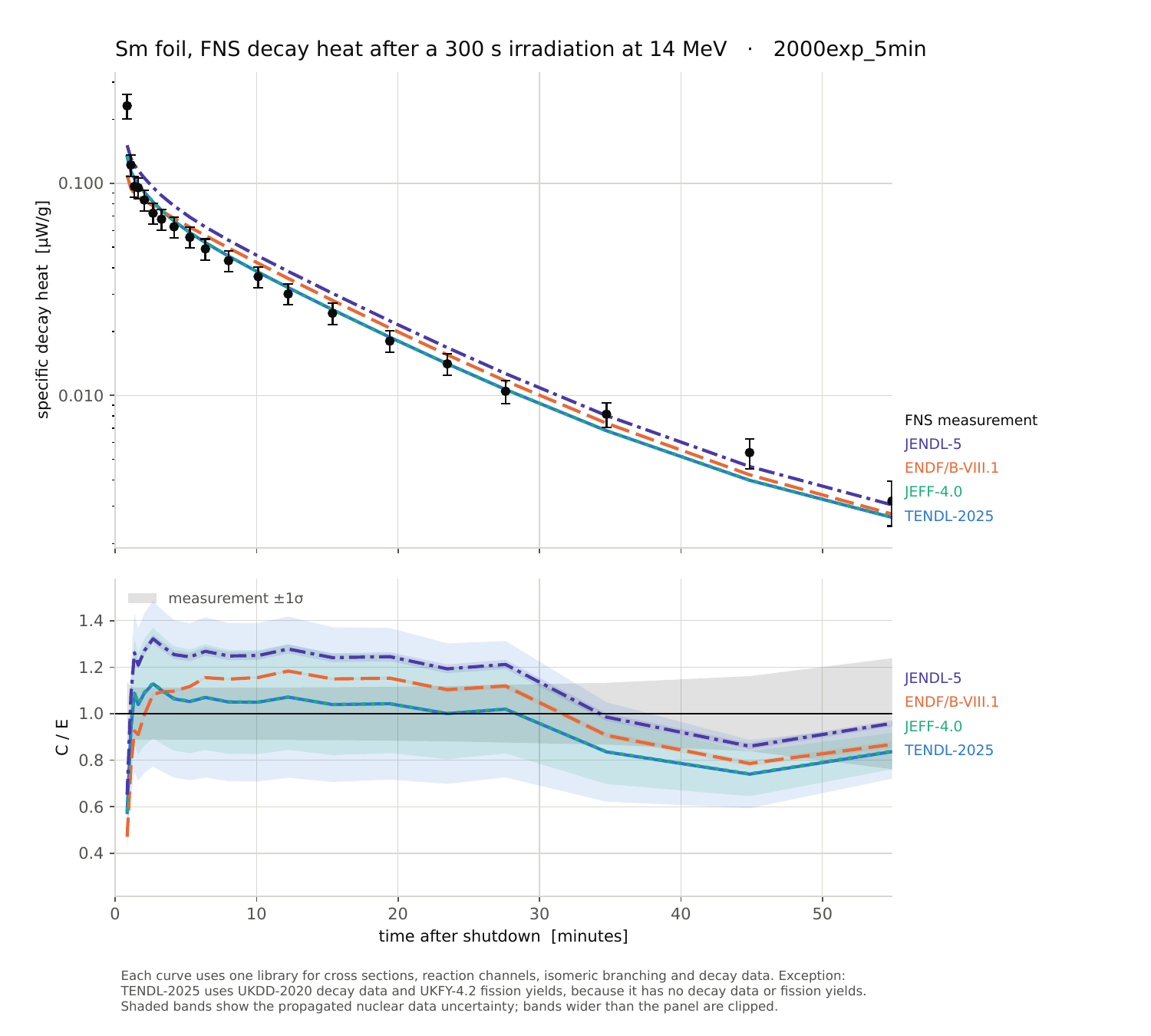}
\caption{Sm: decay heat against the measurement, with C/E beneath. Mean $|C/E-1|$ by library: TENDL-2025 10\%, JEFF-4.0 10\%, ENDF/B-VIII.1 15\%, JENDL-5 22\%.}
\end{figure}
\clearpage

\begin{figure}[H]
\centering
\includegraphics[width=\textwidth,height=0.82\textheight,keepaspectratio]{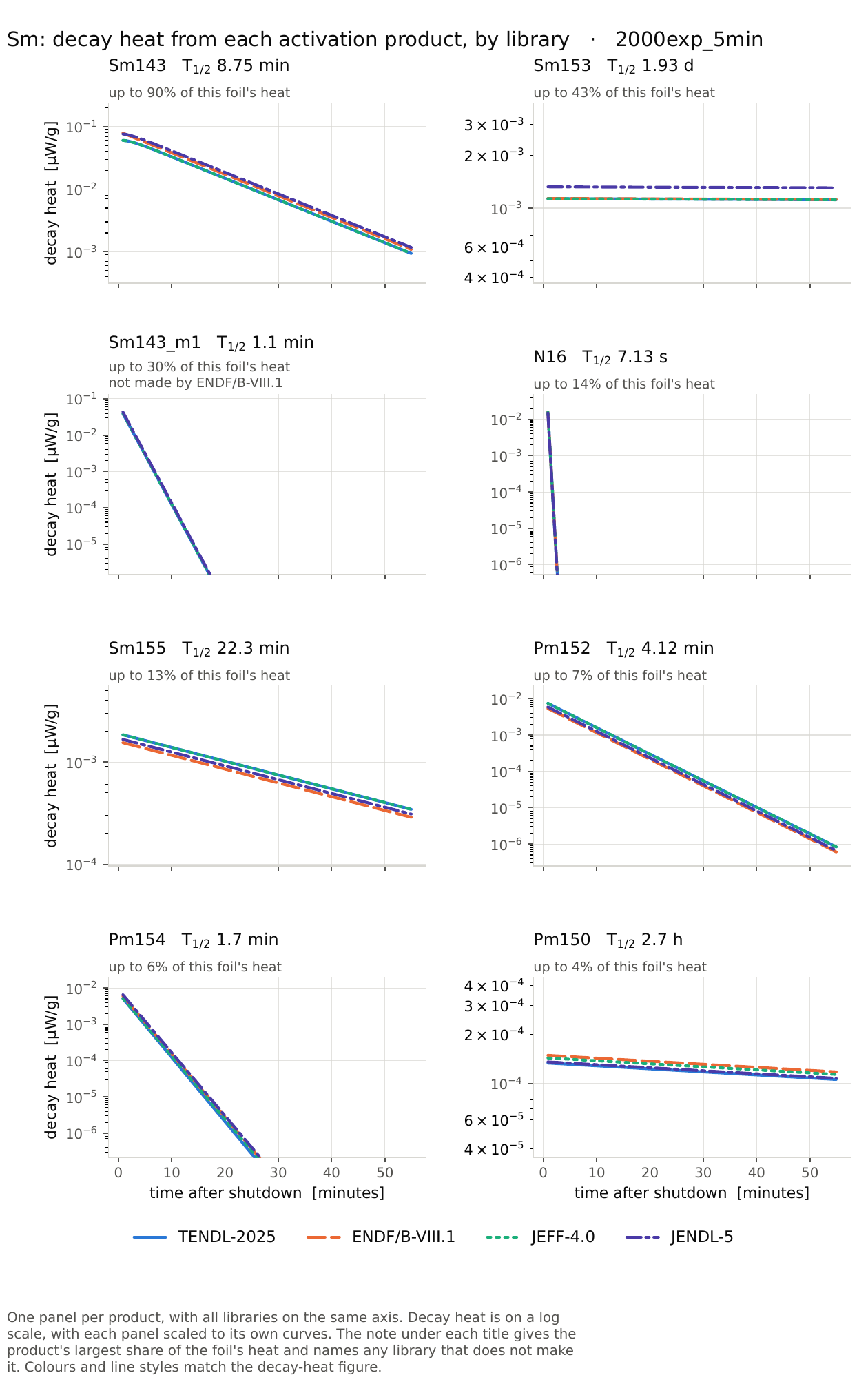}
\caption{Sm: share of the decay heat from each activation product during cooling.}
\end{figure}
\clearpage

\begin{figure}[H]
\centering
\includegraphics[width=\textwidth,height=0.82\textheight,keepaspectratio]{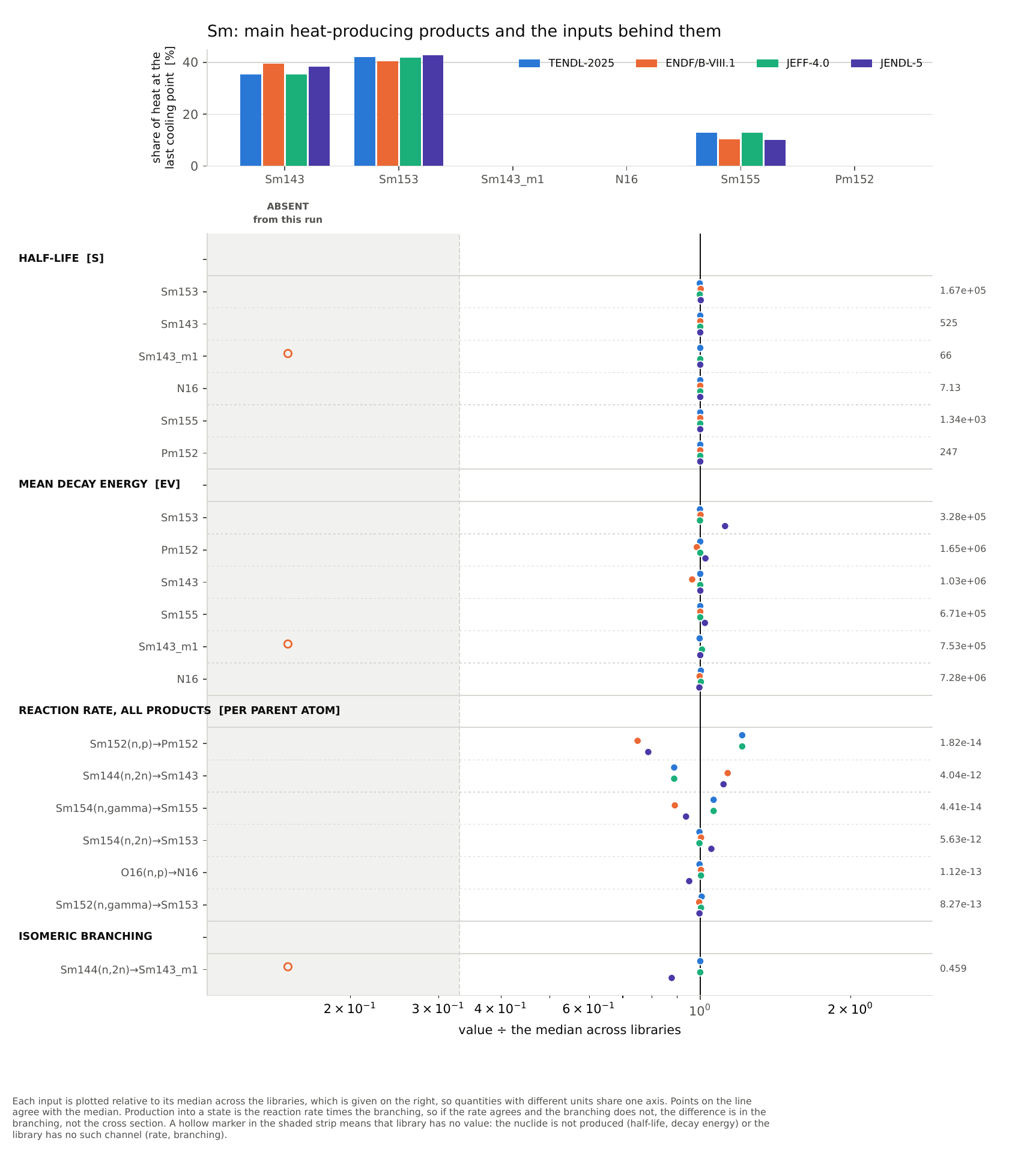}
\caption{Sm: inputs for the main heat-producing products, each relative to its median across the available libraries.}
\end{figure}
\clearpage

\begin{figure}[H]
\centering
\includegraphics[width=\textwidth,height=0.82\textheight,keepaspectratio]{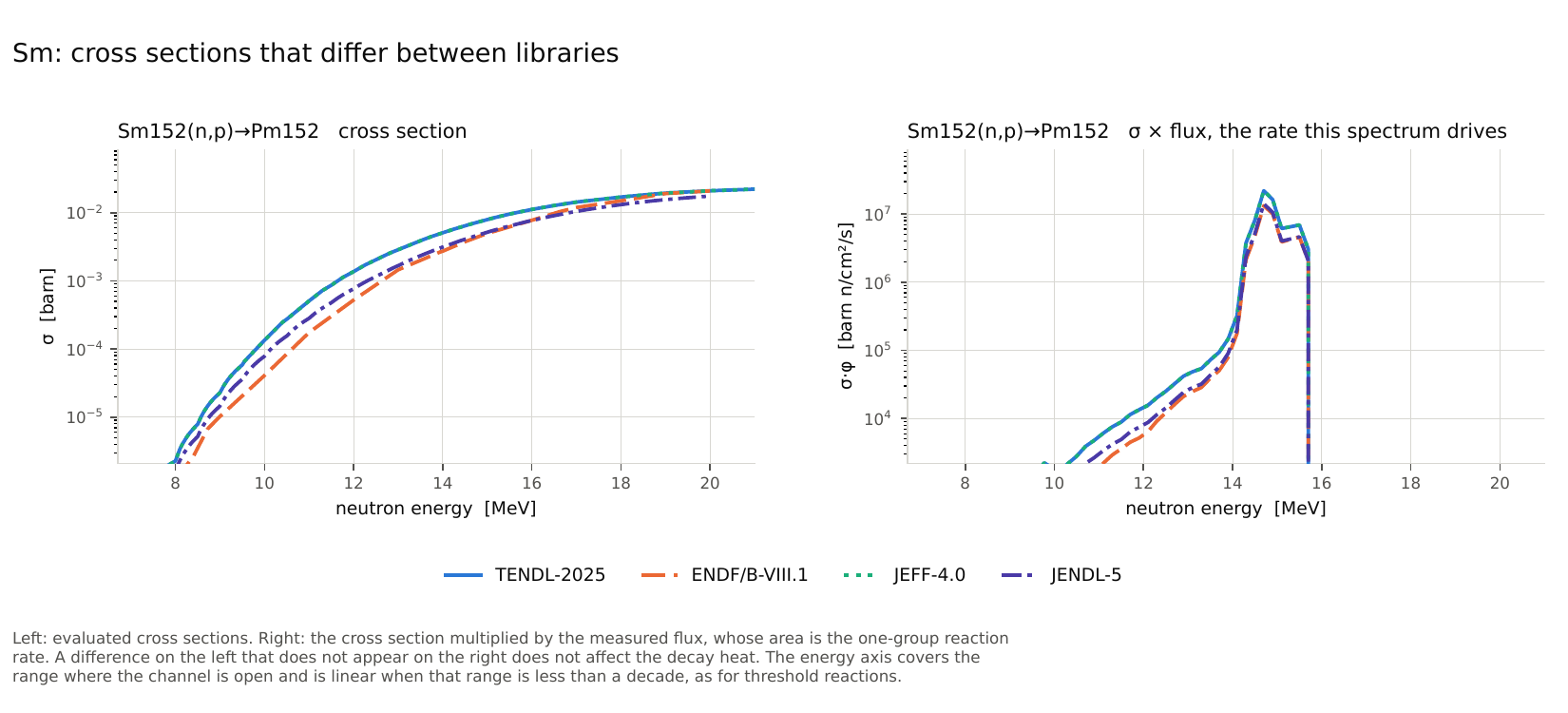}
\caption{Sm: cross sections of the channels whose spectrum-averaged rates differ, with the neutron spectrum and their product.}
\end{figure}
\clearpage

The libraries are between 10 and 22 percent from the measurement, which has a median uncertainty of 11 percent. TENDL-2025 is closest and JENDL-5 furthest.

\subsubsection*{Differences in the nuclear data}
\begin{itemize}\setlength{\itemsep}{6pt}
\item ENDF/B-VIII.1 gives no heat from Sm143\_m1; the other libraries do.\\[2pt]{\itshape The heat breakdown alone does not show why. Possible causes are a missing reaction channel, a different state assignment, different decay data or a reporting threshold.}
\end{itemize}
\subsubsection*{Suggested measurements}
\begin{itemize}\setlength{\itemsep}{4pt}
\item Natural Sm has 7 isotopes, so the measured heat combines several parents. Enriched single-isotope targets would distinguish their contributions. Enrichment alone does not isolate a single reaction channel.
\item The sample is a compound that also contains O. Their activation adds to the measured heat; a metal or purer sample would remove it.
\item This foil makes Sm153 (half-life 1.93 days), which gives up to 43 percent of the calculated heat. A five-minute irradiation and 55 minute measurement only see the start of this contribution. There is no seven-hour FNS measurement of this foil. A longer irradiation followed for days to months would be needed to test this product's production and decay.
\end{itemize}
\clearpage
\subsection{Sn: Tin}\label{foil:Sn}
\begin{figure}[H]
\centering
\includegraphics[width=\textwidth,height=0.82\textheight,keepaspectratio]{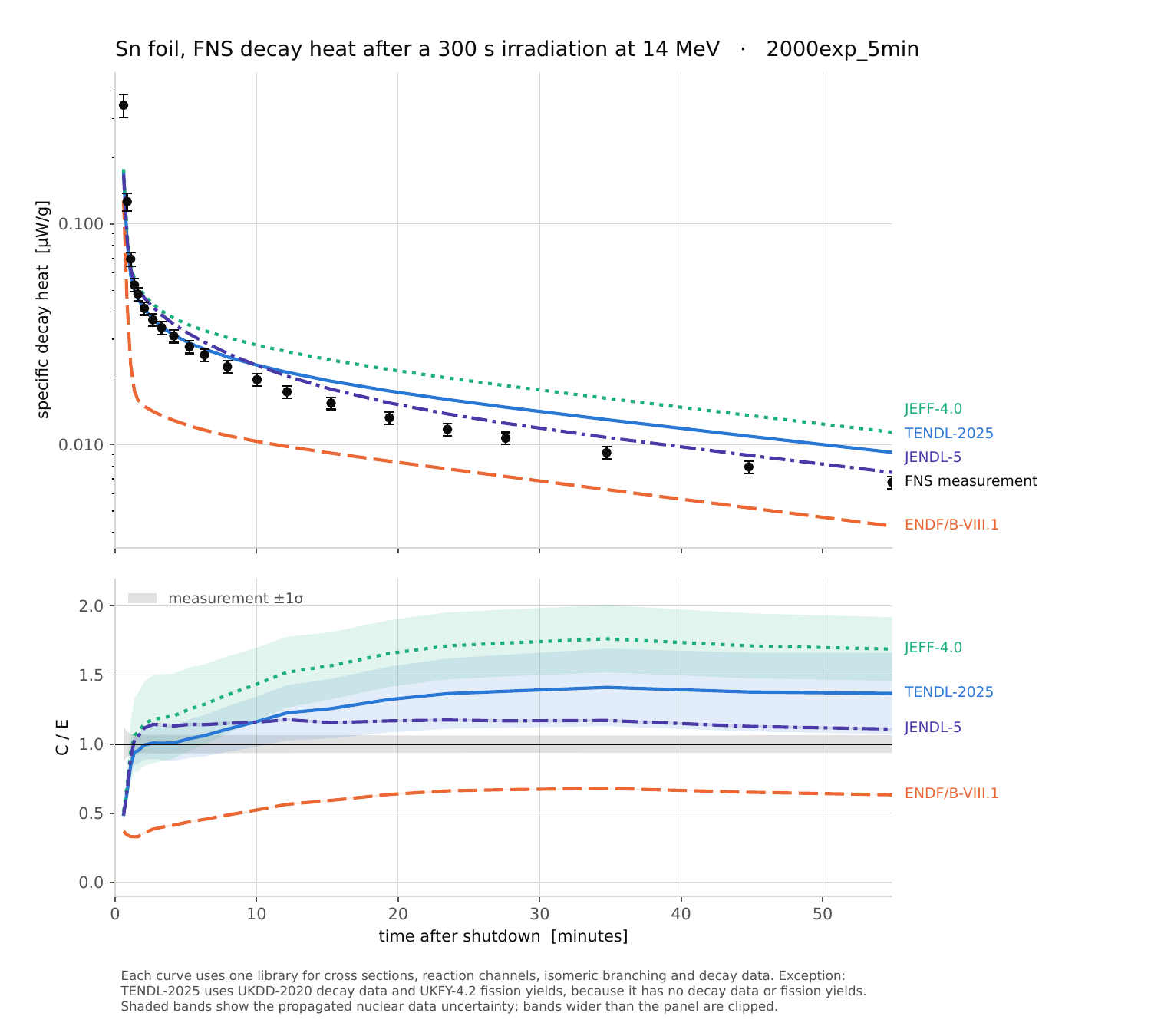}
\caption{Sn: decay heat against the measurement, with C/E beneath. Mean $|C/E-1|$ by library: JENDL-5 16\%, TENDL-2025 20\%, JEFF-4.0 40\%, ENDF/B-VIII.1 51\%.}
\end{figure}
\clearpage

\begin{figure}[H]
\centering
\includegraphics[width=\textwidth,height=0.82\textheight,keepaspectratio]{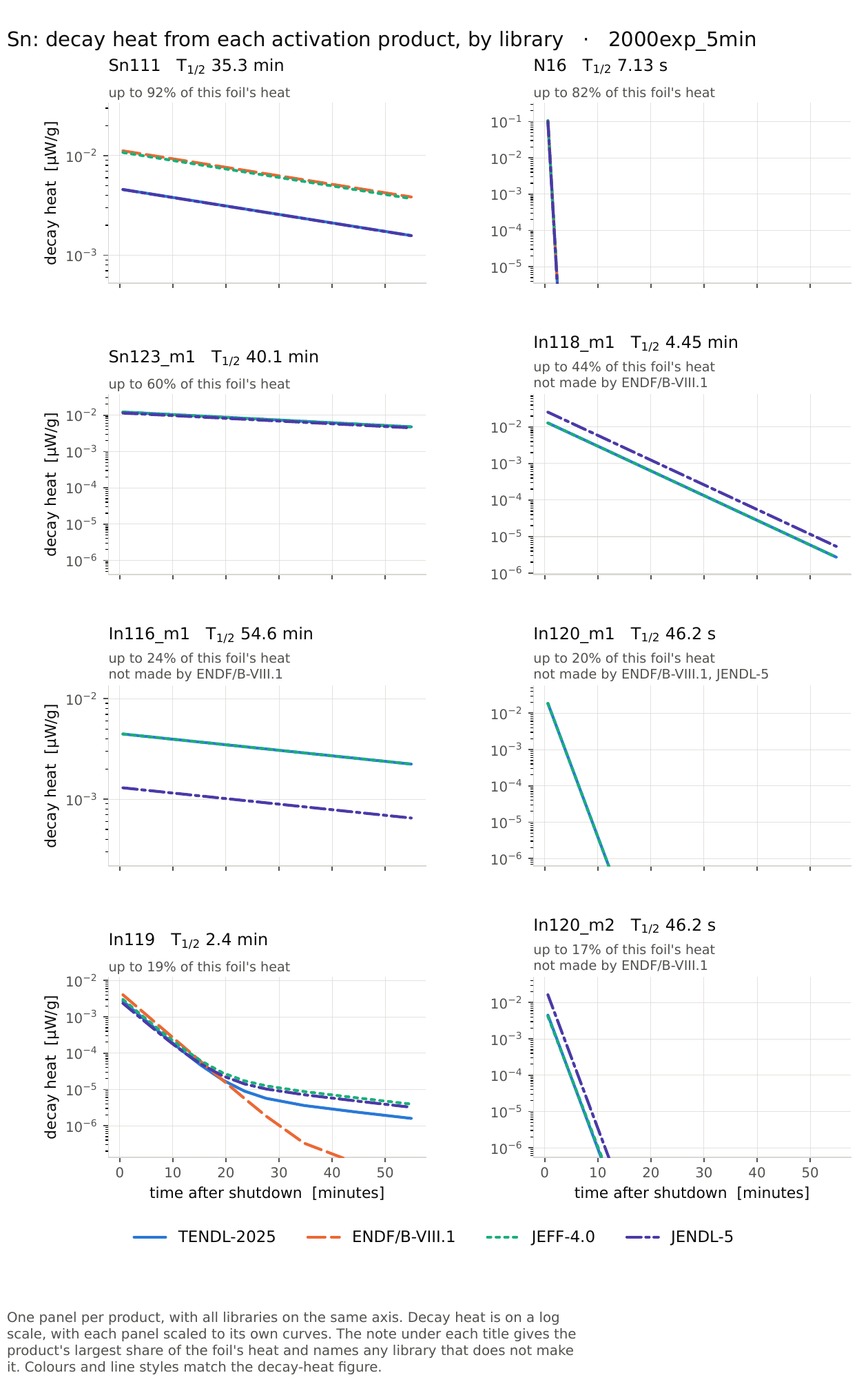}
\caption{Sn: share of the decay heat from each activation product during cooling.}
\end{figure}
\clearpage

\begin{figure}[H]
\centering
\includegraphics[width=\textwidth,height=0.82\textheight,keepaspectratio]{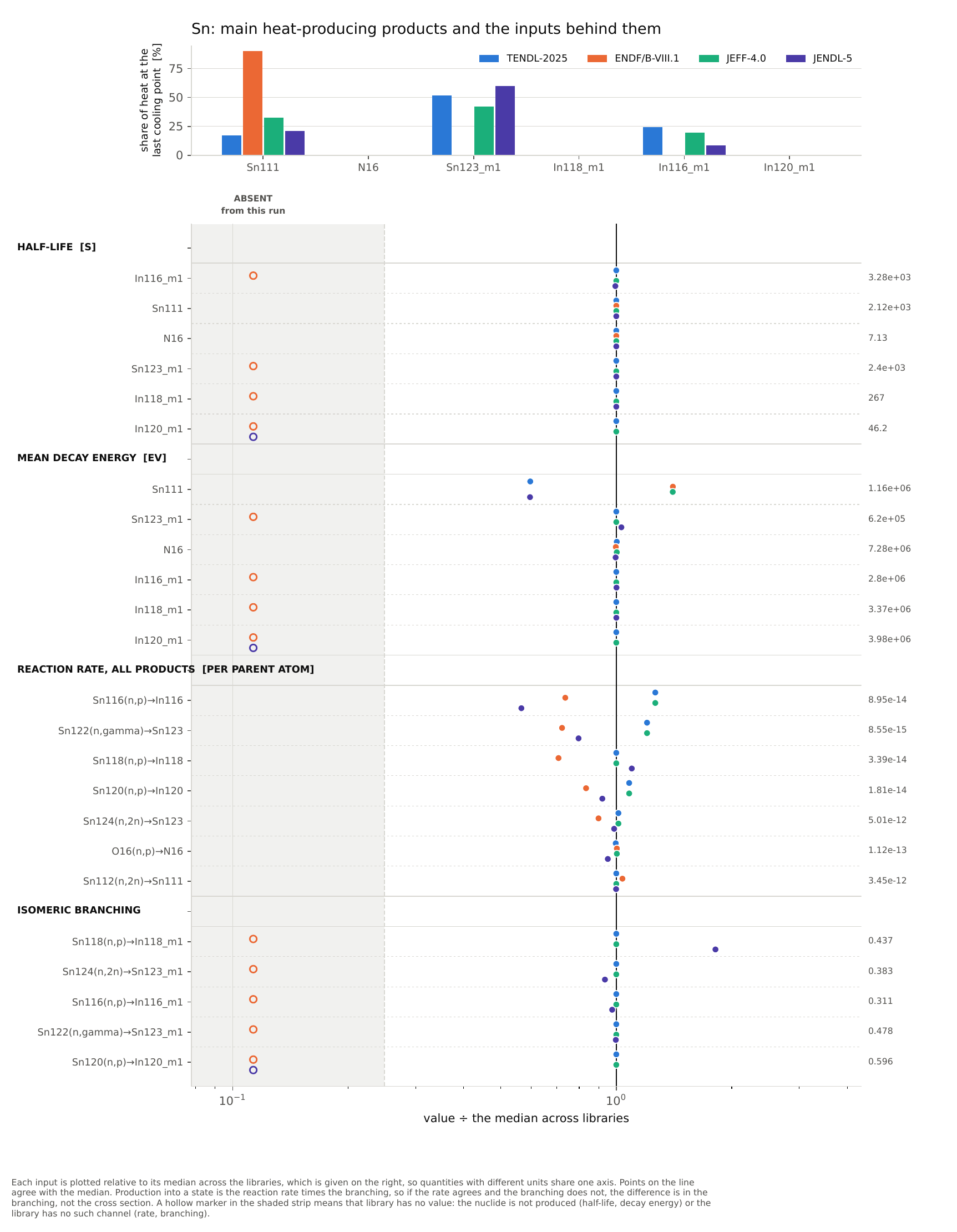}
\caption{Sn: inputs for the main heat-producing products, each relative to its median across the available libraries.}
\end{figure}
\clearpage

\begin{figure}[H]
\centering
\includegraphics[width=\textwidth,height=0.82\textheight,keepaspectratio]{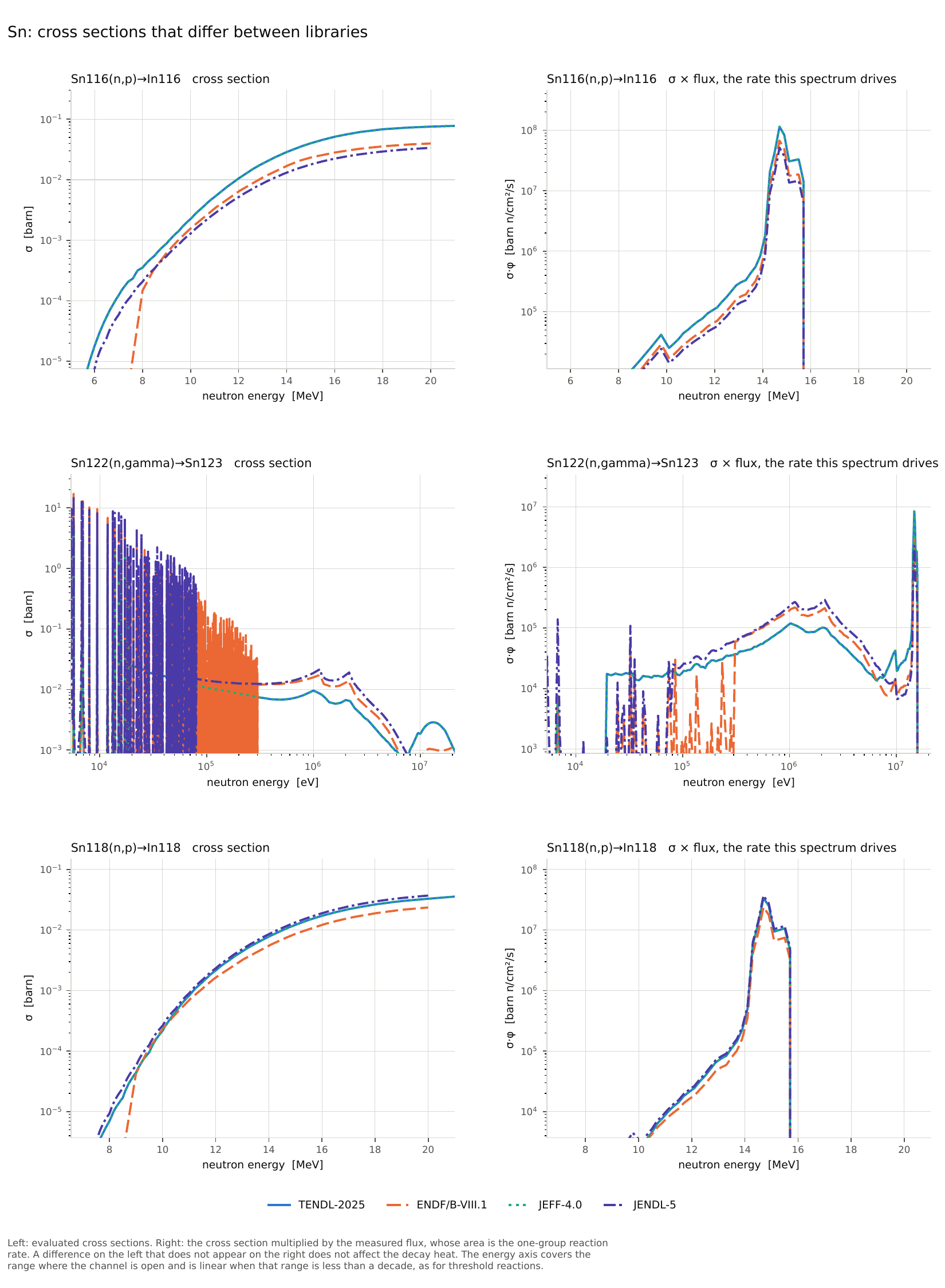}
\caption{Sn: cross sections of the channels whose spectrum-averaged rates differ, with the neutron spectrum and their product.}
\end{figure}
\clearpage

\begin{figure}[H]
\centering
\includegraphics[width=\textwidth,height=0.82\textheight,keepaspectratio]{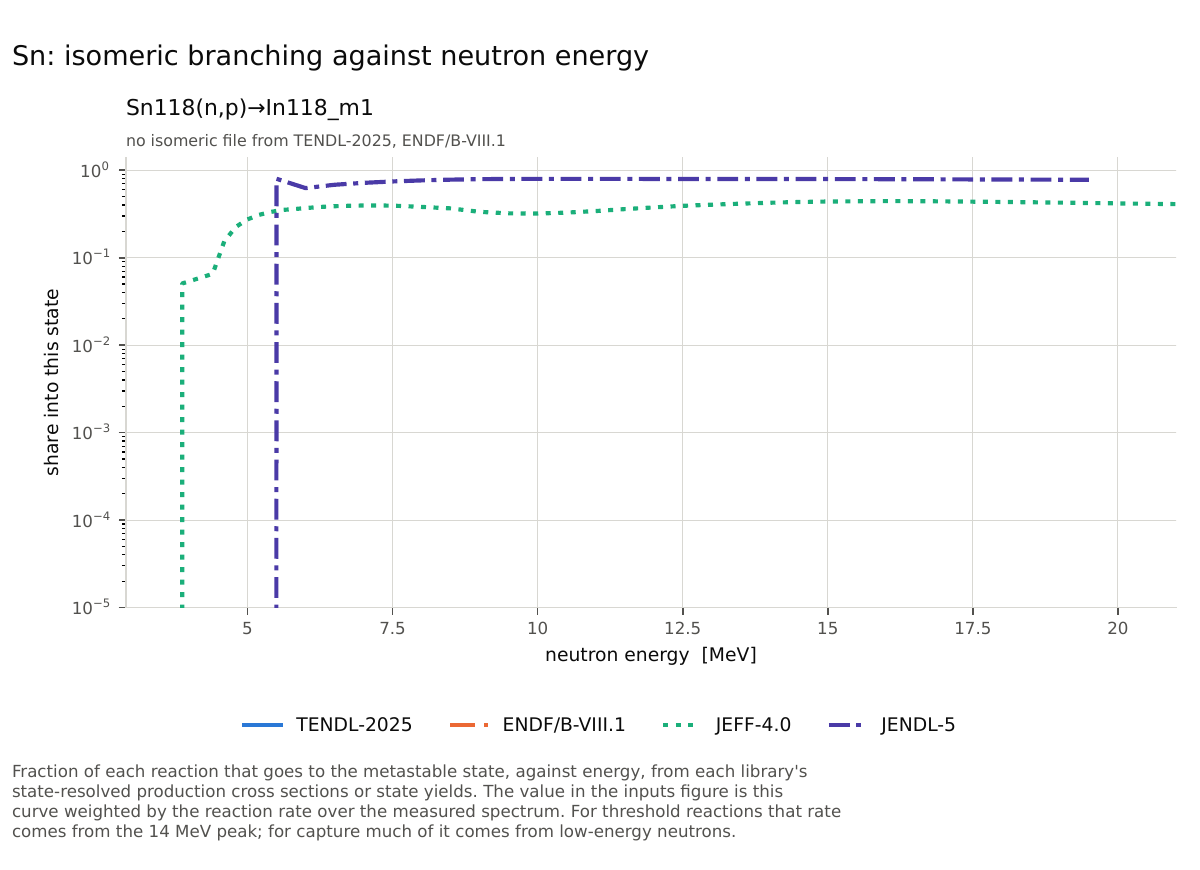}
\caption{Sn: isomeric branching ratio against neutron energy for the channels whose spectrum-averaged ratio differs between libraries.}
\end{figure}
\clearpage

\subsubsection*{1996 seven-hour campaign}
The 1996 campaign irradiated this foil for 7.17 hours at position 7 and measured it out to 402.13 days. In the five-minute calculation Sn121 (half-life 1.13 days) gives up to 3 percent of the heat, and the longer irradiation and cooling test its production and decay data. TENDL-2025 is closest. This campaign uses a different spectrum, so it is not included in the main rankings.

\begin{figure}[H]
\centering
\includegraphics[width=\textwidth,height=0.82\textheight,keepaspectratio]{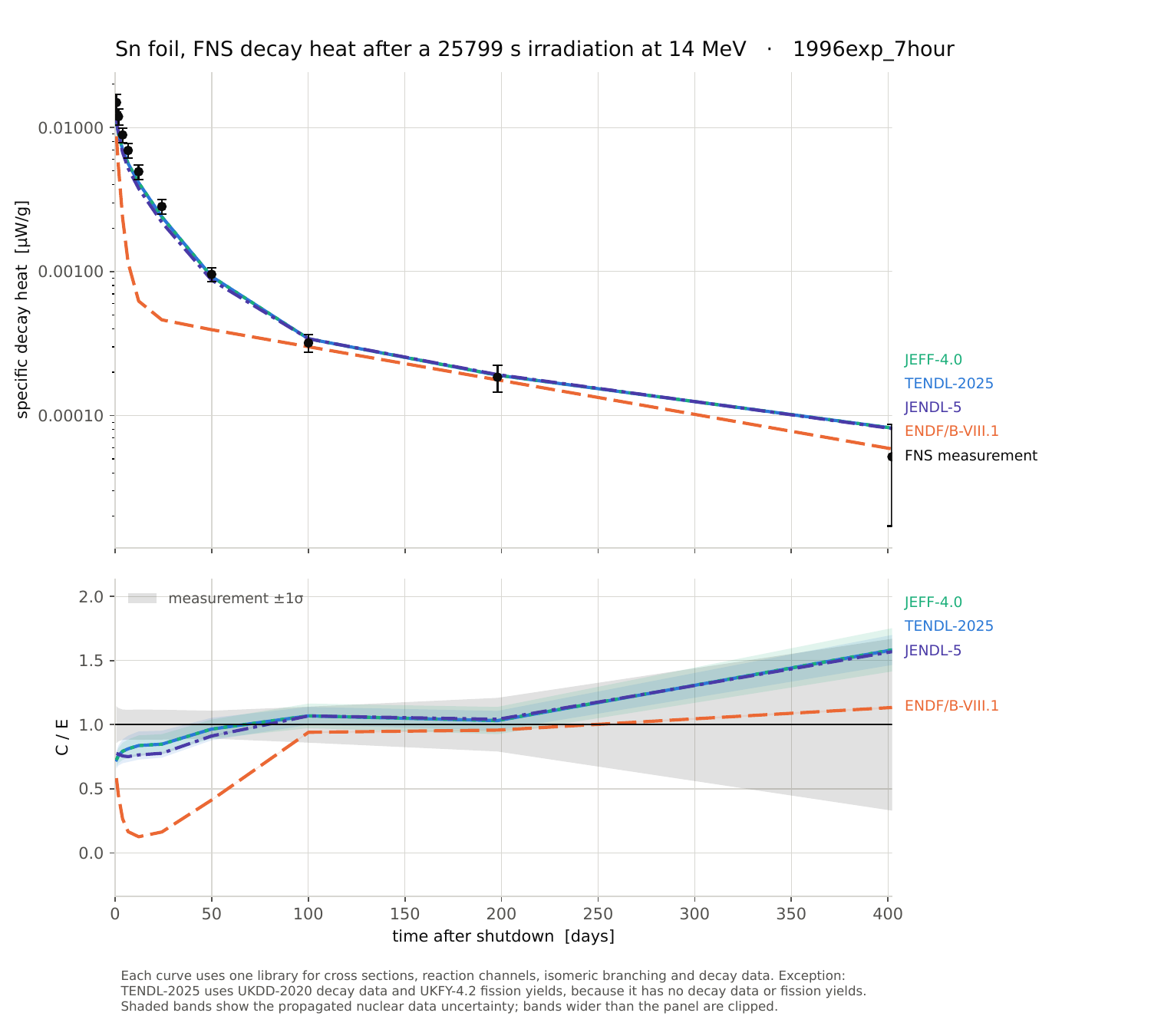}
\caption{Sn: 1996 seven-hour campaign, irradiation position 7. Mean $|C/E-1|$ by library: TENDL-2025 19\%, JEFF-4.0 19\%, JENDL-5 22\%, ENDF/B-VIII.1 51\%. }
\end{figure}
\clearpage

The libraries are between 16 and 51 percent from the measurement, which has a median uncertainty of 7 percent. JENDL-5 is closest and ENDF/B-VIII.1 furthest.

\subsubsection*{Differences in the nuclear data}
\begin{itemize}\setlength{\itemsep}{6pt}
\item JENDL-5 gives the mean decay energy of Sn111 as 6.933e+05 eV, against 1.633e+06 eV for the median of the others, a factor of 2.\\[2pt]{\itshape For a given activity the heat is proportional to the mean decay energy, so this difference goes straight into the heat. Compare the decay schemes and how the emitted energy is counted.}
\item JENDL-5 gives the total reaction rate of Sn116(n,p)→In116 as 5.06e-14, against 1.131e-13 for the median of the others, a factor of 2.\\[2pt]{\itshape This is the total rate of the channel over all its products, so the difference is in the cross section, not in the split between states.}
\item ENDF/B-VIII.1 gives no heat from In118\_m1; the other libraries do.\\[2pt]{\itshape The heat breakdown alone does not show why. Possible causes are a missing reaction channel, a different state assignment, different decay data or a reporting threshold.}
\item ENDF/B-VIII.1 gives no heat from In116\_m1; the other libraries do.
\item ENDF/B-VIII.1, JENDL-5 give no heat from In120\_m1; the other libraries do.
\end{itemize}
\subsubsection*{Suggested measurements}
\begin{itemize}\setlength{\itemsep}{4pt}
\item Natural Sn has 10 isotopes, so the measured heat combines several parents. Enriched single-isotope targets would distinguish their contributions. The candidate parent is Sn116. Enrichment alone does not isolate a single reaction channel.
\item The sample is a compound that also contains O. Their activation adds to the measured heat; a metal or purer sample would remove it.
\item The reaction rates are averaged over the measured spectrum. Activation measurements at several well-defined neutron energies, including below 14 MeV, would show which energy range the differences come from.
\end{itemize}
\clearpage
\subsection{Sr: Strontium}\label{foil:Sr}
\begin{figure}[H]
\centering
\includegraphics[width=\textwidth,height=0.82\textheight,keepaspectratio]{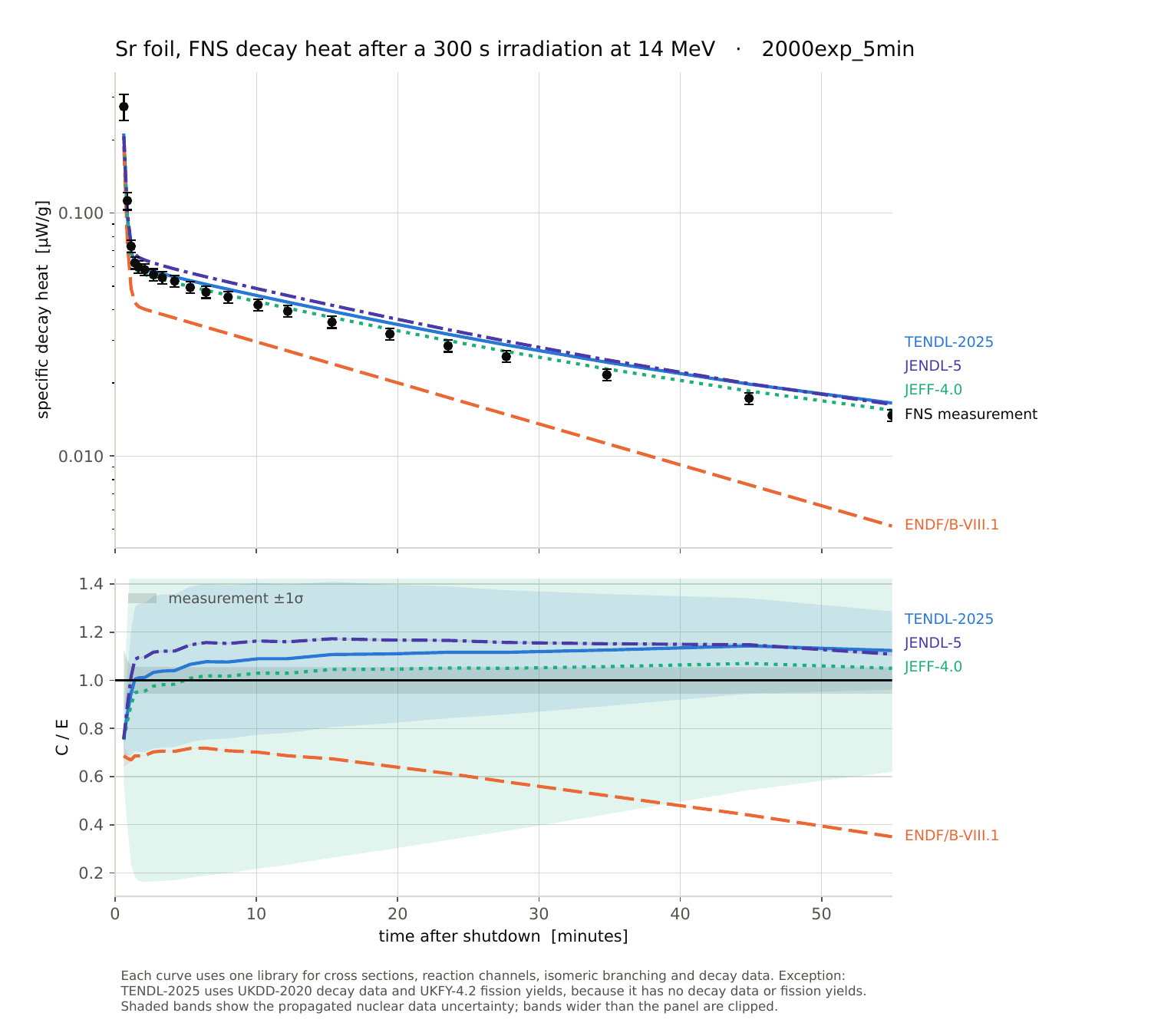}
\caption{Sr: decay heat against the measurement, with C/E beneath. Mean $|C/E-1|$ by library: JEFF-4.0 6\%, TENDL-2025 9\%, JENDL-5 14\%, ENDF/B-VIII.1 35\%.}
\end{figure}
\clearpage

\begin{figure}[H]
\centering
\includegraphics[width=\textwidth,height=0.82\textheight,keepaspectratio]{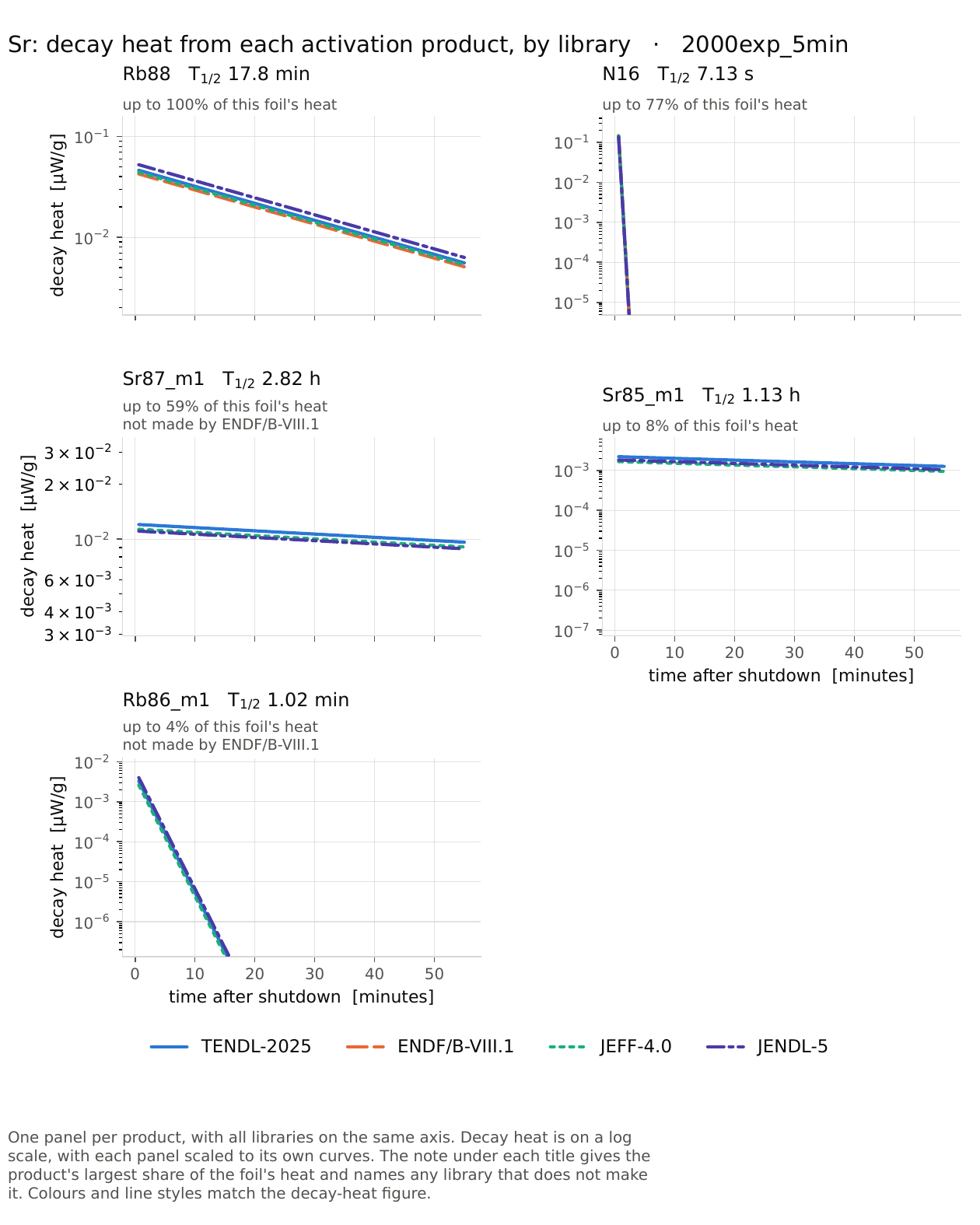}
\caption{Sr: share of the decay heat from each activation product during cooling.}
\end{figure}
\clearpage

\begin{figure}[H]
\centering
\includegraphics[width=\textwidth,height=0.82\textheight,keepaspectratio]{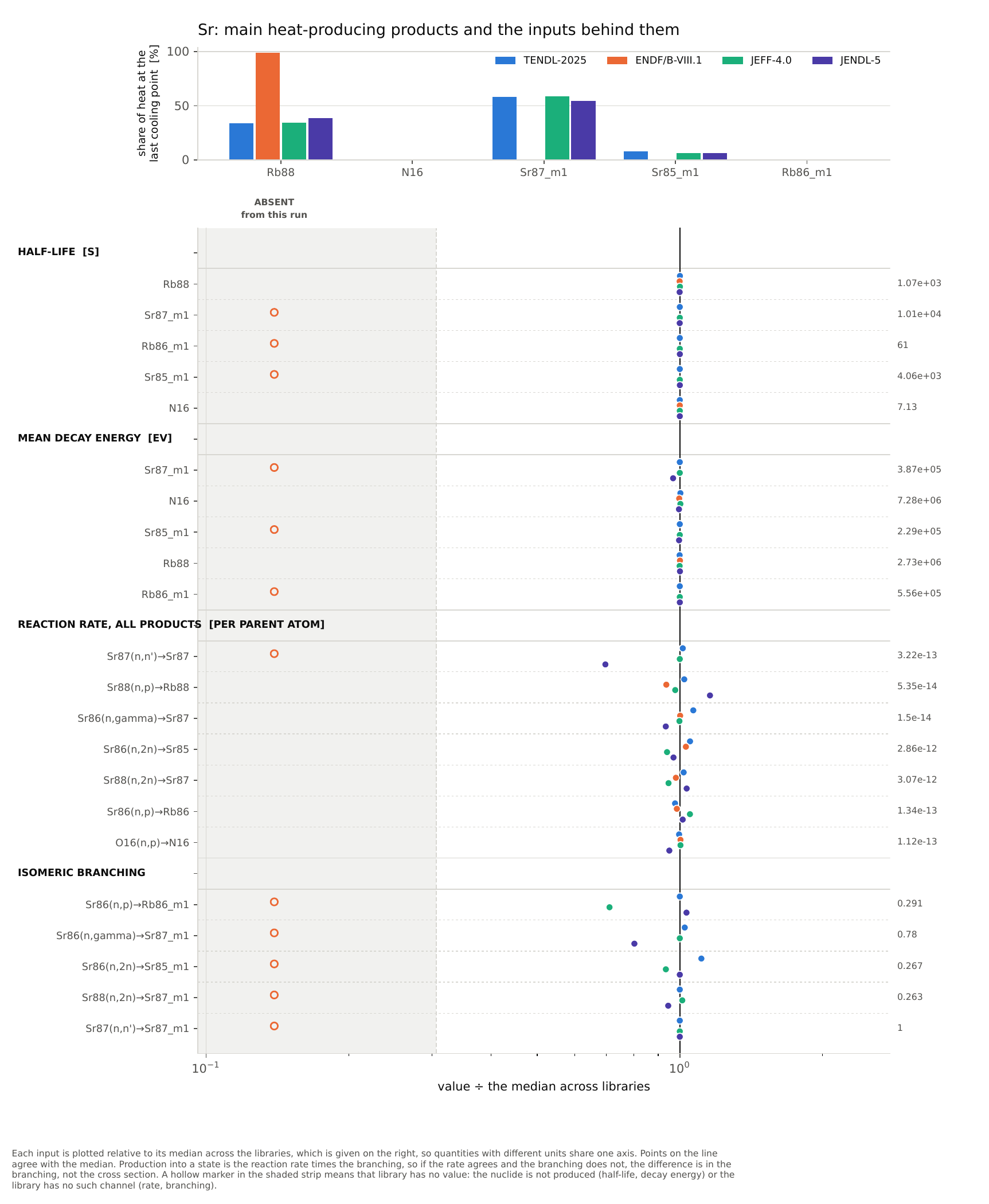}
\caption{Sr: inputs for the main heat-producing products, each relative to its median across the available libraries.}
\end{figure}
\clearpage

ENDF/B-VIII.1 stands out at 35 percent from the measurement, against 14 percent for the next furthest library and 6 percent for JEFF-4.0. The median measurement uncertainty is 6 percent.

\subsubsection*{Differences in the nuclear data}
\begin{itemize}\setlength{\itemsep}{6pt}
\item ENDF/B-VIII.1 gives no heat from Sr87\_m1; the other libraries do.\\[2pt]{\itshape The heat breakdown alone does not show why. Possible causes are a missing reaction channel, a different state assignment, different decay data or a reporting threshold.}
\item ENDF/B-VIII.1 gives no heat from Rb86\_m1; the other libraries do.
\end{itemize}
\subsubsection*{Suggested measurements}
\begin{itemize}\setlength{\itemsep}{4pt}
\item Natural Sr has 4 isotopes, so the measured heat combines several parents. Enriched single-isotope targets would distinguish their contributions. Enrichment alone does not isolate a single reaction channel.
\item The sample is a compound that also contains C, O. Their activation adds to the measured heat; a metal or purer sample would remove it.
\end{itemize}
\clearpage
\subsection{Ta: Tantalum}\label{foil:Ta}
\begin{figure}[H]
\centering
\includegraphics[width=\textwidth,height=0.82\textheight,keepaspectratio]{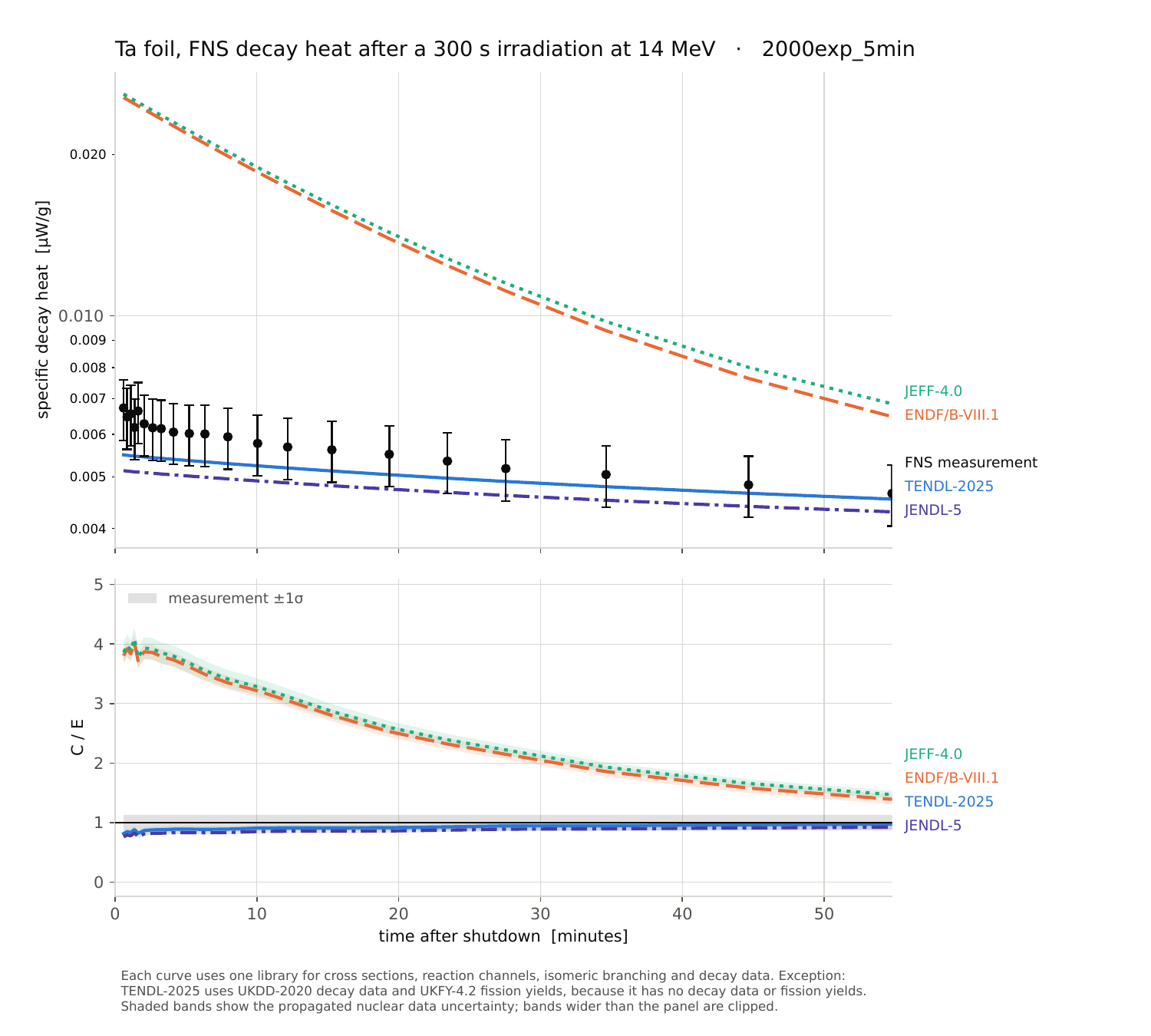}
\caption{Ta: decay heat against the measurement, with C/E beneath. Mean $|C/E-1|$ by library: TENDL-2025 10\%, JENDL-5 16\%, ENDF/B-VIII.1 214\%, JEFF-4.0 220\%.}
\end{figure}
\clearpage

\begin{figure}[H]
\centering
\includegraphics[width=\textwidth,height=0.82\textheight,keepaspectratio]{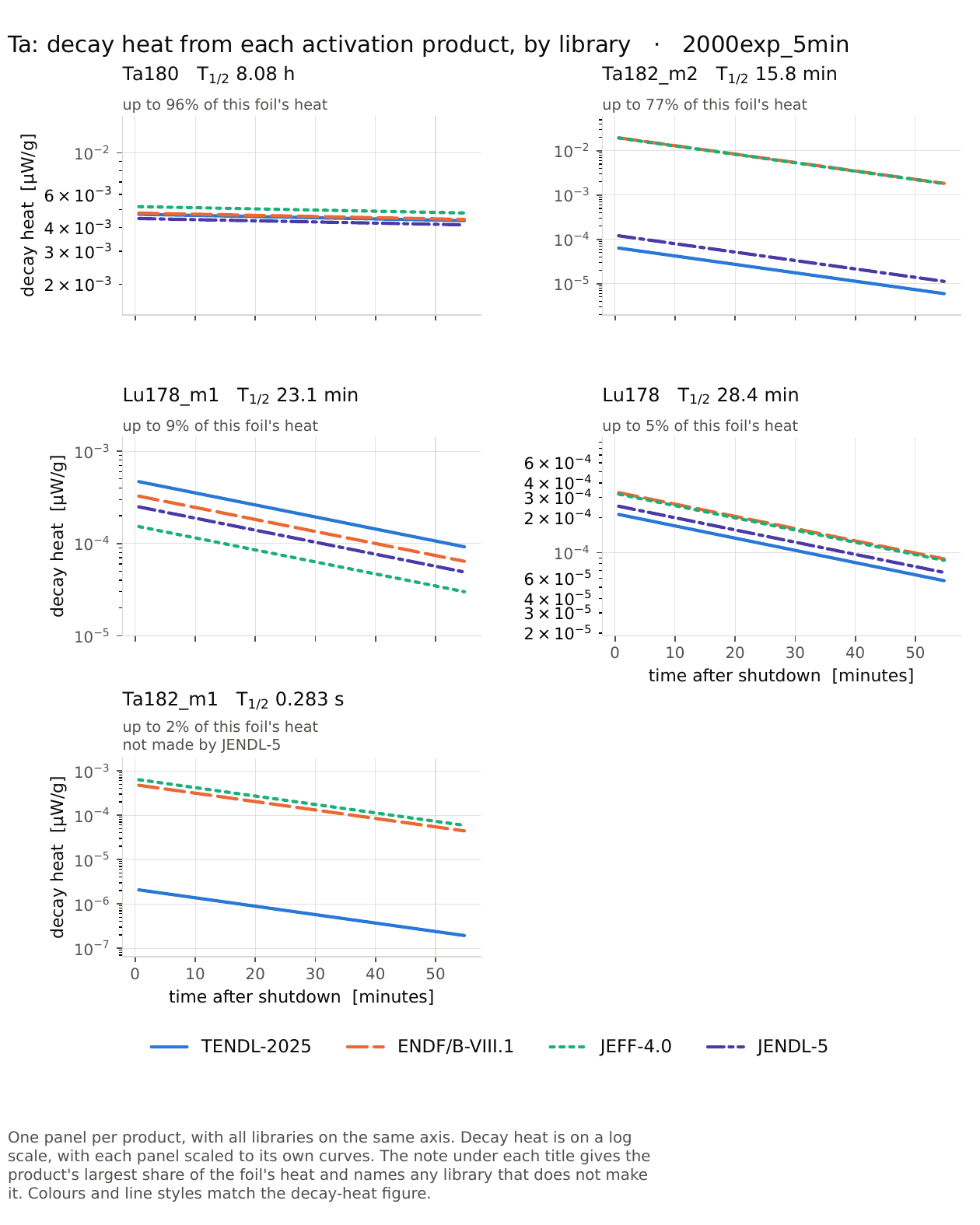}
\caption{Ta: share of the decay heat from each activation product during cooling.}
\end{figure}
\clearpage

\begin{figure}[H]
\centering
\includegraphics[width=\textwidth,height=0.82\textheight,keepaspectratio]{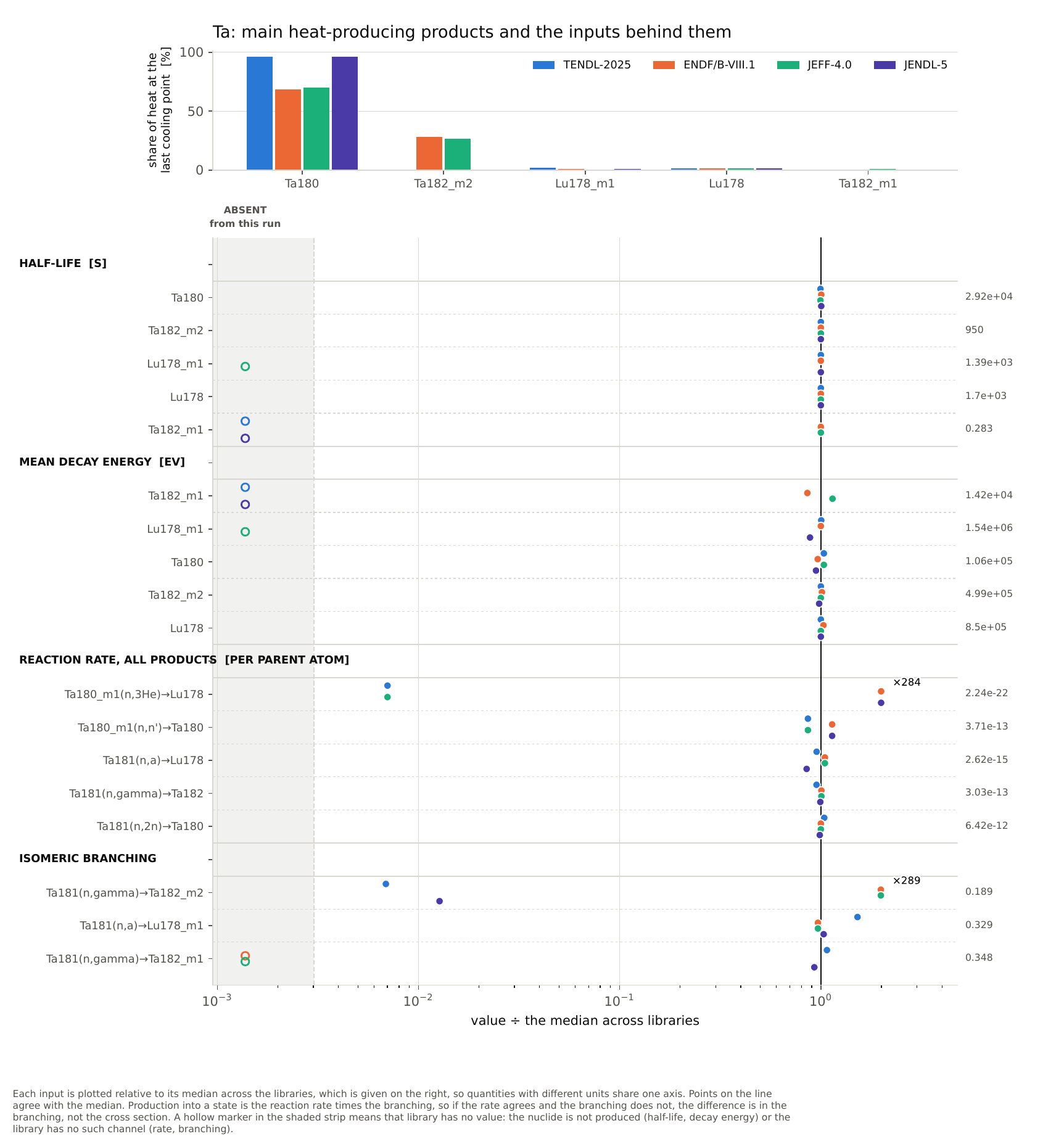}
\caption{Ta: inputs for the main heat-producing products, each relative to its median across the available libraries.}
\end{figure}
\clearpage

\begin{figure}[H]
\centering
\includegraphics[width=\textwidth,height=0.82\textheight,keepaspectratio]{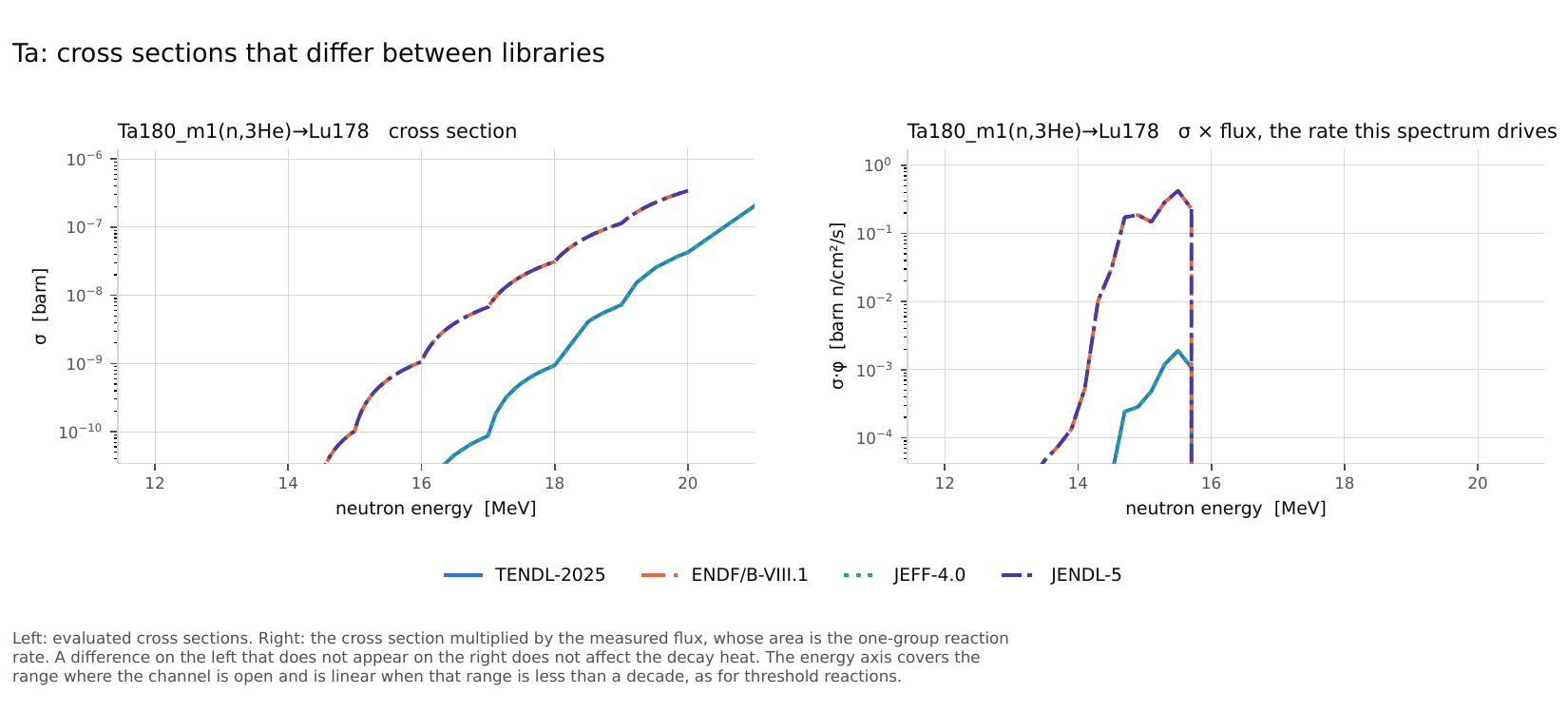}
\caption{Ta: cross sections of the channels whose spectrum-averaged rates differ, with the neutron spectrum and their product.}
\end{figure}
\clearpage

\begin{figure}[H]
\centering
\includegraphics[width=\textwidth,height=0.82\textheight,keepaspectratio]{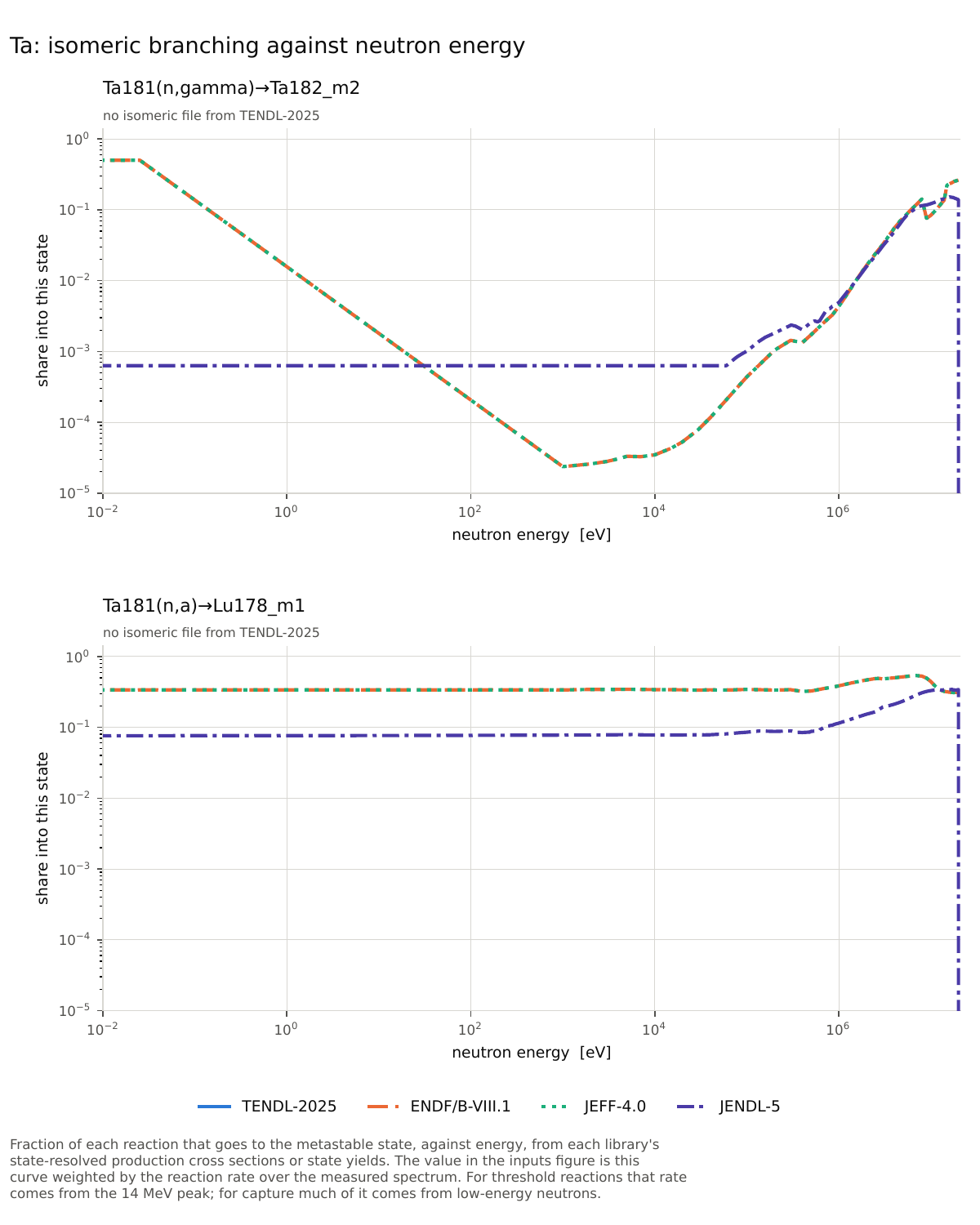}
\caption{Ta: isomeric branching ratio against neutron energy for the channels whose spectrum-averaged ratio differs between libraries.}
\end{figure}
\clearpage

The libraries are between 10 and 220 percent from the measurement, which has a median uncertainty of 13 percent. TENDL-2025 is closest and JEFF-4.0 furthest.

\subsubsection*{Differences in the nuclear data}
\begin{itemize}\setlength{\itemsep}{6pt}
\item TENDL-2025 gives the isomeric branching of Ta181(n,gamma)→Ta182\_m2 as 0.001298, against 0.3748 for the median of the others, a factor of 289.\\[2pt]{\itshape Isomeric branching sets how a channel's production is split between the ground and metastable states. It should be checked together with the total rate of the channel.}
\item TENDL-2025 gives the total reaction rate of Ta180\_m1(n,3He)→Lu178 as 1.572e-24, against 4.469e-22 for the median of the others, a factor of 284.\\[2pt]{\itshape This is the total rate of the channel over all its products, so the difference is in the cross section, not in the split between states.}
\item JENDL-5 gives no heat from Ta182\_m1; the other libraries do.\\[2pt]{\itshape The heat breakdown alone does not show why. Possible causes are a missing reaction channel, a different state assignment, different decay data or a reporting threshold.}
\end{itemize}
\subsubsection*{Suggested measurements}
\begin{itemize}\setlength{\itemsep}{4pt}
\item Natural Ta has 2 isotopes, so the measured heat combines several parents. Enriched single-isotope targets would distinguish their contributions. Candidate parents are Ta180\_m1, Ta181. Enrichment alone does not isolate a single reaction channel.
\item The reaction rates are averaged over the measured spectrum. Activation measurements at several well-defined neutron energies, including below 14 MeV, would show which energy range the differences come from.
\item Measuring the activity of each state separately, for example by gamma spectrometry where the states have distinct lines, would constrain the isomeric ratio. This needs corrections for feeding between states, detector efficiency and irradiation history.
\end{itemize}
\clearpage
\subsection{Tb: Terbium}\label{foil:Tb}
\begin{figure}[H]
\centering
\includegraphics[width=\textwidth,height=0.82\textheight,keepaspectratio]{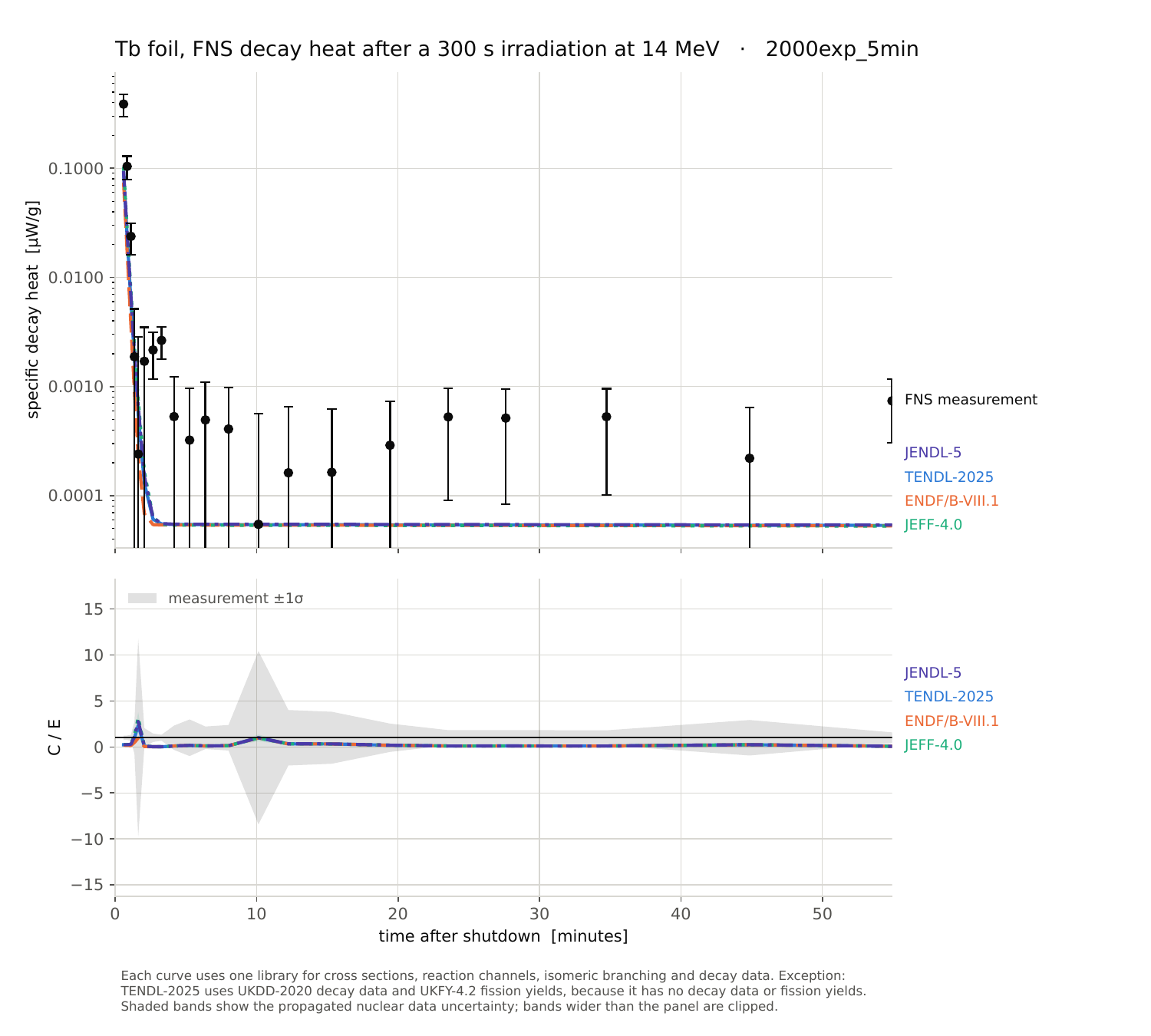}
\caption{Tb: decay heat against the measurement, with C/E beneath. Mean $|C/E-1|$ by library: ENDF/B-VIII.1 76\%, TENDL-2025 79\%, JENDL-5 81\%, JEFF-4.0 83\%.}
\end{figure}
\clearpage

\begin{figure}[H]
\centering
\includegraphics[width=\textwidth,height=0.82\textheight,keepaspectratio]{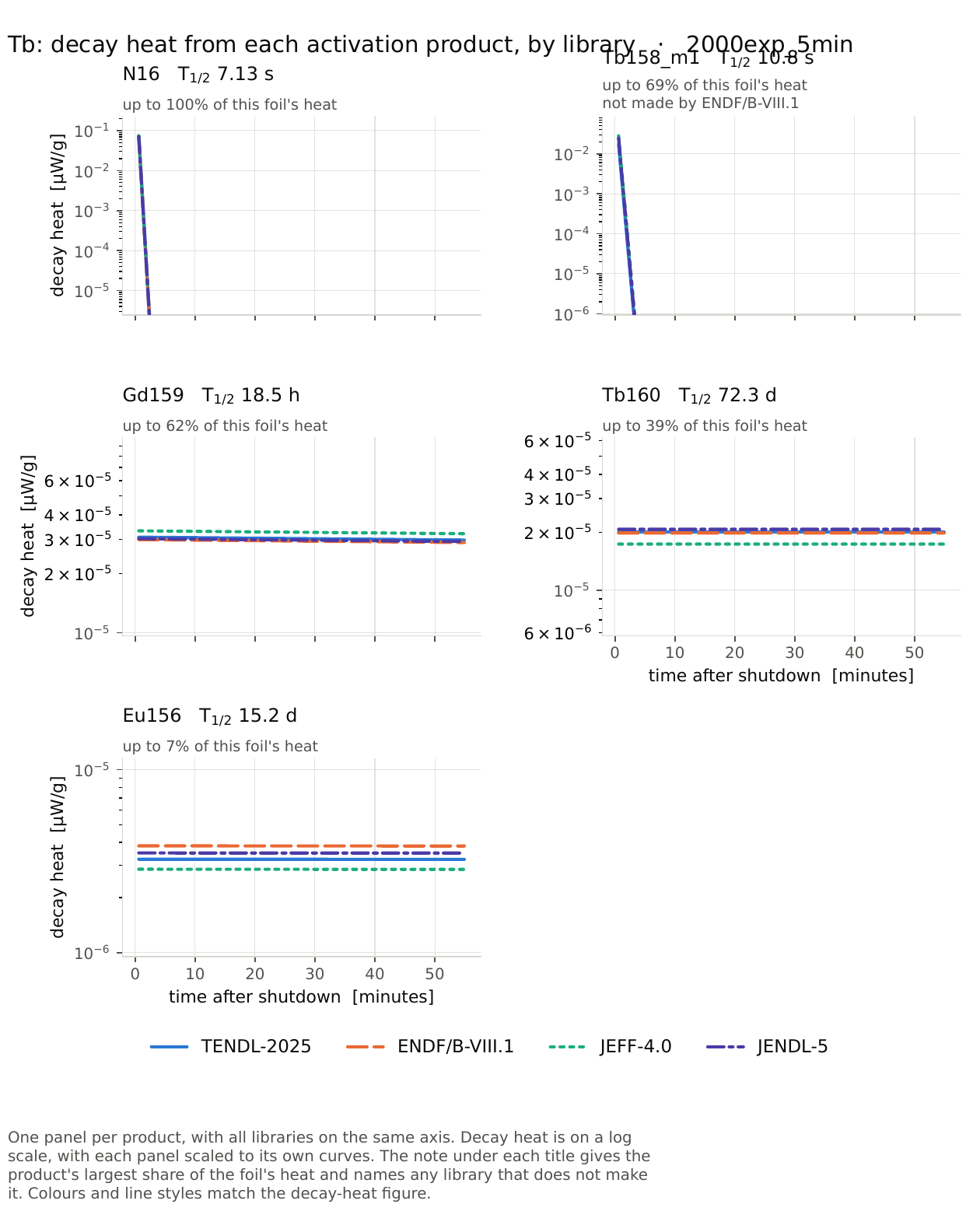}
\caption{Tb: share of the decay heat from each activation product during cooling.}
\end{figure}
\clearpage

\begin{figure}[H]
\centering
\includegraphics[width=\textwidth,height=0.82\textheight,keepaspectratio]{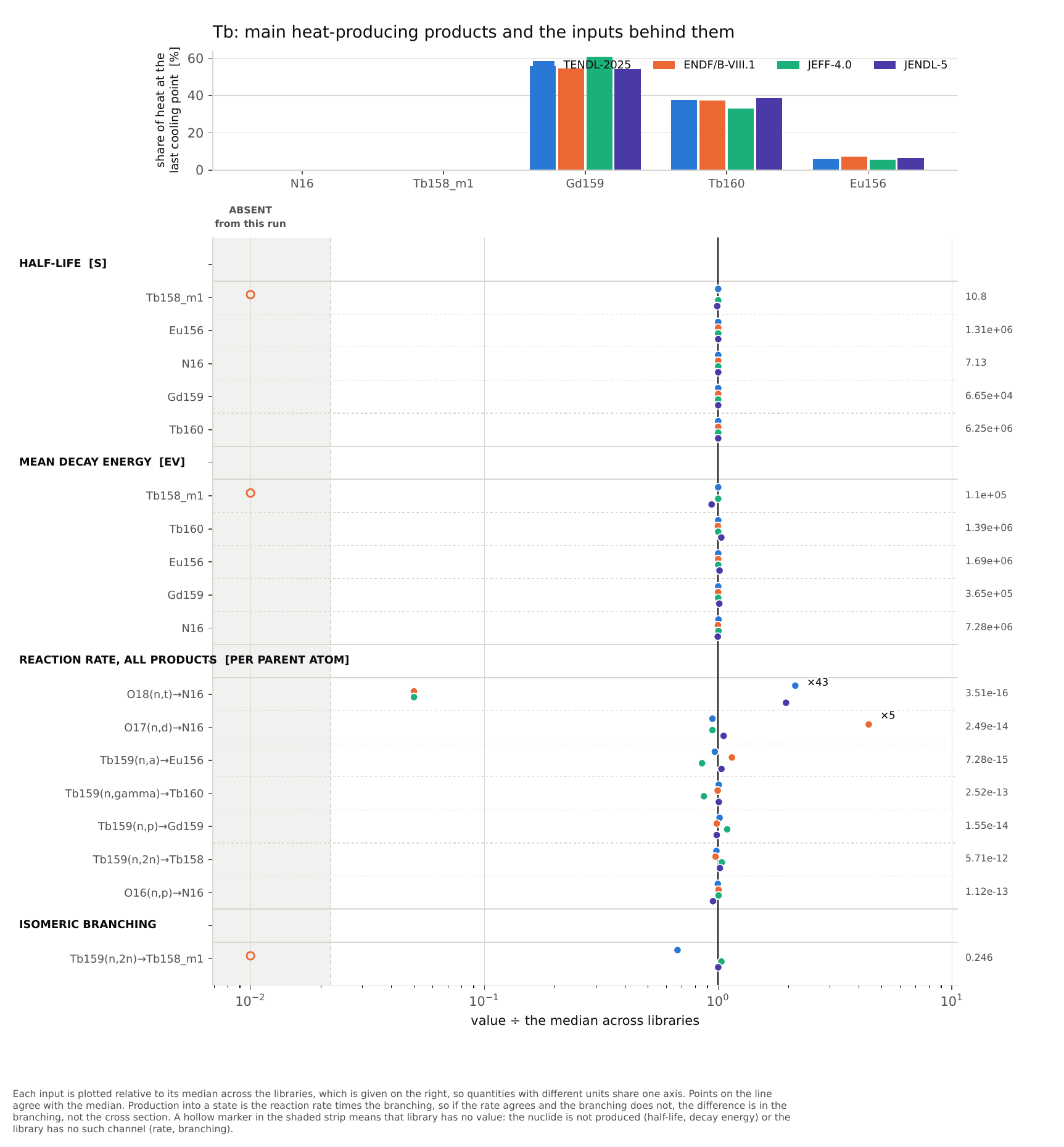}
\caption{Tb: inputs for the main heat-producing products, each relative to its median across the available libraries.}
\end{figure}
\clearpage

\begin{figure}[H]
\centering
\includegraphics[width=\textwidth,height=0.82\textheight,keepaspectratio]{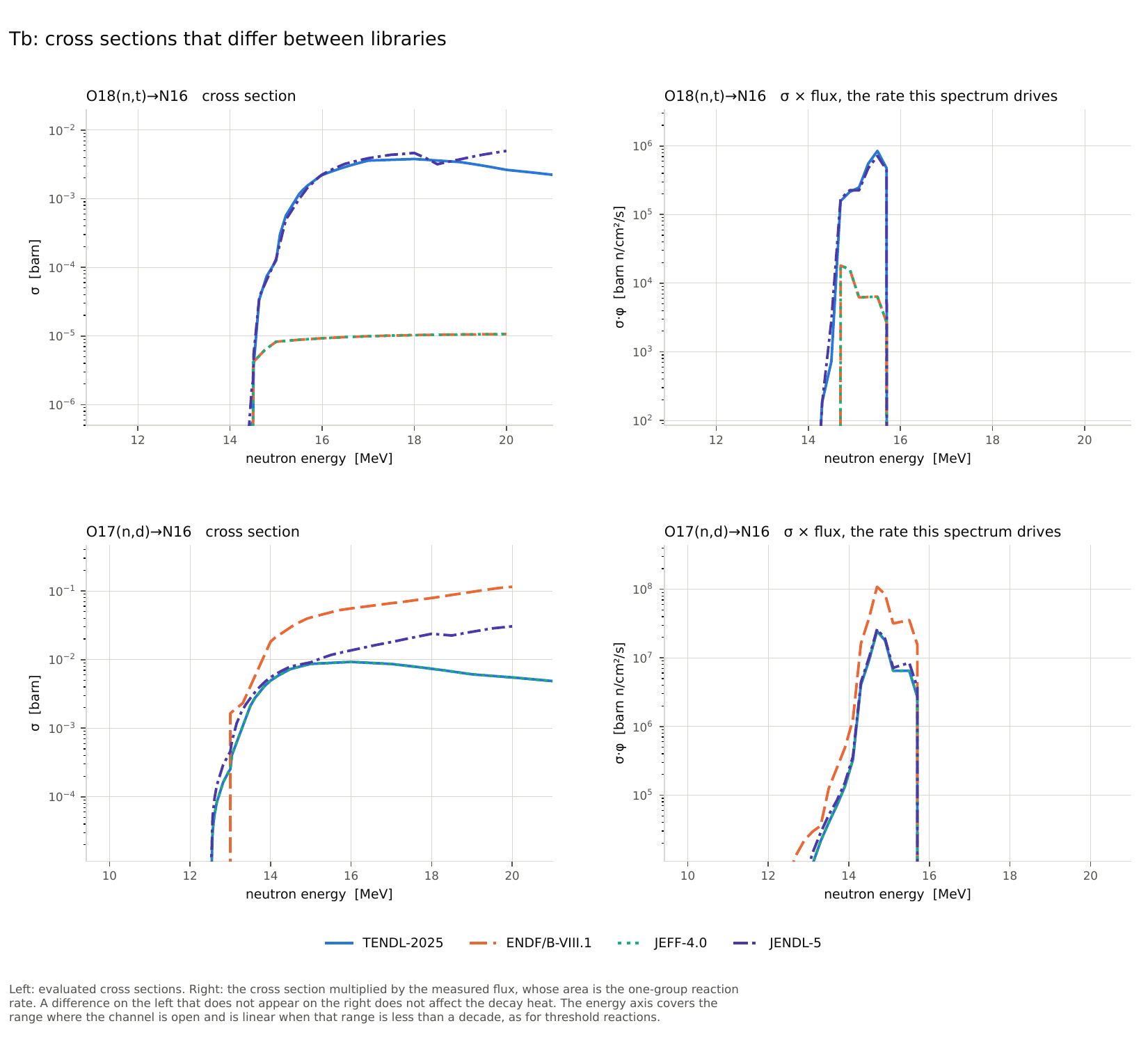}
\caption{Tb: cross sections of the channels whose spectrum-averaged rates differ, with the neutron spectrum and their product.}
\end{figure}
\clearpage

\begin{figure}[H]
\centering
\includegraphics[width=\textwidth,height=0.82\textheight,keepaspectratio]{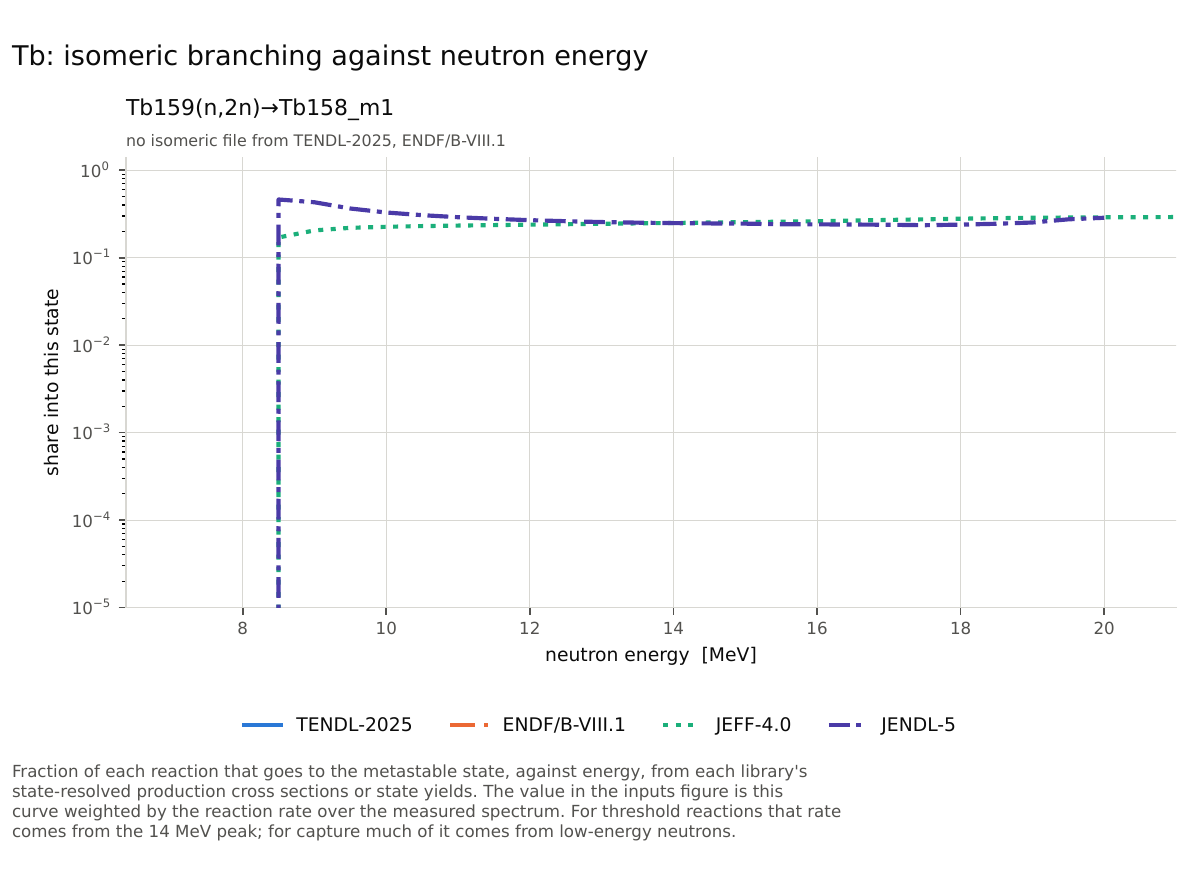}
\caption{Tb: isomeric branching ratio against neutron energy for the channels whose spectrum-averaged ratio differs between libraries.}
\end{figure}
\clearpage

Mean absolute C/E deviations range from 76 to 83 percent. The spread of 7 percentage points is smaller than the median measurement uncertainty of 122 percent, so the measurement cannot separate the libraries.

\subsubsection*{Differences in the nuclear data}
\begin{itemize}\setlength{\itemsep}{6pt}
\item ENDF/B-VIII.1 gives the total reaction rate of O18(n,t)→N16 as 1.751e-17, against 6.84e-16 for the median of the others, a factor of 39.\\[2pt]{\itshape This is the total rate of the channel over all its products, so the difference is in the cross section, not in the split between states.}
\item ENDF/B-VIII.1 gives the total reaction rate of O17(n,d)→N16 as 1.096e-13, against 2.351e-14 for the median of the others, a factor of 5.
\item ENDF/B-VIII.1 gives no heat from Tb158\_m1; the other libraries do.\\[2pt]{\itshape The heat breakdown alone does not show why. Possible causes are a missing reaction channel, a different state assignment, different decay data or a reporting threshold.}
\end{itemize}
\subsubsection*{Suggested measurements}
\begin{itemize}\setlength{\itemsep}{4pt}
\item Tb has a single stable isotope, so the foil is already a single-isotope target and enrichment would not help.
\item The sample is a compound that also contains O. Their activation adds to the measured heat; a metal or purer sample would remove it.
\item This foil makes Tb160 (half-life 72.3 days), which gives up to 39 percent of the calculated heat. A five-minute irradiation and 55 minute measurement only see the start of this contribution. There is no seven-hour FNS measurement of this foil. A longer irradiation followed for days to months would be needed to test this product's production and decay.
\item The reaction rates are averaged over the measured spectrum. Activation measurements at several well-defined neutron energies, including below 14 MeV, would show which energy range the differences come from.
\end{itemize}
\clearpage
\subsection{Te: Tellurium}\label{foil:Te}
\begin{figure}[H]
\centering
\includegraphics[width=\textwidth,height=0.82\textheight,keepaspectratio]{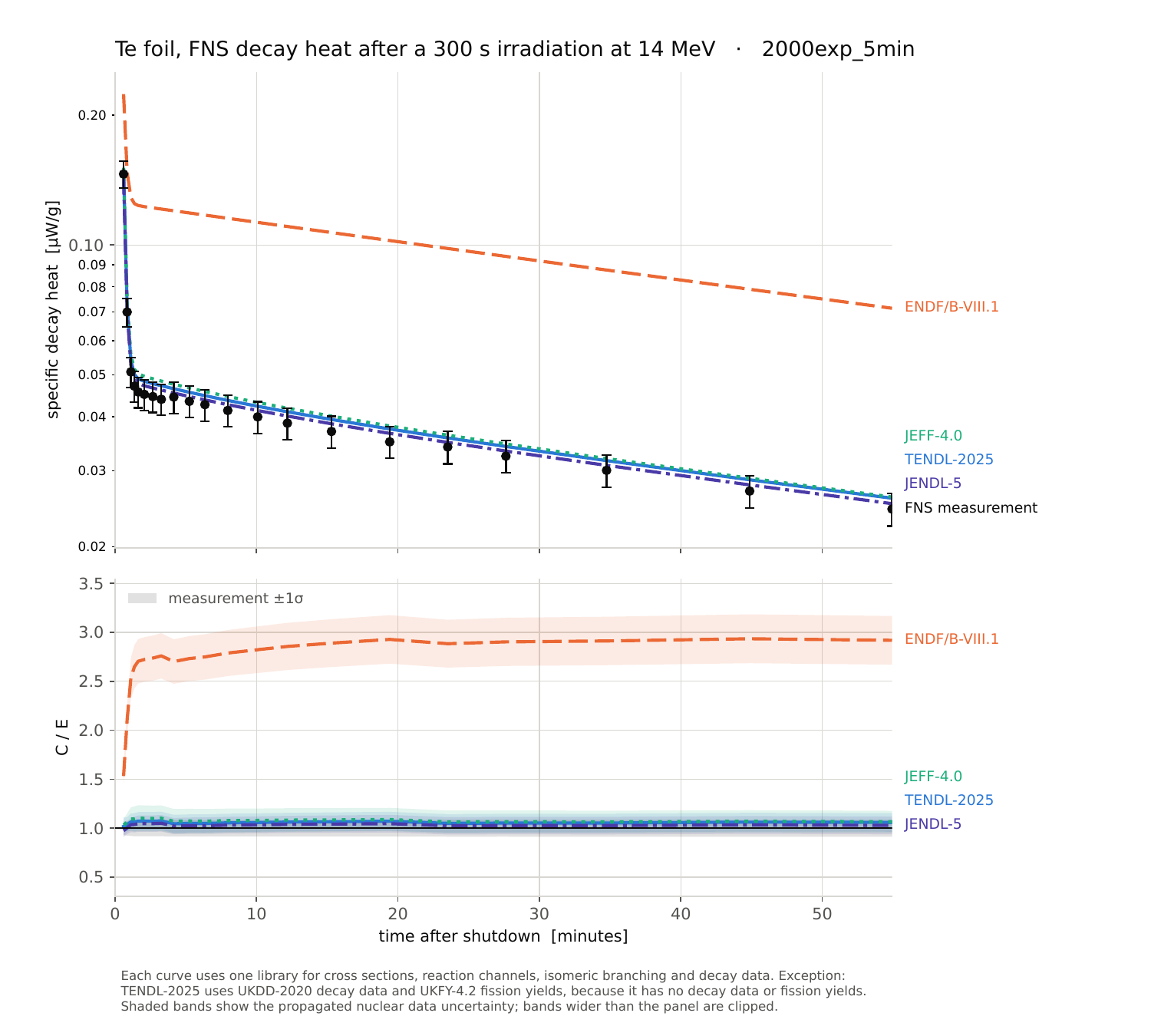}
\caption{Te: decay heat against the measurement, with C/E beneath. Mean $|C/E-1|$ by library: JENDL-5 3\%, TENDL-2025 6\%, JEFF-4.0 8\%, ENDF/B-VIII.1 170\%.}
\end{figure}
\clearpage

\begin{figure}[H]
\centering
\includegraphics[width=\textwidth,height=0.82\textheight,keepaspectratio]{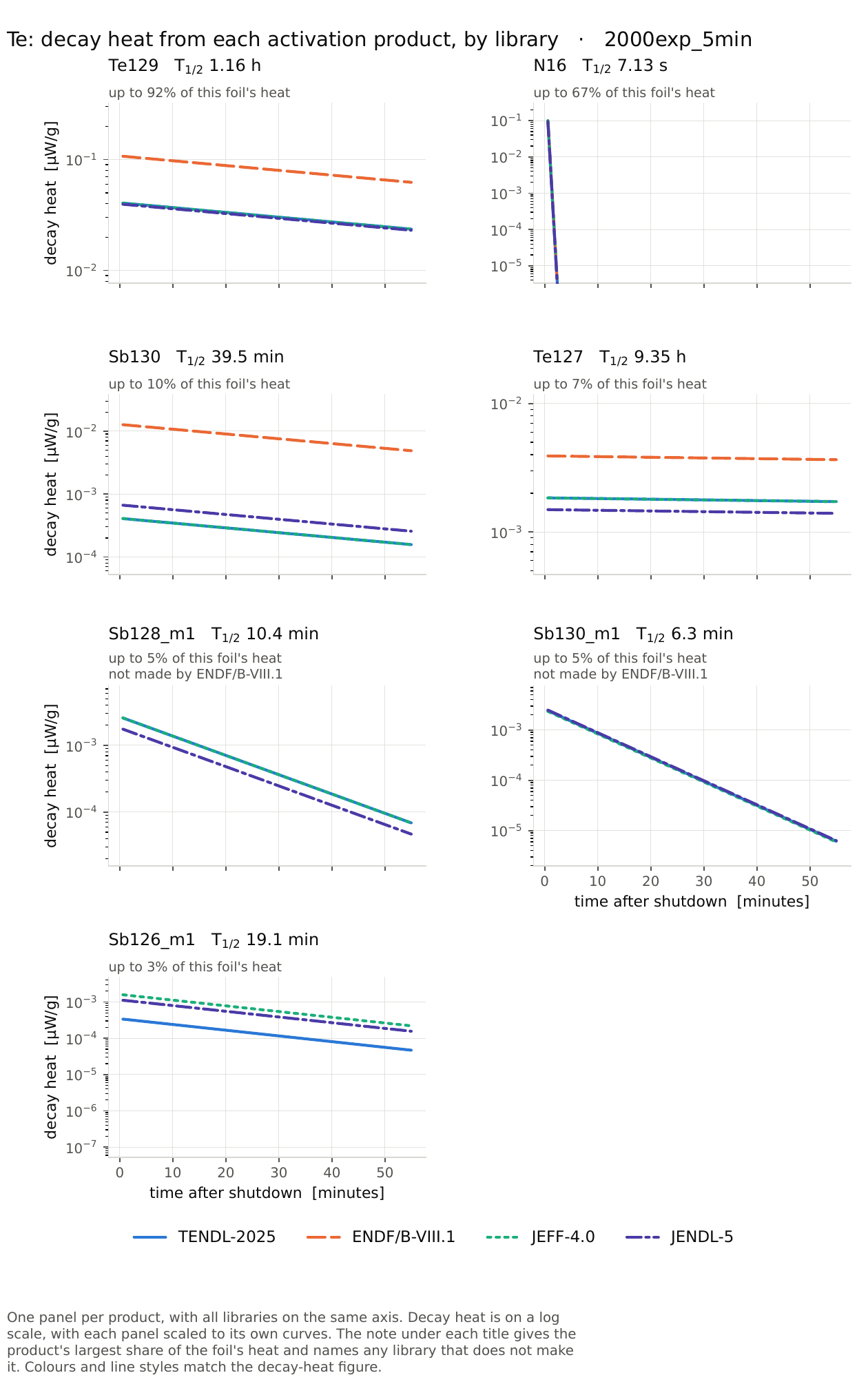}
\caption{Te: share of the decay heat from each activation product during cooling.}
\end{figure}
\clearpage

\begin{figure}[H]
\centering
\includegraphics[width=\textwidth,height=0.82\textheight,keepaspectratio]{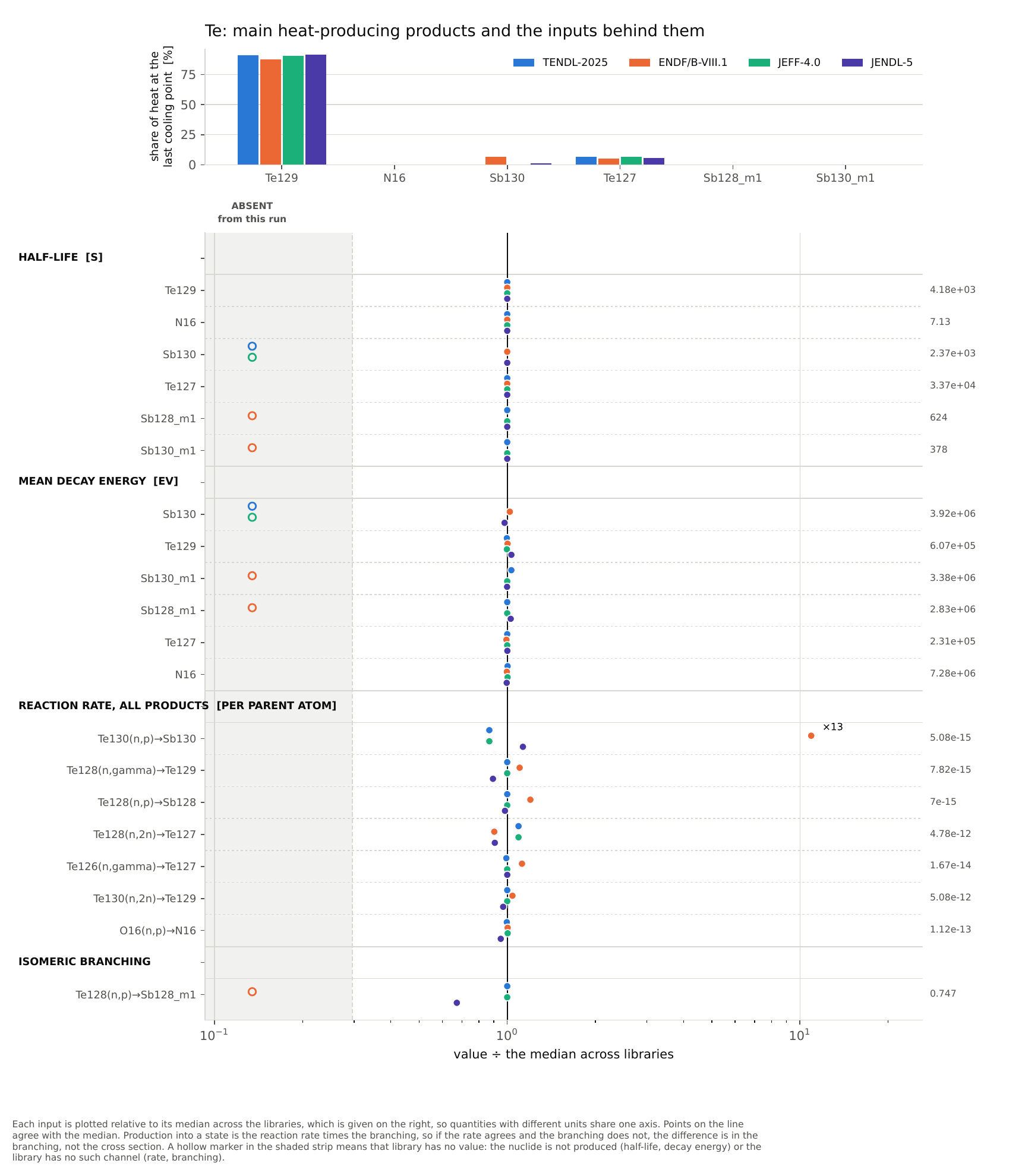}
\caption{Te: inputs for the main heat-producing products, each relative to its median across the available libraries.}
\end{figure}
\clearpage

\begin{figure}[H]
\centering
\includegraphics[width=\textwidth,height=0.82\textheight,keepaspectratio]{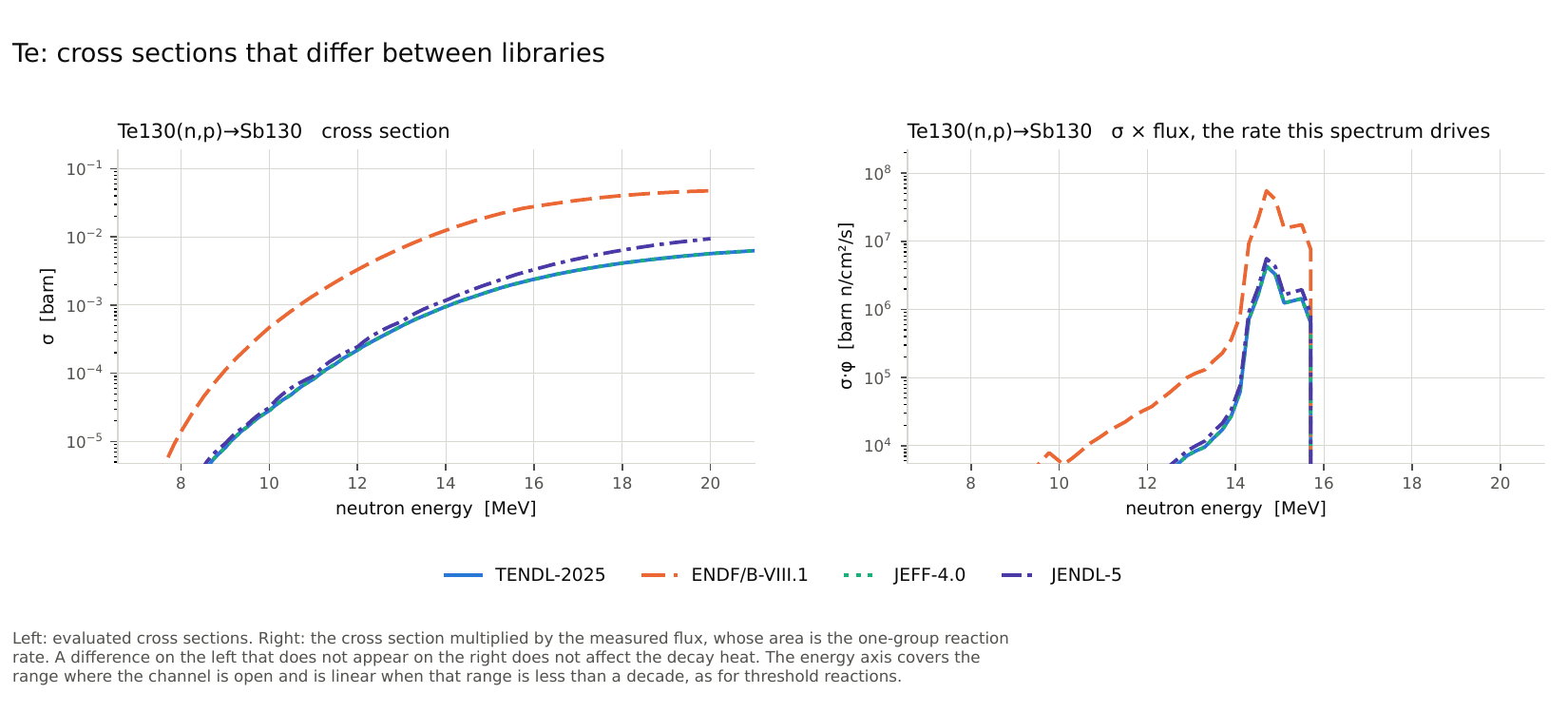}
\caption{Te: cross sections of the channels whose spectrum-averaged rates differ, with the neutron spectrum and their product.}
\end{figure}
\clearpage

ENDF/B-VIII.1 stands out at 170 percent from the measurement, against 8 percent for the next furthest library and 3 percent for JENDL-5. The median measurement uncertainty is 8 percent.

\subsubsection*{Differences in the nuclear data}
\begin{itemize}\setlength{\itemsep}{6pt}
\item ENDF/B-VIII.1 gives the total reaction rate of Te130(n,p)→Sb130 as 5.555e-14, against 4.416e-15 for the median of the others, a factor of 13.\\[2pt]{\itshape This is the total rate of the channel over all its products, so the difference is in the cross section, not in the split between states.}
\item ENDF/B-VIII.1 gives no heat from Sb128\_m1; the other libraries do.\\[2pt]{\itshape The heat breakdown alone does not show why. Possible causes are a missing reaction channel, a different state assignment, different decay data or a reporting threshold.}
\item ENDF/B-VIII.1 gives no heat from Sb130\_m1; the other libraries do.
\end{itemize}
\subsubsection*{Suggested measurements}
\begin{itemize}\setlength{\itemsep}{4pt}
\item Natural Te has 8 isotopes, so the measured heat combines several parents. Enriched single-isotope targets would distinguish their contributions. The candidate parent is Te130. Enrichment alone does not isolate a single reaction channel.
\item The sample is a compound that also contains O. Their activation adds to the measured heat; a metal or purer sample would remove it.
\item The reaction rates are averaged over the measured spectrum. Activation measurements at several well-defined neutron energies, including below 14 MeV, would show which energy range the differences come from.
\end{itemize}
\clearpage
\subsection{Ti: Titanium}\label{foil:Ti}
\begin{figure}[H]
\centering
\includegraphics[width=\textwidth,height=0.82\textheight,keepaspectratio]{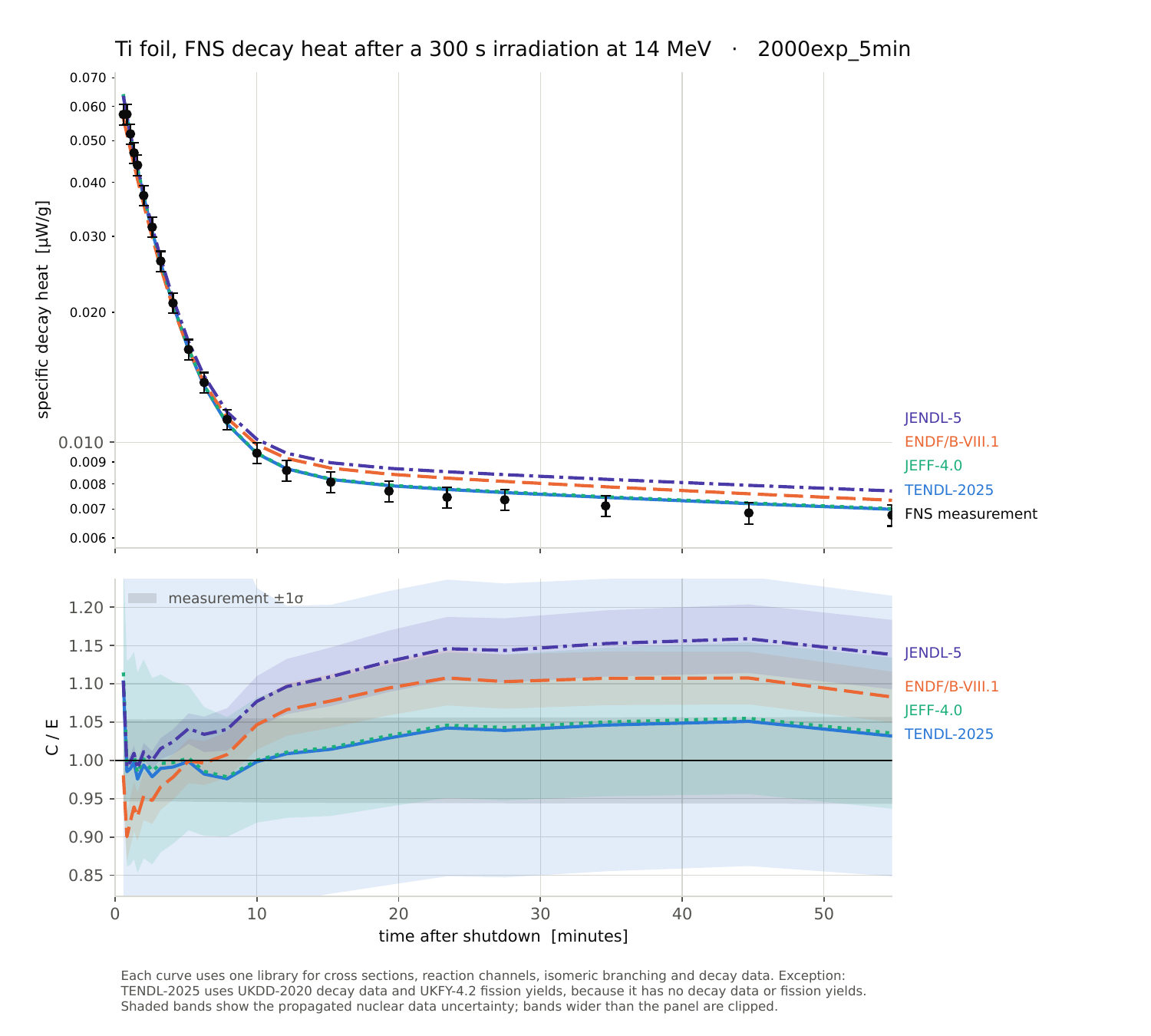}
\caption{Ti: decay heat against the measurement, with C/E beneath. Mean $|C/E-1|$ by library: JEFF-4.0 2\%, TENDL-2025 2\%, ENDF/B-VIII.1 6\%, JENDL-5 7\%.}
\end{figure}
\clearpage

\begin{figure}[H]
\centering
\includegraphics[width=\textwidth,height=0.82\textheight,keepaspectratio]{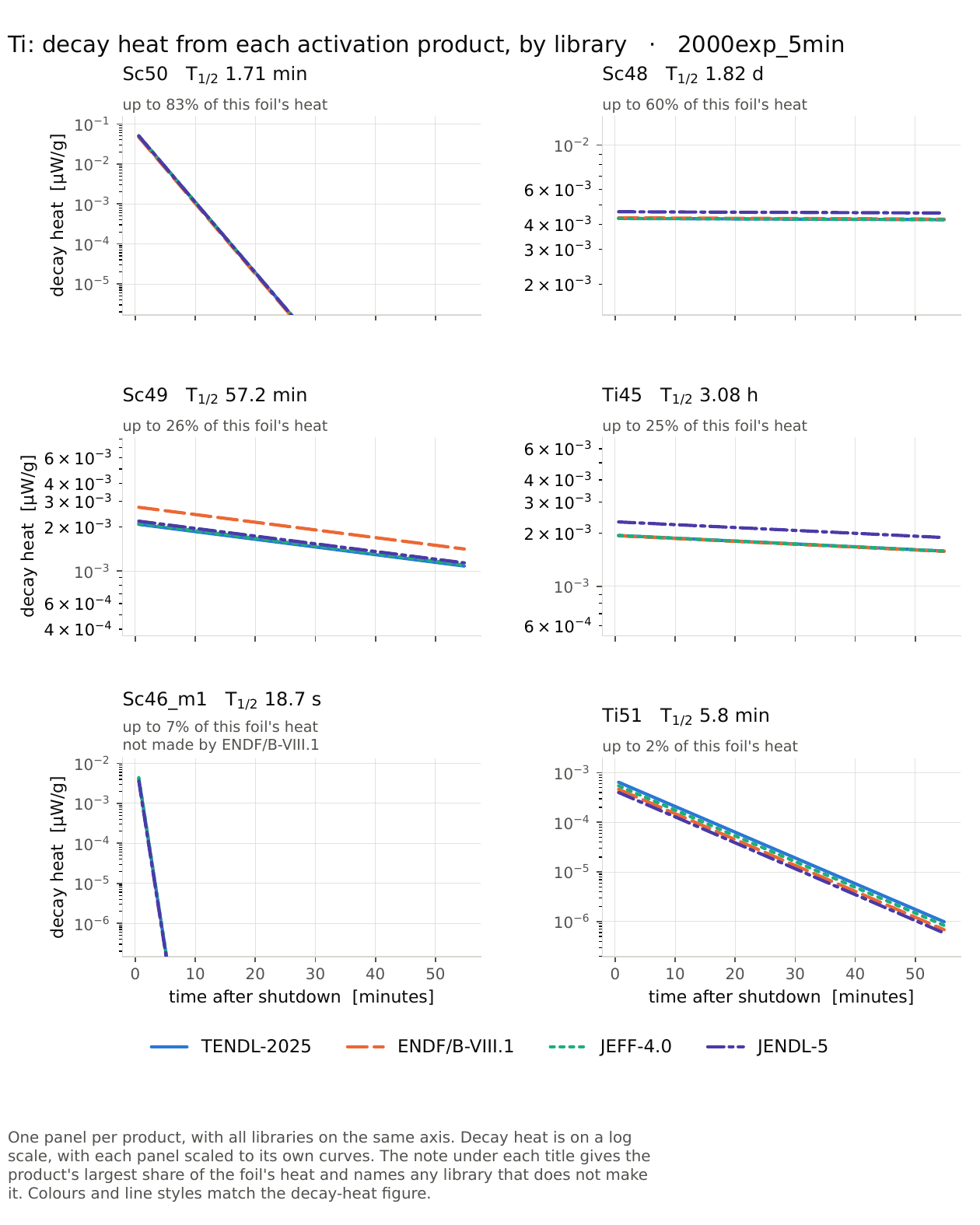}
\caption{Ti: share of the decay heat from each activation product during cooling.}
\end{figure}
\clearpage

\begin{figure}[H]
\centering
\includegraphics[width=\textwidth,height=0.82\textheight,keepaspectratio]{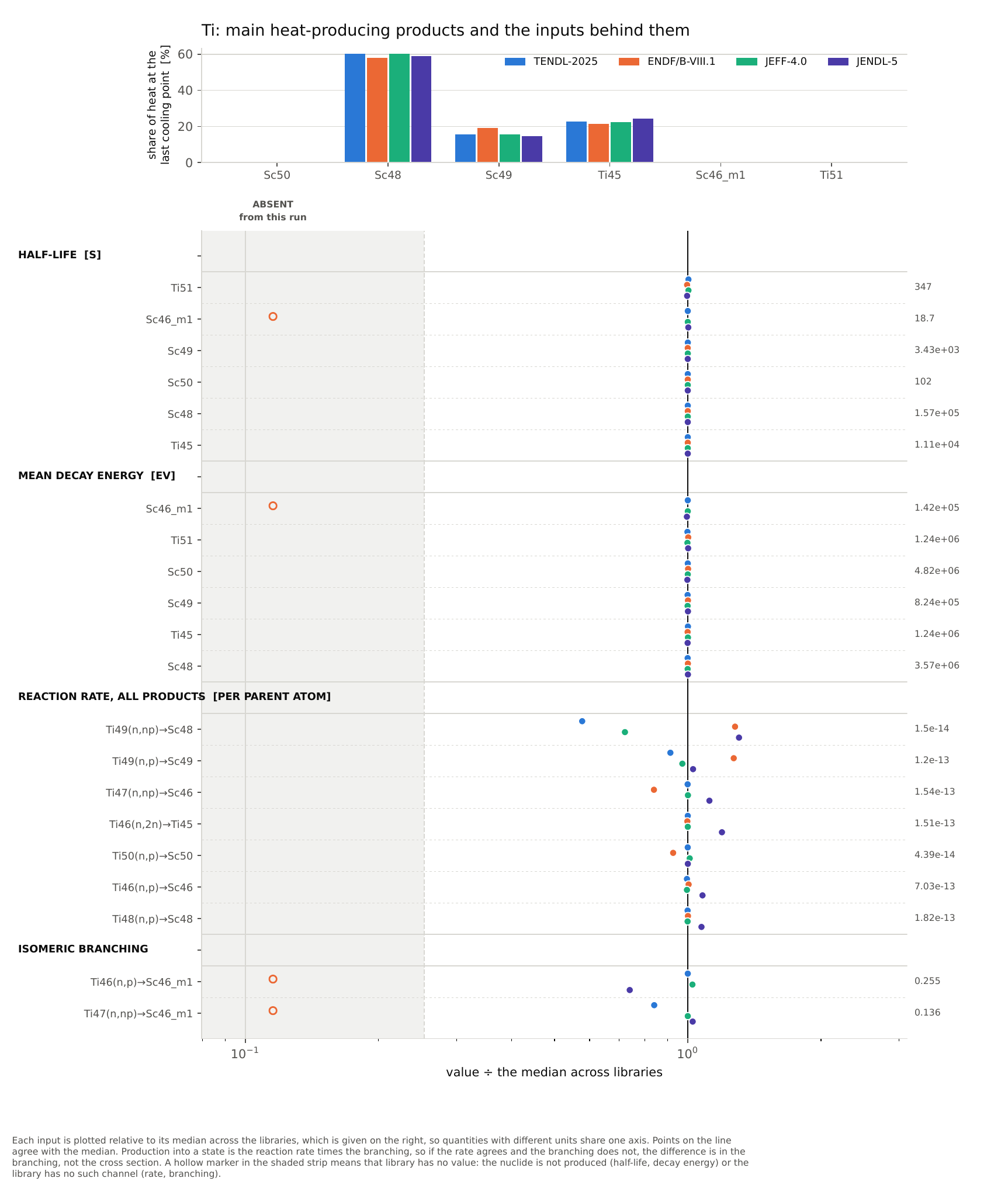}
\caption{Ti: inputs for the main heat-producing products, each relative to its median across the available libraries.}
\end{figure}
\clearpage

\begin{figure}[H]
\centering
\includegraphics[width=\textwidth,height=0.82\textheight,keepaspectratio]{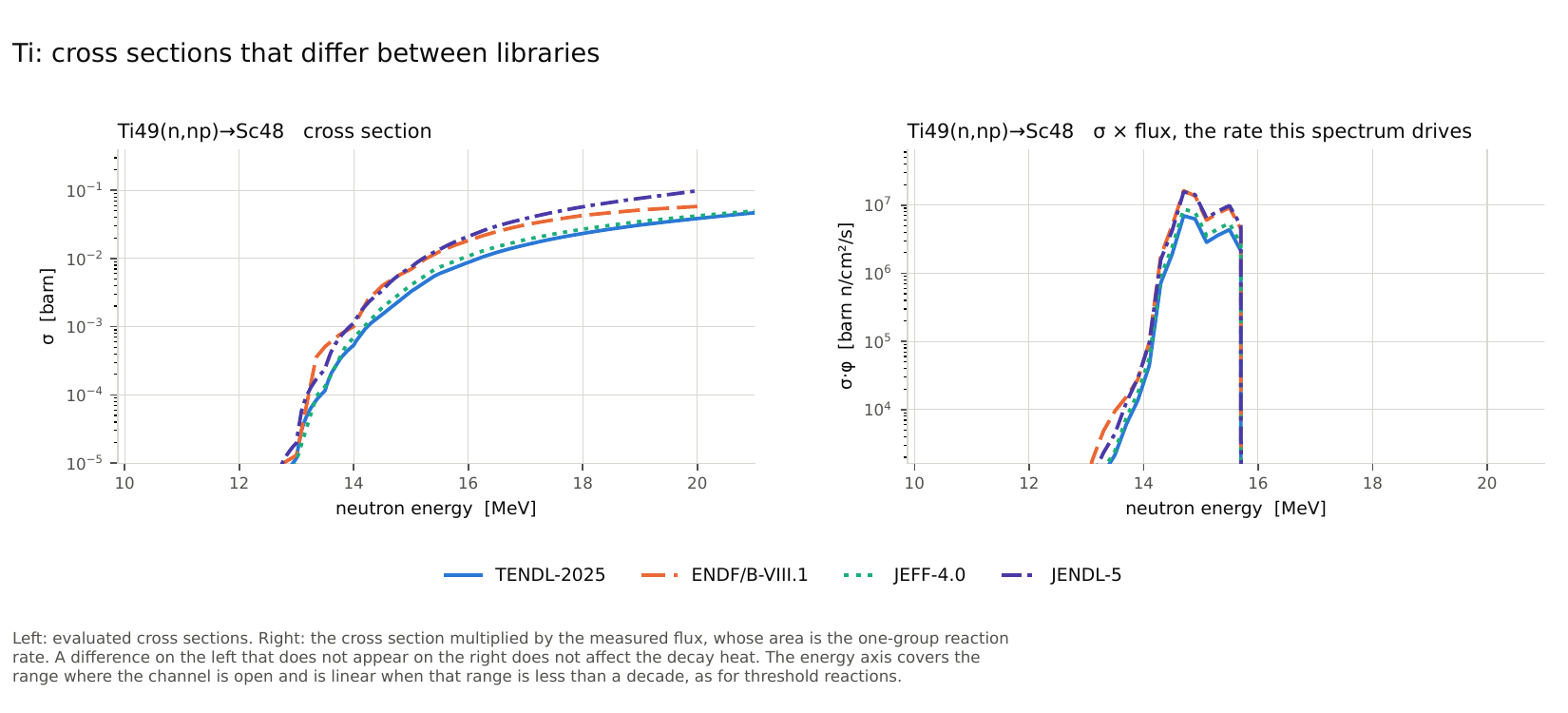}
\caption{Ti: cross sections of the channels whose spectrum-averaged rates differ, with the neutron spectrum and their product.}
\end{figure}
\clearpage

\subsubsection*{1996 seven-hour campaign}
The 1996 campaign irradiated this foil for 7.57 hours at position 7 and measured it out to 402.96 days. In the five-minute calculation Sc48 (half-life 1.82 days) gives up to 60 percent of the heat, and the longer irradiation and cooling test its production and decay data. JENDL-5 is closest. This campaign uses a different spectrum, so it is not included in the main rankings.

\begin{figure}[H]
\centering
\includegraphics[width=\textwidth,height=0.82\textheight,keepaspectratio]{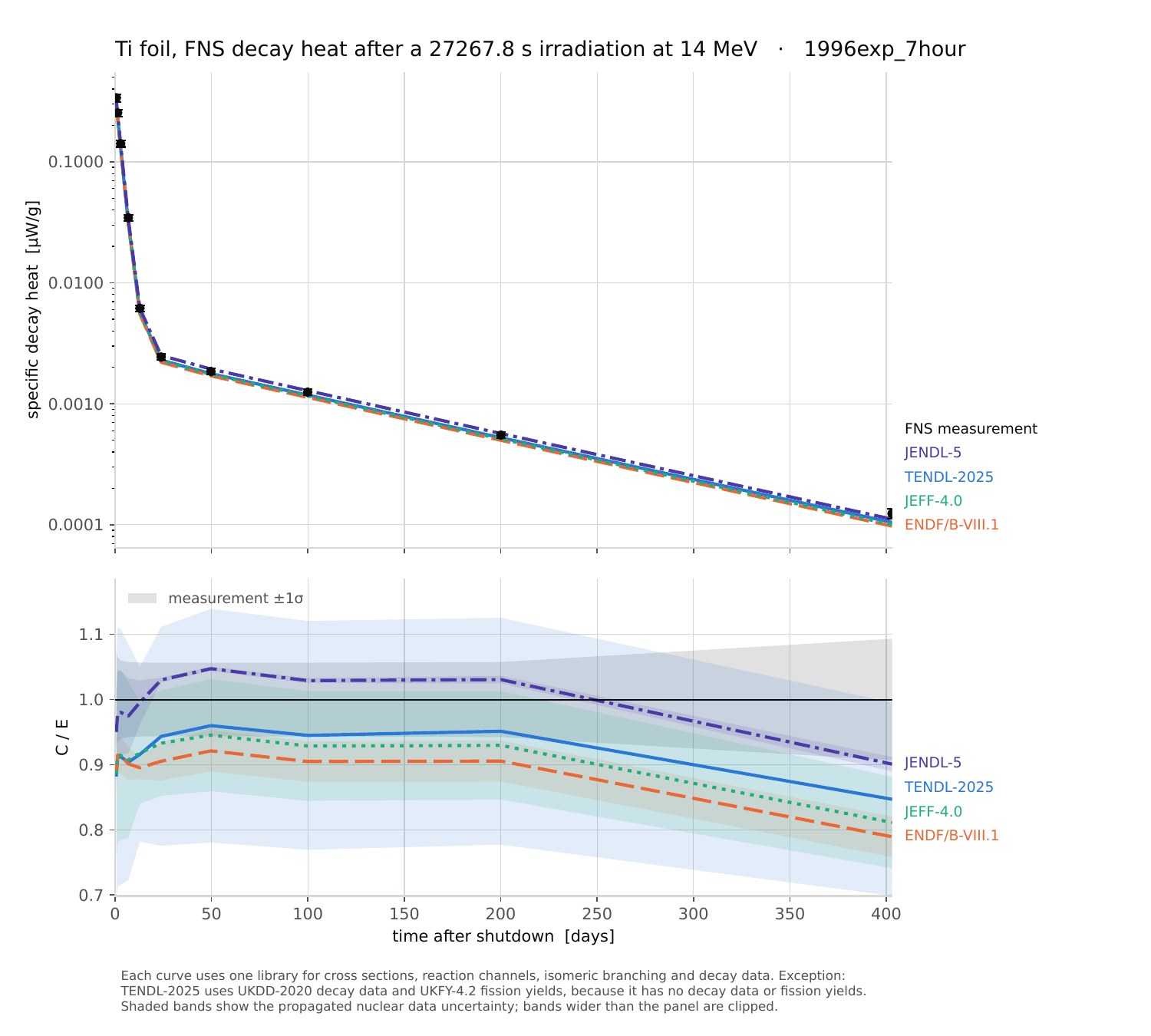}
\caption{Ti: 1996 seven-hour campaign, irradiation position 7. Mean $|C/E-1|$ by library: JENDL-5 4\%, TENDL-2025 8\%, JEFF-4.0 9\%, ENDF/B-VIII.1 11\%. }
\end{figure}
\clearpage

Mean absolute C/E deviations range from 2 to 7 percent. The spread of 5 percentage points is smaller than the median measurement uncertainty of 5 percent, so the measurement cannot separate the libraries.

\subsubsection*{Differences in the nuclear data}
\begin{itemize}\setlength{\itemsep}{6pt}
\item TENDL-2025 gives the total reaction rate of Ti49(n,np)→Sc48 as 8.673e-15, against 1.922e-14 for the median of the others, a factor of 2.\\[2pt]{\itshape This is the total rate of the channel over all its products, so the difference is in the cross section, not in the split between states.}
\item ENDF/B-VIII.1 gives no heat from Sc46\_m1; the other libraries do.\\[2pt]{\itshape The heat breakdown alone does not show why. Possible causes are a missing reaction channel, a different state assignment, different decay data or a reporting threshold.}
\end{itemize}
\subsubsection*{Suggested measurements}
\begin{itemize}\setlength{\itemsep}{4pt}
\item Natural Ti has 5 isotopes, so the measured heat combines several parents. Enriched single-isotope targets would distinguish their contributions. The candidate parent is Ti49. Enrichment alone does not isolate a single reaction channel.
\item The reaction rates are averaged over the measured spectrum. Activation measurements at several well-defined neutron energies, including below 14 MeV, would show which energy range the differences come from.
\end{itemize}
\clearpage
\subsection{Tl: Thallium}\label{foil:Tl}
\begin{figure}[H]
\centering
\includegraphics[width=\textwidth,height=0.82\textheight,keepaspectratio]{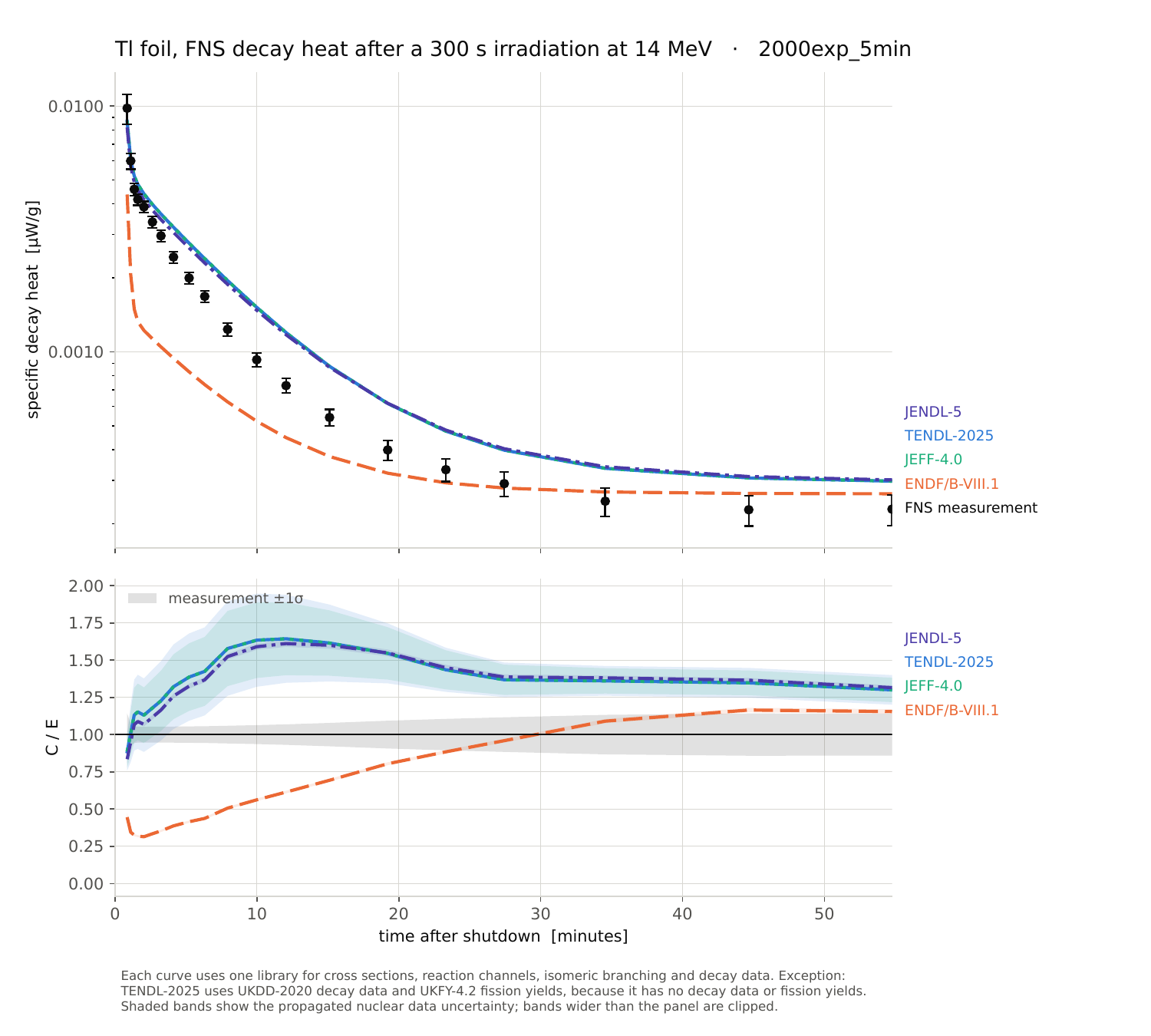}
\caption{Tl: decay heat against the measurement, with C/E beneath. Mean $|C/E-1|$ by library: JENDL-5 32\%, JEFF-4.0 34\%, TENDL-2025 35\%, ENDF/B-VIII.1 44\%.}
\end{figure}
\clearpage

\begin{figure}[H]
\centering
\includegraphics[width=\textwidth,height=0.82\textheight,keepaspectratio]{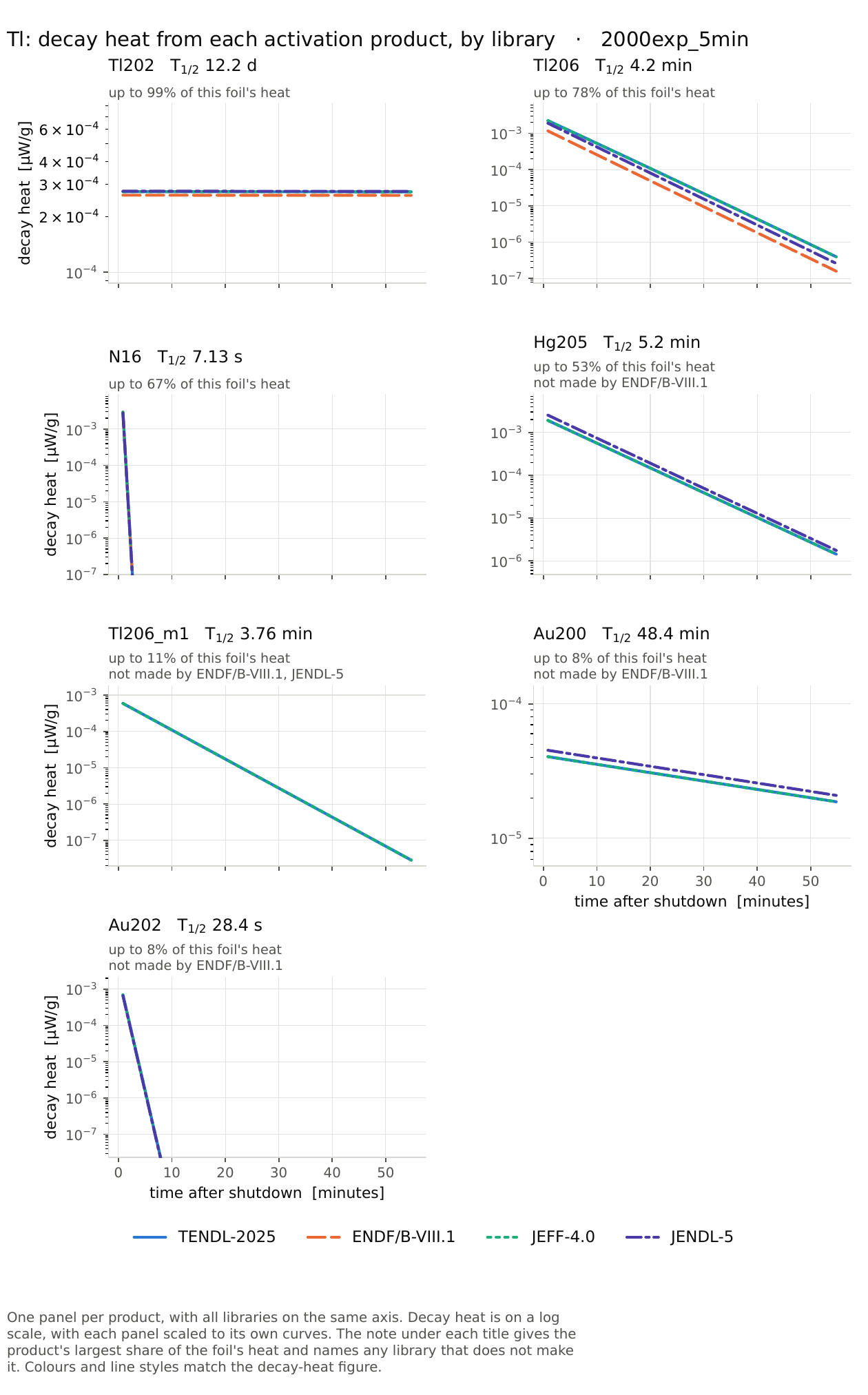}
\caption{Tl: share of the decay heat from each activation product during cooling.}
\end{figure}
\clearpage

\begin{figure}[H]
\centering
\includegraphics[width=\textwidth,height=0.82\textheight,keepaspectratio]{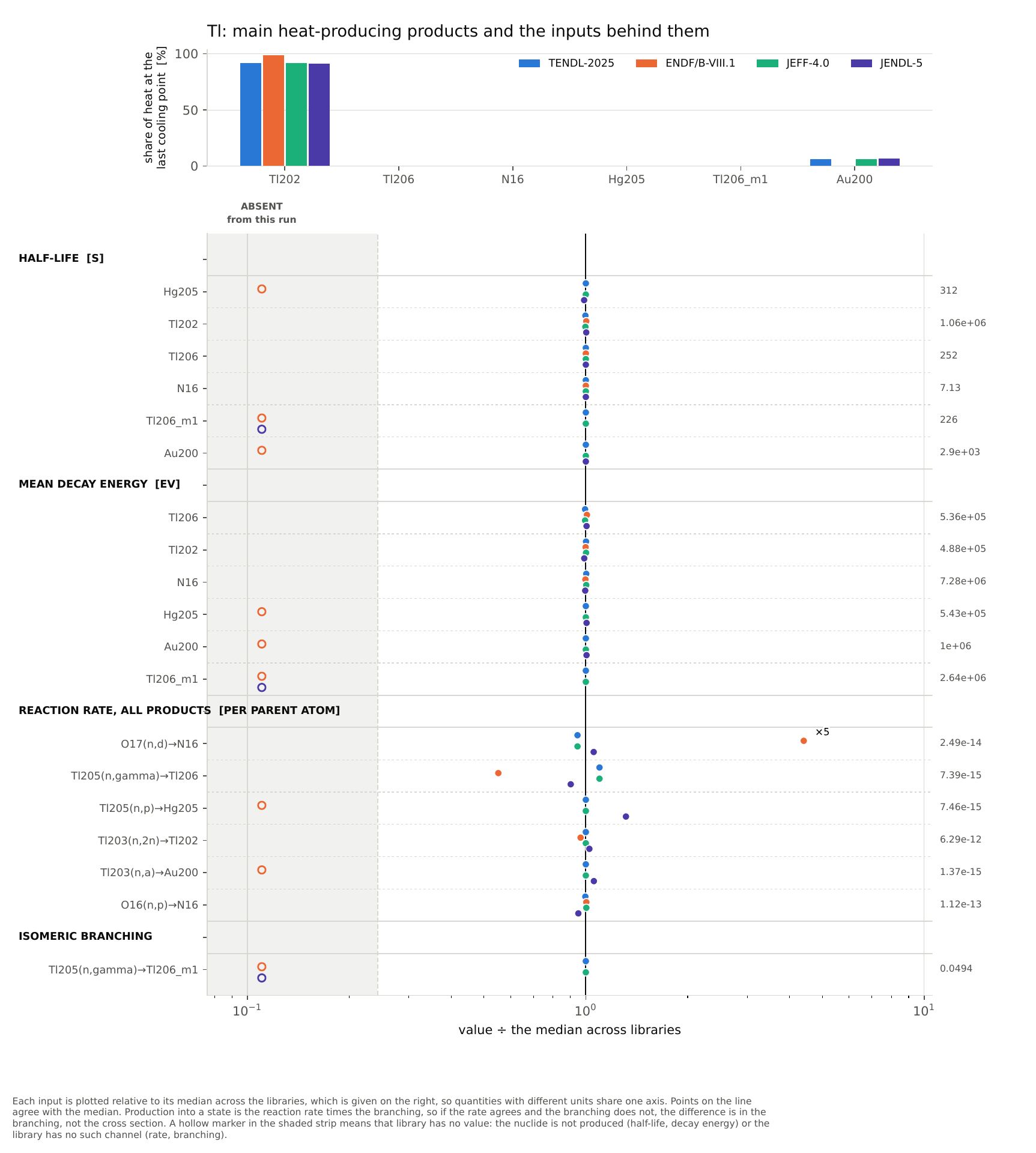}
\caption{Tl: inputs for the main heat-producing products, each relative to its median across the available libraries.}
\end{figure}
\clearpage

\begin{figure}[H]
\centering
\includegraphics[width=\textwidth,height=0.82\textheight,keepaspectratio]{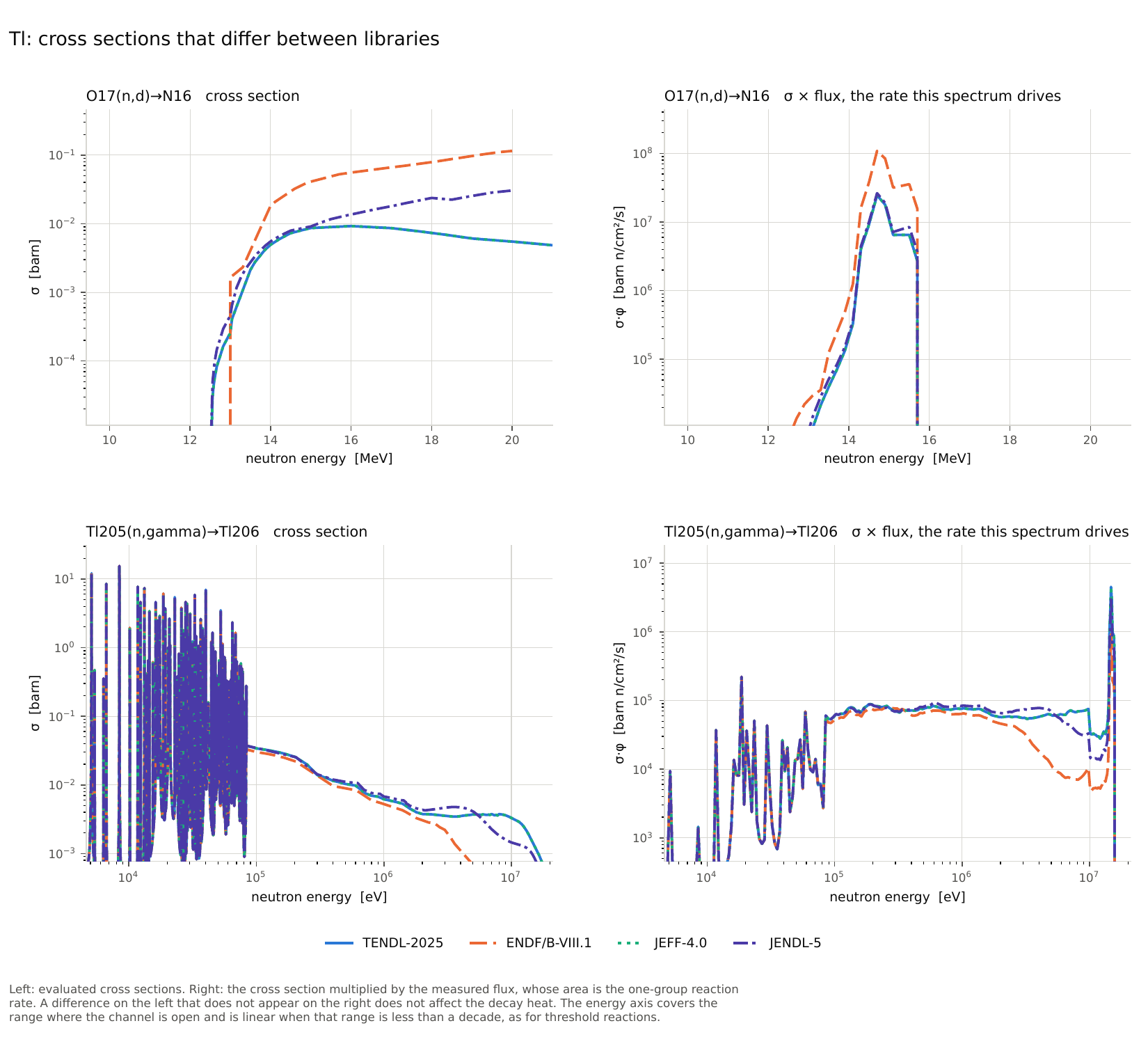}
\caption{Tl: cross sections of the channels whose spectrum-averaged rates differ, with the neutron spectrum and their product.}
\end{figure}
\clearpage

The libraries are between 32 and 44 percent from the measurement, which has a median uncertainty of 7 percent. JENDL-5 is closest and ENDF/B-VIII.1 furthest.

\subsubsection*{Differences in the nuclear data}
\begin{itemize}\setlength{\itemsep}{6pt}
\item ENDF/B-VIII.1 gives the total reaction rate of O17(n,d)→N16 as 1.096e-13, against 2.351e-14 for the median of the others, a factor of 5.\\[2pt]{\itshape This is the total rate of the channel over all its products, so the difference is in the cross section, not in the split between states.}
\item ENDF/B-VIII.1 gives no heat from Hg205; the other libraries do.\\[2pt]{\itshape The heat breakdown alone does not show why. Possible causes are a missing reaction channel, a different state assignment, different decay data or a reporting threshold.}
\item ENDF/B-VIII.1, JENDL-5 give no heat from Tl206\_m1; the other libraries do.
\item ENDF/B-VIII.1 gives no heat from Au200; the other libraries do.
\end{itemize}
\subsubsection*{Suggested measurements}
\begin{itemize}\setlength{\itemsep}{4pt}
\item Natural Tl has 2 isotopes, so the measured heat combines several parents. Enriched single-isotope targets would distinguish their contributions. The candidate parent is O17. Enrichment alone does not isolate a single reaction channel.
\item The sample is a compound that also contains O. Their activation adds to the measured heat; a metal or purer sample would remove it.
\item This foil makes Tl202 (half-life 12.2 days), which gives up to 99 percent of the calculated heat. A five-minute irradiation and 55 minute measurement only see the start of this contribution. There is no seven-hour FNS measurement of this foil. A longer irradiation followed for days to months would be needed to test this product's production and decay.
\item The reaction rates are averaged over the measured spectrum. Activation measurements at several well-defined neutron energies, including below 14 MeV, would show which energy range the differences come from.
\end{itemize}
\clearpage
\subsection{Tm: Thulium}\label{foil:Tm}
\begin{figure}[H]
\centering
\includegraphics[width=\textwidth,height=0.82\textheight,keepaspectratio]{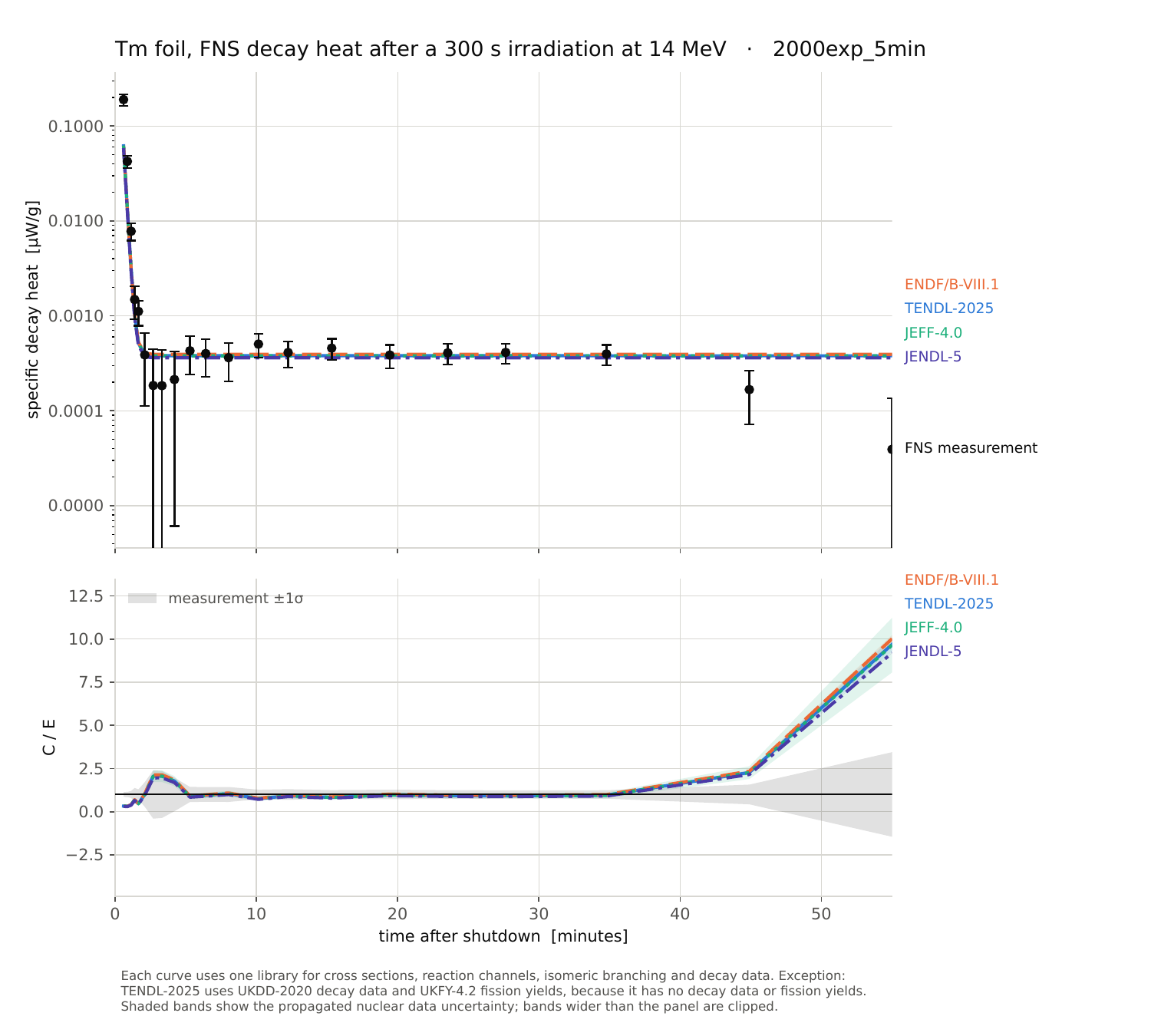}
\caption{Tm: decay heat against the measurement, with C/E beneath. Mean $|C/E-1|$ by library: JENDL-5 77\%, JEFF-4.0 79\%, TENDL-2025 79\%, ENDF/B-VIII.1 81\%.}
\end{figure}
\clearpage

\begin{figure}[H]
\centering
\includegraphics[width=\textwidth,height=0.82\textheight,keepaspectratio]{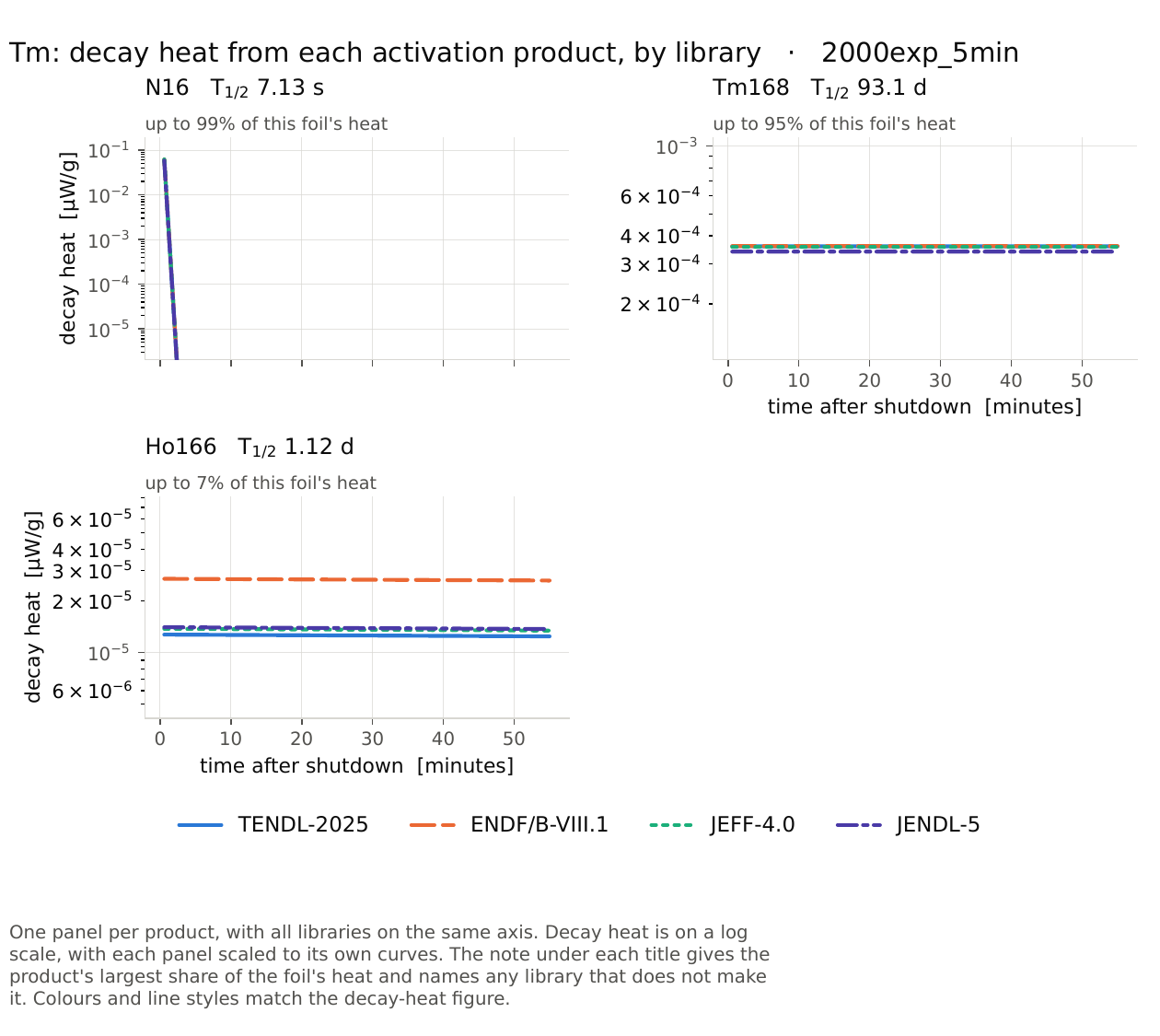}
\caption{Tm: share of the decay heat from each activation product during cooling.}
\end{figure}
\clearpage

\begin{figure}[H]
\centering
\includegraphics[width=\textwidth,height=0.82\textheight,keepaspectratio]{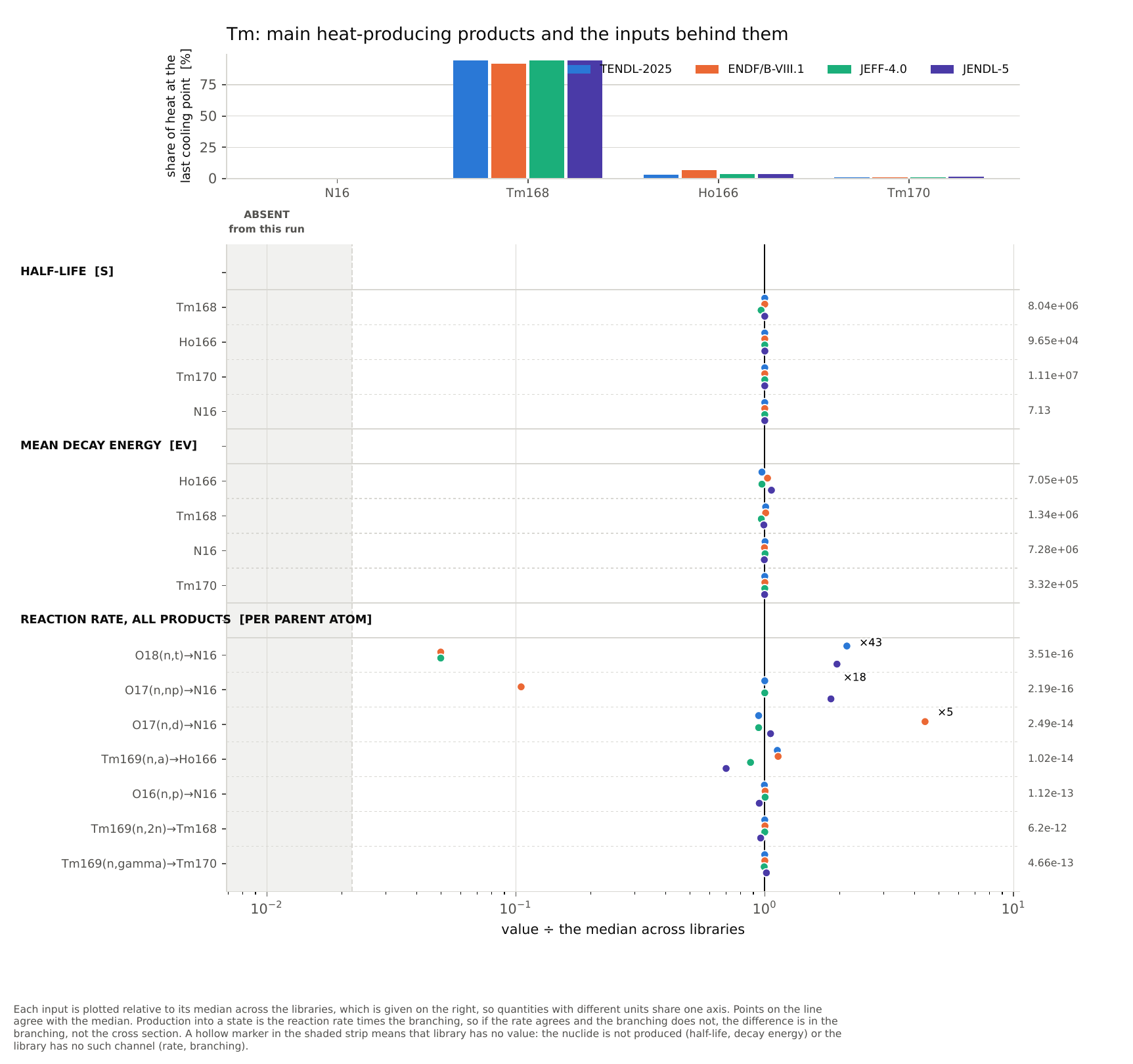}
\caption{Tm: inputs for the main heat-producing products, each relative to its median across the available libraries.}
\end{figure}
\clearpage

\begin{figure}[H]
\centering
\includegraphics[width=\textwidth,height=0.82\textheight,keepaspectratio]{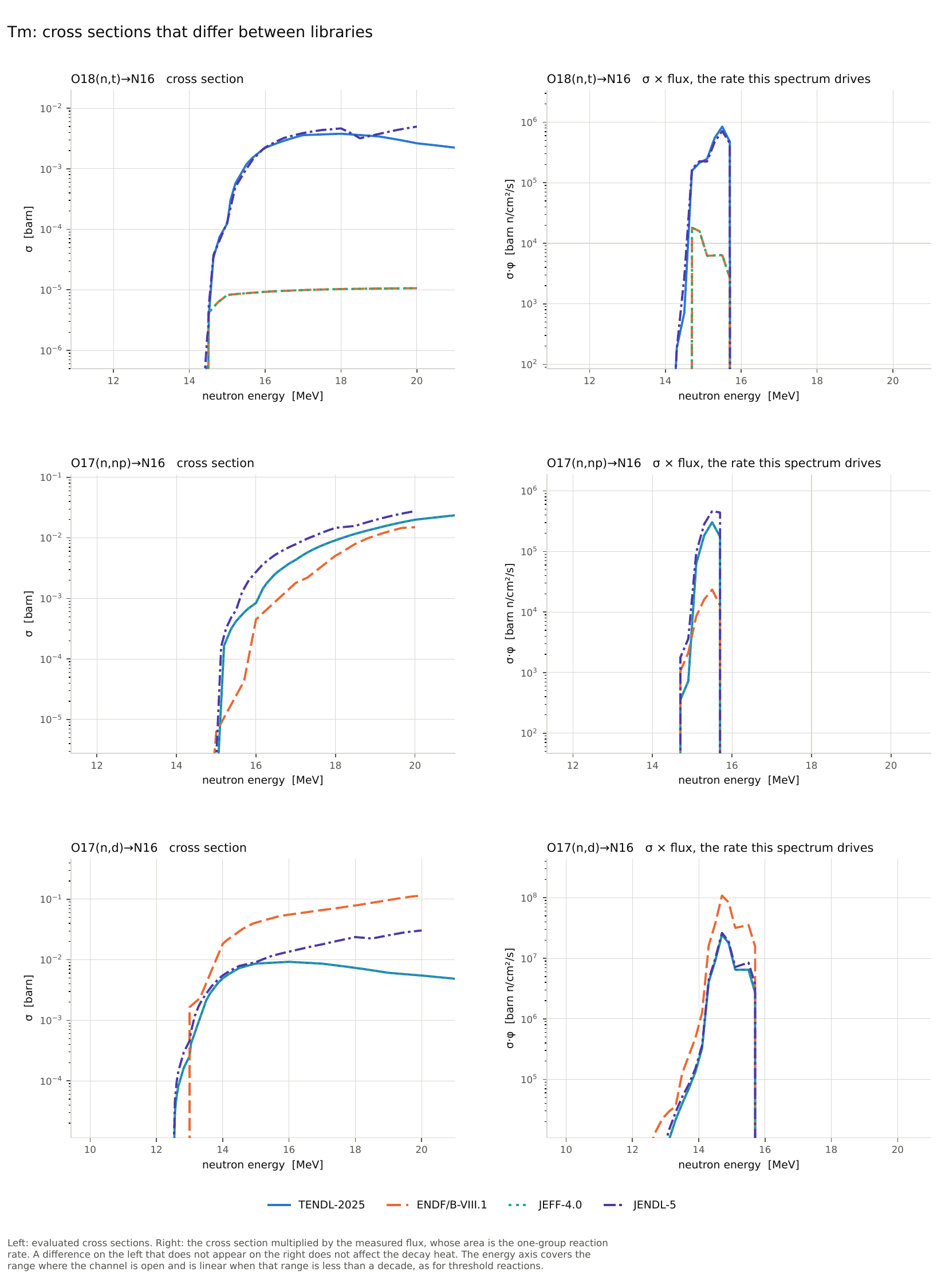}
\caption{Tm: cross sections of the channels whose spectrum-averaged rates differ, with the neutron spectrum and their product.}
\end{figure}
\clearpage

Mean absolute C/E deviations range from 77 to 81 percent. The spread of 4 percentage points is smaller than the median measurement uncertainty of 31 percent, so the measurement cannot separate the libraries.

\subsubsection*{Differences in the nuclear data}
\begin{itemize}\setlength{\itemsep}{6pt}
\item ENDF/B-VIII.1 gives the total reaction rate of O18(n,t)→N16 as 1.751e-17, against 6.84e-16 for the median of the others, a factor of 39.\\[2pt]{\itshape This is the total rate of the channel over all its products, so the difference is in the cross section, not in the split between states.}
\item ENDF/B-VIII.1 gives the total reaction rate of O17(n,np)→N16 as 2.295e-17, against 2.186e-16 for the median of the others, a factor of 10.
\item ENDF/B-VIII.1 gives the total reaction rate of O17(n,d)→N16 as 1.096e-13, against 2.351e-14 for the median of the others, a factor of 5.
\end{itemize}
\subsubsection*{Suggested measurements}
\begin{itemize}\setlength{\itemsep}{4pt}
\item Tm has a single stable isotope, so the foil is already a single-isotope target and enrichment would not help.
\item The sample is a compound that also contains O. Their activation adds to the measured heat; a metal or purer sample would remove it.
\item This foil makes Tm168 (half-life 93.1 days), which gives up to 95 percent of the calculated heat. A five-minute irradiation and 55 minute measurement only see the start of this contribution. There is no seven-hour FNS measurement of this foil. A longer irradiation followed for days to months would be needed to test this product's production and decay.
\item The reaction rates are averaged over the measured spectrum. Activation measurements at several well-defined neutron energies, including below 14 MeV, would show which energy range the differences come from.
\end{itemize}
\clearpage
\subsection{V: Vanadium}\label{foil:V}
\begin{figure}[H]
\centering
\includegraphics[width=\textwidth,height=0.82\textheight,keepaspectratio]{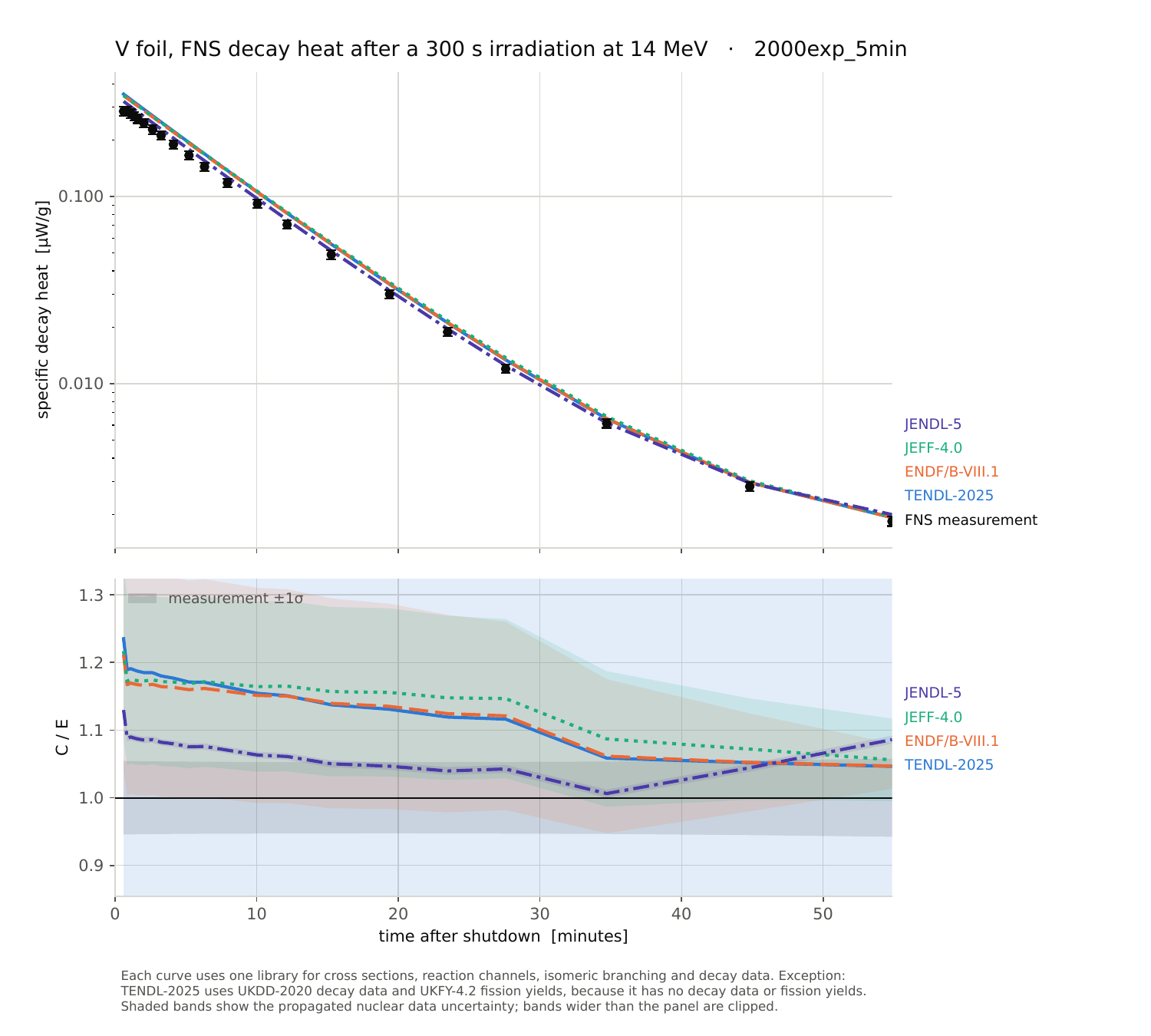}
\caption{V: decay heat against the measurement, with C/E beneath. Mean $|C/E-1|$ by library: JENDL-5 7\%, ENDF/B-VIII.1 14\%, TENDL-2025 15\%, JEFF-4.0 16\%.}
\end{figure}
\clearpage

\begin{figure}[H]
\centering
\includegraphics[width=\textwidth,height=0.82\textheight,keepaspectratio]{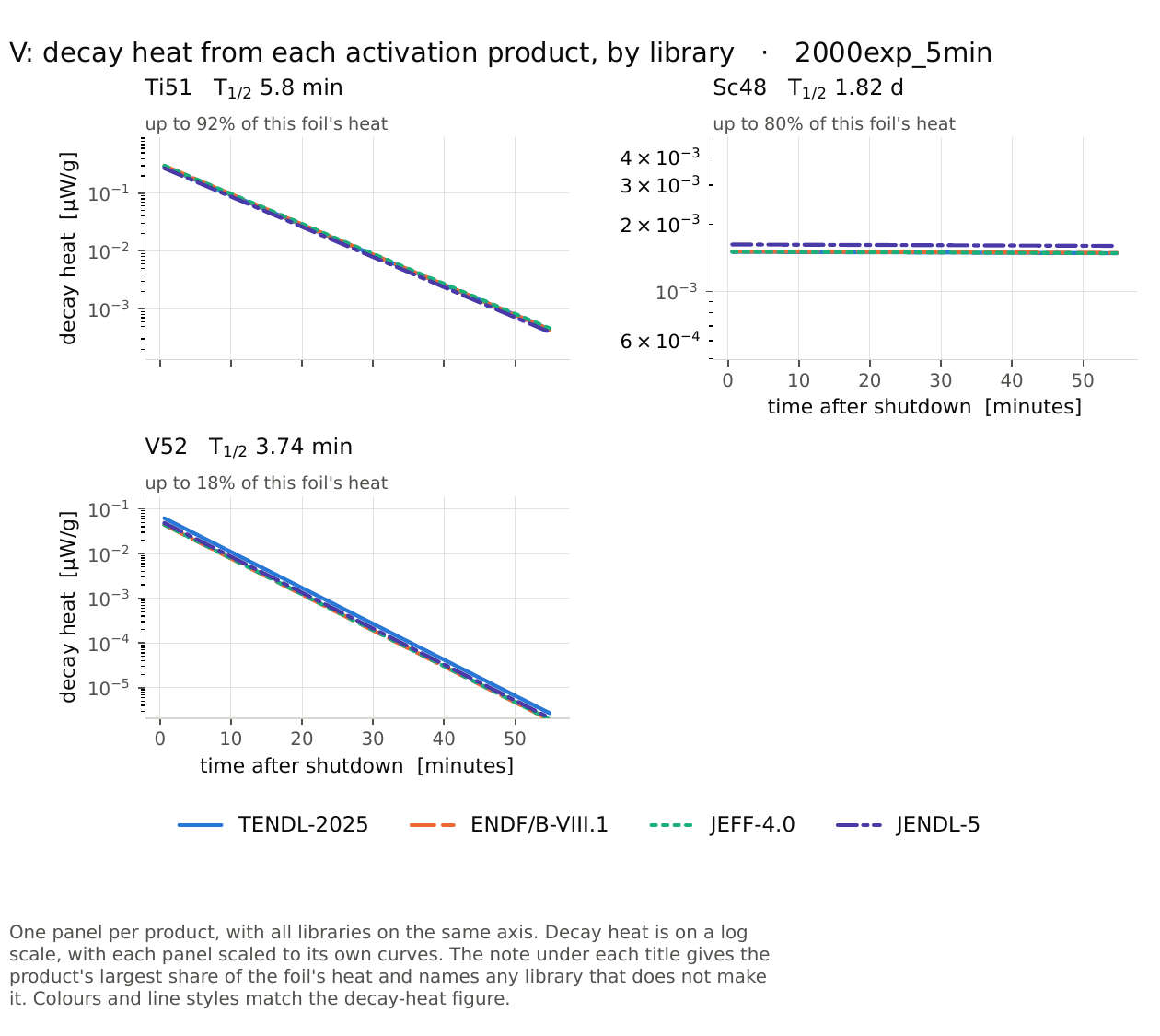}
\caption{V: share of the decay heat from each activation product during cooling.}
\end{figure}
\clearpage

\begin{figure}[H]
\centering
\includegraphics[width=\textwidth,height=0.82\textheight,keepaspectratio]{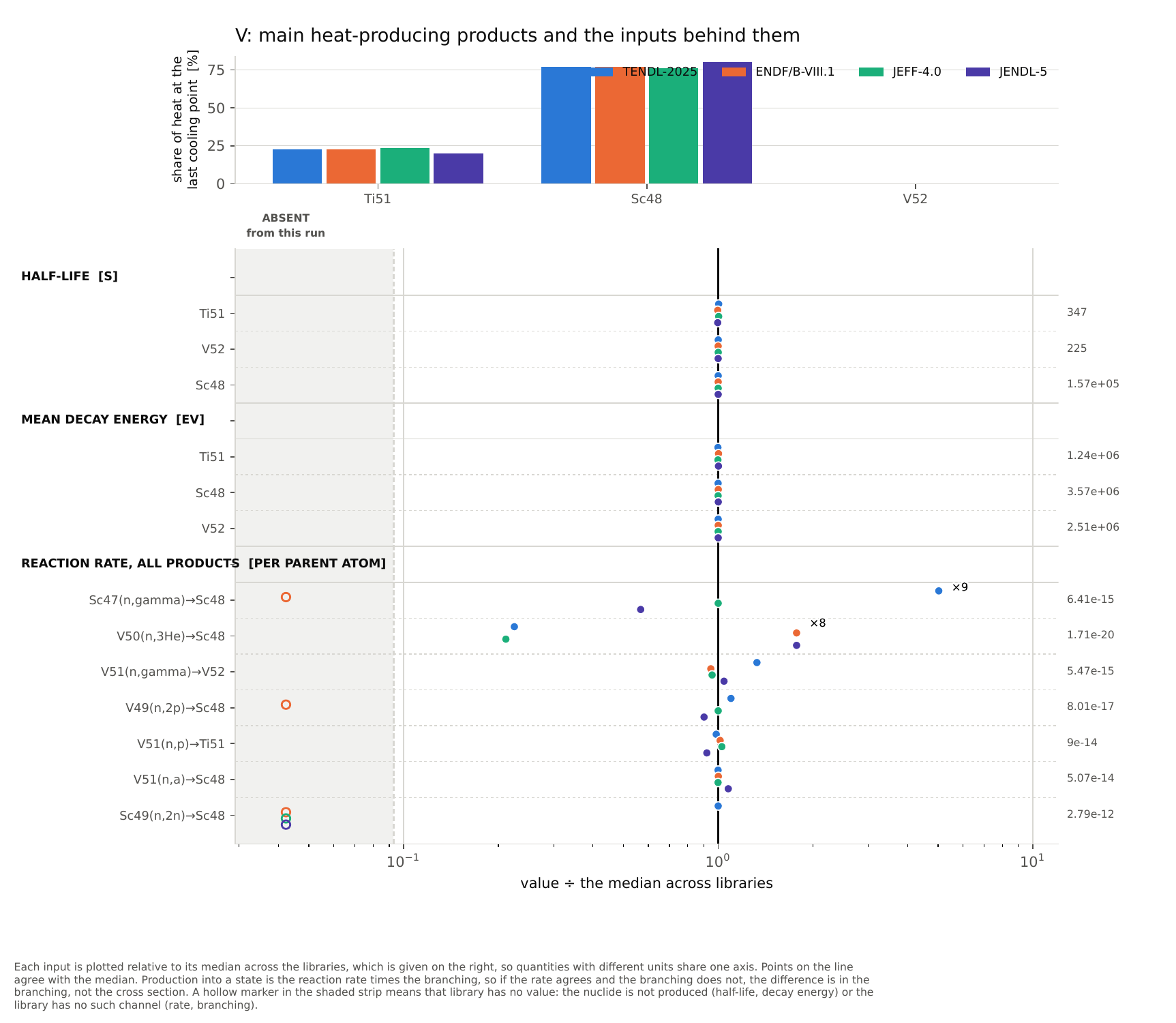}
\caption{V: inputs for the main heat-producing products, each relative to its median across the available libraries.}
\end{figure}
\clearpage

\begin{figure}[H]
\centering
\includegraphics[width=\textwidth,height=0.82\textheight,keepaspectratio]{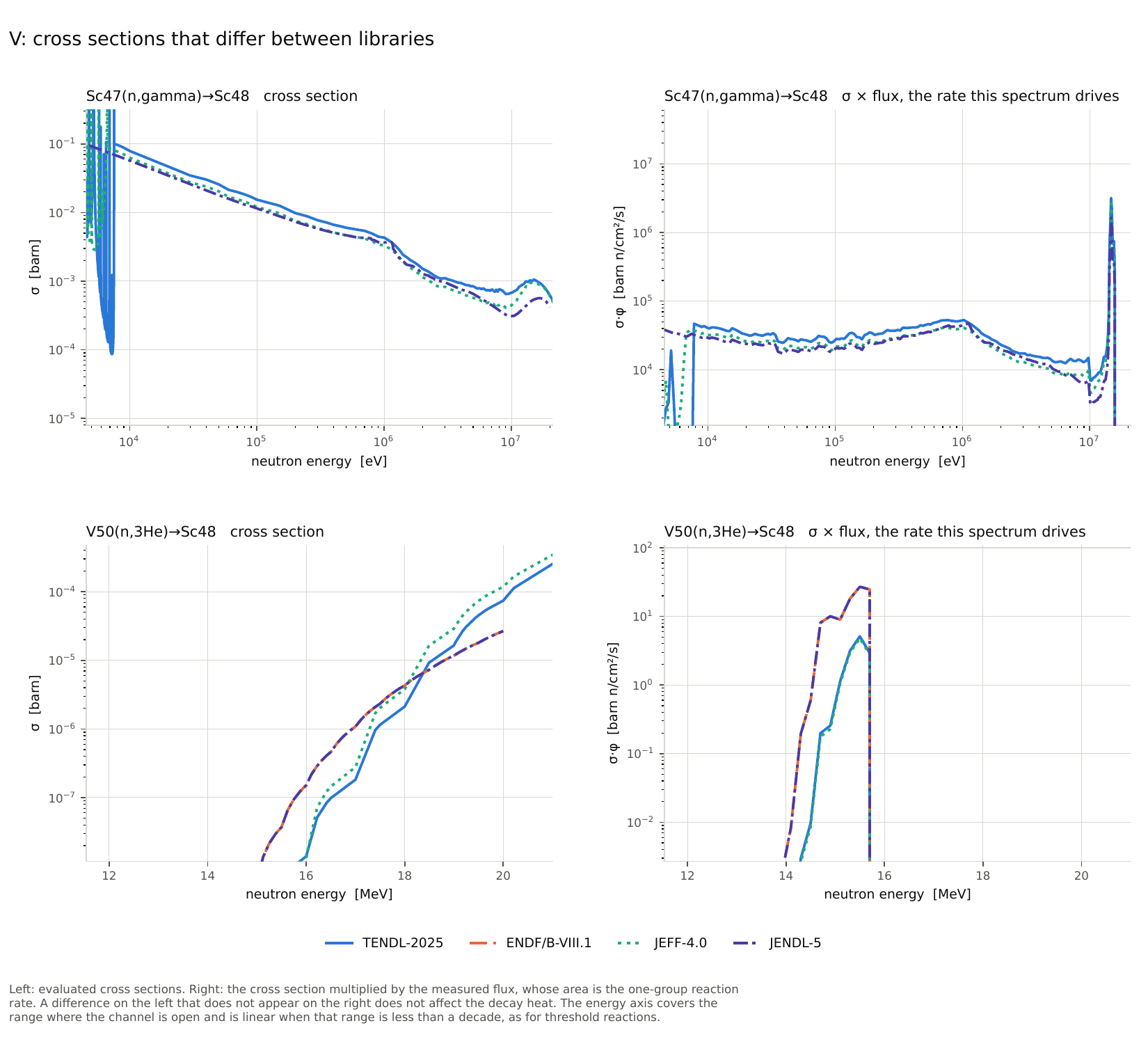}
\caption{V: cross sections of the channels whose spectrum-averaged rates differ, with the neutron spectrum and their product.}
\end{figure}
\clearpage

\subsubsection*{1996 seven-hour campaign}
The 1996 campaign irradiated this foil for 7.57 hours at position 7 and measured it out to 49.73 days. In the five-minute calculation Sc48 (half-life 1.82 days) gives up to 80 percent of the heat, and the longer irradiation and cooling test its production and decay data. JENDL-5 is closest. This campaign uses a different spectrum, so it is not included in the main rankings.

\begin{figure}[H]
\centering
\includegraphics[width=\textwidth,height=0.82\textheight,keepaspectratio]{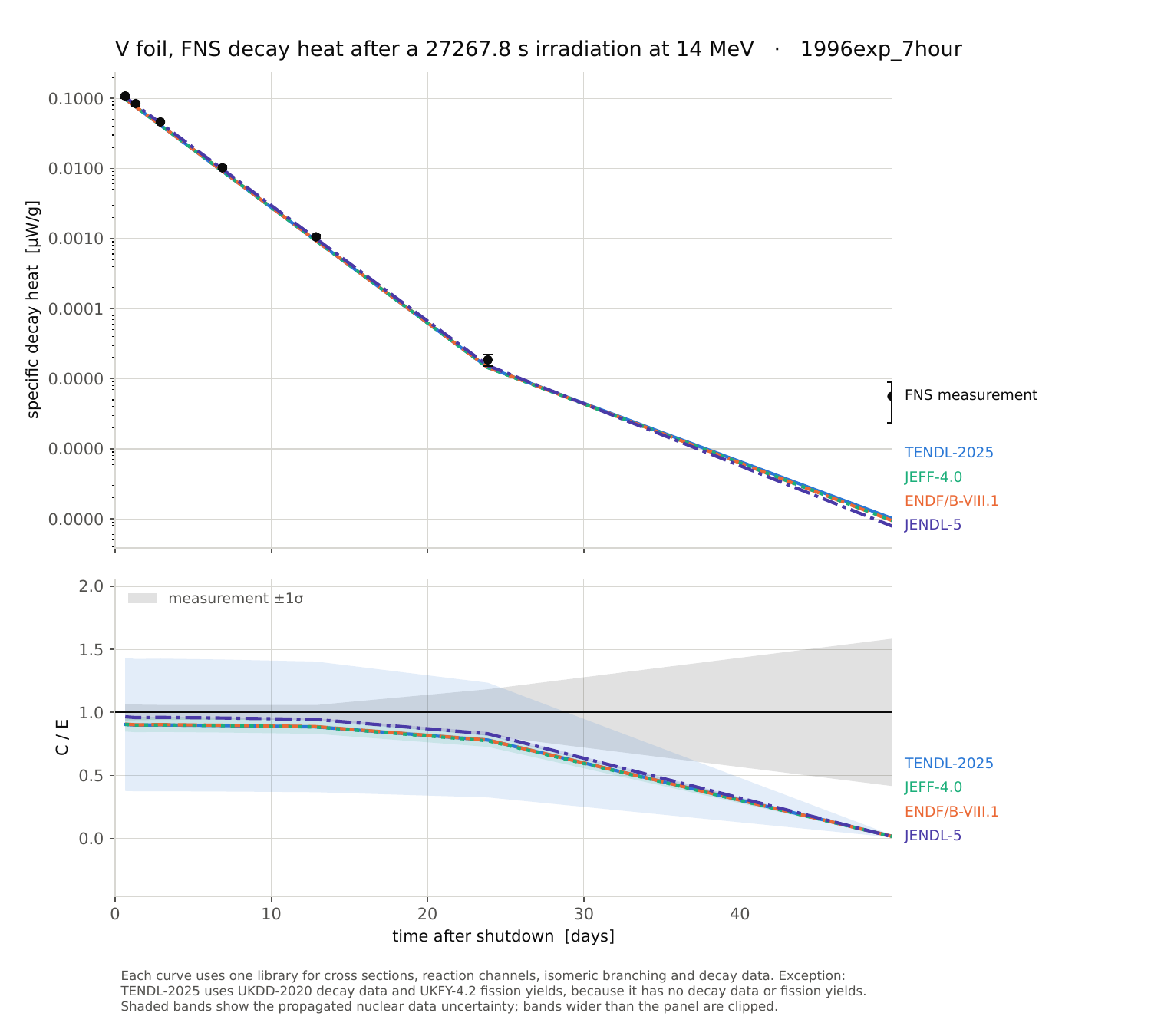}
\caption{V: 1996 seven-hour campaign, irradiation position 7. Mean $|C/E-1|$ by library: JENDL-5 20\%, ENDF/B-VIII.1 24\%, TENDL-2025 25\%, JEFF-4.0 25\%. }
\end{figure}
\clearpage

The libraries are between 7 and 16 percent from the measurement, which has a median uncertainty of 5 percent. JENDL-5 is closest and JEFF-4.0 furthest.

\subsubsection*{Differences in the nuclear data}
\begin{itemize}\setlength{\itemsep}{6pt}
\item JEFF-4.0 gives the total reaction rate of V50(n,3He)→Sc48 as 3.609e-21, against 3.03e-20 for the median of the others, a factor of 8.\\[2pt]{\itshape This is the total rate of the channel over all its products, so the difference is in the cross section, not in the split between states.}
\item TENDL-2025 gives the total reaction rate of Sc47(n,gamma)→Sc48 as 3.222e-14, against 5.024e-15 for the median of the others, a factor of 6.
\end{itemize}
\subsubsection*{Suggested measurements}
\begin{itemize}\setlength{\itemsep}{4pt}
\item Natural V has 2 isotopes, so the measured heat combines several parents. Enriched single-isotope targets would distinguish their contributions. Candidate parents are Sc47, V50. Enrichment alone does not isolate a single reaction channel.
\item The reaction rates are averaged over the measured spectrum. Activation measurements at several well-defined neutron energies, including below 14 MeV, would show which energy range the differences come from.
\end{itemize}
\clearpage
\subsection{W: Tungsten}\label{foil:W}
\begin{figure}[H]
\centering
\includegraphics[width=\textwidth,height=0.82\textheight,keepaspectratio]{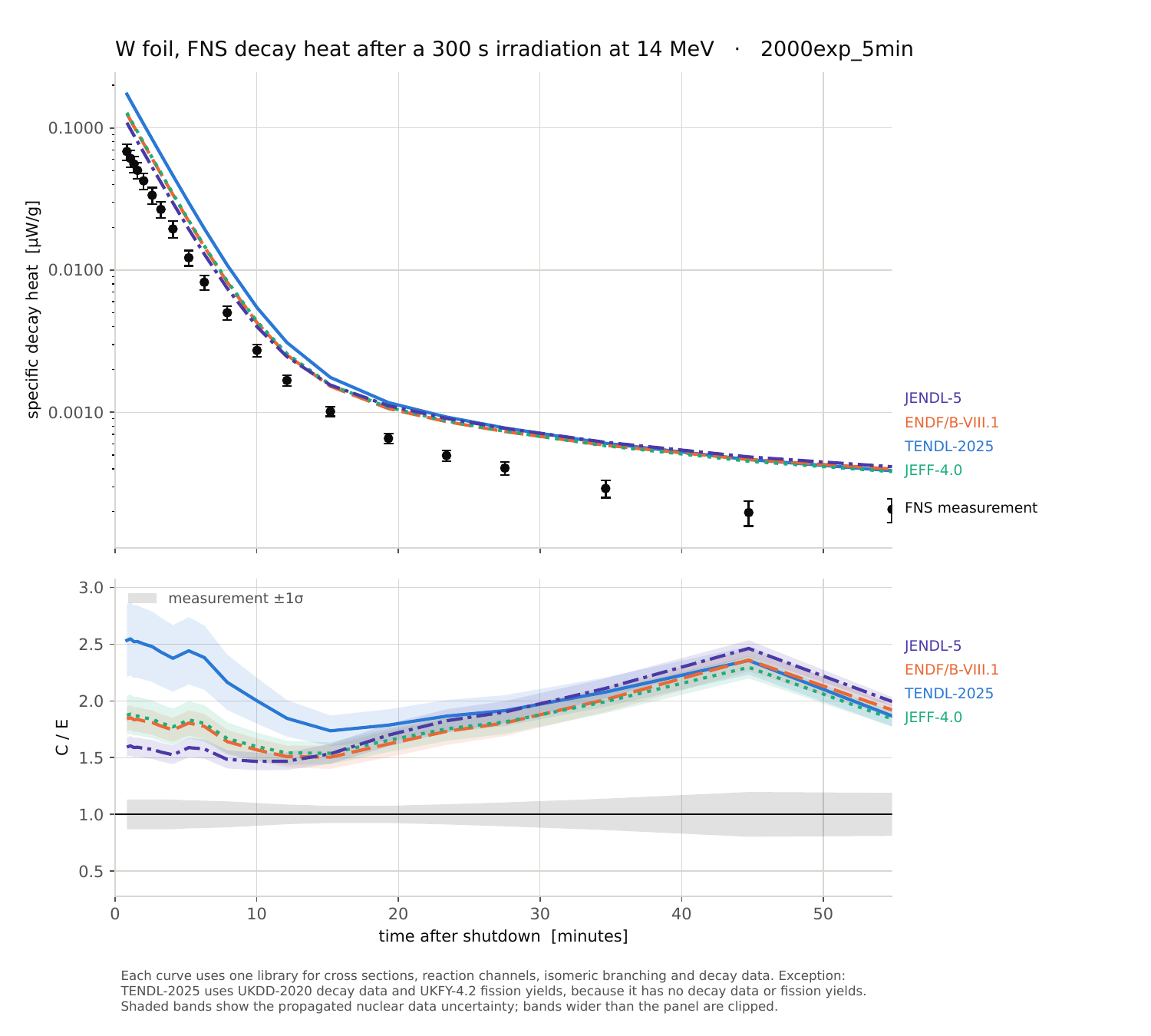}
\caption{W: decay heat against the measurement, with C/E beneath. Mean $|C/E-1|$ by library: JENDL-5 69\%, ENDF/B-VIII.1 79\%, JEFF-4.0 80\%, TENDL-2025 122\%.}
\end{figure}
\clearpage

\begin{figure}[H]
\centering
\includegraphics[width=\textwidth,height=0.82\textheight,keepaspectratio]{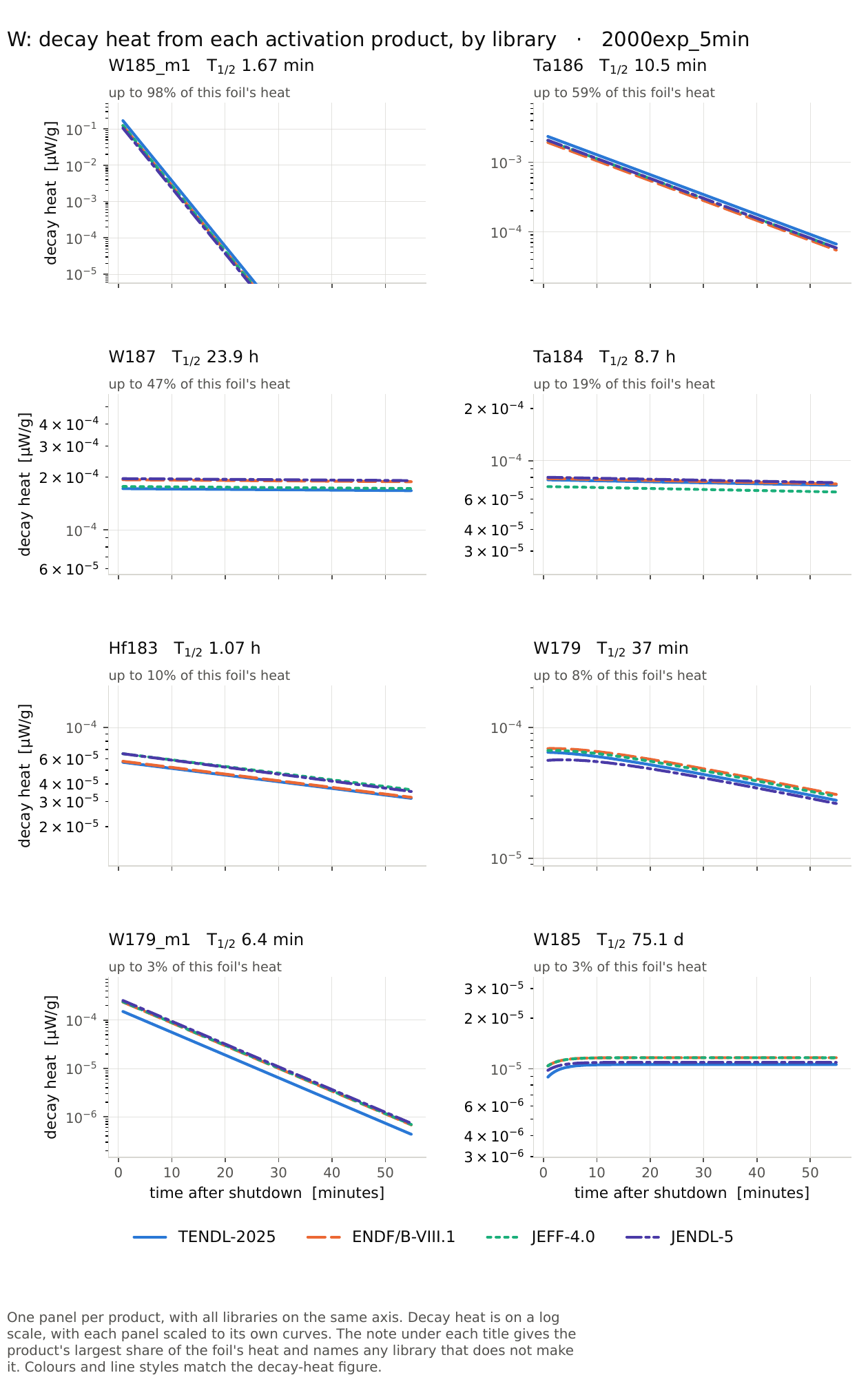}
\caption{W: share of the decay heat from each activation product during cooling.}
\end{figure}
\clearpage

\begin{figure}[H]
\centering
\includegraphics[width=\textwidth,height=0.82\textheight,keepaspectratio]{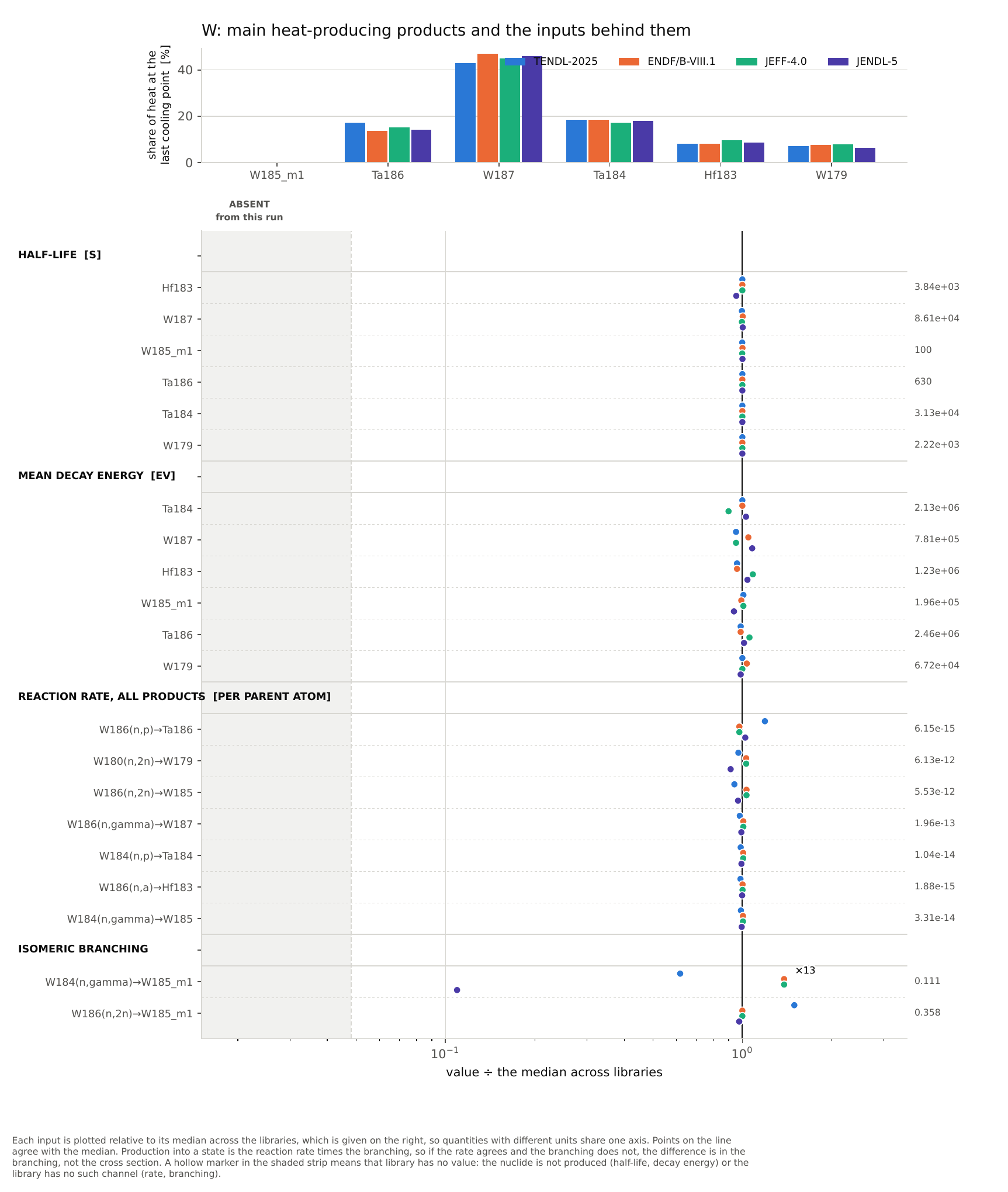}
\caption{W: inputs for the main heat-producing products, each relative to its median across the available libraries.}
\end{figure}
\clearpage

\begin{figure}[H]
\centering
\includegraphics[width=\textwidth,height=0.82\textheight,keepaspectratio]{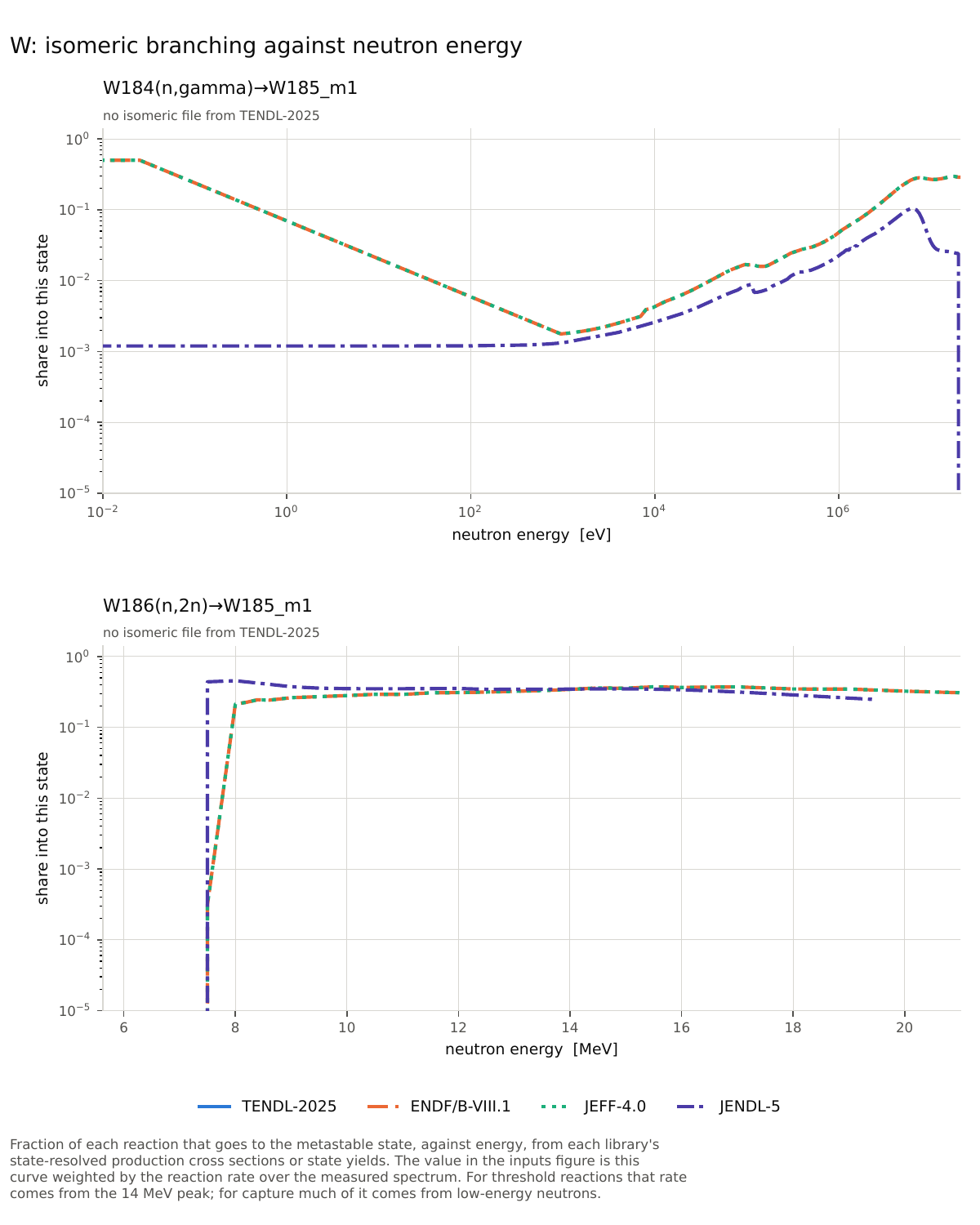}
\caption{W: isomeric branching ratio against neutron energy for the channels whose spectrum-averaged ratio differs between libraries.}
\end{figure}
\clearpage

\subsubsection*{1996 seven-hour campaign}
The 1996 campaign irradiated this foil for 7.57 hours at position 7 and measured it out to 402.95 days. In the five-minute calculation W185 (half-life 75.1 days) gives up to 3 percent of the heat, and the longer irradiation and cooling test its production and decay data. JENDL-5 is closest. This campaign uses a different spectrum, so it is not included in the main rankings.

\begin{figure}[H]
\centering
\includegraphics[width=\textwidth,height=0.82\textheight,keepaspectratio]{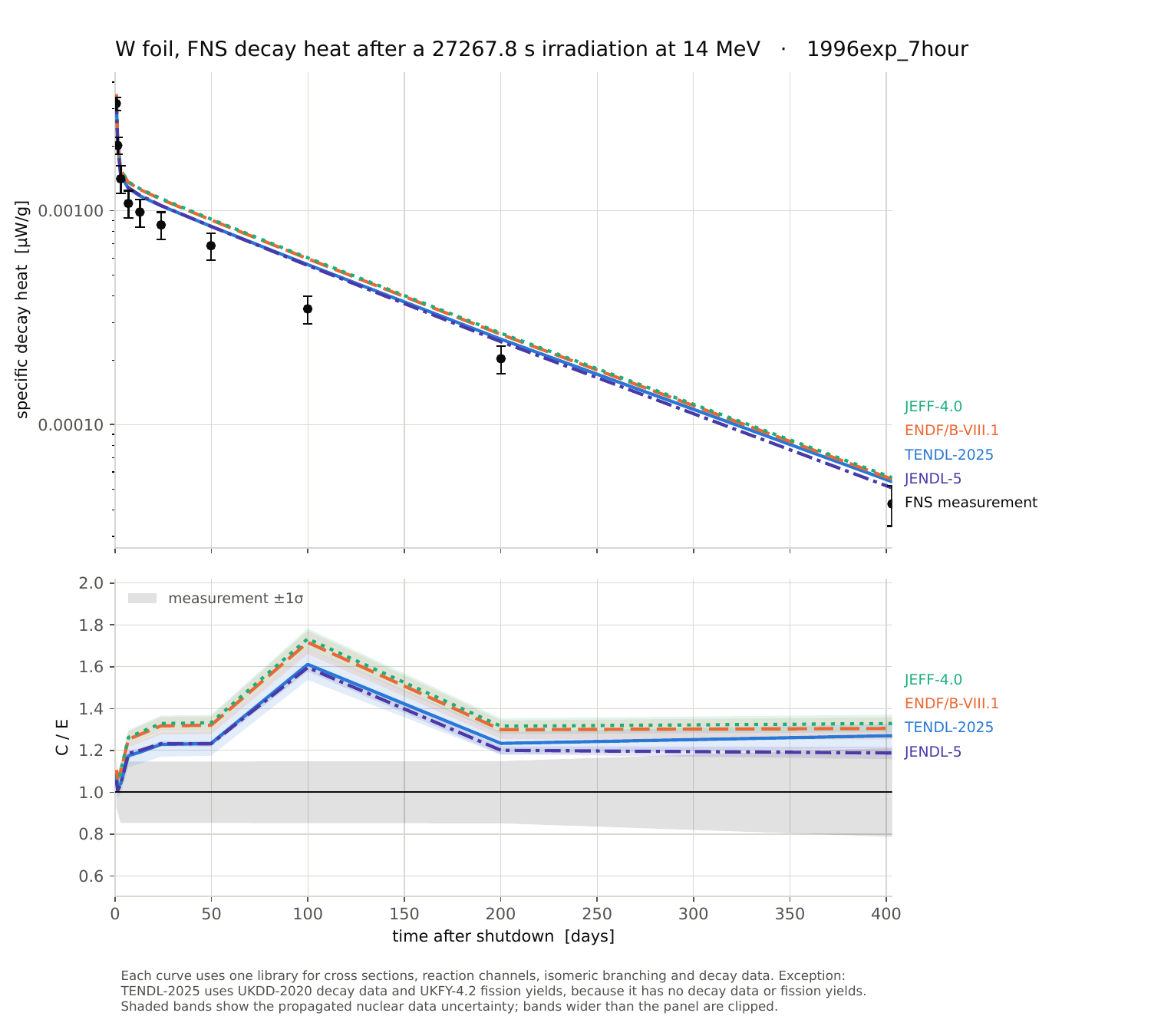}
\caption{W: 1996 seven-hour campaign, irradiation position 7. Mean $|C/E-1|$ by library: JENDL-5 19\%, TENDL-2025 21\%, ENDF/B-VIII.1 28\%, JEFF-4.0 28\%. }
\end{figure}
\clearpage

The libraries are between 69 and 122 percent from the measurement, which has a median uncertainty of 13 percent. JENDL-5 is closest and TENDL-2025 furthest.

\subsubsection*{Differences in the nuclear data}
\begin{itemize}\setlength{\itemsep}{6pt}
\item JENDL-5 gives the isomeric branching of W184(n,gamma)→W185\_m1 as 0.01213, against 0.1532 for the median of the others, a factor of 13.\\[2pt]{\itshape Isomeric branching sets how a channel's production is split between the ground and metastable states. It should be checked together with the total rate of the channel.}
\end{itemize}
\subsubsection*{Suggested measurements}
\begin{itemize}\setlength{\itemsep}{4pt}
\item Natural W has 5 isotopes, so the measured heat combines several parents. Enriched single-isotope targets would distinguish their contributions. The candidate parent is W184. Enrichment alone does not isolate a single reaction channel.
\item Measuring the activity of each state separately, for example by gamma spectrometry where the states have distinct lines, would constrain the isomeric ratio. This needs corrections for feeding between states, detector efficiency and irradiation history.
\end{itemize}
\clearpage
\subsection{Y: Yttrium}\label{foil:Y}
\begin{figure}[H]
\centering
\includegraphics[width=\textwidth,height=0.82\textheight,keepaspectratio]{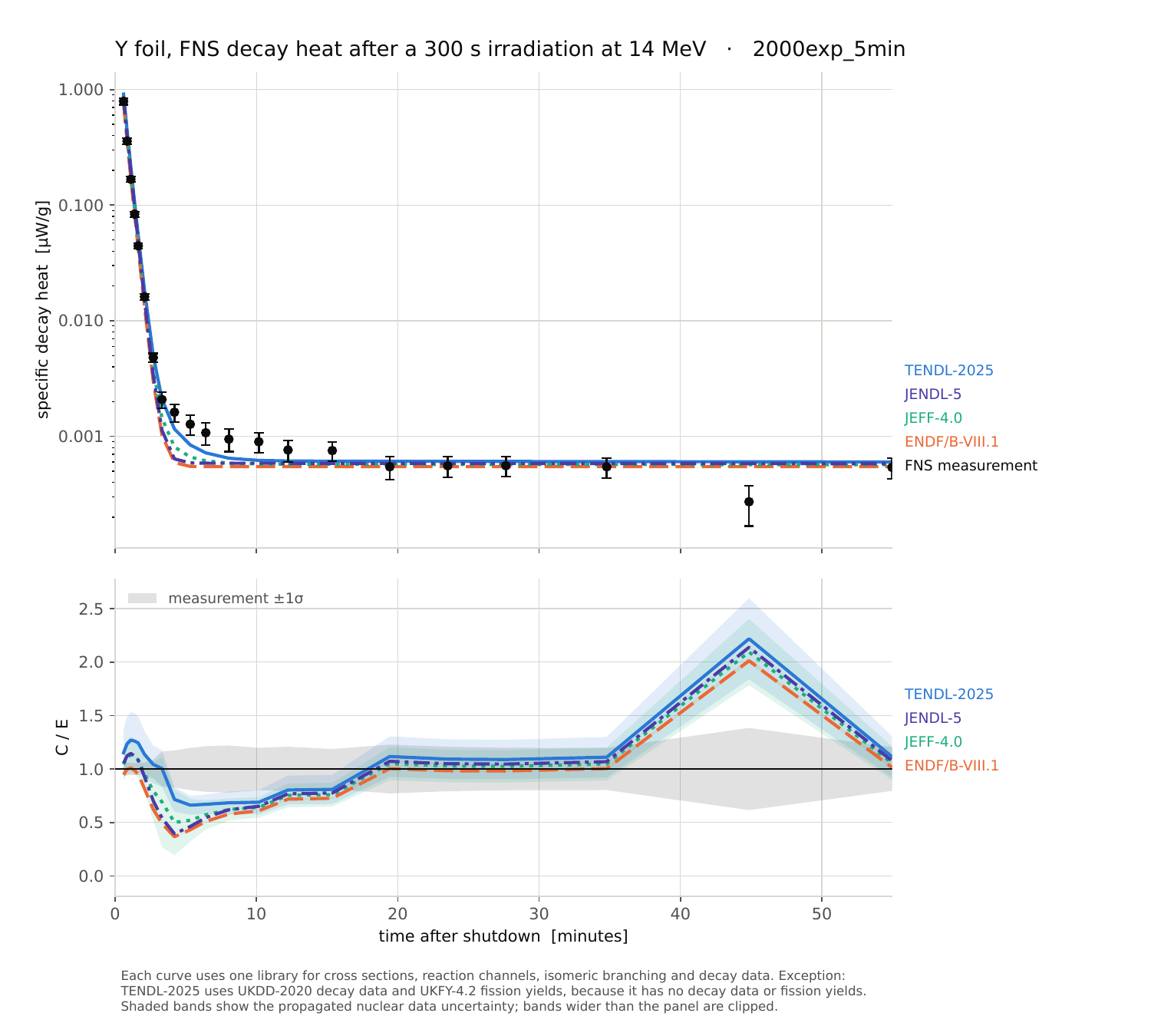}
\caption{Y: decay heat against the measurement, with C/E beneath. Mean $|C/E-1|$ by library: JEFF-4.0 24\%, TENDL-2025 24\%, ENDF/B-VIII.1 25\%, JENDL-5 27\%.}
\end{figure}
\clearpage

\begin{figure}[H]
\centering
\includegraphics[width=\textwidth,height=0.82\textheight,keepaspectratio]{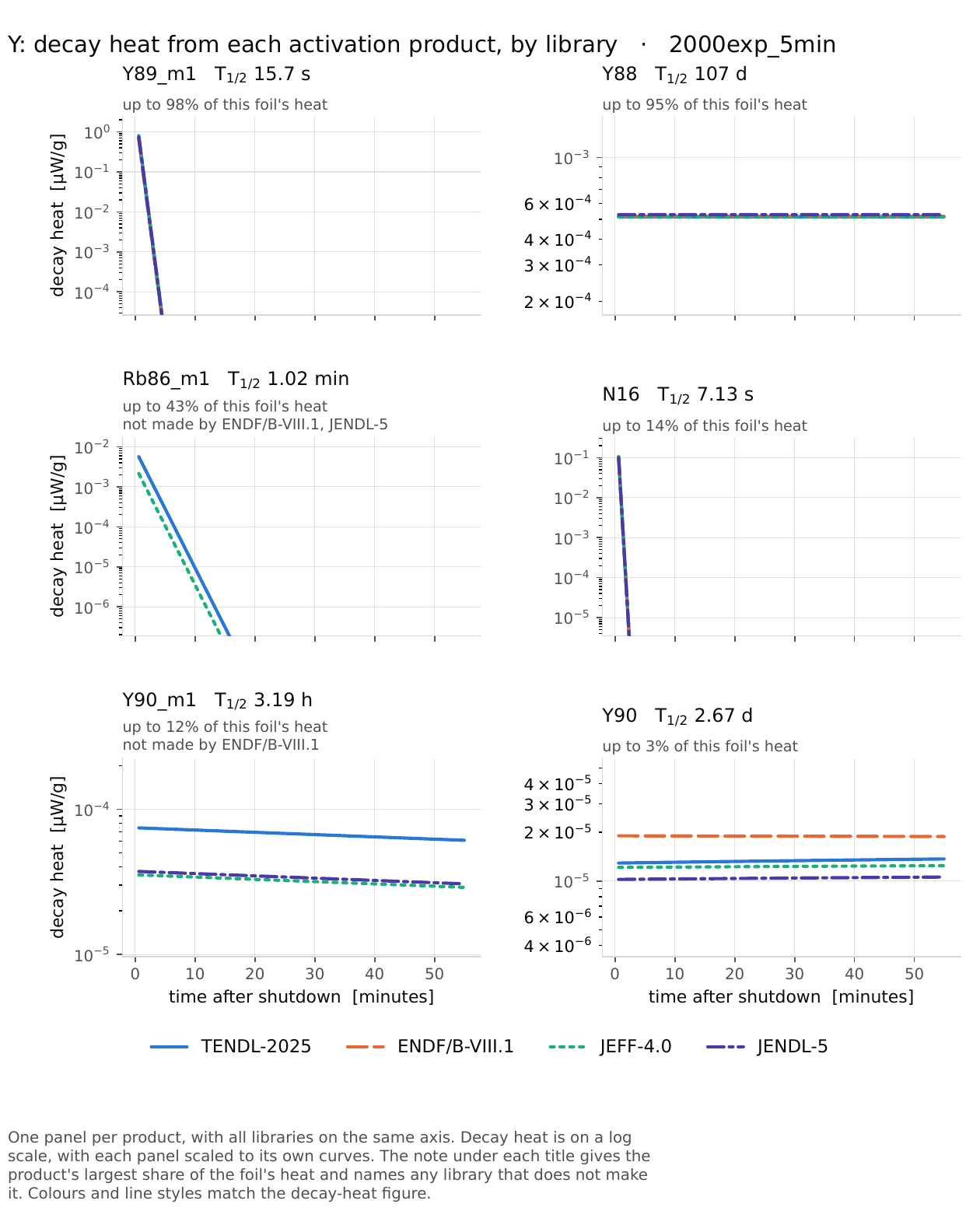}
\caption{Y: share of the decay heat from each activation product during cooling.}
\end{figure}
\clearpage

\begin{figure}[H]
\centering
\includegraphics[width=\textwidth,height=0.82\textheight,keepaspectratio]{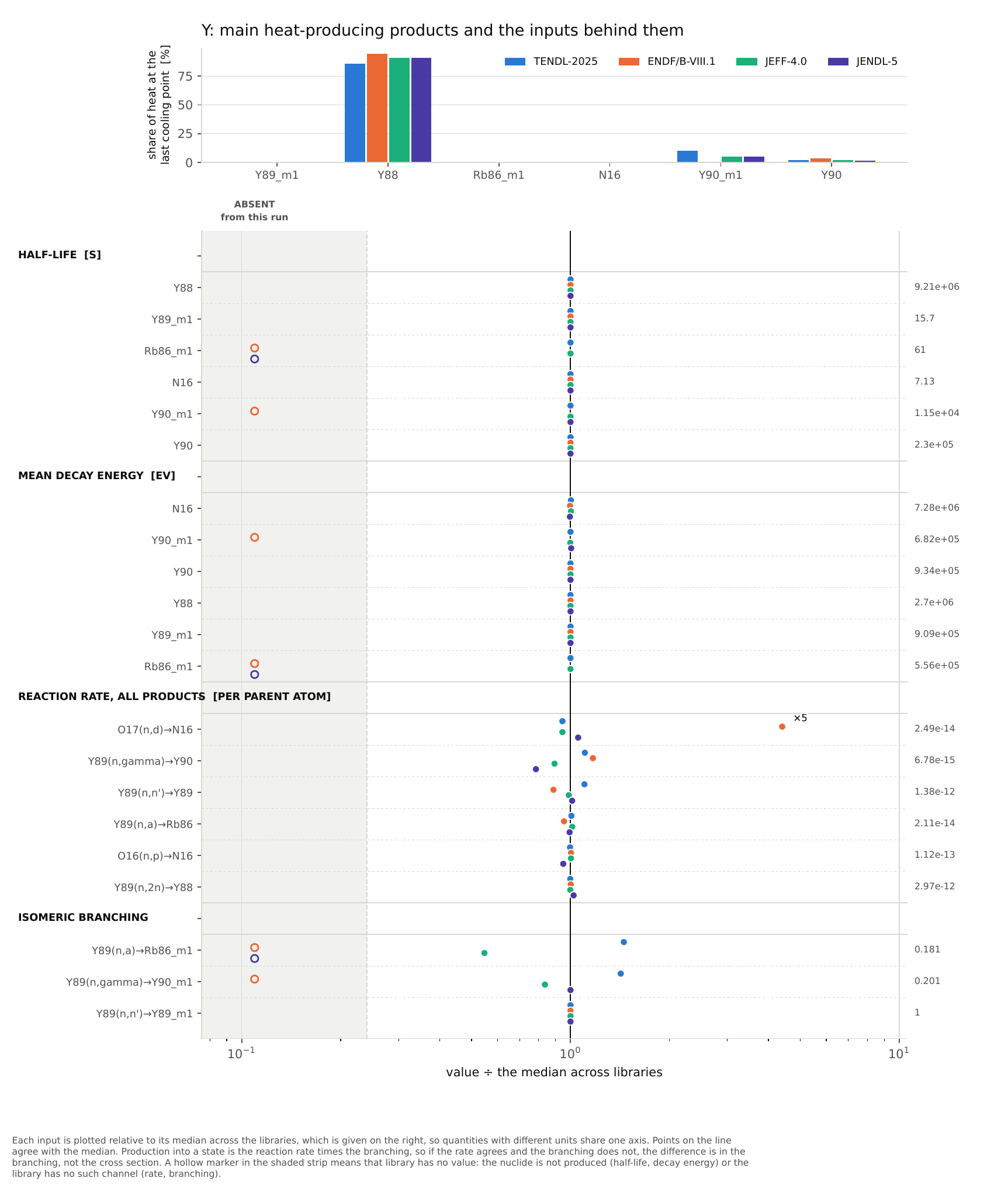}
\caption{Y: inputs for the main heat-producing products, each relative to its median across the available libraries.}
\end{figure}
\clearpage

\begin{figure}[H]
\centering
\includegraphics[width=\textwidth,height=0.82\textheight,keepaspectratio]{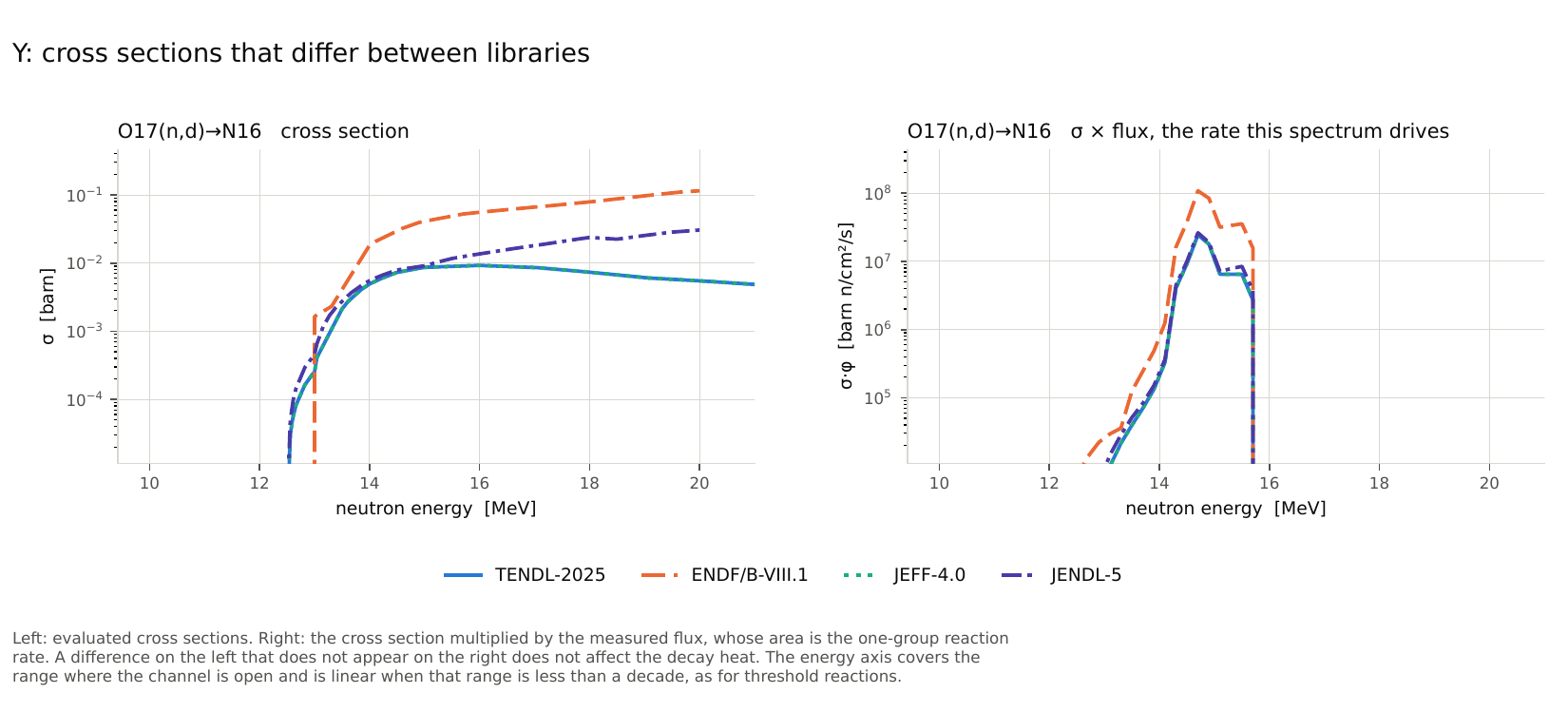}
\caption{Y: cross sections of the channels whose spectrum-averaged rates differ, with the neutron spectrum and their product.}
\end{figure}
\clearpage

\begin{figure}[H]
\centering
\includegraphics[width=\textwidth,height=0.82\textheight,keepaspectratio]{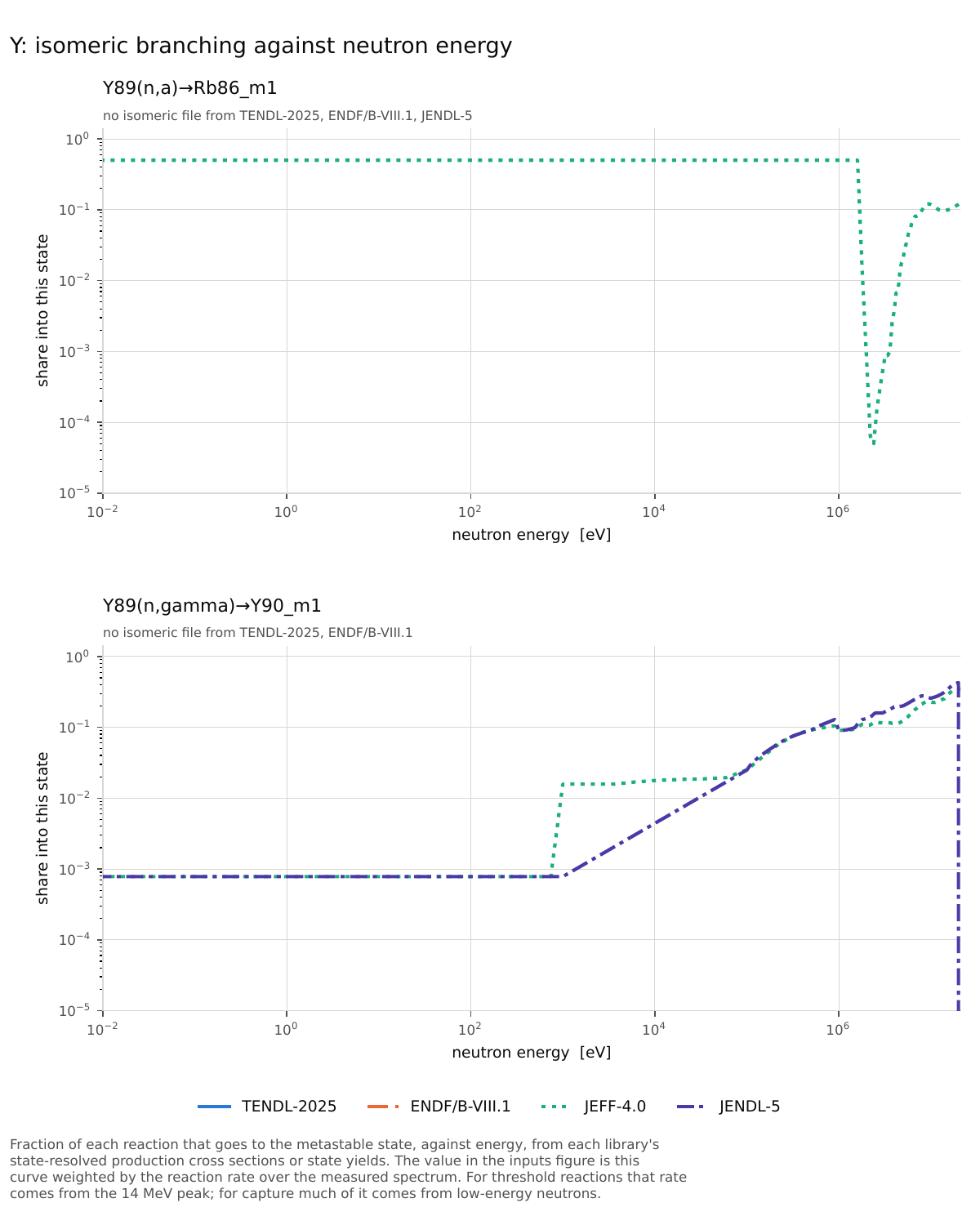}
\caption{Y: isomeric branching ratio against neutron energy for the channels whose spectrum-averaged ratio differs between libraries.}
\end{figure}
\clearpage

\subsubsection*{1996 seven-hour campaign}
The 1996 campaign irradiated this foil for 7.17 hours at position 7 and measured it out to 402.17 days. In the five-minute calculation Y88 (half-life 107 days) gives up to 95 percent of the heat, and the longer irradiation and cooling test its production and decay data. JENDL-5 is closest. This campaign uses a different spectrum, so it is not included in the main rankings.

\begin{figure}[H]
\centering
\includegraphics[width=\textwidth,height=0.82\textheight,keepaspectratio]{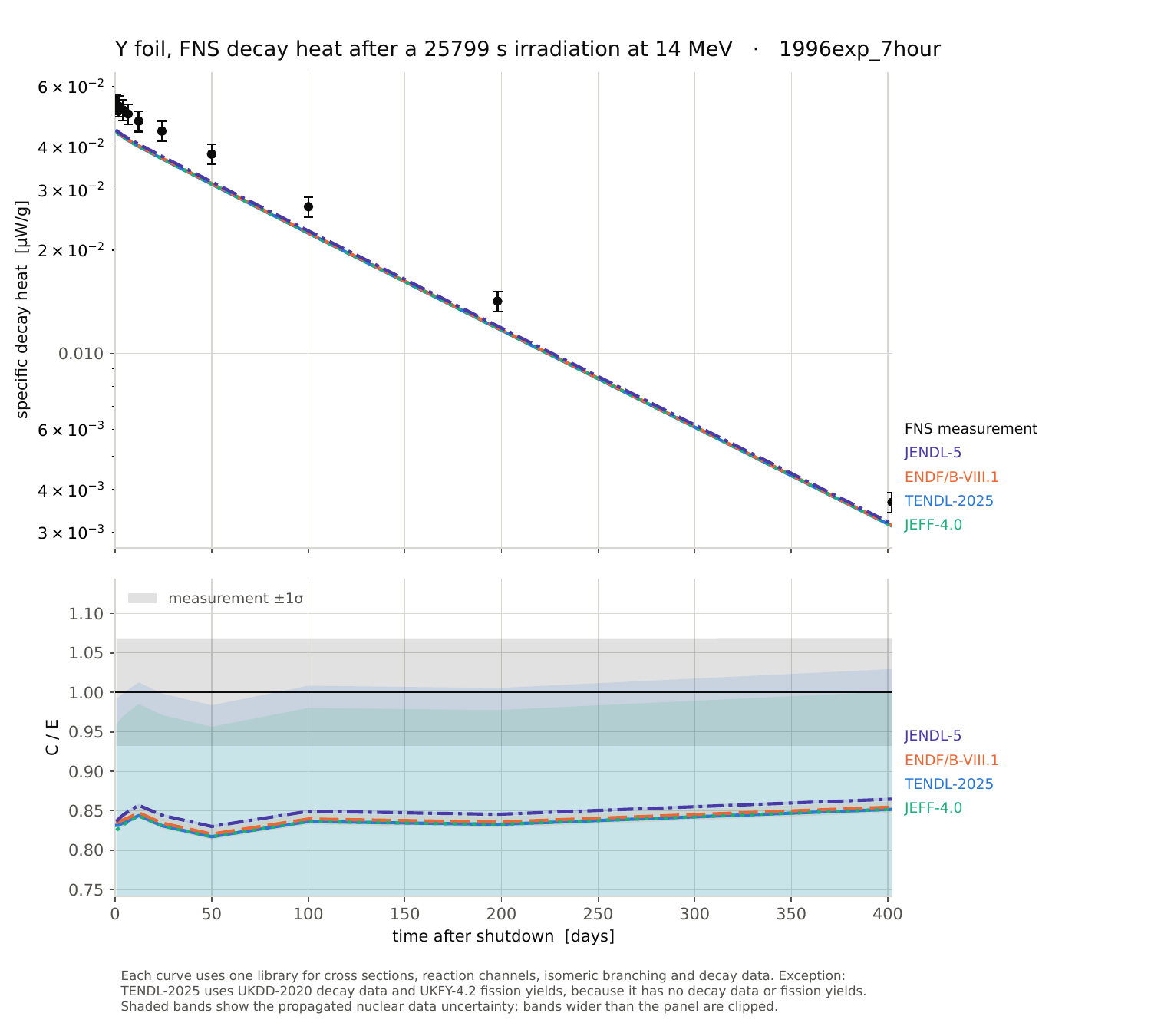}
\caption{Y: 1996 seven-hour campaign, irradiation position 7. Mean $|C/E-1|$ by library: JENDL-5 15\%, ENDF/B-VIII.1 16\%, TENDL-2025 17\%, JEFF-4.0 17\%. }
\end{figure}
\clearpage

Mean absolute C/E deviations range from 24 to 27 percent. The spread of 3 percentage points is smaller than the median measurement uncertainty of 20 percent, so the measurement cannot separate the libraries.

\subsubsection*{Differences in the nuclear data}
\begin{itemize}\setlength{\itemsep}{6pt}
\item ENDF/B-VIII.1 gives the total reaction rate of O17(n,d)→N16 as 1.096e-13, against 2.351e-14 for the median of the others, a factor of 5.\\[2pt]{\itshape This is the total rate of the channel over all its products, so the difference is in the cross section, not in the split between states.}
\item Only TENDL-2025 and JEFF-4.0 have the isomeric branching of Y89(n,a)→Rb86\_m1, and they differ by a factor of 3: 0.263 against 0.09906.\\[2pt]{\itshape Isomeric branching sets how a channel's production is split between the ground and metastable states. It should be checked together with the total rate of the channel.}
\item ENDF/B-VIII.1, JENDL-5 give no heat from Rb86\_m1; the other libraries do.\\[2pt]{\itshape The heat breakdown alone does not show why. Possible causes are a missing reaction channel, a different state assignment, different decay data or a reporting threshold.}
\item ENDF/B-VIII.1 gives no heat from Y90\_m1; the other libraries do.
\end{itemize}
\subsubsection*{Suggested measurements}
\begin{itemize}\setlength{\itemsep}{4pt}
\item Y has a single stable isotope, so the foil is already a single-isotope target and enrichment would not help.
\item The sample is a compound that also contains O. Their activation adds to the measured heat; a metal or purer sample would remove it.
\item The reaction rates are averaged over the measured spectrum. Activation measurements at several well-defined neutron energies, including below 14 MeV, would show which energy range the differences come from.
\item Measuring the activity of each state separately, for example by gamma spectrometry where the states have distinct lines, would constrain the isomeric ratio. This needs corrections for feeding between states, detector efficiency and irradiation history.
\end{itemize}
\clearpage
\subsection{Yb: Ytterbium}\label{foil:Yb}
\begin{figure}[H]
\centering
\includegraphics[width=\textwidth,height=0.82\textheight,keepaspectratio]{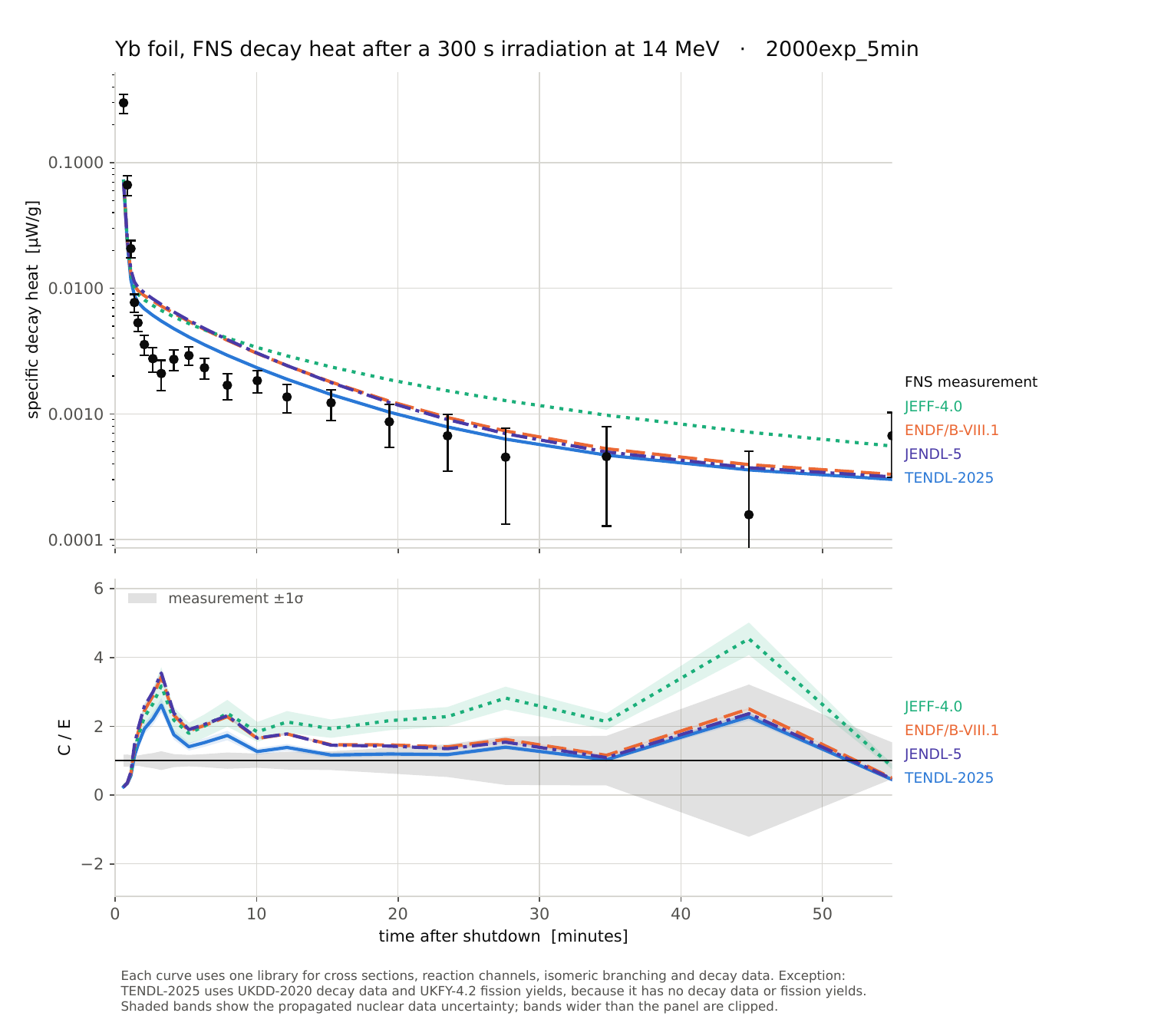}
\caption{Yb: decay heat against the measurement, with C/E beneath. Mean $|C/E-1|$ by library: TENDL-2025 57\%, ENDF/B-VIII.1 89\%, JENDL-5 91\%, JEFF-4.0 115\%.}
\end{figure}
\clearpage

\begin{figure}[H]
\centering
\includegraphics[width=\textwidth,height=0.82\textheight,keepaspectratio]{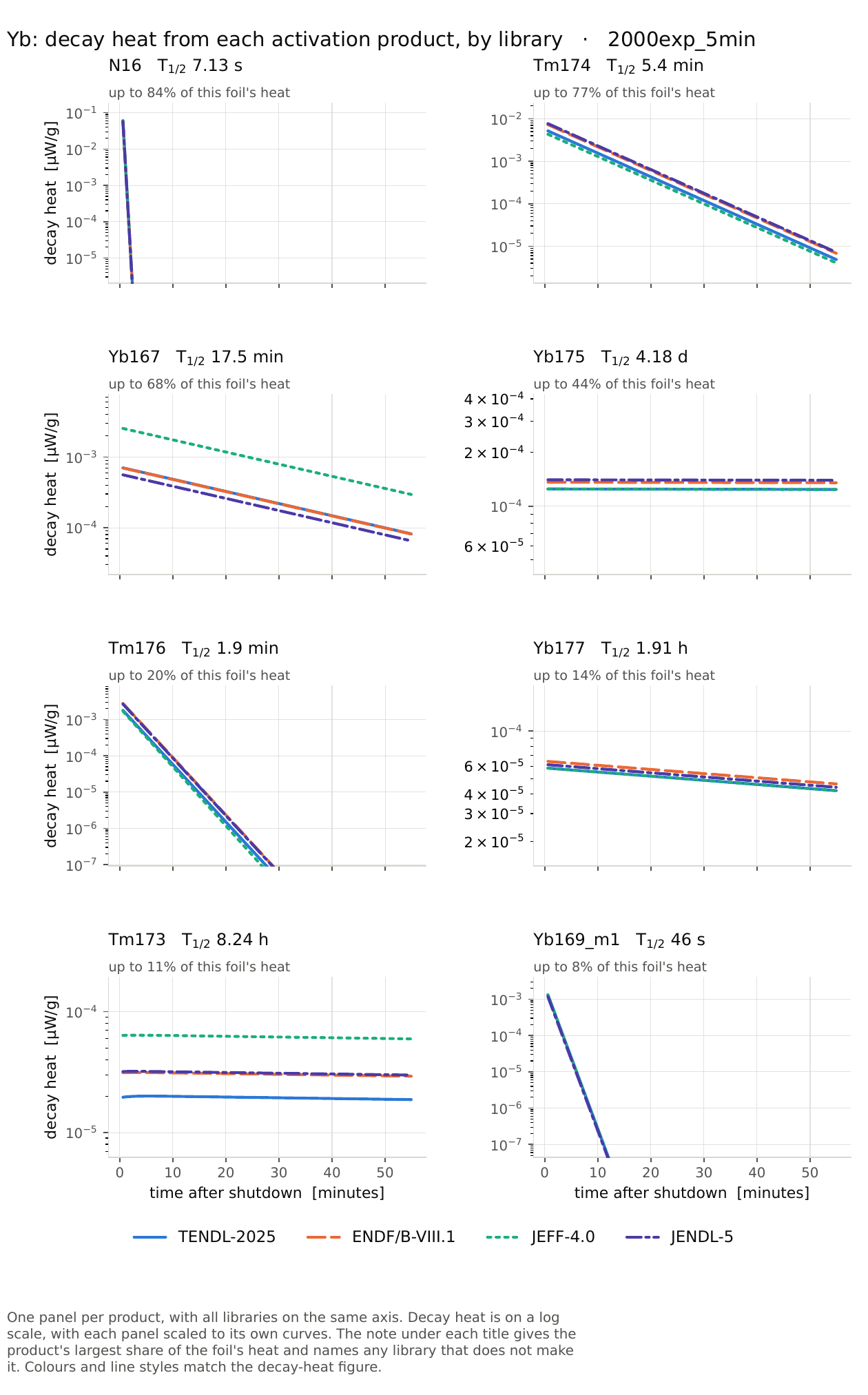}
\caption{Yb: share of the decay heat from each activation product during cooling.}
\end{figure}
\clearpage

\begin{figure}[H]
\centering
\includegraphics[width=\textwidth,height=0.82\textheight,keepaspectratio]{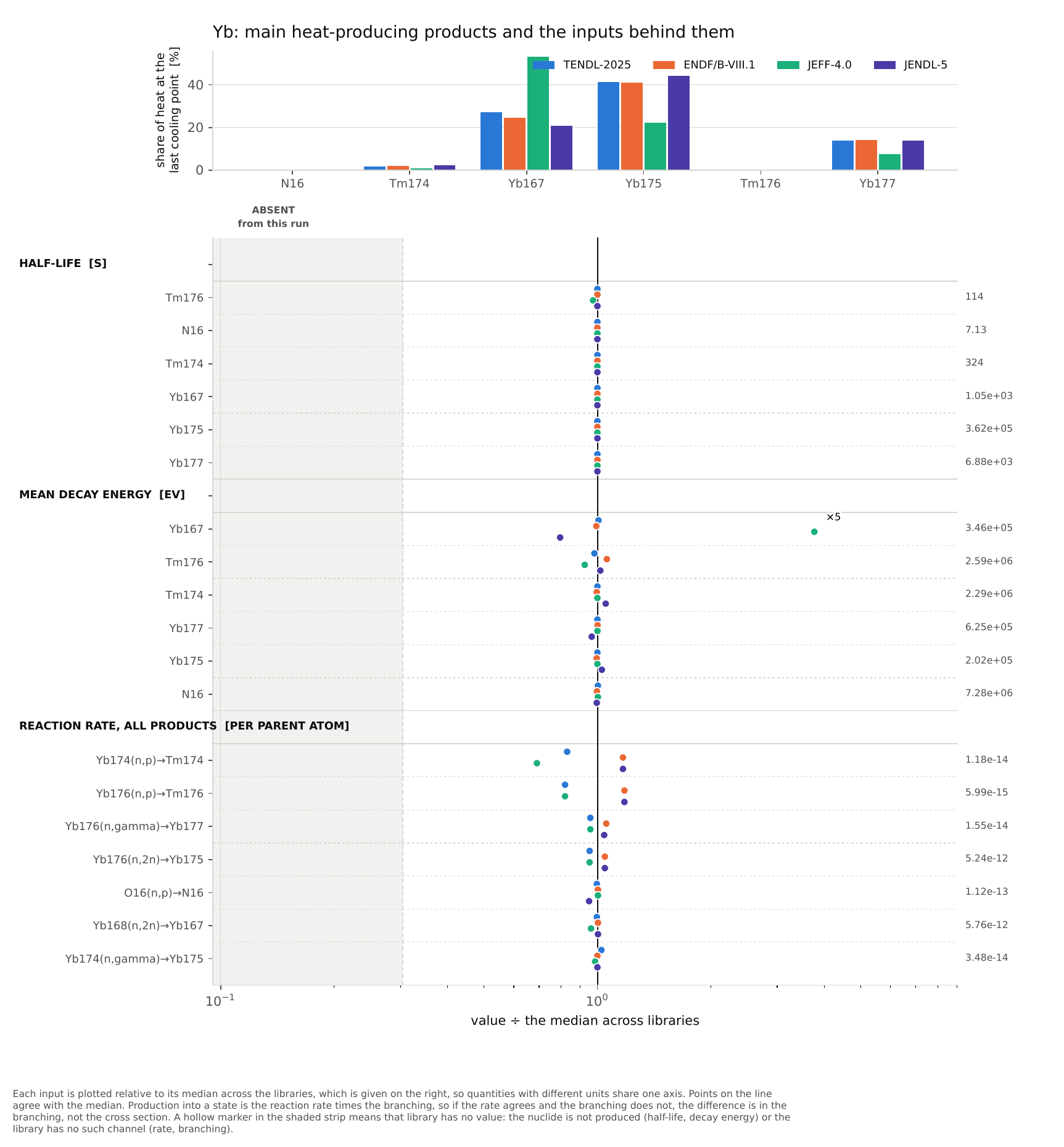}
\caption{Yb: inputs for the main heat-producing products, each relative to its median across the available libraries.}
\end{figure}
\clearpage

\begin{figure}[H]
\centering
\includegraphics[width=\textwidth,height=0.82\textheight,keepaspectratio]{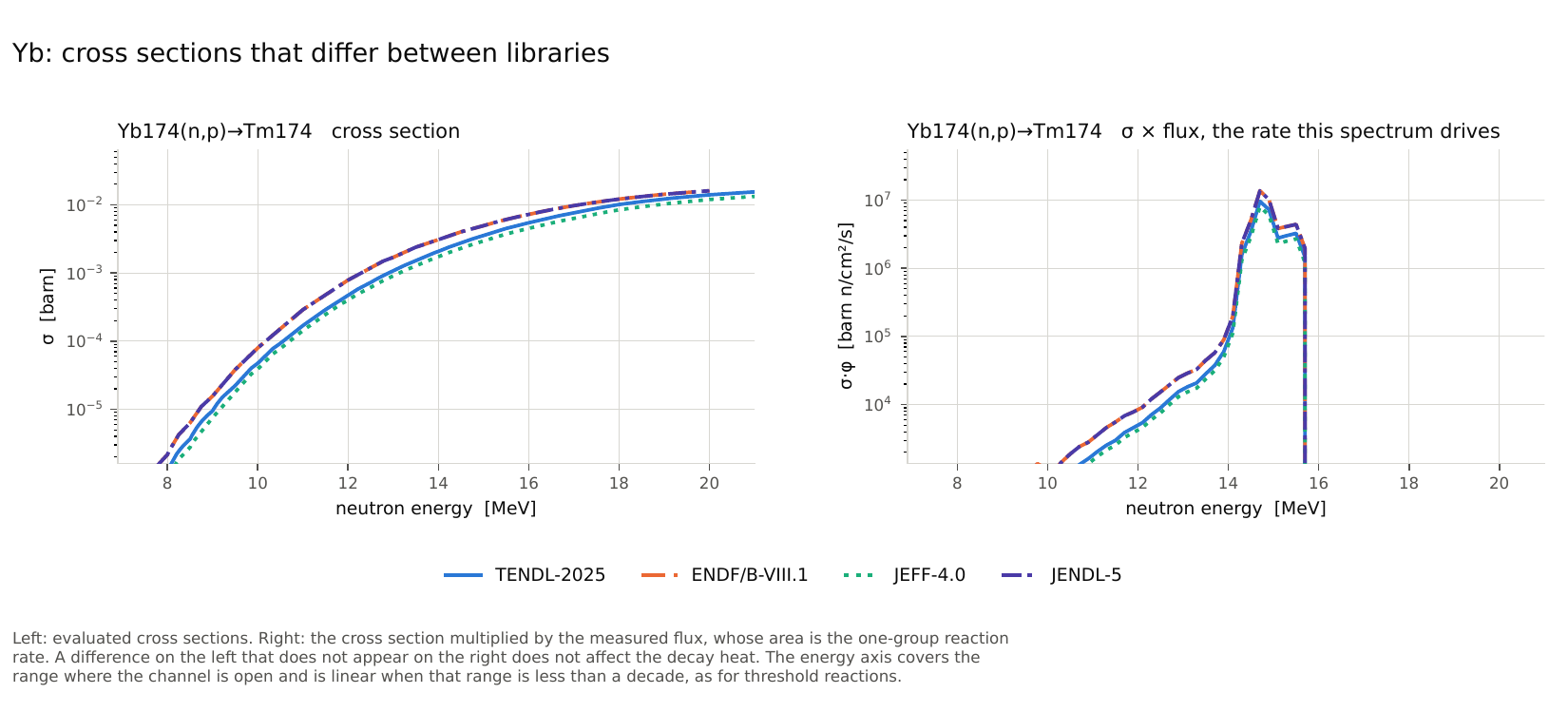}
\caption{Yb: cross sections of the channels whose spectrum-averaged rates differ, with the neutron spectrum and their product.}
\end{figure}
\clearpage

The libraries are between 57 and 115 percent from the measurement, which has a median uncertainty of 22 percent. TENDL-2025 is closest and JEFF-4.0 furthest.

\subsubsection*{Differences in the nuclear data}
\begin{itemize}\setlength{\itemsep}{6pt}
\item JEFF-4.0 gives the mean decay energy of Yb167 as 1.303e+06 eV, against 3.438e+05 eV for the median of the others, a factor of 4.\\[2pt]{\itshape For a given activity the heat is proportional to the mean decay energy, so this difference goes straight into the heat. Compare the decay schemes and how the emitted energy is counted.}
\end{itemize}
\subsubsection*{Suggested measurements}
\begin{itemize}\setlength{\itemsep}{4pt}
\item Natural Yb has 7 isotopes, so the measured heat combines several parents. Enriched single-isotope targets would distinguish their contributions. Enrichment alone does not isolate a single reaction channel.
\item The sample is a compound that also contains O. Their activation adds to the measured heat; a metal or purer sample would remove it.
\item This foil makes Yb175 (half-life 4.18 days), which gives up to 44 percent of the calculated heat. A five-minute irradiation and 55 minute measurement only see the start of this contribution. There is no seven-hour FNS measurement of this foil. A longer irradiation followed for days to months would be needed to test this product's production and decay.
\end{itemize}
\clearpage
\subsection{Zn: Zinc}\label{foil:Zn}
\begin{figure}[H]
\centering
\includegraphics[width=\textwidth,height=0.82\textheight,keepaspectratio]{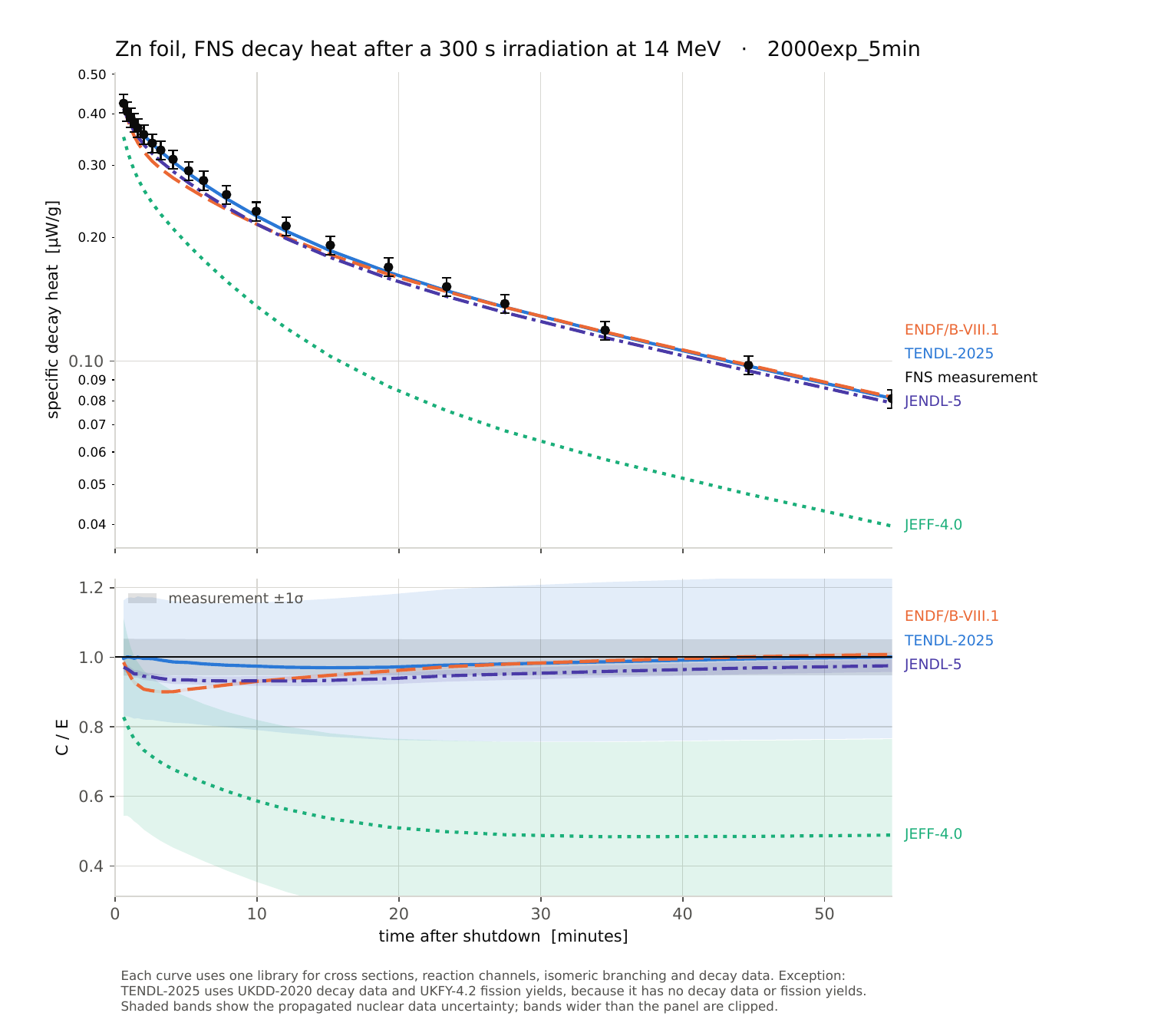}
\caption{Zn: decay heat against the measurement, with C/E beneath. Mean $|C/E-1|$ by library: TENDL-2025 1\%, JENDL-5 5\%, ENDF/B-VIII.1 6\%, JEFF-4.0 37\%.}
\end{figure}
\clearpage

\begin{figure}[H]
\centering
\includegraphics[width=\textwidth,height=0.82\textheight,keepaspectratio]{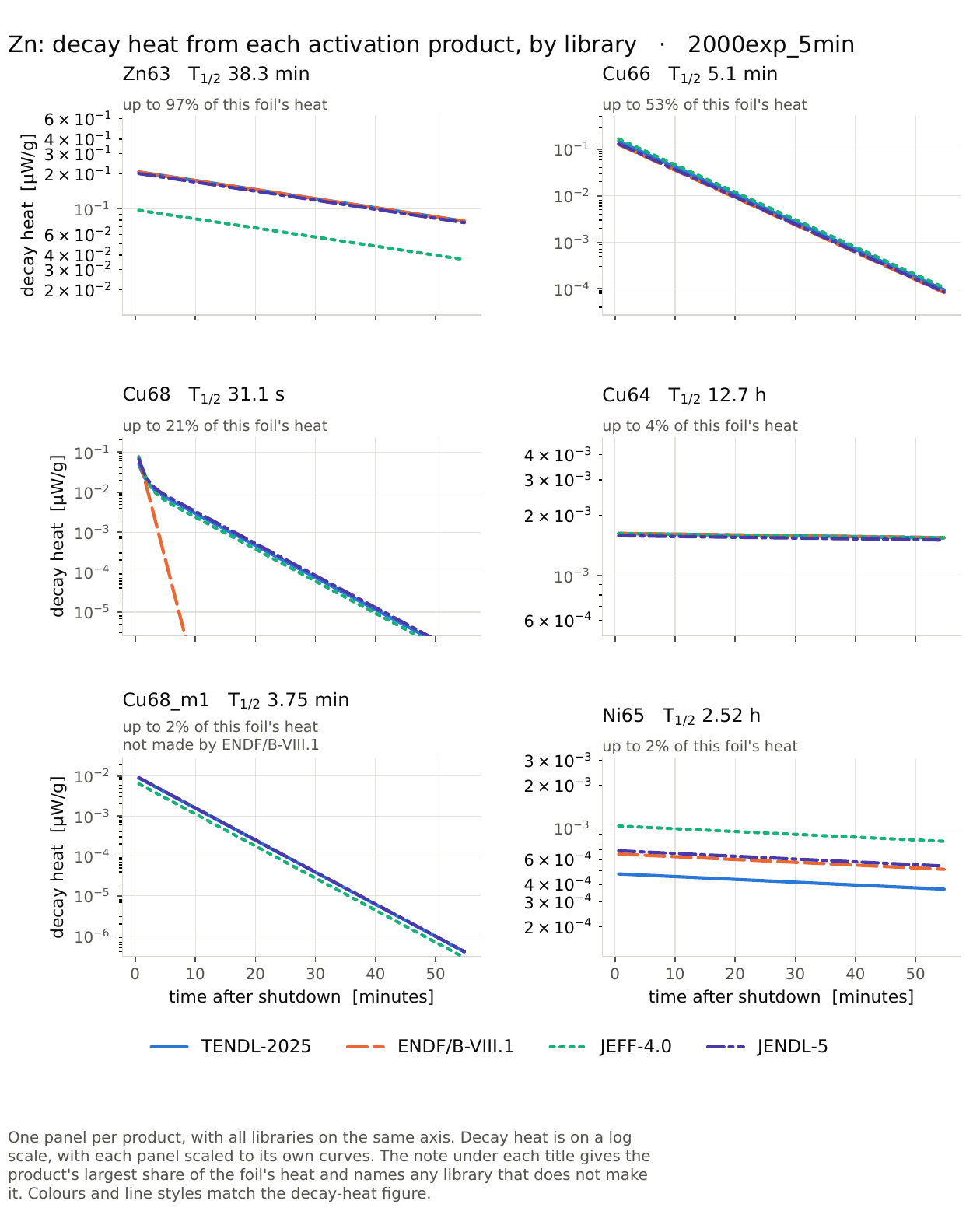}
\caption{Zn: share of the decay heat from each activation product during cooling.}
\end{figure}
\clearpage

\begin{figure}[H]
\centering
\includegraphics[width=\textwidth,height=0.82\textheight,keepaspectratio]{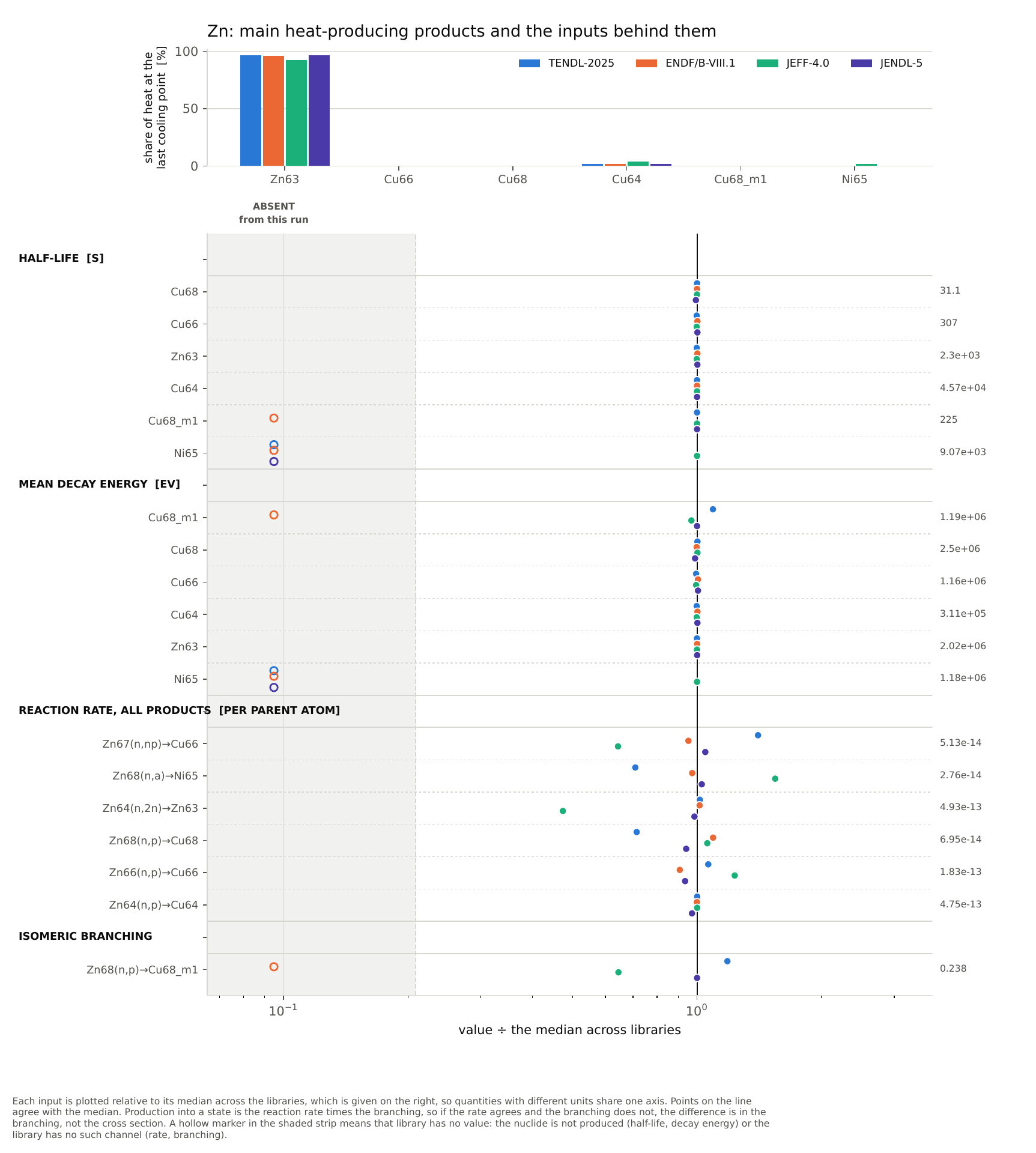}
\caption{Zn: inputs for the main heat-producing products, each relative to its median across the available libraries.}
\end{figure}
\clearpage

\begin{figure}[H]
\centering
\includegraphics[width=\textwidth,height=0.82\textheight,keepaspectratio]{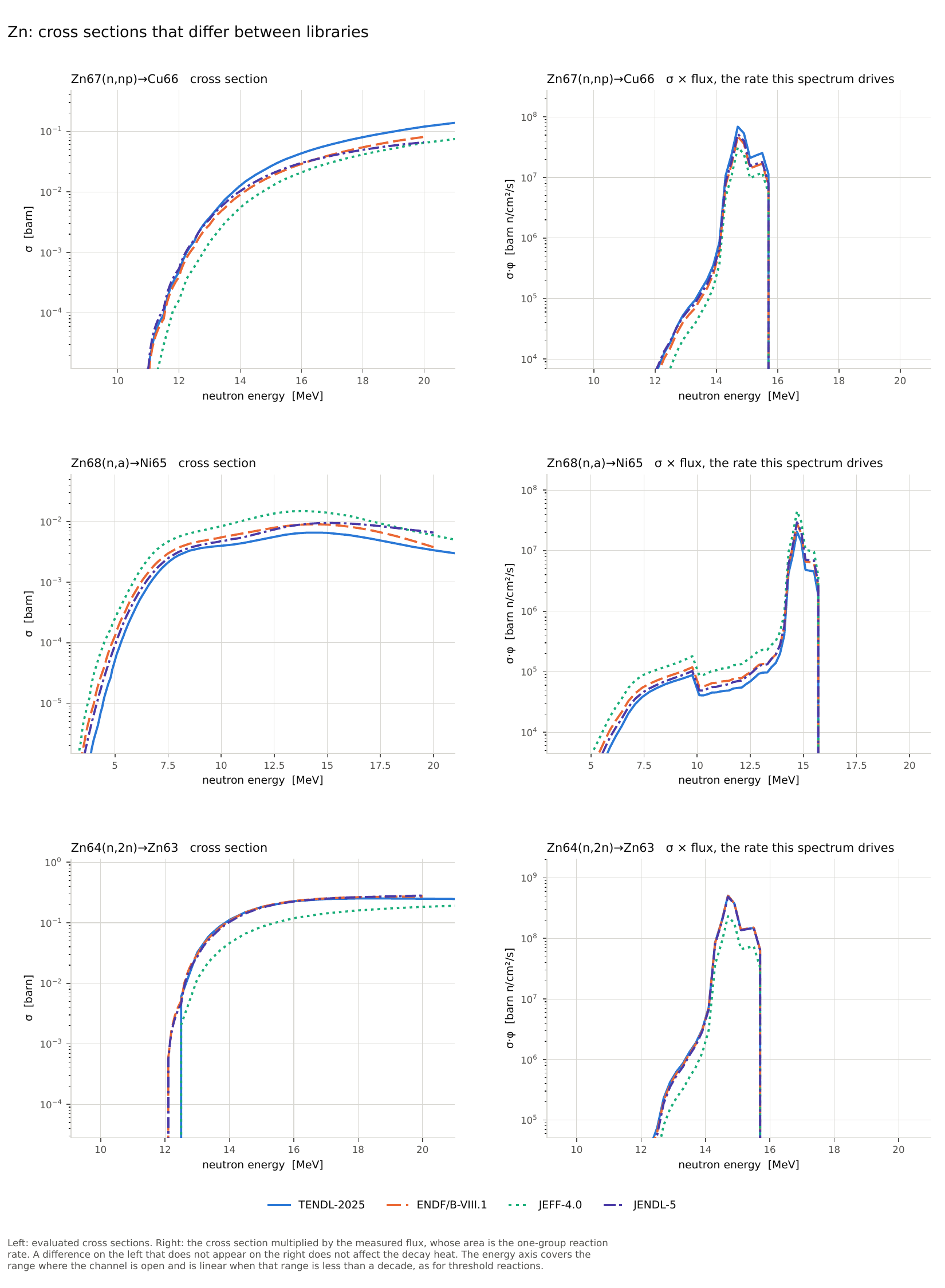}
\caption{Zn: cross sections of the channels whose spectrum-averaged rates differ, with the neutron spectrum and their product.}
\end{figure}
\clearpage

\begin{figure}[H]
\centering
\includegraphics[width=\textwidth,height=0.82\textheight,keepaspectratio]{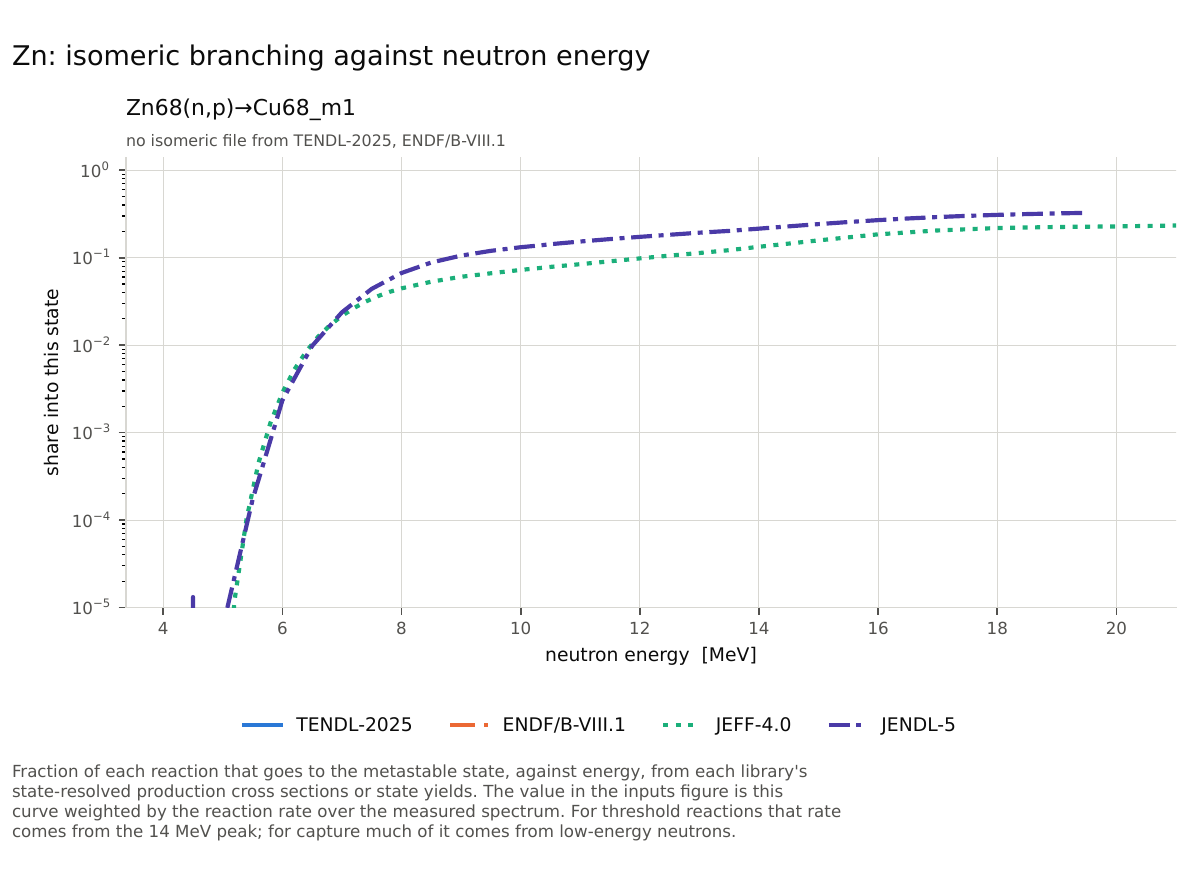}
\caption{Zn: isomeric branching ratio against neutron energy for the channels whose spectrum-averaged ratio differs between libraries.}
\end{figure}
\clearpage

JEFF-4.0 stands out at 37 percent from the measurement, against 6 percent for the next furthest library and 1 percent for TENDL-2025. The median measurement uncertainty is 5 percent.

\subsubsection*{Differences in the nuclear data}
\begin{itemize}\setlength{\itemsep}{6pt}
\item JEFF-4.0 gives the total reaction rate of Zn64(n,2n)→Zn63 as 2.337e-13, against 5.005e-13 for the median of the others, a factor of 2.\\[2pt]{\itshape This is the total rate of the channel over all its products, so the difference is in the cross section, not in the split between states.}
\item ENDF/B-VIII.1 gives no heat from Cu68\_m1; the other libraries do.\\[2pt]{\itshape The heat breakdown alone does not show why. Possible causes are a missing reaction channel, a different state assignment, different decay data or a reporting threshold.}
\end{itemize}
\subsubsection*{Suggested measurements}
\begin{itemize}\setlength{\itemsep}{4pt}
\item Natural Zn has 5 isotopes, so the measured heat combines several parents. Enriched single-isotope targets would distinguish their contributions. The candidate parent is Zn64. Enrichment alone does not isolate a single reaction channel.
\item The reaction rates are averaged over the measured spectrum. Activation measurements at several well-defined neutron energies, including below 14 MeV, would show which energy range the differences come from.
\end{itemize}
\clearpage
\subsection{Zr: Zirconium}\label{foil:Zr}
\begin{figure}[H]
\centering
\includegraphics[width=\textwidth,height=0.82\textheight,keepaspectratio]{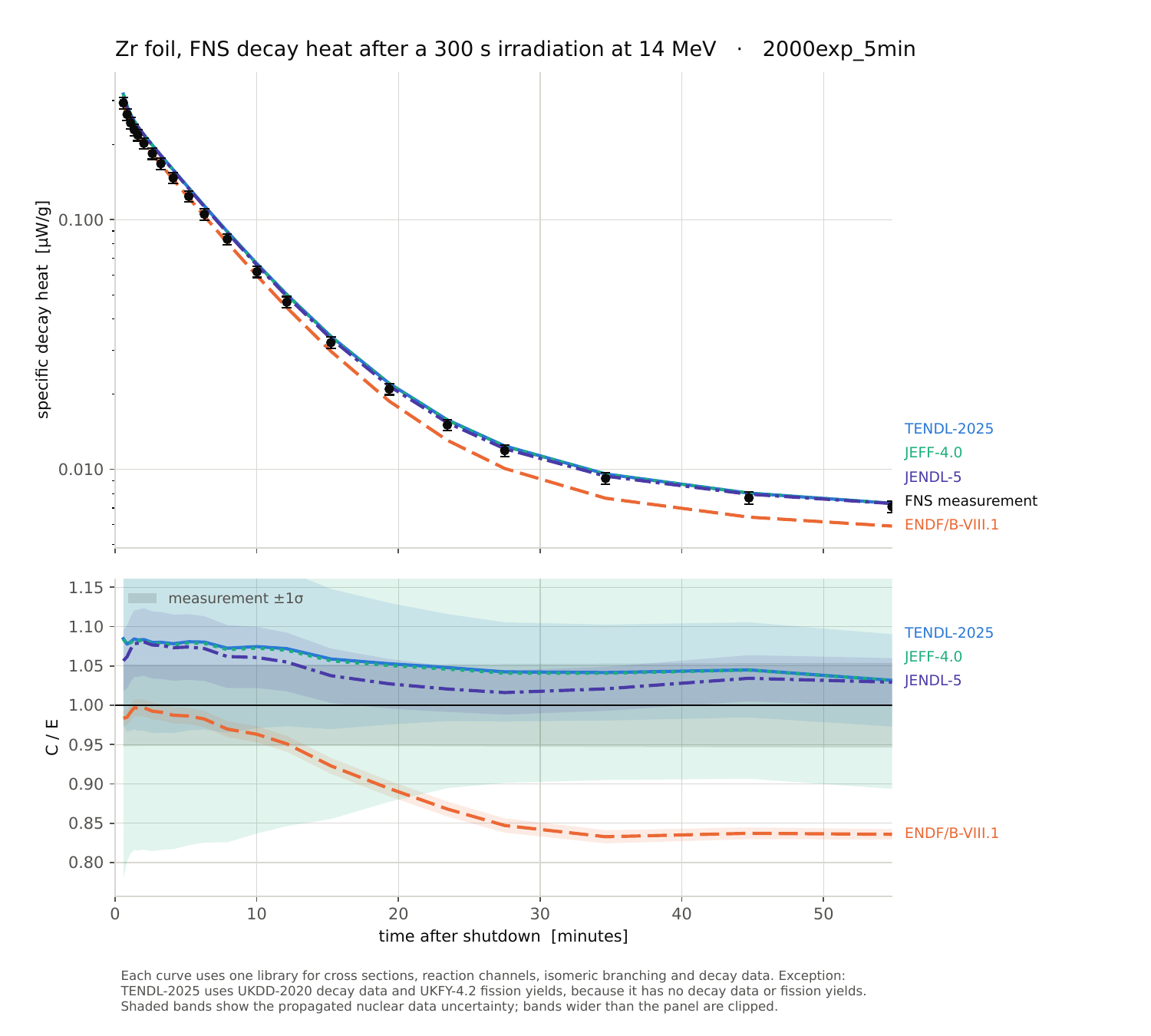}
\caption{Zr: decay heat against the measurement, with C/E beneath. Mean $|C/E-1|$ by library: JENDL-5 6\%, ENDF/B-VIII.1 6\%, JEFF-4.0 7\%, TENDL-2025 7\%.}
\end{figure}
\clearpage

\begin{figure}[H]
\centering
\includegraphics[width=\textwidth,height=0.82\textheight,keepaspectratio]{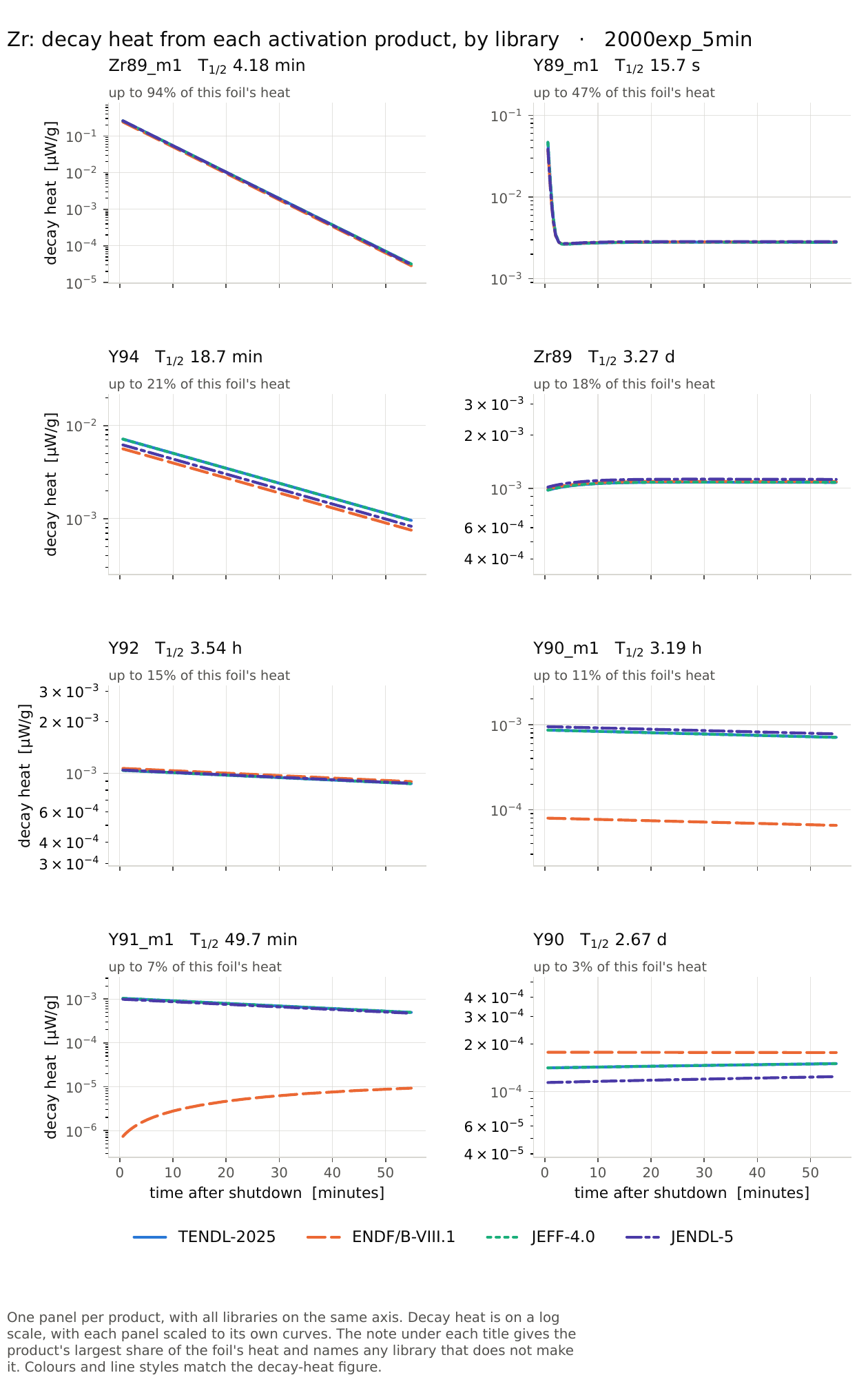}
\caption{Zr: share of the decay heat from each activation product during cooling.}
\end{figure}
\clearpage

\begin{figure}[H]
\centering
\includegraphics[width=\textwidth,height=0.82\textheight,keepaspectratio]{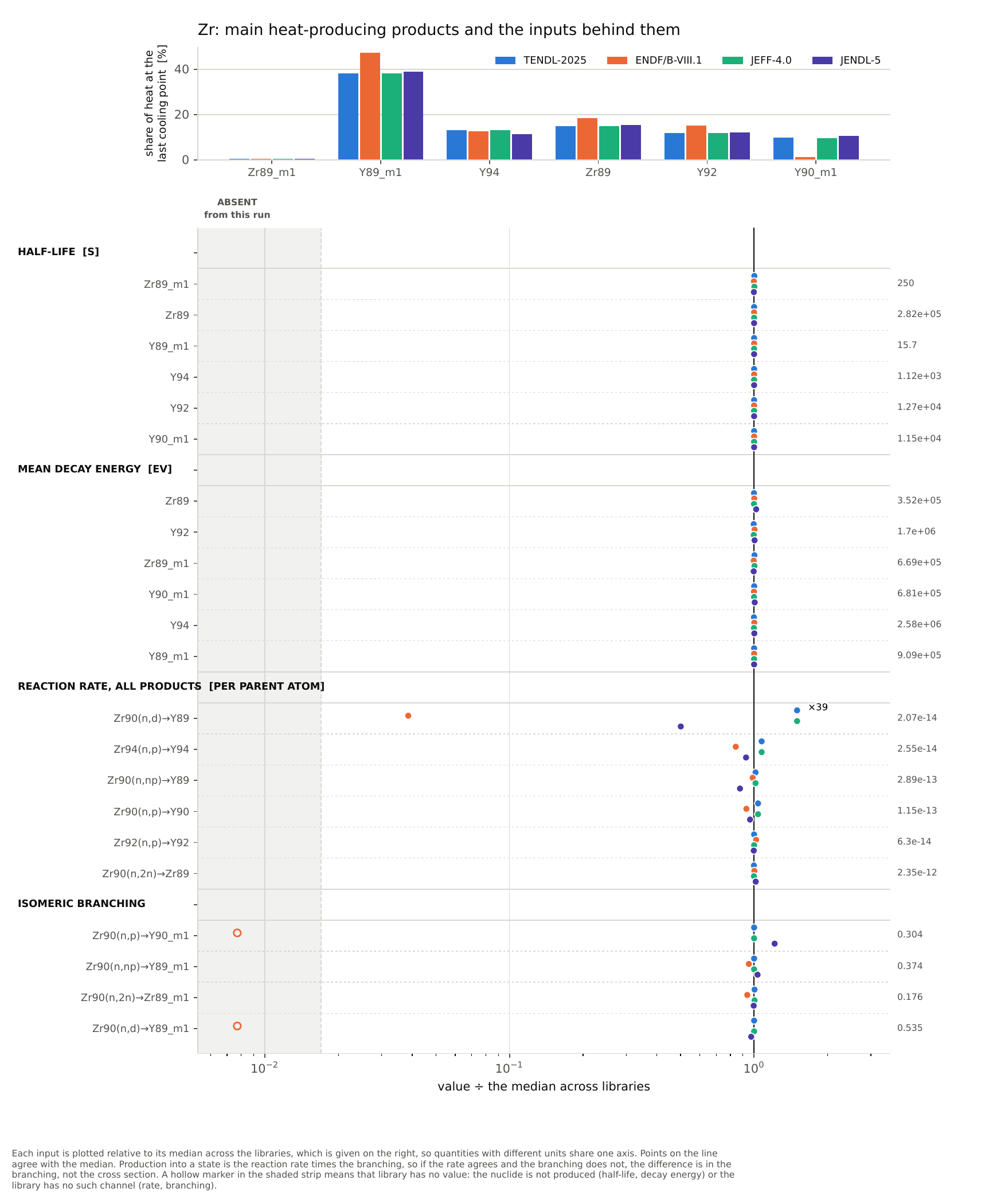}
\caption{Zr: inputs for the main heat-producing products, each relative to its median across the available libraries.}
\end{figure}
\clearpage

\begin{figure}[H]
\centering
\includegraphics[width=\textwidth,height=0.82\textheight,keepaspectratio]{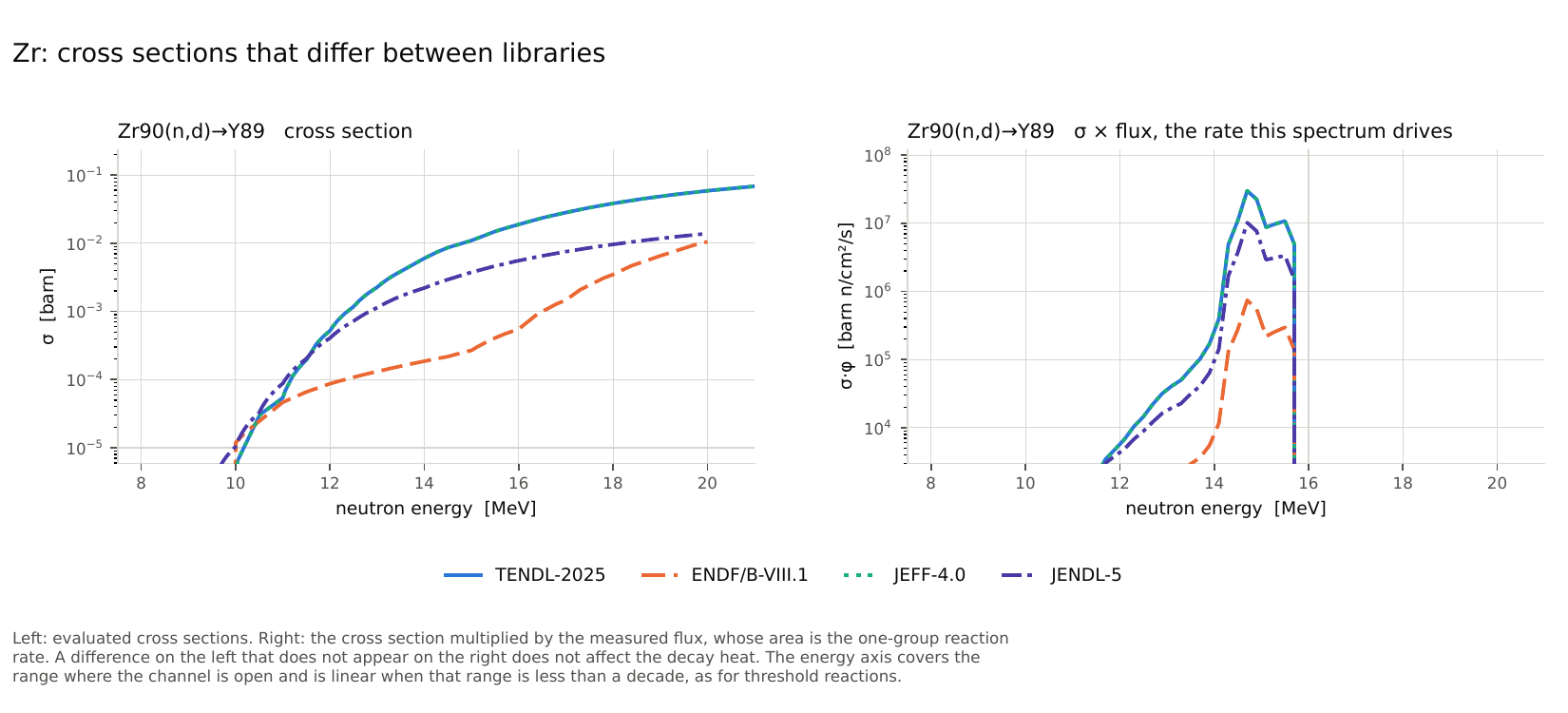}
\caption{Zr: cross sections of the channels whose spectrum-averaged rates differ, with the neutron spectrum and their product.}
\end{figure}
\clearpage

\subsubsection*{1996 seven-hour campaign}
The 1996 campaign irradiated this foil for 7.57 hours at position 7 and measured it out to 402.96 days. In the five-minute calculation Zr89 (half-life 3.27 days) gives up to 18 percent of the heat, and the longer irradiation and cooling test its production and decay data. TENDL-2025 is closest. This campaign uses a different spectrum, so it is not included in the main rankings.

\begin{figure}[H]
\centering
\includegraphics[width=\textwidth,height=0.82\textheight,keepaspectratio]{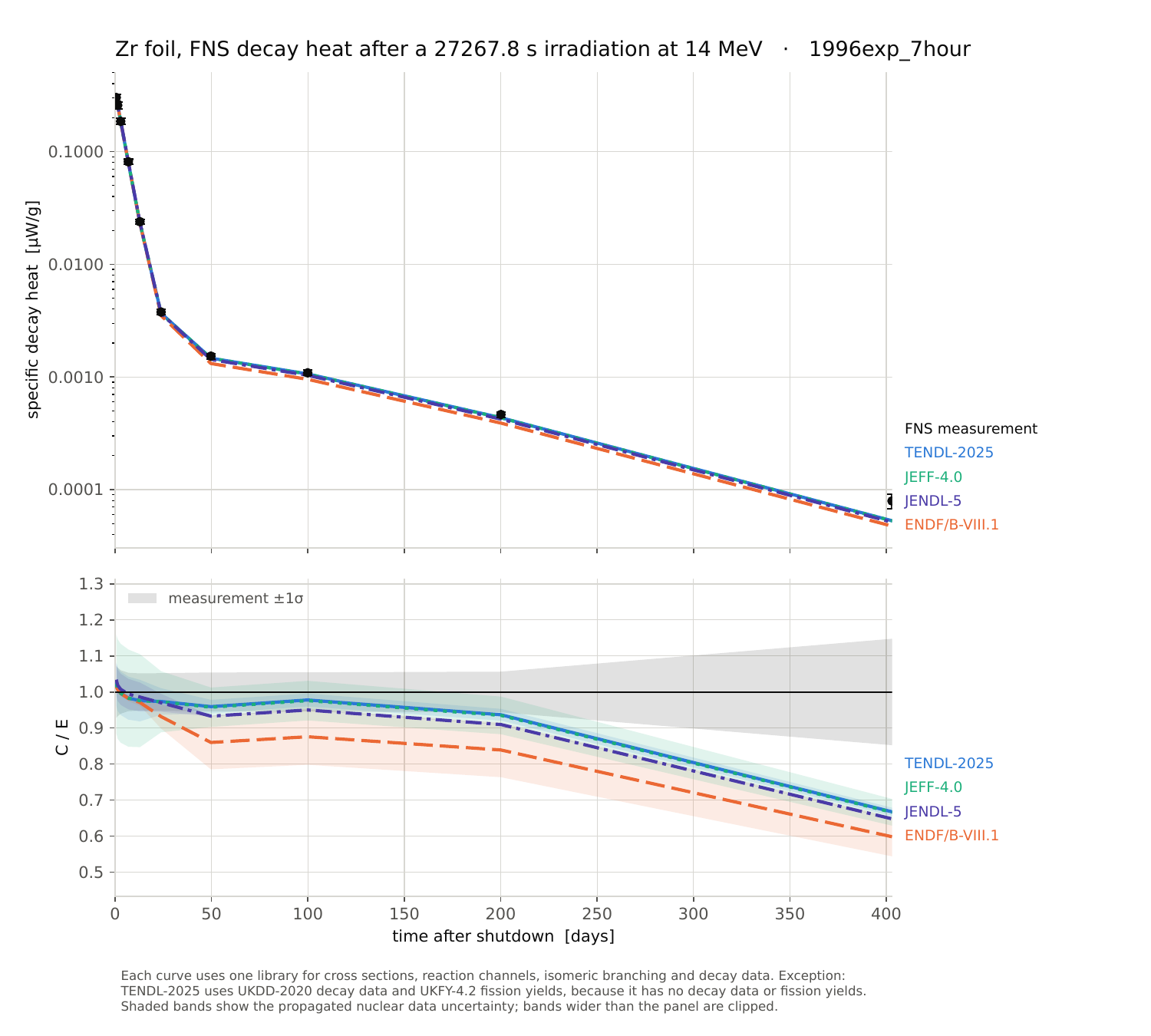}
\caption{Zr: 1996 seven-hour campaign, irradiation position 7. Mean $|C/E-1|$ by library: TENDL-2025 6\%, JEFF-4.0 6\%, JENDL-5 7\%, ENDF/B-VIII.1 10\%. }
\end{figure}
\clearpage

Mean absolute C/E deviations range from 6 to 7 percent. The spread of 1 percentage points is smaller than the median measurement uncertainty of 5 percent, so the measurement cannot separate the libraries.

\subsubsection*{Differences in the nuclear data}
\begin{itemize}\setlength{\itemsep}{6pt}
\item ENDF/B-VIII.1 gives the total reaction rate of Zr90(n,d)→Y89 as 7.966e-16, against 3.097e-14 for the median of the others, a factor of 39.\\[2pt]{\itshape This is the total rate of the channel over all its products, so the difference is in the cross section, not in the split between states.}
\end{itemize}
\subsubsection*{Suggested measurements}
\begin{itemize}\setlength{\itemsep}{4pt}
\item Natural Zr has 5 isotopes, so the measured heat combines several parents. Enriched single-isotope targets would distinguish their contributions. The candidate parent is Zr90. Enrichment alone does not isolate a single reaction channel.
\item The reaction rates are averaged over the measured spectrum. Activation measurements at several well-defined neutron energies, including below 14 MeV, would show which energy range the differences come from.
\end{itemize}
\clearpage
\section{1996 seven-hour campaign}
Every foil measured in the 1996 seven-hour campaign was calculated, using the measured spectrum at irradiation position 7. These results are kept separate from the 2000 five-minute comparison. The element sections show the seven-hour comparison where a product with a half-life over one day gives at least one percent of the five-minute heat. The table lists all 29 foils, including alloys. The libraries column shows how many libraries solved each foil, and the closest library is chosen from those.

\begin{figure}[H]
\centering
\includegraphics[width=\textwidth,height=0.82\textheight,keepaspectratio]{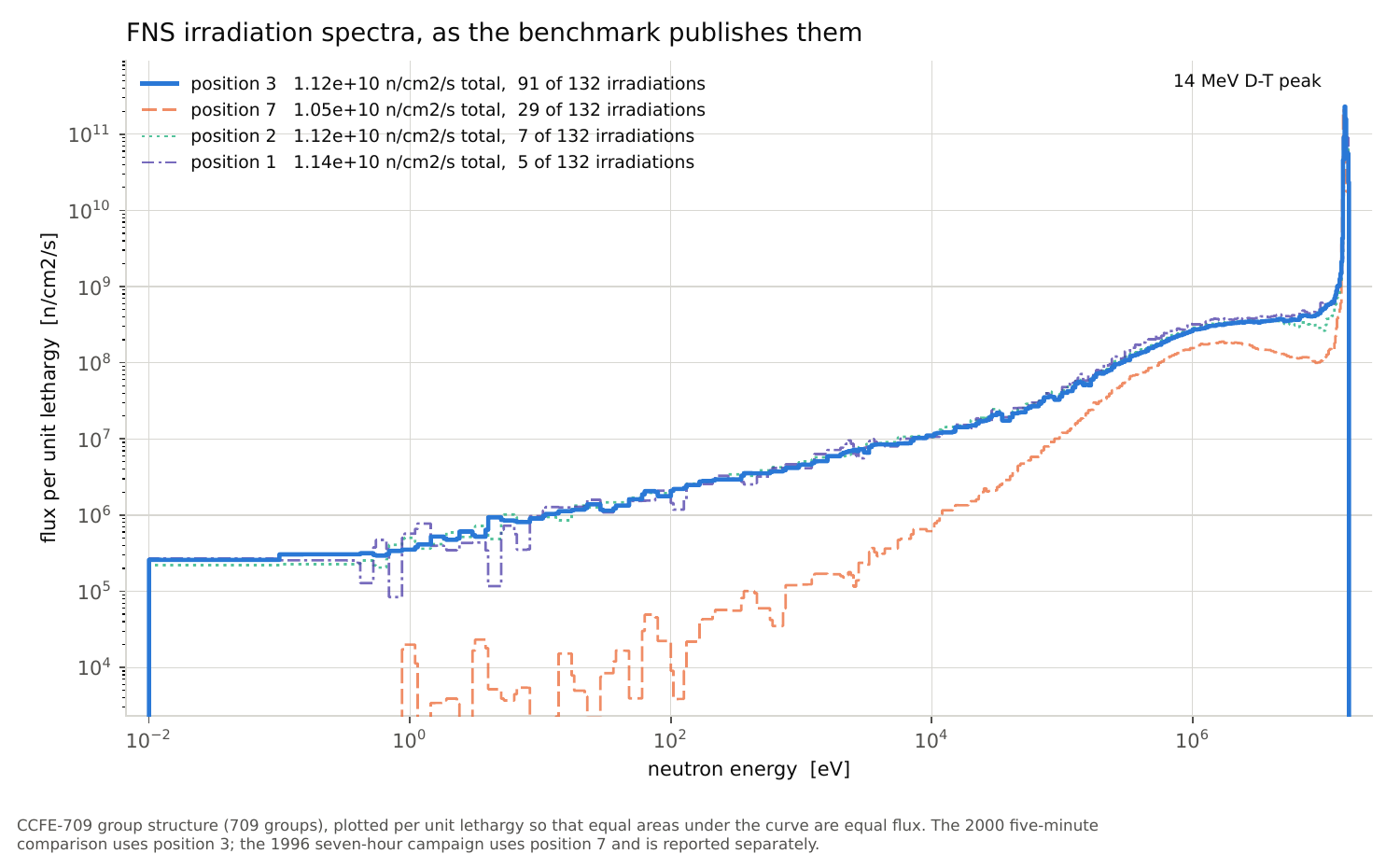}
\caption{The four measured FNS spectra. The main comparison uses position 3 and the seven-hour campaign uses position 7.}
\end{figure}
\clearpage

{\small\setlength{\tabcolsep}{3pt}\begin{longtable}{lrrlrr}\toprule
foil & irradiation & last point & closest library & libraries & mean $|C/E-1|$ range \\ \midrule\endfirsthead
\toprule foil & irradiation & last point & closest library & libraries & mean $|C/E-1|$ range \\ \midrule\endhead
\hyperref[foil:Al]{Al} & 7.57 h & 49.74 days & JENDL-5 & 4/4 & 47--47\% \\
\hyperref[foil:Ba]{Ba} & 7.17 h & 24.18 days & TENDL-2025 & 4/4 & 8--72\% \\
\hyperref[foil:Bi]{Bi} & 7.17 h & 12.17 days & TENDL-2025 & 4/4 & 586--2352\% \\
\hyperref[foil:Ca]{Ca} & 7.17 h & 24.19 days & ENDF/B-VIII.1 & 4/4 & 7--23\% \\
\hyperref[foil:Co]{Co} & 7.57 h & 402.95 days & ENDF/B-VIII.1 & 4/4 & 5--10\% \\
\hyperref[foil:Cr]{Cr} & 7.17 h & 197.96 days & JENDL-5 & 4/4 & 25--26\% \\
\hyperref[foil:Cu]{Cu} & 7.57 h & 402.97 days & ENDF/B-VIII.1 & 4/4 & 8--9\% \\
\hyperref[foil:Fe]{Fe} & 7.57 h & 402.95 days & ENDF/B-VIII.1 & 4/4 & 8--12\% \\
Inc600 & 7.57 h & 402.94 days & JEFF-4.0 & 4/4 & 3--3\% \\
\hyperref[foil:K]{K} & 7.17 h & 100.07 days & TENDL-2025 & 4/4 & 148--1578\% \\
\hyperref[foil:Mn]{Mn} & 7.17 h & 402.18 days & JENDL-5 & 4/4 & 1--6\% \\
\hyperref[foil:Mo]{Mo} & 7.57 h & 200.14 days & JEFF-4.0 & 4/4 & 10--16\% \\
\hyperref[foil:Na]{Na} & 7.17 h & 402.16 days & JEFF-4.0 & 4/4 & 5--30\% \\
\hyperref[foil:Nb]{Nb} & 7.57 h & 99.89 days & JEFF-4.0 & 4/4 & 7--98\% \\
\hyperref[foil:Ni]{Ni} & 7.57 h & 402.94 days & ENDF/B-VIII.1 & 4/4 & 2--7\% \\
NiCr & 7.57 h & 402.94 days & ENDF/B-VIII.1 & 4/4 & 2--6\% \\
\hyperref[foil:Pb]{Pb} & 7.57 h & 49.73 days & ENDF/B-VIII.1 & 4/4 & 27--31\% \\
\hyperref[foil:Re]{Re} & 7.17 h & 402.17 days & JENDL-5 & 4/4 & 7--29\% \\
\hyperref[foil:S]{S} & 7.17 h & 197.94 days & JENDL-5 & 4/4 & 5--6\% \\
SS304 & 7.57 h & 402.93 days & JENDL-5 & 4/4 & 7--8\% \\
SS316 & 7.57 h & 402.93 days & JENDL-5 & 4/4 & 7--8\% \\
\hyperref[foil:Sn]{Sn} & 7.17 h & 402.13 days & TENDL-2025 & 4/4 & 19--51\% \\
\hyperref[foil:Sr]{Sr} & 7.17 h & 402.15 days & TENDL-2025 & 4/4 & 4--13\% \\
\hyperref[foil:Ta]{Ta} & 7.57 h & 402.97 days & JENDL-5 & 4/4 & 43--47\% \\
\hyperref[foil:Ti]{Ti} & 7.57 h & 402.96 days & JENDL-5 & 4/4 & 4--11\% \\
\hyperref[foil:V]{V} & 7.57 h & 49.73 days & JENDL-5 & 4/4 & 20--25\% \\
\hyperref[foil:W]{W} & 7.57 h & 402.95 days & JENDL-5 & 4/4 & 19--28\% \\
\hyperref[foil:Y]{Y} & 7.17 h & 402.17 days & JENDL-5 & 4/4 & 15--17\% \\
\hyperref[foil:Zr]{Zr} & 7.57 h & 402.96 days & TENDL-2025 & 4/4 & 6--10\% \\
\bottomrule\end{longtable}}
The closest library has the smallest mean absolute C/E deviation for that foil in this campaign. These are not counted as wins in the main comparison, and being closest does not mean the measurement can tell the libraries apart.

\clearpage
\section{Results across all foils}
The figures below summarise the 2000 five-minute campaign across all elements.

All 69 common elemental foils are included in the rankings.

\clearpage
\section{Best agreement by element}
\begin{figure}[H]
\centering
\includegraphics[width=\textwidth,height=0.82\textheight,keepaspectratio]{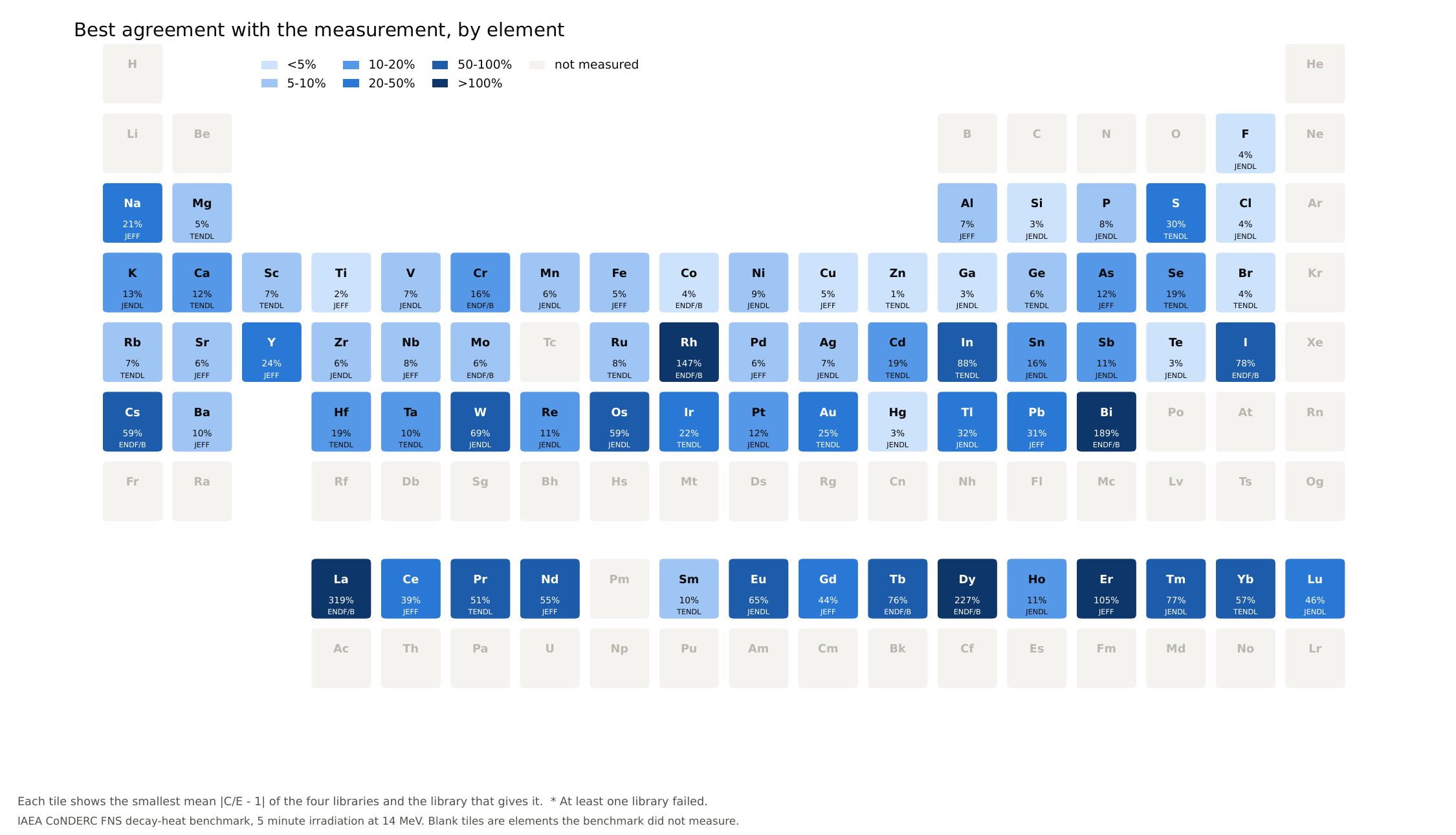}
\caption{The smallest mean absolute C/E deviation reached by any library on each element, and the library that reaches it. Asterisks mark elements with a missing library.}
\end{figure}
\clearpage

\begin{figure}[H]
\centering
\includegraphics[width=\textwidth,height=0.82\textheight,keepaspectratio]{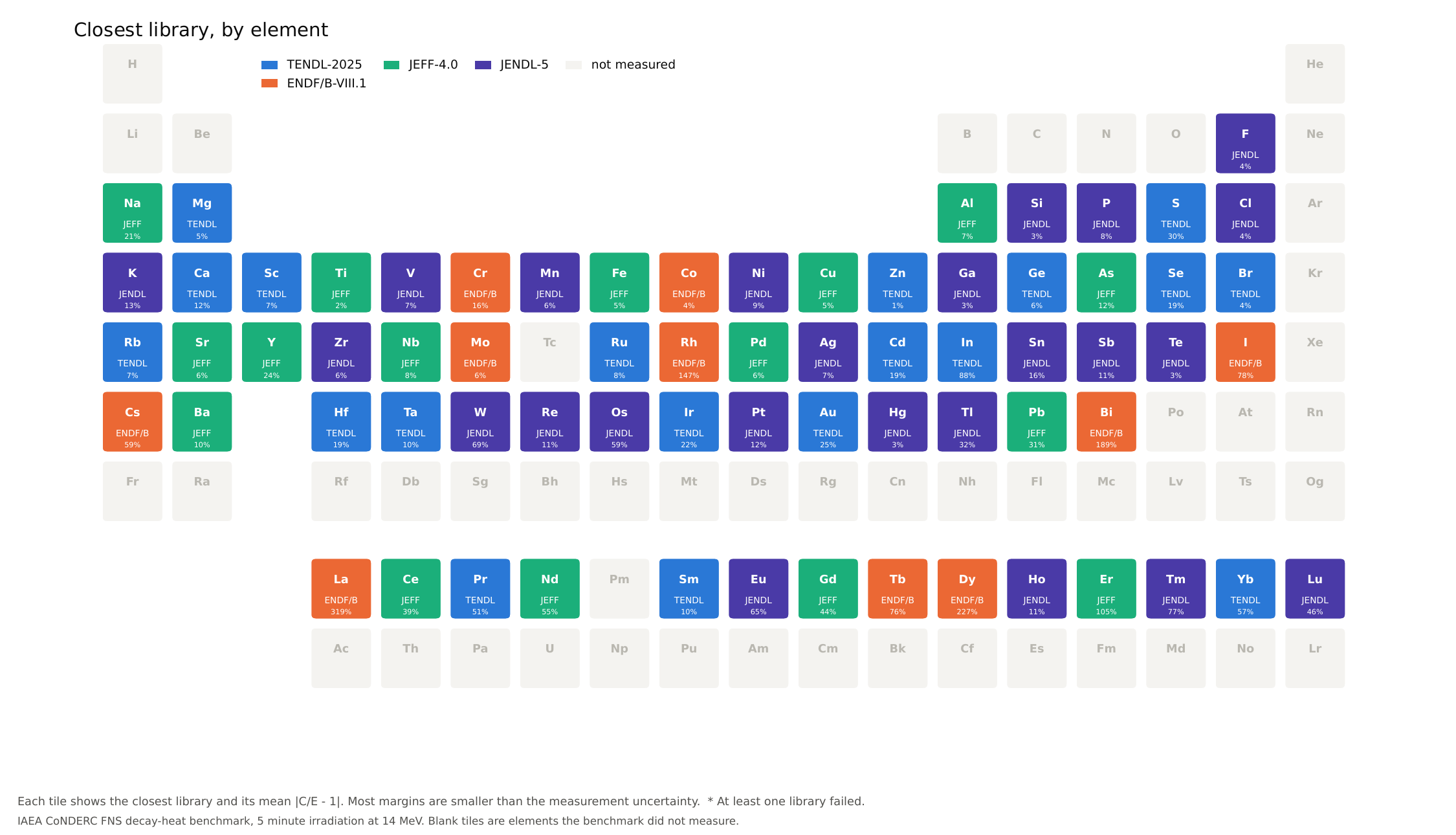}
\caption{The library closest to the measurement on each element. For most elements the margin is smaller than the measurement uncertainty.}
\end{figure}
\clearpage

\begin{figure}[H]
\centering
\includegraphics[width=\textwidth,height=0.82\textheight,keepaspectratio]{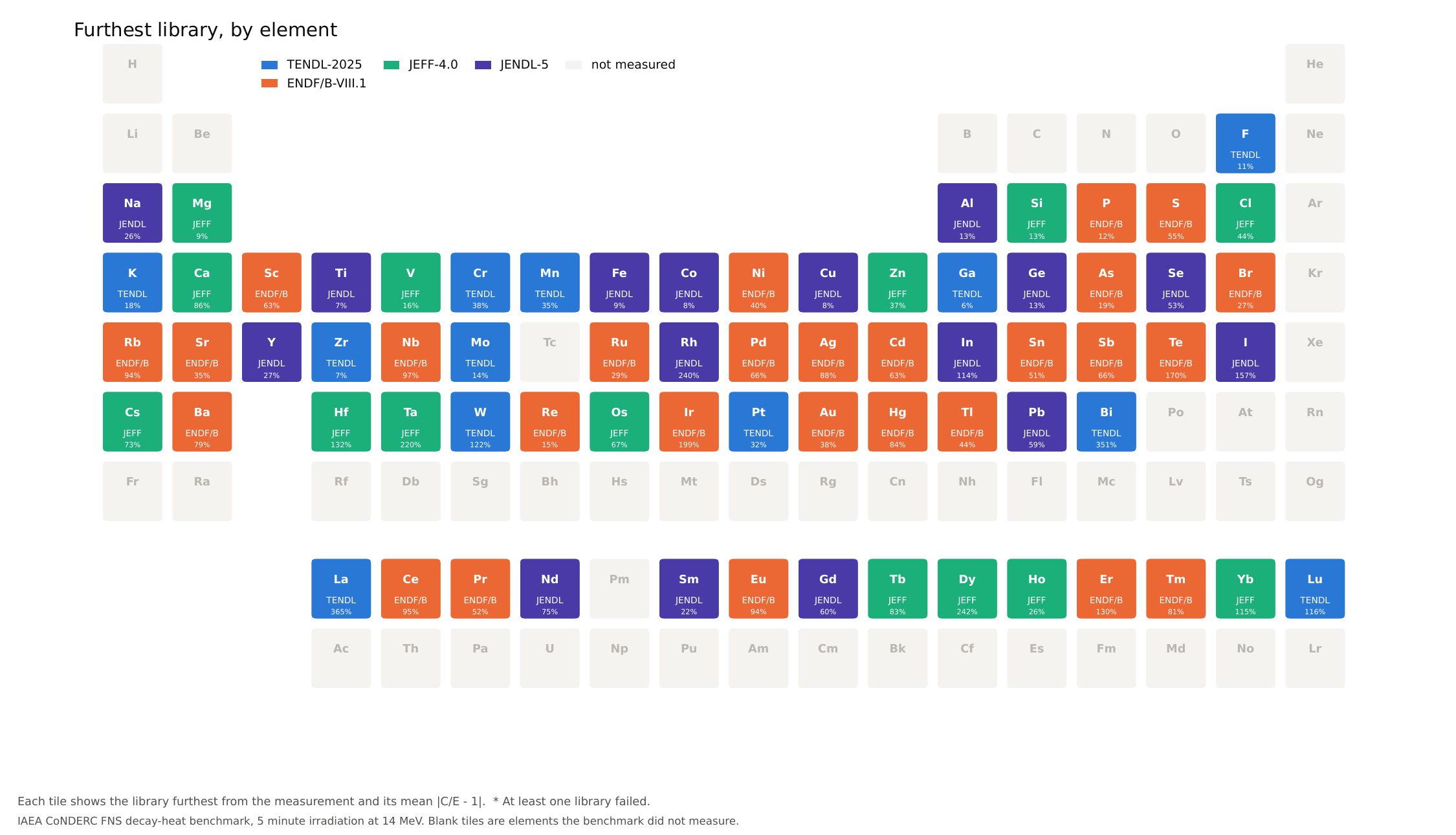}
\caption{The library furthest from the measurement on each element.}
\end{figure}
\clearpage

\section{Library comparison}
\begin{figure}[H]
\centering
\includegraphics[width=\textwidth,height=0.82\textheight,keepaspectratio]{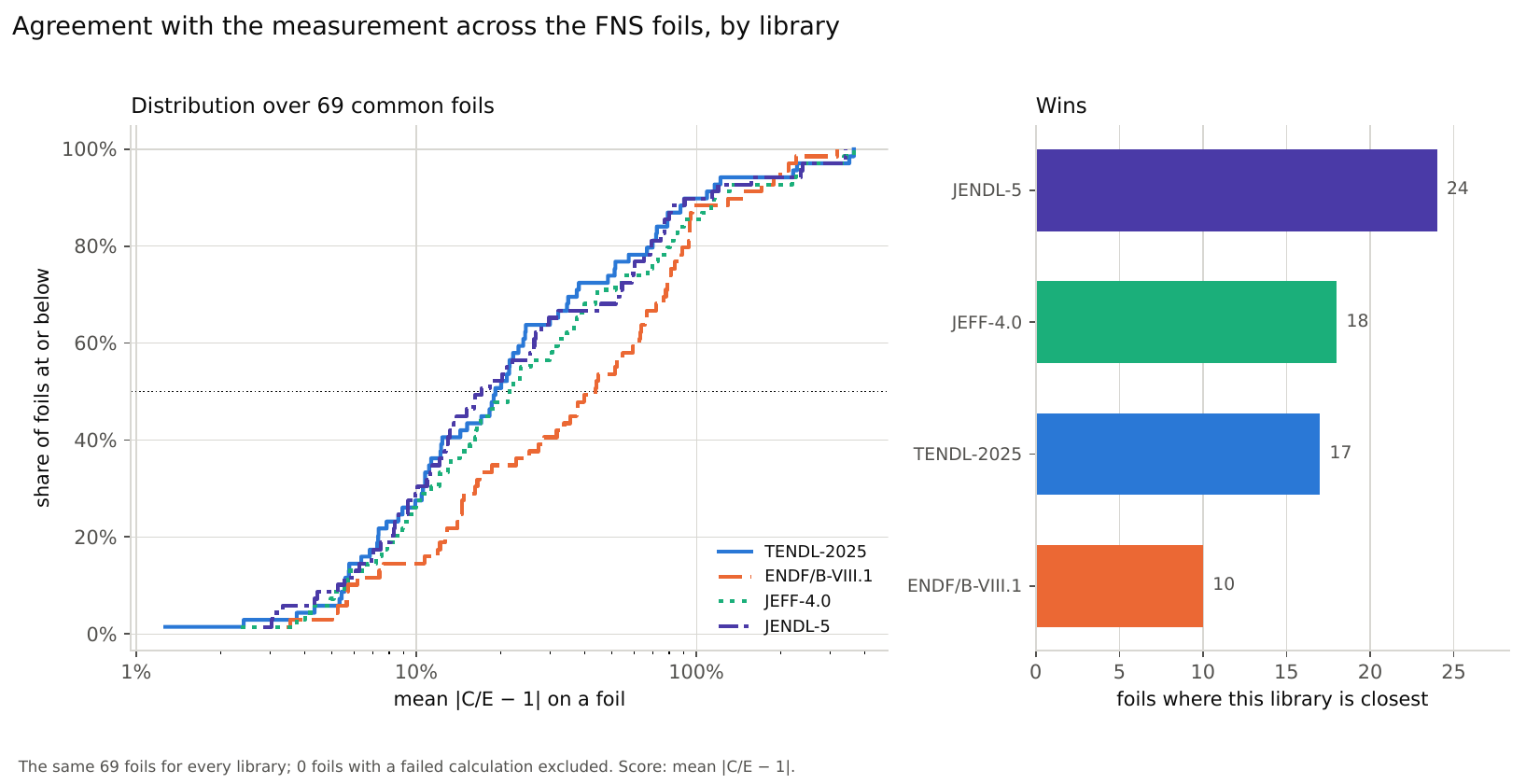}
\caption{Distribution of the per-foil deviation, and the number of foils on which each library is closest, for foils solved by all four libraries.}
\end{figure}
\clearpage

\begin{figure}[H]
\centering
\includegraphics[width=\textwidth,height=0.82\textheight,keepaspectratio]{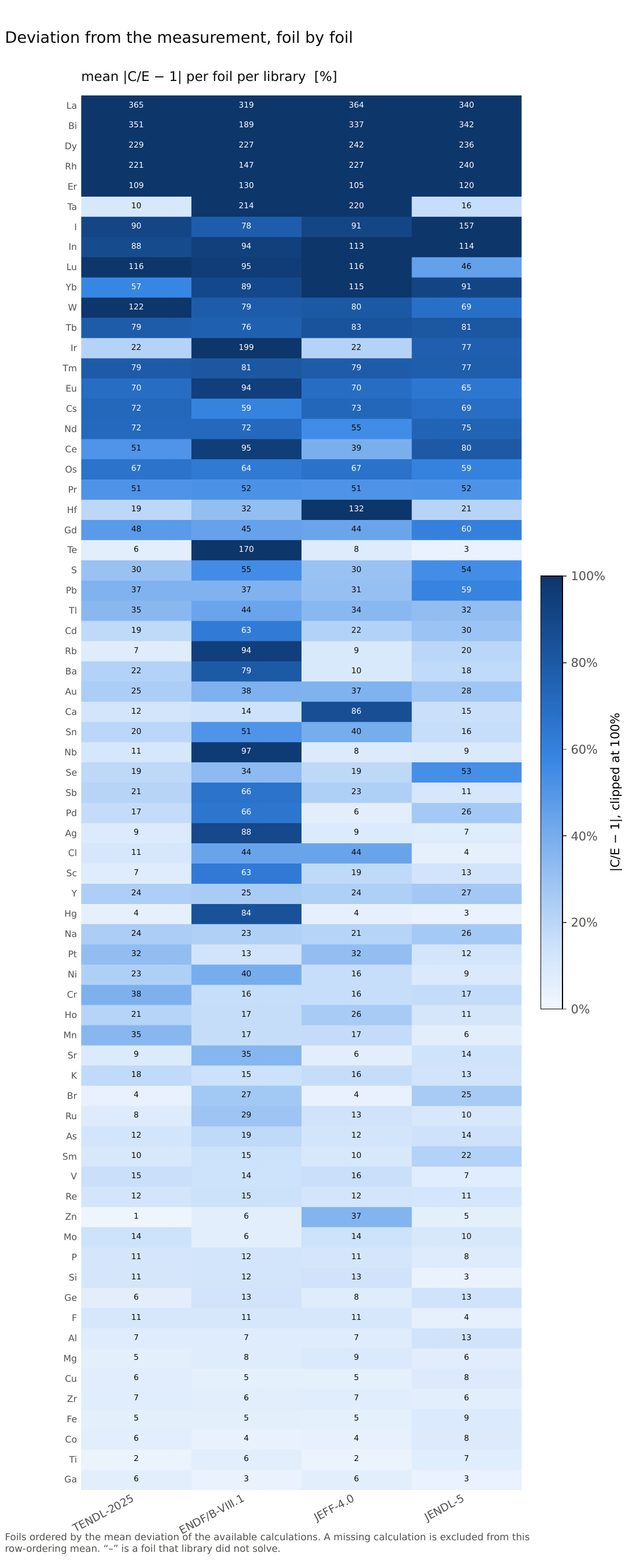}
\caption{Mean deviation for every foil and library, with foils ordered by their mean over the available libraries. Gaps are missing results.}
\end{figure}
\clearpage

\section{Library spread compared with distance from the measurement}
\begin{figure}[H]
\centering
\includegraphics[width=\textwidth,height=0.82\textheight,keepaspectratio]{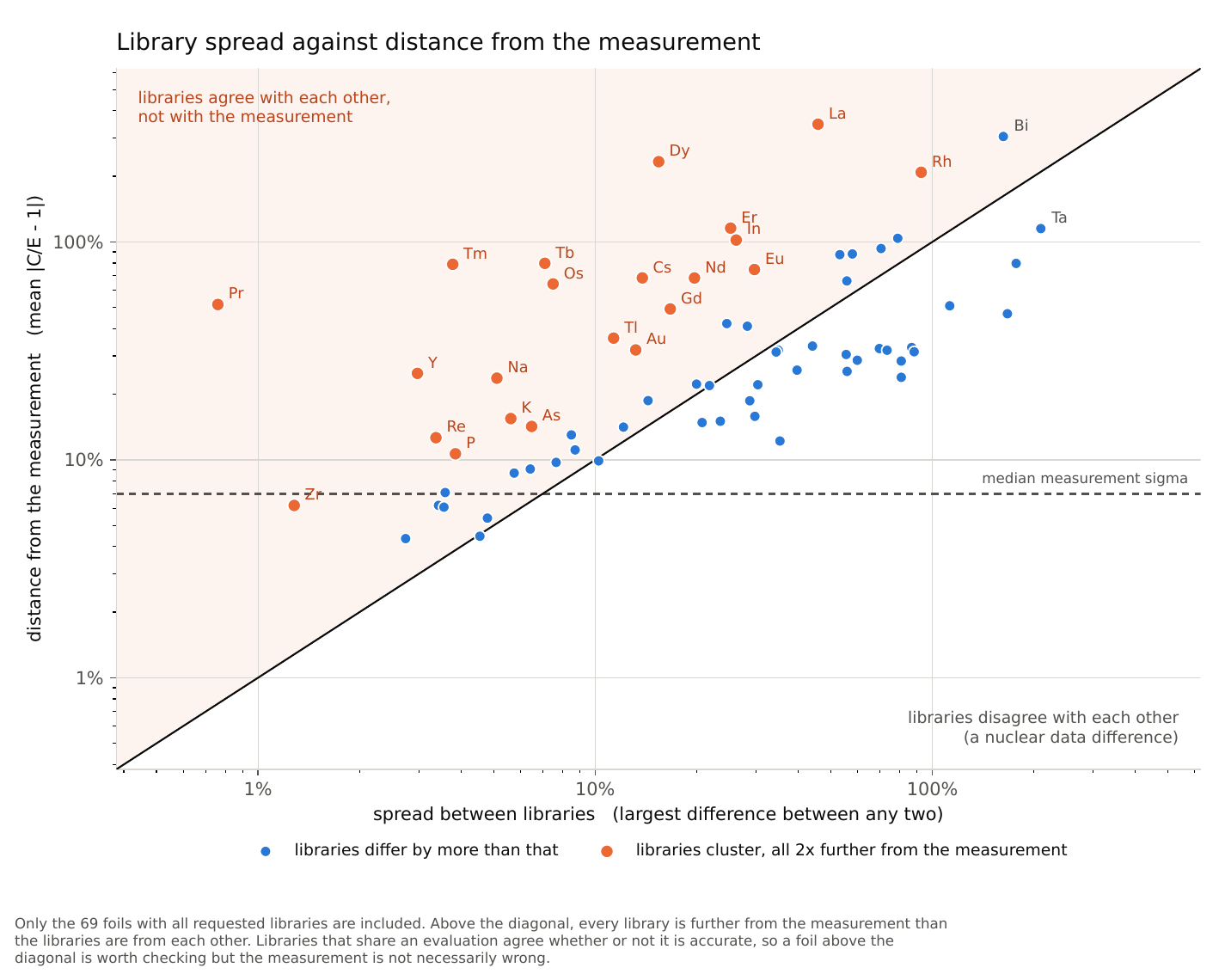}
\caption{Above the diagonal, every library is further from the measurement than the libraries are from each other.}
\end{figure}
\clearpage

\section{Library-specific outliers}
A foil is listed under a library when that library's mean C/E differs from the median of the other three by at least twice the spread of those three, and by more than the median experimental uncertainty. This shows where a library differs, not which library is right.

\subsection*{TENDL-2025}
Closest on 17 of 69 common elemental foils; median deviation 19.1 percent. Listed foils: \hyperref[foil:W]{W} (43\% separation; 19\% span among the other three; 13\% experimental uncertainty); \hyperref[foil:K]{K} (34\% separation; 9\% span among the other three; 6\% experimental uncertainty); \hyperref[foil:Cr]{Cr} (23\% separation; 2\% span among the other three; 6\% experimental uncertainty).

\subsection*{ENDF/B-VIII.1}
Closest on 10 of 69 common elemental foils; median deviation 43.5 percent. Listed foils: \hyperref[foil:Lu]{Lu} (187\% separation; 82\% span among the other three; 24\% experimental uncertainty); \hyperref[foil:Ir]{Ir} (179\% separation; 58\% span among the other three; 7\% experimental uncertainty); \hyperref[foil:Te]{Te} (165\% separation; 5\% span among the other three; 8\% experimental uncertainty); \hyperref[foil:Bi]{Bi} (157\% separation; 27\% span among the other three; 38\% experimental uncertainty); \hyperref[foil:Eu]{Eu} (153\% separation; 5\% span among the other three; 17\% experimental uncertainty); \hyperref[foil:Os]{Os} (130\% separation; 48\% span among the other three; 6\% experimental uncertainty).

\subsection*{JEFF-4.0}
Closest on 18 of 69 common elemental foils; median deviation 21.6 percent. Listed foils: \hyperref[foil:Hf]{Hf} (144\% separation; 47\% span among the other three; 6\% experimental uncertainty); \hyperref[foil:Ca]{Ca} (80\% separation; 11\% span among the other three; 9\% experimental uncertainty); \hyperref[foil:Zn]{Zn} (32\% separation; 5\% span among the other three; 5\% experimental uncertainty).

\subsection*{JENDL-5}
Closest on 24 of 69 common elemental foils; median deviation 17.1 percent. Listed foils: \hyperref[foil:I]{I} (66\% separation; 13\% span among the other three; 7\% experimental uncertainty); \hyperref[foil:Au]{Au} (59\% separation; 20\% span among the other three; 6\% experimental uncertainty); \hyperref[foil:Pb]{Pb} (32\% separation; 8\% span among the other three; 6\% experimental uncertainty); \hyperref[foil:F]{F} (12\% separation; 0\% span among the other three; 5\% experimental uncertainty); \hyperref[foil:Si]{Si} (11\% separation; 2\% span among the other three; 9\% experimental uncertainty); \hyperref[foil:V]{V} (8\% separation; 2\% span among the other three; 5\% experimental uncertainty).

\clearpage
\section{Nuclear data uncertainties}
The uncertainty bands propagate the evaluated uncertainties in cross sections, half-lives, decay branching, decay energies, decay photon lines, fission yields and isomeric branching, where the library provides them. The FNS spectra have no uncertainty per group and there is no transport calculation, so neither contributes. The production coverage column refers to cross-section covariances only. Inputs without evaluated uncertainties do not contribute to the band. No foil fissions, so fission yields carry no decay heat in any calculation.

\begin{center}\begin{tabular}{lrrr}\toprule
library & median stated $\sigma$ & share of production covered & foils \\ \midrule
TENDL-2025 & 14.0\% & 100.0\% & 69 \\
ENDF/B-VIII.1 & 1.8\% & 82.6\% & 69 \\
JEFF-4.0 & 15.3\% & 99.8\% & 69 \\
JENDL-5 & 1.3\% & 9.9\% & 69 \\
\bottomrule\end{tabular}\end{center}
A narrow band can mean missing uncertainty data rather than accurate data, since inputs without uncertainties stay at their nominal values. Each result records the sampled sources, missing inputs, covariance warnings and whether it converged. With a fixed number of replicas the standard deviations are not necessarily converged, and unconverged values need more samples before they are used quantitatively.

\begin{center}\begin{tabular}{lrr}\toprule
library & replicas per foil & converged foils / checked \\ \midrule
TENDL-2025 & 60 & 0 / 69 \\
ENDF/B-VIII.1 & 60 & 0 / 69 \\
JEFF-4.0 & 60 & 0 / 69 \\
JENDL-5 & 60 & 0 / 69 \\
\bottomrule\end{tabular}\end{center}
\clearpage
\section{Recommendations}
\subsection*{Enriched-isotope samples}
The natural samples combine the activation of several isotopes. Irradiating enriched single-isotope targets would show which parent makes a given product. Enrichment does not isolate a single reaction or measure an isomeric ratio on its own. The next section proposes five such experiments, chosen from the reaction routes in these calculations. None of them has been simulated or measured.

\subsection*{Isomeric branching}
For some elements the libraries agree on the total reaction rate but differ by two orders of magnitude in how much goes to a metastable state. For tantalum, the four libraries agree on the Ta181(n,gamma) capture rate to within 6 percent, but the branch to the 950 second isomer Ta182m2 is 37.5 percent in ENDF/B-VIII.1 and JEFF-4.0 and 0.13 and 0.24 percent in TENDL-2025 and JENDL-5. ENDF/B-VIII.1 and JEFF-4.0 overpredict the tantalum heat by a factor of 3.5, which fits a branching error, but total heat alone cannot show which branching is correct. Measuring the products separately would test the branching directly.

\subsection*{Irradiation and cooling times chosen per element}
A five-minute irradiation followed by about an hour of cooling is most sensitive to products with half-lives of minutes. Of the 69 elemental foils, 47 have a product with a half-life over a day among their main heat contributors, and 28 of these have no seven-hour measurement. In some of these the long-lived product gives most of the heat at the last measured point: Cs132 (half-life 6.48 days) gives 94 percent of the cesium heat, Tm168 (half-life 93.1 days) gives 94 percent of the thulium heat, Tl202 (half-life 12.3 days) gives 92 percent of the thallium heat and Au196 (half-life 6.17 days) gives 75 percent of the gold heat. A 55 minute measurement can show how much heat these products give, but not how they decay. Longer irradiations and cooling times chosen for each element would cover them. The 1996 seven-hour results show what this adds for the 29 foils in that campaign.

\subsection*{Samarium-147 background}
Natural samarium is 15 percent Sm147, an alpha emitter with a half-life of 10\textasciicircum{}11 years, which heats the sample without any irradiation. How much this affects the reported measurement depends on how the background heat was subtracted, so it should not be assumed to come from the irradiation.

\subsection*{Foils where the measurement should be checked}
A foil is listed here when its mean deviation from the measurement is at least twice the largest difference between the libraries and is also larger than the experimental uncertainty. This does not mean the measurement is wrong. Data or methods shared by all four libraries can produce the same bias, so both the data and the experiment should be checked.

The six largest cases are: \hyperref[foil:La]{Lanthanum (La)} (347\% mean deviation, 46\% library spread, 56\% experimental uncertainty); \hyperref[foil:Dy]{Dysprosium (Dy)} (234\% mean deviation, 15\% library spread, 13\% experimental uncertainty); \hyperref[foil:Rh]{Rhodium (Rh)} (209\% mean deviation, 93\% library spread, 8\% experimental uncertainty); \hyperref[foil:Er]{Erbium (Er)} (116\% mean deviation, 25\% library spread, 13\% experimental uncertainty); \hyperref[foil:In]{Indium (In)} (102\% mean deviation, 26\% library spread, 6\% experimental uncertainty); \hyperref[foil:Tm]{Thulium (Tm)} (79\% mean deviation, 4\% library spread, 31\% experimental uncertainty).

\clearpage
\section{Proposed enriched-isotope irradiations}\label{isotope-experiments}
The proposal is to irradiate targets enriched in a single isotope instead of the natural element. Where two isotopes are listed, they are two separate targets. This changes which parent nuclei can contribute to the measured products. It is most useful when different isotopes make the same radioactive product, or when the products of a minor isotope are hidden by other activation products in a calorimetric measurement.

The shortlist is based on the natural-sample calculations for the 2000 five-minute campaign. All numbers come from the saved results and diagnostics; no enriched targets have been simulated. Route shares are the fractions of a product made by each reaction during the first irradiation pulse. They are not heat fractions, measured yields or uncertainties. Peak heat shares are the largest fraction of the heat a product gives at any measured cooling time, taken from the full per-nuclide heat curves. Different products peak at different times.

In the notation used, (n,2n) is emission of two neutrons, (n,n') is inelastic scattering, (n,gamma) is neutron capture and (n,p) is proton emission. A suffix m marks a metastable state and m2 the second such state. A single target isotope still undergoes several reactions and makes several radioactive nuclides, so product-resolved measurements are needed alongside the total decay heat.

{\small\setlength{\tabcolsep}{3pt}
\begin{tabular}{@{}p{0.15\linewidth}p{0.28\linewidth}p{0.51\linewidth}@{}}\toprule
Element & Separate enriched targets & Question \\ \midrule
Hf & Hf-177 and Hf-178 & Which parent produces Hf-177m2? \\
Cd & Cd-111 and Cd-112 & Which route produces Cd-111m? \\
Eu & Eu-151 and Eu-153 & How much Eu-152m comes from capture and how much from (n,2n)? \\
Zn & Zn-64 & Is the Zn-64(n,2n)Zn-63 rate right? \\
Ni & Ni-62 & Are Co-62 and Co-62m made in the right amounts? \\
\bottomrule\end{tabular}\par}
\clearpage
\subsection{Isotopic composition of the natural samples}
The percentages are atom fractions within each element, taken from the initial inventories of the calculations. They are the natural abundances assumed in the calculations, not measurements of the original samples or specifications for new targets. Oxygen and other elements are left out, and rounded values may not add up to exactly 100\%. If libraries assumed different abundances, each is labelled.

{\small\setlength{\tabcolsep}{3pt}
\begin{tabular}{@{}p{0.09\linewidth}>{\raggedright\arraybackslash}p{0.51\linewidth}>{\raggedright\arraybackslash}p{0.34\linewidth}@{}}\toprule
Element & Natural isotopes (atom \%) & Proposed targets \\ \midrule
Hf & Hf-174 (0.16\%); Hf-176 (5.26\%); Hf-177 (18.6\%); Hf-178 (27.28\%); Hf-179 (13.62\%); Hf-180 (35.08\%) & Hf-177 and Hf-178 first; Hf-176 later to test capture. \\[5pt]
Cd & Cd-106 (1.245\%); Cd-108 (0.888\%); Cd-110 (12.47\%); Cd-111 (12.795\%); Cd-112 (24.109\%); Cd-113 (12.227\%); Cd-114 (28.754\%); Cd-116 (7.512\%) & Cd-111 and Cd-112 first; Cd-110 later to test capture. \\[5pt]
Eu & Eu-151 (47.81\%); Eu-153 (52.19\%) & Both natural isotopes, Eu-151 and Eu-153, as separate targets. \\[5pt]
Zn & Zn-64 (49.17\%); Zn-66 (27.73\%); Zn-67 (4.04\%); Zn-68 (18.45\%); Zn-70 (0.61\%) & Zn-64 only, for the Zn-63 production test. \\[5pt]
Ni & Ni-58 (68.0769\%); Ni-60 (26.2231\%); Ni-61 (1.1399\%); Ni-62 (3.6345\%); Ni-64 (0.9256\%) & Ni-62 only, for the Co-62 and Co-62m comparison. \\[5pt]
\bottomrule\end{tabular}\par}
Europium is the only element where every natural isotope is proposed, because it has just two and both make the same product. For the other elements a subset is proposed. A natural-element reference sample should still be irradiated and modelled with all its isotopes, since isotopes left off the shortlist still activate.

\clearpage
\subsection{Hafnium: separate Hf-177 and Hf-178 targets}\label{isotope:Hf}
\subsubsection*{Motivation}
Hf-178(n,2n)Hf-177m2 and Hf-177(n,n')Hf-177m2 make the same isomer, so measuring it after irradiating natural hafnium cannot show which parent it came from. The libraries weight the two routes differently, so separate targets would test which production model is right.

\subsubsection*{Target selection}
There is no need to irradiate all six natural hafnium isotopes. Separate Hf-177 and Hf-178 targets are enough to tell the two competing production routes apart. Hf-176 can be added to test the third direct parent, capture on low-energy neutrons. Hf-174 does not make this product directly. Hf-179 and Hf-180 would only make Hf-177m2 at energies above those in this benchmark, although Hf-180 could be used later to study Hf-179m.

\subsubsection*{Possible parents of Hf-177m2}
{\small\setlength{\tabcolsep}{3pt}
\begin{tabular}{@{}p{0.13\linewidth}p{0.23\linewidth}p{0.58\linewidth}@{}}\toprule
Parent & Route to Hf-177m2 & Notes \\ \midrule
Hf-174 & No direct single-neutron route & Reaching mass 177 needs several reactions; not an initial target. \\[4pt]
Hf-176 & (n,gamma) & Capture can make Hf-177m2. Present in the route diagnostics; follow-up target. \\[4pt]
Hf-177 & (n,n') & Inelastic scattering excites the same isotope. Present in the route diagnostics; first target. \\[4pt]
Hf-178 & (n,2n) & Emitting two neutrons leaves mass 177. Present in the route diagnostics; first target. \\[4pt]
Hf-179 & (n,3n) & Only at higher energy: threshold about 16.6 MeV for Hf-177m2. \\[4pt]
Hf-180 & (n,4n) & Only at higher energy: threshold about 24 MeV for Hf-177m2. \\[4pt]
\bottomrule\end{tabular}\par}
These are the direct routes from the natural ground-state hafnium isotopes. Reactions on impurities or on nuclei made during the irradiation are not included. The Hf-177m2 level is 2.740 MeV above the ground state, and the Hf-179 and Hf-180 thresholds include this energy and the recoil, using the TENDL ground-state Q values. A reaction that makes ground-state Hf-177 does not necessarily make the isomer at the same neutron energy. The route diagnostics from the natural-sample calculations find Hf-176, Hf-177 and Hf-178. They follow one neutron reaction and up to three decays, so they do not cover routes with several neutron reactions.

\clearpage
\subsubsection*{Results from the natural-sample calculations}
{\small\setlength{\tabcolsep}{3pt}
\begin{tabular}{@{}p{0.20\linewidth}*{4}{p{0.185\linewidth}}@{}}\toprule
Library & Hf-178 route (\%) & Hf-177 route (\%) & Hf-176 route (\%) & Peak product heat (\%) \\ \midrule
TENDL-2025 & 17.9 & 79.6 & 2.44 & 5.65 \\
ENDF/B-VIII.1 & Not listed & Not listed & Not listed & 0 \\
JEFF-4.0 & 92.8 & 6.99 & 0.181 & 38.1 \\
JENDL-5 & Not listed & Not listed & Not listed & 0 \\
\bottomrule\end{tabular}\par}
The product is Hf-177m2. Route shares are fractions of that product's production, not of the total activation, and need not add up to 100\% because other routes can contribute. Not listed means the route is not among those kept in the diagnostics, not that it is zero. Peak heat comes from the full heat curves, so a zero there is a calculated zero.

JEFF-4.0 attributes 92.8\% of Hf-177m2 production to Hf-178(n,2n), compared with 17.9\% in TENDL-2025. TENDL instead attributes 79.6\% to Hf-177 inelastic scattering. Separate targets would show which is right.

Hf-176 capture supplies 2.44\% of the recorded Hf-177m2 production in TENDL-2025 and 0.181\% in JEFF-4.0. It is a minor route in the natural sample but should stay in the model, and a separate Hf-176 target would measure it directly.

\subsubsection*{Proposed irradiations}
Irradiate Hf-177 and Hf-178 separately in a characterised neutron field. The Hf-177 target tests inelastic production of Hf-177m2 and the Hf-178 target tests the (n,2n) route. Hf-178 also makes Lu-178 through (n,p). The two targets remove the main parent ambiguity without depending on the other hafnium isotopes being modelled correctly.

\subsubsection*{Measurements}
Measure the Hf-177m2 activity and follow the decay heat, with gamma spectrometry to separate other products such as Lu-178. Getting an isomeric branching fraction, rather than just the isomer yield, also needs an independent measurement of the total channel production. Stable ground-state products cannot be measured by gamma counting.

\subsubsection*{Limitations}
The route comparison does not show that this isomer explains the whole hafnium C/E discrepancy, since other reactions and decay data also contribute. Hf-180 could be used later to study Hf-179m, but it does not replace the Hf-177 and Hf-178 comparison.

\clearpage
\subsection{Cadmium: separate Cd-111 and Cd-112 targets}\label{isotope:Cd}
\subsubsection*{Motivation}
Cd-112(n,2n)Cd-111m and Cd-111(n,n')Cd-111m make the same isomer, and natural cadmium contains both parents as well as other isotopes that produce heat. Cd-111m gives about three quarters of the peak heat in TENDL-2025, JEFF-4.0 and JENDL-5 but very little in ENDF/B-VIII.1, so this is a specific product to test.

\subsubsection*{Target selection}
Start with Cd-111 and Cd-112 rather than all eight natural isotopes, with Cd-110 as an optional capture test. Cd-106, Cd-108, Cd-113, Cd-114 and Cd-116 are outside this study of Cd-111m but still activate. Measure any residual Cd-113 in the targets, because its strong low-energy absorption can change the neutron field inside the sample.

\subsubsection*{Results from the natural-sample calculations}
{\small\setlength{\tabcolsep}{3pt}
\begin{tabular}{@{}p{0.22\linewidth}*{3}{p{0.24\linewidth}}@{}}\toprule
Library & Cd-112 route (\%) & Cd-111 route (\%) & Peak product heat (\%) \\ \midrule
TENDL-2025 & 87 & 12.8 & 78.5 \\
ENDF/B-VIII.1 & Not listed & Not listed & 0 \\
JEFF-4.0 & 87 & 12.8 & 75 \\
JENDL-5 & 87.6 & 12.3 & 76.4 \\
\bottomrule\end{tabular}\par}
The product is Cd-111m1. Route shares are fractions of that product's production, not of the total activation, and need not add up to 100\% because other routes can contribute. Not listed means the route is not among those kept in the diagnostics, not that it is zero. Peak heat comes from the full heat curves, so a zero there is a calculated zero.

\subsubsection*{Proposed irradiations}
Separate Cd-112 and Cd-111 targets would measure the isomer yield from (n,2n) and from inelastic scattering. Together they would show which libraries have the Cd-111m production right and which route it comes from. Cd-110 capture is a smaller route that still has to be included when modelling natural cadmium.

\subsubsection*{Measurements}
Measure decay heat together with the Cd-111m activity over a cooling time suited to its 49 minute half-life. Other products, such as Ag-110 from charged-particle reactions, need to be separated from Cd-111m rather than inferred from the total heat.

\subsubsection*{Limitations}
A small contribution is not zero production. The peak heat column uses the full heat breakdown, including products below the diagnostic threshold. A missing route entry does not mean a zero cross section or an error in the evaluation; the processing and the decay chain also need checking.

\clearpage
\subsection{Europium: separate Eu-151 and Eu-153 targets}\label{isotope:Eu}
\subsubsection*{Motivation}
Neutron capture on Eu-151 and Eu-153(n,2n) both make Eu-152m. In the natural sample their contributions add, so the Eu-152m activity alone cannot separate capture from the threshold reaction. Eu-152m also gives a large share of the heat in several libraries.

\subsubsection*{Target selection}
Natural europium has only two isotopes, Eu-151 and Eu-153, and both are needed as separate enriched targets because capture and (n,2n) contribute similar amounts in the natural-sample calculations. Oxygen in an oxide target activates too and has to be modelled separately.

\subsubsection*{Results from the natural-sample calculations}
{\small\setlength{\tabcolsep}{3pt}
\begin{tabular}{@{}p{0.22\linewidth}*{3}{p{0.24\linewidth}}@{}}\toprule
Library & Eu-151 route (\%) & Eu-153 route (\%) & Peak product heat (\%) \\ \midrule
TENDL-2025 & 53.7 & 46.3 & 69 \\
ENDF/B-VIII.1 & Not listed & Not listed & 0 \\
JEFF-4.0 & 53.7 & 46.3 & 69 \\
JENDL-5 & 55.2 & 44.8 & 74.4 \\
\bottomrule\end{tabular}\par}
The product is Eu-152m1. Route shares are fractions of that product's production, not of the total activation, and need not add up to 100\% because other routes can contribute. Not listed means the route is not among those kept in the diagnostics, not that it is zero. Peak heat comes from the full heat curves, so a zero there is a calculated zero.

\subsubsection*{Proposed irradiations}
Irradiate Eu-151 and Eu-153 separately in the same measured spectrum to find each one's Eu-152m production. This separates two different reaction mechanisms, and shows whether the natural-sample calculations get the total yield for the right reasons.

\subsubsection*{Measurements}
Measure the Eu-152m activity and decay heat over several hours, with gamma spectrometry to separate the other europium isomers. The low-energy part of the neutron field needs characterising as well as the 14 MeV peak, because a capture yield depends on it.

\subsubsection*{Limitations}
The benchmark's europium sample is an oxide. Enriching the europium does not remove the activation of oxygen, including early N-16 heat. The chemical form, enrichment and impurities of new targets must be measured, and a metal target would need its own model. The route fractions in the table apply only to the original spectrum.

\clearpage
\subsection{Zinc: a Zn-64 target}\label{isotope:Zn}
\subsubsection*{Motivation}
The diagnostics attribute Zn-63 production to Zn-64(n,2n), and the libraries disagree on its size. The other zinc isotopes make copper and other activation products, so the heat from natural zinc is not a measurement of the Zn-64 channel alone.

\subsubsection*{Target selection}
Only one of the five natural isotopes, Zn-64, is needed, because it is the only direct source of Zn-63. Zn-66, Zn-67, Zn-68 and Zn-70 matter for other products and for the total heat of natural zinc, and would be chosen in a separate study if those products are of interest.

\subsubsection*{Results from the natural-sample calculations}
{\small\setlength{\tabcolsep}{3pt}
\begin{tabular}{@{}p{0.22\linewidth}*{3}{p{0.24\linewidth}}@{}}\toprule
Library & Zn-64 route (\%) & Channel / peer median & Peak product heat (\%) \\ \midrule
TENDL-2025 & 100 & 1.03 & 96.9 \\
ENDF/B-VIII.1 & 100 & 1.03 & 96.2 \\
JEFF-4.0 & 100 & 0.467 & 92.7 \\
JENDL-5 & 100 & 0.972 & 96.6 \\
\bottomrule\end{tabular}\par}
The product is Zn-63. Route shares are fractions of that product's production, not of the total activation, and need not add up to 100\% because other routes can contribute. Not listed means the route is not among those kept in the diagnostics, not that it is zero. Peak heat comes from the full heat curves, so a zero there is a calculated zero.

\subsubsection*{Proposed irradiations}
A Zn-64 target removes most of the other isotopes' contributions and measures the Zn-63 production directly. It is a simple first study to run alongside the hafnium pair, since the question is the size of one channel rather than which of two parents makes a product.

\subsubsection*{Measurements}
Measure the Zn-63 activity, the neutron fluence and spectrum, and the decay heat over its 38 minute half-life. An independent product yield separates an error in the production rate from an error in the decay energy or the calorimetry. Other reactions on Zn-64 still happen and must be included in the model.

\subsubsection*{Limitations}
The relative channel value in the table is the direct Zn-64(n,2n)Zn-63 production per parent, integrated over the irradiation, divided by the median of the other libraries. It is not a C/E. A broad-spectrum irradiation tests the spectrum-averaged rate, not the full excitation function; that needs measurements at other neutron energies.

\clearpage
\subsection{Nickel: a Ni-62 target}\label{isotope:Ni}
\subsubsection*{Motivation}
Ni-62 is a minor isotope of natural nickel, but its (n,p) products Co-62 and Co-62m give a large share of the heat during cooling. The other nickel isotopes make Ni-57 and other cobalt isotopes, which makes the total heat hard to attribute.

\subsubsection*{Target selection}
Only Ni-62 is needed, even though it is a small fraction of natural nickel, because it makes almost all of the Co-62m where the diagnostic is available. Ni-58, Ni-60, Ni-61 and Ni-64 are not needed for this question, but their reactions and residual amounts must still be included in predictions for an enriched target. Validating natural nickel as a whole would need further targets chosen for the other products.

\subsubsection*{Results from the natural-sample calculations}
{\small\setlength{\tabcolsep}{3pt}
\begin{tabular}{@{}p{0.22\linewidth}*{3}{p{0.24\linewidth}}@{}}\toprule
Library & Ni-62 / Ni atoms (\%) & Ni-62 route (\%) & Peak product heat (\%) \\ \midrule
TENDL-2025 & 3.63 & 100 & 51.4 \\
ENDF/B-VIII.1 & 3.63 & Not listed & 0 \\
JEFF-4.0 & 3.63 & 100 & 46.8 \\
JENDL-5 & 3.63 & 100 & 48.4 \\
\bottomrule\end{tabular}\par}
The product is Co-62m1. Route shares are fractions of that product's production, not of the total activation, and need not add up to 100\% because other routes can contribute. Not listed means the route is not among those kept in the diagnostics, not that it is zero. Peak heat comes from the full heat curves, so a zero there is a calculated zero.

\subsubsection*{Proposed irradiations}
A Ni-62 target would concentrate the Ni-62(n,p) signal and remove most of the competing contributions. Measuring Co-62 and Co-62m separately would test both the total production and its split between the ground and metastable states.

\subsubsection*{Measurements}
Use time-resolved activity and decay-heat measurements that cover the 1.5 minute ground state and the 14 minute isomer. Feeding from the isomer to the ground state, transfer time and detector response all have to be accounted for; a single late measurement cannot give both yields.

\subsubsection*{Limitations}
The Ni-62 abundance in the table is its atom fraction in the original natural sample, not the proposed enrichment. New calculations must use the measured composition of the real target. A stronger signal does not guarantee enough counting precision, or that this one product explains the natural-nickel discrepancy.

\clearpage
\subsection{Priorities and practical checks}
\subsubsection*{Priorities}
A first programme would pair the Hf-177 and Hf-178 targets, which test two competing production routes, with Zn-64, which tests one production rate. The Cd-111/Cd-112 and Eu-151/Eu-153 pairs apply the same parent-isotope question to other important isomers. Ni-62 tests a minor isotope whose products are mixed with others in the natural sample. This ordering is based on how clearly each experiment answers its question, not on cost, target availability or expected precision.

\subsubsection*{Limits of enrichment}
A single-isotope target still has several reaction channels, decay chains and metastable products. Gamma spectrometry and calorimetry measure different parts of the problem and should be used together. Niobium, cobalt, rhodium, gold and thulium each have a single stable isotope, so enrichment adds nothing for them.

\subsubsection*{Before choosing samples}
Calculate the enriched compositions with all four libraries before buying or irradiating samples. Include the measured residual isotopes, chemical form and impurities, the intended spectrum and fluence, the sample geometry and self-shielding where relevant, and realistic counting and cooling times. Compare the predicted yields and heat curves with the achievable measurement uncertainty, including the detector and calorimeter response. A natural-element reference sample and flux monitors would separate composition effects from changes in the irradiation or the instruments. Keep every natural isotope in the model of the reference sample, even those not chosen for separate irradiation; a small route share is not a reason to drop an isotope.

\subsubsection*{Status of these proposals}
No enriched-target simulations or new measurements have been made for this section. The 60-replica uncertainty estimates are not converged, and the spread between libraries is not a probability distribution for the true value. TENDL uses ENDF/B decay data here, so agreement between those two is not fully independent. The proposed experiments would test the explanations given above but may not resolve every disagreement.

\clearpage
\section{References}
\subsection*{The benchmark}
\begin{itemize}\setlength{\itemsep}{3pt}
\item \href{https://nds.iaea.org/conderc/fusion/}{IAEA CoNDERC, fusion decay-heat benchmarks}\\{\small\ttfamily https://nds.iaea.org/conderc/fusion/}
\item \href{https://wwwndc.jaea.go.jp/}{JAEA Nuclear Data Center}\\{\small\ttfamily https://wwwndc.jaea.go.jp/}
\end{itemize}
\subsection*{The nuclear data libraries}
\begin{itemize}\setlength{\itemsep}{3pt}
\item \href{https://tendl.web.psi.ch/tendl_2025/tendl2025.html}{TENDL-2025}\\{\small\ttfamily https://tendl.web.psi.ch/tendl\_2025/tendl2025.html}
\item \href{https://www.nndc.bnl.gov/endf/}{ENDF/B-VIII.1, NNDC}\\{\small\ttfamily https://www.nndc.bnl.gov/endf/}
\item \href{https://www.oecd-nea.org/dbdata/jeff/}{JEFF-4.0, OECD NEA}\\{\small\ttfamily https://www.oecd-nea.org/dbdata/jeff/}
\item \href{https://wwwndc.jaea.go.jp/jendl/jendl.html}{JENDL-5, JAEA}\\{\small\ttfamily https://wwwndc.jaea.go.jp/jendl/jendl.html}
\item \href{https://git.oecd-nea.org/fispact/nuclear_data}{UKDD-2020 decay data and UKFY-4.2 fission yields, FISPACT-II nuclear data repository}\\{\small\ttfamily https://git.oecd-nea.org/fispact/nuclear\_data}
\item \href{https://www-nds.iaea.org/public/download-endf/}{IAEA-NDS evaluated data retrieval}\\{\small\ttfamily https://www-nds.iaea.org/public/download-endf/}
\end{itemize}
\subsection*{Nuclear levels for the isotope proposals}
\begin{itemize}\setlength{\itemsep}{3pt}
\item \href{https://www-nds.iaea.org/relnsd/v0/data?fields=levels&nuclides=177hf}{IAEA LiveChart, Hf-177 levels (Hf-177m2: 2740.02 keV, 51.4 min)}\\{\small\ttfamily https://www-nds.iaea.org/relnsd/v0/data?fields=levels\&nuclides=177hf}
\end{itemize}
\subsection*{Previous validation of the same benchmark}
\begin{itemize}\setlength{\itemsep}{3pt}
\item \href{https://www-nds.iaea.org/media/docs/fusion/UKAEA-CCFE-RE-20-004.pdf}{Gilbert, Vilkhivskaya and Sublet, UKAEA-CCFE-RE(20)04, April 2020}\\{\small\ttfamily https://www-nds.iaea.org/media/docs/fusion/UKAEA-CCFE-RE-20-004.pdf}
\end{itemize}
Library names throughout refer to these releases. Decay data and fission yields are taken from the same library as the neutron data wherever it publishes them. TENDL-2025 has neither, so its calculations use UKDD-2020 decay data and UKFY-4.2 fission yields, and every table marks that column.

\end{document}